%% file: These.tex
\documentclass[12pt,twoside,letterpaper]{book}
\usepackage[french, english]{babel}  %support francais et anglais
\usepackage[T1]{fontenc}
\usepackage{amsfonts}
\usepackage[latin1]{inputenc}
\usepackage{dsfont}

\usepackage{times}
\usepackage{mathptmx}

\usepackage{xspace}

\usepackage{lscape}  %for the landscape table

\usepackage{threeparttable} %for footnotes within a table

\usepackage{dcolumn,amsmath}

\usepackage{udem_these_fr}

\usepackage{amsfonts}
\usepackage{amsthm}
\usepackage{amssymb}

\usepackage{flafter}
\usepackage[pdffitwindow=false,
pdfview=FitH,
pdfstartview=FitH,
plainpages=false]{hyperref}

\hypersetup{pdfauthor={Olivier Marchal},
pdftitle={Aspects géométriques et intégrables des modèles de matrices aléatoires},
pdfsubject={Thèse de doctorat},
pdfkeywords={géométrie algébrique, équations de boucles, invariants symplectiques, théorie des cordes topologiques, isomonodromies, polynômes orthogonaux}}

\Auteur{Olivier}{Marchal}

\President{M.}{Yvan}{St Aubin - Université de Montréal}{Pr.}

\Directeur{M.}{Bertrand }{Eynard et John Harnad}{McF}

\Membres{1}{M.}{Marco}{Bertola - Université Concordia}{Ph.D.}
\Membres{2}{M.}{Jacques}{Hurtubise - Université McGill}{Pr.}
\Membres{3}{M.}{Jean-Bernard}{Zuber - Université Paris VI}{Pr.}

\dateaccept{20 décembre 2010}

\PhD{Aspects géométriques et intégrables des modèles de matrices aléatoires}
    {}
    {de mathématiques et de statistique}
   {mathématiques}
    {Décembre}
    {2010}

\newcommand\encadremath[1]{\vbox{\hrule\hbox{\vrule\kern8pt
\vbox{\kern8pt \hbox{$\displaystyle #1$}\kern8pt}
\kern8pt\vrule}\hrule}}
\def\enca#1{\vbox{\hrule\hbox{
\vrule\kern8pt\vbox{\kern8pt \hbox{$\displaystyle #1$}
\kern8pt} \kern8pt\vrule}\hrule}}
\newcommand{\beq}{\begin{equation}}
\newcommand{\eeq}{\end{equation}}
\newcommand{\bea}{\begin{eqnarray}}
\newcommand{\eea}{\end{eqnarray}}

\newcommand{\theoremname}{Theorem}
\addto\captionsfrench{\renewcommand{\theoremname}{Th\'eor\`eme}}
\addto\captionsenglish{\renewcommand{\theoremname}{Theorem}}
\newtheorem{theorem}{\theoremname}[chapter]

\newcommand{\conjecturename}{Conjecture}
\addto\captionsfrench{\renewcommand{\conjecturename}{Conjecture}}
\addto\captionsenglish{\renewcommand{\conjecturename}{Conjecture}}
\newtheorem{conjecture}{\conjecturename}[chapter]

\newcommand{\remarkname}{Remark}
\addto\captionsfrench{\renewcommand{\remarkname}{Remarque}}
\addto\captionsenglish{\renewcommand{\remarkname}{Remark}}
\newtheorem{remark}{\remarkname}[chapter]

\newcommand{\propositionname}{Proposition}
\addto\captionsfrench{\renewcommand{\propositionname}{Proposition}}
\addto\captionsenglish{\renewcommand{\propositionname}{Proposition}}
\newtheorem{proposition}{\propositionname}[chapter]

\newcommand{\lemmename}{Lemma}
\addto\captionsfrench{\renewcommand{\lemmename}{Lemme}}
\addto\captionsenglish{\renewcommand{\lemmename}{Lemma}}
\newtheorem{lemma}{\lemmename}[chapter]

\newcommand{\corollaryname}{Corollary}
\addto\captionsfrench{\renewcommand{\corollaryname}{Corollaire}}
\addto\captionsenglish{\renewcommand{\corollaryname}{Corollary}}
\newtheorem{corollary}{\corollaryname}[chapter]

\newcommand{\definitionname}{Theorem}
\addto\captionsfrench{\renewcommand{\definitionname}{D\'efinition}}
\addto\captionsenglish{\renewcommand{\definitionname}{Definition}}
\newtheorem{definition}{\definitionname}[chapter]

\def \nn{\nonumber}
\def\br{\begin{remark}\rm\small}
\def\er{\end{remark}}
\def\bt{\begin{theorem}}
\def\et{\end{theorem}}
\def\bd{\begin{definition}}
\def\ed{\end{definition}}
\def\bp{\begin{proposition}}
\def\ep{\end{proposition}}
\def\bl{\begin{lemma}}
\def\el{\end{lemma}}
\def\bc{\begin{corollary}}
\def\ec{\end{corollary}}
\def\beaq{\begin{eqnarray}}
\def\eeaq{\end{eqnarray}}
\newcommand{\ba}{\begin{eqnarray*}}
\newcommand{\ea}{\end{eqnarray*}}
\newcommand{\ban}{\begin{eqnarray}}
\newcommand{\ean}{\end{eqnarray}}
\newcommand{\beqn}{\begin{equation}}
\newcommand{\eeqn}{\end{equation}}
\newcommand{\IZ}{\mathbb{Z}}
\newcommand{\IC}{\mathbb{C}}
\newcommand{\IP}{\mathbb{P}}
\newcommand{\IN}{\mathbb{N}}

\newcommand{\curve}{{{\mathcal C}}}
\newcommand{\spcurve}{{{\mathcal S}}}
\newcommand{\Disc}{\mathop{{\rm Disc}\,}}

\newcommand{\cS}{{\cal S}}

\newcommand{\cX}{{\cal X}}

\newcommand{\cO}{{\cal O}}
\newcommand{\cC}{{\cal C}}

\newcommand{\cF}{{\cal F}}

\newcommand{\td}{\tilde}
\def\CC{\mathcal C}

\newcommand{\CYX}{{\mathfrak X}}

\newcommand{\Det}{{\,\rm det}} 
\newcommand{\Tr}{{\,\rm Tr}\:}
\newcommand{\tr}{{\,\rm tr}\:}
\newcommand{\Res}{\mathop{\,\rm Res\,}}
\renewcommand{\l}{\lambda}
\newcommand{\om}{\omega}

\newcommand{\ee}[1]{{{\rm e}^{#1}}}
\newcommand{\eq}[1]{eq.~(\ref{#1})}
\renewcommand{\d}{{{\partial}}}

\newcommand{\Pint}{{\int\kern -1.em -\kern-.25em}}

\renewcommand{\Re}{{\mathrm{Re}}}
\renewcommand{\Im}{{\mathrm{Im}}}

\renewcommand{\l}{\lambda}

\newcommand{\ovl}{\overline}
\newcommand{\genus}{g}
\newcommand{\virg}{{\qquad , \qquad}}
\newcommand{\bcycle}{{\cal B}}
\newcommand{\acycle}{{\cal A}}
\newcommand{\Ycl}{{Y_{\rm cl}}}
\newcommand{\npole}{{\overline{n} }}
\newcommand{\Li}{\rm Li}
\newcommand{\spcurveMM}{{{\mathcal S}_{\rm MM}}}
\def \restr{\mathrm {resTr}}

\def\GC{{\check{\Gamma}}}
\def\G{\Gamma}
\def\Ib{{\mathbf I}}
\def\Hb{{\mathbf H}}

\def\HbT{\hat{\mathbf H}}
\def\eb{{\mathbf e}}
\def\kappab{{ \mathbf \kappa}}
\def\PsiN{\ds{\mathop{\Psi}_N}}

\renewcommand{\theequation}{\arabic{chapter}.\arabic{section}.\arabic{equation}}
\makeatletter
\@addtoreset{equation}{section}

\def \pa{\partial}

\def\Tr{\mathrm {Tr}}
\def\tr{\mathrm {tr}}
\def\det{\mathrm {det}}
\def\ln{\mathrm {ln}}

\def\res{\mathop{\mathrm {res}}\limits_}

\def\ds{\displaystyle}

\def\&{&{\hskip -20pt}}

\def\bea{\begin{eqnarray}}
\def\eea{\end{eqnarray}}

\def \pa{\partial}

\def\Tr{\mathrm {Tr}}
\def\tr{\mathrm {tr}}
\def\det{\mathrm {det}}
\def\ln{\mathrm {ln}}

\def\res{\mathop{\mathrm {res}}\limits_}

\def\ds{\displaystyle}

\def\&{&{\hskip -20pt}}

\def\s{{\sigma}}

\def\GH{{\hat{\Gamma}}}

\def\Cbb{{\mathbb C}}
\def\Nbb{{\mathbb N}}

\def\H{{\cal H}}

\begin{document}

\selectlanguage{french}

%\setcounter{page}{0}
 %\PagesCouverture
%\pagestyle{empty}
\setcounter{page}{1}
 \PagesCouverture
%\newpage~
%\include{FeuilleDegarde}
%\newpage~
%\include{FeuilleDegarde2}
%\newpage~
%\setcounter{page}{5}
\pagestyle{myheadings}
\include{resume}
\include{abstract}
\include{remerciements}

\tabledesmatieres

%\listedestableaux

%\listedesfigures

\include{MesArticles}

\listedesannexes

\include{Notation}

\include{Notation2}

\debutchapitres

\include{intro}
\include{chap1}

\include{chap2}
\include{chap3}
\include{chap4}
\include{conclu}

\bibliographystyle{frplainnat}
%\bibliography{references}

\renewcommand{\theequation}{\Roman{chapter}.\arabic{equation}}
 \debutannexes
 \setcounter{page}{1}
\pagenumbering{roman}
 
\include{ListeFigures}
\include{annexea}

\include{annexeb}
\include{annexec}
\include{annexed}

\include{annexee}
\include{annexef}
\include{annexeg}
\include{annexeh}
\include{annexei}

\end{document}

%% file: resume.tex
\resume
\thispagestyle{myheadings}
\selectlanguage{french}

 %(150 a 250 mots) (1 page)

Cette thèse traite des aspects géométriques et d'intégrabilité associés aux modèles de matrices aléatoires. Son but est de présenter diverses applications des modèles de matrices aléatoires allant de la géométrie algébrique aux équations aux dérivées partielles des systèmes intégrables. Ces différentes applications permettent en particulier de montrer en quoi les modèles de matrices possèdent une grande richesse d'un point de vue mathématique.

Ainsi, cette thèse abordera d'abord l'étude de la jonction de deux intervalles du support de la densité des valeurs propres au voisinage d'un point singulier. On montrera plus précisément en quoi ce régime limite particulier aboutit aux équations universelles de la hiérarchie de Painlevé II des systèmes intégrables. Ensuite, l'approche des polynômes (bi)-orthogonaux, introduite par Mehta pour le calcul des fonctions de partition, permettra d'énoncer des problèmes de Riemann-Hilbert et d'isomonodromies associés aux modèles de matrices, faisant ainsi le lien avec la théorie de Jimbo-Miwa-Ueno. On montrera en particulier que le cas des modèles à deux matrices hermitiens se transpose à un cas dégénéré de la théorie isomonodromique de Jimbo-Miwa-Ueno qui sera alors généralisé. La méthode des équations de boucles avec ses notions centrales de courbe spectrale et de développement topologique permettra quant à elle de faire le lien avec les invariants symplectiques de géométrie algébrique introduits récemment par Eynard et Orantin. Ce dernier point fera également l'objet d'une généralisation aux modèles de matrices non-hermitien ($\beta$ quelconque) ouvrant ainsi la voie à la ``géométrie algébrique quantique'' et à la généralisation de ces invariants symplectiques pour des courbes ``quantiques''. Enfin, une dernière partie sera consacrée aux liens étroits entre les modèles de matrices et les problèmes de combinatoire. En particulier, l'accent sera mis sur les aspects géométriques de la théorie des cordes topologiques avec la construction explicite d'un modèle de matrices aléatoires donnant le dénombrement des invariants de Gromov-Witten pour les variétés de Calabi-Yau toriques de dimension complexe trois utilisées en théorie des cordes topologiques.

L'étendue des domaines abordés étant très vaste, l'objectif de la thèse est de présenter de façon la plus simple possible chacun des domaines mentionnés précédemment et d'analyser en quoi les modèles de matrices peuvent apporter une aide précieuse dans leur résolution. Le fil conducteur étant les modèles matriciels, chaque partie a été conçue pour être abordable pour un spécialiste des modèles de matrices ne connaissant pas forcément tous les domaines d'application présentés ici.

{\bfseries \underline{Mots-clés}: géométrie algébrique, équations de boucles, invariants symplectiques, théorie des cordes topologiques, isomonodromies, polynômes orthogonaux.}

%% file: abstract.tex
\abstract
\thispagestyle{myheadings}
\selectlanguage{english}

 %(150 a 250 mots) (1 page)

This thesis deals with the geometric and integrable aspects associated with random matrix models. Its purpose is to provide various applications of random matrix theory, from algebraic geometry to partial differential equations of integrable systems. The variety of these applications shows why matrix models are important from a mathematical point of view.

First, the thesis will focus on the study of the merging of two intervals of the eigenvalues density near a singular point. Specifically, we will show why this special limit gives universal equations from the Painlevé II hierarchy of integrable systems theory. Then, following the approach of (bi) orthogonal polynomials introduced by Mehta to compute partition functions, we will find Riemann-Hilbert and isomonodromic problems connected to matrix models, making the link with the theory of Jimbo, Miwa and Ueno. In particular, we will describe how the hermitian two-matrix models provide a degenerate case of Jimbo-Miwa-Ueno's theory that we will generalize in this context. Furthermore, the loop equations method, with its central notions of spectral curve and topological expansion, will lead to the symplectic invariants of algebraic geometry recently proposed by Eynard and Orantin. This last point will be generalized to the case of non-hermitian matrix models (arbitrary $\beta$) paving the way to ``quantum algebraic geometry'' and to the generalization of symplectic invariants to ``quantum curves''. Finally, this set up will be applied to combinatorics in the context of topological string theory, with the explicit computation of an hermitian random matrix model enumerating the Gromov-Witten invariants of a toric Calabi-Yau threefold.

Since the range of the applications encountered is large, we try to present every domain in a simple way and explain how random matrix models can bring new insights to those fields. The common element of the thesis being matrix models, each part has been written so that readers unfamiliar with the domains of application but familiar with matrix models should be able to understand it.

%(max. of 10, not words in the title)

%% file: remerciements.tex
\remerciements
\thispagestyle{myheadings}
\selectlanguage{french}
Je tiens à remercier celles et ceux sans lesquels cette thèse n'aurait jamais vu le jour. Tout d'abord mes deux directeurs pour leur grande patience avec une mention spéciale pour les mini-cours personnalisés de Bertrand qui m'ont fait gagner un temps précieux. Je remercie également Michel Bergère pour la voiture rouge et les repas partagés plein de discussions. Egalement, un grand merci à mes amis qui ont toujours su me supporter et me donner de la motivation, du réconfort et du plaisir. Mes pensées sont particulièrement destinées à Guillaume pour les après-midi consoles endiablés, Romain pour m'avoir fait découvrir le squash, Cécilie pour tous les bons moments passés à jouer au volley-ball ensemble, Sophie pour ses dîners savoureux, Hélène pour avoir supporté le début de le thèse, Lucien pour les discussions d'informatique (entre autres), Camille pour m'avoir fait découvrir à distance de lointaines contrées, Mathilde et Elodie pour les randonnées Grenobloises, ainsi que tous mes partenaires de volley-ball pour m'avoir offert des moments de détente. Finalement, mes pensées les plus fortes vont envers mes deux frères et mes parents à qui je dois énormément pour leur soutien constant, surtout dans les périodes les plus difficiles. Enfin, une pensée spéciale va à Audrey qui verra l'aboutissement de cette thèse et je l'espère ce qui suivra ensuite.

%% file: MesArticles.tex
\Articles
\thispagestyle{myheadings}
\selectlanguage{french}

%\underline{\bf{Liste des articles utilisés pour cette thèse:}} (Voir Annexes)

Les articles mentionnés ici sont disponibles dans les annexes numérotés de V à X.
 
\bigskip
\bigskip
\bigskip

$\left[ \textbf{I} \right]$   O. Marchal, M. Cafasso, ``Double scaling limits of random matrices and minimal $(2m,1)$ models: the merging of two cuts in a degenerate case'', \textit{arXiv}:1002.3347v2 [math-ph] (Egalement en annexe \ref{Article[I]})

\bigskip

$\left[ \textbf{II} \right]$ M. Bertola, O. Marchal ``The partition function of the two-matrix model as an isomonodromic tau-function'', \textit{J. Math. Phys.} \textbf{50}, 013529, 2009. (Egalement en annexe \ref{Article[II]}) 

\bigskip

$\left[ \textbf{III} \right]$ B. Eynard, O. Marchal, ``Topological expansion of the Bethe ansatz, and non-commutative algebraic geometry'', \textit{JHEP} 0903:094, 2009, arXiv:0809.3367 [math-ph]. (Egalement en annexe \ref{Article[III]})

\bigskip

$\left[ \textbf{IV} \right]$ L. Chekov, B. Eynard, O. Marchal, ``Topological expansion of the Bethe ansatz, and quantum algebraic geometry'', \textit{arXiv}:0911.1664v2 [math-ph] (Egalement en annexe \ref{Article[IV]})

\bigskip

$\left[ \textbf{V} \right]$ B. Eynard, A. Kashani-Poor, O. Marchal, ``A matrix model for the topological string I: Deriving the matrix model'', \textit{arXiv}:1003.1737v2 [hep-th] (Egalement en annexe \ref{Article[V]})

\bigskip

$\left[ \textbf{VI} \right]$ B. Eynard, A. Kashani-Poor, O. Marchal ``A matrix model for the topological string II: The spectral curve and mirror geometry'', \textit{arXiv}:1007.2194v1 [hep-th] (Egalement en annexe \ref{Article[VI]})

%% file: Notation.tex
\notation
\thispagestyle{myheadings}
\selectlanguage{french}

\begin{center}
\begin{tabular}{r p{12cm} }
$\mathbb{N}$ & Ensemble des nombres naturels \\
$\mathbb{Q}$ & Ensemble des nombres rationels \\
$[a,b]$ & Intervalle fermé des nombre réels compris entre $a$ et $b$ avec $a\leq b$\\
$(a,b)$ ou $]a,b[$ & Intervalle ouvert des nombre réels compris entre $a$ et $b$ avec $a\leq b$\\
$\mathbb{R}$ & Ensemble des nombres réels \\ 
$\mathbb{C}$ & Ensemble des nombres complexes \\
$\mathcal{C}^\infty(E)$ & Ensemble des fonctions infiniment dérivables sur $E$\\ 
$\mathcal{C}_c(E)$ & Ensemble des fonctions continues à support compact sur $E$\\ 
$\mathbb{H}_n$ & Ensemble des matrices carrées hermitiennes de taille $n$   \\
$U(n)$  &  Ensemble des matrices unitaires de taille $n$\\
$Sp(2n)$ & Ensemble des matrices symplectiques de taille $2n$\\
$O(n)$ & Ensemble des matrices orthogonales de taille $n$ \\
$Sym(n)$ & Ensemble des matrices réelles symétriques de taille $n$ \\
$\lambda$ & Vecteur composé de $\lambda_1,...\lambda_n$. Le nombre de composantes ne sera precisé qu'en cas d'ambiguïté  \\
$\Delta(\lambda)$ & Déterminant de Vandermonde des valeurs $\lambda_i$: $\Delta(\lambda)=\prod_{i<j} (\lambda_i-\lambda_j)$\\
$\Tr(A)$ & Trace de la matrice $A$ \\
$\Det(A)$ & Déterminant de la matrice $A$\\
$\Re(z)$ & Partie réelle du nombre complexe $z$\\
$\Im(z)$ & Partie imaginaire du nombre complexe $z$\\

\end{tabular}
\end{center}

%% file: Notation2.tex
\selectlanguage{french}
\begin{center}
\begin{tabular}{l p{12cm} }
$\frac{\partial}{\partial x}f$ & Dérivée partielle de la fonction $f$ par rapport à la variable $x$\\
$\frac{d}{d x}f=f'(x)$ & Dérivée de la fonction $f$ par rapport à son unique variable \\
&réelle $x$\\
$exp(x)=e^x$, $ln(x)$ & Fonctions exponentielles, logarithme Népérien usuelles.\\
&La variable $x$ peut être réelle ou complexe. (Dans le cas \\
& de $ln$ la coupure est supposée implicitement sur l'axe $\mathbb{R}^-$)\\
$cos(x)$, $sin(x)$, $tan(x)$ & Fonctions trigonométriques usuelles\\
$\delta(x)$ & Distribution de Dirac vérifiant $\forall f \in \mathcal{C}_c^\infty(\mathbb{R}):$\\
&$ \int_{\mathbb{R}}f(x)\delta(x)dx=f(0)$\\
$E(x)$ & Partie entière du nombre réel $x$\\
$\text{Pol}(f(z))$ & Partie polynomiale du développement en série de la \\
& fonction $f(z)$ au voisinage de l'infini.
\end{tabular}
\end{center}

%% file: intro.tex
%\introduction
\chapter{Introduction}
 \setcounter{page}{21}
\selectlanguage{french}
\thispagestyle{empty}
\section{Définition d'une matrice aléatoire}

Historiquement \cite{Hist,Hist2}, dans le domaine des matrices de corrélation en statistiques multi-variables, le développement de la théorie des matrices aléatoires a connu sa première avancée majeure dans les annéees 1930 grâce à Wishart. A l'époque, leur essor était encore relativement faible et consistait à s'intéresser aux valeurs propres et aux vecteurs propres de certaines matrices dont les entrées obéissaient à différentes distributions de probabilités. L'introduction de la théorie des matrices aléatoires en physique nucléaire eut lieu en 1951 avec Wigner \cite{VieuxWigner} qui eut l'idée d'utiliser ces matrices pour caractériser la statistique des spectres d'excitation des noyaux lourds. Citons ensuite les travaux de Dyson \cite{VieuxDyson} dans les années 1950-1960 puis de Mehta \cite{Mehta} de 1960 aux années 2000 qui ont contribué à faire avancer la théorie des matrices aléatoires jusqu'à leur niveau actuel.
Ainsi, dans les années 1930, Wishart s'intéressa aux matrices:
\beq A=\begin{pmatrix}
a_{1,1} &\dots &a_{n,n}\\
\vdots & & \vdots\\
a_{n,1} &\dots& a_{n,n}
\end{pmatrix}
\eeq
où les composantes $a_{i,j}$ sont des variables aléatoires réelles indépendantes et identiquement distribuées données par une loi de probabilités $p(x)$:
\beq Prob(a_{i,j} \in [a,b])= \int_a^b p(x)dx\eeq
On peut alors définir une mesure de probabilités sur l'ensemble des matrices réelles $A$ en prenant le produit des lois de probabilités des composantes indépendantes:
\beq Prob(A / a_{i,j} \in [\alpha_{i,j},\beta_{i,j}])=\prod_{i,j=1}^n \int_{\alpha_{i,j}}^{\beta_{i,j}} p(x)dx\eeq
Le premier calcul de Wigner a été de montrer que pour des entrées Gaussiennes, la répartition des valeurs propres (divisées par $\sqrt{ n})$ ainsi que la loi de la plus grande (ou plus petite) valeur propre de ces matrices vont tendre asymptotiquement dans la limite $n \to \infty$ vers des lois de probabilités explicites qui dépendent du type de symétrie de la matrice. Ainsi, historiquement on distingue trois ensembles de matrices différents: les matrices hermitiennes (qui possèdent $n^2$ composantes réelles indépendantes),  qui sont invariantes sous l'action du groupe unitaire $U(n)$, les matrices réelles symétriques (qui possèdent $\frac{n(n+1)}{2}$ composantes réelles indépendantes) qui sont invariantes sous l'action du groupe orthogonal $O(n)$ et les matrices quaternioniques réelles self-duales (qui possèdent $n(2n-1)$ composantes réelles indépendantes, Cf. appendice \ref{annexe2}) qui sont invariantes sous l'action du groupe symplectique $Sp(2n)$. Ces trois ensembles possèdent la propriété que les matrices y sont toujours diagonalisables avec des valeurs propres réelles et que la mesure induite sur l'espace des valeurs propres (c'est à dire apres intégration sur le groupe d'invariance correspondant) peut être mise sous la forme commune:
\beq Z=\int_{\mathbb{R}^n} d\lambda_1\dots d\lambda_n \, \left(\prod_{1\leq i<j\leq n} |\lambda_i-\lambda_j|^{2\beta}\right) e^{-\beta \overset{n}{\underset{i=1}{\sum}} \lambda_i^2}
\eeq 
avec $\beta = 1,\, 1/2,\, 2$ pour respectivement l'ensemble hermitien, symétrique et quaternionique self-dual. Néanmoins beaucoup d'autres ensembles peuvent être également envisagés: matrices unitaires, orthogonales, ou normales avec des valeurs propres localisées sur un contour fixé.

Une deuxième contribution a ensuite été d'observer que pour des entrées non gaussiennes et potentiellement corrélées, certaines lois obtenues pour le cas d'entrées i.i.d. gaussiennes se maintiennent dans la limite $n \to \infty$ sous certaines hypothèses concernant les lois de probabilités des entrées (décroissance exponentielle, indépendance, etc.).  A l'heure actuelle, beaucoup de personnes cherchent à affaiblir les restrictions imposées sur les entrées des matrices (entrées corrélées, distribution de probabilité avec des longues queues, etc.) ou de trouver d'autres lois pour les valeurs propres lorsque les entrées sont distribuées suivant d'autres conditions. On citera entre autres les travaux récents de L. Erdos, A. Guionnet et de K. Johansson (\cite{Erdos,Erdos2,Guillonet,Johansson}) sur ces sujets. Ainsi, on sait désormais que si les entrées indépendantes sont i.i.d. de moyenne nulle et de variance $\sigma^2$ finie, alors la distribution de probabilités des valeurs propres normalisées aura comme limite la loi du demi-cercle de Wigner lorsque la taille $n$ des matrices tend vers l'infini. On peut illustrer ce résultat avec des entrées gaussiennes:

\begin{center} \label{TracyWidom}
	\includegraphics[height=5cm]{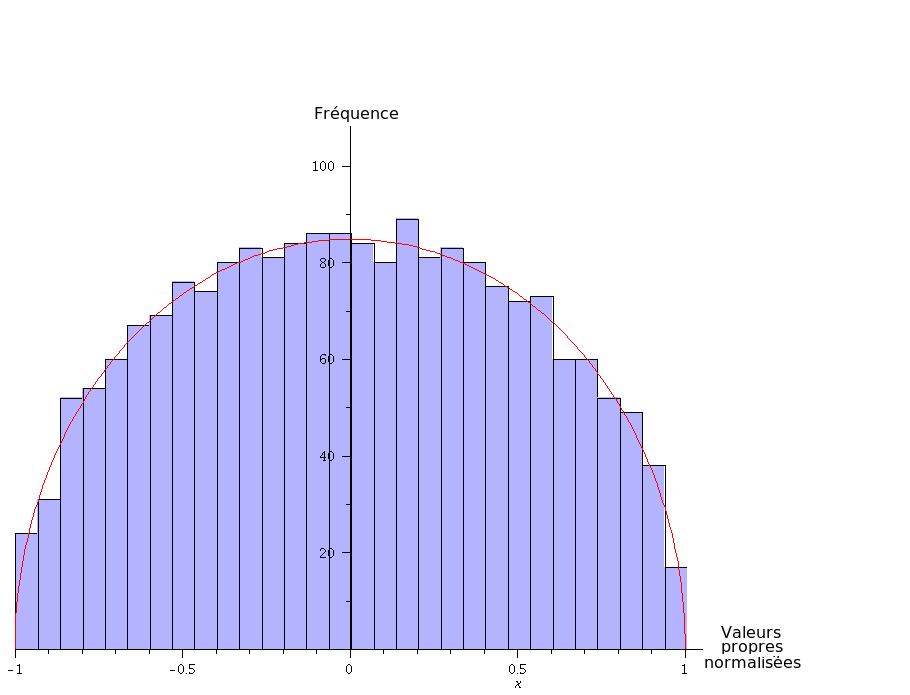}

	\underline{Figure 1}: Histogramme des valeurs propres (divisées par $\sqrt{n}$) d'une matrice $100 \times 100$. La courbe rouge représente la loi théorique du demi-cercle de Wigner.
\end{center}

D'un point de vue mathématique, le résultat peut être exprimé ainsi par:
\beq\lim_{n \to \infty} \mu_{A\in \text{Sym}(n)}(x)=\frac{2}{\pi}\sqrt{1-x^2} \eeq
où $\mu_{A\in \text{Sym}(n)}(x)$ est la mesure de probabilité empirique des valeurs propres:
\beq \mu_{A\in \text{Sym}(n)}(x)=\frac{1}{n}\sum_{i=1}^n \delta(x-\frac{\lambda_i}{2\sqrt{n}})\eeq

En physique, il est souvent plus intéressant de regarder la répartition entre les valeurs propres (niveaux d'énergie) consécutives dans le coeur de la distribution. Cette répartition est connue théoriquement comme nous le verrons par la suite (Chapitre 2, section 6, équation \ref{levelspacingdistribtion}) pour les modèles hermitiens, symétriques réels et quaternioniques self-duaux et donne lieu à un phénomène d'universalité. Pour le cas des matrices symétriques réelles, on obtient la figure suivante: 

\begin{center} \label{LevelSpacingDistribution}
	\includegraphics[height=5cm]{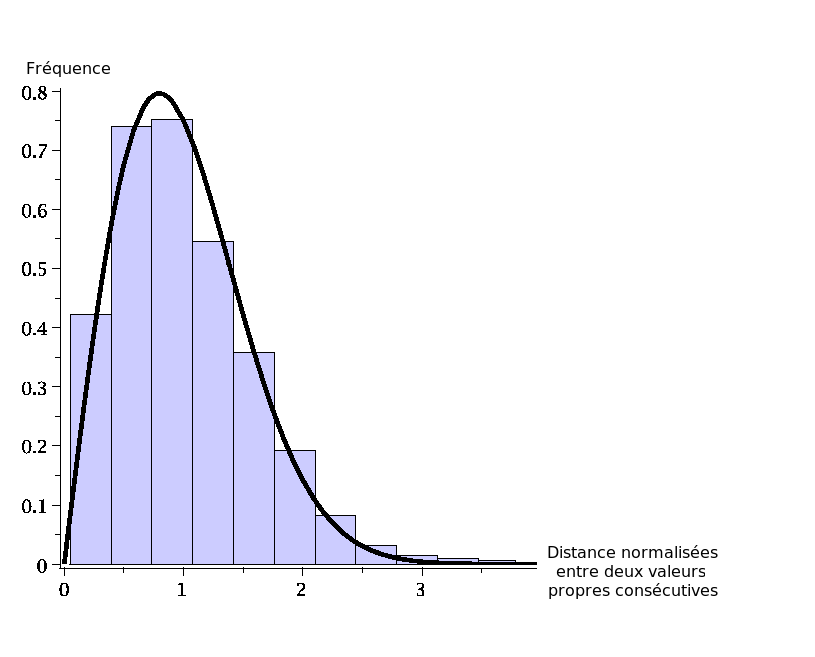}

	\underline{Figure 2}: Histogramme des écarts ($\sqrt{n}(\lambda_{i+1}-\lambda_{i})$) entre les valeurs propres consécutives d'une matrice symétrique réelle gaussienne $1000 \times 1000$. La courbe noire représente la loi théorique.
\end{center}

Cela dit, si la loi de Wigner regroupe beaucoup de lois de distribution pour les entrées, elle n'est aucunement universelle puisque des entrées avec des lois de probabilité n'obéissant pas aux règles énoncées ci-dessus vont donner des distributions de valeurs propres bien différentes. Par exemple, si les entrées sont des variables de Cauchy (dont la loi est $\frac{1}{\pi (x^2+1)}$) la distribution est supportée sur $\mathbb{R}$ tout entier, et peut être illustrée par:

\begin{center} \label{Cauchy}
	\includegraphics[height=5cm]{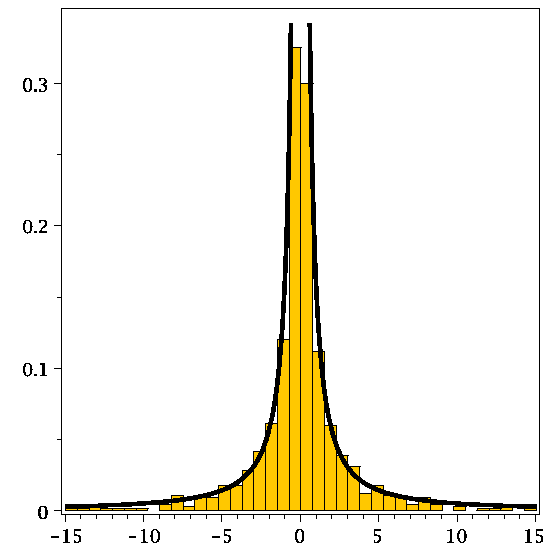}

	\underline{Figure 3}: Histogramme des valeurs propres (normalisées par $1/n$) d'une matrice symétrique $100 \times 100$ avec des i.i.d. suivant la distribution de Cauchy. La courbe noire représente la loi $x \mapsto \frac{1}{2\pi x^{3/2}}$.
\end{center}

On voit donc que l'on est très loin de la loi du demi-cercle, en particulier, la distribution limite obtenue n'est pas normalisable sur $\mathbb{R}$, ne définissant pas mathématiquement une mesure de probabilités.

\section{Lien historique entre les matrices aléatoires et la physique nucléaire}

Les matrices aléatoires ont été introduites par Wigner dans les années 1960 pour expliquer le spectre des noyaux lourds d'uranium qui apparait incroyablement complexe et difficilement résoluble de façon analytique au vu de la complexité et du grand nombre d'interactions présentes au sein d'un noyau. Ainsi pour l'uranium, qui contient plus de 200 protons et neutrons obéissant aux règles complexes des interactions nucléaires, un calcul des différents états d'énergie est impossible analytiquement et compliqué numériquement (surtout en 1960). En revanche, dès les années 1950, la construction d'accélérateurs de particules de hautes énergies permettait l'exploration partielle expérimentale de ces niveaux en bombardant un atome d'uranium avec un neutron accéléré, et des résultats expérimentaux étaient déjà disponibles. A l'époque, la grande majorité des physiciens pensait que les différents écarts entre niveaux de résonance consécutifs devaient être répartis selon une distribution de Poisson:
\beq x=\rho S=\frac{S}{<S>}\virg p(x)dx=e^{-x}dx\eeq
où $\rho$ représente la densité d'état et $<S>$ l'écart moyen. Mais les mesures expérimentales imprécises ne permettaient pas de valider ou d'invalider une telle distribution. Wigner eut alors l'idée de proposer son modèle aléatoire qui donne lieu dans le cas des matrices réelles symétriques à la loi approchée:
\beq \label{rhoS} x=\rho S=\frac{S}{<S>}\virg p(x)dx\sim\frac{\pi}{2}xe^{-\frac{\pi}{4}x^2}dx\eeq
Vers les années 1960, l'amélioration des accélérateurs permit des expériences plus précises qui tranchèrent entre les différentes lois proposées et donna raison aux modèles matriciels de Wigner. Cela est illustré dans le graphique suivant, tiré de \cite{FirkMiller} et de \cite{DerrienLeal}.

\begin{center} \label{NucleiDistribution}
	\includegraphics[height=5cm]{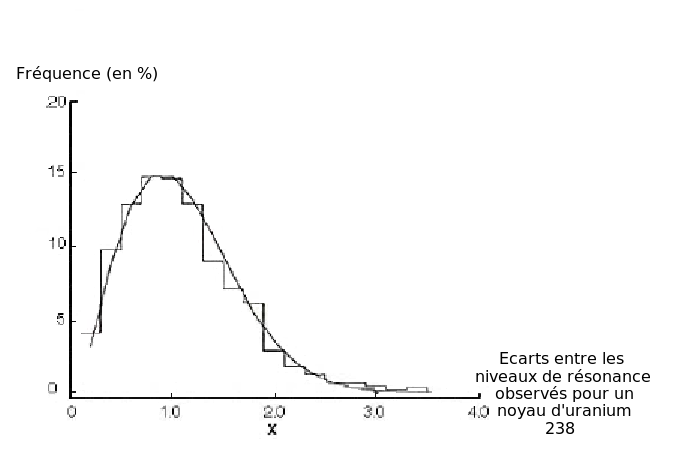}

	\underline{Figure 4}: Une distribution de Wigner avec la distribution des écarts entre niveaux de résonances d'un atome d'uranium 238 pour des énergies allant jusqu'à 20 keV. En abscisse se trouve l'énergie considérée et en ordonnées la probabilité d'avoir une résonance pour cette énergie donnée. 
\end{center}

Depuis Wigner, il est connu que suivant le type de symétries du système étudié, la distribution provient de différents ensembles de matrices. Par exemple, des systèmes présentant une invariance par rotation et une invariance par renversement du temps se verront attachés aux modèles de matrices réelles symétriques (Gaussian orthogonal ensemble) tandis que ceux pour lesquels l'invariance par renversement du temps n'est pas valable sont attachés aux modèles de matrices hermitiennes (Gaussian unitary ensemble). Notons que beaucoup d'autres ensembles ont depuis été étudiés, comme par exemple les ensembles des matrices unitaires ou orthogonales dont les lois de distribution des valeurs propres sont également bien connues.

\section{Matrices aléatoires et autres domaines des mathématiques}

Bien qu'historiquement introduites pour la physique nucléaire, les matrices aléatoires n'ont cessé de se retrouver dans un nombre croissant de domaines à la fois appliqués et théoriques. Coté applications, on peut ainsi mentionner le repliement des brins d'ARN ou de protéines (\cite{RNA}, \cite{RNA2}, \cite{RNA3}, \cite{protein}) et de nombreuses applications en traitement du signal \cite{signal} et dans la théorie des cordes topologiques (Cf. chapitre \ref{chap4}). Dans le domaine des mathématiques, les intégrales matricielles qui font l'objet de cette thèse sont reliées à de nombreux problèmes: systèmes intégrables, polynômes orthogonaux, problèmes de Riemann-Hilbert, combinatoires de cartes et même de manière assez surprenante théorie des nombres. En effet, il semblerait qu'il existe un lien incompris entre les matrices aléatoires et les zéros non triviaux de la fonction $\zeta$ de Riemann. Ce lien est illustré par la ressemblance frappante entre la distribution de Wigner \ref{NucleiDistribution} et celles des zéros de la fonction de Riemann sur la droite $\Re(z)=1/2$:

\begin{center} \label{ZetaZeros}
	\includegraphics[height=5cm]{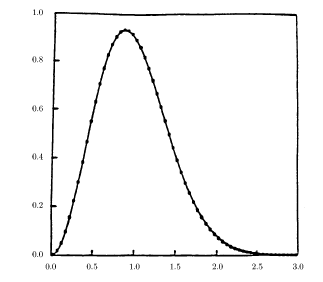}

	\underline{Figure 5}: Distribution des écarts de 70 millions de zéros consécutifs de la fonction $\zeta(s)$ de Riemann (partant du $1020^{ième}$) zéro). Graphique extrait de \cite{FirkMiller}, \cite{Od1}, \cite{Od2}
\end{center}

Ainsi, il semblerait que la distribution des zéros non triviaux de la fonction de Riemann obéissent à une loi des matrices aléatoires bien que le lien entre les deux théories soit encore aujourd'hui mystérieux. Citons également d'autres domaines dans lesquels des lois de matrices aléatoires ont pu être observées: la percolation, l'atome d'hydrogène dans un champ magnétique intense, le dénombrement de certaines familles de graphes, la chromodynamique quantique (QCD), l'étude des partitions planes,... Ainsi les matrices aléatoires, par leurs applications dans de nombreux domaines connaissent à l'heure actuelle un développement important et dans de nombreuses directions. Cette thèse sera le reflet de cette diversité puisqu'elle abordera plusieurs méthodes spécifiques permettant de traiter le problème complexe des intégrales de matrices aléatoires. 

%% file: chap1.tex
\chapter{Intégrales de matrices et densité de valeurs propres }
\pagestyle{myheadings}
\thispagestyle{empty}
 \label{chap1}
\selectlanguage{french}
\section{Définition des intégrales de matrices}

Dans l'introduction, nous avons vu que les valeurs propres de matrices aléatoires de certains ensembles obéissent à des lois simples lorsque la taille de la matrice devient grande. Cette première étape intéressante est néanmoins limitée par le fait qu'en physique statistique, les configurations d'un système sont souvent contraintes par un potentiel d'interaction. Il est donc logique, par analogie avec la physique statistique de Boltzmann, d'introduire les intégrales de matrices suivantes ou fonctions de partition:

\beq \label{IntegralesMatricielles} Z_N(T)=\int_{\mathbb{E}_N} dM e^{-\frac{N}{T} \Tr(V(M))}\eeq
où $\mathbb{E}_N$ désigne un ensemble de matrices de taille $N$ (par exemple hermitiennes, symétriques réelles,...), $T$ désigne la température du modèle, $dM$ correspond aux produits des mesures de Lebesgue des entrées réelles indépendantes et $V(x)$ est le potentiel associé au modèle de matrice étudié.

D'un point de vue probabiliste, toutes les matrices $M$ n'ont désormais plus la même probabilité d'apparition, les matrices telles que $\Tr(V(M))$ est minimal devenant ainsi bien plus probables que les autres, correspondant ainsi à une centralisation préférentielle autour des configurations d'énergies minimales. Par analogie avec les systèmes statistiques, la quantité $\frac{N}{T} \Tr(V(M))$ est appelée l'action du modèle et les contributions les plus importantes à l'intégrale sont donc les matrices qui minimisent cette action. On peut ainsi résumer les probabilités comme:
\beq \text{Prob}\left(M / m_{i,j}\in [a_{i,j},a_{i,j}+d a_{i,j}]\right)=\frac{1}{Z_N}e^{-\frac{N}{T} \underset{i=1}{\overset{N}{\sum}} (V(A))_{i,i} } \left(\prod_{i,j} d a_{i,j}\right)\eeq
ou encore:
\beq \text{Prob}\left(M / m_{i,j}\in [a_{i,j},b_{i,j}]\right)=\frac{1}{Z_N}\int_{a_{1,1}}^{b_{1,1}}\dots \int_{a_{n,n}}^{b_{n,n}}dm_{1,1}\dots dm_{n,n}e^{-\frac{N}{T} \underset{i=1}{\overset{N}{\sum}} (V(M))_{i,i}} \eeq

Notons que selon l'ensemble des matrices considéré, le nombre de composantes réelles indépendantes sur lesquelles on réalise l'intégration peut changer.

\section{Diagonalisation, problème aux valeurs propres}

Dans le cas des ensembles de matrices hermitiennes, symétriques réelles ou quaternioniques self-duales, l'invariance de l'action (grâce à la présence de la trace) sous le groupe unitaire, orthogonal ou symplectique permet d'effectuer l'intégration sur les variables ``angulaires'' et de ramener le problème à celui des valeurs propres des matrices. Cette diagonalisation n'est cependant pas complètement triviale car le Jacobien de la transformation n'est pas évident a priori. Ces diagonalisations, connues depuis longtemps, peuvent être trouvés dans \cite{Mehta} pour chacun des trois cas et peuvent se résumer ainsi:
\beq \label{Diagonalisation} M=U\Lambda U^{-1} \, \Rightarrow \, Z_N\propto \int_{\mathbb{R}^N} d\lambda_1 \dots d\lambda_N \,  \Delta(\lambda)^{2\beta} e^{-\frac{N}{T}\underset{i=1}{\overset{N}{\sum}} V(\lambda_i)}\eeq
où $\Lambda=diag(\lambda_1,\dots,\lambda_N)$, $U$ est une matrice unitaire, orthogonale ou symplectique suivant l'ensemble initial considéré (respectivement hermitiens, symétriques réels et quaternioniques self-duaux). Le paramètre $\beta$ vaut respectivement $1$, $\frac{1}{2}$ et $2$ suivant les ensembles initiaux considérés (respectivement hermitien, symétrique réel et quaternionique). Enfin, les coefficients de proportionnalité peuvent être exprimés facilement dans les trois cas en prenant le potentiel $V(x)$ quadratique. Ces coefficients ne dépendent que de $N$ et peuvent être trouvés explicitement dans \cite{Mehta} par l'intermédiaire de formules exactes dans le formalisme des polynômes (skew) orthogonaux (le cas hermitien se réduisant ainsi aux polynômes de Hermite). Rappelons également que $\Delta(\lambda)$ désigne le déterminant de Vandermonde associé aux valeurs propres $(\lambda_i)_{i=1\dots N}$.

\medskip

\underline{Note}: Nous utiliserons dans cette thèse les conventions dites du "gaz de Coulomb" pour l'exposant $\beta$, utilisées notamment par Laughlin. Cette convention diffère ainsi d'un facteur $2$ par rapport à la notation plus courante de Wigner et de Mehta  dans la littérature. Ainsi, dans la notation de Wigner, l'exposant du déterminant de Vandermonde n'est pas précédé d'un facteur $2$, le cas hermitien correspondant alors à $\beta_\text{lit}=2$. L'intérêt principal de notre convention apparaîtra plus tard lors de l'étude des modèles de matrices où le paramètre $\beta$ est quelconque.

\medskip

Si le cas des matrices hermitiennes, symétriques réelles et quaternioniques self-duales est intéressant, il ne constitue cependant pas le cas le plus général. En effet, il est facile de généraliser les cas ci-dessus pour des ensembles de matrices normales (i.e. qui commutent avec leur adjoint et qui sont donc diagonalisables sur une base orthonormale de vecteurs propres) dont les valeurs propres sont assujeties à être situées sur un contour $\mathcal{C}$ fixé du plan complexe. Enfin, on peut également choisir d'étudier directement une version de \ref{Diagonalisation} dans lequel le paramètre $\beta$ est arbitraire, même si pour le cas où $\beta$ est quelconque, il n'existe pas d'ensemble de matrices simples connus à ce jour qui reproduisent une telle mesure de probabilité pour les valeurs propres. Ainsi, la version plus générale du modèle diagonalisé que nous étudierons dans cette thèse est:
\beq  \label{DiagonalisationGenerale} Z_N= \int_{\mathcal{C}^N} d\lambda_1 \dots d\lambda_N \,  \Delta(\lambda)^{2\beta} e^{-\frac{N}{T}\underset{i=1}{\overset{N}{\sum}} V(\lambda_i)}\eeq

\section{Distribution des valeurs propres: mesure d'équilibre}

La première question qui vient à l'esprit lorsque l'on regarde le cas des matrices hermitiennes:
\beq \label{DiagonalisationHerm} Z_N= \int_{\mathbb{R}^N} d\lambda_1 \dots d\lambda_N \,  \Delta(\lambda)^{2} e^{-\frac{N}{T}\underset{i=1}{\overset{N}{\sum}} V(\lambda_i)}=\int_{\mathbb{R}^N} d\lambda_1 \dots d\lambda_N \,  e^{-\frac{N}{T}\underset{i=1}{\overset{N}{\sum}} V(\lambda_i)+2\underset{i<j}{\sum}\ln(|\lambda_i-\lambda_j|)}\eeq
est de se demander si la distribution des valeurs propres correctement normalisées et soumises à l'action $-\frac{N}{T}\underset{i=1}{\overset{N}{\sum}} V(\lambda_i)+2\underset{i<j}{\sum}\ln(|\lambda_i-\lambda_j|)$ va suivre une distribution de probabilité simple lorsque $N \to +\infty$ comme dans le cas de la loi du demi-cercle de Wigner. D'un point de vue physique, l'action effective précédente subie par les valeurs propres possède deux contributions évoluant en sens opposés: une force de type Coulombienne $\frac{1}{(\lambda_i-\lambda_j)}$ provoquant une répulsion à courte distance entre les valeurs propres (analogue à celle de charges ponctuelles de même signe en électrostatique) et un terme potentiel $-V(\lambda_i)$ poussant les valeurs propres vers le minimum ou les minima du potentiel $V$. Si intuitivement, on pressent qu'un équilibre entre l'attraction par le puit de potentiel et la répulsion à courte distance va aboutir à une configuration stable et prédéfinie, la réponse définitive à cette question a été apportée par \cite{BPS} et \cite{JOH} et se formule ainsi:
\begin{theorem} Soit $V(x)$ un potentiel polynômial de degré pair. Soit $d\nu_N(x)=\rho_N(x)dx$ la distribution des valeurs propres sur l'axe réel donnée par: $$\forall \phi(x) \in \mathcal{C}^\infty_c(\mathbb{R}):\, \int_{\mathbb{R}} \phi(x)d \nu_N(x)=\int_{\mathbb{R}^N} \left[\frac{1}{N} \underset{j=1}{\overset{N}{\sum}} \phi(\lambda_j)\right] \frac{1}{Z_N}e^{-\frac{N}{T}\underset{i=1}{\overset{N}{\sum}} V(\lambda_i)+2\underset{i<j}{\sum}\ln(|\lambda_i-\lambda_j|)} \prod_{i=1}^N d\lambda_i$$
Soit $d\nu(x;\lambda)$ la mesure discrète localisée aux $\lambda_j$:
$$ d\nu(x;\lambda)=\frac{1}{N}\sum_{j=1}^N \delta(x-\lambda_j)dx$$
Alors ces deux mesures admettent une limite commune notée $d \nu_{eq}(x)$ lorsque $N\to \infty$. Cette mesure appelée \textit{mesure d'équilibre} est supportée par un nombre fini d'intervalles $[a_i,b_i]$ et est absolument continue par rapport à la mesure de Lebesgue: $d\nu_{eq}(x)=\rho_{eq}(x)dx$ avec
\beq \label{densiteequilibre} \rho_{eq}(x)=\frac{1}{2\pi} h(x)R^{\frac{1}{2}}(x)\prod_{i=1}^q\mathds{1}_{[a_i,b_i]}(x) \,\,\,\, ,\,\,\,\, R(x)=\prod_{i=1}^q (x-a_i)(b_i-x) \eeq
et $h(x)$ est un polynôme de degré $deg(h)=deg(V)-q-1$. Le support de la distribution ainsi que la densité $\rho_{eq}(x)$ sont entièrement déterminés par les contraintes:
\beq \label{Contraintes} V'(z)=\underset{z\to \infty}{Pol}(h(z)R^{\frac{1}{2}}(z))\,\,,\,\, \Res_{z \to \infty} h(z)R^{\frac{1}{2}}(z)=-2\eeq
et 
\beq \label{Contraintes2} \int_{b_i}^{a_{i+1}} h(z)R^{\frac{1}{2}}(z)dz=0\,\,\,,\,\,\,  \forall \, 1\leq i\leq q-1\eeq
où la notation $\underset{z\to \infty}{Pol}(f(z))$ signifie la partie polynômiale du développement en série de Laurent de la fonction $f(z)$ à l'infini.
\end{theorem}

Notons que \ref{Contraintes} est équivalent à:
\beq \label{ContraintesInversees} h(z)=\underset{z\to \infty}{Pol}(\frac{V'(z)}{R^{\frac{1}{2}}(z)})\eeq
Remarquons également que les contraintes \ref{Contraintes} et \ref{Contraintes2} restent difficiles à utiliser en pratique. D'abord, elles ne déterminent pas le nombre d'intervalles $q$ de façon immédiate. Il faut ainsi faire une hypothèse sur la valeur de $q$ puis tenter de satisfaire \ref{Contraintes} et \ref{Contraintes2} et si cela n'est pas possible, postuler une autre valeur de $q$ et recommencer. D'autre part, ces contraintes sont fortement non-linéaires et étant donné un potentiel $V(x)$, il est quasiment impossible de déterminer analytiquement la distribution $\rho_{eq}(x)$ ou les extrémités des intervalles. En revanche, il est très facile en utilisant \ref{Contraintes} de trouver un potentiel associé à une distribution d'équilibre $\rho_{eq}(x)$ donnée.

Enfin, il n'est pas évident a priori que les conditions \ref{Contraintes} et \ref{Contraintes2} aboutissent à une densité de probabilité (qui, rappelons le, doit être positive et d'intégrale totale égal à un). Il est également habituel de distinguer les cas où la mesure d'équilibre s'annule sur son support:

\begin{definition} La mesure d'équilibre $d \nu_{eq}(x)$ est dite \textit{régulière} (sinon \textit{singulière}) si elle est strictement positive sur chacun de ses intervalles $]a_i,b_i[$ et si $\forall \, 1\leq i\leq q:  \, \rho_{eq}(x)\underset{x\to a_i}{\sim} \sqrt{x-a_i}$ et $\rho_{eq}(x)\underset{x\to b_i}{\sim} \sqrt{b_i-x}$ . Dans le cas où la mesure est singulière, le potentiel $V(x)$ associé est dit \textit{critique}.
\end{definition}

\section{Simulations et exemples}

Le résultat précédent peut se visualiser très bien avec des simulations numériques pour le cas des matrices hermitiennes. En effet, bien qu'il soit impossible de calculer la fonction de partition $Z_N$ pour de grandes valeurs de $N$, il est en revanche possible de simuler des tirages de valeurs propres $(\lambda_1,\dots,\lambda_N)$ suivant la loi \ref{DiagonalisationHerm} par l'algorithme de Metropolis-Hastings ou par des méthodes de Monte-Carlo. Ainsi, on peut simuler le comportement de la densité d'équilibre associée au potentiel (le paramètre $\epsilon$ mesure l'écart par rapport au cas quartique donné par $\epsilon=1/2$)
\beq \label{potentielcritique} V(x,T,\epsilon)=\frac{1}{T}\left(\frac{x^4}{4}-\frac{4 \cos(\pi\epsilon)x^3}{3}+\cos(2\pi\epsilon)x^2+8\cos(\pi\epsilon)x\right)\eeq
Ce potentiel a été étudié dans \cite{BleherEynard} et devient critique pour la valeur $T_c=1+4\cos(\pi\epsilon)$ où la densité est alors explicitement connue:
\beq \label{densitecritique}\rho_c(x)=\frac{1}{2\pi T_c} \left(x-2\cos\,\pi \epsilon\right)^2\sqrt{4-x^2}\eeq
Pour $T>T_c$, il peut être montré que la densité d'équilibre a un support réduit à un intervalle, alors que pour $T<T_c$, son support est constitué de deux intervalles disjoints. En utilisant l'algorithme présenté en annexe \ref{MetropolisHastings} sur un ordinateur portable standard équipé du logiciel Maple 13, on peut ainsi obtenir pour $\epsilon=1/2$ et $N=200$:

\begin{center} \label{MetropolisCritique}
	\includegraphics[height=5cm]{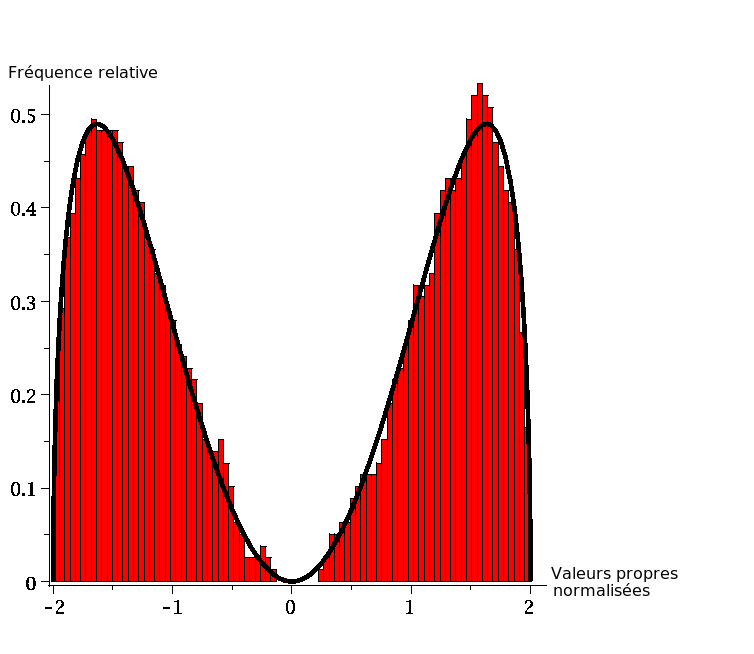}

	\underline{Figure 6}: Histogramme des valeurs propres obtenues par l'algorithme de Metropolis-Hastings pour le potentiel \ref{potentielcritique} à $T=T_c$. La courbe noire représente la densité \ref{densitecritique}. L'échelle de l'axe vertical est choisie pour que l'aire sous l'histogramme soit égale à un (fréquence relative). 
\end{center}

\begin{center} \label{MetropolisCritique2}
	\includegraphics[height=5cm]{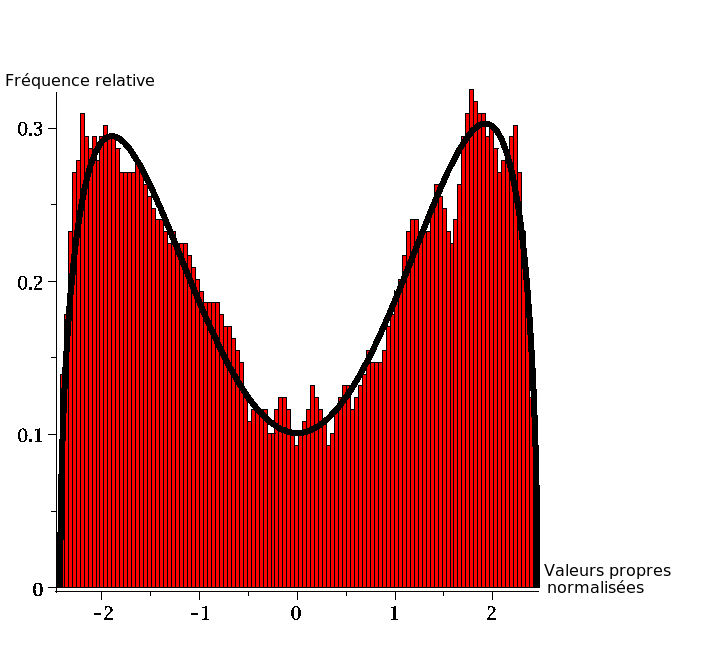}

	\underline{Figure 7}: Histogramme des valeurs propres obtenues par l'algorithme de Metropolis-Hastings pour le potentiel \ref{potentielcritique} à $T=2T_c$. La courbe noire représente la densité donnée par \ref{densiteequilibre} où les extrémités $a_1$ et $b_1$ sont déduites de la simulation. 
\end{center}

\begin{center} \label{MetropolisCritique3}
	\includegraphics[height=5cm]{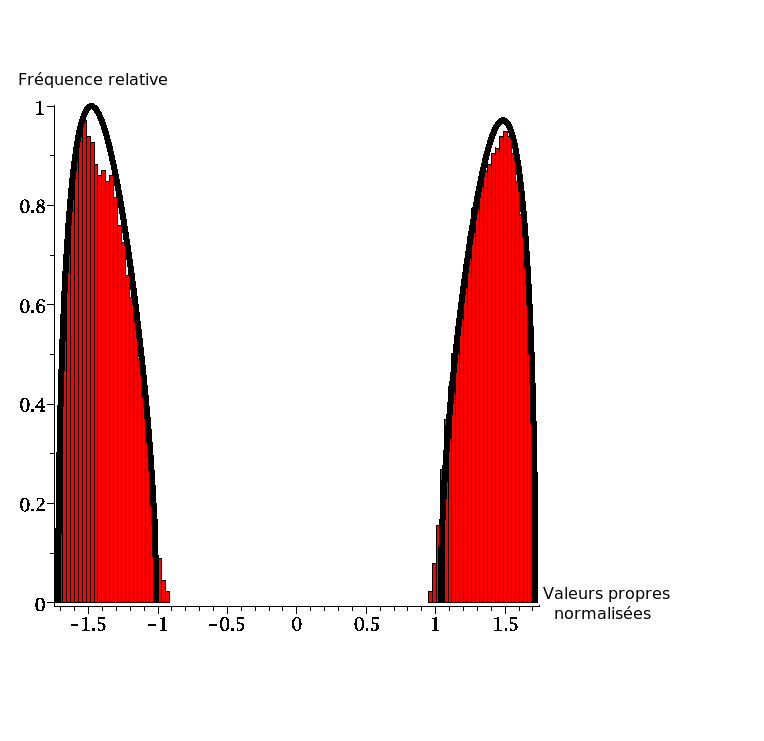}

	\underline{Figure 8}: Histogramme des valeurs propres obtenues par l'algorithme de Metropolis-Hastings pour le potentiel \ref{potentielcritique} à $T=0.5 T_c$. La courbe noire représente la densité donnée par \ref{densiteequilibre} où les extrémités $a_1,\,a_2,\,b_1,\,b_2$ sont déduites de la simulation. 
\end{center}

Cette méthode de simulation permet d'obtenir rapidement les histogrammes des valeurs propres des modèles à une matrice étant donné n'importe quel potentiel. En particulier, il est alors extrêmement facile de trouver numériquement les extrémités $a_i$ et $b_i$ des intervalles supportant la distribution et d'en déduire alors grâce à \ref{ContraintesInversees} la densité associée.

\underline{Note}: Si l'algorithme de Metropolis-Hastings peut être utilisé pour simuler la répartition des valeurs propres, il ne peut pas être utilisé pour calculer directement la fonction de partition $Z_N$ (qui se simplifie à chaque étape de l'algorithme). Le calcul numérique de $Z_n$ se révèle être lui particulièrement difficile dès que $N>2$ à cause de la ``malédiction des dimensions'' (``curse of dimensionnality'') qui demande alors une puissance de calcul très importante.

\underline{Note 2}: Dans le cas d'un potentiel quadratique $V(z)=\frac{z^2}{2}$, la densité des valeurs propres obtenue correspond à la loi du demi-cercle de Wigner. 

\begin{center} \label{Quadratique}
	\includegraphics[height=5cm]{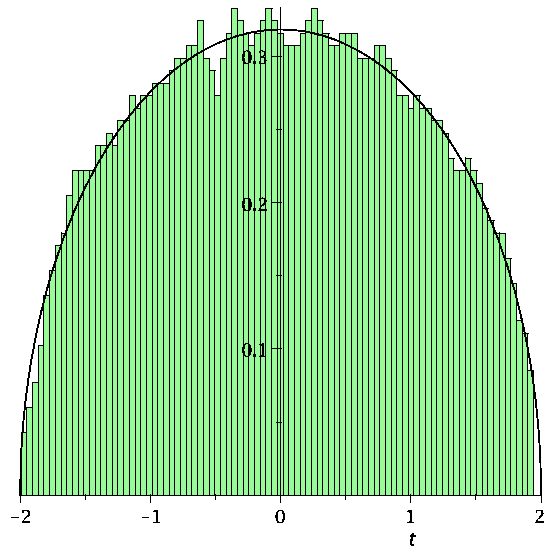}

	\underline{Figure 9}: Histogramme des valeurs propres obtenues par l'algorithme de Metropolis-Hastings pour le potentiel $V(z)=\frac{z^2}{2}$ et $T=1$. En noir, est représentée la loi du demi-cercle $\frac{1}{2\pi}\sqrt{4-x^2}$
\end{center}

\section{Les fonctions de corrélation à $n$-points et l'universalité}

Au delà de la fonction de partition ou de la densité d'équilibre des valeurs propres, il est intéressant de connaître les corrélations entre les valeurs propres lorsque la taille des matrices devient grande. On définit ainsi les fonctions de corrélations à $n$-points dans le cas hermitien (Dyson 1962) par:
\begin{definition} Les fonctions de corrélation non-connexes entre les valeurs propres sont définies par:
\beq \label{FonctionsCorrelations}\rho_n(\lambda_1,\dots,\lambda_n)= \frac{N!}{Z_N (N-n)!}\int_\mathbb{R}\dots\int_\mathbb{R} d\lambda_{n+1} \dots d\lambda_N \,  e^{-\frac{N}{T}\underset{i=1}{\overset{N}{\sum}} V(\lambda_i)+2\underset{i<j}{\sum}\ln(|\lambda_i-\lambda_j|)}\eeq
\end{definition}
Ces fonctions représentent la densité de probabilité de trouver des valeurs propres en $\lambda_1,\dots,\lambda_n$, la position des autres valeurs propres restant non-observées (libres). En particulier, la fonction $\rho_1(x)$ redonne la densité des valeurs propres étudiée dans la section précédente (et dont la limite $N \to +\infty$ est donnée par la mesure d'équilibre \ref{densiteequilibre}). Pour $n>1$, ces fonctions traduisent les corrélations existantes entre les valeurs propres et il est intéressant de se demander si ces fonctions admettent une expression particulière dans la limite où $N \to +\infty$ ou dans la limite $n,N \to \infty$ avec $\frac{n}{N}$ fixé.

Le résultat majeur, connu sous le résultat d'\textbf{universalité} est alors le suivant:
\begin{theorem}
 Pour les modèles de matrices hermitiennes, symétriques réelles et quaternioniques self-duales, les fonctions de corrélations non-connexes à $n$-points ($n>1$)  à petite distance (i.e. d'ordre $1/N$) sont indépendantes du potentiel polynomial pair $V(z)$. En particulier, elles peuvent être calculées par le potentiel gaussien $V(z)=z^2$. Par ailleurs, la connaissance de la fonction de corrélation à $2$-points est suffisante pour déterminer les autres fonctions de corrélations à l'aide de formules déterminantales. (Cf. \cite{Mehta} pour les formules déterminantales spécifiques des trois ensembles) Ainsi, les fonctions à $2$-points non-connexes dans le coeur de la distribution sont données en notant $r=N|\lambda_1-\lambda_2|\rho(\lambda_1)$ par:
\bea \text{Hermitien}: W_2(r)&=&1-\left(\frac{\sin(\pi r)}{\pi r}\right)^2\cr
\text{Réel symétrique}:  W_2(r)&=&1-\left(\frac{\sin(\pi r)}{\pi r}\right)^2-\left(\int_r^\infty\,\frac{\sin(\pi s)}{\pi s} \,ds \right) \frac{d}{dr} \frac{\sin(\pi r)}{\pi r} \cr
\text{Quaternionique}:  W_2(r)&=&1-\left(\frac{\sin(2\pi r)}{2\pi r}\right)^2+\left(\int_0^r\,\frac{\sin(2\pi s)}{2\pi s} \,ds \right) \frac{d}{dr} \frac{\sin(2\pi r)}{2\pi r}\cr
\eea
\end{theorem}
qui peuvent être représentées graphiquement:

\begin{center} \label{FonctionsDeuxPoints}
	\includegraphics[height=8cm]{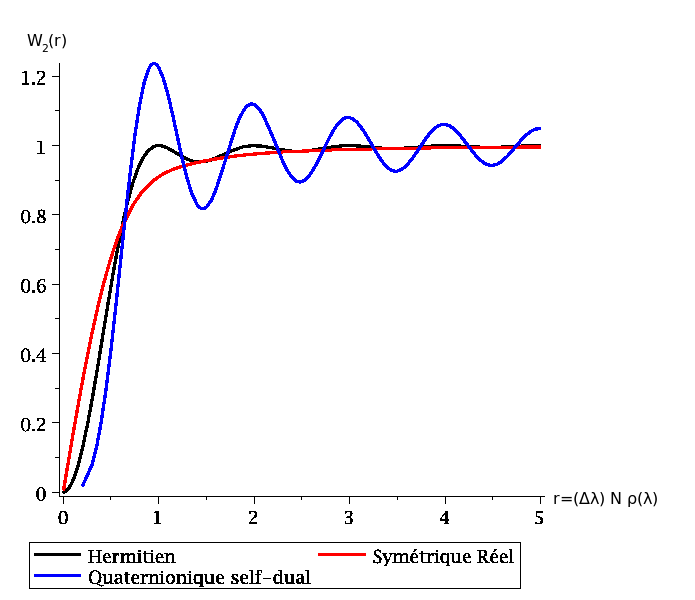}

	\underline{Figure 10}: Graphe des fonctions universelles à deux points pour le cas des ensembles de matrices hermitiennes, symétriques réelles et quaternioniques self-duales. 
\end{center}

En ce qui concerne les modèles de matrices généraux où l'exposant $\beta$ est arbitraire, le théorème précédent n'est pas acquis. En particulier, l'existence d'un phénomène d'universalité ou de formules permettant de déduire les fonctions de corrélation à $n$-points à partir de celle à $2$-points n'est pas connu à l'heure actuelle.

\section{Universalité et premier lien avec les systèmes intégrables}

Soit $E_\beta(J)$ la probabilité qu'aucune valeur propre ne soit dans l'intervalle $J$, alors dans le cas des matrices hermitiennes, symétriques réelles et quaternioniques self-duales, il est connu (Gaudin 1961 pour les matrices hermitiennes et des entrées gaussiennes, Mehta 1971 pour les matrices réelles symétriques et quaternioniques self-duales) que $E_\beta(J)$ peut s'exprimer à l'aide d'un déterminant de Fredholm (Cf. annexe \ref{annexe3} pour la définition générale d'un déterminant de Fredholm):
\beq E_\beta(J)=\det(Id-K_\beta(J)) \eeq
où $K_\beta(J)$ est un opérateur intégral agissant sur $J$ exprimé sous la forme:
\beq \label{rrrrr}\frac{\phi(y)\psi(x)-\psi(y)\phi(x)}{y-x}\eeq
avec $\psi(x)$ et $\phi(x)$ les polynômes orthogonaux d'ordre $n$ et $n-1$ du système (Cf. Chapitre \ref{chap2} pour la définition des polynômes orthogonaux).
Notons également qu'à partir de la connaissance de $E_\beta(J)$, on peut facilement par dérivation par rapport aux bords de l'intervalle $J$ obtenir les densités de valeurs propres ainsi que diverses autres quantités à divers endroits de la distribution (le coeur ou une des extrémités de la distribution). Le second intérêt de pouvoir exprimer $E_\beta(J)$ à l'aide d'un déterminant de Fredholm est qu'il devient possible d'en extraire des limites lorsque $n$ devient grand. Ainsi, dans le cas des matrices hermitiennes, le noyau $K_\beta(J)$ tend asymptotiquement dans le coeur de la distribution vers le noyau "sinus" ("sine kernel"):
\beq K_{\beta=2}(J)\mathop{\to}_{n \to \infty} \frac{sin\, \pi(x-y)}{\pi(x-y)}\eeq
En particulier, on peut alors obtenir la loi des écarts relatifs entre les valeurs propres (normalisées) par:
\beq \label{levelspacingdistribtion} p_{\text{écarts relatifs}}=\frac {d^2}{d s^2} E_{\beta}(J=[0,s]) \eeq
connu sous le nom de distribution de Gaudin.
En ce qui concerne le voisinage des extrémités du support de la distribution (par exemple la plus grande valeur propre), le noyau a cette fois-ci pour limite le noyau d'Airy. En posant:
\beq \lambda_{max}=2\sqrt{T}\sqrt{n} +\frac{\hat{\lambda}}{n^{1/6}}\eeq
alors la loi de $\hat{\lambda}$ tend asymptotiquement vers la loi dite d'Airy:
\beq \label{plusgrandevptheo}\text{prob}(\hat{\lambda}\leq s)\mathop{\to}_{n \to \infty}\frac{d}{ds}\Big[ \det(Id-K_{Airy}([s,+\infty[))\Big] \eeq
où $K_{Airy}(J)$ est donné par \ref{rrrrr} avec $\psi(x)=Airy(x)$ et $\phi(x)=Airy'(x)$.

\medskip
\medskip

Si l'expression en termes de déterminant de Fredholm est intéressante, elle n'est en général pas facilement manipulable et se prête difficilement à des analyses numériques. Heureusement en 1980, dans leur célèbre article \cite{JMMS}, Jimbo, Miwa, Môry et Sato ont obtenu une représentation du noyau sinus en termes de solution d'une équation de Painlevé, faisant ainsi le lien avec les systèmes intégrables. Ainsi, on a:
\begin{theorem} Représentation du noyau sinus à l'aide de l'équation de Painlevé V (\cite{JMMS}):
\beq \det(Id-\lambda K_\beta([0,s[)) = exp(\int_{0}^{\pi s} \frac{\sigma(x,\lambda)}{x} dx)\eeq
où $\sigma(x,\lambda)$ est l'unique solution de l'équation (cas particulier de l'équation de Painlevé V) différentielle:
\beq (x \sigma''(x,\lambda))^2+4\left(x\sigma'(x,\lambda)-\sigma(x,\lambda)\right) \left(x\sigma'(x,\lambda)-\sigma(x,\lambda)+(\sigma'(x,\lambda))^2 \right)=0\eeq
avec $\sigma(x,\lambda)\underset{x\to 0}{\to}-\frac{\lambda}{\pi}x -\frac{\lambda^2}{\pi^2}x^2$.
\end{theorem}

De la même façon, on sait désormais que le noyau d'Airy est relié à l'équation de Painlevé II \cite{TW, TW2}:

\begin{theorem} Représentation du noyau d'Airy à l'aide de l'équation de Painlevé II (\cite{TW}):
\beq 
\det(Id-K_{Airy}([s,+\infty[)=exp(-\int_s^\infty (x-s)q(x)dx)\eeq
où $\sigma(x)$ est l'unique solution de l'équation de Painlevé II:
\beq q''(x)=xq(x)+2q(x)^3 \virg q(x)\underset{x\to \infty}{\sim} Ai(x)\eeq
\end{theorem}

Grâce à ces représentations différentielles, il est alors possible (bien que numériquement les équations de Painlevé soient assez instables) de comparer les lois théoriques des écarts relatifs entre les valeurs propres dans le coeur ainsi que la distribution de la plus grande valeur propre à des simulations numériques:

\begin{center} \label{GaudinDistribution}
	\includegraphics[height=8cm]{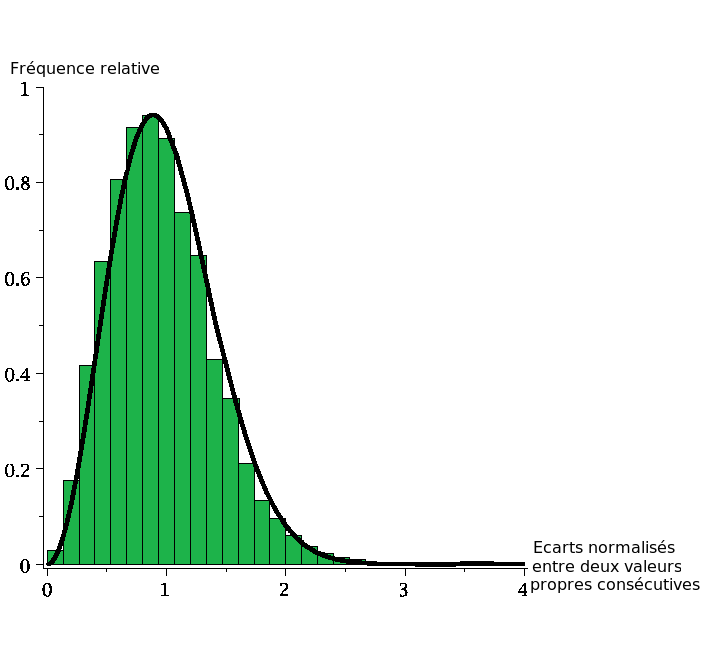}

	\underline{Figure 11}: Histogramme des écarts normalisés de deux valeurs propres consécutives de $500$ matrices hermitiennes de taille $300 \times 300$ (entrées gaussiennes) avec la distribution théorique de Gaudin \ref{levelspacingdistribtion}. 
\end{center}

\begin{center} \label{Plusgrandevp}
	\includegraphics[height=8cm]{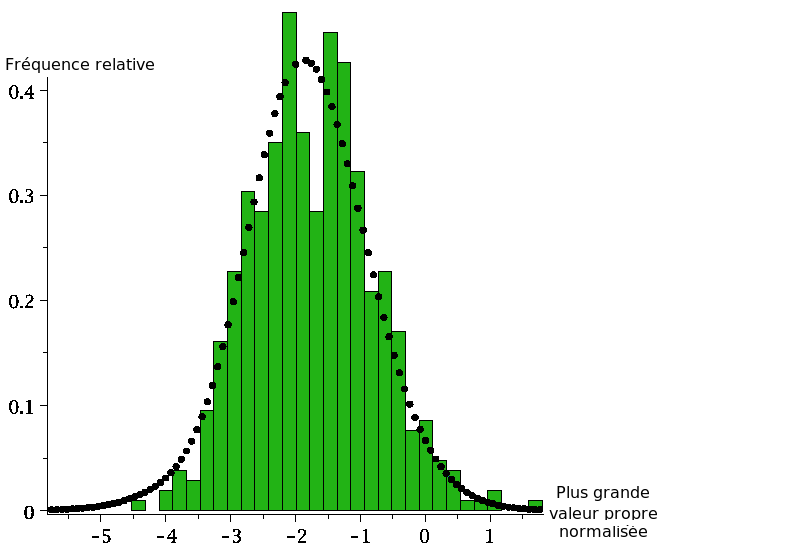}

	\underline{Figure 12}: Histogramme des plus grandes valeurs propres normalisées de $500$ matrices hermitiennes de taille $300 \times 300$ (entrées gaussiennes) avec la distribution théorique \ref{plusgrandevptheo}.
\end{center}

Ou bien encore avec la fonction de répartition (pour éviter d'avoir à dériver):

\begin{center} \label{Plusgrandevp2}
	\includegraphics[height=8cm]{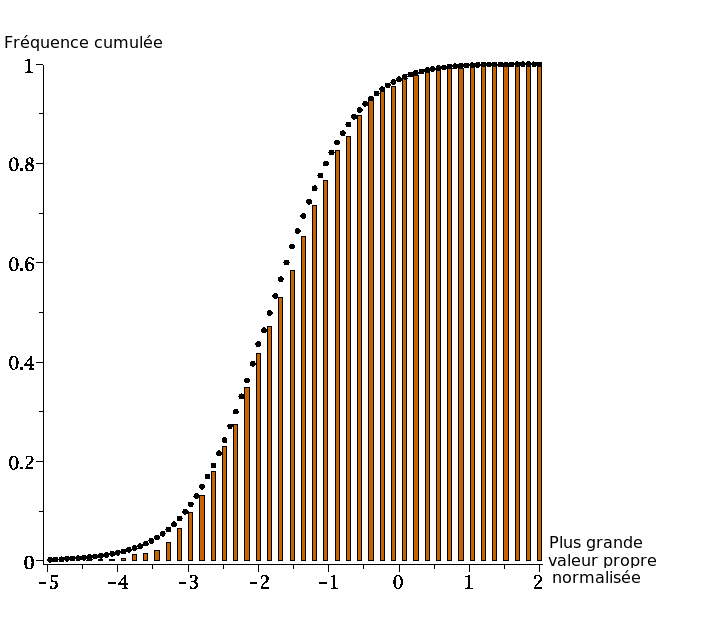}

	\underline{Figure 13}: Histogramme cumulé des plus grandes valeurs propres normalisées de $500$ matrices hermitiennes de taille $300 \times 300$ (entrées gaussiennes) avec la fonction de répartition théorique \ref{plusgrandevptheo}.
\end{center}

Notons également que dans le cas des matrices réelles symétriques et quaternioniques self-duales, des expressions similaires en termes de déterminant de Fredholm et de solutions de Painlevé sont connues \cite{TracyWidom, Mehta} donnant en particulier les densités au voisinage de la plus grande valeur propre ainsi que dans le coeur de la distribution. Enfin, le point le plus important est que ces expressions satisfont à un théorème d'universalité (\cite{TracyWidom, Mehta}):

\begin{theorem}
 La loi de Gaudin \ref{levelspacingdistribtion} reste valable quels que soient les potentiels polynomiaux pairs $V(x)$ en tout point situé dans le coeur du support de la distribution des valeurs propres (i.e. loin des extrémités du support) où la densité de probabilité est strictement positive. La loi d'Airy reste également valide quels que soient les potentiels polynomiaux pairs $V(x)$ pour les valeurs propres situées dans un voisinage d'une des extrémités, notée $a$, du support de la distribution des valeurs propres, sous la condition que la densité s'y comporte localement comme $\rho(x) \underset{x \to a}{\sim} \sqrt{x-a}$ (extrémité dite régulière).
\end{theorem}

Le cas des extrémités non régulières ou de points où la densité s'annule sont des cas singuliers et ont constitué une partie de mon travail de thèse (Cf. \textbf{[I]} à l'annexe \ref{Article[I]}). Les résultats à ce sujet seront présentés dans le prochain chapitre où l'on verra que les points $a$ où la densité s'annule comme $(x-a)^{2m}$ sont en relation avec d'autres solutions d'équations de Painlevé.

\section{Résolvantes et développement topologique dans le cas hermitien}

Afin de calculer les fonctions de corrélations à $n$-points, il est utile de définir les \textit{résolvantes} par:
\begin{definition} Les résolvantes sont définies par:
\beq \label{resolvents}  \hat{\om}_n(x_1,\dots,x_n)=\left< \Tr\frac{1}{x_1-M}\dots\Tr\frac{1}{x_n-M}\right>=\left<\sum_{i_1,\dots,i_n=1}^N \frac{1}{x_1-\lambda_{i_1}}\dots\frac{1}{x_n-\lambda_{i_n}}\right>\eeq
\end{definition}

Ici, la notation $<>$ désigne la valeur moyenne selon la distribution de probabilité: 
\beq \label{ValeurMoyenne}\forall f: \, \mathbb{E}_N \rightarrow \mathbb{C}\,\, :\,\,\, <f(M)>=\frac{1}{Z_N}\int_{\mathbb{E}_N} dM f(M) e^{-\frac{N}{T} \Tr(V(M))}\eeq
ce qui se traduit dans le problème aux valeurs propres par (pour le cas hermitien): 
\beq \label{ValeurMoyenneReduite}
\forall g:\, \mathbb{R}^N \rightarrow \mathbb{C}\,\, :\,\, <g(\lambda)>= \frac{1}{Z_N} \int_{\mathbb{R}^N} d\lambda_1 \dots d\lambda_N \, g(\lambda) \Delta(\lambda)^{2} e^{-\frac{N}{T}\sum_{i=1}^N V(\lambda_i)}\eeq

Il est également intéressant d'introduire les cumulants des expressions \ref{resolvents} par:
\begin{definition} Les cumulants des résolvantes sont définis par:
\beq \label{cumulants}\om_n(x_1,\dots,x_n)=\left< \Tr\frac{1}{x_1-M}\dots\Tr\frac{1}{x_n-M}\right>_c\eeq
où l'indice $_c$ désigne les cumulants de la valeur moyenne des produits, c'est à dire en notant $J=\{1,\dots,n\}$:
\bea <A_1> &=& <A_1>_c \cr
<A_1 A_2> &=& <A_1 A_2>_c + <A_1>_c <A_2>_c \cr
<A_1 A_2 A_3> &=& <A_1 A_2 A_3>_c + <A_1 A_2>_c <A_3> + <A_1 A_3>_c <A_2> \cr
&&+ <A_2 A_3>_c <A_3> + <A_1> <A_2> <A_3>\cr
<A_1\dots A_n>&=&<A_J>=\sum_{k=1}^n\sum_{I_1 \bigsqcup I_2 \dots \bigsqcup I_k=J}\prod_{i=1}^k <A_{I_i}>_c\cr 
\eea
où la somme a lieu sur une décomposition de l'ensemble $J$ en une réunion disjointe d'ensembles non vides.
\end{definition}

Les densités des fonctions de corrélations $\rho_n(x_1,\dots,x_n)$ définies préalablement (\ref{FonctionsCorrelations}) s'obtiennent alors comme les discontinuités des résolvantes $ \hat{\om}_n(x_1,\dots,x_n)$ et inversement les résolvantes $ \hat{\om}_n(x_1,\dots,x_n)$ sont les transformées de Stieljes des densités. Par exemple la densité des valeurs propres $\rho(x)=\rho_1(x)$ s'obtient par:
\beq  \hat{\om}_1(x)=\int \frac{\rho(x')}{x-x'}dx' \Longleftrightarrow \rho(x)=\frac{1}{2i\pi} \left(\om_1(x-i0)-\om_1(x+i0)\right)\eeq

\textbf{Ainsi connaître les densités de corrélations des valeurs propres est équivalent à connaître les résolvantes $ \hat{\om}_n(x_1,\dots,x_n)$ ou leurs cumulants $\om_n(x_1,\dots,x_n)$.}

Hélas, le calcul des fonctions de corrélation ou des résolvantes n'est pas en général possible analytiquement. Une solution est alors de rechercher un développement perturbatif en série de puissances de $\frac{1}{N}$ et d'écrire:

\begin{definition} Le développement perturbatif (topologique) dans le cas des modèles de matrices hermitiennes est défini de façon formelle par \cite{OE, Hooft, Hooft2}:
\bea \label{devtopo}
F(T)=\ln(Z_N(T))&=&\sum_{g=0}^\infty\left(\frac{N}{T}\right)^{2-2g}F_g(T)\cr
\hat{\om}_n(x_1,\dots,x_n,T)&=& \sum_{g=0}^\infty\left(\frac{N}{T}\right)^{2-2g-n}\hat{\om}_n^{(g)}(x_1,\dots,x_n,T)\cr
\eea
\end{definition}

Il est important de préciser qu'un tel développement n'existe pas toujours (Si par exemple $F(T)$ possèdait un terme en $e^{-\frac{N}{T}}$, ce dernier ne contribuerait pas au développement formel car il est exponentiellement petit dans la limite $\frac{N}{T} \to \infty$ ) mais existe systématiquement dans des cas particuliers où le contour des valeurs propres correspond à un contour de ``steepest descent'' associé au potentiel $V$. De plus, cette série est toujours divergente et il ne s'agit là que d'un développement asymptotique que l'on peut manipuler de façon formelle. Quoi qu'il en soit, il est toujours possible de supposer l'existence d'un tel développement et de le manipuler de façon formelle. Cela est en particulier utile pour les problèmes de dénombrements où les $F_g$ comptent le nombre de surfaces triangulées de genre $g$ (et qui justifie le nom de développement topologique). Le lecteur intéressé pourra se référer entre autres à \cite{countingsurface, OE}. Cela peut également être utile si l'on s'intéresse à la limite $N \to +\infty$ ou $T\to 0$ puisqu'alors seuls les premiers termes de la série contribuent de façon significative. On notera enfin que \cite{Nonperturbative} permet de calculer la partie non-perturbative de la fonction de partition lorsque le développement perturbatif est connu.

\section{Les équations de boucles du modèle hermitien}

L'introduction du développement topologique (\ref{devtopo}), permet de trouver des relations entre les différentes résolvantes $\om_n^{(g)}(x_1,\dots,x_n)$ et de résoudre par récurrence le problème. Les relations entre les différentes résolvantes sont données par la méthode des équations de boucles, connues aussi comme équations de Schwinger-Dyson qui consistent en de simples intégrations par parties judicieuses dans l'intégrale matricielle. La démonstration de ces équations est devenue classique pour le cas du modèle à une matrice et peut être trouvée par exemple dans \cite{Mehta}, \cite{OE}. Par ailleurs dans le chapitre \ref{chap3}, nous présentons en détail la dérivation des équations de boucles dans le cas du modèle à deux matrices avec $\beta$ quelconque, à partir desquelles on peut facilement obtenir les équations de boucles qui nous intéressent présentement.
Introduisons donc les fonctions:
\begin{definition}
\beq U_n(x_1;x_2\dots, x_n)=\left<\Tr \frac{V'(x_1)-V'(M)}{x_1-M}\Tr \frac{1}{x_2-M}\dots \Tr \frac{1}{x_n-M}\right>_c\eeq 
ainsi que leur développement topologique (formel ou non):
\beq U_n(x_1;x_2\dots, x_n)=\sum_{g=0}^\infty \left(\frac{N}{T}\right)^{2-2g-n}U_k^{(g)}(x_1;x_2\dots, x_n)\eeq
\end{definition}

alors les équations de boucles donnent les relations suivantes \cite{OE}:
\beq \label{eqboucle1}\om_1(x_1)^2+\frac{T^2}{N^2}\om_2(x_1,x_2)=V'(x_1)\om_1(x_1)-U_1(x_1)\eeq
puis en définissant la notation $J=\{x_2,\dots,x_n\}$:
\bea \label{eqboucles2} &&\left(V'(x_1)-2\om_1(x_1)\right)\om_n(x_1,\dots,x_n)\cr
&&=U_n(x_1;x_2,\dots,x_n)+\frac{T^2}{N^2}\om_{n+1}(x_1,x_1,x_2,\dots,x_n)\cr
&&+\sum_{I\subset J} \om_{j+1}(x_1,x_I) \om_{n-j}(x_1,x_{J/I})\cr
&&+\sum_{j=1}^k \frac{\partial}{\partial x_j} \left( \frac{ \om_{n-1}(x_1,\dots,x_j,\dots,x_n)-\om_{n-1}(x_1,\dots,x_1,\dots,x_n)}{x_j-x_1}\right)
\cr
\eea
On peut alors projeter ces équations dans le développement topologique et identifier les puissances terme à terme. Cela donne:
\beq \om_1^{(0)}(x_1)^2=V'(x_1)\om_1^{(0)}(x_1)-U_1^{(0)}(x_1)\eeq
et 
\bea &&\left(V'(x_1)-2\om_1^{(0)}(x_1)\right)\om_k^{(g)}(x_1,\dots,x_k)\cr
&&=U_k^{(g)}(x_1;x_2,\dots,x_k)+\om_{k+1}^{(g-1)}(x_1,x_1,x_2,\dots,x_k)\cr
&&+\sum_{I\in J}\sum_{h=0}^g \om_{j+1}^{(h)}(x_1,x_I) \om_{k-j}^{(g-h)}(x_1,x_{J/I})\cr
&&+\sum_{j=1}^k \frac{\partial}{\partial x_j} \left( \frac{ \om_{k-1}^{(g)}(x_1,\dots,x_j,\dots,x_k)-\om_{k-1}^{g)}(x_1,\dots,x_1,\dots,x_k)}{x_j-x_1}\right)
\cr
\eea

En rebaptisant les fonctions de la façon suivante:
\beq\label{oups} w(x)=\om_1^{(0)}(x)\,\,\,\,\, , \,\,\,\,\, y(x)=w(x)-\frac{V'(x)}{2}\,\,\,\,\, , \,\,\,\,\, P(x)=\frac{V'(x)^2}{4}-U_1^{(0)}(x)\eeq
et 
\beq P_k^{(g)}(x_1;x_2,\dots,x_k)=U_k^{(g)}(x_1;x_2,\dots,x_k)+\sum_{j=2}^k \frac{\partial}{\partial x_j} \left( \frac{ \om_{k-1}^{(g)}(x_1,\dots,x_j,\dots,x_k)-\om_{k-1}^{(g)}(x_1,\dots,x_1,\dots,x_k)}{x_j-x_1}\right)\eeq
on trouve les équations de boucles sous leur forme standard \cite{OE}:

\begin{theorem} Les équations de boucles peuvent être mises sous leur forme standard:
\beq \label{courbe spectrale w(x)} y^2(x)=P(x) \eeq
puis la formule de récurrence topologique:
\bea-2y(x)\om_k^{(g)}(x,x_2,\dots,x_k)
&=& P_k^{(g)}(x;x_2,\dots,x_k)+\om_{k+1}^{(g-1)}(x,x,x_2,\dots,x_k)\cr
&&+\sum_{I\in J}\sum_{h=0}^g \om_{j+1}^{(h)}(x,x_I) \om_{k-j}^{(g-h)}(x,x_{J/I})\cr \label{Formule de recurrence topologique}
\eea
\end{theorem}

L'intérêt de la méthode des équations de boucles apparaît alors. En effet, les fonctions $x\to U_n(x;x_2,\dots,x_n)$, $x \to P(x)$ et $x \to P_n(x;x_2,\dots,x_n)$ n'ont de singularités qu'aux singularités du potentiel $V(x)$. En particulier lorsque ce dernier est polynômial, ces fonctions sont également polynômiales en $x$. L'intérêt de la méthode provient également du fait qu'une fois la courbe $y^2(x)=P(x)$ déterminée (c'est-à-dire une fois l'ordre dominant de la densité des valeurs propres connu, par exemple à l'aide de \ref{densiteequilibre} ou de la donnée des fractions de remplissage), toutes les autres fonctions $\om_n^{(g)}(x_1,\dots,x_n)$ peuvent être calculées à l'aide des travaux de \cite{OE} et de résultats de géométrie algébrique résumés dans le paragraphe ci-dessous. Notons que la détermination du polynôme $P(x)$ peut varier suivant le contexte. En effet, par sa définition \ref{oups}, il est clair que seule la moitié supérieure des coefficients de $P(x)$ peut être directement déterminée par le potentiel $V'(x)$. Pour la moitié inférieure, cela depend du contexte qui est étudié: si l'on s'intéresse au modèle convergent, alors la densité des valeurs propres est donnée par \ref{densiteequilibre} qui est intégralement connue par les contraintes \ref{Contraintes} \ref{Contraintes2}. En revanche, on peut également s'intéresser à des modèles formels où une fraction donnée $\epsilon_i$ (appelées couramment ``fraction de remplissage'') des valeurs propres se trouve dans la coupure $[a_i,b_i]$ de \ref{densiteequilibre}. Dans ce cas, les contraintes \ref{Contraintes2} ne sont plus exigibles mais doivent être remplacées par les conditions:
\beq \epsilon_i=\oint_{\mathcal{A}_i} ydx\eeq
où $\mathcal{A}_i$ est le $\mathcal{A}_i$-cycle contour entourant la coupure $[a_i,b_i]$.

\section{Invariants symplectiques et géométrie algébrique}

Les résultats concernant les modèles de matrices se généralisent en fait à des courbes algébriques quelconques \cite{OE} appelées \textit{courbe spectrale}:
\beq \label{E} E(x,y)=0=\sum_{i=0}^{d_x}\sum_{j=0}^{d_y}E_{i,j}x^iy^j\eeq
Le cas particulier des modèles à une matrice hermitienne correspond ainsi au cas où $E(x,y)$ est de degré $2$ en $y$ et est donné par \ref{courbe spectrale w(x)} (et dépend donc d'un paramètre $T$). L'équation \ref{E} définit une surface de Riemann à partir de laquelle des quantités vont être calculées (Cf. \cite{OE} pour plus de détails). Soient $\{a_i\}_{i=1..n}$ les points de branchements supposés \textbf{simples} (sinon la construction échoue) de la courbe spectrale $E(x,y)=0$ que l'on suppose de genre $\genus$ et sur laquelle on a défini une base de cycles non contractibles indépendants: $(\mathcal{A}_i, \mathcal{B}_i)_{i=1..\genus}$ avec:
 $$\mathcal{A}_i \cap \mathcal{B}_j=\delta_{i,j}$$ 
Sur cette courbe algébrique, on peut également définir une base de formes holomorphes indépendantes $du_i(x)$ que l'on normalise comme suit: $$\oint_{\mathcal{A}_j} du_i =\delta_{i,j}$$
Un résultat de géométrie algébrique nous dit alors que la matrice des périodes de Riemann $\tau_{i,j}=\oint_{\mathcal{B}_{j}} du_i$ est symétrique. 

Sur cette courbe algébrique, on peut enfin définir un noyau de Bergman $B(p,q)$ comme unique forme bilinéaire ayant un pôle double sans résidu en $p=q$ et normalisée de la façon suivante: $B(p,q) \sim \frac{dz(p)dz(q)}{(z(p)-z(q))^2}$. Il possède les propriétés d'être symétrique et de satisfaire $\oint_{q \in\mathcal{B}_i} B(p,q)=2i\pi\, du_i(p)$ et $\oint_{q\in \mathcal{A}_i} B(p,q)=0$

A partir de ces définitions classiques de géométrie algébrique on peut définir ($a$ est un point de base qui peut être choisi arbitrairement):
$$\Phi(p)=\int_a^p y\,dx$$
$$W_2^{(0)}(p_1,p_2)=B(p_1,p_2)$$
\bea W^{(g)}_{k+1}(p,p_K) &=& \sum_i\Res_{q\to a_i}\frac{dE_q(p)}{y(q)-y(\overline{q})}\cr
&&\left(\sum_{h=0}^{g}\sum_{J\in K}
W_{|J|+1}^{(h)}(q, p_J)W^{(g-h)}_{k-|J|+1}(q, p_{K/J}) +W^{(g-1)}_{k+2} (q, q, p_K)\label{Recurrence}
\right)\cr
\eea
où $dE_q(p)=G(p,q)=\int^q B(p,q')$ est la forme de troisième espèce. Ces fonctions satisfont les relations suivantes:
\begin{enumerate}
 \item Elles satisfont des ``équations de boucles'' identiques à \ref{Formule de recurrence topologique}. Les quantités:
\bea P^{(g)}_k (x(p), p_K) &=& \sum_i \big[ 2y(p^i)W_{k+1}^{(g)}(p^i,p_K)+W_{k+2}^{(g-1)}(p^i,p^i,p_K)\cr
&&+\sum_{J\in K}\sum_{h=0}^g W_{j+1}^{(h)}(p^i,p_J) W_{k-j-1}^{(g-h)}(p^i,p_{K/J})\big]
\eea
sont des fonctions rationnelles en $x(p)$ sans pôle aux points de branchements.
\item Les $W_n^{(g)}$ sont des fonctions symétriques de leurs arguments, de résidus nuls aux points de branchements et d'intégrales nulles sur les $\mathcal{A}$-cycles.
\item $\forall k \geq 1\,:$
 \beq \Res_{p_{k+1} \to a,p_1,...,p_k}\Phi(p_{k+1})W^{(g)}_{k+1}(p_K, p_{k+1})
= (2g + k - 2)W^{(g)}_k (p_K) + \delta_{g,0}\delta_{k,1}y(p_1)dx(p_1)\eeq
\item Les invariants sont alors définis par extension de la relation précédente pour $k=0$:
\beq F^{(g)}=\frac{1}{2-2g} \sum_i \Res_{q \to a_i} W_1^{(g)}(q)\Phi(q) \label{InvSymp}\eeq
Ils sont invariants par transformations symplectiques de la courbe spectrale.
\end{enumerate}

Le lien avec les modèles de matrices est alors le suivant: les fonctions $W_n^{(g)}(x_1,\dots,x_n)$ définies par la récurrence topologique \ref{Recurrence} correspondent aux résolvantes $\om_n^{(g)}(x_1,\dots,x_n)$ des modèles de matrices hermitiennes \ref{devtopo}. Il en est de même avec les invariants symplectiques \ref{InvSymp} qui sont les mêmes que les énergies libres $F_g(T)$ des modèles de matrices \ref{devtopo}. Ce résultat se généralise systématiquement lorsque la courbe spectrale \ref{E} provient d'un modèle de matrices hermitiennes. (Les courbes de degrés plus élevés pouvant provenir de modèles à deux matrices hermitiennes). Il est à noter que la récurrence topologique \ref{Recurrence} est facilement applicable en pratique puisqu'elle consiste uniquement à prendre des résidus aux points de branchements. \textbf{En résumé, les travaux de Eynard et Orantin \cite{OE} permettent, étant donnée la courbe spectrale, (i.e. le terme dominant de la densité des valeurs propres) de déterminer le développement perturbatif de toutes les résolvantes (i.e. de toutes les densités de corrélations des valeurs propres)}.

\section{Potentiel singulier et double limite d'échelle}

Lorsque le potentiel $V(x)$ est singulier (i.e. lorsque la densité limite des valeurs propres s'annule en au moins un point intérieur strictement à son support), le modèle de matrice associé présente des caractéristiques intéressantes dans le régime appelé \textit{double limite d'échelle} \cite{Kazakov, doublelimiteechelle, doublelimiteechelle}. En effet, il est alors possible \cite{BleherEynard} de plonger le potentiel critique $V_c(x)$ dans un ensemble de potentiels dépendant d'un paramètre supplémentaire $s$: $V(x,s)$ tel que, pour une certaine valeur du paramètre $s$, on obtient le potentiel critique:
\beq V(x,s=s_c)=V_c(x)\eeq
Bien que le choix du plongement soit a priori arbitraire, le paramètre le plus naturel pour plonger le potentiel critique est la température $T$. Par analogie avec la physique statistique traditionnelle, le modèle de matrice est dit présenter une \textit{transition de phase} pour la valeur critique $T=T_c$. Ainsi nous avons vu par exemple précédemment \ref{potentielcritique} que le potentiel:
\beq V(x,T)=\frac{1}{T}\left(\frac{x^4}{4}-\frac{4 \cos(\pi\epsilon)x^3}{3}+\cos(2\pi\epsilon)x^2+8\cos(\pi\epsilon)x\right)\eeq
présente une transition de phase pour la valeur $T=T_c=1+4\cos(\pi\epsilon)$.

L'étude des potentiels singuliers et des transitions de phase est intéressante car elle permet de faire le lien avec les systèmes intégrables dans le cadre de la double limite d'échelle. La double limite d'échelle consiste à prendre simultanément la limite $N \to +\infty$ et $T \to T_c$ de telle sorte que pour une certaine valeur bien choisie $\alpha$, le produit $(T-T_c)N^{-\alpha}$ reste d'ordre un. D'une façon générale, les transitions de phase peuvent être étudiées pour des singularités de la densité des valeurs propres $\rho(x)$ de type $(p,q)\in \mathbb{N}^2$:
\beq \rho(x)\underset{x\to a}{\sim} (x-a)^{\frac{p}{q}}\eeq
Si $a$ est une extrémité du support de $\rho(x)$, les valeurs $(p=1,q=2)$ ne donnent pas lieu à un point critique. En revanche, dans tous les autres cas, la densité de valeurs propres $\rho(x)$ présente un point singulier en $x=a$. Dans le cas des modèles à une matrice hermitienne, le fait que la densité d'équilibre des valeurs propres soit donnée par une courbe hyperelliptique n'autorise que deux valeurs de $q$: $q=1$ et $q=2$. Le cas $q=1$ correspond à un point intérieur au support de la densité où cette dernière s'annule. Le cas $q=2$ et $p\neq 1$ correspond quant à lui à une extrémité singulière où la densité ne se comporte plus localement comme une racine carrée. Ces deux situations correspondent aux deux images suivantes:

\begin{center} \label{DensiteCritiqueZoomee}
	\includegraphics[height=5cm]{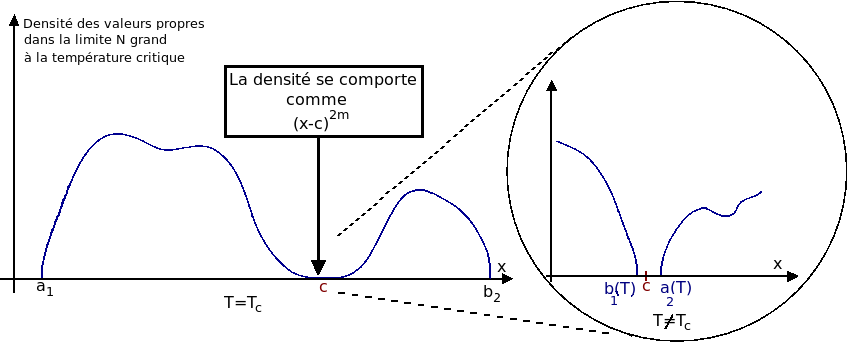}

	\underline{Figure 14}: Double limite d'échelle d'une densité critique dont le point critique est intérieur au support. La singularité est de type $(p=2m,q=1)$.
\end{center}  

\begin{center}
\label{DensiteCritiqueZoomee2}
	\includegraphics[height=5cm]{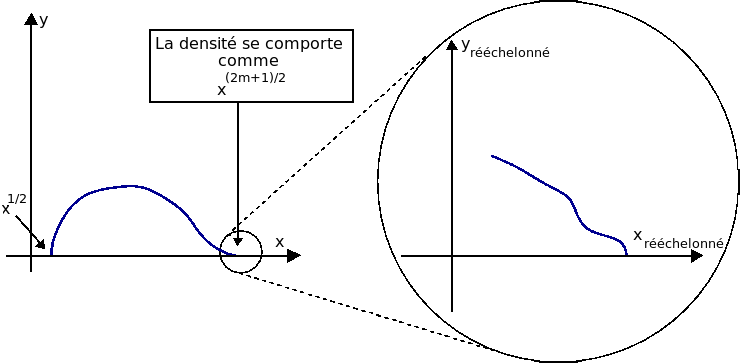}

	\underline{Figure 15}: Double limite d'échelle d'une densité critique dont le point critique est à une extrémité du support. La singularité est de type $(p=2m+1,q=2)$. \cite{BergereEynard}
\end{center}

De nombreux résultats concernant les doubles limites d'échelle sont connus aujourd'hui. Ainsi, d'après \cite{OE}, il est connu que les doubles limites d'échelle pour une singularité de type $(p,q)$ sont reliées aux modèles minimaux $(p,q)$ de la théorie conforme (CFT). Un résultat majeur de \cite{Kazakov} montre ainsi que si les invariants symplectiques $F_g(T)$ et les fonctions de corrélations $\om_n^{(g)}(x_1,\dots,x_n,T)$ peuvent être calculés pour des valeurs régulières de la température $T$, ces quantités divergent lorsque $T\to T_c$ et ne peuvent être définies par la récurrence topologique \ref{Recurrence} habituelle pour $T=T_c$ puisque des points de branchements ne sont plus simples. En revanche, il est montré dans \cite{OE} le résultat suivant:

\begin{theorem}
Sous le changement d'échelle général (point singulier $(p,q)$ en $x=a$):

\beq x_i=a+(T-T_c)^{\frac{1}{p+q-1}}\xi_i\eeq
tel que:
\beq y_{\text{rescaled}}(\xi)=\lim_{T\to T_c} (T-T_c)^{\frac{p+q}{p+q-1}} y( x_1, T)\eeq
et
\beq \label{rescaling} \om_{\text{rescaled},n}^{(g)}(\xi_1,\dots,\xi_n)=\lim_{T\to T_c}(T-T_c)^{(2-2g-n)\frac{p+q}{p+q-1}} \om_n^{(g)}( x_1,\dots,x_n,T)\eeq
avec
\beq F_{\text{rescaled},g}=\lim_{T\to T_c} (T-T_c)^{(2-2g)\frac{p+q}{p+q-1}}F_g(T)\eeq

Les quantités, $y_{\text{rescaled}}(\xi)$, $\om_{\text{rescaled},n}^{(g)}(\xi_1,\dots,\xi_n)$ et $F_{\text{rescaled},g}$ sont bien définies (au sens où la limite existe et est finie) et correspondent aux invariants symplectiques et aux résolvantes de la courbe $y_{\text{rescaled}}(\xi)$. 
\end{theorem}

Ce résultat peut être utilisé dans les deux cas provenant des modèles hermitiens à une matrice. Ainsi, le cas $(2m+1,q=2)$ a été traité par M. Bergère et B. Eynard dans \cite{BergereEynard} alors que j'ai traité le cas $(p=2m,q=1)$ dans l'article \textbf{[I]} présenté dans l'annexe \ref{Article[I]} avec Mattia Cafasso. Dans notre cas, la densité critique est:  
\beq  \label{SingularDensity}
\rho(x,T_c)=\rho_c(x)=\frac{1}{2\pi} (x-b\epsilon)^{2m}\sqrt{b^2-x^2}\eeq
où le point critique est $x=b\epsilon$ de type $(2m,1)$.

\begin{center} \label{SingularDensityPicture}
	\includegraphics[height=5cm]{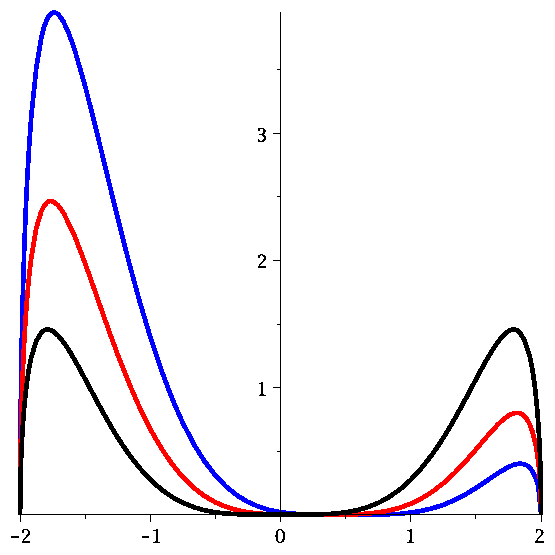}

	\underline{Figure 16}: Image de la fonction de densité critique donnée par (\ref{SingularDensity}) pour les valeurs $b=2$, $m=2$. La courbe noire représente le cas où $\epsilon=0$, la rouge celle où $\epsilon=\frac{1}{4}$ et la bleue celle où $\epsilon=\frac{1}{8}$.
\end{center} 

Nous avons montré que (Cf. \textbf{[I]}, Section 2.4 de l'annexe \ref{Article[I]}) cette densité critique correspond au potentiel critique suivant:
\begin{theorem}
 La densité critique \ref{SingularDensity} correspond au potentiel:
\bea V'(x,T)&=&\frac{1}{T}\Big(\sum_{j=0}^{2m+1} \Big( \dbinom{2m}{j-1} (-b\epsilon)^{2m+1-j} \cr
&&+\sum_{n=1}^{E(\frac{2m+1-j}{2})} \dbinom{2m}{2n+j-1} \frac{(-1)^j(2n-2)!\epsilon^{2(m-n)+1-j}b^{2m+1-j}}{n!(n-1)!2^{2n-1}} \Big) x^j\Big)\cr 
\eea
avec une température critique:
\beq T_c=\frac{b^{2m+2}}{2}\sum_{n=1}^{m+1} \frac{\epsilon^{2m-2n+2}(2m)! }{n!(2m-2n+2)!(n-1)!2^{2n-1}}\eeq
\end{theorem}

On peut alors montrer (Cf. \textbf{[I]}, Section 2.4 de l'annexe \ref{Article[I]}) que le changement d'échelle $x=b\epsilon+\xi (T-T_c)^{\frac{1}{2m}}$ donne alors la courbe réduite suivante:
\begin{theorem} Après changement d'échelle, la courbe réduite est donnée par:
 \beq \label{RescaledCurve}\encadremath{  y_{\text{rescaled}}(\xi)=b\pi\sqrt{1-\epsilon^2}\sqrt{(\gamma^2-\xi^2)}\left( \xi^{2m-1}+ \sum_{n=1}^{m-1} \frac{(2n)!}{(n!)^22^{2n}} \gamma^{2n}\xi^{2m-1-2n} \right)}\eeq
 où le paramètre $\gamma$ intervenant dans l'expression de la courbe est donné par:
\beq\label{gamma} \gamma^{2m}=-\frac{(m!)^22^{2m+1}}{b^2(1-\epsilon^2)(2m)!}\eeq
\end{theorem}

Le calcul des fonctions $\om_{\text{rescaled},n}^{(g)}(\xi_1,\dots,\xi_n)$ et des invariants symplectiques $F_{\text{rescaled},g}$ peut alors être effectué par la méthode générale de Eynard et Orantin pour la courbe \ref{RescaledCurve}. Nous avons ainsi trouvé dans \textbf{[I]} présenté dans l'annexe \ref{Article[I]} que:

\begin{theorem} La fonction de corrélation à deux points est donnée par (Cf. \textbf{[I]}, Section 2.4 de l'annexe \ref{Article[I]}):
\beq \label{W20xi}\om_2^{(0)}(\xi_1,\xi_2)=\frac{1}{4(\xi_1-\xi_2)^2}\left(-2+ \sqrt{ \frac{(\gamma+\xi_1)(\gamma-\xi_2)}{(\gamma-\xi_1)(\gamma+\xi_2)} }+\sqrt{ \frac{(\gamma-\xi_1)(\gamma+\xi_2)}{(\gamma+\xi_1)(\gamma-\xi_2)} } \right)\eeq
\end{theorem}

Notons que l'écriture en terme des variables $\xi$ n'est pas optimale, puisqu'elle fait intervenir des racines carrées et plus généralement des fonctions multivariées. Cela provient du fait que la courbe spectrale \ref{RescaledCurve} présente elle-même une racine carrée. Ainsi, il est souvent plus agréable de travailler avec une représentation paramétrique de la courbe de la forme $(\xi(z),y_{\text{rescaled}}(z))$ où $z$ est un point courant d'une surface de Riemann et $\xi(z)$ et $y_{\text{rescaled}}(z))$ sont cette fois-ci des fonctions univariées. Dans notre cas, la courbe spectrale \ref{RescaledCurve} est de genre $0$ donc une bonne paramétrisation est d'utiliser la transformation de Joukovski:
\beq \label{Para2} \xi=\frac{\gamma}{2}\left( z+\frac{1}{z}\right)=\frac{\gamma(z^2+1)}{2z}  \Leftrightarrow z=\frac{1+\sqrt{\xi^2-\gamma^2}}{\gamma}\eeq
En plus de ce changement de paramétrisation, il est souvent préférable (et c'est également ces quantités que le formalisme d'Eynard et Orantin calcule) de considérer des formes différentielles plutôt que des fonctions. En effet, les formes différentielles sont des objets plus intrinsèques que les fonctions dans le sens où elles ne dépendent plus du choix de la paramétrisation $\xi(z)$ retenue. Ainsi, on peut définir plus intrinsèquement les formes différentielles:

\begin{definition} Les formes différentielles associées aux résolvantes sont définies par:
\beq \label{formesdiff}\mathcal{W}_n^{(g)}(z_1,\dots,z_n)dz_1\dots dz_n=\om_n^{(g)}(\xi(z_1),\dots,\xi(z_n))d\xi_1\dots d\xi_n +\delta_{n,2}\delta_{g,0} \frac{d\xi(z_1)d\xi(z_2)}{(\xi(z_1)-\xi(z_2))^2} \eeq
\end{definition}

Dans notre cas, on peut alors prouver que :
\begin{theorem}
  Après utilisation de la transformation de Joukovski \ref{Para2}, la 2-forme $\mathcal{W}_2^{(0)}(z_1,z_2)dz_1dz_2$ s'écrit:
\beq \mathcal{W}_2^{(0)}(z_1,z_2)dz_1dz_2=\frac{dz_1dz_2}{(z_2-z_1)^2}\eeq
\end{theorem}

\section{Modèles $(p,q)$ de la théorie conforme couplée à la gravité}

Les modèles minimaux interviennent dans l'étude des représentations du groupe conforme en dimension $2$. Pour $n \geq 2$, le groupe des transformations conforme dans $\mathbb{R}^n$ est le groupe des transformations de $\mathbb{R}^n$ dans lui-même qui conserve les angles. Ce groupe est engendré par l'inversion: $\vec{x}\mapsto \frac{\vec{x}}{||x||}$ ainsi que par le groupe de Poincaré (translations, rotations/boosts, dilatations). Il est isomorphe à $SO(n+1,1)$ et ses représentations peuvent être étudiées, y compris lorsque $\mathbb{R}^n$ est muni d'une métrique non-euclidienne. Dans le cas de la dimension $2$, il est remarquable que l'algèbre de Lie du groupe conforme est de dimension infinie, ce qui n'est plus le cas dans les dimensions supérieures où le groupe conforme présente beaucoup moins d'intérêt. Il est maintenant connu que les représentations irréductibles de charge centrale $c<1$ du groupe conforme en dimension $2$ (dont l'algèbre de Lie est de dimension infinie) peuvent être classifiées par deux paramètres entiers $(p,q)$ dont la charge centrale correspondante \cite{Conformal} est donnée par:
\beq c=1-6\frac{(p-q)^2}{pq}\eeq
L'étude de ces représentations irréductibles est d'une grande importance en physique puisque la symétrie conforme est une symétrie que l'on rencontre dans beaucoup de domaines. On citera ainsi les modèles connus suivants \cite{conffieldtheory}:
\begin{enumerate}
 \item $(1,2)$: Airy ($c=-2$)
 \item $(3,2)$: Modèle de gravité pure ($c=0$)
 \item $(4,3)$: Modèle d'Ising ($c=\frac{1}{2}$)
 \item $(6,5)$: Modèle de Potts à $3$ états ($c=\frac{4}{5}$)
\end{enumerate}

Il existe beaucoup d'approches dans la présentation des modèles minimaux $(p,q)$. En particulier, dans le cadre de cette thèse, il est utile de voir les modèles $(p,q)$ comme une réduction de la hiérarchie Kadamtsev-Petviashvili (KP) des systèmes intégrables. Ainsi, dans \cite{BleherEynard}, les auteurs ont montré qu'une représentation en termes de paire de Lax du modèle $(2m,1)$ est:
\begin{theorem}
 Le modèle de théorie conforme couplée à la gravité $(2m,1)$ peut être mis sous la forme d'une paire de Lax (\cite{BleherEynard}): 

\bea \label{LaxPair}
\frac{1}{N}\frac{\partial}{\partial x} \Psi(x,t)&=& \mathcal{D}(x,t) \Psi(x,t)\cr
\frac{1}{N}\frac{\partial}{\partial t} \Psi(x,t)&=& \mathcal{R}(x,t) \Psi(x,t)
\eea
où $\Psi(x,t)$ est une matrice $2 \times 2$ dont les entrées peuvent être écrites comme:
\beq \Psi(x,t)=\begin{pmatrix}
                \psi(x,t) & \phi(x,t)\\
		\td{\psi}(x,t)& \td{\psi}(x,t)\\
               \end{pmatrix}
\eeq
et satisfaisant la normalisation $\det\left(\Psi(x,t)\right)=1$ (pour que le wronskien du système différentiel soit égal à l'unité).
Les matrices $\mathcal{D}(x,t)$ et $\mathcal{R}(x,t)$ sont données par:
\beq 
\mathcal{R}(x,t)=\begin{pmatrix}
                  0 & x+u(t)\\
		  -x+u(t) &0\\
                 \end{pmatrix}
\eeq
et
\beq \label{tuka}
\mathcal{D}(x,t)=\sum_{k=0}^m t_k \mathcal{D}_k(x,t)
\eeq
avec
\beq
\mathcal{D}_k(x,t)=\begin{pmatrix}
                  -A_k(x,t) & xB_k(x,t)+C_k(x,t)\\
		  xB_k(x,t)-C_k(x,t) &A_k(x,t)\\
                 \end{pmatrix}
\eeq
où $A_k$, $B_k$, $C_k$ sont des polynômes en $x$ de degrés respectifs $2k-2$, $2k-2$ et $2k$.
\end{theorem}

Les paires de Lax sont des blocs fondamentaux des systèmes intégrables (KdV, KP, etc.) puisqu'elles permettent l'intégration explicite des équations aux dérivées partielles auxquelles elles sont reliées. En particulier, on peut montrer que les systèmes différentiels exprimés sous la forme d'une paire de Lax, possèdent une infinité de quantités conservées, ce qui rend l'intégration possible. Dans notre cas, la paire de Lax \ref{LaxPair} permet de fournir une représentation de la hiérarchie de Painlevé II. Dans l'article \textbf{[I]}, présenté à l'annexe \ref{Article[I]}), on montre ainsi à la section 3.2:

\begin{theorem}
 La relation de compatiblité ($\left[\frac{1}{N} \frac{\partial}{\partial x} -\mathcal{D}(x,t), \mathcal{R}(x,t)-\frac{1}{N} \frac{\partial}{\partial t}\right]=0$)
 implique que la fonction inconnue $u(x,t)$ satisfait l'équation des cordes («string equation»):
\beq \label{StringEquation} 
\sum_{k=0}^m t_k \hat{R}_k(u(t))=-tu(t)\eeq
où les $\hat{R}_k(x)$ sont les polynômes de Gelfand-Dikii associés à la hiérarchie de Painlevé II (dont la récurrence est donnée dans \textbf{[I]} dans l'annexe \ref{Article[I]}). En particulier pour $m=1$, la fonction $u(x,t)$ doit satisfaire l'équation de Painlevé II:
\beq \frac{d^2 u}{d t^2}(t)=2u^3(t)+4(t+t_0)u(t)\eeq
\end{theorem}

Notons qu'un travail similaire a été fait dans le cas $(p=2m+1,q=2)$ dans \cite{BergereEynard} à la différence près que la hiérarchie trouvée dans ce cas est celle de KdV et non celle de Painlevé II. Rappelons que les équations de Painlevé, au nombre de six, sont les seules équations différentielles ordinaires du second ordre dont les singularités possèdent la propriété de Painlevé, c'est-à-dire que les singularités mobiles (dépendant des conditions initiales) ne peuvent être que des pôles (pas de singularités essentielles mobiles). Ces équations présentent ainsi des propriétés très particulières de symétrie (transformation de Backlund) et peuvent être exprimées dans un formalisme Hamiltonien.

Pour l'instant, le lecteur peut se demander le rapport entre ces équations intégrables, les modèles de matrices hermitiennes et leurs doubles limites d'échelle. En fait, le c\oe ur de la réponse à cette question est l'existence d'une courbe spectrale naturelle associée à une paire de Lax de type (\ref{LaxPair}), et de façon plus générale à n'importe quelle paire de Lax dans la limite semi-classique. En effet, la présence du facteur $\frac{1}{N}$ dans (\ref{LaxPair}) permet de définir un développement en puissances de $\frac{1}{N}$ pour n'importe quelle quantité (noté ici génériquement $\mathcal{F}(x,t)$) de la forme:
\beq \mathcal{F}(x,t)= \sum_{j=0}^\infty \frac{\mathcal{F}_j(x,t)}{N^{j}}\eeq
On peut alors définir la courbe spectrale associée par:

\begin{definition} La courbe spectrale naturellement associée à une paire de Lax est définie par:
 \beq\label{courbe} \det(\,yId-\mathcal{D}_\infty(x,t))=0\eeq
où $\mathcal{D}_\infty(x,t)$ est donc la limite $N \to +\infty$ de la matrice $\mathcal{D}$ de l'équation \ref{tuka}. 
\end{definition}

Le résultat important est alors que (Cf. \textbf{[I]} Section 3.3 de l'annexe \ref{Article[I]}) :
\begin{theorem}
 La courbe spectrale (\ref{courbe}) coïncide avec la courbe spectrale obtenue lors de la double limite d'échelle (\ref{RescaledCurve}) avec l'identification $\gamma=u_0(t)$.
\end{theorem}

Ce résultat non trivial traduit un lien entre la hiérarchie de Painlevé II et les doubles limites de modèles de matrices aléatoires. Mais ce lien va au delà des simples courbes spectrales, puisque l'on peut définir de façon naturelle dans le contexte des paires de Lax, des fonctions $W_n^{(g)}(x_1,\dots,x_n)$ qui correspondent exactement aux résolvantes $\om_{\text{rescaled},n}^{(g)}(x_1,\dots,x_n)$ associées à la double limite d'échelle (\ref{RescaledCurve} et \ref{rescaling}).
La définition ``naturelle'' de ces fonctions dans le cadre des paires de Lax est la suivante \cite{determinantalformulae}:

\begin{definition} 
Soit le noyau $K(x_1,x_2)$ défini à l'aide des fonctions de Baker-Akhiezer:
\beq \label{K}
K(x_1,x_2)=\frac{\psi(x_1)\td{\phi}(x_2)-\td{\psi}(x_1)\phi(x_2)}{x_1-x_2}\eeq
Les fonctions de corrélations (connexes) sont alors définies par:
\beq W_1(x)=\psi'(x)\td{\phi}(x)-\td{\psi}'(x)\phi(x)\eeq
\beq \label{defcorr} W_n(x_1,\dots,x_n)=-\frac{\delta_{n,2}}{(x_1-x_2)^2}- (-1)^n\sum_{\sigma=cycles} \prod_{i=1}^n K(x_{\sigma(i)},x_{\sigma(i+1)})\eeq
Les fonctions non-connexes correspondantes prennent une forme déterminantale typique des modèles de matrices (\cite{determinantalformulae}): 
\beq \label{defnoncorr}W_{n,n-c}(x_1,\dots,x_n)=\mathop{det}^{'}(K(x_i,x_j))\eeq
où la notation $\mathop{det}^{'}$ signifie que le déterminant doit être calculé comme habituellement par une somme sur les permutations $\sigma$ de produits $(-1)^\sigma\underset{i=1}{\overset{n}{\prod}} K(x_i,K_{\sigma_i})$, à l'exception des termes $i=\sigma(i)$ et $i=\sigma(j) \,, j=\sigma(i)$ où l'on doit remplacer les termes $K(x_i,x_i)$ et $K(x_i,x_j)K(x_j,x_i)$ par respectivement $W_1(x_i)$ et $-W_2(x_i,x_j)$. 
\end{definition}

Le théorème principal de \textbf{[I]} présenté en annexe \ref{Article[I]} établit alors l'égalité des fonctions de corrélation de la double limite d'échelle de modèle de matrice singuliers de type $(2m,1)$ avec les fonctions de corrélations issues de la paire de Lax du modèle de théorie conforme couplée à la gravité $(2m,1)$ qui peut se résumer ainsi (Cf. \textbf{[I]}, Section 3.6) :
\begin{theorem} Lien entre la hiérarchie de Painlevé II et la double limite d'échelle d'un modèle de matrices:
 \beq\forall \, (n,g):\, W_n^{(g)}(x_1,\dots,x_n)=\om_{\text{rescaled,n}}^{(g)}(x_1,\dots,x_n)\eeq
\end{theorem}

\section{Conclusion et perspectives}

Dans ce chapitre, nous avons esquissé le lien profond entre les doubles limites d'échelle de type $(p=2m,q=1)$ et les paires de Lax des systèmes intégrables correspondant au modèle de la théorie conforme $(2m,1)$ à tous les ordres du développement topologique. Afin de rendre ce lien complet, les calculs techniques (et longs) des preuves peuvent être trouvés dans l'article \textbf{[I]} présenté en annexe \ref{Article[I]} de ce mémoire. Dans leur article \cite{BergereEynard}, Bergère et Eynard montrent qu'un lien identique existe pour le cas des modèles $(p=2m+1,q=2)$ aussi bien pour les courbes spectrales que pour les résolvantes, à tous les niveaux du développement topologique. Des résultats similaires sont également vérifiés par Alvarez, Alonso et Medina dans \cite{AlvarezAlonsoMedina} dans le cas de densités limites présentant un support composé de plusieurs segments. Moyennant la généralisation d'un résultat technique (le fait que les fonctions de corrélation définies par \ref{defcorr} satisfassent les équations de boucles du modèle à deux matrices) valable pour l'instant pour $q<3$, il est attendu prochainement que ce type de résultats se généralise pour toutes les valeurs de $(p,q)$ avec les doubles limites d'échelle du modèle hermitien à deux matrices. D'un point de vue physique, il est intéressant de constater que dans la limite de double échelle, qui correspond à une sorte de transition de phase du modèle de matrices, on retombe, après des changements d'échelle, sur des lois universelles (KdV, équations de Painlevé,...). Ces lois universelles sont en analogie avec les exposants critiques universels des transitions de phase pour différents systèmes (modèle d'Ising, percolation, etc.). Même si à l'heure actuelle, l'étendue des phénomènes physiques ou biologiques pouvant être modélisés par des modèles de matrices aléatoires reste inconnue, l'universalité très spécifique de ces régimes limites permettra sans doute d'éclairer dans le futur cette question.

%% file: chap2.tex
\chapter{Modèles matriciels et polynômes orthogonaux}
 \label{chap2}
\thispagestyle{empty}
 \selectlanguage{french}
\section{Introduction des polynômes orthogonaux}

Une méthode très efficace pour calculer la fonction de partition des modèles de matrices hermitiens:
\beq  \label{DiagonalisationGenerale2} Z_N= \int_{\Gamma^N} d\lambda_1 \dots d\lambda_N \,  \Delta(\lambda)^{2} e^{-\frac{N}{T}\underset{i=1}{\overset{N}{\sum}}V(\lambda_i)}\eeq
est d'introduire la base des polynômes orthogonaux moniques $P_n(x)$:

\begin{definition}
Les polynômes orthogonaux sont définis par:
\beq \int_{\Gamma} P_n(x)P_m(x)e^{-\frac{N}{T} V(x)}dx=h_n\delta_{n,m}\eeq
où $P_n(x)$ est un polynôme monique (i.e. $P_n(x)=x^n+\dots$).
\end{definition}

En général, il est aussi pratique d'introduire de véritables fonctions orthonormales pour la mesure de Lebesgue par:

\begin{definition}
Définition des fonctions orthonormales:
\beq \psi_n(x)=\frac{1}{\sqrt{h_n}}p_n(x)e^{-\frac{1}{2}V(x)} \label{DefPsin}\eeq
qui vérifient:
\beq \int_{\Gamma} \psi_n(x)\psi_m(x)dx=\delta_{n,m}\eeq
\end{definition}

Le lien entre les modèles de matrices et les polynômes orthogonaux est alors le suivant: en écrivant le déterminant de Vandermonde comme un déterminant puis en pratiquant des combinaisons linéaires sur les lignes et les colonnes, il est bien connu (\cite{Bessis, Mehta}) que la fonction de partition du modèle à une matrice hermitienne $Z_N$ peut être réécrite comme:
\beq \label{poly} Z_N=N! \prod_{j=0}^{N-1} h_j\eeq
Le choix des matrices hermitiennes n'est pas le seul possible, on pourrait par exemple choisir d'étudier des matrices normales (i.e. commutant avec leur adjoint) dont les valeurs propres seraient imposées sur une certaine courbe $\Gamma$ du plan complexe. Notons dans ce cas que, si le support imposé des valeurs propres part à l'infini, cette direction doit être compatible avec le choix du potentiel $V(x)$ pour que l'intégrale matricielle converge (et que les polynômes orthogonaux existent).

Relier le calcul de la fonction de partition des modèles matriciels hermitiens aux polynômes orthogonaux permet d'appliquer les nombreux résultats connus sur les polynômes orthogonaux aux modèles matriciels (en particulier des cas connus comme les polynômes de Legendre, Laguerre, Hermite, Jacobi et bien d'autres). On citera en particulier les résultats:
\begin{enumerate}
 \item Pour toute suite de polynômes orthogonaux, il existe une relation de récurrence relativement à trois indices consécutifs.
\beq \label{rec3termes} x\psi_n(x)=\gamma_{n+1}\psi_{n+1}+\beta_n\psi_n+\gamma_n \psi_{n-1}\eeq
où $\gamma_n=\sqrt{\frac{h_n}{n_{n-1}}}$ et les coefficients $\beta_n$ dépendent du potentiel de départ.
 \item Les zéros des polynômes orthogonaux sont toujours situés sur le contour d'intégration $\Gamma$ et les racines des polynômes se trouvent strictement entre les racines du polynôme de degré supérieur dans la suite (entrelacement).
\end{enumerate}
En pratique, dès que le potentiel $V(x)$ dépasse le second ordre, le calcul analytique des polynômes orthogonaux devient difficile. Certes la relation de récurrence à trois termes \ref{rec3termes} ramène le problème à la connaissance des coefficients $\beta_n$ et $\gamma_n$ qui peuvent être évalués numériquement. Ainsi, on sait par exemple (\cite{NotesBleher}) que si l'on définit:
\beq Q=\begin{pmatrix}
\beta_0 &\gamma_1& 0&0 &\dots\\
\gamma_1 & \beta_1 &\gamma_2 &0 &\dots\\
0 &\gamma_2 & \beta_2 & \gamma_3 & \dots\\
0 &0 &\gamma_3 &\beta_3 &\dots\\
\vdots & \vdots & \vdots & \vdots& \ddots
\end{pmatrix}
\eeq
alors les coefficients $\beta_n$ et $\gamma_n$ obéissent aux équations:
\bea \gamma_n [V'(Q)]_{n,n-1}&=&\frac{nT}{N}\cr
[V'(Q)]_{n,n}&=&0\cr
\eea
Hélas, si cette derniere formule est compacte, elle donne lieu à des formules extrêmement compliquées lorsque le potentiel $V$ possède un degré élevé, si bien qu'en pratique le calcul des polynômes orthogonaux reste difficile en dehors de quelques cas connus.

\section{Ecriture du problème de Riemann-Hilbert}

L'écriture d'un problème de Riemann-Hilbert nécessite l'introduction des transformées de Cauchy $\td{\psi}_n(x)$ des polynômes orthogonaux précédents:

\begin{definition} Transformée de Cauchy des polynômes orthogonaux:
\beq \td{\psi}_n(x)=e^{\frac{1}{2} V(x)} \int_{\td{\Gamma}} dx \frac{e^{-\frac{1}{2}V(z)}\psi_n(z)}{x-z} \label{CauhyPsin}\eeq
Le contour $\td{\Gamma}$ doit être choisi dual (au sens de \textbf{[II]}) de $\Gamma$ définissant les polynômes orthogonaux.
\end{definition}

Le problème de Riemann-Hilbert, pour un potentiel $V(x)=\underset{i=0}{\overset{d}{\sum}} \frac{t_i}{i} x^i$, peut alors être formulé sur une matrice $2\times 2$. Les polynômes orthogonaux et leurs transformées de Cauchy regroupés sous la forme de la matrice $2 \times 2$:
 
\begin{theorem}
Soit la matrice
\beq \Psi_n(x)=\begin{pmatrix}
                    \psi_{n-1}(x) & \td{\psi}_{n-1}(x)\\
\psi_n(x) & \td{\psi}_n(x)
                   \end{pmatrix}
\eeq
Alors $\Psi_n(x)$ est \textbf{l'unique} solution du problème de Riemann-Hilbert suivant (Cf. \textbf{[II]}, Section 2.2 de l'annexe \ref{Article[II]}):
\begin{enumerate}
\item \underline{Régularité}: $\forall\,  z \notin \Gamma, \, z \mapsto \Psi_n(z)$ est analytique. 
\item \underline{Système différentiel}:
 \beq \frac{d}{dx}\Psi_n(x)=\frac{d}{dx} \begin{pmatrix} 
                    \psi_{n-1} &\td{\psi}_{n-1} \cr
\psi_{n}&\td{\psi}_{n}
                   \end{pmatrix}
= \mathcal{D}_n(x)  \begin{pmatrix} 
                    \psi_{n-1} &\td{\psi}_{n-1} \cr
\psi_{n}&\td{\psi}_{n}
                   \end{pmatrix}\eeq
avec $\mathcal{D}_n(x)=\frac{1}{2} t_dx^{d-1} \sigma_3+O\left(x^{d-2}\right)$ est une matrice polynômiale de degré $d-1$ et $\sigma_3=\text{diag}(1,-1)$
\item \underline{Saut constant sur $\Gamma$}:
\beq \forall\,  x \in \Gamma: \,\, \Psi_n(x)_+=\Psi_n(x)_- \, J_n\eeq
où $J_n$ est une matrice constante (indépendante de $x$)
\item \underline{Asymptotique à l'infini}: 
Si l'on définit la matrice $T_n(x)=(\frac{1}{2}V(x)-n\ln(x))\sigma_3$ alors
\beq \Psi_n(x)\sim C_n \left(Id+ \frac{Y_{n,1}}{x}+\frac{Y_{n,2}}{x^2}+\dots\right)e^{T_n(x)}\eeq
où les $C_n$ et $Y_{n,i}$ sont des matrices constantes (indépendantes de $x$)
\end{enumerate}
\end{theorem}

Le problème de Riemann-Hilbert n'apporte pas en soi de nouvelles informations par rapport au calcul des polynômes orthogonaux. En revanche, il en constitue une reformulation pratique et standard sur laquelle beaucoup de méthodes sont applicables, en particulier la méthode de diffusion inverse («inverse scattering method») introduite par Ablowitz et Segur permettant d'obtenir des asymptotiques exacts des polynômes orthogonaux et des fonctions de partition (Cf. \cite{NotesBleher}). L'écriture d'un modèle hermitien à une matrice est un élément intéressant mais qui peut s'insérer plus généralement dans l'écriture d'un problème de Riemann-Hilbert pour un modèle à deux matrices hermitiennes. C'est dans ce contexte plus général, mais aussi plus technique, que j'ai réalisé mon travail (\textbf{[II]} présenté en annexe \ref{Article[II]}) avec Marco Bertola sur les fonctions tau-isomonodromiques. Avant de parler de ces fonctions tau qui sont un élément majeur dans la théorie de l'intégrabilité, il est préférable de définir et d'énoncer quelques propriétés du modèle à deux matrices.

\section{Modèle à deux matrices hermitiennes et problème de Riemann-Hilbert associé}

Le modèle à deux matrices hermitiennes se caractérise par la fonction de partition suivante:
\beq Z_{2MM}=\int\int_{E_N} dM_1 dM_2 e^{-\Tr(V_1(M_1)+V_2(M_2)-M_1M_2)}  \label{2MM}\eeq
où $V_1(x)$ et $V_2(y)$ sont deux potentiels polynômiaux et $M_1$ et $M_2$ sont des matrices hermitiennes (ou normales) :
\beq V_1(x)=\sum_{j=1}^{d_1+1} \frac{u_j}{j}x^j \, \, , \, \, V_2(y)=\sum_{j=1}^{d_2+1} \frac{v_j}{j}y^j \eeq
Ce problème peut également être rapporté à un problème aux valeurs propres moyennant l'utilisation de l'intégrale d'Itzykson-Zuber-Harish-Chandra (\cite{HarishChandra}, \cite{ItzyksonZuber}). Il se ramène alors à:
\beq Z\sim  \int\int_{\kappa}\left(\prod_{j=1}^N dx_jdy_j\right) \Delta^2(X)\Delta^2(Y)e^{-\underset{j=1}{\overset{N}{\sum}} V_1(x_j)+V_2(y_j)-x_jy_j}I(x_1,\dots,x_N,y_1,\dots,y_N)  \label{2MMvp}\eeq
où $I(x_1,\dots,x_N,y_1,\dots,y_N)=I(X,Y)$ est l'intégrale d'Itzykson-Zuber définie par:
\beq \label{IZhermitien} I(X,Y)=\int_{\mathcal{U}_N} dU e^{\frac{N}{T}\Tr(XUYU^{-1})}=\left(\frac{N}{T}\right)^{-\frac{N(N-1)}{2}}\left(\prod_{p=1}^{N-1}p!\right)\frac{det(e^{\frac{N}{T}X_{i}Y_{j}})_{i,j}}{\Delta(X)\Delta(Y)}\eeq

Le contour d'intégration des valeurs propres noté génériquement $\int\int_\kappa$ signifie en fait n'importe quelle combinaison linéaire de chemins admissibles (au sens où l'intégrale converge):
\beq \int\int_\kappa=\sum_{i,j} \kappa_{i,j} \int_{\Gamma_i} dx_i \int_{\hat{\Gamma}_j} dy_j\eeq
Comme dans le cas à une matrice, on peut alors introduire des polynômes bi-orthogonaux pour écrire cette intégrale. Ces polynômes moniques $\pi_n(x)$ et $\sigma_m(y)$ de degrés respectifs $n$ et $m$, sont définis par \cite{Bessis, Mehta1}:
\begin{definition} Les polynômes bi-orthogonaux sont définis par:
\beq \int\int_{\kappa} \pi_n(x)\sigma_m(y)e^{-V_1(x)-V_2(y)+xy} dx dy=h_n \delta_{n,m}\label{ortho2M} \eeq
\end{definition}
Notons que cette fois-ci, les polynômes sont orthogonaux pour une ``double intégration'' ce qui rend leur calcul numérique encore plus délicat que pour les polynômes orthogonaux. La principale différence avec le cas des modèles à une matrice est que les matrices ne vont plus être de taille $2\times 2$, mais vont être de taille $d_1 \times d_1$ ou $d_2 \times d_2$. En cela, le cas à deux matrices est plus difficile d'un point de vue technique, mais la majeure partie des résultats du cas à une matrice s'étend pour le cas à deux matrices.
 
\begin{definition}
Les pseudo-polynômes orthonormaux sont définis par:
$$\psi_n(x)=\frac{1}{\sqrt{h_n}}\pi_n(x)e^{-V_1(x)} \, \, , \, \,  \phi_m(y)=\frac{1}{\sqrt{h_m}}\sigma_m(y)e^{-V_2(y)}$$
On définit également les vecteurs:
\beq
\Psi_N(x)=\begin{pmatrix} 
        \psi_{N-d_2}(x)\\
\vdots \\
\psi_N(x)
       \end{pmatrix}
\,\, , \,\, 
\Phi_N(x)=\begin{pmatrix} 
        \phi_{N-d_1}(x)\\
\vdots \\
\phi_N(x)
       \end{pmatrix}
\eeq
\end{definition}

Alors on a les résultats suivants (Cf. \textbf{[II]}, Section 3.2 de l'annexe \ref{Article[II]}):
\begin{theorem}
\beq \Psi_{N+1}(x)= \begin{pmatrix}
             0&1&0&0\\
0&0&\ddots &0\\
0&0&0&1\\
\frac{-\alpha_{d_2}(N)}{\gamma(N)} &\dots&\frac{-\alpha_{1}(N)}{\gamma(N)} & \frac{x-\alpha_{0}(N))}{\gamma(N)}
            \end{pmatrix} \Psi_N(x)\eeq
\beq 
\Phi_{N+1}(y)= \begin{pmatrix}
             0&1&0&0\\
0&0&\ddots &0\\
0&0&0&1\\
\frac{-\beta_{d_1}(N)}{\gamma(N)} &\dots&\frac{-\beta_{1}(N)}{\gamma(N)} & \frac{y-\beta_{0}(N))}{\gamma(N)}
            \end{pmatrix} \Phi_N(y)
\eeq
où, par définition, les nombres $\alpha_i(N)$, $\beta_i(N)$ et $\gamma(N)$ définissent les coefficients de la récurrence à $d_1$ et $d_2$ termes satisfaite par les $\psi_n(x)$ et les $\phi_m(y)$:
\bea x\psi_n(x)&=&\sum_{i=0}^{d_2} \alpha_i(n)\psi_{n-i}(x) +\gamma(n)\psi_{n+1}(x)\cr
y\phi_n(x)&=&\sum_{i=0}^{d_1} \beta_i(n)\phi_{n-i}(x) +\gamma(n)\phi_{n+1}(x) \label{relationrecurrence}
\eea
\end{theorem}

Il est également possible d'obtenir une relation matricielle reliant les dérivées:
\beq \frac{d}{dx} \Psi_N(x) =\mathcal{D}_N(x) \Psi_N(x) \,\, , \,\, \frac{d}{dx} \Phi_N(y) =\td{\mathcal{D}}_N(y) \Phi_N(y)\eeq
où les matrices $\mathcal{D}_N(x)$ et $\td{\mathcal{D}}_N(y)$ peuvent s'exprimer en fonction des coefficients $\alpha_i(N)$, $\beta_i(N)$ et $\gamma(N)$. Nous renvoyons à \cite{BEH} pour une formule exacte. Comme pour le cas à une matrice, il est également possible de trouver un problème de Riemann-Hilbert satisfait par les polynômes bi-orthogonaux. Pour cela, il est nécessaire de définir l'équivalent de la transformée de Cauchy et d'obtenir des matrices de taille $d_1$ ou $d_2$. Ces résultats ont été établis dans \cite{KMcL} et peuvent se résumer de la manière suivante.

\begin{theorem}
Définissons:
\beq \mathcal{C}_i[f](x)=\frac{1}{2i\pi} \int\int_{\hat{\kappa}} dz dy \frac{f(z)}{z-x} y^j e^{- V_1(z)-V_2(y)+zy}\eeq
alors la matrice 
\beq \Gamma_N(x)=\begin{pmatrix}
        \pi_N(x) & \mathcal{C}_0[\pi_N](x) & \dots &\mathcal{C}_{d_2-1}[\pi_N](x)\\ \\
\pi_{N-1}(x) & \mathcal{C}_0[\pi_{N-1}](x) & \dots &\mathcal{C}_{d_2-1}[\pi_{N-1}](x)\\
\vdots &\vdots &\vdots&\vdots\\
\pi_{N-d_2}(x) & \mathcal{C}_0[\pi_{N-d_2}](x) & \dots &\mathcal{C}_{d_2-1}[\pi_{N-d_2}](x)\\
       \end{pmatrix}
\eeq
satisfait le problème de Riemann-Hilbert suivant (résultat généralisé de \cite{KMcL} qui se limite à des polynômes de degré $4$ ):
\begin{enumerate}
 \item 
$z \mapsto \Gamma_N(z)$ est analytique sur $\mathbb{C}$ sauf sur les contours $\Gamma_j$ (introduits au départ dans la définition des polynômes orthogonaux \ref{ortho2M}) où elle présente un saut:
\beq 
\Gamma_{N,+}(z)=\Gamma_{N,-}(z)\begin{pmatrix}
                        1 &w_{j,1} &\dots& w_{j,d_2}\\
0&1 &0 &0\\
\vdots &\ddots& & \vdots\\
0 & \dots &0 & 1
                       \end{pmatrix}
\label{RH2}
\eeq
où
$$w_{j,\nu}=e^{-V_1(x)}\sum_{k=1}^{d_2}\kappa_{j,k} \int_{\hat{\Gamma}_k} dy \,y^{\nu-1} e^{-V_2(y)+xy}$$
\item
Son asymptotique à l'infini est donné par:
\beq \Gamma_N(x)\sim \left(Id+\frac{Y_{1,N}}{x} +O\left(\frac{1}{x^2}\right)\right) \begin{pmatrix}
                                                                                   x^N&0&0\\
0& x^{-m_N-1} Id_{r_N}&0\\
0&0& x^{-m_N} Id_{d_2-r_N}
                                                                                  \end{pmatrix}\eeq
où l'on a défini $m_N$ et $r_N$ comme respectivement le quotient et le reste de la division Euclidienne de $N$ par $d_2$:
$$N=m_N d_2+r_N$$
\end{enumerate}
\end{theorem}

En général, pour utiliser les techniques connues des problèmes de Riemann-Hilbert, en particulier les techniques d'isomonodromies vers lesquelles on se destine, il est préférable d'avoir un problème de Riemann-Hilbert dans lequel les sauts sont constants. Dans notre cas, cela peut être fait en multipliant à droite par une matrice bien choisie. L'inconvénient est alors que l'asymptotique à l'infini devient beaucoup plus complexe. On trouve (les détails sont dans \textbf{[II]}, Section 2.2 de l'annexe \ref{Article[II]})

\begin{theorem} Soit:
$\forall \, 1\leq k \leq d_2$:
\beq
 \psi_m^{(k)}(x) \mathop{=}^{\text{def}} \frac{1}{2\pi i}\int_{\Gamma_k}ds \int \!\!\!\int_{\kappa}  dz dw 
\frac{\pi_m(z)e^{-V_1(z)}}{x-z}\frac{V_2'(s)-V_2'(w)}{s-w}  e^{-V_2(w)+V_2(s) +zw - xs }, \quad 1\leq k \leq d_2,
\label{psikmdef}
\eeq
et
\beq
\psi^{(0)}_m(x)\mathop{=}^{\text{def}} \pi_m(x)e^{-V_1(x)}
\eeq

ainsi que la matrice de taille $(d_2 + 1) \times (d_2 +1)$ $\PsiN(x)$ ($N \ge d_2$):
\beq
\PsiN(x):= \left[
\begin{array}{ccc}
\psi_N^{(0)}(x) & \dots& \psi_N^{(d_2)}(x)\\
\vdots &&\vdots\\
\psi_{N-d_2}^{(0)}(x) & \dots& \psi_{N-d_2}^{(d_2)}(x)
\end{array}\right]
\label{hatpsi}
\eeq

Le problème de Riemann-Hilbert satisfait par cette matrice est alors le suivant:

\label{24}
La matrice $\PsiN$ est l'unique solution du problème de Riemann-Hilbert:
\begin{enumerate} 
\item  \underline{Régularité}
$z \mapsto \PsiN(z)$ est analytique sur $\mathbb{C}$ sauf sur les contours $\Gamma_{j}$
\item \underline{Sauts constants}: \beq
\PsiN{}_{+}(x) \,= \PsiN{}_{-} (x)\Hb^{(j)}
\eeq avec 
\beq \Hb^{(j)} \, :=  \Ib  - 2\pi i \eb_0  \kappab^T \,\,\,\, ,\,\,\,\,
 \eb_0\, := \begin{pmatrix}1 \\ 0 \\ \vdots \\  0 \end{pmatrix}
\,\,\,\, ,\,\,\,\, \kappab:= \begin{pmatrix} 0 \\ \kappa_{j 1} \\ \vdots \\ \kappa_{j d_2}\end{pmatrix}
\eeq
\item \underline{Asymptotique à l'infini}:
\beq \PsiN(x)\sim\Gamma_N \begin{pmatrix}
x^{N}e^{-V_1(x)} & 0 &  0\\
0 & x^{-m_N-1}Id_{r_N}& 0\\
0&0& x^{-m_N}Id_{d_2-r_N} \\
\end{pmatrix}  \Psi_0(x) 
\eeq
où
\beq \label{eee} {\Gamma_N}=Id+\frac{{Y}_{N,1}}{x}+...\eeq 
et où $\Psi_0(x)$ est la solution ``nue'' dont l'asymptotique à l'infini peut être calculée par la méthode de «steepest descent». 
\item $\Psi_{N}'(x)\Psi_N^{-1}=D_N(x)$ où $D_N(x)$ est polynômiale en $x$
\item $\partial_{u_K}\Psi_{N}(x)\Psi_N^{-1}=U_{K,N}(x)$ est polynômiale en $x$.
\item $\partial_{v_J}\Psi_{N}(x)\Psi_N^{-1}=V_{J,N}(x)$ est polynômiale en $x$.
\item $\det(\Psi_{N+1}\Psi_N^{-1})=Cste$
\end{enumerate}
\end{theorem}

L'intérêt de pouvoir réécrire le problème sous la forme d'un problème de Riemann-Hilbert avec des sauts constants est qu'il permet de faire le lien avec la théorie des isomonodromies développées par Jimbo-Miwa-Ueno dans leur série d'articles \cite{JMI} \cite{JMII} \cite{JMIII}.

\section{Les transformations isomonodromiques}

\subsection{Systèmes Fuchsiens et équations de Schlesinger}

Considérons un système d'équations différentielles du type:
\beq \label{Fuch}\frac{dY}{dx}=AY=\sum_{i=1}^{n}\frac{A_i}{x-\lambda_i}Y \eeq
où $x \in \mathbb{C}$ et les $A_i$ sont des matrices $n \times n$ indépendantes de $x$. Les points $\lambda_i$ peuvent être vus comme des pôles de l'équation différentielle dont les matrices $A_i$ correspondantes seraient les résidus. Si l'on considère $Y$ une solution de \ref{Fuch}, alors on peut produire d'autres solutions en partant d'un point de base $b$ et en faisant le prolongement analytique des solutions le long d'une courbe qui entoure un des pôles $\lambda_i$. De cette façon, lorsque l'on est retourné au point de base $b$ après ce cheminement, on obtient une nouvelle solution $Y'$ différente de $Y$. Ces deux solutions sont reliées par une matrice de monodromies $M_i$:
   \beq Y'=YM_i \eeq
Ainsi on peut établir un morphisme entre le groupe fondamental de $\mathbb{C}\mathbb{P}\setminus \{\lambda_1\dots \lambda_n\}$ (i.e. les lacets entourant les $\lambda_i$) et le groupe des matrices inversibles $GL_n(\mathbb{C})$ (les matrices de monodromies $M_i$). Il est clair que cette construction dépend du point de base $b$ d'où partent et où arrivent les lacets. Un changement de point de base correspond pour les matrices de monodromies à une conjugaison globale par une matrice fixée (caractérisant le changement de point de base). Si l'on veut donc s'affranchir du choix du point de base on ne s'intéressera donc aux matrices de monodromies qu'à une conjugaison globale près.
La question légitime qui vient alors à l'esprit est de se demander ce que déterminent exactement ces matrices de monodromies. En particulier, la connaissance des matrices de monodromies suffit-elle à caractériser entièrement le système différentiel \ref{Fuch}? La réponse à cette question est négative: il existe des systèmes Fuchsiens qui admettent des matrices de monodromies identiques et d'une façon générale un jeu de matrices de monodromies fixé correspond à plusieurs systèmes Fuchsiens. Notons ici que l'on ne tient pas compte des reformulations possibles d'un même système différentiel par simple changement de coordonnées (qui ne changera donc pas les monodromies), on se placera donc dans le cas où $A$ et: 
\beq  g^{-1}(x)Ag(x)-g^{-1}(x)\frac{dg(x)}{dx} \eeq
sont considérées équivalents pour toute transformation $\td{x}=g(x)$ de coordonnées.
Une autre question naturelle qui vient à l'esprit est de savoir si pour des matrices de monodromies données, il existe toujours un système fuchsien \ref{Fuch} qui redonne ces matrices. Il est connu depuis Plemelj que, sauf certains cas dégénérés dans lequel la réponse est négative, la réponse à cette question est affirmative résolvant ainsi le vingt-et-unième problème de Hilbert.

\subsection{Les transformations isomonodromiques}

Si pour des matrices de monodromies données il existe en général beaucoup de systèmes Fuchsiens correspondants, on peut alors se demander quelques types de transformations ``isomonodromiques'' permettent de connecter ces différents systèmes. Si l'on suppose que les matrices $A_i$ dépendent de la position des pôles $\lambda_j$, il a été montré en 1912 par Schlesinger que dans le cas générique, les transformations isomonodromiques (i.e. ne changeant pas les matrices de monodromies) doivent satisfaire les équations d'holonomies intégrables connues maintenant sous le nom d'équations de Schlesinger \cite{blouh}:
\bea \frac{\partial A_i}{\partial \lambda_j} &=& \frac{[A_i,A_j]}{\lambda_i-\lambda_j} \qquad \qquad j\neq i \cr
 \frac{\partial A_i}{\partial \lambda_i} &=& -\sum_{j\neq i}\frac{[A_i,A_j]}{\lambda_i-\lambda_j} \eea

Notons que ces équations peuvent être interprétées comme des équations de courbure nulle sur l'espace des paramètres de déformations $\lambda_j$.

\subsection{Les singularités d'ordre supérieur, la contribution de l'école japonaise} 

Un des buts de l'école japonaise de Jimbo-Miwa-Ueno a été de généraliser les résultats précédents dans le cadre de singularités d'ordre supérieur \cite{JMI,JMII,JMIII}:
 \beq \label{Fuchsger} \frac{dY}{dx}=AY=\sum_{i=1}^{n}\sum_{j=1}^{r_i+1}\frac{A^{(i)}_j}{(x-\lambda_i)^j}Y \eeq 
où les matrices $A_j^{(i)}$ sont indépendantes de $x$. Cette fois-ci les données de monodromies sont plus délicates à définir. En effet, en plus des matrices de monodromies, il faut cette fois-ci ajouter des matrices de Stokes reliant des solutions entre deux secteurs de Stokes d'un même pôle. Enfin, il faut également rajouter des matrices de connexion reliant les solutions canoniques de différents secteurs de différents pôles. Les solutions canoniques sont définies grâce à un théorème de Birkhoff. En effet, une solution simple mais purement formelle consiste à résoudre terme à terme en $x_i=x-\lambda_i$, les équations donnant la connexion $g_i$ ($T_j^{(i)}$ et $M^{(i)}$ sont diagonales) :
    \beq \frac{d(g_i^{-1}Z_i)}{dx_i} = \left(\sum_{j=1}^{r_i} \frac{(-j)T^{(i)}_j}{x_i^{j+1}}+\frac{M^{(i)}}{x_i}\right)(g_i^{-1}Z_i) \eeq
qui donneraient alors localement:
    \beq Z_i = g_i \exp\left(M^{(i)} \log(x_i)+\sum_{j=1}^{r_i}\frac{T^{(i)}_j}{x_i^{j}}\right) \eeq 
Malheureusement, la résolution terme à terme en puissances de $x_i$ donne en général lieu à une série divergente. Cependant, le théorème de Birkhoff assure l'existence d'une unique solution convergente $G_i$ qui est asymptotiquement équivalente à $g_i$ dans un secteur du pôle $\lambda_i$. Dès lors, la solution:
  \beq  Z_i = G_i \exp\left(M^{(i)} \log(x_i)+\sum_{j=1}^{r_i}\frac{T^{(i)}_j}{x_i^{j}}\right) \eeq 
est bien définie dans un des secteurs du pôle $\lambda_i$ et constitue une solution canonique de \ref{Fuchsger} dans ce secteur. Les données de monodromies consistent alors en des matrices reliant les différents secteurs d'un même pôle (matrices de Stokes) ou entre différents pôles (matrices de monodromies ou matrices de connexion).

Les transformations isomonodromiques peuvent alors être définies commes les transformations préservant les données de monodromies. Si l'on s'autorise à varier les quantités suivant la position des pôles $\lambda_i$ et suivant les résidus diagonaux $T^{(i)}_j$ alors les transformations isomonodromiques d'un système (\ref{Fuchsger}) caractérisé par $A$ doivent satisfaire l'équation:
   \beq \label{rr} dA + [\Omega,A] + \frac{d\Omega}{dx} = 0 \eeq
où la $1$-forme $\Omega$ est définie par:
    \beq \label{forme}\Omega = \sum_{i=1}^{n}\left(A d\lambda_i - g_i D \left( \sum_{j=1}^{r_i}T^{(i)}_j \right)g_i^{-1} \right)  \eeq
et $D$ représente la dérivation extérieure sur les $T^{(i)}_j$. Notons à nouveau que \ref{rr} possède l'interprétation géométrique d'une courbure nulle. On peut alors prouver que \ref{rr} permet de montrer que \ref{forme} est une forme fermée donc localement exacte, c'est-à-dire que l'on peut définir une fonction $\tau$-isomonodromique (à une constante multiplicative près) par:
\beq \Omega=d (\ln \tau )\eeq 

\subsection{Propriétés d'intégrabilité des transformations isomonodromiques}

Une des propriétés les plus intéressantes des transformations isomonodromiques (demontrée par Malgrange dans le cas Fuchsien et par Miwa dans le cas général) est que toutes les singularités essentielles des solutions sont fixées, bien que la position des pôles $\lambda_j$ puisse bouger. En d'autres termes, cela veut dire que les solutions satisfont automatiquement la propriété de Painlevé (singularités essentielles fixées) signifiant que l'on retrouve l'aspect des systèmes intégrables.

\section{Fonctions de partition des modèles matriciels et fonction tau}

La principale nouveauté apportée dans l'article \textbf{[II]}, Section 3 de l'annexe \ref{Article[II]} a été de généraliser la définition d'une fonction $\tau$-isomonodromique dans le cadre dégénéré du problème de Riemann-Hilbert \ref{24}. En effet dans ce cas la matrice:
\beq S=\begin{pmatrix} \label{tabarnak}
x^{N}e^{-V_1(x)} & 0 &  0\\
0 & x^{-m_N-1}Id_{r_N}& 0\\
0&0& x^{-m_N}Id_{d_2-r_N} \\
\end{pmatrix}\eeq
possède des valeurs propres dégénérées, et la théorie générique de Jimbo-Miwa-Ueno, n'est donc a priori plus valide. Néanmoins, en définissant convenablement une $1$-forme similairement à \ref{forme}, on peut prouver que cette nouvelle $1$-forme reste fermée et définie donc également une fonction $\tau$-isomonodromique. En particulier dans \textbf{[II]} (annexe \ref{Article[II]}) on montre que la bonne définition de la fonction tau s'inscrit dans le cadre très général suivant:

Soit une matrice $ \Psi(x)$ vérifiant l'asymptotique:
\beq
 \Psi(x) \sim  Y(x)\,  \Xi(x)\ ,
 \ \ Y(x):= \left(\mathbf 1 + \frac{Y_1}x + \frac {Y_2}{x^2}+ \dots \right) x^S
 \eeq 
 où $\Xi(x) = \Xi(x; \mathbf t)$ est une expression explicite supposée connue et $S$ est une matrice diagonale indépendante des temps d'isomonodromies et dont les valeurs propres peuvent être multiples (et qui dans l'application aux modèles de matrice sera donnée par \ref{tabarnak}). Cela implique en particulier que si l'on définit la 1-forme matricielle $\mathcal{H}(x; \mathbf t)$ par: 
  \beq
 \mathcal{H}(x; \mathbf t) = d \Xi(x; \mathbf t)\, \Xi(x;\mathbf t)^{-1}
 \eeq
alors $\mathcal{H}(x) = \sum \mathcal{H}_a d t_a$ est solution d'équation de courbure nulle:
\beq
\pa_a \mathcal{H}_b - \pa _b \mathcal{H}_a = [\mathcal{H}_a,\mathcal{H}_b]
\label{barezcc}
\eeq
Dans ce contexte général, la fonction tau s'obtient alors par la formule:

\begin{definition}
La fonction tau est la 1-forme définie par:
\beq
\omega:= \sum \omega_a d t^a:=\sum_a \res{} \tr \left(Y^{-1}Y' \mathcal{H}_a \right) d t^a
\eeq
\end{definition}

 Les résultats principaux de l'article ont consisté en la démonstration de la fermeture de la 1-forme et de l'obtention de l'égalité entre cette fonction tau et la fonction de partition du modèle de matrices. En particulier une partie importante de la démonstration a consisté à utiliser des transformations de Schlesinger discrètes sur le paramètre $N$ (taille des matrices) pour expliciter le rapport $\frac{\tau_{N+1}}{\tau_N}$ en fonction d'un des coefficients de $Y_{N,1}$ (Cf. équation \ref{eee}). Une fois cela établi, la définition même de ce coefficient ainsi que les propriétés d'orthogonalité des polynômes bi-orthogonaux permettent de constater que ce même coefficient est en fait le rapport $\frac{Z_{N+1}}{Z_N}$. Finalement, l'étude pour $N=1$ permet de montrer le théorème suivant (Cf. \textbf{[II]}, Section 3.2 de l'annexe \ref{Article[II]}):

\begin{theorem} La fonction de partition du modèle à deux matrices et la fonction tau-isomonodromique vérifie l'égalité (\textbf{[II]} présenté en annexe \ref{Article[II]}):
\beq \forall N \in \mathbb{N} \,: \, Z_N = (v_{d_2+1})^{\frac{d_2\alpha_N(\alpha_N+1)}{2}+\alpha_N(N-d_2\alpha_N)} \, \tau_N\eeq
avec $\alpha_N=E\left(\frac{N}{d_2}\right)$ (l'égalité étant entendue à une constante multiplicative près indépendante de $N$ et des potentiels. $E(x)$ désigant la partie entière de $x$).
\end{theorem}

En d'autres termes, à l'exception d'une puissance multiplicative en $v_{d_2+1}$ (qui provient d'une mauvaise normalisation de la fonction $\tau$) \textbf{on retrouve, après avoir généralisé la définition de Jimbo-Miwa-Ueno à un cas où l'asymptotique est dégénéré, le fait que la fonction de partition du modèle à deux matrices hermitiennes est une fonction $\tau$-isomonodromique.} Ce résultat était déjà connu pour le cas à une matrice hermitienne après le travail de M.Bertola, B.Eynard et J. Harnad \cite{Harnad} et permet de renforcer le lien profond entre les modèles de matrices aléatoires (au moins ceux où les matrices sont hermitiennes) et la théorie de l'intégrabilité. Nous renvoyons le lecteur intéressé par la démonstration complète à l'article \textbf{[II]} situé en annexe \ref{Article[II]}.

\section{Cas des modèles non-hermitiens}

Après avoir vu que les fonctions de partition des modèles à une puis deux matrices hermitiennes donnent lieu à des fonctions $\tau$-isomonodromiques, il est naturel de se demander si le résultat s'étend à d'autres ensembles de matrices non-hermitiennes. La réponse à cette question n'est pas connue à l'heure actuelle. En effet, un ingrédient crucial dès le départ a été de réduire le calcul de la fonction de partition à une intégrale sur les valeurs propres puis à un problème de polynômes (bi)-orthogonaux grâce aux propriétés du déterminant de Vandermonde (\ref{poly}). Mais, dans le cas où la puissance du déterminant de Vandermonde n'est pas deux, l'usage des polynômes (bi)-orthogonaux n'est plus possible (pour le cas symétrique réel et quaternionique self-dual il est néanmoins possible de définir des polynômes skew-orthogonaux \cite{Mehta}). Dans le cas où l'exposant est $\beta$-quelconque, des liens avec les polynômes de Jack ou de McDonald peuvent être espérés, mais les propriétés de ces polynômes restent quasiment inconnues et aucune reformulation en termes de problème de Riemann-Hilbert n'est connue à ce jour. Sans cette reformulation, la question de la définition d'une fonction $\tau$ et de son éventuel lien avec la fonction de partition reste sans objet. En revanche, comme nous le verrons par la suite, d'autres méthodes, en particulier la méthode des équations de boucles, se généralisent plus directement à des matrices non-hermitiennes où la puissance du déterminant de Vandermonde est arbitraire.

%% file: chap3.tex
\chapter{Modèles de matrices pour $\beta$-arbitraire}
 \label{chap3}
\thispagestyle{empty}
 \selectlanguage{french}
\section{Généralisation des modèles matriciels aux ``ensembles $\beta$''}

Jusqu'à présent nous avons considéré des modèles de matrices hermitiennes, c'est-à-dire que l'intégrale était définie sur des matrices hermitiennes. Cependant, comme nous l'avons mentionné auparavant, d'autres ensembles de matrices peuvent avoir un intérêt physique, comme les matrices symétriques réelles ou les matrices quaternioniques self-duales. Dans ces deux cas précis, la diagonalisation du problème en un problème aux valeurs propres est connue et est donnée par:
\beq Z_N\propto \int_{\mathbb{C}^N} d\lambda_1 \dots d\lambda_N \,  \Delta(\lambda)^{2\beta} e^{-\frac{N}{T}\underset{i=1}{\overset{N}{\sum}} V(\lambda_i)}\eeq
où la valeur du paramètre $\beta$ vaut $1$, $1/2$ ou $2$ selon l'ensemble de matrice choisie (Cf. \ref{Diagonalisation}). Dès lors, il est intéressant de se demander si l'on ne pourrait pas étudier directement ces trois ensembles de matrices en conservant ce paramètre $\beta$ arbitraire, afin de traiter les trois modèles d'un seul coup. Cette approche constitue ce que l'on appelle dans la littérature \textit{le modèle à une matrice avec $\beta$-quelconque}. La généralisation au modèle à deux matrices est plus délicate. En effet, pour le cas hermitien $\beta=1$, la diagonalisation en un problème aux valeurs propres fait intervenir l'intégrale d'Itzykson-Zuber (\ref{IZhermitien}) sur le groupe unitaire, spécifique aux matrices hermitiennes. Il est donc nécessaire de trouver une généralisation naturelle de cette intégration. La généralisation pour $\beta$-quelconque du modèle à deux matrices est donnée par:
\beq \label{Diagonalized2betaMM} Z_{\beta}\mathop{=}^{def}\int dX dY \Delta(X)^{2\beta} \Delta(Y)^{2\beta} e^{-\frac{N\beta}{T}\left[\underset{i=1}{\overset{N}{\sum}}V_1(x_i)+\underset{j=1}{\overset{N}{\sum}}V_2(y_j)\right]}I_\beta(X,Y)\eeq
où $I_\beta(X,Y)$ est la généralisation ``naturelle'' de l'intégrale d'Itzykson-Zuber que l'on va maintenant décrire plus en détail.

\section{L'intégrale d'Itzykson-Zuber généralisée}

Dans le cas hermitien, l'intégrale d'Itzykson-Zuber est définie par:
\beq  I_\text{herm}(X,Y)=\int_{\mathcal{U}_N} dU e^{\frac{N}{T}\Tr(XUYU^{-1})}\eeq
où $\mathcal{U}_N$ est le groupe unitaire équipé de la mesure de Haar.
Il est également intéressant de définir les quantités suivantes:
\beq M_{i,j}=\int_{\mathcal{U}_N} dU \| U_{i,j} \|^2 e^{\frac{N}{T}\Tr(XUYU^{-1})}\eeq
qui peuvent être utilisées pour déterminer l'intégrale d'Itzykson-Zuber par la formule:
\beq \label{OO}I_\text{herm}(X,Y)=\sum_{i=1}^N M_{i,j}=\sum_{j=1}^N M_{i,j}\eeq
Cette dernière formule est évidente puisque $\underset{i=1}{\overset{N}{\sum}}\| U_{i,j} \|^2=1=\underset{j=1}{\overset{N}{\sum}}\| U_{i,j} \|^2$ sur le groupe unitaire.
Dans le cas hermitien, les $M_{i,j}$ sont connus pour satisfaire l'équation de Dunkl \cite{Angular}:
\beq \label{CalogeroMoser} 
\forall\,  1\leq i,j\leq N\, :\, \frac{\partial}{\partial x_i} M_{i,j} + \sum_{k\neq i} \frac{M_{i,j}-M_{k,j}}{x_i-x_k}=y_j M_{i,j}
\eeq
qui sera à la base de la généralisation au cas où $\beta$ est quelconque. Dans \cite{Angular}, les auteurs montrent que l'on peut généraliser les $M_{i,j}$ au cas où $\beta$ est arbitraire (que l'on notera $M_{i,j}^{(\beta)}$) par les conditions suivantes:
\begin{definition}
Les $M_{i,j}^{(\beta)}$ sont définis par les propriétés suivantes:
\begin{enumerate}
 \item Les $M_{i,j}^{(\beta)}$ satisfont l'équation de Calogero-Moser-Dunkl généralisée:
\beq \label{GeneralizedCM} \forall\,  1\leq i,j\leq N\, :\, \frac{\partial}{\partial x_i} M_{i,j}^{(\beta)}+ \beta\sum_{k\neq i} \frac{M_{i,j}^{(\beta)}-M_{k,j}^{(\beta)}}{x_i-x_k}=\frac{N\beta}{T} y_j M_{i,j}^{(\beta)}
\eeq (Le facteur $\frac{N\beta}{T}$, absent de \cite{Angular}, provient de la présence dans l'exponentielle du préfacteur $\frac{N\beta}{T}$ qui peut être absorbé par le changement de variables $Y \leftrightarrow \frac{N\beta}{T}Y$).
\item $M_{i,j}^{(\beta)}$ doivent être des matrices stochastiques: $I_\beta=\underset{i=1}{\overset{N}{\sum}} M_{i,j}^{(\beta)}$ doit être indépendant de $j$ et $I_\beta=\underset{j=1}{\overset{N}{\sum}} M_{i,j}^{(\beta)}$ doit être indépendant de $i$.
\item La fonction $(X,Y) \mapsto I_\beta(X,Y)=\underset{i=1}{\overset{N}{\sum}} M_{i,j}^{(\beta)}(X,Y)=\underset{j=1}{\overset{N}{\sum}} M_{i,j}^{(\beta)}(X,Y)$ doit être une fonction symétrique de ses variables.
\end{enumerate}
\end{definition}

Ces conditions permettent de définir les $M_{i,j}^{(\beta)}$ de façon unique à l'exception d'une constante multiplicative globale sans intérêt. De plus, ces conditions sont vérifiées pour les trois cas connus: $\beta=1, \frac{1}{2}$ et $2$ pour lesquels des démonstrations spécifiques existent (utilisant les propriétés spécifiques de ces ensembles de matrices). Il est alors logique de définir l'intégrale d'Itzykson-Zuber généralisée par la formule connue pour le cas hermitien:

\begin{definition} L'intégrale d'Itzykson-Zuber généralisée est définie par:
\beq \label{I} I_\beta(X,Y)\mathop{=}^{\text{def}} \sum_{i=1}^N M_{i,j}^{(\beta)}(X,Y)=\sum_{j=1}^N M_{i,j}^{(\beta)}(X,Y)\eeq
\end{definition}

Les indices $i$ et $j$ dans les sommes peuvent être choisis de façon arbitraire puisque les sommes en sont indépendantes: notons qu'à nouveau, cette définition recouvre les trois cas connus $\beta=1, \frac{1}{2}, 2$. Cette définition permet alors de montrer que $I_\beta(X,Y)$ satisfait l'équation suivante:
\begin{theorem}
L'intégrale d'Itzykson-Zuber généralisée vérifie l'équation :
\beq \label{Hamilton} H_X^{(\beta)}I_\beta\mathop{=}^{def}\sum_{i=1}^N \frac{\partial^2 I_\beta}{\partial x_i^2} +\beta \sum_{i\neq j} \frac{ 1}{x_i-x_j} \left( \frac{ \partial I_\beta}{\partial x_i} - \frac{ \partial I_\beta}{\partial x_j}\right) =\left(\frac{N\beta}{T}\right)^2\left(\sum_{j=1}^N y_j^2\right) I_\beta \eeq
où $H_X^{(\beta)}$ est l'Hamiltonien de Calogero-Moser ce qui légitime le choix de la généralisation.
\end{theorem}

La démonstration de cette identité est facile:

\medskip
\underline{Preuve de \ref{Hamilton}}:

Observons d'abord que:
$$ \frac{\partial I_\beta}{\partial x_i}=\frac{\partial }{\partial x_i}\left(\underset{j=1}{\overset{N}{\sum}} M_{i,j}^{(\beta)}\right) $$ en prenant la somme sur $j$ de \ref{GeneralizedCM} et en observant que $\underset{j=1}{\overset{N}{\sum}} \left(M_{i,j}^{(\beta)}-M_{k,j}^{(\beta)}\right)=0$ pour tout $i$ et $k$, on trouve que:
$$ \frac{\partial I_\beta}{\partial x_i}=\frac{N\beta}{T}\sum_{j=1}^N y_j M_{i,j}^{(\beta)}$$
En dérivant cette égalité, on obtient alors:
\bea \frac{\partial^2 I_\beta}{\partial x_i^2}&=&\frac{N\beta}{T}\sum_{j=1}^N y_j \frac{\partial M_{i,j}^{(\beta)}}{\partial x_i}\cr
&=&\frac{N\beta}{T}\sum_{j=1}^N y_j \left(\frac{N\beta}{T}y_j M_{i,j}^{(\beta)}-\beta\sum_{k\neq i} \frac{ M_{i,j}^{(\beta)}-M_{k,j}^{(\beta)}}{x_i-x_k}\right)\cr
&=&\left(\frac{N\beta}{T}\right)^2\sum_{j=1}^N y_j^2 M_{i,j}^{(\beta)} -\beta\sum_{k\neq i} \frac{\frac{\partial I_\beta}{\partial x_i} - \frac{\partial I_\beta}{\partial x_k}}{x_i-x_k}
\eea
qui donne exactement \ref{Hamilton}.

\medskip
\medskip

Ainsi, il est alors naturel de définir le modèle à deux matrices avec $\beta$-quelconque de la façon suivante:
\begin{definition}
 \textit{le modèle à deux matrices avec $\beta$-quelconque} est défini par:
\beq  Z_{\beta}\mathop{=}^{def}\int\int dX dY \Delta(X)^{2\beta} \Delta(Y)^{2\beta} e^{-\frac{N\beta}{T}\left[\underset{i=1}{\overset{N}{\sum}}V_1(x_i)+\underset{j=1}{\overset{N}{\sum}}V_2(y_j)\right]}I_\beta(X,Y)\eeq
avec l'intégrale d'Itzykson-Zuber généralisée $I_\beta(X,Y)$ définie par (\ref{GeneralizedCM}) et (\ref{I}).
\end{definition}

Notons en particulier que dans le cas où $V_2(y)$ est un potentiel quadratique on retombe sur le modèle à une matrice avec $\beta$-quelconque précédemment décrit.

\section{Equations de boucles pour le modèle à deux matrices et $\beta$-quelconque}

Maintenant que le modèle à deux matrices est généralisé pour des valeurs de $\beta$ quelconque, il faut se demander quelles méthodes employer pour le résoudre. D'après ce que l'on a vu précédemment, il est clair que pour une valeur arbitraire de $\beta$, la méthode des polynômes orthogonaux ou bi-orthogonaux ne pourra pas fonctionner car elle est spécifique de la puissance $2$ du déterminant de Vandermonde. Cela dit, l'utilisation de certains types de polynômes pourrait peut être permettre la résolution de ce modèle comme c'est le cas dans le cas hermitien, mais à l'heure actuelle, aucune réponse générale n'a été trouvée bien que l'utilisation des polynômes de Jack ou de MacDonald semble être une possibilité. La seconde alternative consiste alors à utiliser l'approche des équations de boucles. Comme nous allons le voir, l'écriture des équations de boucles possède l'avantage de se généraliser relativement facilement au cas où $\beta$ est arbitraire. En revanche, la résolution de ces équations de boucles devient beaucoup plus délicate en dehors du cas hermitien dont la spécificité ressort nettement dans les équations. Cette section aura donc pour objet de décrire en détail l'obtention des équations de boucles pour le modèle à deux matrices avec $\beta$-quelconque. Le lecteur intéressé pourra alors retrouver le cas hermitien en prenant $\beta=1$, ainsi que le cas à une matrice en prenant $V_2(y)=\frac{y^2}{2}$.

\subsection{Notations}

La principale difficulté dans l'écriture des équations de boucles se situe dans le nombre important d'indices et de fonctions à définir préalablement. Nous utiliserons donc les notations suivantes:
\begin{definition}
 \begin{itemize}
 \item Les potentiels sont supposés polynômiaux:
\beq V_1'(x)=\sum_{k=0}^{d_1} t_k x^k \virg V_2'(x)=\sum_{k=0}^{d_2} \tilde{t}_k x^k \eeq
\item Les résolvantes sont définies par:
\beq \label{Wn} W_n(z_1,\dots,z_n)=\left<\sum_{i_1,\dots,i_n=1}^N  \frac{1}{z_1-x_{i_1}}\dots \frac{1}{z_n-x_{i_n}}\right>_c \eeq
où les crochets $\left<\right>$ indiquent que l'on prend la valeur moyenne relativement à la mesure définie par  \ref{Diagonalized2betaMM}. L'indice $_c$ indique que l'on prend la partie connexe lors d'un produit de traces. Par exemple si l'on note: $X=(x_1,\dots,x_N)$ et Y=$(y_1,\dots,y_N$), alors
\beq \left<A(X,Y)\right>\mathop{=}^{def}\frac{1}{Z_\beta} \int dX dY A(X,Y) e^{-\frac{N\beta}{T} \left(\tr V_1(X)+V_2(Y)\right)}\Delta(X)^{2\beta}\Delta(Y)^{2\beta} I_\beta(X,Y)\eeq
et la partie connexe indique que l'on doit calculer:
\beq \left< A(X,Y) B(X,Y)\right>_c=\left< A(X,Y)B(X,Y) \right>-\left< A(X,Y)\right> \left< B(X,Y)\right>\eeq
Pour simplifier un peu les notations, nous noterons $W(x)=W_1(x)$ pour la première résolvante qui joue un rôle particulier.
\item Afin de fermer les équations de boucles, nous allons avoir besoin des fonctions suivantes:
\beq \label{Un} U_n(x,y;z_1,\dots,z_n)=\sum_{i,j,i_1,\dots,i_n=1}^N\left<  \frac{1}{x-x_i} \frac{M_{i,j}^{(\beta)}}{I_\beta}\frac{V'_2(y)-V'_2(y_j)}{y-y_j}\frac{1}{z_1-x_{i_1}} \dots  \frac{1}{z_n-x_{i_n}}\right>_c
 \eeq
qui est un polynôme en $y$. Finalement nous aurons besoin également de:
 \beq \label{Pn} P_n(x,y;z_1,\dots,z_n)=\sum_{i,j,i_1,\dots,i_n=1}^N\left<  \frac{V'_1(x)-V'_2(x_i)}{x-x_i} \frac{M_{i,j}^{(\beta)}}{I_\beta}\frac{V'_2(y)-V'_2(y_j)}{y-y_j} \frac{1}{z_1-x_{i_1}} \dots \frac{1}{z_n-x_{i_n}}\right>_c \eeq
qui est un polynôme à la fois en $x$ et en $y$.
\end{itemize}
\end{definition}

\medskip
\medskip

Pour résoudre les futures équations de boucles, nous allons avoir besoin d'écrire le développement topologique (identique à celui de \cite{OE}) des fonctions précédentes . Afin de garantir l'existence de tels développements, nous nous plaçons dans le cas d'un modèle ``formel'' de matrices, c'est-à-dire que nous supposons l'existence d'un développement perturbatif en puissances de $\frac{1}{N}$, sans nous préocupper de la convergence des séries (considérées comme formelles).

\begin{definition} Le développement topologique des fonctions de corrélation est défini par: 
\bea\label{topexp} W_n(x_1,\dots,x_n)&=&\beta^{-\frac{n}{2}}\sum_{g=0}^\infty \left(\frac{N\sqrt{\beta}}{T}\right)^{2-2g-n} W_n^{(g)}(x_1,\dots,x_n)\cr
U_0(x,y)&=&\frac{N}{T}\left(U_0^{(0)}(x,y)-x+V_2'(y)\right)+\beta^{-\frac{1}{2}}\sum_{g=1}^\infty \left(\frac{N\sqrt{\beta}}{T}\right)^{1-2g} U_0^{(g)}(x,y)\cr
U_n(x,y;x_1,\dots,x_n)&=&\beta^{-\frac{n+1}{2}}\sum_{g=0}^\infty \left(\frac{N\sqrt{\beta}}{T}\right)^{2-2g-(n+1)} U_n^{(g)}(x,y;x_1,\dots,x_n)\cr
P_0(x,y)&=&\frac{N}{T}\left(P_0^{(0)}(x,y)+\hbar-T\right)+\beta^{-\frac{1}{2}}\sum_{g=1}^\infty \left(\frac{N\sqrt{\beta}}{T}\right)^{1-2g} P_0^{(g)}(x,y)\cr
P_n(x,y;x_1,\dots,x_n)&=&\beta^{-\frac{n+1}{2}}\sum_{g=0}^\infty \left(\frac{N\sqrt{\beta}}{T}\right)^{2-2g-(n+1)} P_n^{(g)}(x,y;x_1,\dots,x_n)\cr
\eea
\end{definition}

Une remarque importante est que nous avons choisi ici de translater les fonctions $U_0^{(0)}(x,y)$ et $P_0^{(0)}(x,y)$. Bien que cela puisse paraître étrange, cela permettra par la suite de simplifier légèrement l'écriture des équations de boucles. Par ailleurs, nous avons omis pour des raisons de simplicité d'écriture la dépendance des fonctions ($W_n$, $W_n^{(g)}$, $P_n^{(g)}$, $U_n^{(g)}$ etc.) dans le paramètre $\beta$.

\medskip
Finalement, il est aussi utile de définir les nombres $F_g$ comme le développement topologique de la fonction de partition elle-même:
\begin{definition} La fonction de partition s'écrit formellement:
\beq Z_\beta=e^F \, \, \virg \,\, F=\sum_{g=0}^\infty \left(\frac{N\sqrt{\beta}}{T}\right)^{2-2g} F_g\eeq
\end{definition}
ainsi que le paramètre $\hbar$ qui jouera un rôle crucial dans la suite et que l'on obtient à partir de $\beta$ par:

\begin{definition} Le paramètre $\hbar$ est relié au paramètre $\beta$ par la relation: 
\beq \encadremath{\hbar=\frac{T}{N} \left(1-\frac{1}{\beta}\right)=\frac{T}{N \sqrt{\beta}} \left(\sqrt{\beta}-\frac{1}{\sqrt{\beta}}\right) } \eeq
\end{definition}

Tout comme pour le cas hermitien, nous introduisons également les opérateurs ``d'insertion de boucles'' définis par:
\begin{definition} Les opérateurs d'insertion sont définis par:
\bea \label{insertionop}
 \frac{\d}{\d V_1(x)}&:=&-\sum_{k=1}^\infty \frac{1}{x^{k+1}}\,k\frac{\d}{\d t_{k-1}} \cr
\frac{\d}{\d V_2(x)}&:=&-\sum_{k=1}^\infty \frac{1}{y^{k+1}}\,k\frac{\d}{\d \td{t}_{k-1}}
\eea
\end{definition}

Ils possèdent les propriétés:
\beq
\frac{\d V_j(x)}{\d V_l(x')}=\delta_{j,l}\,\frac{1}{x-x'}\virg \frac{\d V_j'(x)}{\d V_l(x')}=-\delta_{j,l}\,\frac{1}{(x-x')^2} 
\eeq
Ces opérateurs sont particulièrement intéressants car ils permettent de passer d'une résolvante à la suivante:
\beq \frac{N\beta}{T}\, W(x)=\frac{\d F}{\d V_1(x)} \,\, \virg \,\,  \frac{\d F_g}{\d V_1(x)}=W_1^{(g)}(x)\eeq
et également:
\bea
\frac{N\beta}{T}\,W_n(x_1,\dots,x_n) &=& \frac{\d W_{n-1}(x_1\dots,x_{n-1})}{\d V_1(x_n)}\cr
\frac{N\beta}{T}\,U_n(x,y;x_1,\dots,x_n) &=& \frac{\d U_{n-1}(x,y;x_1\dots,x_{n-1})}{\d V_1(x_n)}
\eea
ce qui donne dans les développements topologiques:
\bea \label{der}
W_n^{(g)}(x_1,\dots,x_n) &=& \frac{\d W_{n-1}^{(g)}(x_1\dots,x_{n-1})} {\d V_1(x_n)}\cr
U_n^{(g)}(x,y;x_1,\dots,x_n) &=& \frac{\d U_{n-1}^{(g)}(x,y;x_1\dots,x_{n-1})}{\d V_1(x_n)}
\eea
On voit donc que la connaisance des $F_g$ permet ensuite par simple application de ces opérateurs de dérivation de trouver tous les $W_n^{(g)}(x_1,\dots,x_n)$ correspondants.

Enfin, afin d'avoir des notations plus compactes, nous introduisons les fonctions translatées:
\begin{definition}
Soit la fonction translatée:
\beq
Y(x):=(V'_1(x)-W_1^{(0)}(x))
\eeq
On définit la courbe spectrale par:
\beq \label{spectr}
E(x,y):=(V'_1(x)-y)(V'_2(y)-x)-P_0^{(0)}(x,y)
\eeq
\end{definition}

Notons que la plupart de nos fonctions étant polynômiales en $y$, nous pouvons les développer sur la base des $y^k$ de la façon suivante (en prenant en compte le degré):
\begin{definition}
Le développement en puissances de $y^k$ donnent les identités formelles suivantes:
\bea \label{yprojection}
U_n(x,y;x_1,\dots,x_n)&=& \sum_{k=0}^{d_2} U_{n,k}(x;x_1,\dots,x_n) y^k \cr
P_n(x,y;x_1,\dots,x_n)&=& \sum_{k=0}^{d_2-1} P_{n,k}(x;x_1,\dots,x_n) y^k \cr
E(x,y)&=& \sum_{k=0}^{d_2+1} E_k(x) y^k \cr
U_n^{(g)}(x,y;x_1,\dots,x_n)&=& \sum_{k=0}^{d_2} U_{n,k}^{(g)}(x;x_1,\dots,x_n) y^k \cr
P_n^{(g)}(x,y;x_1,\dots,x_n)&=& \sum_{k=0}^{d_2-1} P_{n,k}^{(g)}(x;x_1,\dots,x_n) y^k \cr
\eea
D'une façon générale, un indice $_k$ supplémentaire et l'absence de variable $y$ signifie que l'on a pris la projection sur $y^k$.
\end{definition}

\medskip
\medskip

Une fois toutes ces notations introduites, nous pouvons passer à l'écriture des équations de boucles. Celle-ci se fera en deux temps par l'écriture de deux intégrales triviales.

\subsection{Etape une: un résultat préliminaire}

Intéressons-nous tout d'abord à l'intégrale nulle (car on peut intégrer d'abord sur $y_j$ une dérivée totale et le contour d'intégration est supposé sans bords) suivante:
\beq 0=\sum_{i,j=1}^N \int dXdY \, \frac{\partial}{\partial y_j} \left(e^{-\frac{N\beta}{T} \tr (V_1(X)+V_2(Y) )} \Delta(X)^{2\beta} \Delta(Y)^{2\beta} \frac{1}{x-x_i} M_{i,j}^{(\beta)}\right)\eeq 

On peut faire agir la dérivée sur chaque terme, ce qui donne trois contributions différentes:
\begin{itemize}
 \item Agissant sur l'exponentielle on trouve:
\beq -\frac{N\beta}{T}\sum_{i,j=1}^N\left< V_2'(y_j)\frac{1}{x-x_i}\frac{M_{i,j}^{(\beta)}}{I_\beta}\right>\eeq
\item Agissant sur le déterminant de Vandermonde, on trouve:
\beq \label{oo} 2\beta \left< \sum_{i,j=1}^N \sum_{k \neq i} \frac{1}{y_j-y_k} \frac{1}{x-x_i} \frac{M_{i,j}^{(\beta)}}{I_\beta}\right>\eeq
\item Finalement, agissant sur $M_{i,j}^{(\beta)}$ et en utilisant l'équation différentielle satisfaite par les $M_{i,j}^{(\beta)}$ (\ref{GeneralizedCM}) on trouve:
\beq \label{ooo} \left< \sum_{i,j=1}^N\frac{1}{x-x_i} \left( \frac{N\beta}{T} x_iM_{i,j}^{(\beta)}-\beta \sum_{k\neq j} \frac{M_{i,j}^{(\beta)}-M_{i,k}^{(\beta)}}{y_j-y_k}\right) \frac{1}{I_\beta}\right> \eeq
\end{itemize}

On voit alors que \ref{oo} s'annule avec la dernière partie de \ref{ooo} ce qui donne :
\beq \sum_{i,j=1}^N\left< V_2'(y_j)\frac{1}{x-x_i}\frac{M_{i,j}^{(\beta)}}{I_\beta}\right>=\sum_{i,j=1}^N\left< \frac{x_i}{x-x_i}\frac{M_{i,j}^{(\beta)}}{I_\beta}\right> \eeq
Comme $\underset{j=1}{\overset{N}{\sum}} M_{i,j}^{(\beta)}=I_\beta$ on trouve finalement:
\beq \label{firstloop} \encadremath{\sum_{i,j=1}^N\left< V_2'(y_j)\frac{1}{x-x_i}\frac{M_{i,j}^{(\beta)}}{I_\beta}\right>=-N+xW(x)} \eeq
concluant ainsi la première étape.

\subsection{Etape deux: les équations de boucles}

La deuxième étape consiste à regarder l'intégrale nulle (pour des raisons similaires à la précédente) suivante:
\beq 
0=\sum_{i,j=1}^N \int dXdY \, \frac{\partial}{\partial x_i} \left(e^{-\frac{N\beta}{T} \tr (V_1(X)+V_2(Y) )} \Delta(X)^{2\beta} \Delta(Y)^{2\beta} \frac{1}{x-x_i} M_{i,j}^{(\beta)}\frac{V'_2(y)-V'_2(y_j)}{y-y_j}\right)\eeq 
Le lecteur remarquera que cette intégrale est très similaire aux définitions des fonctions $U_0(x,y)$ et $P_0(x,y)$ (\ref{Un} et \ref{Pn}). A nouveau, on peut faire agir la dérivation sur chacun des termes du produit. Il y a cette fois-ci quatre contributions:
\begin{itemize}
 \item Agissant sur l'exponentielle on trouve:
\beq \label{i}-\frac{N\beta}{T} \sum_{i,j=1}^N \left< \frac{V'_1(x_i)}{x-x_i} \frac{M_{i,j}^{(\beta)}}{I_\beta} \frac{V'_2(y)-V'_2(y_j)}{y-y_j}\right> \, (i)\eeq 
\item Agissant sur le déterminant de Vandermonde, on trouve:
\beq \label{ii}2\beta\sum_{i,j=1}^N \left<  \sum_{k \neq i} \frac{1}{x_i-x_k}\frac{1}{x-x_i} \frac{M_{i,j}^{(\beta)}}{I_\beta}\frac{V'_2(y)-V'_2(y_j)}{y-y_j}\right> \, (ii)\eeq
\item Agissant sur $\frac{1}{x-x_i}$ on trouve:
\beq \label{iii}\sum_{i,j=1}^N \left<  \frac{1}{(x-x_i)^2} \frac{M_{i,j}^{(\beta)}}{I_\beta}\frac{V'_2(y)-V'_2(y_j)}{y-y_j}\right> \, (iii)\eeq
\item Enfin, agissant sur $M_{i,j}^{(\beta)}$ et en utilisant de nouveau \ref{GeneralizedCM} on trouve:
\beq \label{iv} \sum_{i,j=1}^N \left< \frac{1}{x-x_i} \frac{1}{I_\beta}\left(\frac{N\beta}{T}y_j M_{i,j}^{(\beta)}-\beta\sum_{k \neq i}\frac{M_{i,j}^{(\beta)}-M_{k,j}^{(\beta)}}{x_i-x_k}\right) \frac{V'_2(y)-V'_2(y_j)}{y-y_j}\right> \, (iv)\eeq 
\end{itemize}
Maintenant, il convient de remarquer les identités suivantes. Tout d'abord, dans \ref{i}, on peut séparer $V'_1(x_i)=V'_1(x_i)-V'_1(x)+V'_1(x)$ de telle sorte que:
\beq (i) \, \Leftrightarrow -\frac{N\beta}{T} \left( V_1'(x)U_0(x,y)-P_0(x,y)\right)\eeq
Deuxièmement, on peut couper $(ii)$ de la façon suivante:
\bea
(ii)&=& \beta\sum_{i,j=1}^N \left<  \sum_{k \neq i} \frac{1}{x_i-x_k}\frac{1}{x-x_i} \frac{M_{i,j}^{(\beta)}}{I_\beta}\frac{V'_2(y)-V'_2(y_j)}{y-y_j}\right>\cr
&&+ \beta \sum_{i,j=1}^N \left<  \sum_{k \neq i} \frac{1}{x-x_i}\frac{1}{x-x_k} \frac{M_{i,j}^{(\beta)}}{I_\beta}\frac{V'_2(y)-V'_2(y_j)}{y-y_j}\right>\cr
&&+\beta \sum_{i,j=1}^N \left<  \sum_{k \neq i} \frac{1}{x-x_k}\frac{1}{x_i-x_k} \frac{M_{i,j}^{(\beta)}}{I_\beta}\frac{V'_2(y)-V'_2(y_j)}{y-y_j}\right>\cr
&=& \beta \sum_{i,j=1}^N \left<  \sum_{k \neq i} \frac{1}{x-x_i}\frac{1}{x-x_k} \frac{M_{i,j}^{(\beta)}}{I_\beta}\frac{V'_2(y)-V'_2(y_j)}{y-y_j}\right> \,(ii)'\cr
&&+\beta\sum_{i,j=1}^N \left<\sum_{k \neq i}\frac{M_{i,j}^{(\beta)}-M_{k,j}^{(\beta)}}{(x_i-x_k)(x-x_i)} \frac{V'_2(y)-V'_2(y_j)}{y-y_j} \right>\, (ii)''\cr
\eea
Remarquons alors que $(ii)''$ est identique au dernier terme de $(iv)$ ce qui provoquera leurs annulations respectives. Ensuite, on peut couper $(ii)'$ en une somme sur $i,k$ moins le cas où $i=k$ qui est quant à lui identique à $(iii)$, à un facteur $\beta$ près. Ainsi, on peut regrouper $(ii)$, $(iii)$ et $(iv)$ pour obtenir:
\bea  \label{v} (ii)+(iii)+(iv)&=&(1-\beta) \sum_{i,j=1}^N \left<  \frac{1}{(x-x_i)^2} \frac{M_{i,j}^{(\beta)}}{I_\beta}\frac{V'_2(y)-V'_2(y_j)}{y-y_j}\right> \, (1)\cr
&&+\beta \sum_{i,j,k=1}^N \left<\frac{1}{x-x_i}\frac{1}{x-x_k} \frac{M_{i,j}^{(\beta)}}{I_\beta}\frac{V'_2(y)-V'_2(y_j)}{y-y_j}\right> \, (2)\cr
&&+\frac{N\beta}{T}\sum_{i,j=1}^N \left< \frac{y_j}{x-x_i} \frac{M_{i,j}^{(\beta)}}{I_\beta}\frac{V'_2(y)-V'_2(y_j)}{y-y_j} \right> \, (3)
\eea
Observons maintenant l'identité:
\beq (1)=(1-\beta)\sum_{i,j=1}^N \left<  \frac{1}{(x-x_i)^2} \frac{M_{i,j}^{(\beta)}}{I_\beta}\frac{V'_2(y)-V'_2(y_j)}{y-y_j}\right>=(\beta-1)\frac{\partial}{\partial x} U_0(x,y)\eeq
et également:
\bea \frac{T}{N\beta}\left< \frac{\partial}{ \partial V_1(x)} U_0(x,y)\right>_c&=&\sum_{i,k,j=1}^N\left<\frac{1}{x-x_i}\frac{1}{x-x_k} \frac{M_{i,j}^{(\beta)}}{I_\beta}\frac{V'_2(y)-V'_2(y_j)}{y-y_j}\right>\cr
&&-\sum_{i,j=1}^N\left<\frac{1}{x-x_i}\frac{M_{i,j}^{(\beta)}}{I_\beta}\frac{V'_2(y)-V'_2(y_j)}{y-y_j}\right>\left< \frac{1}{x-x_k}\frac{M_{i,j}^{(\beta)}}{I_\beta}\right>\cr
&&=\sum_{i,k,j=1}^N\left<\frac{1}{x-x_i}\frac{1}{x-x_k} \frac{M_{i,j}^{(\beta)}}{I_\beta}\frac{V'_2(y)-V'_2(y_j)}{y-y_j}\right>\cr
&&-W(x)U_0(x,y)
\eea
On reconnait ici le second terme de \ref{v}:
\bea (2) &\Leftrightarrow& \frac{T}{N}\left< \frac{\partial}{ \partial V_1(x)} U_0(x,y)\right>_c+\beta W(x)U_0(x,y)\cr
 &\Leftrightarrow& \beta U_1(x,y;x)+\beta W(x)U_0(x,y)
\eea
Finalement, il nous reste à traiter $(3)$. On peut alors réécrire $y_j \leftrightarrow y_j-y+y$ et le couper en deux pour avoir:
\bea
(3) &\Leftrightarrow& \frac{N\beta}{T}y U_0(x,y) -\frac{N\beta}{T}V'_2(y)\sum_{i,j=1}^N\left<\frac{1}{x-x_i}\frac{M_{i,j}^{(\beta)}}{I_\beta}\right>\cr
&&+ \frac{N\beta}{T}\sum_{i,j=1}^N\left<\frac{V'_2(y_j)}{x-x_i}\frac{M_{i,j}^{(\beta)}}{I_\beta}\right>
\eea

Mais rappelons que $\underset{j=1}{\overset{N}{\sum}} M_{i,j}^{(\beta)}=I_\beta$. et que d'après notre résultat préliminaire nous avons \ref{firstloop} ce qui donne finalement:
\beq (3) \Leftrightarrow \frac{N\beta}{T}y U_0(x,y)-\frac{N\beta}{T}V'_2(y)W(x)+\frac{N\beta}{T}(-N +x W(x))\eeq

En regroupant ensemble toutes les contributions, nous arrivons donc à l'équation:
\bea 0&=&-\frac{N\beta}{T} \left( V_1'(x)U_0(x,y)-P_0(x,y)\right)-(1-\beta)\frac{\partial}{\partial x} U_0(x,y)\cr
&&+\beta U_1(x,y;x)+\beta W(x)U_0(x,y)+\frac{N\beta}{T}yU_0(x,y)\cr
&&-\frac{N\beta}{T}V'_2(y)W(x)+\frac{N\beta}{T}(-N +x W(x))
\eea
qui peut être réécrite (en multipliant par $-\frac{T}{N\beta}$) pour donner \textit{l'équation de boucle maîtresse}:
\begin{theorem} La fonction $W(x)$ satisfait l'équation maîtresse suivante:
\beq \label{loopequation} \encadremath{\left(y-V'_1(x)+\frac{T}{N}W(x)+\hbar \partial_x\right)U_0(x,y)=(V'_2(y)-x)W(x)-P_0(x,y)
+N-\frac{T}{N} U_1(x,y;x)}\eeq
\end{theorem}

Afin de résoudre cette équation de boucle maîtresse, on peut la projeter sur le développement topologique et obtenir le théorème suivant:

\begin{theorem}
Les fonctions de corrélations satisfont les équations de boucles à $\beta$ quelconque:

\medskip
\underline{Equation de boucles à l'ordre dominant}:
 
\beq \label{eqboucles0} \encadremath{
\left(y-Y(x)+\hbar\partial_x\right)U_0^{(0)}(x,y)=E(x,y)}
\eeq

\medskip
\underline{Equations de boucles aux ordres supérieurs}:
\begin{equation}
\label{loopequationg}
\fbox{$
   \begin{array}{rcl}
\left(y-Y(x)+\hbar\partial_x\right)U_0^{(g)}(x,y)&=&-W_1^{(g)}(x)U_0^{(0)}(x,y)-\overset{g-1}{\underset{h=1}{\sum}}
 W_1^{(g-h)}(x)U_0^{(h)}(x,y)\cr
&&- P_0^{(g)}(x,y)-U_1^{(g-1)}(x,y;x)\end{array}
   $}
\end{equation}
\end{theorem}

\subsection{Analyse des équations de boucles: singularité du cas hermitien}

Les équations de boucles (\ref{eqboucles0}) et (\ref{loopequationg}) permettent d'obtenir le cas à une matrice (prendre $V_2'(y)=y$) et/ou le cas hermitien (prendre $\beta=1 \Leftrightarrow \hbar=0$). Il est alors évident de constater que le cas hermitien constitue un cas très particulier, puisqu'alors les équations de boucles deviennent purement algébriques (le facteur $\hbar$ devant les dérivées devenant nul). Cette simplification spécifique a permis à B. Eynard et N. Orantin \cite{OE} de résoudre ces équations de boucles par des méthodes de géométrie algébrique, et même de généraliser, dans le cadre d'une courbe algébrique quelconque (appelée courbe spectrale) $E(x,y)=0$ la définition d'invariants symplectiques $F_g$ qui résolvent les équations de boucles hermitiennes (Cf. chapitre \ref{chap1}). En revanche, dans le cas où $\beta$ est quelconque (i.e. $\hbar \neq 0$), la nature des équations de boucles change, puisqu'elles deviennnent non plus algébriques, mais différentielles. Dès lors, la résolution, qui à ce jour n'est pas encore complètement explicite, change de nature également. Cela dit, un point important est que la limite $\hbar \to 0$ doit redonner les résultats du cas hermitien, c'est-à-dire de la théorie des invariants symplectiques correspondants. Le travail réalisé en collaboration avec B. Eynard et L. Chekhov a été de développer un formalisme de résolution de ces équations de boucles dans le cas où $\hbar \neq 0$ en réalisant une ``déformation quantique'' de la théorie des invariants symplectiques de B. Eynard et N. Orantin. Cette résolution, encore partielle à l'heure actuelle, consiste à s'intéresser à une ``courbe quantique'':
\beq P(x,y)=0 \virg [y,x]=\hbar \eeq
et à définir à partir de cette ``courbe'' des fonctions qui résolvent les équations de boucles \ref{eqboucles0} et \ref{loopequationg}. Bien que certains résultats concernant le modèle à deux matrices soient en cours de réalisation, nous nous contenterons dans cette thèse de ne traiter que des modèles à une matrice présentés dans les articles \textbf{[III]} et \textbf{[IV]} présentés respectivement en annexe \ref{Article[III]} et \ref{Article[IV]}.

\section{Le modèle à une matrice pour $\beta$-quelconque et la géométrie algébrique quantique}

Le cas du modèle à une matrice possède l'avantage d'être beaucoup plus simple d'un point de vue technique que le cas à deux matrices. En effet, dans le cas du modèle à une matrice, les équations de boucles se réécrivent sous la forme:
\beq \label{loop1}
{W^{(0)}_1}(x)^2 - V_1'(x)W^{(0)}_1(x) + \hbar\,{\partial_x} W_1^{(0)}(x)=  -P^{(0)}_1(x)
\eeq
et en posant $J=(x_1,\dots,x_n)$:
\bea\label{loopeqPng}
 P_{n+1}^{(g)}(x;x_1...,x_n)
 &=&
\left(2W_1^{(0)}(x)-V_1'(x)\right)\overline{W}_{n+1}^{(g)}(x,x_1,...,x_n) + \hbar \partial_{x}{\overline{W}_{n+1}^{(g)}(x,x_1...,x_n)} \cr
&& + \sum_{I\subset J} \ovl{W}_{|I|+1}^{(h)}(x,x_I) \ovl{W}_{n-|I|+1}^{(g-h)}(x,J/I) +
\ovl{W}_{n+2}^{(g-1)}(x,x,J)  \cr
& &+ \sum_{j}
\partial_{x_j} \left( {{\ovl{W}_n^{(g)}(x,J/\{j\})-{\ovl{W}_n^{(g)}(x_j,J/\{j\})}} \over {(x-x_j)}}\right)
\eea
où la notation $\ovl{W}_n^{(g)}$ signifie:
\beq
\ovl{W}_{n}^{(g)}(x_1,...,x_n) = W_{n}^{(g)}(x_1,...,x_n) - {\delta_{n,2}\delta_{g,0}\over 2}\, {1\over (x_1-x_2)^2}
\eeq

Comme nous allons le voir, le modèle à une matrice est relié à l'équation de Schrödinger, c'est-à-dire à une équation différentielle ordinaire de degré $2$. Dans le cas du modèle à deux matrices, l'équation différentielle est de degré $d_2$, (degré du potentiel $V_2(y)$) ce qui rend les calcus plus compliqués.

\subsection{Lien entre les équations de boucles et la géométrie algébrique quantique}

\begin{definition}
On définit la fonction $\psi(x)$ par:
\beq \label{l}\hbar \frac{\psi'(x)}{\psi(x)}=W_1^{(0)}(x)-\frac{V_1'(x)}{2}\eeq
\end{definition}
On remarque que la première équation de boucles se réécrit (Cf. \textbf{[IV]} en annexe \ref{Article[IV]}):
\beq \hbar^2 \psi''(x)=U(x)\psi(x)\virg U(x)=\frac{V'(x)^2}{4}-\hbar\frac{V''(x)}{2} -P_1^{(0)}(x)\eeq
c'est-à-dire que la fonction $\psi(x)$ satisfait une équation de Schrödinger. Dans le cas du modèle à deux matrices, cette équation se généralise à une équation d'ordre plus élevée:
\bea
W_1^{(0)}(x)&=&\hbar\frac{\psi'(x)}{\psi(x)}\cr
0&=& \left(V_1'(x)-\hbar \partial_x\right)^{d_2+1}\psi(x)-\sum_{k=0}^{d_2}\left(V_1'(x)-\hbar \partial_x\right)^k \left(E_{k}(x) \psi(x)\right)\eea
où les $E_k(x)$ sont donnés par $E(x,y)=\underset{k=0}{\overset{d_2+1}{\sum}} E_k(x)y^k$ et où nous rappelons que $E(x,y)$ est la courbe spectrale donnée par \ref{spectr}. (Notons que l'on retrouve bien le cas à une matrice en prenant $d_2=1$ et $E_0(x)=-2V_1'(x)$ comme prévu). Ainsi on peut récrire ces deux modèles sous la forme suivante:
\beq \label{ODE}\encadremath{ \hat{y}=V_1'(x)-\hbar \partial_x \virg  E(x,\hat{y})\psi(x)\mathop{=}^{\text{def}}\sum_{k=0}^{d_2+1} \hat{y}^k E_k^{(0)}(x) \psi(x)=0} \eeq
avec $[\hat{y},x]=\hbar$. Notons en particulier, que les variables $x$ et $\hat{y}$ ne commutant plus, il est nécessaire de préciser la position de l'une par rapport à l'autre (les variables $\hat{y}$ se retrouvant toujours à gauche). Dans le cas à une matrice, cela se réécrit avec les notations spécifiques \ref{l} comme:
\beq \hat{y}=\hbar \partial_x \virg \left(\hat{y}^2-U(x)\right)\psi(x)=0\eeq

On voit donc émerger une courbe ``quantique'' (au sens où les variables ne commutent plus: $[\hat{y},x]=\hbar$) donnée par $E(x,\hat{y})\psi(x)=0$, où $\frac{\psi'(x)}{\psi(x)}$ représente $W_1^{(0)}(x)$ à une translation triviale par $V_1'(x)$ près. A noter que dans le cas hermitien, on retombe sur une courbe algébrique ``classique'' $E(x,y)=0$ où $x$ et $y$ commutent de nouveau. On voit donc toute la singularité du cas hermitien, puisque l'on passe alors du domaine différentiel au domaine algébrique ou de façon équivalente de variables non-commutantes à des variables commutantes. Dans le cas hermitien, les travaux de B. Eynard et N. Orantin \cite{OE}, \cite{countingsurface}, \cite{toprecint} permettent de construire à partir de la courbe algébrique $E(x,y)=0$ toutes les autres fonctions de corrélation $W_n^{(g)}$ ainsi que les invariants symplectiques $F_g$. Cette construction, présentée brièvement au chapitre \ref{chap1} utilise des notions avancées de géométrie algébrique: genre, noyau de Bergmann, intégration sur une surface de Riemann, formes holomorphes, etc.

Notre démarche a alors été de partir de la courbe ``quantique'' et d'une solution $\psi(x)$ associée, et de généraliser les notions développées par B. Eynard et N. Orantin pour le cas des courbes algébriques ``classiques''. En particulier, nous nous sommes intéressés à la généralisation de la notion de genre, de formes holomorphes, de noyau de Bergman sur notre courbe ``quantique`` dans le but de résoudre les équations de boucles.

\subsection{La géométrie algébrique ``quantique'' dans le cas d'équations hyper-elliptiques}

Donnons-nous donc une courbe quantique hyper-elliptique (i.e. de degré $2$ en $y$) de la forme:
\begin{definition}
Une courbe quantique hyper-elliptique consiste en la donnée d'une équation différentielle et d'une solution $\psi$:
\beq E(x,\hat{y})\psi(x)=\sum_{j=0}^2\hat{y}^jE_j(x)\psi(x)=0\eeq
\end{definition}
En divisant par le coefficient dominant devant $\psi''(x)$ et en translatant convenablement la fonction $\psi(x)$, on peut se ramener à une équation de type Schrödinger:
\beq \label{Schro}\hbar^2\psi''(x)=U(x)\psi(x)\eeq
Il est à noter que les solutions $\psi(x)$ de cette équation différentielle ne sont pas uniques, tout comme les solutions d'une équation algébrique $y^2=U(x) \Leftrightarrow u=\pm \sqrt{U(x)}$ ne le sont pas également. Néanmoins, il est immédiat de constater que la dimension de l'espace vectoriel des solutions (ici $2$) correspond toujours au degré en $y$ de la courbe, c'est-à-dire également au nombre de solutions de l'équation algébrique classique associée. En supposant que $U(x)$ est un polynôme de degré $2d$, on peut alors définir:
\begin{definition} Le potentiel associée à la courbe est défini par:
\beq\label{defV'}
V'(x) = 2\,(\sqrt{U})_+ = \sum_{k=0}^d t_{k+1}\,x^k
\eeq
où $_+$ désigne la partie polynômiale au voisinage de l'infini. On peut également définir:
\beq\label{defP}
P(x) = \frac{V'^2(x)}{4} - U(x) -\hbar \frac{V''(x)}{2}
\eeq
qui est un polynôme de degré $d-1$.
On définit également le paramètre $t_0$ par:
\beq\label{deft0}
t_0 = \mathop{{\rm lim}}_{x\to\infty}\, \frac{xP(x)}{V'(x)}
\eeq
\end{definition}

Traditionnellement, les c\oe{}fficients $t_i$ sont appelés les ``Casimirs'' et les autres c\oe{}fficients de $P(x)$ les ``charges'' qui jouent un rôle particulier. La fonction $\psi(x)$ étant solution d'une équation de type Schrödinger et $U(x)$ étant supposée polynômiale, elle présente donc un phénomène de Stokes, c'est-à-dire que bien qu'étant analytique sur $\mathbb{C}$, son asymptotique à l'infini possède des discontinuités (singularité essentielle) suivant certaines directions: 
\begin{center} \label{Stokes}
	\includegraphics[height=5cm]{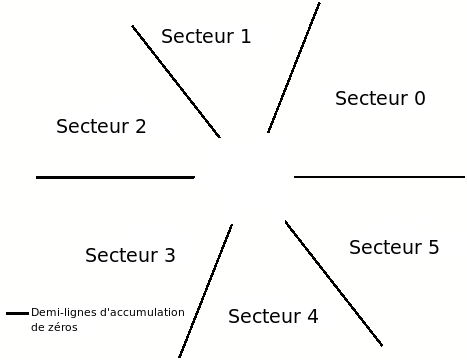}

	\underline{Figure 17}: Exemple de secteurs pour un potentiel de degré $\deg V=3$, i.e. $d=2$. Dans le cas général où  $\deg V=d+1$, il y a $2d+2$ secteurs. 
\end{center}
 Dans chaque secteur $S_k$, l'asymptotique de $\psi(x)$ est donné sous la forme:
\beq \label{asymp}\psi(x)\mathop{{\sim}}_{S_k} \ee{\pm\,\frac{1}{2\hbar}V(x)}\,x^{C_{k}}\,\, (A_k+\frac{B_k}{x}+\dots)\eeq
Une fois ces considérations prises en compte, il est alors possible de généraliser des notions de géométrie algébrique dans le cadre de notre courbe ``quantique'':

\begin{enumerate}
 \item \textbf{Les différents feuillets}

Dans le cadre d'une équation algébrique du second degré, il existe deux solutions distinctes qui correspondent à deux feuillets en géométrie algébrique. Ces feuillets sont reliés par des points de branchements correspondants aux points où les solutions sont identiques. Dans le cadre de notre courbe quantique, notre solution $\psi(x)$ présente deux comportements asymptotiques différents par le choix du signe $\pm$ dans la formule \ref{asymp}. Dès lors, nous pouvons séparer les secteurs de Stokes en deux feuillets: le feuillet ``physique'', où $\psi(x)$ se comporte comme $\psi(x)\mathop{{\sim}}_{S_k} \ee{\textbf{+}\,\frac{1}{2\hbar}V(x)}\,x^{C_{k}}$ et le feuillet non-physique, où $\psi(x)\mathop{{\sim}}_{S_k} \ee{\textbf{-}\,\frac{1}{2\hbar}V(x)}\,x^{C_{k}}$. Notons que les secteurs de Stokes n'ont de sens que dans un voisinage de l'infini, ce qui rend leur interprétation plus délicate que dans le cas algébrique. Par ailleurs, afin de fixer la solution $\psi(x)$, nous choisirons la solution qui est exponentiellement décroissante dans le secteur $0$. A noter que dans des cas très spécifiques, l'équation de Schrödinger \ref{Schro} peut avoir des solutions polynômiales ne présentant pas de phénomène de Stokes. Ces cas singuliers ont été traités dans l'article \textbf{[III]} présenté en annexe \ref{Article[III]}.

\item \textbf{Coupures et points de branchement}

Dans le cadre de la géométrie algébrique, une équation du second degré peut être vue comme une surface de Riemann de genre $g_{\text{alg}}$, c'est-à-dire comme deux copies du plan complexe reliées par $g_{\text{alg}}+1$ coupures. Dans le cas où la courbe est donnée par $$y^2=U(x)=\prod_{i=1}^{2d}(x-a_i)$$
avec des $a_i$ distincts, les points de branchement sont les racines $a_i$ et les coupures peuvent être prises comme reliant $\underset{p=1}{\overset{d}{\bigcup}}[a_{2p-1},a_{2p}]$:
 \begin{center} \label{Coupures}
	\includegraphics[height=5cm]{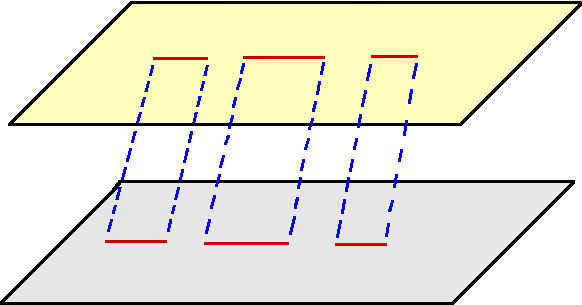}

	\underline{Figure 18}: Exemple de coupures dans le cas d'une équation algébrique. 
\end{center}
Lorsque deux $a_i$ coïncident, le point de branchement devient alors dégénéré et le nombre de coupures diminue. Dans le cas général, on peut récrire:
$$y^2=U(x)=Q^2(x)\prod_{i=1}^{2m}(x-a_i)$$
où $Q(x)$ est un polynôme. Les points de branchements sont de nouveau les $a_i$ restants et les coupures peuvent être choisies comme:
$\underset{p=1}{\overset{m}{\bigcup}}[a_{2p-1},a_{2p}]$. Le genre de la courbe reste quant à lui toujours donné par $m-1$. Il est alors facile de remarquer que $0 \leq g_{\text{alg}} \leq d-1$.

Les notions de genre, de coupures et de points de branchements se généralisent alors de la façon suivante. Notons $s_i$ les zéros de la fonction $\psi(x)$. Alors si $\psi$ présente un phénomène de Stokes, elle possède une infinité de zéros qui ne peuvent s'accumuler que le long des demi-lignes de Stokes où l'asymptotique est discontinu. Dans le cas générique, seul le secteur $0$ est singulier pour $\psi_0$ (car sinon, on se retrouve dans la situation où $\psi_0$ est également sous-dominante dans un autre secteur, disons $i$. Dans ce cas cela signifie qu'il existe une solution intégrable du secteur $0$ au secteur $i$ à l'équation de Schrödinger ce qui n'est pas le cas pour une équation de Schrödinger générique), et la situation se présente sous la forme:
  \begin{center} \label{Coupures2}
	\includegraphics[height=5cm]{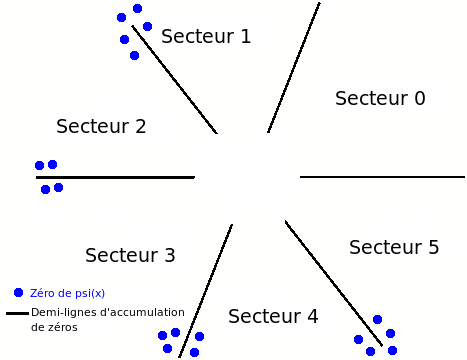}

	\underline{Figure 19}: Demi-lignes d'accumulation de zéros dans le cas générique 
\end{center}
Dès lors, il est possible de définir \textbf{les coupures comme une paire de deux demi-lignes d'accumulation de zéros}. Cette appariement présente un caractère arbitraire, qui correspond dans le cas algébrique au choix de regrouper les points de branchements pour créer les coupures. Le choix le plus naturel est alors de regrouper par paires deux demi-lignes d'accumulation de zéros consécutives. Tout comme dans le cas algébrique, \textbf{le genre est alors défini comme le nombre de coupures moins un}. Dans le cas générique, toutes les demi-lignes de Stokes accumulent des zéros (sauf celle délimitant le secteur $0$) et donc le genre est maximal. Cela dit, tout comme dans le cas algébrique, il se peut que l'équation de départ soit singulière et que $\psi(x)$ présente d'autres demi-lignes de Stokes n'accumulant pas de zéros. Le genre de la courbe quantique diminue alors d'une unité à chaque fois comme dans le cas algébrique. Quoiqu'il en soit, le genre de la courbe quantique satisfait les inégalités:
\beq -1 \leq g \leq d-1 \eeq
Le cas où $g=-1$ correspond au cas singulier où $\psi(x)$ est polynômiale (i.e. n'a pas de phénomène de Stokes), il correspond au cas algébrique où $y^2=Q^2(x)$ qui ne présente alors pas beaucoup d'intérêt. Néanmoins, dans le cas quantique, ce cas existe et demande un traitement particulier donné dans l'article \textbf{[III]} présenté en annexe \ref{Article[III]}. La notion de points de branchements est quant à elle plus floue dans le cas quantique. En effet, seule une des extrémités des demi-lignes d'accumulation de zéros (celle en direction de l'infini) est bien définie. Toutefois, on peut interpréter les points de branchements comme les directions asymptotiques dans lesquelles les zéros de $\psi$ s'accumulent.

\item \textbf{Cycles}

Dans le cadre de la géométrie algébrique, on sait que l'on peut choisir une base de $2g_\text{alg}$ cycles d'homologie indépendants $\mathcal{A}_i$ et $\mathcal{B}_j$ sur la surface de Riemann de genre $g_\text{alg}$ décrivant notre équation algébrique. Cette base vérifie les propriétés de croisement:
\bea \mathcal{A}_i\cap\mathcal{B}_j&=&\delta_{i,j}\cr
\mathcal{A}_i\cap\mathcal{A}_j&=&\emptyset\cr
\mathcal{B}_i\cap\mathcal{B}_j&=&\emptyset\eea
Le choix canonique est de prendre les cycles $\mathcal{A}_i$ entourant chacune des $g-1$ premières coupures, tandis que les cycles $\mathcal{B}_i$ traversent les feuillets et se rejoignent dans la dernière coupure. Ce choix est bien sûr arbitraire, et n'importe quel autre choix de base indépendante et respectant les conditions de croisement est possible:
 \begin{center} \label{Cycles}
	\includegraphics[height=5cm]{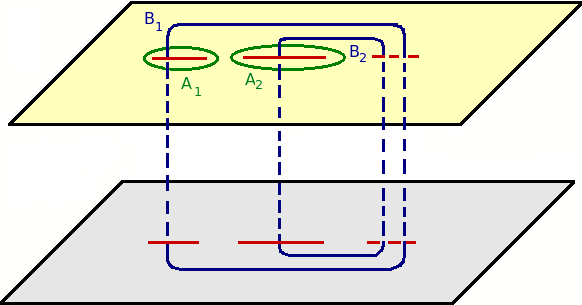}

	\underline{Figure 20}: Exemple de cycles dans le cas d'une équation algébrique hyperelliptique. 
\end{center}

Dans le cas quantique, le choix des $\mathcal{A}$-cycles est similaire: il consiste à choisir $g$ chemins reliant les différents infinis physiques entre eux. A nouveau, le choix canonique est d'entourer les coupures en restant dans le feuillet physique. Si les coupures ont été choisies consécutivement, la situation est alors décrite par:
 \begin{center} \label{Cycles2}
	\includegraphics[height=8cm]{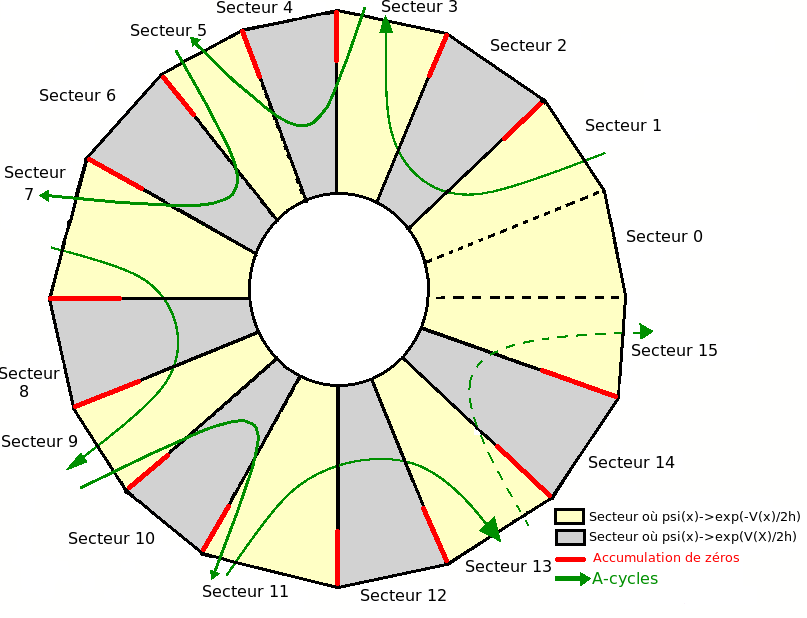}

	\underline{Figure 21}:Exemple de $\mathcal{A}$-cycles dans le cas d'une courbe quantique. 
\end{center}

Le choix des $\mathcal{B}$-cycles suit la même construction que dans le cas algébrique, le chemin part de la dernière coupure, i.e. du secteur $0$, traverse le $\mathcal{A}$-cycle correspondant pour passer dans l'autre feuillet, puis revient. La seule différence est qu'ici le retour impose une nouvelle fois de traverser le $\mathcal{A}$-cycle. Finalement la situation peut être visualisée comme:
 \begin{center} \label{Cycles3}
	\includegraphics[height=8cm]{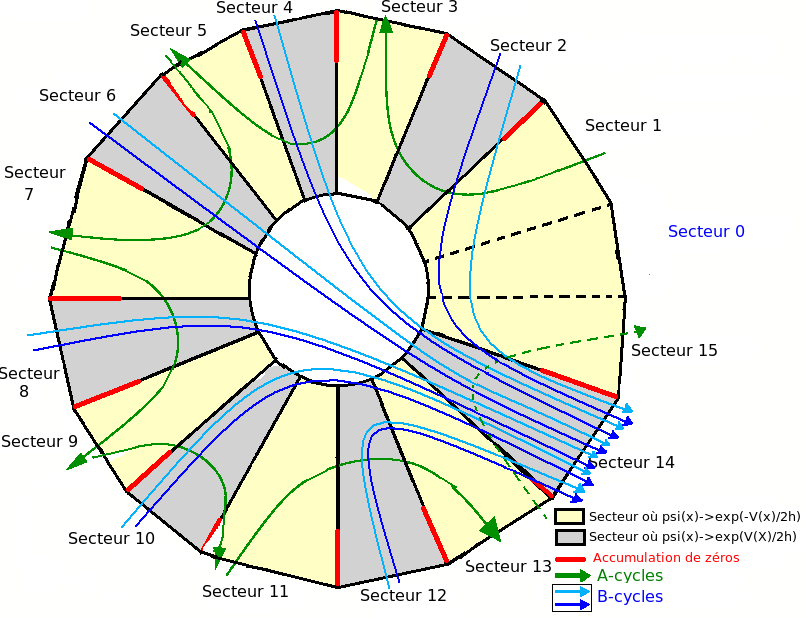}

	\underline{Figure 22}: Exemple de $\mathcal{B}$-cycles dans le cas d'une courbe quantique. 
\end{center} 
Notons que cette fois-ci on a les relations:
\bea \mathcal{A}_i\cap\mathcal{B}_j&=&2\delta_{i,j}\cr
\mathcal{A}_i\cap\mathcal{A}_j&=&\emptyset\cr
\mathcal{B}_i\cap\mathcal{B}_j&=&\emptyset\eea

Par ailleurs, dans le cas algébrique comme dans le cas quantique, il est possible de définir des $\hat{\mathcal{A}}$-cycles et des $\hat{\mathcal{B}}$-cycles pour des coupures dégénérées (i.e. des demi-lignes sans accumulation de zéros ou des zéros multiples dans le cas algébrique). Dans le cas algébrique, ils correspondent à des cycles pincés:
\begin{center}\label{Cycles33}
	\includegraphics[height=5cm]{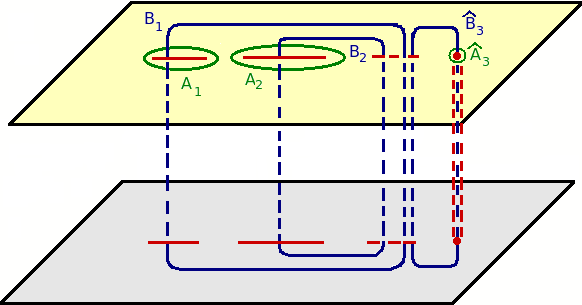}

	\underline{Figure 23}: Exemple de cycles pincés dans le cas d'une équation algébrique hyperelliptique. 
\end{center}

Dans le cas quantique, les $\hat{\mathcal{A}}_i$-cycles dégénérés sont des chemins partant de l'infini $0$ et allant dans le secteur dégénéré. Ils peuvent être visualisés de la manière suivante:
 \begin{center} \label{Cycles4}
	\includegraphics[height=8cm]{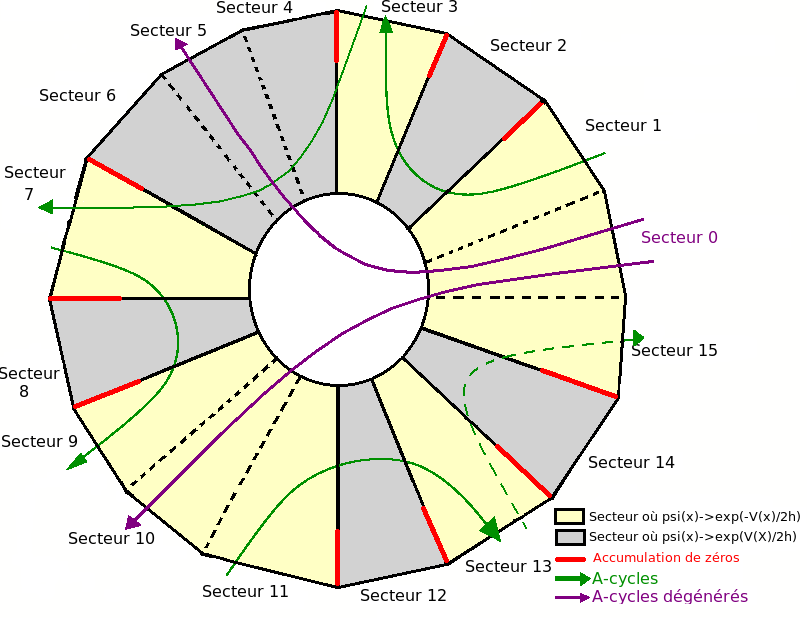}

	\underline{Figure 24}: Exemple de $\hat{\mathcal{A}}$-cycles dans le cas d'une courbe quantique. 
\end{center} 
Notons qu'il y a toujours exactement $g$ $\mathcal{A}$-cycles non-dégénérés et au total $d-1$ $\mathcal{A}$-cycles (dégénérés ou non) indépendants.

\item \textbf{Formes holomorphes}

En géométrie algébrique, les formes holomorphes sur une surface de Riemann de genre $g_\text{alg}$ sont les fonctions ne présentant aucun pôle sur la surface. Nous allons généraliser cette notion dans notre cas quantique de la façon suivante:
soit $h_k(x)$ une base des polynômes de degré inférieur à $d-2$. Intéressons nous d'abord aux fonctions:
\beq \label{tito}
v_k(x) = \frac{1}{\hbar\, \psi^2(x)}\,\int_{\infty_0}^x h_k(x')\,\psi^2(x')\,dx'
\virg
\deg h_k\leq d-2.
\eeq
Grâce aux propriétés dans les différents secteurs à l'infini de la fonction $\psi(x)$, on peut montrer (\textbf{[IV]} en annexe \ref{Article[IV]}) que les intégrales suivantes sont bien définies, et ne dépendent que de la classe d'homologie des contours $\mathcal{A}$.
\beq
I_{k,\alpha} = \oint_{\acycle_\alpha} v_k(x)\,dx
\virg \alpha=1,\dots,g, \, k=1,\dots,d-1.
\eeq

Pour les cas de cycles dégénérés $\hat{\cal{A}}_\alpha$, de telles intégrales divergeraient, on préférera donc prendre:
\beq
I_{k,\alpha} = \oint_{\hat{\acycle}_\alpha} h_k(x)\, \psi^2(x)\,dx
\virg \alpha=g+1,\dots,d-1, \, k=1,\dots,d-1.
\eeq

La matrice $I_{k,\alpha}$ avec $k,\alpha=1,\dots,d-1$ est une matrice carrée donnant une connexion entre l'ensemble des chemins \{ $\acycle_\alpha,\hat\acycle_\alpha$ \} et l'espace vectoriel des polynômes de degré inférieur à $d-2$. On peut donc choisir la base $h_k$, duale des $\acycle$-cycles, c'est-à-dire satisfaisant les relations:
\beq\label{eqdualvkacycle}
I_{k,\alpha}=\delta_{k,\alpha}.
\eeq

\bigskip

En choisissant cette base, on obtient alors les relations:
\beq \label{orthogonalbasis}
\forall\, i=1,\dots,\genus ,j=1,\dots,d-1\, , \qquad \quad
 \oint_{\mathcal{A}_i} v_j(x)\,\, dx = \delta_{i,j}
\eeq
\beq
\forall i=g+1,\dots,d-1, \, j=1,\dots,d-1 : \oint_{\hat{\mathcal{A}}_i} h_j(x)\, \psi^2(x)\,dx =\delta_{j,i}
\eeq
Par ailleurs, par sa définition même, on peut montrer aisément que les fonctions $v_k(x)$ avec $k\leq g$ se comportent comme
\beq
\forall \, k=1,\dots, g, \qquad v_k(x)=O(x^{-2})
\eeq
dans tous les secteurs à l'infini. Elles possèdent donc toutes les propriétés requises pour être la généralisation des formes holomorphes. Enfin, on peut facilement montrer que dans la limite $\hbar \to 0$ on retrouve:
\beq
v_k(x) \mathop{\sim}_{\hbar \to 0} \frac{\pm h_k(x)}{ \sqrt{U(x)}} 
\eeq
qui sont effectivement les formes holomorphes de la géométrie algébrique pour la courbe $y^2=U(x)$

\item \textbf{Matrice des périodes de Riemann}

En géométrie algébrique, une fois les cycles et les formes holomorphes $v_k$ définis, la matrice des périodes de Riemann de taille $g \times g$ est définie par:
\beq \label{tau}\forall 1 \leq i,j \leq g:
\tau_{i,j} \stackrel{{\rm def}}{=} \oint_{\bcycle_i} v_j(x)\,\, dx.
\eeq 
En effet, les formes holomorphes étant normalisées sur les $\mathcal{A}$-cycles, (\ref{orthogonalbasis}), il est naturel de s'intéresser à leurs intégrales sur les cycles duaux $\mathcal{B}_j$. Le théorème de Riemann sur les surfaces de Riemann énonce alors que la matrice des périodes $\tau$ est symétrique. Notons qu'étant donnée la définition de la matrice $\tau$, ce résultat n'est pas du tout évident. Dès lors, si notre généralisation quantique se veut correcte, elle se doit de maintenir un tel résultat. Dans l'article \textbf{[IV]} (section 3.4 de l'annexe \ref{Article[IV]}), il est montré que:
\begin{theorem} 
La matrice $\tau$ quantique définie par $\tau_{i,j} \stackrel{{\rm def}}{=} \oint_{\bcycle_i} v_j(x)\,\, dx$ (où les $v_j(x)$ sont définis par \ref{tito}), est symétrique.
\end{theorem}

Notons que ce résultat est hautement non-trivial compte-tenu des définitions précédemment introduites et constitue donc un premier pas important de la théorie.

\item  \textbf{Les fractions de remplissage}

Les fractions de remplissage jouent un rôle essentiel dans la théorie des matrices aléatoires, car elles indiquent les différentes proportions de valeurs propres se retrouvant sur chaque coupure (intervalles) dans la limite $N \to \infty$ de la mesure d'équilibre (\ref{densiteequilibre}). En dehors du cadre des modèles de matrices, elles sont définies dans le contexte de la géométrie algébrique par:
\beq \forall\, 1\leq i\leq g_\text{alg}\,:\, \epsilon_i\mathop{=}^{\text{def}}\frac{1}{2i\pi}\oint_{\mathcal{A}_i}ydx\virg \forall\, i>g_\text{alg}\,: \, \epsilon_i=0 \eeq
Dans notre cas quantique, les fractions de remplissage $\epsilon_1,\dots,\epsilon_d$ sont définies par:
\begin{definition} Les fractions de remplissage sont définies par:
\beq
\forall\, 1\leq \alpha\leq g\,:\,
\epsilon_\alpha = \frac{1}{2i\pi}\,\oint_{\acycle_\alpha} \left(\om(x)-\frac{t_0}{x}\right) +\frac{t_0 n_\alpha}{(d+1)} 
\eeq
où l'entier $n_\alpha$ compte la moitié des demi-lignes de Stokes entourées par le cycle $\mathcal{A}_\alpha$, ce qui est équivalent à dire que $\frac{n_\alpha}{d+1}$ représente la fraction angulaire du plan complexe défini par le cycle $\mathcal{A}_\alpha$.

Pour $g+1\leq \alpha \leq d-1$ la définition est:
\beq
g+1\leq \alpha \leq d-1\,:\,
\epsilon_\alpha =0
\eeq
et pour $\alpha=d$, nous choisissons un cycle  $\mathcal{A}_d$, (non-indépendant des autres $\mathcal{A}_i$) qui entoure les autres zéros $s_i$ qui ne sont pas entourés par les cycles $\acycle_1,\dots,\acycle_g$. Une fois ce contour choisi, nous définissons alors:
\beq\epsilon_d =\frac{1}{2i\pi}\,\oint_{\acycle_d}\left(\hbar\frac{\psi'(x)}{\psi(x)}+\frac{V'(x)}{2}-\frac{t_0}{x}\right) +\frac{t_0 n_d}{(d+1)} 
\eeq
\end{definition}

Ces fractions de remplissage satisfont trivialement à la relation $\underset{\alpha=1}{\overset{d}{\sum}} \epsilon_\alpha=t_0$ qui est également vraie dans le cas algébrique. Notons cependant que dans le cas quantique, les fractions de remplissage apparaissent arbitraires car dépendantes du choix des contours $\mathcal{A}$ relativement aux zéros de $\psi(x)$. En effet, si l'on déforme le contour $\mathcal{A}_\alpha$ pour englober un zéro $s_i$ supplémentaire, la fraction de remplissage correspondante $\epsilon_\alpha$ augmentera de $\hbar$ (mais une autre baissera de $\hbar$). Leur interprétation pour les modèles de matrices devant être indépendante du choix précis des contours, elle ne peut donc avoir de sens qu'à un multiple entier de $\hbar$ près. 

\item \textbf{Le noyau de récurrence $K(x_0,x)$}

Le noyau de récurrence $K(x_0,x)$ n'est pas une quantité standard de géométrie algébrique, mais est un ingrédient essentiel dans la définition des invariants symplectiques $F_g$ d'Eynard et Orantin. En effet, comme nous le verrons par la suite, il permet l'écriture d'une récurrence ``topologique'' permettant de calculer les fonctions de corrélation pour un ordre donné à partir des autres fonctions d'ordre inférieur. Pour le cas algébrique, nous avons vu précédemment que ce noyau est défini par $\frac{dE_q(p)}{y(q)-y(\overline{q})}=\frac{\int_q^{\overline{q}}B(\xi,p)}{2(y(q)-y(\overline{q}))}$ (Cf. \ref{Recurrence}), c'est-à-dire à partir (par primitivation au voisinage d'un point de branchement) du noyau de Bergman $B(\xi,p)$ de la surface de Riemann.
Dans le cas quantique, il est plus naturel de définir d'abord le noyau de récurrence puis ensuite par dérivation de construire l'équivalent du noyau de Bergman (dont il faudra naturellement vérifier les propriétés). 
\begin{definition}
Le noyau de récurrence est défini par:
\beq
K(x,z) = \frac{1}{\hbar\psi^2(x)}\, \int_{\infty_0}^x \psi^2(x')\,{dx'\over x'-z} - \sum_{\alpha=1}^{d-1} v_\alpha(x)\,C_\alpha(z)
\eeq
avec:$ \forall \, \alpha=1,\dots,g$:
\beq
\hbar C_\alpha(z) 
=\oint_{\acycle_\alpha}\, \frac{dx''}{\psi^2(x'')}\,\int_{\infty_0}^{x''} \psi^2(x')\,\frac{dx'}{x'-z}
\eeq
et $ \forall \, \alpha=g+1,\dots,d-1$:
\beq
C_\alpha(z) 
=\oint_{\hat{\cal{A}}_\alpha}\,  \psi^2(x')\,\frac{dx'}{x'-z}.
\eeq
\end{definition}

Nous renvoyons le lecteur à \textbf{[IV]} situé en annexe \ref{Article[IV]} pour les preuves d'existence des intégrales, pour les subtilités de définition des fonctions $C_\alpha(z)$ ainsi que pour les démonstrations techniques des propriétés à venir. Ce noyau possède les propriétés suivantes:

\begin{theorem}
Le noyau $K(x,z)$ possède les propriété suivantes:

$\bullet$ Il possède une discontinuité le long d'un chemin reliant $\infty_0$ a $x$ dont le saut est donné par: 
\beq
\delta K(x,z) = \frac{2i\pi}{\hbar} \,\,\frac{\psi^2(z)}{\psi^2(x)}
\eeq

$\bullet$ Pour $\alpha=1,\dots g$, il possède une discontinuité de $\infty_0$ à un point $P_\alpha \in \mathcal{A}_\alpha$ donnée par le saut:
\beq
\delta K(x,z) = \frac{2i\pi\,\,\psi^2(z)}{ \hbar}\,\,\oint_{\acycle_\alpha} \frac{dx''}{\psi^2(x'')}v_\alpha(x)
\eeq
et une discontinuité sur le cycle $\mathcal{A}_\alpha$ donnée par le saut:
\beq\label{discKxzAalpha}
\delta K(x,z) = \frac{2i\pi\,\,\psi^2(z)}{ \hbar}\,\,\int_{P_\alpha}^z \frac{dx''}{ \psi^2(x'')} v_\alpha(x)
\eeq

$\bullet$ Pour $\alpha=g+1,\dots,d-1$, il possède une discontinuité sur le cycle $\hat{\mathcal{A}}_\alpha$ donnée par:
\beq
\delta K(x,z) = {2i\pi\,\,\psi^2(z)}v_\alpha(x)
\eeq

$\bullet$ Le comportement au voisinage de l'infini est donné par:
\bea
K(x,z) \mathop{\sim}_{x\to\infty}\, O(x^{-2}) \,\,\, \text{dans tous les secteurs}\cr
K(x,z) \mathop{\sim}_{z\to\infty}\, O(z^{-d}) \,\,\, \text{dans tous les secteurs}
\eea

$\bullet$ Pour $\alpha=1,\dots,g$, et $z$ du côté des cycles $\acycle_\alpha$ ne contenant pas $\infty_0$:
\beq
\oint_{\acycle_\alpha} K(x,z)\,dx = 0
\eeq
\end{theorem}

On voit donc que ce noyau possède beaucoup de lignes de discontinuité. Cela n'est pas sans rappeler le fait que dans le cas algébrique,
la quantité $\frac{dE_q(p)}{y(q)-y(\overline{q})}$ n'est définie qu'au voisinage des points de branchement mais pas de façon globale. A partir de ce noyau $K(x_0,x)$, on peut définir l'équivalent des formes de $3^\text{ième}$ espèce, ainsi que l'équivalent du noyau de Bergmann pour le cas quantique.

\item \textbf{Les formes de troisième espèce: le noyau $G(x_0,x)$}

\begin{definition}
A partir du noyau précédent, on définit le noyau $G(x_0,s)$ par la formule:
\beq\label{eqdefG}
G(x,z) =- \hbar\,\psi^2(z)\, \partial_z\, \frac{K(x,z)}{\psi^2(z)}=2\hbar\frac{\psi'(z)}{\psi(z)}K(x,z)-\hbar \partial_z\,K(x,z)
\eeq
\end{definition}

Ce nouveau $G(x_0,s)$ possède de meilleures propriétés de régularité que le noyau précédent $K(x_0,x)$. En effet, il est facile de voir que toutes les discontinuités proportionnelles à $\psi^2(z)$ vont être annulées lors de la dérivation. 
\begin{theorem}
Les propriétés du noyau $G(x_0,s)$ sont alors les suivantes (Cf. \textbf{[IV]}, section 4.2, annexe \ref{Article[IV]}):

$\bullet$ $G(x,z)$ est une fonction analytique de $x$, avec un pôle simple en $x=z$ de résidu $-1$, et des pôles doubles aux $s_{j}$ (zéros de $\psi(x)$) sans résidu, et potentiellement une singularité essentielle à l'infini.

$\bullet$ $G(x,z)$ est une fonction analytique de $z$, avec un pôle simple en $z=x$ de résidu $+1$, des pôles simples pour $z=s_{j}$, et une discontinuité le long des $\acycle_\alpha$-cycles pour $\alpha=1,\dots,g$.
\beq
\delta G(x,z) = -2i\pi \, v_\alpha(x)
\eeq
En particulier, elle n'a pas de discontinuité le long des autres cycles $\hat{\cal{A}}_\alpha$.

$\bullet$ Les limites à l'infini sont données par: 
\beq
G(x,z)=O(1/x^2)
\eeq

$\bullet$ Pour $\alpha=1,\dots,g$, et $z$ du côté des cycles $\acycle_\alpha$ ne contenant pas $\infty_0$:
\beq
\oint_{\acycle_\alpha} G(x,z)\,dx = 0
\eeq
\end{theorem}

Ce noyau constitue une bonne généralisation des formes de troisième espèce en géométrie algébrique. En effet, en géométrie algébrique, ces formes possèdent les propriétés d'être analytiques partout sur la surface de Riemann (c'est-à-dire en dehors des coupures) et de posséder un unique pôle simple. Par ailleurs, ces formes sont également normalisées sur les $\mathcal{A}$-cycles. Dans notre cas ``quantique'', $z \mapsto G(x,z)$ est analytique partout à l'exception des $\mathcal{A}$-cycles et des $s_i$ qui constituent justement les coupures ``quantiques''. Enfin, elle possède également un pôle simple en $z=x$ et elle est correctement normalisée sur les $\mathcal{A}$-cycles. Par ailleurs, si l'on effectue la limite $\hbar \to 0$ (i.e. on remplace $\psi \sim e^{\frac{\pm 1}{ \hbar}\int\sqrt{U}}$) on retrouve:
\beq
G(x,z) \sim 2\sqrt{U(z)} K(x,z) \sim \frac{1}{ x-z}\,\frac{\sqrt{U(z)}}{ \sqrt{U(x)}} - 2\sum_\alpha\, v_\alpha(x)\,C_\alpha(z)\sqrt{U(z)}
\eeq
Dans ce cas, la forme $G(x,z)dx$ possède donc un pôle simple en $x=z$, de résidu $+1$ dans le feuillet physique, et de résidu $-1$ dans l'autre feuillet. Elle est également normalisée sur les $\acycle$-cycles: $\oint_{\acycle_i} G(x,z)dx=0$. On retrouve ainsi les toutes les propriétés usuelles des formes de $3^\text{ième}$ espèce de la géométrie algébrique.

\item \textbf{Le noyau de Bergman}

Le noyau de Bergman (ou différentielle fondamentale de deuxième espèce) est une quantité fondamentale en géométrie algébrique. En effet, il constitue une fonction intrinsèque pouvant être définie sur n'importe quelle surface de Riemann. De plus, dans la démarche de Eynard et Orantin \cite{OE}, il sert à construire les deux noyaux précédents. Ici, notre démarche étant inverse, le noyau de Bergman est défini à partir du noyau $G(x,z)$ de la même façon que le noyau de Bergman est relié aux formes de troisième espèce en géométrie algébrique:
\begin{definition} Le noyau de Bergman est défini par la formule:
\beq
B(x,z) \mathop{=}^{\text{def}} -\frac{1}{ 2}\, \partial_z\, G(x,z).
\eeq
\end{definition}
Pour pouvoir prétendre être une bonne généralisation du noyau de Bergman de la géométrie algébrique, il faut que ce nouveau noyau en vérifie les propriétés fondamentales. La vérification est donnée dans \textbf{[IV]} (section 4.3, annexe \ref{Article[IV]}).

\begin{theorem} Le noyau de Bergman satisfait les propriétés suivantes:

$\bullet$ $B(x,z)$ est une fonction analytique en $x$ et en $z$, avec un pôle double en $x=z$ sans résidu, et des pôles doubles aux $s_{j}$ sans résidu, ainsi que potentiellement une singularité essentielle à l'infini. On notera tout particulièrement qu'il n'a pas de discontinuité sur les $\mathcal{A}$-cycles.

$\bullet$ $B(x,z)$ est une fonction symétrique de ses variables: $B(x,z)=B(z,x)$. Cette propriété est hautement non-triviale compte-tenu de sa définition, et constitue une propriété fondamentale du noyau de Bergman.

$\bullet$ $B(x,z)$ se comporte dans tous les secteurs de l'infini comme:
\beq
B(x,z) \mathop{=}_{x \to \infty} O(1/x^2) \virg  B(x,z) \mathop{=}_{x \to \infty} O(1/z^2)
\eeq
dans tous les secteurs.

$\bullet$ Le noyau $B(x,z)$ est normalisé convenablement sur les cycles:
\beq
\forall \, \alpha=1,\dots,g \, :\, \oint_{\acycle_\alpha} B(x,z)\,dx = 0
\virg
\oint_{\acycle_\alpha} B(x,z)\,dz = 0
\eeq
et
\beq \forall \, \alpha=1,\dots,g \, :\,
\oint_{\bcycle_\alpha} B(x,z)\,dz = 2i\pi v_\alpha(x)
\eeq
\end{theorem}

$\bullet$ Enfin une dernière propriété en lien avec les modèles de matrice (et qui d'après les travaux de Eynard et Orantin \cite{OE} et également vraie en géométrie algébrique) est que ce noyau $B(x,z)$ satisfait les équations de boucles suivantes:
\begin{theorem} Le noyau de Bergman satisfait les équations (la preuve est dans \textbf{[IV]}, annexe A):
\beq\label{loopeqBx}
(2\frac{\psi'(x)}{ \psi(x)}+\partial_x)\,\left(B(x,z)-\frac{1}{ 2(x-z)^2}\right) + \partial_z\,\frac{\frac{\psi'(x)}{ \psi(x)}-\frac{\psi'(z)}{\psi(z)}}{x-z} = P_2^{(0)}(x,z)
\eeq
où $P_2^{(0)}(x,z)$ est un polynôme en $x$ de degré au plus $d-2$.
\beq\label{loopeqBz}
(2\frac{\psi'(z)}{ \psi(z)}+\partial_z)\,\left(B(x,z)-\frac{1}{ 2(x-z)^2}\right) + \partial_x\,\frac{\frac{\psi'(x)}{ \psi(x)}-\frac{\psi'(z)}{\psi(z)}}{x-z} =\td{P}_2^{(0)}(z,x)
\eeq
où  $\td{P}_2^{(0)}(z,x)$ est en polynôme en $z$ de degré au plus $d-2$.
\end{theorem}

Ces équations de boucles nous permettront par la suite d'identifier $B(x,z)$ avec la résolvante $W_2^{(0)}(x,z)$ des modèles de matrices qui satisfait cette équation (\ref{loopeqPng}).

\end{enumerate}

\section{Une solution des équations de boucles pour $\beta$-quelconque}

$\bullet$ Une fois la généralisation des quantités fondamentales de géométrie algébrique effectuée, il devient facile de trouver une solution des équations de boucles pour $\beta$-quelconque. Il suffit pour cela de reprendre la même formule de récurrence topologique que celle développée par Eynard et Orantin (\cite{OE}) en remplaçant chacune des quantités par sa généralisation ``quantique''. Ainsi, on définit les résolvantes de la façon suivante:
\begin{definition} Les résolvantes sont définies par la récurrence:
\beq
W_1^{(0)}(x) = \om(x)
\virg
W_2^{(0)}(x_1,x_2)=B(x_1,x_2)
\eeq
\bea\label{mainrecformula}
 W^{(g)}_{n+1}(x_0,J)  
&=&   \frac{1}{ 2i\pi}\,\sum_i \int_{{\cal C}_i}\, dx\,\,  K(x_0,x)\, \Big( \ovl{W}_{n+2}^{(g-1)}(x,x,J) \cr
&& + \sum_{h=0}^g\sum'_{I\subset J} {W}_{|I|+1}^{(h)}(x,x_I) {W}_{n-|I|+1}^{(g-h)}(x,J/I) \Big)\cr
\eea
où $J$ est une notation compacte pour les variables $J=\{ x_{1},\dots,x_{n} \}$, et où $\sum\sum'$ signifie que l'on exclut les termes $(h=0,I=\emptyset)$ et $(h=g,I=J)$ (pour obtenir une véritable relation de récurrence). Par ailleurs, nous avons effectué une translation des fonctions de corrélation:
\beq
\ovl{W}_{n}^{(g)}(x_1,...,x_n) = W_{n}^{(g)}(x_1,...,x_n) - \frac{\delta_{n,2}\delta_{g,0}}{2}\, \frac{1}{ (x_1-x_2)^2}
\eeq
Enfin, les points $x_0$ ainsi que tous les autres $x_i$ sont supposés être du même côté des $\acycle$-cycles que $\infty_0$. Le contour
${\cal C}_i$ est un contour qui entoure les demi-lignes d'accumulation de zéros $L_i$ (i.e. les points de branchements quantiques), et qui est choisi de telle façon que chaque $s_j$ soit entouré exactement une seule fois et que le contour n'intersecte aucun des $\mathcal{A}_\alpha$-cycles pour $1 \leq \alpha \leq g$.
\end{definition}

\medskip

Cette formule est à comparer avec le cas de la géométrie algébrique classique \cite{OE} où:
$$\Phi(p)=\int^p y\,dx$$
$$W_2^{(0)}(p_1,p_2)=B(p_1,p_2)$$
\bea W^{(g)}_{k+1}(p,p_K) &=& \sum_i\Res_{q\to a_i}\frac{dE_q(p)}{y(q)-y(\overline{q})}\cr
&&\left(\sum_{h=0}^{g}\sum_{I\in J}'
W_{|I|+1}^{(h)}(q, p_I)W^{(g-m)}_{k-|I|+1}(q, p_{J/I}) +W^{(g-1)}_{k+2} (q, q, p_J)
\right)\cr
\eea

On constate donc que les deux formules sont identiques et sont formées d'une somme de résidus autour des points de branchements. En d'autres termes, on constate donc que la forme de la récurrence topologique est indépendante de $\hbar$. De plus, il est montré dans \textbf{[IV]}, annexe C, le théorème suivant: \begin{theorem}
Les fonctions de corrélation $W_n^{(g)}$ définies par \ref{mainrecformula} vérifient les équations de boucles du modèle à une matrice avec $\beta$-quelconque \ref{loopeqPng} et constituent donc une solution aux équations de boucles du modèle à une matrice avec $\beta$-quelconque.
\end{theorem}

\medskip
\medskip

$\bullet$ Dans les deux cas, les démonstrations (\cite{OE}) font appel à l'identité bilinéaire de Riemann qui peut être énoncée de la façon suivante:

\begin{theorem} Identité bilinéaire en géométrie algébrique:

 Soient $\om_1$ et $\om_2$ deux formes méromorphes et $\mathcal{A}_1,\dots,\mathcal{A}_g, \mathcal{B}_1,\dots,\mathcal{B}_g$ une base de cycles d'une surface de Riemann de genre $g$, alors en définissant la fonction $\Phi_1(p)=\int_{p_0}^p \om_1$ où $p_0$ est un point arbitraire de la surface de Riemann:
\beq \Res_{p \to \text{tous les poles}} \Phi_1 \om_2 =\frac{1}{2i\pi}\sum_{j=1}^g\int_{\mathcal{A}_j}\om_1 \int_{\mathcal{B}_j} \om_2 -\int_{\mathcal{A}_j}\om_2 \int_{\mathcal{B}_j} \om_1 \eeq 
\end{theorem}
Dans le cas quantique, cette identité peut être généralisée de la manière suivante (Cf. \textbf{[IV]}, section 4.4.4 de l'annexe \ref{Article[IV]}):
\begin{theorem} Considérons une fonction $f(x)$ de la forme:
\beq
f(x) = \frac{1}{ \hbar\psi^2(x)}\,\int_{\infty_0}^x \psi^2(x')\,g(x')\, dx' \,\, \Leftrightarrow \,\, \hbar(2\frac{\psi'(x)}{ \psi(x)}+\partial_x)\, f(x) = g(x)
\eeq
où $g(x')$ est un polynôme de degré inférieur à $d-1$. Alors on a l'identité bilinéaire de Riemann généralisée:
\bea
f(x) &=& \frac{1}{ 2i\pi}\,\oint_{{\cal C}} K(x,z)\,g(z)\,dz + \sum_\alpha v_\alpha(x) \oint_{\acycle_\alpha} f(z)\,dz\cr
&=&\frac{1}{ 2i\pi}\,\oint_{{\cal C}} G(x,z)\,f(z)\,dz + \sum_\alpha v_\alpha(x) \oint_{\acycle_\alpha} f(z)\,dz
\eea
où $\mathcal{C}$ est le contour de récurrence introduit dans \ref{mainrecformula}.
\end{theorem} 

Notons que cette identité généralise bien l'identité bilinéaire de Riemann traditionnelle de géométrie algébrique que l'on peut énoncer comme suit:
\beq \om (q)=\Res_{p \to poles \, de\, \om} dS_{p,p_0}(q)\om(p) +\sum_{i=1}^g d u_i(q)\oint_{\mathcal{A}_i} \om \eeq
où $dS_{p,p_0}(q)$ est la forme de troisième espèce et $d u_i$ sont les formes holomorphes. Dans notre formalisme, $G(x,z)$ joue le role de cette forme de troisième espèce et les $v_\alpha(x)$ jouent le role des formes holomorphes. Au final, la seule différence d'écriture se situe dans la présence d'une intégrale de contour $\mathcal{C}$ au lieu d'une somme sur les résidus.
\medskip
\medskip

$\bullet$ On peut également montrer que ces fonctions de corrélation définies par \ref{mainrecformula} satisfont les propriétés attendues pour des fonctions de corrélations. En particulier, on a le résultat suivant (Cf. \textbf{[IV]}, annexe B):
\begin{theorem} $W_n^{(g)}(x_1,\dots,x_n)$ est une fonction analytique avec des pôles sans résidu aux $s_i$, est une fonction symétrique de ses variables et se comporte en $O\left(\frac{1}{x_i^2}\right)$ dans tous les secteurs à l'infini.
\end{theorem}

\section{Vers des invariants symplectiques $F_g$ généralisés?}

Un résultat crucial de la récurrence topologique d'Eynard et Orantin est de pouvoir inverser la récurrence topologique de manière extrêmement simple par la formule:
\beq \label{recinverse} (2-2g-n)W_n^{(g)}(p_1,\dots,p_n)=\sum_{i}\Res_{q\to a_i}\Phi(q)W_{n+1}^{(g)}(p_1,\dots,p_n,q)\eeq
qui permet ``de remonter d'une variable''. En étendant cette propriété au cas $n=0$, les auteurs ont ainsi défini les nombres $F^{(g)}_{\beta=1}$:
\beq F^{(g)}_{\beta=1}(p_1,\dots,p_n)=\frac{1}{2-2g}\sum_{i}\Res_{q\to a_i}\Phi(q)W_{1}^{(g)}(q)\eeq
et ainsi pu montrer qu'il s'agissait d'invariants symplectiques, c'est-à-dire qu'ils ne changent pas sous une transformation de la courbe algébrique de départ $E(x,y)$ lors d'une transformation symplectique, (i.e. toute transformation des coordonnées $\td{x}=f(x,y) \,\,,\,\, \td{y}=g(x,y)$ laissant invariant la forme $dx \wedge dy$: $dx \wedge dy= d\td{x} \wedge d\td{y}$). Depuis leur découverte, ces invariants symplectiques ont connu de nombreuses applications en dénombrement, en théorie des cordes topologiques et permettent de définir une fonction $\tau$ et une hiérarchie intégrable \cite{OE}. Notons également que par définition, ils permettent d'écrire un développement perturbatif de la fonction de partition par la formule:
\beq \ln\,\, Z_{\beta=1}= \sum_{g=0}^\infty \left(\frac{N}{T}\right)^{2-2g}F^{(g)}_{\beta=1}\eeq
Dès lors, il apparait intéressant de pouvoir généraliser de tels nombres au cas quantique. Malheureusement, plusieurs étapes manquent à l'heure actuelle. Tout d'abord, la généralisation de \ref{recinverse} n'est que partielle à ce jour pour le cas à une matrice puisque l'on a seulement (Cf. \textbf{[IV]}, section 7, présenté en annexe \ref{Article[IV]}):
\begin{theorem}
Les fonctions $W_n^{(g)}$ satisfont la formule:
\beq \label{recinversebeta}
(2-2g-n-\hbar\,\partial_{\hbar})\,W_n^{(g)} = \left(t_0\,\partial_{t_0}+\sum_{k=1}^{d+1}\,t_k\,\partial_{t_k}+\sum_{i=1}^g \epsilon_i \partial_{\epsilon_i}\right)\,W_n^{(g)}=\hat H.\,W_{n+1}^{(g)}
\eeq
où $\hat H$ est un opérateur linéaire qui agit comme (et dont on peut montrer qu'il redonne le membre de droite de \ref{recinverse} dans la limite où $\hbar \to 0$):
\beq
\hat H.f(x) = t_0\,\int_{\infty_0}^{\infty_-} f +\sum_{j=1}^{d+1} \Res_{\infty_0}\, \frac{t_{j}\,x^j}{ j}\,f + \sum_{i=1}^g \epsilon_i\,\oint_{\bcycle_i} f.
\eeq
\end{theorem}

La différence principale avec le cas hermitien est alors la présence supplémentaire du facteur $\hbar \partial_\hbar$ dans le membre de gauche de \ref{recinversebeta} qui ne permet pas de définir les $F_g$ de façon unique dans le cas $n=0$. Pour pouvoir résoudre cette difficulté, il faudrait pouvoir exprimer $\hbar \partial_\hbar W_n^{(g)}$ comme un opérateur intégral agissant sur $W_{n+1}^{(g)}$ ce qui n'est pas connu à l'heure actuelle.  La seconde difficulté consiste ensuite à généraliser les résultats obtenus pour le cas du modèle à une matrice avec $\beta$ quelconque au cas du modèle à deux matrices avec $\beta$ quelconque. Nous travaillons à l'heure actuelle à la réalisation de cette étape et les résultats préliminaires semblent indiquer que cette généralisation est possible bien que le formalisme devienne plus technique. Enfin, la dernière étape, conditionnée à la réussite des deux précédentes, serait, une fois la définition des $F_g$ obtenue dans le cas des modèles à une et deux matrices, de trouver l'équivalent de la propriété d'invariance symplectique et de la démontrer. Il serait assez logique que cette propriété consiste en l'invariance des $F_g$ sous n'importe quelle reparamétrisation $\td{x}=f(x,\hat{y}) \,\,,\,\, \td{y}=g(x,\hat{y})$ conservant le commutateur $[\hat{y},x]=\hbar$ bien qu'il ne s'agisse ici que d'une conjecture extrêmement lointaine.

\section{Conclusions}

Dans ce chapitre, nous avons montré comment les quantités de géométrie algébrique ainsi que la récurrence topologique d'Eynard et Orantin utilisées pour résoudre les équations de boucles du modèle hermitien, peuvent être généralisées dans ce qui pourrait devenir de la géométrie algébrique ``quantique'' pour le cas du modèle à une matrice avec $\beta$ quelconque. Cependant, beaucoup de choses restent à faire dans ce nouveau domaine, en particulier trouver une généralisation des invariants symplectiques, traiter le cas des modèles à deux matrices, trouver et démontrer une généralisation de la propriété d'invariance symplectique. Par ailleurs, dans la méthode proposée, les fonctions de corrélations $W_n^{(g)}$ dépendent explicitement du choix de la solution $\psi(x)$ de l'équation différentielle \ref{ODE} qui constitue la courbe ``quantique''. Comprendre cette dépendance et le rôle de la solution $\psi(x)$ constitue donc une étape supplémentaire importante pour la résolution explicite des modèles de matrices avec $\beta$-quelconque les plus simples. Des applications de cette théorie sont également en cours de développement, comme par exemple: le dénombrement des surfaces non-orientables (le cas hermitien donnant celui des surfaces orientables tel que données dans (\cite{David}, \cite{David2}, \cite{BiPZ}, \cite{countingsurface}, \cite{countingsurface2})), et la théorie des cordes topologiques, en particulier de la fonction de partition de Nekrasov (\cite{Nekrasov} \cite{Nekrasov}, \cite{Nekrasov3}) et de la conjecture AGT (\cite{AGT}). Des liens avec les systèmes intégrables et la théorie des équations différentielles (puisque la méthode permettrait d'associer à toute équation différentielle ordinaire linéaire (\ref{ODE}) des invariants $F_g$) seraient également possibles bien qu'ils soient à l'heure actuelle encore hypothétiques. 

%% file: chap4.tex
\chapter{Théorie des cordes topologiques et modèles de matrices}
 \label{chap4}
\thispagestyle{empty}
 \selectlanguage{french}
\section{La théorie des cordes topologiques}

La théorie des cordes s'est développée en physique fondamentale dans le but de concilier les deux grandes théories de la physique moderne: la mécanique quantique (et la théorie des champs qu'elle implique) et la relativité générale. En effet, durant les cinquante dernières années, de nombreuses expériences ont permis de vérifier l'exactitude des deux théories: la théorie quantique des champs permettant de faire des prédictions sur le monde microscopique (laser, collision de particules, modèle standard,...) tandis que la relativité générale permet elle de faire des prédictions sur des corps en interaction gravitationnelle (avancée du périhélie de Mercure, correction pour le GPS, etc.). Cela dit, malgré la réussite de chacune des théories pour prédire efficacement les résultats dans leur domaine respectif, les deux formalismes mathématiques sont incompatibles lorsque l'on tente de quantifier la gravitation comme les autres champs. Certes, il est possible d'adopter une position pragmatique et de n'utiliser chacune des théories que dans son domaine respectif (le monde microscopique pour la mécanique quantique et les gros objets célestes pour la relativité générale). Après tout, la science n'a pas pour but d'être ``unifiée'' ou d'être ``belle'', mais juste de fournir des modèles explicatifs et prédictifs, ce que chaque branche réalise parfaitement pour les échelles d'énergie expérimentées jusqu'ici. Toutefois, il reste que certains phénomènes impliquent les deux échelles. Ainsi, les trous noirs, l'univers primordial, l'anomalie des sondes Pioneer, ou les futures collisions de particules à très hautes énergies du LHC sont autant d'expériences qui poussent (ou pousseront) chacune des théories à ses limites. Dès lors, il est nécessaire de trouver une théorie permettant de regrouper sous un même formalisme toutes les interactions, y compris l'interaction gravitationnelle. Cette quête de la théorie de l'unification est un des sujets qui anime la physique théorique depuis plus de cinquante ans, mais qui n'a actuellement toujours pas de solution acceptable. Si la théorie des cordes constitue un candidat pour réaliser l'unification, elle n'est plus aujourd'hui l'unique théorie en lice. Ainsi, la gravitation quantique à boucles («loop quantum gravity»), la géométrie non-commutative, la dynamique Newtonienne modifiée (MOND) figurent parmi les adversaires les plus avancés et sont en développement rapide. D'autres plus exotiques sont également en cours d'élaboration: temps discret, particules supplémentaires aux propriétés étranges, modification de la théorie des champs, etc. Hélas aucune d'entre elles n'a pour l'instant résolu le problème de la ``grande unification'' de façon satisfaisante. Parmi toutes ces théories, la théorie des cordes semble être l'un des candidats les plus sérieux. L'idée de départ repose sur le fait que ce que nous appelons particules élémentaires (électrons, quarks, neutrinos,...) ne seraient pas des particules ponctuelles mais de minuscules cordes vibrantes. Si l'idée de base semble séduisante, la cohérence mathématique impose immédiatement que le nombre de dimensions spatiales de notre univers n'est plus de $3$ mais de $10$. Dès lors, comment expliquer que nous ne ressentons pas les $6$ dimensions spatiales supplémentaires? La réponse à cette question est relativement simple: nous ne les ressentons pas car elles sont de tailles minuscules: échelle de Planck ou tout du moins une échelle suffisamment petite pour que les énergies accessibles par les technologies actuelles n'aient pas permis de les détecter. En revanche, lors de l'étude de phénomènes extrêmes (univers primordial), l'influence de ces dimensions supplémentaires pourrait être détectée. D'un point de vue mathématique, la question qui apparaît naturellement est de savoir comment décrire des dimensions ``de taille minuscule'', c'est-à-dire comment compactifier les dimensions supplémentaires? Pour visualiser les choses, le lecteur non spécialiste peut s'imaginer qu'un segment peut être replié en un cercle si l'on joint les deux bouts. Un ruban peut quant à lui être replié en un cylindre puis un tore, mais peut également être replié pour former un ruban de Moebius puis une bouteille de Klein. Le concept naturel en mathématique pour décrire la compactification est la notion de variétés. Ainsi, la sphère, le tore, la bouteille de Klein sont des variétés compactes, et l'on pourrait imaginer que les $6$ dimensions manquantes correspondent en fait à une variété de dimension $6$ compacte très enroulée sur elle-même. L'image la plus utilisée pour décrire ce type de variété est la suivante:

\begin{center} \label{CalabiYau}
	\includegraphics[height=5cm]{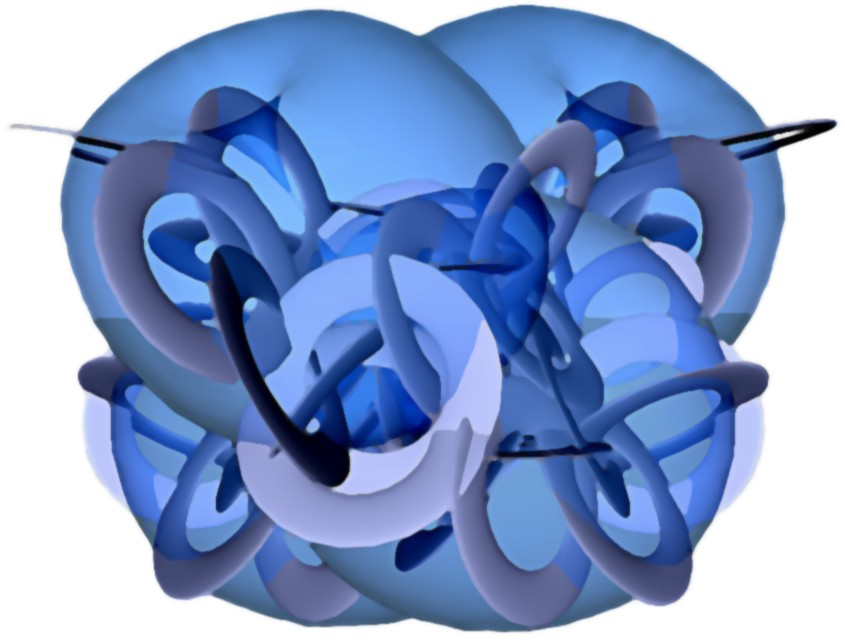}

\underline{Figure 25}: Représentation imagée d'une variété de Calabi-Yau
\end{center}

Le choix du type de variété utilisé pour la compactification se pose alors de manière cruciale. Sans rentrer dans les détails techniques (la théorie des cordes faisant à elle seule l'objet de thèses ou de livres entiers), les conditions impliquées par la physique et le formalisme considéré imposent de choisir une variété de type Calabi-Yau que nous allons décrire rapidement dans la prochaine section.

\section{Variété de Calabi-Yau}

Le but de cette section n'est pas de rentrer dans les détails géométriques de la construction des variétés de Calabi-Yau. Le lecteur interessé trouvera dans \cite{Klemm} une excellente introduction aux variétés de Calabi-Yau. Dans ce paragraphe, nous nous contenterons de donner les définitions et quelques propriétés pour aboutir à la symétrie miroir et à la conjecture de Bouchard, Klemm, Marino et Pasquetti (BKMP) et aux modèles de matrices.

\begin{definition}
 Une variété de Calabi-Yau est définie comme une variété kählérienne (i.e. une variété hermitienne M, c'est-à-dire une variété complexe munie d'une métrique hermitienne h, telle que la 2-forme $\om = - \Im\, h$ soit fermée) dont la première classe de Chern est nulle. De façon encore équivalente, un espace de Calabi-Yau de dimension complexe $n$\, (ce qui correspond à une dimension réelle $2n$\,) peut être vu comme une variété riemannienne d'holonomie réduite à $SU(n)$\, (le groupe d'holonomie d'une variété riemannienne de dimension réelle $2n$\, étant génériquement le groupe $SO(2n)$\,).
\end{definition}

Définir précisément les notions de classe de Chern et de groupe d'holonomie ainsi que leurs propriétés amènerait la discussion bien au delà du domaine des matrices aléatoires, nous nous contentons donc de citer \cite{corde1} pour informations. 

Un des théorèmes importants des variétés de Calabi-Yau a été la démonstration de la conjecture d'Eugène Calabi, formulée en 1957, par Shing-Tung Yau en 1977 (d'où le nom de variété de Calabi-Yau) de l'existence sur de telles variétés d'une métrique dont le tenseur de Ricci s'annule (et qui constitue un élément nécessaire pour la cohérence de la théorie physique). L'étude des variétés de Calabi-Yau en basse dimension a été également réalisée. On sait désormais que:
\begin{enumerate}
 \item En dimension complexe $1$, la seule variété Calabi-Yau est le 2-tore.
 \item En dimension complexe $2$, il n'existe que deux variétés Calabi-Yau à un isomorphisme près. Il s'agit du $4$-tore et de l'espace $K3$. Sur ce dernier, aucune métrique Ricci-plate explicite n'est connue, bien que l'existence soit assurée par le théorème de Yau. Il en va de même pour toutes les variétés de Calabi-Yau non triviales de dimensions supérieures.
\item A partir de la dimension complexe $3$ (dimension réelle $6$) le nombre de variétés de Calabi-Yau devient infini et il n'existe pas encore de classification générale. On sait toutefois en construire beaucoup qui possèdent en plus la propriété d'être des variétés toriques.
\end{enumerate}

La notion de variété de Calabi-Yau torique est ainsi donnée par la définition suivante:
\begin{definition}
 Par définition, une variété de Kähler $\mathcal{M}$, $n$ dimensionnelle est dite ``torique'' s'il existe un tore maximal $T$ inclus dans les automorphismes bi-holomorphes de $\mathcal{M}$ tel que $T$ soit isomorphe à $(\mathbb{C}^*)^n=\mathbb{R}^n \times (\mathbb{S}^1)^n$ etque l'action du tore $T$ sur lui même s'étendent à toute la variété $\mathcal{M}$.
\end{definition}

Enfin, puisque nous sommes intéressés à compactifier un espace de dimension réelle $6$, nous allons donc nous intéresser à des variétés de Calabi-Yau de dimension complexe $3$. Grâce à leurs applications en théorie des cordes, ces variétés ont été abondamment étudiées. En particulier, il a été découvert récemment que pour ces variétés, il existe une symétrie ``miroir'', c'est-à-dire une dualité entre familles de variétés de Calabi-Yau de dimension 3. Du point de vue mathématique, cette symétrie exprime une relation entre les nombres de courbes rationnelles sur une telle variété et les périodes des structures de Hodge sur la variété ``miroir'' associée. La théorie des cordes topologiques s'intéresse au dénombrement d'applications pseudoholomorphes d'une surface de Riemann de genre $g$ vers une variété de Calabi-Yau donnée (donc au premier aspect de la dualité précédente). On peut montrer (\cite{Marino}) que cela revient plus ou moins à trouver les invariants de Gromov-Witten (qui sont des nombres rationnels) de cette variété de Calabi-Yau. Les invariants de Gromov-Witten étant relativement éloignés des matrices aléatoires, ils ne seront pas abordés en détail ici et nous renvoyons le lecteur à \cite{GromovWitten} pour une introduction. Lorsque la variété est torique, on peut lui associer grâce à la symétrie miroir une variété duale qui peut être décrite par une équation du type (le second aspect de la dualité précédente):
\beq \label{miroireq} H(e^x,e^y)=0\eeq
où $H$ est un polynôme dont les coefficients codent les propriétés géométriques de la variété torique de départ. L'intérêt de cette formulation est qu'elle permet de faire un lien avec les matrices aléatoires et en particulier le développement topologique des modèles hermitiens et qu'elle offre des meilleures perspectives de calculs pratiques. (Car d'un point de vue calculatoire, les invariants de Gromov-Witten sont en général très difficiles à calculer.) En effet, dans ce cas, on a vu dans la première partie que les modèles hermitiens donnent lieu à une courbe spectrale algébrique ainsi qu'à une collection d'invariants $F^{(g)}$ de cette courbe. La conjecture BKMP (\cite{BKMP}) peut alors être énoncée de la façon suivante:
\begin{conjecture}
``Les invariants de Gromov Witten d'une variété torique de Calabi-Yau $\mathcal{M}$ de dimension trois sont les invariants symplectiques (au sens d'Eynard et Orantin) $F^{(g)}$ de la courbe spectrale $H(e^x,e^y)=0$ de sa variété miroir''.
\end{conjecture}

Cette conjecture présente un double intérêt. Tout d'abord, elle permet de faire le lien entre la théorie des cordes topologiques et les modèles de matrices hermitiens, permettant en particulier d'appliquer les techniques des modèles de matrice pour la théorie des cordes topologiques. Ensuite, si la conjecture s'avère exacte, elle permettrait, dans le cas des variétés de Calabi-Yau toriques de dimension $3$, d'avoir un algorithme explicite pour calculer les invariants de Gromov-Witten par le formalisme de récurrence topologique d'Eynard et Orantin. Il est à noter que dans de nombreux cas simples ou plus compliqués (\cite{BKMPcase1,BKMPChen}) cette conjecture a pu être explicitement vérifiée.

\section{La formule du vertex topologique}

Le point de départ pour relier la théorie des matrices aléatoires avec les invariants de Gromov-Witten est la formule du vertex topologique \cite{TopologicalVertex,TopologicalVertex2}. Rappelons que les invariants de Gromov-Witten, notés $\mathcal{N}_{g,D}$, d'une variété torique de Calabi-Yau $\mathcal{M}$ de dimension $3$, comptent le nombre d'applications de surfaces connexes de genre $g$ dans $\mathcal{M}$ étant donné une classe d'homologie $D=(D_1,\dots,D_k)$. On peut alors rassembler ces invariants sous la forme d'une double série génératrice (en notant $Q^D\overset{def}{=}\prod_{i=1}^k Q_i^{D_i}$):
\beq  
GW(\mathcal{M},Q,g_s)=\sum_{g=0}^\infty \sum_D Q^D g_s^{2g-2}\mathcal{N}_{g,D}(\mathcal{M})\eeq
On peut rajouter, comme habituellement dans ce type de dénombrement, les surfaces non-connexes en prenant l'exponentielle:
\beq Z_{GW}(\mathcal{M},Q,g_s)=e^{GW(\mathcal{M},Q,g_s)}=\sum_{g=0}^\infty \sum_D Q^D g_s^{2g-2}\mathcal{N}^*_{g,D}(\mathcal{M})\eeq

Il a été montré \cite{TopologicalVertex} que cette dernière fonction de partition est identique à celle donnée par la formule du vertex topologique qui peut être exprimée dans un cas particulier dit ``fiducial'' (le cas général peut également être exprimé, mais comme je le mentionne plus bas, on peut toujours se ramener au cas fiducial par des opérations appelées ``transitions de flop'' ce qui facilite les calculs) comme :
\beq Z_{vertex}(\mathcal{M})=\sum_{j=0..n, i=1..m-1} \sum_{\alpha_{i,j}}\prod_{i=1}^m Z_{ligne}(\vec{\alpha}_i,\vec{\alpha}^T_{i+1})\prod_{i,j} q^{s_{j,i}|\alpha_{i,j}|} \label{Zvertex} \eeq

où 
\bea
Z_{ligne}(\alpha;\beta^T)  \label{z_strip}
&=&   \prod_{i=0}^n \left(\frac { [\alpha_i] [\beta^T_i] }{\quad \quad [\beta_i,\alpha_i^T]_{Q_{\beta_i,\alpha_i}}}\right) \,  \frac{\underset{i<j}{\prod}\left([\alpha_i,\alpha_j^T]_{Q_{\alpha_i,\alpha_j}}\right) \,\,\underset{i<j}{\prod} \left([\beta_i,\beta_j^T]_{Q_{\beta_i,\beta_j}}\right) }{
 \underset{i<j}{\prod} \left([\alpha_i,\beta_j^T]_{Q_{\alpha_i,\beta_j}} [\beta_i,\alpha_j^T]_{Q_{\beta_i,\alpha_j}}\right) }
  \cr  
\eea
et les $\alpha_i$ sont des partitions planes et les crochets appliqués à une partition plane $\gamma=(\gamma_1,\gamma_2,\dots)$ sont définis par:
\bea \label{toutoumou}
[\gamma] 
&=& (-1)^d  q^{\frac{1}{4} \kappa(\gamma)} \prod_{1 \le i < j \le d} \frac{[\gamma_i - \gamma_j + j - i]}{[j-i]} \prod_{i=1}^{d} \prod_{j=1}^{\gamma_i} \frac{1}{[d + j - i]} \eea
avec $q=e^{-g_s}$ faisant le lien avec la série génératrice des invariants de Gromov-Witten, $\kappa(\gamma)=\underset{k}{\sum} \gamma_k(\gamma_k-2k+1)$ le second Casimir de la partition plane $\gamma$ et par definition le crochet appliqué à un nombre entier est le ``q-nombre'': 
\beq \forall\, n \in \,\mathbb{N}\,: \,\,[n]=q^{-\frac{n}{2}}-q^{\frac{n}{2}}\eeq
Ici nous utilisons les notations françaises des partitions planes décrites par la figure:

\begin{center} \label{partitionsplanes}
	\includegraphics[height=5cm]{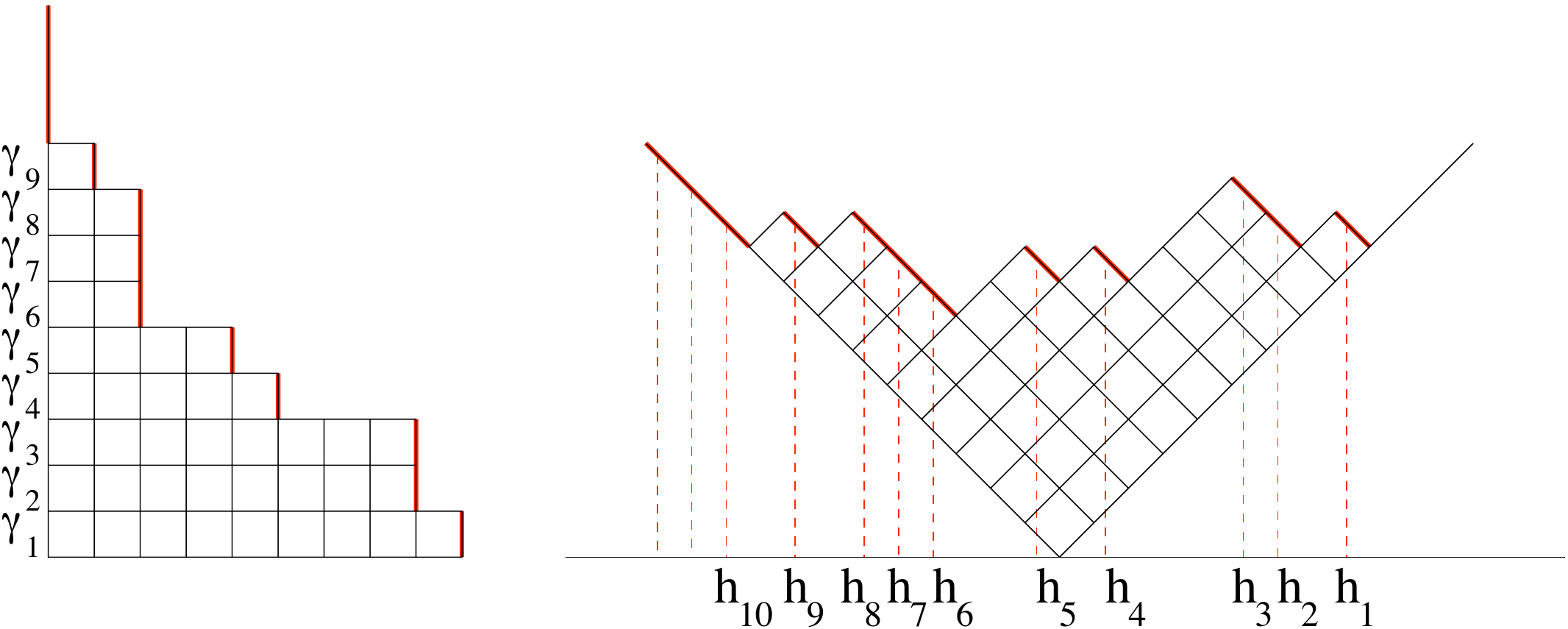}

	\underline{Figure 26}: Exemple de représentation de partitions planes. On remarque ainsi comment on peut passer par une simple rotation des entiers $\gamma_i$ aux entiers $h_i$.
\end{center}

 La transposée $\gamma^T$ d'une partition $\gamma$ est définie comme la partition plane dans laquelle on a inversé les lignes et les colonnes. La notation $|\gamma|$ désigne le nombre total de boîtes de la partition plane. Le terme de couplage entre deux partitions est donné par:
\bea[\gamma, \delta^T]
&=&  Q_{\gamma,\delta}^{-\frac{|\gamma| + |\delta|}{2}} q^{-\frac{\kappa(\gamma) - \kappa(\delta) }{4}} \prod_{i=1}^{d} \prod_{j=1}^{d} \frac{  [ h_i(\gamma)-h_j(\delta)] }{ [ a_\gamma-a_\delta  + j - i] }  \cr
&& \times \prod_{i=1}^{d} \prod_{j=1}^{\gamma_i} \frac{1}{[a_\gamma-a_\delta + j - i + d]} \prod_{i=1}^{d} \prod_{j=1}^{\delta_i} \frac{1}{[a_\gamma-a_\delta - j + i - d]} \,\cr
&&  \prod_{k=0}^\infty g(Q_{\gamma,\delta}^{-1}q^{-k})\,
\eea
avec $h_i(\gamma)=\gamma_i-i+d+a$ et $g(x)=\underset{n=1}{\overset{\infty}{\prod}} \left(1-\frac{1}{x}q^n\right)$. Le paramètre $Q_{\gamma,\delta}$ reflète les paramètres de Kähler de notre variété de départ à partir desquels on peut définir les paramètres $a_\gamma$ de façon équivalente (dans le formalisme du vertex topologique il est beaucoup plus pratique d'utiliser les paramètres $a_\gamma$ que les $Q_{\gamma,\delta}$) par:
\beq Q_{\gamma,\delta}=q^{a_\gamma-a_\delta}\label{Qenfonctiondesa}\eeq
Les nombres $s_{i,j}$ décrivant l'interaction entre deux lignes dans \ref{Zvertex} peuvent être reliés à des différences de $a_{\alpha_{i,j}}$ (Cf. \textbf{[V]} présenté en annexe \ref{Article[V]}).

\begin{center} \label{boiteimage}
	\includegraphics[height=5cm]{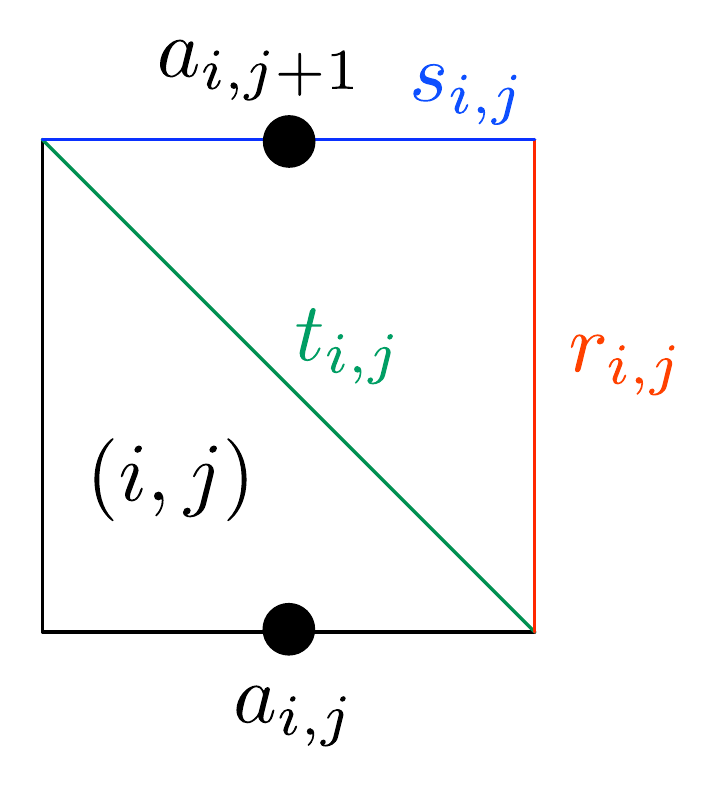}

	\underline{Figure 27}: Présentation des conventions de notation pour les paramètres de  Kähler 
\end{center}

Traditionnellement, dans le formalisme du vertex topologique, la géométrie d'une variété de Calabi-Yau torique de dimension trois est représentée par un diagramme du type:

\begin{center}  \label{diagrammevertex}
	\includegraphics[height=5cm]{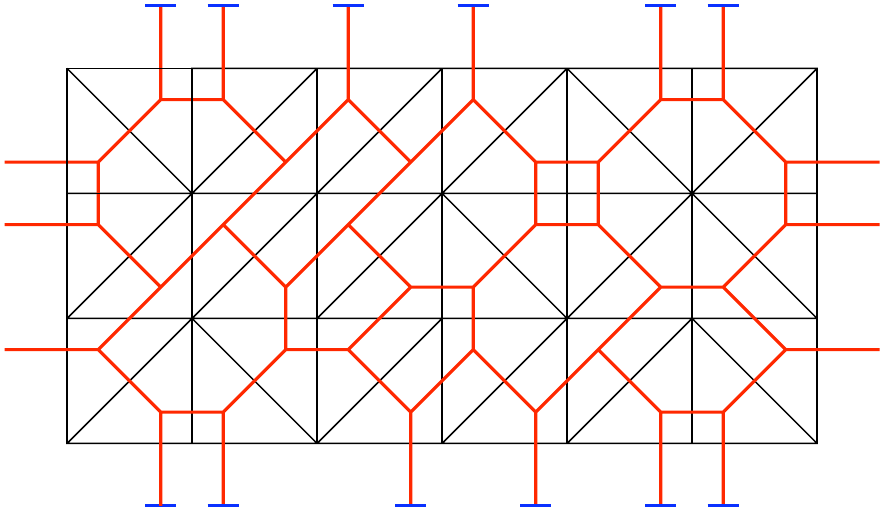}

	\underline{Figure 28}: Exemple de représentations d'une variété de Calabi-Yau torique de dimension $3$ par le formalisme du vertex topologique. Le diagramme dual est représenté en rouge. Le choix de l'orientation des diagonales de chaque carré reflète les propriétés géométriques de la variété.
\end{center}

Le passage d'une boîte diagonale haute à une boîte diagonale basse est appelé ``transition de flop''. 

\begin{center}\label{Transitiondeflop}
	\includegraphics[height=5cm]{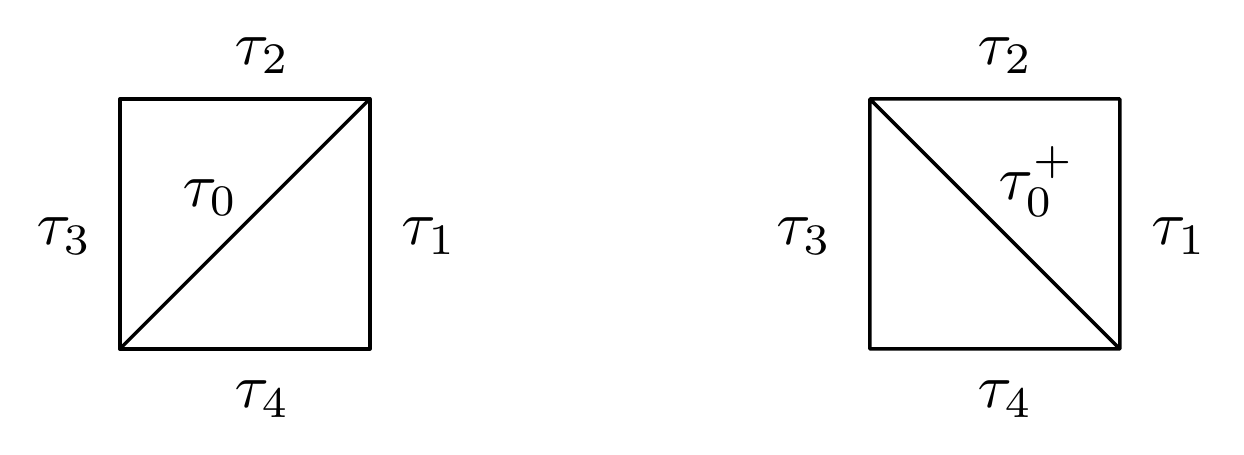}

	\underline{Figure 29}: Illustration d'une transition de flop.
\end{center}

Il est connu que les invariants de Gromov-Witten sont invariants sous les transitions de flop, ce qui signifie que n'importe quelle configuration peut être choisie pour le calcul de ces invariants. En particulier, notre choix fiducial est possible et se décrit par le diagramme:

\begin{center} \label{fiducialimage}
	\includegraphics[height=5cm]{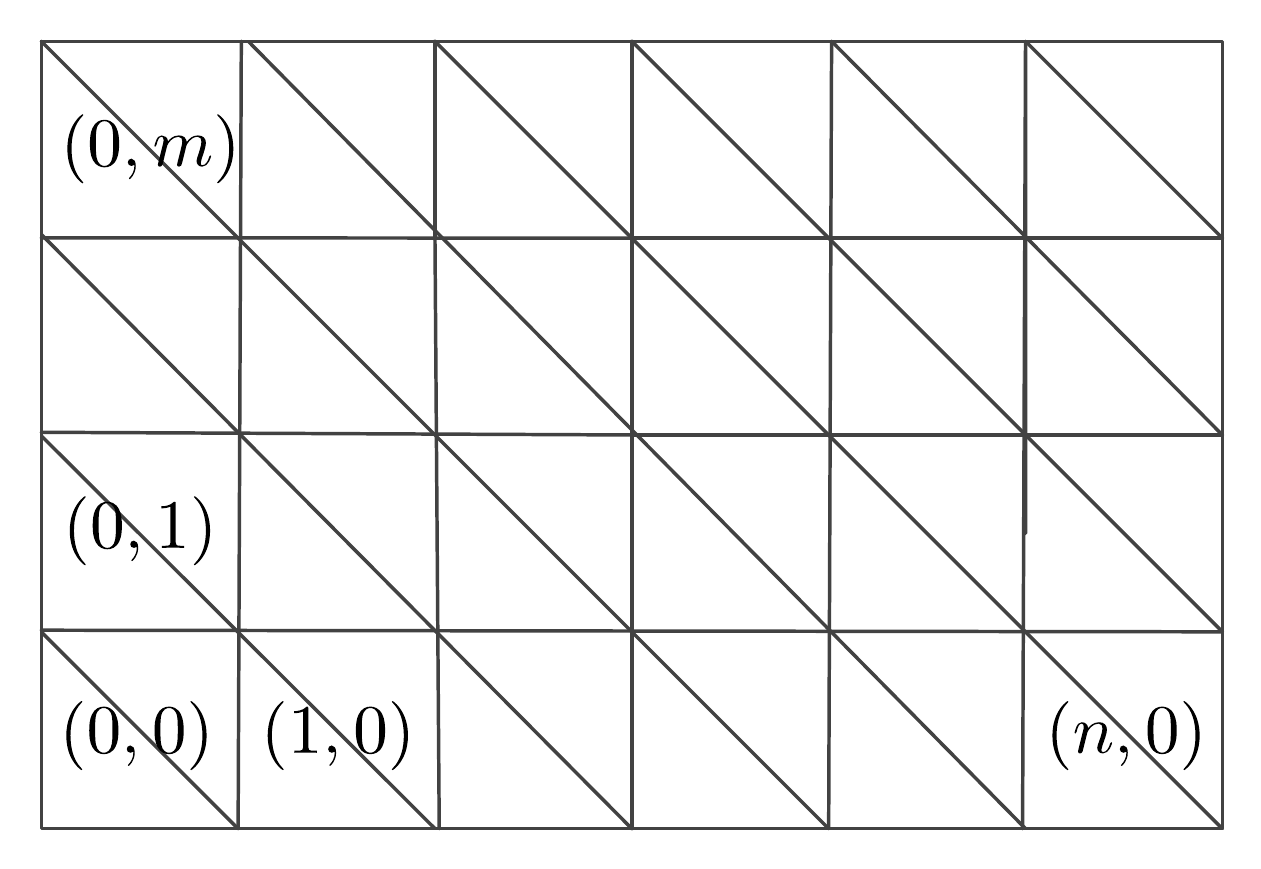}

	\underline{Figure 30}: Représentation de la géométrie fiduciale.
\end{center}

Ainsi, grâce aux résultats précédents, on voit que la fonction de partition des invariants de Gromov-Witten peut être ramenée à une fonction de partition exprimée en termes de partitions planes, qui sont connues pour avoir des liens avec les matrices aléatoires \cite{Vershik, Douglas, Okunkov, Partitions, Partitions2}. Notons que si la conjecture s'avère exacte, l'invariance des nombres de Gromov-Witten sous les transitions de flop doit être retrouvée du côté des invariants symplectiques $F^{(g)}$. Dans \textbf{[V]} (Cf. annexe \ref{Article[V]}), il est montré qu'une transition de flop correspond à une transformation symplectique de la courbe spectrale $E(x,y)$ du modèle de matrice décrit dans les prochains paragraphes, qui comme nous l'avons vu précédemment laisse invariants les nombres $F^{(g)}$.

\section{Reformulation en termes de modèles de matrices}

Grâce à la formule du vertex topologique et à l'invariance des nombres de Gromov-Witten par transitions de flop, nous avons vu dans le paragraphe précédent qu'il était possible de reformuler la série génératrice des invariants de Gromov-Witten sous la forme d'une fonction de partition impliquant des partitions planes (\ref{Zvertex}). Dans notre travail \textbf{[V]} (Cf. annexe \ref{Article[V]}), nous avons montré comment il est possible de réécrire cette fonction de partition sous la formule d'un modèle de matrice hermitien. Compte-tenu de la longueur des calculs nécessaires, nous nous contererons ici de mentionner le résultat obtenu, laissant au lecteur le soin de consulter \textbf{[V]} en annexe \ref{Article[V]} de ce mémoire pour les détails de la dérivation. Le modèle de matrice obtenu est:

\bea \label{ZMM}
Z_{\rm MM}(Q,g_s,\vec\alpha_m,\vec\alpha_0^T)
&=& \Delta(X(\vec \alpha_m))\, \Delta(X(\vec \alpha_0)) \, 
\prod_{i=0}^m \int_{H_N(\Gamma_i)} dM_i \,
 \prod_{i=1}^{m}\int_{H_N({\mathbb R}_+)}\,dR_i \cr
&& \prod_{i=1}^{m} e^{\frac{-1}{g_s}\,\tr\, V_{\vec a_i}(M_i)-V_{\vec a_{i-1}}(M_i) 
} \,\,\,
 \prod_{i=1}^{m} e^{\frac{-1}{g_s}\,\tr\, V_{\vec a_{i-1}}(M_{i-1})-V_{\vec a_{i}}(M_{i-1}) 
} \cr
&& \prod_{i=1}^m e^{\frac{1}{g_s} \tr\, (M_i-M_{i-1})R_i} \,
 \prod_{i=1}^{m-1} e^{(S_i+i\pi)\,\tr\, \ln\, M_i}\, \cr
&& e^{\tr\, \ln\, f_{0}(M_0)}\,e^{\tr\, \ln\, f_{m}(M_m)}\, \prod_{i=1}^{m-1} e^{\tr\, \ln f_{i}(M_i)}.
\eea 
où les matrices sont de taille $N=(n+1)\, d$ et où les matrices de début et de fin de chaîne sont données par:
\beq
X(\vec \alpha_m)  = {\rm diag} (X(\vec \alpha_m)_i)_{i=1,\dots,N}
\,\, , \qquad
X(\vec \alpha_m)_{(j-1)d+i} = q^{h_i(\alpha_{j,m})},
\eeq
\beq
X(\vec \alpha_0)  = {\rm diag} (X(\vec \alpha_0)_i)_{i=1,\dots,N}
\,\, , \qquad
X(\vec \alpha_0)_{(j-1)d+i} = q^{h_i(\alpha_{j,0})},
\eeq
avec la notation habituelle $\Delta(X)=\underset{i<j}{\prod}(X_i-X_j)$ du déterminant de Vandermonde et où enfin les potentiels sont définis par:
\beq
V_{\vec a_i}(x) = -g_s \sum_{j=1}^n \ln \, {g(q^{a_{j,i}}/x)}
\eeq
avec $ \forall \, i=1,\dots,m-1$ l'introduction des fonctions $f_i(x)$ par:
\beq
f_i(x) =  \prod_{j=0}^n \frac{{g(1)^2\,\,e^{(\frac{1}{2}+\frac{i\pi}{ \ln q})\, \ln{(x q^{1-a_{j,i}})}}\, \,e^{\frac{(\ln{(x q^{1-a_{j,i}})})^2}{  2 g_s}}}}{ g(x\,q^{1-a_{j,i}})\, g(q^{a_{j,i}}/x)\,}.
\eeq
et les cas spéciaux:
\beq
f_0(x) = \prod_{j=0}^n {g(x\,q^{1-a_{j,0}-d})\over x^d\, g(x\,q^{1-a_{j,0}}) }\,,\hspace{1cm} f_{m+1}(x) = \prod_{j=0}^n {g(x\,q^{1-a_{j,m+1}-d})\over x^d\, g(x\,q^{1-a_{j,m+1}}) }
\eeq
Les paramètres de Kähler de la fonction de partition du vertex topologique \ref{Zvertex} s'identifient alors comme $a_{\alpha_{i,j}}=a_{i,j}$ et les couplages $s_{i,j}$ s'identifient avec les nombres $S_i$  par la formule (valable quel que soit le choix de $j$):
\beq
S_i =s_{j,i} + \sum_{k\leq j} (a_{k,i-1}-a_{k,i}) + \sum_{k<j} (a_{k,i+1}-a_{k,i})
\eeq
Rappelons également que le paramètre $Q$ est contenu dans les $a_j$ par la formule \ref{Qenfonctiondesa}. Le théorème fondamental est alors le suivant (Cf. \textbf{[V]}, section 4.4 de l'annexe \ref{Article[V]}):

\begin{theorem} La fonction de partition du vertex topologique \ref{Zvertex} est identique à celle du modele de matrices (à des facteurs de proportionnalité triviaux près):
 \beq Z_{vertex}(\mathcal{M})= Z_{\rm MM}(Q,g_s,\vec\alpha_m,\vec\alpha_0^T)\eeq
\end{theorem}

\medskip
\underline{\textbf{Idée de la preuve:}}
\medskip

La preuve de ce théorème est présentée dans mon article \textbf{[VI]} réalisé en collaboration avec B. Eynard et A. Kashani Poor et présenté en annexe \ref{Article[VI]}. Compte tenu de la longueur importante de la preuve, nous nous contenterons ici de n'en expliciter que les grandes lignes, en laissant au lecteur la possibilité de se référer à l'annexe \ref{Article[VI]} pour les détails. L'idée de la preuve est la suivante: Si l'on part du modèle de matrices \ref{ZMM}, on voit que les potentiels $f_i$ ont des pôles simples et que l'intégration correspond donc juste à prendre des résidus en ces pôles. En particulier elle localise les valeurs propres des matrices $M_i$ aux entiers de la forme $q^{h_i}$ et se transforme ainsi en une somme sur des entiers que l'on peut écrire en termes d'une somme sur des partitions. Les intégrales sur les matrices $R_i$, réalisent les transformées de Laplace d'intégrales d'Itzykson-Zuber, c'est à dire des déterminants de Cauchy, i.e. les dénominateurs dans la formule \ref{z_strip}. Les déterminants de Vandermonde proviennent eux de la diagonalisation de l'intégrale sur les matrices $M_i$ et réalisent les numérateurs de la formule \ref{z_strip}. Enfin, les potentiels $f_j$ contenant les fonctions $g$, réalisent les poids de la formule \ref{toutoumou}. On retrouve alors la fonction de partition du vertex topologique \ref{Zvertex} établissant ainsi l'égalité.

\section{Analyse du modèle de matrices et conjecture BKMP}

D'après le paragraphe précédent, nous avons vu qu'il était possible de reformuler la fonction de partition donnant les invariants de Gromov-Witten d'une variété torique de Calabi-Yau de dimension $3$ en un modèle de matrice hermitien donné par \ref{ZMM}. Hélas, le modèle de matrice obtenu est une chaîne de matrices possédant des restrictions de positivité sur les valeurs propres des matrices $R_i$. Cette restriction correspond à la présence de bords dits durs en $0$ sur les valeurs propres des matrices $R_i$ et ce type de modèle de matrice (chaîne+ bords durs) n'a pas été étudié en détail pour l'instant. Néanmoins, il n'y a quasiment aucun doute sur la possibilité d'extension des résultats connus sur les chaînes de matrices sans bords durs dans le cas des bords durs moyennant des modifications habituelles liées aux bords durs. En effet, le cas des chaînes de matrices sans bords durs a déjà été traité dans \cite{Chain}. De même, le cas à une et deux matrices (correspondant à des chaines de longueurs $0$ et $1$) a été traité avec des bords durs dans \cite{Hardedges, Hardedges2}. Une fois ce point purement technique établi, un autre problème se présente alors dans la résolution de la conjecture BKMP. En effet, lorsque l'on résout les équations de boucles, la solution $W_1^{(0)}(x)$, ou de manière équivalente la courbe spectrale, n'est pas unique puisqu'il y a autant de solutions que d'extrema du potentiel (il faut alors spécifier l'extremum autour duquel on se situe ou des fractions de remplissage). Or dans le cas de la chaîne de matrices obtenue, le potentiel extrêmement compliqué présente une infinité d'extrema et donc une infinité de solutions. Dans ce cas, il est connu \cite{OE} que la véritable solution au problème est alors de rechercher la courbe spectrale qui minimise, parmis les courbes solutions, le premier invariant $F_0$, ce qui en pratique s'avère être extrêmement difficile. En particulier dans notre article $\textbf{[VI]}$, nous montrons seulement que la courbe miroir satisfait effectivement les équations de boucles, mais pas qu'elle minimise globalement $F_0$. Ce point reste donc à éclaircir pour aboutir à une démonstration finale de la conjecture BKMP. Néanmoins dans notre article $\textbf{[VI]}$, nous montrons que la courbe spectrale ``minimale'', c'est à dire celle de plus petits degré et genre possibles, correspond bien à la courbe miroir recherchée. Malheureusement il n'existe pas à l'heure actuelle de démonstration générale permettant d'affirmer que la courbe spectrale ``minimale'' correspond toujours à la courbe spectrale minimisant le prépotentiel $F_0$.

%% file: conclu.tex
\conclusion
%\chapter{Conclusion}
\selectlanguage{french}

Cette thèse a ainsi présenté différentes méthodes utiles pour résoudre les modèles de matrices aléatoires ainsi qu'une application possible en théorie des cordes. A la frontière entre de nombreux domaines et grandes questions des mathématiques actuelles, il apparaît clairement que les matrices aléatoires nous réservent encore bien des surprises. En particulier, la notion d'universalité évoquée pour les cas hermitiens, réels symétriques et quaternioniques self-duaux est une propriété surprenante qui laisse à penser qu'une sorte de théorème central limite encore incompris serait à l'oeuvre dans les modèles de matrices. Si cette propriété surprenante pouvait être généralisée au cas où l'exposant $\beta$ est quelconque, elle ouvrirait sans doute de grandes possibilités tant fondamentales, avec la possibilité de développer la théorie des polynômes orthogonaux, qu'appliquées avec la théorie des cordes topologiques. 

L'extension de la notion d'intégrabilité sous-jacente au cas hermitien est également une voie à suivre très prometteuse. En effet, si la méthode, présentée dans cette thèse, des équations de boucles pour le cas où l'exposant $\beta$ est arbitraire aboutissait à des résultats similaires à ceux du cas hermitien, cela ouvrirait de grandes perspectives quant à une meilleure compréhension de l'intégrabilité au sens quantique et des équations différentielles linéaires. 

Néanmoins, beaucoup de travail reste encore à accomplir pour parvenir à de tels débouchés, qui hélas, par l'incertitude même de toute recherche, pourraient également s'avérer inacessibles ou chimériques. Espérons donc que l'effervescence de ces dernières années se poursuive et que la théorie des matrices aléatoires continue d'alimenter des domaines variés des mathématiques et de la physique en nous réservant, qui sait, peut être quelques autres grandes découvertes.

%% file: ListeFigures.tex
\annexe{Liste des figures} \label{ListeFigures}
\selectlanguage{french}
%\begin{center}
% \textbf{LISTE DES FIGURES}
%\end{center}

\medskip
\medskip
\medskip
\medskip

\begin{enumerate}
\item  Loi de Tracy-Widom, Page 23 
\item  Histogramme des écarts de valeurs propres consécutives pour des matrices symétriques, Page 24
\item  Histogramme des valeurs propres pour des entrées de Cauchy, Page 24 
\item  Distribution de Wigner pour un atome d'uranium, Page 26
\item  Distribution des zéros de la fonction $\zeta$ de Riemann, Page 27 
\item  Histogramme des valeurs propres par l'algorithme de Metropolis à la température critique, Page 34 
\item  Histogramme des valeurs propres par l'algorithme de Metropolis au delà de la température critique, Page 35  
\item  Histogramme des valeurs propres par l'algorithme de Metropolis en dessous de la température critique, Page 35  
\item  Histogramme des valeurs propres par l'algorithme de Metropolis pour le cas Gaussien, Page 36 
\item  Tracé des fonctions universelles à deux points pour les ensembles hermitiens, symétriques réels et quaternioniques self-duaux, Page 38 
\item  Distribution de Gaudin et histogramme des écarts des valeurs propres normalisés dans le bulk, Page 41
\item  Histogramme des plus grandes valeurs propres normalisées et densité théorique correspondante, Page 41
\item  Histogramme cumulé des plus grandes valeurs propres normalisées et fonction de répartition théorique correspondante, Page 42
\item  Double limite d'échelle pour un point critique à l'intérieur du support, Page 52 
\item  Double limite d'échelle pour un point critique à l'extrémité du support, Page 52 
\item  Image d'une densité critique selon différents choix du paramètre $\epsilon$, Page 54 
\item  Exemple de secteurs de Stokes, Page 100  
\item  Illustration de coupures en géométrie algébrique, Page 102 
\item  Illustration de demi-lignes d'accumulation de zéros, Page 103 
\item  Exemple de cycles d'holonomie en géométrie algébrique, Page 104 
\item  Exemple de $\mathcal{A}$-cycles pour une courbe quantique, Page 105 
\item  Exemple de $\mathcal{B}$-cycles pour une courbe quantique, Page 105 
\item  Exemple de cycles d'holonomie dégénérés en géométrie algébrique, Page 106 
\item  Exemple de cycles dégénérés pour une courbe quantique, Page 107
\item  Représentation imagée d'une variété de Calabi-Yau, Page 125 
\item  Exemple de représentation de partitions planes, Page 129 
\item  Présentation des conventions de notation pour les paramètres de Kähler, Page 130 
\item  Diagramme d'une variété de Calabi-Yau par la formule du vertex topologique, Page 131 
\item  Illustration d'une transition de flop, Page 131 
\item  Représentation de la géométrie fiduciale, Page 132 
\end{enumerate}

%% file: annexea.tex
\annexe{Algorithme de Metropolis} \label{MetropolisHastings}
\selectlanguage{french}

L'algorithme de Metropolis-Hastings est un outil puissant en physique statistique et permet de simuler n'importe quel type de distribution de probabilités $f(x_1,\dots,x_n)$. Un avantage important de l'algorithme est que la connaissance de la fonction de distribution $f(x_1,\dots,x_n)$ à une constante multiplicative près est suffisante pour implémenter l'algorithme. En particulier, dans notre cas, l'impossibilité de calculer la normalisation $Z_N$ n'est pas un obstacle à l'implémentation concrète de l'algorithme. Le principe de l'algorithme est le suivant: on part d'un état $(x_1,\dots,x_n)$ quelconque puis on propose un nouvel état $(\td{x}_1,\dots, \td{x}_n)$ en fonction de l'état précédent $(x_1,\dots,x_n)$. Dans notre cas précis, seule une des composantes (tirée aléatoirement) $x_i$ sera changée et sera obtenue par un tirage aléatoire d'une variable normale centrée en $x_i$ et de variance $0.1$. On procède alors au calcul du rapport: 
\beq
r=\frac{f(\td{x}_1,\dots, \td{x}_n)}{f(x_1,\dots,x_n)}\eeq
On tire ensuite une variable aléatoire $\alpha$ suivant une loi uniforme sur $[0,1]$. Si $\alpha<r$, on conserve alors le nouvel état $(\td{x}_1,\dots, \td{x}_n)$, sinon on conserve l'ancien. On itère enfin ce processus un très grand nombre de fois et l'état final obtenu donne alors un échantillon distribué suivant la loi de probabilité $f(x_1,\dots,x_n)$.

Dans le cas présenté en \ref{potentielcritique}, le code en Maple est:

\underline{Initialisation des paramètres}:

\begin{tabular}{l}
$N:=200:$ \\
$\epsilon:=0.5:$\\
$ c_1:=\cos(\pi*\epsilon): c_2:=\cos(2*\pi*\epsilon):$\\
$ T_c:=1+4*c_1^2:$\\
$ T:=2*T_c:$\\
$V:=z->1/T*(z^4/4 - 4*c_1*z^3/3 + 2*c_2*z^2/2 + 8*c_1*z):$\\
\end{tabular}

\underline{Définition de l'état initial}:

\begin{tabular}{l}
etatini:=Array(1..N):\\
for k from 1 to N do etatini[k]:=$(-1)^k*2*k/N$: od:\\

\end{tabular}

\underline{Algorithme de Metropolis-Hastings-Gibbs}:

\begin{tabular}{l}
Gibbs:=proc(e,NN,sigma)\\
etatcourant:=e:\\
for k from 1 to N*NN do\\
valeurpropreencours:=Generate(integer(range = 1 .. N)):\\
oldVP:=etatcourant[valeurpropreencours]:\\
nouvelleVP:=Sample(RandomVariable(Normal(oldVP,sigma)),1)[1]:\\
diffpotentiel:=-N/T*V(nouvelleVP)+N/T*V(oldVP):\\
for j from 1 to N do\\
if j<>valeurpropreencours then\\
diffpotentiel:=diffpotentiel+evalf(2*ln(abs(nouvelleVP-etatcourant[j])))\\
-evalf(2*ln(abs(oldVP-etatcourant[j]))):\\
fi:\\
od:\\
ratio:=exp(diffpotentiel):\\
alpha:=GenerateFloat():\\
if alpha< min(ratio,1) then etatcourant[valeurpropreencours]:=nouvelleVP:\\
fi:\\
od:\\
return(etatcourant):\\
end proc:\\
\end{tabular}

\underline{Commande d'exécution}:

Histogram(Gibbs(etatini,50,0.1), averageshifted = 4);

%% file: annexeb.tex
\annexe{Quaternions et matrices quaternioniques} \label{annexe2}
\selectlanguage{french}

Les matrices quaternioniques self-duales, constituent le troisième ensemble standard dans l'étude des matrices aléatoires. Il est connu depuis les années 1950 avec Wigner, et correspond à un ensemble de matrices diagonalisables par un élément du groupe symplectique $Sp(2n)$, généralisant ainsi les matrices hermitiennes et réelles symétriques (respectivement diagonalisables par un élément du groupe unitaire $U(n)$ et orthogonal $O(n)$). Néanmoins, cet ensemble de matrice étant moins connu, nous rappelons dans cet annexe sa définition.

\medskip
\medskip
\textbf{1. Les quaternions}

\medskip
\medskip

Les quaternions constituent une généralisation bien connue des nombres complexes. Ils sont définis par une algèbre complexe (non commutative) de dimension $4$:

\beq q=q^{(0)}+q^{(1)}e_1+q^{(2)}e_2+q^{(3)}e_3=q^{(0)}+\vec{q}.\vec{e}\eeq
où les $q^{(i)}$ sont des nombres complexes et les vecteurs de l'algèbre $e_i$ obéissent aux opérations (définissant le produit sur l'algèbre):
\bea 
&& e_1^2=e_2^2=e_3^2=-1 \cr
&& e_1.e_2=-e_2.e_1=e_3 \cr
&& e_2.e_3=-e_3.e_2=e_1 \cr
&& e_3.e_1=-e_1.e_3=e_2\eea

Les quaternions possèdent une représentation naturelle en termes de matrices $2 \times 2$:
\beq e_1=\begin{pmatrix}
          i& 0\\
0&-i
         \end{pmatrix} \virg e_2=\begin{pmatrix}
          0& 1\\
-1&0
         \end{pmatrix} \virg e_2=\begin{pmatrix}
          0& i\\
i&0
         \end{pmatrix} \eeq
avec en plus la matrice unité:
\beq 1=\begin{pmatrix}
        1& 0\\
0& 1
       \end{pmatrix}\eeq

Le conjugué d'un quaternion est alors défini par:
\beq \ovl{q}=q^{(0)}-\vec{q}.\vec{e}\eeq
qu'il ne faut pas confondre avec son complexe conjugué:
\beq q^*=q^{(0)\,*}-\vec{q}^*.\vec{e}\eeq
Un quaternion vérifiant $q^*=q$ est dit réel, un quaternion avec $q^*=-q$ est dit imaginaire pur et enfin un quaternion avec $\ovl{q}=q$ est dit scalaire. Finalement, le conjugué hermitien d'un quaternion est défini comme:
\beq q^\dagger=\ovl{q}^*\eeq
Un quaternion avec $q^\dagger=q$ est dit hermitien (et il correspond dans sa représentation en termes de matrices à une matrice $2 \times 2$ hermitienne)

\medskip
\medskip
\textbf{2. Les matrices ``quaternioniques''}

\medskip
\medskip

Les matrices ``quaternioniques'' sont des matrices complexes $A$ de taille paire $2n \times 2n$ qui peuvent s'écrire comme une matrice quaternionique $Q$ de taille $n \times n$ grâce à la représentation de dimension deux des quaternions. Par exemple:
\beq A=\begin{pmatrix}
      i&0&  0 &i &1 &0\\
0 &-i & i &0 &0&1\\
0&1 &1 &0 &0 &i\\
-1 &0 & 0&1 &i &0\\
2+3i &4+5i &3& 0 &3 & 5i\\
-4+5i &2-3i &0 &3 &5i &3\\
     \end{pmatrix}
\Leftrightarrow Q=\begin{pmatrix}
     e_1 &e_3 &1\\
e_2 & 1& e_3\\
2+3e_1+4e_2+5e_3 & 3&3 +5e_2\\
     \end{pmatrix}
\eeq

Les opérations habituelles sur la matrice $A$ se reflètent sur la matrice $Q$ de la façon suivante:
\begin{enumerate}
\item  Transposition: $(Q^T)_{i,j}=-e_2.\ovl{q}_{j,i}.e_2$
\item  Conjugué hermitien: $(Q^\dagger)_{i,j}=q^\dagger_{j,i}$
\item  Renversement du temps: $(Q^R)_{i,j}=\ovl{q}_{i,j}$
\end{enumerate}
La matrice $Q^R$ est dite duale de $Q$ et une matrice vérifiant $Q^R=Q$ est dite self-duale. Une matrice $A$ de taille $2n \times 2n$ 
dont la matrice quaternionique $Q$ est self-duale possède $n(2n-1)$ composantes réelles indépendantes.

\medskip
\medskip
\textbf{3. Groupe symplectique}

\medskip
\medskip

Soit la matrice de taille $2n \times 2n$:
\beq Z=\begin{pmatrix}
 0&1&0&0 &\dots&\dots\\
-1 &0 &0 &0 &\dots&\dots\\
0&0&0 &1 & \dots &\dots\\
0&0& -1&0 &\dots&\dots \\
\dots & \dots &\dots&\dots& \ddots &\ddots \\
\dots & \dots &\dots&\dots &\ddots &\ddots \\
       \end{pmatrix}\eeq
qui est composée de blocs $2 \times 2$ sur la diagonale et nulle ailleurs. Le groupe symplectique $Sp(2n)$ est alors défini comme l'ensemble des matrices $B$ de taille $2n \times 2n$ telles que:
\beq Z=B\, Z \, B^T\eeq
Le groupe symplectique est connu depuis Weyl (1946) et se retrouve très utilisé en physique puisque le formalisme hamiltonien possède intrinsèquement une structure symplectique. La propriété principale dans notre contexte est que l'ensemble des matrices quaternioniques self-duales est invariant sous les transformations $C\mapsto B^R CB$ où $B$ est n'importe quelle matrice symplectique. De plus, n'importe quelle matrice quaternionique self-duale peut être diagonalisée en une matrice quaternionique $D$ réelle et scalaire par une transformation symplectique:
\beq A=B \,D \,B^R\eeq
et la mesure de probabilité sur les matrices quaternioniques self-duales est invariante par transformation symplectique.

\medskip
\medskip
\medskip

Toutes ces propriétés permettent de passer du modèle des intégrales de matrices quaternioniques self-duales au problème aux valeurs propres correspondant avec un Jacobien égal au determinant de Vandermonde à la puissance $4$. (Cf. \cite{Mehta} pour la démonstration) En particulier, on voit la similitude très forte entre cet ensemble de matrice un peu exotique et les ensembles de matrices hermitiennes et réelles symétriques plus habituels: toutes les matrices de ces ensembles peuvent être diagonalisées par un élément du groupe symplectique, unitaire ou orthogonal et les mesures de probabilité sont invariantes par les transformations correspondantes. Seul le nombre de composantes réelles indépendantes de ces matrices change ce qui aboutit à des exposants du déterminant de Vandermonde différents selon les trois ensembles. 

%% file: annexec.tex
\annexe{Déterminants de Fredholm} \label{annexe3}
\selectlanguage{french}

%\begin{center}
%\textbf{Déterminants de Fredholm}
%\end{center}

Le déterminant de Fredholm est une notion qui généralise le déterminant d'une matrice dans le cadre d'opérateurs d'un espace de Hilbert qui ne diffèrent de l'identité que par un opérateur vérifiant la propriété dite de trace:
\beq \label{TraceClass} \sum_{k=0}^\infty \left< (A^{*}A)^{\frac{1}{2}} e_k , e_k\right> <\infty \eeq
où $<,>$ désigne le produit scalaire de l'espace de Hilbert et $e_k$ est une base orthonormale.

Soit donc $A$ un tel opérateur, on définit alors son déterminant de Fredholm par:
\beq \det(Id+A)\overset{\text{def}}{=}\sum_{k=0}^\infty \Tr(\lambda^k(A))\eeq
où par définition $\forall v_i \, \in H$:
\beq \lambda^k(A) \,\, v_1 \wedge v_2 \wedge \dots \wedge v_k= Av_1 \wedge Av_2 \wedge \dots \wedge Av_k\eeq

Une telle définition permet de généraliser les propriétés habituelles du déterminant des matrices. Si $A$ et $B$ sont des opérateurs vérifiant \ref{TraceClass} et si $T$ est un opérateur inversible alors:
\bea \det( (Id+A)(Id+B))&=&\det(Id+A) \, \det(I+B)\cr
\det(T (Id+A)T^{-1})&=&\det(Id+A)\cr
\det( e^A)&=&\exp(Tr(A))
 \eea
 
Le cas matriciel est alors un cas particulier des déterminants de Fredholm où $H$ est de dimension finie $m$:
\beq \Tr(\lambda^k(A))=\sum_{i_1,\dots,i_n}\frac{(\det(A_{i_p,i_q} ) )_{p,q=1..n}}{n!}\eeq
ce qui donne la formule de von Koch (1892) du déterminant:
\beq \det(I + zA) =\sum_{n=0}^m \frac{z^n}{n!} \sum_{i_1,\dots,i_n=1}^m (\det(A_{i_p,i_q} ) )_{p,q=1..n} \eeq

Un deuxième cas, tout spécialement intéressant pour la théorie des matrices aléatoires est celui où l'espace de Hilbert $H$ est l'espace des fonctions de carrés intégrables $\mathcal{L}^2(a,b)$. L'opérateur de trace est alors défini pour un noyau $K(x,y)$ continu sur $(a,b)$ par:
\beq \Tr(\lambda^k(K))=\frac{1}{k!} \int_{(a,b)^k} (\det(K(x_p,x_q)))_{p,q=1..k} \,\, dx_1\dots dx_k \eeq
ce qui aboutit au déterminant de Fredholm:
\beq \det(I + zK) =\sum_{n=0}^m \frac{z^n}{n!} \int_{(a,b)^n} (\det(K(x_p,x_q)))_{p,q=1..n} \,\, dx_1\dots dx_n \eeq
qui peut également se réécrire comme:
\beq \det(I-zK)=\exp(-\sum_{k=0}^\infty \frac{z^n}{n}\Tr K^n) \eeq
où la "trace" de $K^n$ est définie comme:
\bea \Tr(K)&=&\int_a^b K(x,x)dx\cr
\Tr(K^2)&=&\int_a^b\int_a^b K(x,y)K(y,x)\,dx \,dy\cr
&\dots&\cr\eea

Notons qu'il est en général rare de pouvoir calculer explicitement les déterminants de Fredholm de façon exacte, même pour des noyaux $K(x,y)$ simples. 

%% file: annexed.tex
\annexe{Double scaling limits of random matrices and minimal $(2m,1)$ models: the merging of two cuts in a degenerate case} \label{Article[I]}
\selectlanguage{french}

\begin{center}
\vspace{26pt}
\vspace{20pt}

{\sl O.\ Marchal}\hspace*{0.05cm}\footnote{ E-mail: olivier.marchal@polytechnique.org }\\
D\'{e}partement de math\'{e}matiques et de statistique\\
Universit\'{e} de Montr\'{e}al, Canada\\
Institut de physique th\'{e}orique\\
F-91191 Gif-sur-Yvette Cedex, France.\\
\vspace{6pt}
{\sl M.\ Cafasso}\hspace*{0.05cm}\footnote{ E-mail: mattia.cafasso@gmail.com }\\
Centre de recherche math\'{e}matique\\
Concordia University\\
Montr\'{e}al, Canada\\

\vspace{6pt}
\end{center}

\vspace{20pt}
\begin{center}
{\bf Abstract}: 
In this article, we show that the double scaling limit correlation functions of a random matrix model when two cuts merge with degeneracy $2m$ (i.e. when $y\sim x^{2m}$ for arbitrary values of the integer $m$) are the same as the determinantal formulae defined by conformal $(2m,1)$ models. Our approach follows the one developed by Berg\`{e}re and Eynard in \cite{BergereEynard} and uses a Lax pair representation of the conformal $(2m,1)$ models (giving Painlev\'e II integrable hierarchy) as suggested by Bleher and Eynard in \cite{BleherEynard}. In particular we define Baker-Akhiezer functions associated to the Lax pair to construct a kernel which is then used to compute determinantal formulae giving the correlation functions of the double scaling limit of a matrix model near the merging of two cuts. 
\end{center}

%-----------------------------ABSTRACT--------------------------------------
\vspace{26pt}

%*********************************************************************
%==================== ARTICLE =======================================%******************************************

\section*{1 Introduction}

It has been known for a long time that the study of random matrix models in different scaling limits gives rise to a great number of famous integrable equations; both PDEs of solitonic type (KdV and, more generally, Gelfand-Dikii equations) and ODEs arising from isomonodromic systems (like Painlev\'e equations). A key idea in these studies is the notion of spectral curves attached to algebraic equations $P(x,y)=0$. The genus of the curve gives the number of intervals on which the eigenvalues of the matrices will accumulate when their size tends to infinity. It is well known that, in the generic case, the curve behaves like $y\sim \sqrt{x-a}$ near a branchpoint $a$ (an extremity of an interval); the appropriate double scaling limit gives the celebrated Airy kernel in connection with $(1,2)$ minimal models. But it may happen by taking a fine-tuned limit (see for instance \cite{BergereEynard}), that the behavior near a branchpoint differs from the generic case and takes the form of $y^p \sim (x-a)^q$. In such a case, it is expected that the double scaling limit is related to the conformal $(p,q)$ minimal model. In their articles \cite{BergereEynard} and \cite{determinantalformulae}, the authors opened the way to rigorous mathematical proofs in order to establish that the correlation functions of the double-scaling limit of a matrix model are the same as the ones defined by determinantal formulae arising from $(p,q)$ models. In their articles, they apply this method to all $(2m+1,2)$ models, i.e. suitable limits of matrix models where the spectral curve behaves like $y^2 \sim x^{2m+1}$ near an endpoint. In this article, we will use the same method for the $(2m,1)$ case which corresponds to a point where two cuts are merging with a degeneracy $2m$. For a generic merging, i.e. $m=1$ it has been proven in \cite{BleherEynard} that the suitable double scaling limit of the matrix model is connected to the Painlev\'e II equation. Some similar results have been established with the study of a suitable Riemann-Hilbert problem. For example the case of an even-quartic polynomial has been studied in \cite{RHP}. It would be also interesting to derive similar results, for these kernels, as the ones proved in \cite{CIK}. Here, using the approach of \cite{BergereEynard}, we find, as expected, that the correlation functions of the double scaling limit of the merging of two cuts with degeneracy $2m$ are expressed through the Lax system of the Painlev\'{e} II hierarchy (see \cite{mKDV} and \cite{MM}).

\section*{2 Double scaling limit in random matrices: the merging of two cuts}
\subsection*{2.1 Hermitian matrix models and equilibrium density} 
It is well known in the literature that the study of the Hermitian matrix model with partition function:
\beq \label{HermIntegral} Z_N=\int_{\mathbb{H}_N} \exp(-N\Tr(V(M)))dM\eeq
with an even polynomial potential
\beq \label{defPotential}
V(x)=\sum_{i=1}^{2d} t_i x^i
\eeq
can be reduced into an eigenvalue problem: $\lambda=\{ \lambda_j,j=1,...,N\}$ for the matrix $M$ with distribution:
\beq \label{DiagProblem}
\td{Z}_N=\int_{\mathbb{R}^N} \exp\left(2\sum_{1\leq j<k\leq N} \log|\lambda_j-\lambda_k|-N\sum_{i=1}^N V(\lambda_j)\right)
\eeq
When $N \to \infty$, the distribution of the eigenvalues on the line $d\nu_N(x)=\rho_N(x)dx$ is defined (in the distribution theory sense) by the formula
\beq \label{distribdensity}
\int_{\mathbb{R}}\phi(x)d\nu_N(x)=\frac{1}{\td{Z}_N}\int_{\mathbb{R}^N}\left(\frac{1}{N}\sum_{j=1}^n \phi(\lambda_j)\right) \exp\left(2\sum_{1\leq j<k\leq N} \log|\lambda_j-\lambda_k|-2N\sum_{i=1}^N V(\lambda_j)\right).
\eeq
For any test function $\phi(x)$ there is a weak limit $d\nu_\infty (x):=\underset{N \to \infty}{\lim} d\nu_N(x)$ which is the same as the equilibrium density $d\nu_{eq}(x)$ given by the limit of the empirical density:
\beq \label{empdistribution}
d \nu_{eq}(x)=\mathop{\lim}_{N \to \infty} \frac{1}{N} \sum_{j=1}^N \delta(x-\lambda_j)
\eeq
For details about the existence of the distribution limits; the equality between the equilibrium density $d\nu_{eq}(x)$ and $d\nu_\infty(x)$ and the following characterizations we refer the reader to \cite{BPS}, \cite{JOH}. Nowadays, many properties of the equilibrium density are known. For example, we know that the equilibrium density is supported by a finite number of intervals $[a_j,b_j], \, j=1,\dots,q$ and that it is absolutely continuous with respect to the Lebesgue measure:
\beq d\nu_{eq}(x)=\rho(x)dx=\frac{1}{2i\pi} h(x) R^{1/2}(x), \,\, \, R(x)=\prod_{j=1}^q (x-a_j)(x-b_j)\eeq
where $h(x)$ is a polynomial of degree $2d-q-1$ and $R^{1/2}(x)$ is to be understood as the value on the upper cut of the principal sheet of the complex-valued function $R^{1/2}(z)$ with cuts on $J=\bigcup_{j=1}^q[a_j,b_j]$. Eventually, the equilibrium density $d\nu_{eq}(x)$ is completely defined by the knowledge of the extremities $a_j$'s and $b_j$'s and the unknown coefficients of the polynomial $h(x)$. It has been proved that such quantities are uniquely determined by the following set of equations:
\begin{enumerate}
 \item Connexion between $h(z)$ and the potential $V(z)$:
\beq V'(z)=\underset{z \to \infty}{\text{Pol}}\left( h(z) R^{1/2}(z)\right) \label{Vconnexionh}\eeq
 \item Residue constraint:
\beq \label{ResidueConstraint}\Res_{z \to \infty} \left(h(z)R^{1/2}(z)\right)=-2\eeq
 \item Integrals constraints:
\beq \label{IntegralsConstraints}\int_{b_j}^{a_j+1}h(z)R^{1/2}(z)dz=0\, ,\,\, \forall j\in \{1,\dots,q-1\}
\eeq
\end{enumerate}
Note also that the relation between $h(z)$ and $V(z)$ \ref{Vconnexionh} can be inverted by:
\beq \label{hconnexionV} h(z)=\underset{z \to \infty}{\text{Pol}}\left(\frac{V'(z)}{R^{1/2}(z)}\right)\eeq

In theory, the previous set of equation is sufficient to determine the whole solution $d\nu_{eq}(x)$ but, practically, since the equations are highly non-linear, it becomes very hard to compute the unknown coefficients for two or more intervals or for potentials of degree higher than $4$. Moreover, in some exceptional situations, the previous set of equations has multiple solutions. In such situations, the good solution is determined by a positivity condition:
\beq \label{positivity} h(x)\geq0, \,\, \forall x \in J=\bigcup_{j=1}^q[a_j,b_j]\eeq
When $\forall x \in \bigcup_{j=1}^q]a_j,b_j[: \,\, h(x)>0$, the potential $V(x)$ and the equilibrium measure $d\nu_{eq}(x)$ are called \textit{regular}. Otherwise the equilibrium density is called \textit{singular} and the corresponding potential is called \textit{critical}, meaning that there is at least one point on $J$ where the equilibrium measure vanishes. For a regular potential, the situation can be summarized with the following picture:

\begin{center}
	\includegraphics[height=5cm]{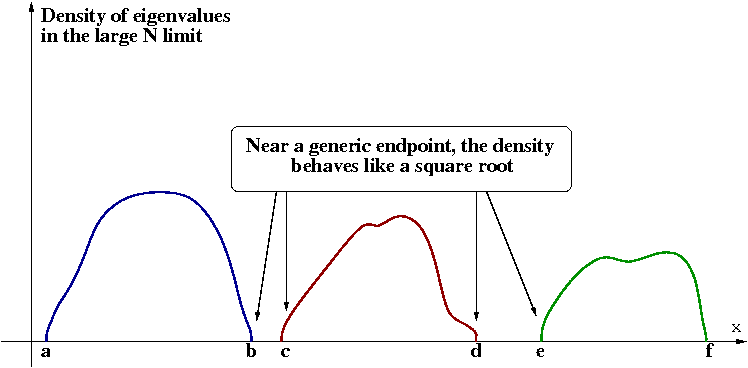}

	\underline{Figure 1}: Example of a typical eigenvalue density for a regular potential. The density is spread here in three intervals
\end{center}

\subsection*{2.2 Singular densities for the $(2m,1)$ case}

In order to study what happens at a singular density, one embeds the potential $V(x)$ into a parametric family $V(x,t)$ so that for some $t=t_c$ the problem is at the critical potential: $V(x,t_c)=V_c(x)$. Then the interesting questions are to determine the asymptotics of the eigenvalues correlation functions when $t \to t_c$. Indeed for $t \neq t_c$ the potential is regular and all the previous results stand. Therefore one can define $a_j(t)$, $b_j(t)$ and $h(x,t)$ determining completely the equilibrium density for $t\neq t_c$ and study their limits when $t \to t_c$.
In matrix models, it is often interesting to study a modified version of the integral \ref{HermIntegral} by introducing a parameter $T$ often referred as ``\textit{the temperature}'':
\beq \label{HermIntegralwithT} Z_N=\int_{\mathbb{H}_N} \exp(-\frac{N}{T}\Tr(V(M)))dM\eeq
It turns out that $T$ can be used as a parameter for the study of singular densities. In order to fit into our previous description, we need to introduce the following notation:
\beq V(x,T)=\frac{V(x)}{T}\eeq
In the study of the $(2m,1)$ model, we assume that at $T=T_c$ the potential $V(x,T_c)=V_c(x)$ becomes singular and gives rise to a singular density defined by the following $2m$ singular density:
\beq \label{SingularDensityAnn}
\rho(x,T_c)=\rho_c(x)=\frac{1}{2i\pi} (x-b\epsilon)^{2m}\sqrt{b^2-x^2}=\frac{1}{2i\pi}h_c(x)\sqrt{b^2-x^2}\eeq
with $\epsilon \in ]-1,1[$ representing the position of the singular point in the interval $]-b,b[$ supporting the distribution.
For $T\neq T_c$, we assume that the density is supported by two intervals $]a_1(T),b_1(T)[$ and $]a_2(T),b_2(T)[$ and define (note the normalization with $\frac{1}{T}$):
\beq \label{Density}
\rho(x,T)= \frac{1}{2i\pi T} h(x,T)\sqrt{(x-a_1(T))(x-b_1(T))(x-a_2(T))(x-b_2(T))}\eeq

Note that in order to recover our singular density at $T=T_c$ we must have:
\begin{center}
\begin{enumerate}
 \item $a_1(T)\underset{T\to T_c} {\to} -b$
 \item $b_1\underset{T\to T_c} {\to} b\epsilon$
 \item $a_2(T)\underset{T\to T_c} {\to} b\epsilon$
 \item $b_2(T) \underset{T\to T_c} {\to} b$
 \item $h(x,T)\underset{T\to T_c} {\to} h(x)$
\end{enumerate}
\end{center}

The previous assumptions correspond to the merging to two cuts with degeneracy $2m$ (order of the singularity). The most general case would be a singular point $a$ with $\rho_c^q(x)\underset{T\to T_c}{\sim}(x-a)^p, \, (p,q) \in \mathbb{N}^2$, which is expected to correspond to the $(p,q)$ minimal model (for $q>2$ we are speaking about multi-matrix models). In our case the situation can be summarized with the following pictures:

\begin{center}
	\includegraphics[height=6cm]{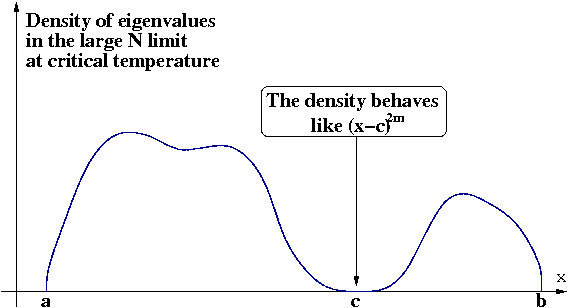}

	\underline{Figure 2}: Example of a critical eigenvalue density for a critical potential. At point $c$, the density is singular and behaves like $x^{2m}$

\bigskip
\bigskip

\includegraphics[height=6cm]{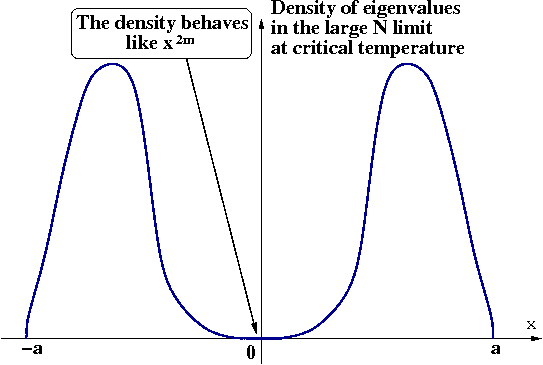}

	\underline{Figure 3}: Example of a critical eigenvalue density for a critical even potential. At the origin, the density is singular and behaves like $x^{2m}$
\end{center}

In \cite{BleherEynard}, the authors studied the case $m=1$ in details and conjectured some connections with Painlev\'{e} II hierarchy for higher $m$.

\subsection*{2.3 Double scaling limits in matrix models}

In the study of matrix models, one is usually interested in the following functions called \textit{resolvents}:
\bea \label{Corrfunctions}\hat{w}_n(x_1,\dots,x_n)&=&\big{<} \Tr\left(\frac{1}{x_1-M}\right)\dots \Tr\left(\frac{1}{x_n-M}\right)\big{>}\cr
&=&\big{<}\sum_{i_1,\dots,i_n} \Tr\left(\frac{1}{x_1-\lambda_{i_1}}\right)\dots \Tr\left(\frac{1}{x_n-\lambda_{i_n}}\right)\big{>}
\eea
and in their cumulants, also known as \textit{correlation functions}:
\bea \label{Corrfunctions2}\hat{w}_n(x_1,\dots,x_n)&=&\big{<} \Tr\left(\frac{1}{x_1-M}\right)\dots \Tr\left(\frac{1}{x_n-M}\right)\big{>}_{c}\cr
&=&\big{<}\sum_{i_1,\dots,i_n} \Tr\left(\frac{1}{x_1-\lambda_{i_1}}\right)\dots \Tr\left(\frac{1}{x_n-\lambda_{i_n}}\right)\big{>}_{c}
\eea
Here, the brackets stand for the integration relatively to the probability measure $Z_N^{-1}d\nu_N(x)$, the $\lambda_i$'s are the eigenvalues of the matrices and the index $c$ stands for the cumulants part (for example: $<AB>_c=<AB>-<A><B>$ and so on). The joint density correlation functions $\rho_n(x_1,\dots,x_n)$ can easily be deduced from the former correlation functions: densities are discontinuities of the resolvents and resolvents are Stieljes transforms of densities. For example:
\beq \hat{w}_1(x)=\int \frac{\rho_1(x')}{x-x'}dx' \Longleftrightarrow \rho_1(x)=\frac{1}{2i\pi} \left(\hat{w}_1(x-i0)-\hat{w}_1(x+i0)\right)\eeq
Then we want to use a formal $\frac{1}{T}$ power-series development which unfortunately is not necessarily well-defined for all matrix models. Indeed, if one is interested in \textit{convergent matrix models}, then one must be sure that such a series expansion commutes with integrations. In general, this does not happen and solutions of the convergent matrix model differ from the solutions of the \textit{formal matrix model} (where by definition the development is assumed to exist and to commute with integrations). The explanation of this phenomenon is simple: when we use a series expansion, it automatically ignores the exponentially small factors (one can think, for example, to $\exp(-x^2)$ which has at $x=\infty$ the same asymptotic expansion as the zero function). To sum up, formal matrix models are easier to handle, because by definition the formal expansion exists and we can perform formal operations on it; but the price to pay is that we only get a part of the convergent solutions (we miss the exponentially decreasing terms). It could appear disappointing to consider just formal matrix models, since they do not carry the whole convergent solutions (and thus leads only to a significative but incomplete part of the convergent solutions), but fortunately differences between formal and convergent matrix models have been well studied, and in \cite{ConvForm1}, \cite{Nonperturbative}, the authors show how to reconstruct with theta functions the convergent solutions from the formal ones. From now on, we will place ourselves in the case of formal matrix models, i.e. we assume that there automatically exists an expansion of type:
\beq \label{TopDev}\ln Z_N=\sum_{g=0}^\infty \left(\frac{N}{T}\right)^{2-2g}\hat{f}_g\eeq
and 
\beq \hat{w}_n(x_1,\dots,x_n)=\sum_{g=0}^\infty \left(\frac{N}{T}\right)^{2-2g-n}\hat{w}_n^{(g)}(x_1,\dots,x_n)\eeq
The numbers $\hat{f}_g$ are called \textit{symplectic or spectral invariants} of the model (invariant relatively to symplectic transformations of the spectral curve). The previous expansion can be understood as a large $N$ expansion and therefore in the limit $N\to \infty$ one expects that the leading value ($g=0$) corresponds to the ''real'' large $N$ limit of the model. In fact this intuition is correct and it has been proved that
\beq \hat{y}(x)=i\pi\rho_{eq}(x)=\frac{1}{2}V'(x)-\hat{w}_1^{(0)}(x).\eeq 
This formula establish a direct link between the equilibrium density and the leading order of the first correlation function. The function $\hat{y}(x)$ (which is up to a trivial rescaling the equilibrium density) is often named the \textit{spectral curve} of the problem. In our case, it satisfies:
\beq \hat{y}^2(x)=\text{Polynomial}(x)=\frac{1}{2T} h^2(x,T)(x-a_1(T))(x-b_1(T))(x-a_2(T))(x-b_2(T))\eeq
This identity defines the algebraic spectral curve $\hat{y}^2=P(x)$ where $P$ is a polynomial. We remind the reader that Eynard and Orantin showed in \cite{OE} that for any algebraic curve $P(x,y)=0$ we can associate some symplectic invariants $f_g$ and $w_n^{(g)}(x_1,\dots,x_n)$. Moreover, when the algebraic curve comes from a matrix model, these invariants are the same as the one we defined earlier in \ref{Corrfunctions} and \ref{Corrfunctions2}. 

In our case, the function $\hat{y}(x,T)=\frac{1}{2T} h^2(x,T)(x-a_1(T))(x-b_1(T))(x-a_2(T))(x-b_2(T))$ depends on the temperature $T$ and so are the corresponding invariants $\hat{w}_n(x_1,\dots,x_n,T)$ and $f_g(T)$. When $T\to T_c$ it is known that $ \forall g>1, \, f_g\to \infty$ and that the correlation functions diverges. This is so because the expansion \ref{TopDev} reaches its radius of convergence in $T$. In order to recover finite quantities, one has to rescale properly the variables at $T\sim T_c$. In our case we will prove that the good rescaling is given by:
\beq x_i=b\epsilon+(T-T_c)^{\frac{1}{2m}}\xi_i\eeq
so that
\beq \hat{y}_{\text{rescaled}}(\xi)=\lim_{T\to T_c} \frac{\hat{y}( b\epsilon+(T-T_c)^{\frac{1}{2m}}\xi, T)}{T-T_c}\eeq
and
\beq \hat{w}_{\text{rescaled},n}^{(g)}(\xi_1,\dots,\xi_n)=\lim_{T\to T_c} \frac{\hat{w}_n^{(g)}( b\epsilon+(T-T_c)^{\frac{1}{2m}}\xi_1,\dots,b\epsilon+(T-T_c)^{\frac{1}{2m}}\xi_n,T)}{(T-T_c)^n}\eeq
and
\beq \hat{f}_{\text{rescaled},g}=\lim_{T\to T_c} (T-T_c)^{-(2-2g)}\hat{f}_g\eeq
are finite quantities and that the new $\hat{w}_{\text{rescaled},n}^{(g)}(\xi_1,\dots,\xi_n)$ and $\hat{f}_{\text{rescaled},g}$ are the spectral invariants of the rescaled curve $\hat{y}_{\text{rescaled}}(\xi)$. In the general context of matrix model, such a rescaling is called a \textit{double scaling limit} since we have performed a double limit $N\to \infty$ and $T\to T_c$ so that $N(T-T_c)^{2m}$ remains finite:
\beq \ln Z_N=\sum_{g=0}^\infty \left(\frac{N}{T}\right)^{2-2g} \hat{f}_g\sim \sum_{g=0}^\infty \left(\frac{N}{T_c}\right)^{2-2g}(T-T_c)^{(2-2g)}\hat{f}_{\text{rescaled},g}\eeq
From a geometric point of view, this double scaling limit corresponds to a local zoom in the region of the degenerate point $b\epsilon$. The rate of the zoom depends on both the temperature $T$ and the size of the matrices $N$ so that $N(T-T_c)$ remains finite. It can be illustrated in the following picture:

\bigskip
\begin{center}
\includegraphics[height=6cm]{zoom.png}

	\underline{Figure 4}: Example of a critical eigenvalue density near the critical temperature
\end{center} 

In the context of matrix models, double scaling limits are often very important because they are expected to give universal (independent of the potential) rescaled spectral curve and correlation functions related to $(p,q)$ minimal models (and thus in our case the $(2m,1)$ minimal model). On the other hand, $(p,q)$ minimal models are studied through string reductions of some well known integrable systems. In the rest of the paper, we will prove that, in the case of the merging of two cuts, the rescaled spectral curve corresponds to the spectral curve of the $(2m,1)$ minimal model. Then, using the method introduced by Berg\`{e}re and Eynard in \cite{BergereEynard}, we prove that the rescaled correlation functions and the spectral invariants correspond to some ``correlation'' functions expressed with some  determinantal formulae \cite{determinantalformulae} for the $(2m,1)$ minimal model.

\subsection*{2.4 The rescaled spectral curve in our $2m$ degenerate matrix model case} 

In order to get the rescaled spectral curve, we need to perform a few consecutive steps. First we can express explicitly the corresponding critical potential corresponding to $\rho_c(x)$ in \ref{SingularDensityAnn} using \ref{Vconnexionh}.
The computation is straightforward and uses only the general Taylor expansion of:
\beq\sqrt{1+x}=1+\sum_{n=1}^\infty (-1)^{n+1} \frac{ (2n-2)!}{n!(n-1)!2^{2n-1}} x^n\eeq
It gives:
\beq V'_c(x)=\sum_{j=0}^{2m+1} \left( \dbinom{2m}{j-1} (-b\epsilon)^{2m+1-j} +\sum_{n=1}^{E(\frac{2m+1-j}{2})} \dbinom{2m}{2n+j-1} \frac{(-1)^j(2n-2)!\epsilon^{2(m-n)+1-j}b^{2m+1-j}}{n!(n-1)!2^{2n-1}} \right) x^j 
\eeq
where $E(\frac{2m+1-j}{2})$ stands for the greatest integer lower or equal to $\frac{2m+1-j}{2}$. The critical temperature is given by:
\beq T_c=\frac{b^{2m+2}}{2}\sum_{n=1}^{m+1} \frac{\epsilon^{2m-2n+2}(2m)! }{n!(2m-2n+2)!(n-1)!2^{2n-1}}\eeq

Then, we need to use some reformulations of conditions \ref{ResidueConstraint} and \ref{hconnexionV}. Indeed, it is known for a long time (a proof can be found in appendix A of \cite{BergereEynard} but the results were derived much before) and has been used intensively in \cite{AAM} that the set of equations \ref{ResidueConstraint} and \ref{hconnexionV} leads to the following ordinary differential equations (sometimes called \textit{hodograph equations}):
\bea \label{Diffset}
     \frac{d }{dT}a_1(T)&=& \frac{4 (a_1(T)-x_0(T))}{h(a_1(T),T)(a_1(T)-b_1(T))(a_1(T)-a_2(T))(a_1(T)-b_2(T))}\cr
     \frac{d }{dT}b_1(T)&=& \frac{4 (b_1(T)-x_0(T))}{h(b_1(T),T)(b_1(T)-a_1(T))(b_1(T)-a_2(T))(b_1(T)-b_2(T))}\cr
     \frac{d }{dT}a_2(T)&=& \frac{4 (a_2(T)-x_0(T))}{h(a_2(T),T)(a_2(T)-b_2(T))(a_2(T)-a_1(T))(a_2(T)-b_1(T))}\cr
     \frac{d }{dT}b_2(T)&=& \frac{4 (b_2(T)-x_0(T))}{h(b_2(T),T)(b_2(T)-a_2(T))(b_2(T)-a_1(T))(b_2(T)-b_1(T))}\cr
\eea
where the point $x_0(T)$ ($b_1(T)\leq x_0(T)\leq a_2(T)$ ) is determined by:
\beq \int_{b_1(T)}^{a_2(T)} \frac{z-x_0(T)}{\sqrt{(b_1(T)-z)(z-a_1(T))(b_2(T)-z)(z-a_2(T))}}dz=0 \label{Definition x_0}\eeq 
This set of equations taken at $T=T_c$ for $a_1$ and $b_2$ gives:
\bea \frac{d a_1(T)}{dT}_{|T=T_c}&=&-\frac{2}{(1+\epsilon)^{2m}b^{2m+1}}\cr
\frac{d b_2(T)}{dT}_{|T=T_c}&=&\frac{2}{(1-\epsilon)^{2m}b^{2m+1}}\cr\eea
so that in a neighbourhood of $T_c$:
\beq \label{a1}\encadremath{a_1(T)\underset{T=T_c}{\sim}-b-\frac{2}{(1+\epsilon)^{2m}b^{2m+1}}(T-T_c)+o(T-T_c)}\eeq
\beq \encadremath{b_2(T)\underset{T=T_c}{\sim}b+\frac{2}{(1-\epsilon)^{2m}b^{2m+1}}(T-T_c)+o(T-T_c)} \eeq
As mentioned earlier, we expect that the functions $a_j(T)$ and $b_j(T)$ will be analytic functions of $\Delta=(T-T_c)^\nu$, where $\nu$ is an exponent that we will determine later. Therefore we introduce the following notations:
\bea  \label{Texpansion}
b_1(T)&=&b\epsilon+\alpha\Delta+ \sum_{n=1}^\infty b_{1,n}\Delta^n\cr
a_2(T)&=&b\epsilon+\gamma\Delta+ \sum_{n=1}^\infty a_{2,n}\Delta^n\cr
x_0(T)&=&b\epsilon+X_0\Delta+\sum_{n=1}^\infty x_{n}\Delta^n\cr
h(z,T)&=&(z-b\epsilon)^{2m-1}+ P(z)\Delta+\sum_{n=1}^\infty h_{n}(z)\Delta^n\eea
where $P(z)$ and $h_n(z)$ are polynomials of degree at most $2m-2$. In equations \ref{Diffset} for $a_2(T)$ and $b_1(T)$ we see that the l.h.s. is of order $(T-T_c)^{\nu-1}$ whereas the r.h.s. is of order $(T-T_c)^{-(2m-1)\nu}$. Hence, to have compatible equations we must have, as announced in the previous subsection, that
\beq \label{Exponent} \encadremath{ \nu=\frac{1}{2m}}\eeq
The next step is purely technical and consists in proving that $\alpha=-\gamma$. Since it is only a technical point, we postpone this discussion in Appendix \ref{AppendixB}. With the help of this relation we can now determine the rescaled spectral curve.

First remember that for $T=T_c$ , we have \ref{SingularDensityAnn}:
\beq (z-b\epsilon)^{2m}=h_c(z)=\underset{z\to \infty}{\text{Pol}} \left(\frac{V'_c(z)}{\sqrt{z^2-b^2}}\right) \label{Definitionhc}\eeq
For $T\neq T_c$, reminding that $V(z,T)=\frac{V'(z)}{T}$ and that $\rho(z,T)$ is defined with a factor $\frac{1}{T}$ in \ref{Density} (which will cancel the one of $V(x,T)$) we have:
\beq h(z,T)=\underset{z\to \infty}{\text{Pol}} \left(\frac{V'(z)}{\sqrt{(z-a_1)(z-b_2)(z-a_2)(z-b_1)}}\right)\eeq
We now use the fact that up to order $\Delta^{2m-1}$, both $a_1$ and $b_2$ are respectively equal to $-b$ and $b$ (\ref{a1}). Therefore we get:
\beq h(z,T)=\underset{z\to \infty}{\text{Pol}} \left(\frac{V'(z)}{\sqrt{(z^2-b^2)(z-a_2)(z-b_1)}} +O\left(\Delta^{2m}\right)\right) \label{intermediate}\eeq
Then, from the definition of $h_c(x)$ we have that:
$$ (z-b\epsilon)^{2m}=\frac{1}{T_c}\underset{z\to \infty}{Pol} \left(\frac{V'(z)}{\sqrt{z^2-b^2}}\right)$$
so that:
\beq \frac{V'(z)}{\sqrt{z^2-b^2}}=T_c (z-b\epsilon)^{2m} + O\left(\frac{1}{z}\right)\eeq
Putting back this identity into \ref{intermediate} and noticing that $\frac{1}{\sqrt{(z^2-b^2)(z-a_2)(z-b_1)}}$ only gives negative powers of $z$ that will disappear when taking the polynomial part, we find that:
\bea h(z,T)&=&T_c \underset{z\to \infty}{\text{Pol}} \left(\frac{(z-b\epsilon)^{2m}}{\sqrt{(z-a_2)(z-b_1)}} +O\left(\Delta^{2m}\right)\right)\cr
&=& T_c \underset{z\to \infty}{\text{Pol}} \left(\frac{(z-b\epsilon)^{2m-1}}{\sqrt{1+\frac{2b\epsilon-a_2-b_1}{z-b\epsilon}+ \frac{(b\epsilon-a_2)(b\epsilon-b_1)}{(z-b\epsilon)^2}}} +O\left(\Delta^{2m}\right)\right)\cr
\eea
We can now insert the Taylor series of the square-root:
\beq (1+x)^{-\frac{1}{2}}=\sum_{n=0}^\infty \frac{(-1)^n(2n)!}{(n!)^22^{2n}}x^n\eeq
to get:
\bea  \frac{h(z,T)}{T_c}&=&\underset{z\to \infty}{\text{Pol}} \left((z-b\epsilon)^{2m-1}\left(1+\sum_{n=1}^\infty \frac{(-1)^n(2n)!}{(n!)^22^{2n}}\left(\frac{2b\epsilon-a_2-b_1}{z-b\epsilon}+ \frac{(b\epsilon-a_2)(b\epsilon-b_1)}{(z-b\epsilon)^2}\right)^n \right) \right)\cr
&&+O\left(\Delta^{2m}\right)\cr
&=&(z-b\epsilon)^{2m-1} +\cr
&&\underset{z\to \infty}{\text{Pol}} \Big(\sum_{n=1}^\infty\sum_{k=0}^n \frac{(-1)^n(2n)!}{(n!)^22^{2n}} \dbinom{n}{k}\left(2b\epsilon-a_2-b_1\right)^k \big((b\epsilon-a_2)(b\epsilon-b_1)\big)^{n-k} \cr
&&(z-b\epsilon)^{2m-1+k-2n} \Big) +O\left(\Delta^{2m}\right) \cr
\eea
Let's now introduce the following ensemble :\beq
\label{set}I_m=\{ (n,k)\in (\mathbb{N}^* \times \mathbb{N}) \, / \, 2n-k\leq 2m-1 \text{ and } k\leq n\}\eeq
Clearly $I_m$ is a finite set and we can rewrite the previous identity as:
 \bea  \frac{h(z,T)}{T_c}&=&(z-b\epsilon)^{2m-1} + \cr
&&\Big(\sum_{(n,k)\in I_m} \frac{(-1)^n(2n)!}{(n!)^22^{2n}} \dbinom{n}{k}\left(2b\epsilon-a_2-b_1\right)^k \big((b\epsilon-a_2)(b\epsilon-b_1)\big)^{n-k} \cr
&&(z-b\epsilon)^{2m-1+k-2n} \Big)+O\left(\Delta^{2m}\right) \cr
\eea
We can now introduce the series expansion in $\Delta$: 
$$2b\epsilon-a_2-b_1=-(\alpha+\gamma)\Delta+O\left(\Delta^2\right)$$ 
and
$$(b\epsilon-a_2)(b\epsilon-b_1)=\left(\alpha\Delta+\sum_{n=2}^\infty a_{2,n}\Delta^n\right)\left(\gamma\Delta+\sum_{n=2}^\infty b_{1,n}\Delta^n\right)=\Delta^2 \alpha\gamma+ O\left(\Delta^3\right)$$
Then we perform the rescaling \beq z=b\epsilon+\Delta\xi \label{rescalingAnn}\eeq
We only need to take into account terms with degree strictly less than $\Delta^{2m}$ so that only a few terms remain:

\bea
\frac{h(\xi,\Delta)}{T_c}&=&\left( \xi^{2m-1}+ \sum_{(n,k)\in I_m} \frac{(-1)^n(2n)!}{(n!)^22^{2n}} \dbinom{n}{k}(-1)^k (\alpha+\gamma)^k (\alpha\gamma)^{n-k}\xi^{2m-1+k-2n} \right)\Delta^{2m-1}\cr
&&+O\left(\Delta^{2m}\right)
\eea
so that:
\beq h_{\text{rescaled}}(\xi)=T_c\left( \xi^{2m-1}+ \sum_{(n,k)\in I_m} \frac{(-1)^n(2n)!}{(n!)^22^{2n}} \dbinom{n}{k}(-1)^k (\alpha+\gamma)^k (\alpha\gamma)^{n-k}\xi^{2m-1+k-2n} \right)\eeq

Eventually we get the rescaled spectral curve by taking into account the trivial term $R^{1/2}(z,T)=\sqrt{(z-a_1(T))(z-a_2(T))(z-b_1(T))(z-b_2(T))}$ with the rescaling \ref{rescalingAnn}:
\bea R^{\frac{1}{2}}(b\epsilon+\xi\Delta,\Delta)&=&\sqrt{(b\epsilon+\xi\Delta-a_1(\Delta))(b\epsilon+\xi\Delta-a_2(\Delta))(b\epsilon+\xi\Delta-b_1(\Delta))(b\epsilon+\xi\Delta-b_2(\Delta))}\cr
&=&b\sqrt{\epsilon^2-1}\sqrt{(b\epsilon+\xi\Delta-a_2(\Delta))(b\epsilon+\xi\Delta-b_1(\Delta))}+O\left(\Delta^{2m}\right)\cr
&=&ib\Delta\sqrt{1-\epsilon^2}\sqrt{(\xi-\alpha)(\xi-\gamma)}+O\left(\Delta^2\right)\cr
\eea
so that:
\bea \rho(b\epsilon+\xi\Delta,\Delta)&=& \frac{b\sqrt{1-\epsilon^2}}{2\pi}\sqrt{(\xi-\alpha)(\xi-\gamma)}\cr
&&\left( \xi^{2m-1}+ \sum_{(n,k)\in I_m} \frac{(-1)^n(2n)!}{(n!)^22^{2n}} \dbinom{n}{k}(-1)^k (\alpha+\gamma)^k (\alpha\gamma)^{n-k}\xi^{2m-1+k-2n} \right)\Delta^{2m} \cr
&&+O\left(\Delta^{2m+1}\right)\eea
giving that:
\bea \hat{y}_{\text{rescaled}}(\xi)&=&b\pi\sqrt{1-\epsilon^2}\sqrt{(\xi-\alpha)(\gamma-\xi)}\cr
&&\left( \xi^{2m-1}+ \sum_{(n,k)\in I_m} \frac{(-1)^n(2n)!}{(n!)^22^{2n}} \dbinom{n}{k}(-1)^k (\alpha+\gamma)^k (\alpha\gamma)^{n-k}\xi^{2m-1+k-2n} \right)\cr\eea
In the appendix \ref{AppendixB}, we prove that $\alpha=-\gamma$ so that it eventually leads to:
\beq \label{RescaledCurveAnn}\encadremath{\alpha=-\gamma \,\, ,\,\, :\,  \hat{y}_{\text{rescaled}}(\xi)=b\pi\sqrt{1-\epsilon^2}\sqrt{(\gamma^2-\xi^2)}\left( \xi^{2m-1}+ \sum_{n=1}^{m-1} \frac{(2n)!}{(n!)^22^{2n}} \gamma^{2n}\xi^{2m-1-2n} \right)}\eeq

We can even compute the precise value of $\gamma$. Indeed, using \ref{set} to compute the leading term of the $\Delta$-expansion of $h(a_2,T)$ and putting it back into \ref{Diffset} (and using the fact that with the definition of $x_0$ \ref{Definition x_0} we have $X_0=0$ when $\alpha+\gamma=0$) we have:

\beq \encadremath{ \label{gammaAnn}\alpha=-\gamma\,\,\, \text{with} \,\,\, \gamma^{2m}=\alpha^{2m}= -\frac{4m}{b^2(1-\epsilon^2)\left(\underset{n=0}{\overset{m-1}{\sum}} \frac{(2n)!}{(n!)^22^{2n}}\right) }=-\frac{(m!)^22^{2m+1}}{b^2(1-\epsilon^2)(2m)!} }\eeq 
In this case, introducing the new variable $s$ by $\xi=\gamma s$ or equivalently 
\beq z=b\epsilon+\gamma\Delta s\eeq
we get:
\beq \label{RescaledCurve2}\encadremath{\alpha=-\gamma \,\, ,\,\, :\,  \hat{y}_{\text{rescaled}}(s)=b\pi\gamma^{2m}\sqrt{1-\epsilon^2}\sqrt{(1-s^2)}\left( s^{2m-1}+ \sum_{n=1}^{m-1} \frac{(2n)!}{(n!)^22^{2n}} s^{2m-1-2n} \right)}\eeq

Eventually \ref{RescaledCurve2} shows as expected that when performing a double scaling limit $z=b\epsilon+\gamma\Delta s$ (with $\gamma$ a complex number given by \ref{gammaAnn} whose argument gives oscillations in the ($\Re(z)$,$\Im(z)$) plane), we recover a universal curve. In the next section, we will see that this rescaled spectral curve \ref{RescaledCurveAnn} is exactly, (up to the trivial normalization factor $b\sqrt{1-\epsilon^2}$) the spectral curve arising in the Lax pair representation of the Painlev\'{e} II hierarchy with $t_m=1$, all other $t_j$'s (See next section for a definition) taken to zero and the identification $u_0(t)=\gamma$ (coherently with \ref{u0Equation}). Before proceeding in the study of the Lax pair representation, we remind the reader that from general results of Eynard and Orantin \cite{OE}, the rescaled invariants and correlation functions $\hat{w}_{\text{rescaled},n}^{(g)}$ and $\hat{f}_{\text{rescaled},g}$ are automatically the symplectic invariants and correlation functions of the new rescaled spectral curve $\hat{y}_{\text{rescaled}}(\xi)$ and thus do automatically satisfied the famous loop equations \cite{OE}.

\section*{3 Correlation functions and invariants arising in the Lax pair representation of the $(2m,1)$ minimal model}

In the previous section, we have found the rescaled spectral curve for a double scaling limit of a $2m$ degenerate merging of two cuts in matrix models. As conjectured in \cite{BleherEynard}, we expect that this universal double scaling limit is connected to the Painlev\'{e} II hierarchy. In order to prove this result, we will follow the approach \cite{BergereEynard} developed and successfully applied for the $(2m+1,2)$ models. It consists in finding a natural spectral curve $y_{\text{Lax}}(x)$ from a Lax pair representation of the hierarchy and check that it is equal to our rescaled curve defined in the previous section. Then from another work of Berg\`{e}re and Eynard,\cite{determinantalformulae} we can define from the Lax pair representation some new correlation functions $W_n^{(g)}(x_1,\dots,x_n)$ and invariants $F_g$ by some determinantal formulae and a suitable kernel. In particular, they proved that these new functions do satisfy the same loop equations as our correlations functions. Eventually, with the study of the pole structure and $W_2^{(0)}$ we will end by proving that our new correlation functions $W_n^{(g)}(x_1,\dots,x_n)$ and invariants $F_g$ are identical to the rescaled ones defined in the previous section.

\subsection*{3.1 A Lax pair representation for the $(2m,1)$ minimal model}

In their paper \cite{BleherEynard}, the authors claimed that a good Lax pair representation for the $(2m,1)$ minimal model should be given by a set of two $2\times 2$ matrices $\mathcal{R}(x,t)$ and $\mathcal{D}(x,t)$ satisfying the following Lax pair representation:
\bea \label{LaxPairAnn}
\frac{1}{N}\frac{\partial}{\partial x} \Psi(x,t)&=& \mathcal{D}(x,t) \Psi(x,t)\cr
\frac{1}{N}\frac{\partial}{\partial t} \Psi(x,t)&=& \mathcal{R}(x,t) \Psi(x,t)
\eea
where $\Psi(x,t)$ is a two by two matrix whose entries will be written as:
\beq \Psi(x,t)=\begin{pmatrix}
                \psi(x,t) & \phi(x,t)\\
		\td{\psi}(x,t)& \td{\psi}(x,t)\\
               \end{pmatrix}
\eeq
and satisfies the normalization $\det\Psi(x,t)=1$.

The compatibility condition of the Lax pair is then:
\beq \label{CompatibilityCondition}
\left[\frac{1}{N} \frac{\partial}{\partial x} -\mathcal{D}(x,t), \mathcal{R}(x,t)-\frac{1}{N} \frac{\partial}{\partial t}\right]=0
\eeq
In order to specify completely the Lax pair, we need to impose some conditions about the shape of the matrices $\mathcal{R}(x,t)$ and $\mathcal{D}(x,t)$. In our case we will assume:
\beq 
\mathcal{R}(x,t)=\begin{pmatrix}
                  0 & x+u(t)\\
		  -x+u(t) &0\\
                 \end{pmatrix}
\eeq
and 
\beq 
\mathcal{D}(x,t)=\sum_{k=0}^m t_k \mathcal{D}_k(x,t)
\eeq
with
\beq
\mathcal{D}_k(x,t)=\begin{pmatrix}
                  -A_k(x,t) & xB_k(x,t)+C_k(x,t)\\
		  xB_k(x,t)-C_k(x,t) &A_k(x,t)\\
                 \end{pmatrix}
\eeq
and $A_k$, $B_k$, $C_k$ are polynomials of $x$ of degree respectively $2k-2$, $2k-2$, $2k$. Note that in the literature one can find several different Lax pair corresponding to the same problem. Indeed any conjugation (change of basis) give equivalent matrices that describe the same problem but in different coordinates (see section \ref{SectionLax}). In fact any equivalent Lax pair can be used since the quantities we will define later will be invariant from this choice.
In order to have more compact notation, we will use the following convention: a dot will indicate a derivative relatively to $t$ normalized by a coefficient $1/N$, namely:
\beq \label{Notation} \dot{f}(x,t)\mathop{=}^{\text{def}}\frac{1}{N} \frac{\partial f(x,t)}{\partial t}\eeq
Putting back this specific shape of matrices into the compatibility equation gives the following recursion:
\bea A_0&=&0, B_0=0, C_0=1\cr
C_{k+1}&=&x^2C_k+ \check{R}_k(u)\cr
B_{k+1}&=&x^2 B_k +\hat{R}_k(u)\cr
A_{k+1}&=&x^2+\frac{1}{2} \dot{\hat{R}}(u)
\eea 
where $\check{R}_k$ and $\hat{R}_k$ are the modified Gelfand-Dikii polynomials given by the following recursion:
\bea \label{recursionGD}
\hat{R}_0(u)&=&u \,\, \check{R}_0(u)=\frac{u^2}{2}\cr
\hat{R}_{k+1}(u)&=&u\check{R}_k(u)-\frac{1}{4}\frac{d^2}{d t^2} \hat{R}_k(u)\cr
\frac{d}{d t}\check{R}_k(u)&=&u \frac{d}{d t}\hat{R}_k(u)
\eea

It is then easy to see that the matrices $\mathcal{R}(x,t)$ and $\mathcal{D}(x,t)$ satisfy \ref{LaxPairAnn} if and only if $u(t)$ satisfies the string equation (see details in \cite{BleherEynard}.)
\beq \label{StringEquationAnn} \encadremath{
\sum_{k=0}^m t_k \hat{R}_k(u(t))=-tu(t)}\eeq
which gives an explicit differential equation of order $m$ satisfied by $u(t)$ (since the polynomials $\hat{R}_k$ can be explicitly computed from the recursion \ref{recursionGD}). In particular the case $m=1$ gives Painlev\'e II equation:
\beq \frac{d^2 u}{d t^2}(t)=2u^3(t)+4(t+t_0)u(t)\eeq
where $t_0$ is a free parameter that can be set to $0$ by a time-translation $\td{t}=t+t_0$.

\medskip

\underline{Remark}: Seculiar equations

As it is always the case for a linear differential equation, we can get a seculiar equation on $\psi(x,t)$ by combining the two components of the differential equation in $t$ given by \ref{LaxPairAnn}. In our case, we find that both $\psi(x,t)$ and $\phi(x,t)$ are solution of the seculiar equation:
\beq \label{Seculiar Equation}
\ddot{\psi}(x,t) -\frac{\dot{u}(t) \dot{\psi}(x,t)}{x+u(t)}=\left(u^2(t)-x^2\right)\psi(x,t)\eeq 
which by a simple standard change of variable can be transformed into a Schrodinger-like equation.

\subsection*{3.2 Large $N$ development}

From the fact that a dot derivative contributes with a factor $\frac{1}{N}$, it is easy to see from the string equation \ref{StringEquationAnn} that $u(t)$ admits a series development at large $N$:
\beq u(t)=\sum_{j=0}^\infty \frac{u_j(t)}{N^{2j}}= u_0(t)+\frac{u_1(t)}{N^2}+\dots \label{UExpansion}\eeq

\underline{Note}: The fact that $u(t)$ admits such a development in $\frac{1}{N^2}$ and not $\frac{1}{N}$ comes from the fact that the modified Gelfand-Dikii polynomials $\hat{R}_k$'s are a sum of terms involving only even numbers of dots-derivatives (i.e. even power of $\frac{1}{N}$).

Putting back this expansion into the string equation \ref{StringEquationAnn} and looking at the power of $N^0$ of the series gives us that $u_0(t)$ must satisfy the following algebraic relation:
\beq \encadremath{
\label{u0Equation}
-t=\sum_{j=1}^m t_j \frac{(2j)!}{2^{2j} (j!)^2} u_0(t)^{2j}}
\eeq

From that result, it is then easy to see that the matrices $\mathcal{R}(x,t)$ and $\mathcal{D}(x,t)$ also admit a large $N$ expansion:
\beq \mathcal{R}(x,t)=\begin{pmatrix}
                  0 & x+u_0(t)\\
		  -x+u_0(t) &0\\
                 \end{pmatrix} 
+\frac{1}{N^2}\begin{pmatrix}
                  0 & u_1(t)\\
		  u_1(t) &0\\
                 \end{pmatrix}
+\dots= \sum_{j=0}^\infty \frac{\mathcal{R}_j(x,t)}{N^{2j}}\eeq
and
\beq \mathcal{D}(x,t)= \sum_{j=0}^\infty \frac{\mathcal{D}_j(x,t)}{N^{j}}\eeq
where the first matrix can be explicitly computed:

\beq \label{MatriceD0} 
\mathcal{D}_0(x,t)=\begin{pmatrix} 0& t+B_0+C_0\\
                    -t +B_0-C_0 &0\\
                   \end{pmatrix}
\eeq
with
\bea
B_0&=&\sum_{j=1}^m t_j\sum_{k=0}^{j-1} x^{2(j-k)-1}\frac{(2k)!}{2^{2k} (k!)^2}u_0(t)^{2k+1}\cr
C_0 &=&\sum_{j=1}^m t_j\left(\sum_{k=0}^{j}x^{2j}+\sum_{k=0}^{j} x^{2(j-k)}\frac{(2k)!}{2^{2k} (k!)^2}u_0(t)^{2k}\right)
\eea

It should also be possible to find equations defining recursively the next matrices $\mathcal{R}_j(x,t)$ and $\mathcal{D}_j(x,t)$ by looking at the next orders in the series expansion. But since we will have no use of such results we do not mention them here.

\subsection*{3.3 Spectral Curve attached to the Lax pair}

By definition, the spectral curve of a differential system like \ref{LaxPairAnn} is given by $\det(\,yId-\mathcal{D}_0(x,t))=0$, that is to say by the large $N$ limit of the eigenvalues of the spectral problem (which we expect to give the large $N$ limit of our matrix model). Note in particular that this definition is independent of a change of basis (conjugation by a matrix). From all the previous results, we can compute this two by two determinant and get:
\bea
y^2
&=&\left[ \sum_{j=1}^m t_j\sum_{k=0}^{j-1} x^{2(j-k)-1}\frac{(2k)!}{2^{2k} (k!)^2}u_0(t)^{2k+1}+\sum_{j=1}^m t_j\left(\sum_{k=0}^{j}x^{2j}+\sum_{k=0}^{j} x^{2(j-k)}\frac{(2k)!}{2^{2k} (k!)^2}u_0(t)^{2k}\right)\right]\cr
&& \left[ \sum_{j=1}^m t_j\sum_{k=0}^{j-1} x^{2(j-k)-1}\frac{(2k)!}{2^{2k} (k!)^2}u_0(t)^{2k+1}-\sum_{j=1}^m t_j\left(\sum_{k=0}^{j}x^{2j}+\sum_{k=0}^{j} x^{2(j-k)}\frac{(2k)!}{2^{2k} (k!)^2}u_0(t)^{2k}\right)\right]\cr
&=&\left[ \sum_{j=1}^m t_j\sum_{k=0}^{j-1} x^{2(j-k)-1}\frac{(2k)!}{2^{2k} (k!)^2}u_0(t)^{2k+1}+\sum_{j=1}^m t_j\left(\sum_{k=0}^{j-1}x^{2j}+\sum_{k=0}^{j} x^{2(j-k)}\frac{(2k)!}{2^{2k} (k!)^2}u_0(t)^{2k}\right)\right]\cr
&& \left[ \sum_{j=1}^m t_j\sum_{k=0}^{j-1} x^{2(j-k)-1}\frac{(2k)!}{2^{2k} (k!)^2}u_0(t)^{2k+1}-\sum_{j=1}^m t_j\left(\sum_{k=0}^{j}x^{2j}+\sum_{k=0}^{j-1} x^{2(j-k)}\frac{(2k)!}{2^{2k} (k!)^2}u_0(t)^{2k}\right)\right]\cr
\eea 
where in the last identity we have use the algebraic equation satisfied by $u_0(t)$ \ref{u0Equation}. Then, it is then a straightforward computation to see that the product can be rewritten as:
\beq \label{SpectralCurve}\encadremath{
y_{\text{Lax}}^2=P(x,t)=\left(u_0(t)^2-x^2\right)\left(\sum_{j=1}^m t_j\sum_{k=0}^{j-1}\frac{x^{2(j-k)-1}(2k)!}{2^{2k} (k!)^2}u_0(t)^{2k}\right)^2 }\eeq

In particular in the specific case where $\forall j<m:\,t_j=0, $ and $t_m=1$, we find that the spectral curve reduces to:
\beq \label{SpectralCurveRed}\encadremath{
 \forall j<m:\, t_j=0,\, t_m=1 \, \,\Rightarrow \, y_{\text{Lax}}(x)=\sqrt{u_0(t)^2-x^2}\sum_{k=0}^{m-1}\frac{x^{2(m-k)-1}(2k)!}{2^{2k} (k!)^2}u_0(t)^{2k} }\eeq

\textbf{As expected, with the identification $u_0(t)=\gamma$ we recover exactly the rescaled-spectral curve of our matrix model \ref{RescaledCurve}}.

\underline{Note:}
In \ref{SpectralCurve} we can see that the only simple zeros of $P(x,t)$ are at $x=\pm u_0(t)$. Moreover since the polynomial $P(x,t)$ is obviously even and that there is no constant term in $x$ in the sum, we get that $P(x,t)$ has a double zero at $x=0$ and has double roots at some points $\pm \lambda_i, \, i=1,\dots,m-1$

\subsection*{3.4 Asymptotics of the matrix $\Psi(x,t)$}

The next step in the method of \cite{BergereEynard} is to determine an asymptotic of the functions $\psi(x,t)$ and $\phi(x,t)$. From the Schrodinger-like equation \ref{Seculiar Equation}, we have a BKW expansion:
$$\psi(x,t)=g(x,t)e^{N h(x,t)}\left(1+ \frac{\psi_1(x,t)}{N}+\frac{\psi_2(x,t)}{N^2}+\dots\right)$$
Putting back into the seculiar equation gives the following result:
\bea \label{Asymptotics}
\psi(x,t)&=&\frac{1}{\sqrt{2}} \left(\frac{u_0(t)+x}{u_0(t)-x}\right)^{\frac{1}{4}}e^{N\int^t\sqrt{u_0^2(t')-x^2}dt'}\left(1+\frac{\psi_1(x,t)}{N}+\dots\right)\cr
\phi(x,t)&=&-\frac{1}{\sqrt{2}} \left(\frac{u_0(t)+x}{u_0(t)-x}\right)^{\frac{1}{4}}e^{-N\int^t\sqrt{u_0^2(t')-x^2}dt'}\left(1+\frac{\phi_1(x,t)}{N}+\dots\right)\cr
\td{\psi}(x,t)&=&\frac{1}{\sqrt{2}} \left(\frac{u_0(t)-x}{u_0(t)+x}\right)^{\frac{1}{4}}e^{N\int^t\sqrt{u_0^2(t')-x^2}dt'}\left(1+\frac{\td{\psi}_1(x,t)}{N}+\dots\right)\cr
\td{\phi}(x,t)&=&\frac{1}{\sqrt{2}} \left(\frac{u_0(t)-x}{u_0(t)-x}\right)^{\frac{1}{4}}e^{-N\int^t\sqrt{u_0^2(t')-x^2}dt'}\left(1+\frac{\td{\phi}_1(x,t)}{N}+\dots\right)
\eea

One can easily check that at dominant order in $N$ the previous asymptotics gives $\det(\Psi(x,t))=1+O\left(\frac{1}{N}\right)$. The next step is to transform the integration over $t$ in the exponential as a integral over $x$ by using the property of the spectral curve. Indeed, the spectral curve defines a Riemann surface which can be parametrized locally by $x(z,t)$ and $y(z,t)$ where $z$ is a running point on the Riemann surface. Thus, the function $y$ can be seen as both a function of $(z,t)$ or $(x,t)$. In order to avoid confusion here, we will write differently the function when it is seen as a function of $(z,t)$ or as a function of $(x,t)$ (we put a tilda for the function in $(x,t)$ and keep $y$ for the function of $(z,t)$):
\beq \td{y}(x,t)= \sqrt{P(x,t)}=y(z(x,t),t)\eeq
Then, using standard chain rule derivation, one can compute:
\beq \frac{\partial y}{\partial z}\frac{\partial x}{\partial t}-\frac{\partial y}{\partial t}\frac{\partial x}{\partial z}=-\frac{\partial \td{y}}{\partial t} \frac{\partial x}{\partial z}\eeq
From the expression of the spectral curve \ref{SpectralCurve} (which gives explicitly $\td{y}(x,t)$)  one can compute $\frac{\partial \td{y}}{\partial t}$:
\bea \frac{\partial \td{y}}{\partial t}&=&\frac{x u_0(\partial_t{u}_0)}{\sqrt{u_0^2-x^2}}\left(\sum_{j=1}^m t_j\sum_{k=0}^{j-1}\frac{x^{2(j-1-k)}(2k)!}{2^{2k} (k!)^2}u_0(t)^{2k}\right) \cr
&&+x(\partial_t{u}_0) \sqrt{u_0^2-x^2}\left(\sum_{j=1}^m t_j\sum_{k=1}^{j-1}\frac{x^{2(j-1-k)}(2k)! 2k}{2^{2k} (k!)^2}u_0(t)^{2k-1}\right)\cr
&=&\frac{x(\partial_t{u}_0)}{\sqrt{u_0^2-x^2}}\sum_{j=1}^mt_j\frac{(2j)! (2j)}{2^{2j} (j!)^2}u_0(t)^{2j-1}\cr
&=&-\frac{x}{\sqrt{u_0^2-x^2}}
\eea
To get the last identity, we have used the string equation \ref{u0Equation} for $u_0(t)$. Therefore by introducing the parametrization:
\beq \label{Xpara} z^2=u_0(t)^2-x^2 \Leftrightarrow x^2=u_0(t)^2-z^2\eeq
one finds that:
\beq x'(z,t)=\frac{\partial x}{\partial z}=-\frac{\sqrt{u_0^2-x^2}}{x}\,\,\, , \,\,\, \frac{\partial x}{\partial t}=-\frac{u_0(\partial_t u_0)}{x} \eeq
so that eventually:
\beq \frac{\partial y}{\partial z}\frac{\partial x}{\partial t}-\frac{\partial y}{\partial t}\frac{\partial x}{\partial z}=-\frac{\partial \td{y}}{\partial t} \frac{\partial x}{\partial z}=-\frac{x}{\sqrt{u_0^2-x^2}}\frac{\sqrt{u_0^2-x^2}}{x}=-1\eeq
The last identity can be rewritten as:
\beq \label{PoissonBracket} \encadremath{\frac{\partial y}{\partial t}\frac{\partial x}{\partial z}-\frac{\partial y}{\partial z}\frac{\partial x}{\partial t}=1}\eeq
and interpreted as the remaining of a non-commutative structure of $[P,Q]=\frac{1}{N}$ in the limit $N \to \infty$ which in such situations often transform into a Poisson structure for $y(z,t) \leftrightarrow P$ and $x(z,t)\leftrightarrow Q$ by simply replacing the commutator with a Lie bracket: \beq\{y(z,t),y(z,t)\}=1 \label{CommutationStructure}\eeq

With the help of this structure, we can get a reformulation of the integral:
\beq \frac{\partial \td{y}}{\partial t}=\frac{1}{x'(z)}\eeq
hence:
\beq\frac{\partial \int^x \td{y}dx}{\partial t}=z\eeq
and
\beq \int^t \sqrt{u_0^2(t')-x^2}dt'=\int^t zdt  =\int^x \td{y}dx\eeq
Eventually we have the following large $N$ developments:
\bea \label{Asymptotics2}
\psi(x,t)&=&\frac{1}{\sqrt{2}} \left(\frac{u_0(t)+x}{u_0(t)-x}\right)^{\frac{1}{4}}e^{N\int^x\td{y}dx}\left(1+\frac{\psi_1(x,t)}{N}+\dots\right)\cr
\phi(x,t)&=&-\frac{1}{\sqrt{2}} \left(\frac{u_0(t)+x}{u_0(t)-x}\right)^{\frac{1}{4}}e^{-N\int^x\td{y}dx}\left(1+\frac{\phi_1(x,t)}{N}+\dots\right)\cr
\td{\psi}(x,t)&=&\frac{1}{\sqrt{2}} \left(\frac{u_0(t)-x}{u_0(t)+x}\right)^{\frac{1}{4}}e^{N\int^x\td{y}dx}\left(1+\frac{\td{\psi}_1(x,t)}{N}+\dots\right)\cr
\td{\phi}(x,t)&=&\frac{1}{\sqrt{2}} \left(\frac{u_0(t)-x}{u_0(t)-x}\right)^{\frac{1}{4}}e^{-N\int^x\td{y}dx}\left(1+\frac{\td{\phi}_1(x,t)}{N}+\dots\right)
\eea

\subsection*{3.5 Kernels and correlation functions in the Lax pair formalism}

It was established in \cite{determinantalformulae} that one can define a kernel $K(x_1,x_2)$ and define from it (through determinantal formulae) some functions $W_n(x_1,\dots,x_n)$ that have nice properties. In particular the authors showed in \cite{determinantalformulae} that these functions do satisfy some loop equations and thus are likely to correspond to our matrix model correlation functions. Following \cite{determinantalformulae} we define the kernel by:
\beq \label{KAnn}
K(x_1,x_2)=\frac{\psi(x_1)\td{\phi}(x_2)-\td{\psi}(x_1)\phi(x_2)}{x_1-x_2}\eeq
Then we define the (connected) correlation functions by:
\beq W_1(x)=\psi'(x)\td{\phi}(x)-\td{\psi}'(x)\phi(x)\eeq
\beq \label{defcorrAnn} W_n(x_1,\dots,x_n)=-\frac{\delta_{n,2}}{(x_1-x_2)^2}- (-1)^n\sum_{\sigma=cycles} \prod_{i=1}^n K(x_{\sigma(i)},x_{\sigma(i+1)})\eeq
and eventually we define non-connected functions $W_{n,n-c}$ by determinantal formulae:
\beq \label{defnoncorrAnn}W_{n,n-c}(x_1,\dots,x_n)=\mathop{det}^{'}(K(x_i,x_j))\eeq
where the notation $\mathop{det}^{'}$ means that the determinant is computed in the usual way as a sum over permutations $\sigma$ of products $(-1)^\sigma\prod_{i=1}^n K(x_i,K_{\sigma_i})$, except for terms when $i=\sigma(i)$ and  when when $i=\sigma(j) \,, j=\sigma(i)$. In such cases, one must replace $K(x_i,x_i)$ by $W_1(x_i)$ and $K(x_i,x_j)K(x_j,x_i)$ by $-W_2(x_i,x_j)$. For additional details, we invite the reader to look at (\cite{determinantalformulae}) 

\medskip

As in our problem we will need the large $N$ developments of these functions, we introduce the notations:
\bea \label{NexpansionKernelsCorrFuncts}K(x_1,x_2)&=&K_0(x_1,x_2)e^{N\int_{x_2}^{x_1}\td{y}dx}\left(1+\sum_{g=1}^\infty N^{-g}K^{(g)}(x_1,x_2)\right)\cr
W_n(x_1,\dots,x_n)&=&\sum_{g=0}^\infty N^{2-2g-n}W_n^{(g)}(x_1,\dots,x_n)\cr
W_{n,n-c}(x_1,\dots,x_n)&=&\sum_{g=0}^\infty N^{n-2g}W_{n,n-c}^{(g)}(x_1,\dots,x_n)\eea
Then, we can insert all our previous results concerning the leading terms of the series expansion \ref{Asymptotics2},\ref{defcorrAnn} and  \ref{NexpansionKernelsCorrFuncts}. It gives:
\beq K_0(x_1,x_2)=\frac{1}{2(x_1-x_2)}\left(\left(\frac{u_0+x_1}{u_0-x_1}\right)^{\frac{1}{4}}\left(\frac{u_0-x_2}{u_0+x_2}\right)^{\frac{1}{4}}+\left(\frac{u_0-x_1}{u_0+x_1}\right)^{\frac{1}{4}}\left(\frac{u_0+x_2}{u_0-x_2}\right)^{\frac{1}{4}}\right)
\eeq
\beq W_1^{(0)}(x)=\td{y}(x)\eeq
and
\beq W_2^{(0)}(x_1,x_2)=\frac{1}{4(x_1-x_2)^2}\left(-2+ \sqrt{ \frac{(u_0+x_1)(u_0-x_2)}{(u_0-x_1)(u_0+x_2)} }+\sqrt{ \frac{(u_0-x_1)(u_0+x_2)}{(u_0+x_1)(u_0-x_2)} } \right)\eeq 

In order to get rid of the square-roots in the expressions above, it is better to introduce a proper parametrization of our spectral curve \ref{SpectralCurve}. Let us define:
\beq \label{Para2Ann} \encadremath{x=\frac{u_0}{2}\left( z+\frac{1}{z}\right)=\frac{u_0(z^2+1)}{2z}  \Leftrightarrow z=\frac{1+\sqrt{x^2-u_0^2}}{u_0}}\eeq
In particular, under such a change of variables we obtain several useful identities:
\bea \sqrt{\frac{u_0-x}{u_0+x}}&=&i\frac{z-1}{z+1}\cr
 u_0-x&=&-\frac{u_0(z-1)^2}{2z}\cr
 u_0-x&=&\frac{u_0(z+1)^2}{2z}\cr
 \sqrt{u_0^2-x^2}&=&\frac{iu_0}{2z}(z+1)(z-1)\cr
\frac{d x(z)}{dz}&=& \frac{u_0(z^2-1)}{2z^2}
\eea

\medskip

Eventually we can rewrite $W_2^{(0)}$ in terms of the new variable $z$:
\beq W_2^{(0)}(z_1,z_2)=\frac{4z_1^2z_2^2}{u_0^2 (z_1^2-1)(z_2^2-1)(z_1z_2-1)^2}\label{W20}\eeq

Although these functions have some interesting features, they still depend on the choice of coordinates on the Riemann surface defined by the spectral curve. Therefore, we introduce similarly to \cite{BergereEynard} and \cite{OE} the corresponding differential forms:

\beq \mathcal{W}_n^{(g)}(z_1,\dots,z_n)=W_n^{(g)}(x(z_1),\dots,x(z_n))x'(z_1)\dots x'(z_n) +\delta_{n,2}\delta_{g,0} \frac{x'(z_1)x'(z_2)}{(x(z_1)-x(z_2))^2} \eeq
These differentials are symmetric rational functions of all their variables. Moreover as proved in the crucial theorem \ref{Pole Structure} these functions only have poles at $z_i=\pm 1$ (except again $\mathcal{W}_2^{(0)}(z_1,z_2)$ which may have a pole at $x(z_1)=x(z_2)$). Eventually, a direct computation from \ref{W20} gives:
\beq \encadremath{\mathcal{W}_2^{(0)}(z_1,z_2)=\frac{1}{(z_2-z_1)^2} }\eeq

\subsection*{3.6 Loop equations, determinantal formulae, pole structure and unicity}

The previous determinantal definitions may seem rather arbitrary, but as we mention before they have the interesting property (proved in \cite{determinantalformulae}) to satisfy the following loop equations.
\begin{theorem} Loop equations satisfied by the determinantal functions:
 \bea \label{loopnc} P_n(x;x_1,\dots,x_n)&=& W_{n+2,n-c}(x,x,x_1,\dots,x_n)\cr
&&+\sum_{j=1}^n \frac{\partial}{\partial x_j} \frac{W_n(x,x_1,\dots,x_{j-1},x_{j+1},\dots,x_n)-W_n(x_1,\dots,x_n)}{x-x_j}\cr\eea
is a polynomial of the variable $x$. The previous theorem is equivalently reformulated for the standard connected functions:
\bea \label{loop} P_n(x;x_1,\dots,x_n)&=& \sum_{h=0}^g \sum_{I \subset J} W_{1+|I|}^{(h)}(x,I)W_{1+n-|I|}^{(g-h)}(x,J/I) \cr
&&+\sum_{j=1}^n \frac{\partial}{\partial x_j} \frac{W_n(x,J/\{x_j \} )-W_n(x_1,\dots,x_n)}{x-x_j}\cr\eea
is a polynomial of the variable $x$.
\end{theorem}

We emphasize again that loop equations are an essential step because it is well known in the matrix model world \cite{Mehta} that the correlation functions introduced in our first section do satisfy these loop equations. Unfortunately, loop equations generally admit several solutions encoded essentially in the unknown coefficients of the polynomial $P_n$. Therefore we need some additional results to get unicity. The first one deals with the pole structure:

\begin{theorem}\label{Pole Structure}  \underline{Pole Structure}:

The functions $z\to \psi_k(z,t)$ are rational functions with poles only at $z\in \{ \pm i,0,\infty\}$. The coefficients of these fractions depend on $u_0(t)$ and its derivatives. Hence the determinantal correlation functions $W_n^{(g)}$ are symmetric and rational functions in the variables $z_i$ with poles only at $z_i=\pm 1$.
\end{theorem}

\underline{Proof}: The last part of the theorem is obvious from the definitions as soon as the results regarding the $\psi_k(z,t)$'s are established. This proof is presented in Appendix \ref{AppPoleStructure} and is highly non-trivial. It uses the whole structure of integrability (i.e. the two differential equations \ref{LaxPairAnn}) to eliminate other possible poles (at the other zeros of $y_{text{Lax}}(x)$).

With the knowledge of the pole structure of the $W_n^{(g)}$, the fact that they satisfy the loop equations and the knowledge of $W_2^{(0)}$ we have a unicity theorem. In fact under these conditions we can identify our differentials $\mathcal{W}_{n}^{(g)}$'s with the ones defined by the standard recursion relation introduced by Eynard and Orantin in \cite{OE}:

\begin{theorem}
The differentials $\mathcal{W}_{n}^{(g)}$ satisfy the following recursion:
\bea \label{recursion}
  \mathcal{W}_{n+1}^{(g)}(z_1,\dots,z_n,z_{n+1})&=&\Res_{z \to \pm 1} \frac{dz}{2u_0 y(z)(1-\frac{z_{n+1}}{z})(\frac{1}{z}-z_{n+1})}\big[ \mathcal{W}_{n+2}^{(g-1)}(z,\overline{z},z_1,\dots,z_n)\cr
&&+\sum_{h=0}^g \mathop{\sum}_{I \in J}^{'} \mathcal{W}_{1+|I|}^{(h)}(z,I)\mathcal{W}_{n+1-|I|}^{(g-h)}(\overline{z},J/I)\big]\cr
\eea
where $J$ is a short-writing for $J=(z_1,\dots,z_n)$ and $\overset{g}{\underset{{h=0}}{\sum}} \overset{'}{\underset{I \in J}{\sum}}$ means that we exclude the terms $(h,I)=(0,\emptyset)$ and $(h,I)=(g,J)$ in the sum. The notation $\overline{z}$ stands for the conjugate point of $z$ near the poles where the residue is taken. In our case: $\overline{z}=\frac{1}{z}$
\end{theorem}

\underline{Note}: It is worth noticing that in Eynard and Orantin's notation we have in our case (we omit the dependance in the $t$ parameter):
\bea \om(z)&=&y(z)\frac{u_0 (z^2-1)}{z^2}\cr
 y(\overline{z})&=&y(1/z)=-y(z)\cr
dE_z(p)&=&\frac{1}{2}\int_z^{\frac{1}{z}} \frac{ds}{(s-p)^2}=\frac{1-z^2}{2(z-p)(pz-1)}\eea
so that:
\beq
\frac{dE_z(z_{n+1})}{\om(z)}=\frac{z^2}{2u_0 y(z)(z-z_{n+1})(1-z_{n+1}z)}=\frac{1}{2u_0 y(z)(1-\frac{z_{n+1}}{z})(\frac{1}{z}-z_{n+1})}\eeq
 
\underline{Proof of \ref{recursion}}: The unicity proof has been done in various article but for completeness we rederive it here with our notations. First of all Cauchy's theorem states that:
\beq \mathcal{W}_{n+1}^{(g)}(z_1,\dots,z_{n+1})= \Res_{z \to z_{n+1}} \frac{dz}{z-z_{n+1}}\mathcal{W}_{n+1}^{(g)}(z_1,\dots,z_{n+1})\eeq
We can move the integration contour to enclose all other poles, i.e. only $\pm 1$ in our case:
\bea \mathcal{W}_{n+1}^{(g)}(z_1,\dots,z_{n+1})&=& \Res_{z \to \pm 1} \frac{dz}{z_{n+1}-z}\mathcal{W}_{n+1}^{(g)}(z_1,\dots,z_{n+1})\cr
&=&\Res_{z \to \pm 1} \frac{x'(z)dz}{z_{n+1}-z}W_{n+1}^{(g)}(x(z_1),\dots,x(z_{n+1}))\eea
Then using the loop equations \ref{loop} and separating the coefficients $W_1^{(0)}$ in the sum gives:
\bea
-2W_1^{(0)}(x)W_{n+1}^{(g)}(x_1,\dots,x_n,x)&=&\sum_{h=0}^g \mathop{\sum}^{'}_{I \subset J} W_{1+|I|}^{(h)}(x,I)W_{1+n-|I|}^{(g-h)}(x,J/I)\cr
&&\sum_{j=1}^n \frac{\partial}{\partial x_j} \frac{W_n(x,J/\{x_j \} )-W_n(x_j,J/\{x_j \})}{x-x_j}\cr
&&-P_n^{(g)}(x,x_1,\dots,x_n)\eea

The polynomial $P_n^{(g)}(x,x_1,\dots,x_n)$ does not contribute to the residue, and after using the relation between $x$ and $z$ we are left with \ref{recursion}.

\section*{4 Lax pairs for the (2m,1) minimal model and for the Painlev\'e II hierarchy \label{SectionLax}}

\subsection*{4.1 The (2m,1) minimal model and the Flashka-Newell Lax pair}

As observed in \cite{BleherEynard} the string equation \ref{StringEquationAnn} is nothing but the $m^{\text{th}}$ member of the so-called Painlev\'e II hierarchy. The Painlev\'e II (PII) hierarchy, a collection of ODEs of order $2m$, arises as a self-similar reduction of the mKdV hierarchy. In the papers \cite{mKDV} and \cite{MM} this relationship has been used to construct a Lax pair for the PII hierarchy starting from the relevant Lax pair for the modified KdV hierarchy. We call this PII Lax pair the Flashka-Newell Lax pair since the first member of the hierarchy was find, for the first time, in \cite{FN}. In this subsection we prove that, up to a linear transformation of the wave function and a rescaling of the variables, the Flashka-Newell Lax Pair is equivalent to the $(2m,1)$ minimal model Lax pair. In order to simplify notation we forget, in this section, the rescaling given by $1/N$ over the variables $x$ and $t$. We begin with the case $t_1=0=t_2=\ldots=t_{m-1}$. 

\begin{proposition}
	Define $\tilde\Psi$ as a new wave function
	$$\tilde\Psi:=J\Psi$$ with
	$$J:=\begin{pmatrix}
			1 & i\\
			i & 1
		\end{pmatrix}$$
	and set $t_m\longmapsto (4^{m+1}/2)$ (all other parameters $t_j$ equal to $0$). Then $\tilde\Psi$ satisfies the Flashka-Newell Lax pair as written in \cite{mKDV}.
\end{proposition}
\underline{Proof}
 Since $J$ is constant we observe that $\tilde\Psi$ solve the Lax system
 \bea\label{FNLAxPair}
	\frac{\partial}{\partial x}\tilde{\Psi}(x,t)&=& \tilde{\mathcal{D}}_m(x,t) \tilde{\Psi}(x,t)\cr
	\frac{\partial}{\partial t} \tilde{\Psi}(x,t)&=& \tilde{\mathcal{R}}(x,t) \tilde{\Psi}(x,t)
\eea
 with $ \tilde{\mathcal{D}}(x,t), \tilde{\mathcal{R}}(x,t)$ obtained through conjugation with $J$; i.e.
 $$
 	\tilde{\mathcal{R}}(x,t)=J\mathcal{R}(x,t)J^{-1}=\begin{pmatrix}
											-ix & u\\
											u & ix
										\end{pmatrix}	
 $$
 and
 $$
 	\tilde{\mathcal{D}}_m(x,t)=\frac{4^{m+1}}{2}J\mathcal{D}_m(x,t)J^{-1}=\frac{4^{m+1}}{2}\begin{pmatrix}
					-iC_m(x,t) & iA_m(x,t)+xB_m(x,t)\\
					-iA_m(x,t)+xB_m(x,t) & iC_m(x,t).
				\end{pmatrix}
 $$
These two matrices are exactly the ones appearing in (16a) and (16b) in \cite{mKDV} (modulo the identification $u\longleftrightarrow w, x\longleftrightarrow\lambda, t\longleftrightarrow z$). 
For the matrix $\tilde{\mathcal R}$ this is self-evident. For $\tilde{\mathcal{D}}_m$ we just have to observe that it has the same shape as the matrix written in the right-hand side of (16b) (see eqs (14); in particular the polar part in (16b) is zero thanks to (14b)). On the other hand this conditions, plus compatibility condition, determines uniquely $\tilde{\mathcal{D}}_m$.

\medskip

Of course the result above is extended to the case in which all $t_j$ enter in $\cal D$ just taking linear combinations of the matrices studied in the previous proposition. This has been done, for the Flashka-Newell pair, in \cite{MM} (note, nevertheless, that there the spectral parameter is rotated; $\lambda\rightarrow -i\lambda$). Hence we have the following proposition.

\begin{proposition}
	Under a rescaling of all time variables $t_j\longrightarrow \frac{4^{j+1}}{2} t_j$ the (2m,1)-minimal model Lax pair is equivalent to the Flashka-Newell Lax pair for the PII hierarchy.
\end{proposition}

\section*{5 Conclusion and outlooks}

In section $1$, we have established that the double scaling limit of a matrix model with a $2m$-degenerate point can define a universal rescaled spectral curve $\hat{y}_{\text{rescaled}}(x)$. In section $1$ we also reminded that the correlation functions and symplectic invariants $\hat{w}_n^{(g)}(x_1,\dots,x_n)$ and $\hat{f}_g$ can also be rescaled in a suitable way in order to give some new functions  $\hat{w}_{\text{rescaled,n}}^{(g)}(x_1,\dots,x_n)$  and new symplectic invariants  $\hat{f}_{\text{rescaled,g}}$ corresponding respectively to the correlation functions and symplectic invariants of the rescaled curve $\hat{y}_{\text{rescaled}}(x)$. Then, starting from a Lax pair of the Painlev\'{e} II hierarchy and using the same method as \cite{BergereEynard} we have constructed a spectral curve $y_{\text{Lax}}(x)$ which coincides with $\hat{y}_{\text{rescaled}}(x)$ for a natural choice of the flow parameters $t_j$'s. Finally, with the definition of a suitable kernel and determinantal formulae, we have defined in the same way as \cite{BergereEynard} some functions $W_n^{(g)}$ having interesting properties (loop equations). Studying in details the pole structure and computing $\mathcal{W}_{2}^{(0)}(z_1,z_2)$, we have eventually shown that the function $W_n^{(g)}$'s are in fact exactly the correlation functions of the curve $y_{\text{Lax}}(x)$. Since the two spectral curves are the same, we have proved the statement:
\begin{theorem}
The correlation functions (and spectral curve) of the double scaling limit of a $2m$-degenerate merging of two cuts are the same as the
functions $W_n^{(g)}$ (and spectral curve) defined by determinantal formulae of the integrable Painlev\'{e} II hierarchy's kernel.
\end{theorem}

This result reinforces the links between double scaling limit in matrix models and integrable $(p,q)$ minimal models. With this new result and the one of \cite{BergereEynard}, the two models are shown to be identical for $(p=2m ,q=1)$ and $(p=2m+1,q=2)$ ($m \in \mathbb{N}^*)$. However even if this identity is expected to hold for every $(p,q)$, some complete proofs as the one presented here are still missing. Indeed, if our reasoning may seem easy to generalize for arbitrary value of $p$ and $q$, the crucial theorem \ref{loop} establishing that the functions $W_n^{(g)}$ coming from determinantal formulae do satisfy the loop equations (proved in \cite{determinantalformulae}) is only valid for $q\leq2$ at the moment. Therefore a good approach to the generalization for arbitrary value of $(p,q)$ could be to first extend this theorem for every $(p,q)$ and then to use the method presented here to extend the result.

Another approach could be to use this approach to study other integrable systems whose Lax pairs are known. Indeed, it is possible to perform the same method as the one presented here for any Lax pair. In particular, for every Lax pair, it would be interesting to analyse the associated spectral curve and the corresponding determinantal correlation functions.

\section*{Appendix: Pole structure for $\psi_k(z,t)$}\label{AppPoleStructure}

In order to use the unicity theorem \ref{loop} showing that the $W_n^{(g)}$'s are the expected correlation functions, we need to precise the pole structure of the function $\psi_k(x,t)$'s and $\phi_k(x,t)$'s from which they are defined. In order to determine the functions $\psi_k(x,t)$'s, one can insert the series expansion \ref{Asymptotics2} into the seculiar equations. Since the case $\psi_k(x,t)$ and $\phi_k(x,t)$'s are similar (they satisfy the same seculiar equation), we will focus only on the $\psi_k(x,t)$'s. The main issue of this appendix is that putting the large $N$ asymptotics of $\psi(x,t)$ \ref{Asymptotics2} into the seculiar equation a priori gives unwanted poles at the zeros of $y(x)$ for $\psi_k(x,t)$ that we need to rule out. It is the purpose of this appendix to explain how this can be done.

\subsection*{Study of the differential equation in $t$}

From the fact that $u(t)$ satisfies the string equation we remind the reader that we have \ref{u0Equation}:
\beq t=-\sum_{j=1}^m t_j \frac{(2j)!}{2^{2j} (j!)^2} u_0(t)^{2j}=P_0(u_0)\eeq
From this, it follows that $\frac{du_0}{dt}$ is:
\beq  \frac{du_0}{dt}=\frac{1}{P_0'(u_0)}\eeq
Performing more derivations relatively to $t$ can give the derivatives of $u_0(t)$ to any order as a fraction whose denominator is always a power of $P_0'(u_0)$. For example:
\bea \frac{d^2u_0}{dt^2}&=&-\frac{P''_0(u_0)}{(P_0'(u_0))^3}\cr
 \frac{d^3u_0}{dt^3}&=&-\frac{P_0'''(u_0)}{(P_0'(u_0))^4}+3\frac{(P_0''(u_0))^2}{(P_0'(u_0))^5}\eea
and so on.

As a consequence, any power of any derivative of $u_0$ remains a rational function of $u_0$ with poles only at the roots of $P_0'(x)$. For example, expressions like $\frac{d u_0}{dt}\frac{d^3 u_0}{dt^3}+ \left(\frac{d u_0}{dt}\right)^2\frac{d^2 u_0}{dt^2}$ will be rational functions of $u_0$ with poles only at the roots of $P_0'(x)$.

Now, putting back the development of $u(t)=u_0(t)+\frac{u_2(t)}{N^2}+\frac{u_3(t)}{N^3}+\dots$ into the full string equation \ref{StringEquationAnn} gives that any subleading order $u_k$ can be expressed as a rational function of $u_0$ with poles only at the roots of $P'_0(x)$.

Eventually, inserting the shape of the function $\psi(x,t)$ into the seculiar equation and evaluating the order $N^{-k}$ gives the following equation $\forall k\geq 2$:
\bea \label{psik witht}
\partial_t \psi_{k-1}&=&\frac{\partial_{t^2}g}{2gh} \psi_{k-2} -\frac{\partial_{t}g}{gh}\partial_t \psi_{k-2} -\frac{\partial_{t^2}\psi_{k-2}}{2h} +\frac{1}{2} \left(\frac{\partial_t u}{u+x}\right)_{k} +\frac{1}{2} \sum_{i=0}^{k-2} \left(\frac{\partial_t u}{u+x}\right)_{k-i}\psi_i\cr
&&+\frac{\partial_t g}{2gh} \left(\frac{\partial_t u}{u+x}\right)_{k-1}+\frac{1}{2} \sum_{i=0}^{k-2}\left(\frac{\partial_t u}{u+x}\right)_{k-1-i}\left(\psi_i \frac{g_t}{gh} +\frac{\partial_t \psi_i}{h}\right)\cr
&&+\frac{1}{2} \sum_{i=0}^{k-2} \left(u^2\right)_{k-i} \frac{\psi_i}{h}
\eea
where we have written in short:
\bea
\psi_0(x,t)&=&1\cr
h(x,t)&=&\sqrt{u_0(t)^2-x^2}\cr
g(x,t)&=&=\left(\frac{u_0(t)+x}{u_0(t)-x}\right)^{1/4}\cr \label{definitions}
\eea
and the notation $\left(\frac{\partial_t u}{u+x}\right)_{k}$ stands for the term in $N^{-k}$ in the expansion of $\frac{\partial_t u}{u+x}$. Note in particular that these terms can be expressed as a fraction with poles at $u_0(t)+x=0$ and at $P'(u_0(t))=0$ (the last are independent of $x$).
For example the first one is:
\beq \label{psi1 witht} \partial_t \psi_1(x,t)=\frac{ (\partial_t u_0)^2 x^2 (u_0-x)^{\frac{3}{2}}}{4(u_0+x)^{\frac{3}{2}}} +\frac{(u_0+x)^{\frac{1}{2}}}{(u_0-x)^{\frac{1}{2}}} u_2(t)\eeq
where remember that $u_2(t)$ can be expressed as a rational function of $u_0(t)$ whose poles are known are only when $u_0(t)$ is at a root of $P_0'$ (and thus are independent of $x$). From this expression, it is clear that $\psi_1(x,t)$ may only have $x$-dependent singularities at $x=\pm{u_0}$ and at $x=\infty$.

\subsection*{Study of the differential equation in $x$}

The technic presented in the previous subsection can be carried out for the differential equation in $x$. Starting with the second equation of the Lax pair \ref{LaxPairAnn}:
\beq 
\frac{1}{N}\frac{\partial} {\partial x} \Psi(x,t) =
\begin{pmatrix}
                  -A(x,t) & xB(x,t)+C(x,t)\\
		  xB(x,t)-C(x,t) &A(x,t)\\
                 \end{pmatrix}
\Psi(x,t)
\eeq
we can derive another seculiar equation for both $\psi(x,t)$ and $\phi(x,t)$:
\bea \label{seculiar}
0&=&\frac{1}{N^2} \frac{\partial^2}{\partial x^2}\psi(x,t) -\frac{1}{N^2} \left(\frac{ \partial_x (xB+C)}{xB+C}\right)\frac{\partial}{\partial x}\psi(x,t) \cr
&&+\frac{1}{N} \left(\partial_x A -A \frac{ \partial_x (xB+C)}{xB+C} \right) \psi(x,t)-y^2(x,t)\psi(x,t)\eea 
where we have used that:
\beq \det(\Psi)=1 \Leftrightarrow y^2(x,t)=A(x,t)^2+x^2B(x,t)^2-C(x,t)^2\eeq
Note in particular in the last identity that the r.h.s. should have a large $N$ development whereas the l.h.s. $y(x)$ given by \ref{SpectralCurve} does not. Therefore, the l.h.s. must have vanishing subleading orders in $\frac{1}{N^k} ,\forall \, k>0$.

Moreover, reformulating \ref{MatriceD0} give:
\bea\label{identity2}
 A_0&=&0\cr
\left(xB+C\right)_0&=&y(x,t)\sqrt{\frac{u_0+x}{u_0-x}}
\eea
where the subscript $0$ stands for the first order in the large $N$ expansion. Indeed, it comes from the fact that:
\bea y^2(x,t)&=&\left(xB+C\right)_0\left(xB-C\right)_0=P(x,t)=P_1(x,t)P_2(x,t)\cr
 \left(xB+C\right)_0=P_1(x,t)&=&(u_0+x)\left(\sum_{j=1}^m t_j\sum_{k=0}^{j-1}\frac{x^{2(j-1-k)}(2k)!}{2^{2k} (k!)^2}u_0(t)^{2k}\right)\cr
 \left(xB-C\right)_0=P_2(x,t)&=&(u_0-x)\left(\sum_{j=1}^m t_j\sum_{k=0}^{j-1}\frac{x^{2(j-1-k)}(2k)!}{2^{2k} (k!)^2}u_0(t)^{2k}\right)\cr
\eea
and eventually:
\beq P_2(x,t)=P_1(x,t) \frac{u_0-x}{u_0+x} \eeq
With \ref{identity2} it is easy to see that:
\beq \label{identity3} \left(\frac{ \partial_x (xB+C)}{xB+C}\right)_0=\frac{\partial_x y}{y} +\frac{u_0}{u_0^2-x^2}\eeq
which will be crucial for the coherence of the computation. Indeed, putting the large $N$ expansion of $\psi(x,t)$:
$$ \psi(x,t)=g(x,t)e^{N h(x,t)}\left(1+ \frac{\psi_1(x,t)}{N}+\frac{\psi_2(x,t)}{N^2}+\dots\right)$$
into \ref{seculiar} and comparing the first orders in $\frac{1}{N}$ gives:
\bea 
0&=&g(x,t)y^2(x,t)-g(x,t)y^2(x,t)\cr
0&=& \partial_x(g(x,t)y(x,t)) +y(x,t)\partial_xg(x,t) +g(x,t)y(x,t) \left(\frac{ \partial_x (xB+C)}{xB+C}\right)_0
\eea
The second equation with the help of \ref{identity3} determines $g(x,t)$ coherently with \ref{definitions}, that is to say:
$$g(x,t)=\left(\frac{u_0(t)+x}{u_0(t)-x}\right)^{1/4}$$
Note now that $\forall \,k>0$, the function $\left(\frac{ \partial_x (xB+C)}{xB+C}\right)_k$ only has singularities at the singularities of $\frac{ 1}{xB_0+C_0}$ according to the standard rules of Taylor series for a fraction. 
The next order, $\frac{1}{N^2}$, gives us the function $\psi_1(x,t)$ (with the notation that a subscript $k$ defines the term in $N^{-k}$ in the expansion at large $N$):
\bea \label{psi1 withx}
\partial_x \psi_1(x,t)&=&-\frac{\partial_{x^2}g}{2gy}+ \frac{\partial_xg}{2gy}\left(\frac{ \partial_x (xB+C)}{xB+C}\right)_0+\frac{1}{2}\left(\frac{ \partial_x (xB+C)}{xB+C}\right)_1\cr
&&-\frac{1}{2}\left(\partial_xA -A\frac{ \partial_x (xB+C)}{xB+C}\right)_1\cr
\eea
From the definition of $g(x,t)$, it is easy to compute:
\bea \frac{\partial_x g}{g}&=&\frac{1}{2}\frac{u_0}{u_0^2-x^2}\cr
\frac{\partial_{x^2}g}{g}&=&\frac{u_0x}{u_0^2-x^2}+\frac{1}{4} \frac{u_0^2}{(u_0^2-x^2)^2}\eea
and thus to see that $\partial_x \psi_1(x,t)$ is a function of $x$ that may only have singularities at $x=\pm u_0$, at $x=\infty$ and at the others zeros of $y(x)=0$. (it is so because $\left(\frac{ \partial_x (xB+C)}{xB+C}\right)_1$  have the same singularities as $\frac{1}{xB_0+C_0}$ which by \ref{identity2} are only at $x=\pm u_0$, $x=\infty$ and at the zeros of $y(x)$).

It is then possible to extend this result for higher terms in the large $N$ expansion. The power $\frac{1}{N^k}$ gives:
\bea \label{psik withx}
\partial_x \psi_{k-1}&=&-\frac{\partial_{x^2}g}{2gy}\psi_{k-2}-\frac{\partial_xg}{gy}\partial_x \psi_{k-2} -\frac{1}{y}\partial_{x^2}\psi_{k-2}\cr
&&+\frac{1}{2}\sum_{i=0}^{k-2}\left(\frac{ \partial_x (xB+C)}{xB+C}\right)_{k-1-i} \psi_i +\frac{\partial_x g}{2gy}\sum_{i=0}^{k-2}\left(\frac{ \partial_x (xB+C)}{xB+C}\right)_{k-2-i}\psi_i\cr
&&+\frac{1}{2y}\sum_{i=0}^{k-2}\left(\frac{ \partial_x (xB+C)}{xB+C}\right)_{k-2-i}\partial_x\psi_i-\frac{1}{2y}\sum_{i=0}^{k-1-i}\left(\partial_x A- A\frac{ \partial_x (xB+C)}{xB+C}\right)_{k-1-i}\psi_i\cr
\eea
where we have define $\psi_0=1$. The precise form of the relation is mostly irrelevant, but the main fact is that if all the $\psi_i(x,t) $ with $i<k$ are assumed to have singularities only at $x=\pm u_0$, $x=\infty$ and at the other zeros of $y(x)=0$, then the same is true for $\partial_x \psi_{k}$ by a simple recursion.

\subsection*{Pole structure of $\psi_k(x,t)$}

With the help of \ref{psik witht} and \ref{psik withx} we are now able to prove that the only singularities of $x \mapsto \psi_k(x,t)$ are at $x=\pm u_0$ and at $x=\infty$. 

From \ref{psik witht} we have shown that $\partial_t \psi_k(x,t)$ can only have singularities at $x=\pm u_0(t)$, at $x=\infty$ and when $u_0(t)$ is at a root of $P_0'$. But from \ref{psik withx} we have shown that $\partial_x \psi_k(x,t)$ can only have singularities at $x=\pm u_0(t)$, at $x=\infty$ and at the other zeros of $y(x)=0$ given by $x=\lambda_i(t)$ solution of $\overset{m}{\underset{j=1}{\sum}} t_j\overset{j-1}{\underset{k=0}{\sum}}\frac{x^{2(j-k)-1}(2k)!}{2^{2k} (k!)^2}u_0(t)^{2k}=0$ in \ref{SpectralCurve}. But these poles are incompatible with the former result. Indeed if $\psi_k(x,t)$ had a pole at $x=\lambda_i(t)$, then $\partial_t \psi_k(x,t)$ would also have a pole at $x=\lambda_i(t)$, but we have shown that the only $x$-dependent singularities of $\partial_t \psi_k(x,t)$ are at $x\pm u_0(t)$ or $x=\infty$ giving rise to a contradiction. \textbf{Therefore:
 $x \mapsto \psi_k(x,t)$ has only singularities at $x=\pm u_0$ (square-root poles) and $x=\infty$ (poles) and in particular has no pole at the other zeros of $y(x)=0$}. This result is highly non trivial because we need to combine the two differential equations (i.e. the whole integrable structure) to get it. Hence, the structure of integrability seems to play an important underlying role in the pole structure and we can hope that such a result could extend to every integrable system. 

\subsection*{Pole structure in the $z$ variable}

In order to have only poles (and not square root singularities), we want to shift the former result to the $z$ variable defined by:
\beq z^2=\frac{u_0-x}{u_0+x} \Leftrightarrow x=u_0 \frac{1-z^2}{1+z^2}\eeq
Note that we have the identities:
\bea \frac{\partial x}{\partial t}&=&(\partial_t u_0)\frac{1-z^2}{1+z^2}\cr
\frac{\partial x}{\partial z}&=&-\frac{4u_0z}{(1+z^2)^2}\cr
u_0+x&=&\frac{2u_0}{1+z^2}\cr
u_0-x&=&\frac{-2z^2}{1+z^2}\cr
g(z,t)&=&\frac{(-u_0)^{\frac{1}{4}}}{z^{\frac{1}{2}}}\cr
y(z,t)&=&\frac{4z^2u_0^2}{(1+z^2)^2} P_0(\left(\frac{1-z^2}{1+z^2}\right)u_0)\cr
\frac{\partial_x g}{g}(z,t)&=&\frac{(1+z^2)^2}{8u_0^2z^2}\cr
\frac{\partial_{x^2}g}{g}(z,t)&=&\frac{(1+z^2)^2}{4u_0^2z^2}+\frac{(1+z^2)^4}{64u_0^2z^4}\label{computations}\eea

Note also that every polynomial in $x$ will give a polynomial in $\frac{1-z^2}{1+z^2}$, that is to say a rational function in $z$ with poles at $z^2+1=0$.

The rules for derivation gives that:
\beq \partial_t \td{\psi}_k(z,t)=\partial_t \psi_k(x,t)+\frac{\partial x}{\partial t} \frac{\partial \psi_k(x,t)}{\partial x}\eeq
\beq \partial_z \td{\psi}_k(z,t)=\frac{\partial x}{\partial z}\partial_x \psi_1(x,t) \eeq
where all these terms are already known from the previous sections. If one uses \ref{computations} and the remark that a polynomial in $x$ will give a rational function in $z$ with poles at $z^2+1=0$ (and remember that functions $A,B,C$ are polynomials in $x$), one can see that the singularities of $\psi_k(x,t)$ at $x=\pm{u_0}$ (square-root type) and at $x=\infty$ (poles), will transform into poles at $z=0$ ($\Leftrightarrow x=-u_0$), $z=\infty$ ($\Leftrightarrow x=u_0$) and $z=\pm i$ ($\Leftrightarrow x=\infty$).

Hence we have the final result:
\textbf{ \label{poles} $\forall k\geq 0$: the functions $z\to \psi_k(z,t)$ are rational functions with poles only at $z\in \{ \pm i,0,\infty\}$. The coefficients of these fractions depend on $u_0(t)$ and its derivatives.}  

\section*{Appendix: Discussion about $\alpha=-\gamma$}\label{AppendixB}

When computing the rescaled spectral curve in the matrix model double scaling limit, we need to find a relationship between $\alpha$ and $\gamma$ that are given by \ref{Texpansion}:
\bea
b_1(T)&=&b\epsilon+\alpha\Delta+ \sum_{n=1}^\infty b_{1,n}\Delta^n\cr
a_2(T)&=&b\epsilon+\gamma\Delta+ \sum_{n=1}^\infty a_{2,n}\Delta^n\cr
\eea
where we remind that $\Delta=(T-T_c)^{\frac{1}{2m}}$.
A first argument in favour of the fact that $\alpha=-\gamma$ is the case when $\epsilon=0$. Indeed, in such a case, the situation is fully symmetric around the singular point $0$. Therefore, one expects the two endpoints $b_1(T)$ and $a_2(T)$ to be symmetric around $x=0$ for every value of $T$ around $T_c$. In such a case the identity $\forall T \simeq T_c: \, a_2(T)=-b_1(T)$ gives $\alpha=-\gamma$. When $\epsilon\neq0$, we can carry out a similar reasoning at first orders in $\Delta$. Indeed, if we center the origin at $b\epsilon$, then as we observed it several times, the endpoints $a_1$ and $b_2$ can be considered to be respectively $-b$ and $b$ up to order $\Delta^6$. Therefore in the function $R^{\frac{1}{2}}(x)$ they only add a multiplicative trivial factor depending on $\epsilon$ ($\sqrt{1-\epsilon^2}$ to be precise) which will not change the symmetry around $b\epsilon$ of the endpoints $a_2$ and $b_1$ at first orders in $\Delta$.

Eventually, another more explicit approach is to put the developments \ref{Texpansion} into all the equations \ref{hconnexionV}, \ref{ResidueConstraint}, \ref{IntegralsConstraints} and \ref{Definition x_0} determining $h(z,T)$, $x_0(T)$ and the endpoints $a_1(T), b_1(T), a_2(T)$ and $b_2(T)$. Doing so leads to an algebraic equation of degree $2m$ connecting $\alpha$ and $\gamma$:
\beq Q(\alpha,\eta)=0 \eeq
with $Q$ a symmetric, homogeneous polynomial of degree $2m$. Unfortunately the system does not admit a unique solution as soon as $m>1$. Indeed, although the solution $\alpha=-\gamma$ is always there, when $m>1$ there are also other possibilities such as $\alpha=\lambda \gamma, \lambda \in \mathbb{C}$ and $\gamma$ satisfying an equation of degree $2m$ with complex coefficients. Though it might appear surprising that the set of equations may have several distinct solutions (thus giving several eigenvalues density), one must remember that they are some additional constraints for the solution. Indeed, if one wants to have a density distribution, it means that all quantities involved must at least be real and positive. Therefore only the solution $\alpha=-\gamma$ is possible.

\underline{Note}: In fact $\alpha$ and $\gamma$ are not necessarily well defined. Indeed, there are only defined up to a multiplicative $(2m)^{\text{th}}$ root of unity since the equation defining them is homogeneous of degree $2m$. This is because the notion of $\Delta=(T-T_c)^{\frac{1}{2m}}$ is also ambiguous, whereas $\Delta^{2m}$, $\alpha^{2m}$ and $\gamma^{2m}$ are well-defined quantities. (which explain why the development in $a_1(T)$ and $b_2(T)$ is well defined). Indeed, if one changes: 
\beq \forall n \in \{ 1,\dots,2m-1 \}: \,\, \Delta \rightarrow \td{\Delta}=\Delta e^{\frac{2in\pi}{2m}} \, \, \, \text{,} \,\,\,  \alpha \rightarrow \td{\alpha}=\alpha e^{\frac{-2in\pi}{2m}}\, \, \, \text{and} \,\,\, \gamma \rightarrow \td{\gamma}=\gamma e^{-\frac{2in\pi}{2m}}\eeq
then \ref{Texpansion} remains unchanged. With the change $\xi \rightarrow \td{\xi}=\xi e^{-\frac{2in\pi}{2m}}$, the rescaled spectral curve remains unchanged.

%% file: annexee.tex
\annexe{The partition function of  the two-matrix model as an isomonodromic tau-function} \label{Article[II]}
\selectlanguage{english}

\baselineskip 16pt 
\begin{center}
 {M. Bertola}$^{\ddagger,\sharp}$\footnote{bertola@crm.umontreal.ca},  {O. Marchal}$^{\dagger, \sharp}$\footnote{olivier.marchal@polytechnique.org}
\\
\bigskip
\begin{small}
$^{\dagger}$ {\em Institut de Physique Th\'eorique,
CEA, IPhT, F-91191 Gif-sur-Yvette, France
CNRS, URA 2306, F-91191 Gif-sur-Yvette, France

$^\sharp$
Centre de recherches math\'ematiques,
Universit\'e de Montr\'eal\\ C.~P.~6128, succ. centre ville, Montr\'eal,
Qu\'ebec, Canada H3C 3J7} \\
\smallskip
$^{\ddagger}$ {\em Department of Mathematics and
Statistics, Concordia University\\ 1455 de Maisonneuve W., Montr\'eal, Qu\'ebec,
Canada H3G 1M8} \\ 
\end{small}
\end{center}
\bigskip
%\maketitle
\smallskip
%\vspace{20pt}
\bigskip
\begin{center}{\bf Abstract}
\end{center}
\smallskip
We consider the Itzykson-Zuber-Eynard-Mehta two-matrix model and prove that the partition function is an isomonodromic tau function in a sense that generalizes Jimbo-Miwa-Ueno's  \cite{JMI}. 
In order to achieve the generalization we need to define a notion of tau-function for isomonodromic systems where the $ad$--regularity of the leading coefficient is not a necessary requirement.

\bigskip
\bigskip

\section*{1 Introduction}

Random matrices models have been studied for years and have generated important results in many fields of both theoretical physics and mathematics. 

The two-matrix model 
\beq
\d\mu(M_1,M_2) = {\rm e}^{-\Tr (V_1(M_1) + V_2(M_2) -M_1M_2)} \d M_1\d M_2\label{101}
\eeq
was used to model $2D$ quantum gravity \cite{DKK} and was investigated from a more mathematical point of view in \cite{MS, EynMehta, BEH1,BEH2,BEH,BEH4, BE}; the {\em partition function} of the model
\beq
\mathcal Z_N(V_1,V_2) = \int\int \d\mu(M_1,M_2)
\eeq
has important properties in the large $N$--limit for the enumeration of discrete maps on surfaces \cite{ZJDFG} of arbitrary genus and it is also known to be a tau-function for the $2$--Toda hierarchy.
In the case of  the Witten conjecture, proved by Kontsevich \cite{Kontsevich} with the use of matrix integrals not too dissimilar from the above one,  the enumerative properties of the tau function imply some nonlinear (hierarchy of) PDEs (the  KdV hierarchy for the mentioned example). 
On a similar level,  one expects some hierarchy of PDEs  for the case of the two-matrix model and possibly some  Painlev\'e\ property (namely the absence of movable essential singularities). 
The Painlev\'e\ property is characteristic of tau-functions  for isomonodromic families of ODEs that depend on parameters; hence a way of establishing such property for the partition function $\mathcal Z_N$ is that of identifying it with an instance of isomonodromic tau function \cite{JMI, JMII}. 

 This is precisely the  purpose of this article; we capitalize on previous work that showed how to relate the matrix model to certain biorthogonal polynomials \cite{MS, EynMehta} and how these appear in a natural fashion as the solution of certain isomonodromic family \cite{Harnad} .
 
 The paper extends to the case of the two matrix model the work contained in \cite{Harnad, BEH4, BertoGekhtman}; it uses, however, a different approach, closer to the recent \cite{BE_mom}. 
 
 In \cite{Harnad, BEH4, BertoGekhtman, ITW} the partition function of the one--matrix model (and certain shifted T\"oplitz determinants) were identified as isomonodromic tau functions by using {\em spectral residue formul\ae} in terms of the spectral curve of the differential equation. 
 Such spectral curve has interesting properties inasmuch as --in the one-matrix case-- the spectral invariants can be related to the expectation values of the matrix model. Recently the spectral curve of the two matrix model \cite{BE} has been written explicitly in terms of expectation values of the two--matrix model and hence one could use their result and follow a similar path for the proof as the one followed in \cite{BEH4}. 
Whichever one of the two approaches one chooses,  a main obstacle  is that the definition of isomonodromic tau function \cite{JMI, JMII} relies on a genericity assumption for the ODE which fails in the case at hand, thus requiring a generalization in the definition.
 
 According to this logic, one of the purposes of this paper  
  is to extend the notion of tau-function introduced by Jimbo-Miwa-Ueno's  \cite{JMI}, to the two-matrix Itzykson-Zuber model. This task is accomplished in a rather general framework in Sec. 
\ref{taudef}. 

We then show that the partition function has a very precise relationship with the tau-function so introduced, allowing us to (essentially) identify it as an isomonodromic tau function (Thm. \ref{main}).

\section*{2 A Riemann Hilbert formulation of the two-matrix model}
\label{2mmRHP}
According to the seminal work \cite{MS, EynMehta} and 
following the notations and definitions introduced in \cite{BEH1, BEH2}, we consider paired sequences of monic polynomials $\{\pi_m(x), \s_m(y)\}_{m=0\dots \infty}$
 $ (m=\deg{\pi_m} =\deg{\s_m}$), that are biorthogonal in the sense that
 \beq
 \int \!\!\! \int _\varkappa dx dy \pi_m(x)\s_n(y) e^{-V_1(x) -V_2(y) +xy} =  h_m \delta_{mn}, \quad h_m\neq 0.
 \eeq
The functions $V_1(x), V_2(y)$ appearing here are referred to as  {\it potentials}, terminology drawn from random matrix theory, in which such quantities play a fundamental role.
 
  Henceforth, the second potential $V_2(y)$ will be chosen as a polynomial of degree $d_2+1$
  \beq
  V_2(y) = \sum_{j=1}^{d_2+1} \frac{v_j}{j} y^j , \quad v_{d_2+1}  \ \ne 0
  \eeq
  For the purposes of most of the considerations to follow, 
  the first potential $V_1(x)$ may have very general analyticity properties as long as the manipulations make sense, but for definiteness and clarity we choose it to be polynomial as well.
  
 The symbol $\int\int_\varkappa$ stands for any linear combination of integrals of the form
 \beq
\int\!\!\! \int _\varkappa  dx dy  := \sum_{j}\sum_k \varkappa_{jk}\int_{\G_j} dx \int_{\GH_k} dy ,\qquad \varkappa_{ij} \in \mathbb C
\eeq
 where  the  contours $\{\GH_k\}_{k=1\dots d_2}$ will be chosen as follows. In the $y$--plane, define  $d_2+1$ ``wedge sectors'' $\{ \hat{S}_k\}_{k=0\dots d_2}$  such that $\hat{S}_k$ is bounded by the pairs of rays: $r_k:= \{y \vert  \arg{y}= \theta + \frac{2k\pi}{d_2+1}\}$ and $r_{k-1}:= \{y \vert \arg{y}= \theta +\frac{2(k-1)\pi}{d_2 +1}\}$, where $\theta:= \arg{v_{d_2+1}}$. Then $\GH_k$ is any smooth oriented contour  within the sector $\hat{S}_k$ starting from $\infty$ asymptotic to the ray $r_k$ (or any ray within the sector that is at an angle $< \frac{\pi}{2(2d_2+1)}$ to it, which is equivalent for purposes of integration), and returning to $\infty$ asymptotically along $r_{k-1}$ (or at an angle  $ < \frac{\pi}{2(2d_2+1)}$ to it). These will  be referred to as the ``wedge contours''. We also define a set of smooth oriented contours $\{\GC_k\}_{k=1, \dots d_2}$,  that have intersection matrix $\GC_j \cap \GH_k= \delta_{jk}$ with the $\GH_k$'s, 
  such that $\GC_k$ starts from $\infty$ in sector $\hat{S}_0$, asymptotic to the ray $\check{r}_0 :=\{y \vert \arg(y) = \theta - \frac{\pi}{d_2+1}$  and returns to $\infty$  in sector $\hat{S}_k$ asymptotically along the ray $\check{r}_k := \{y \vert \arg(y) = \theta + \frac{2(k-\frac{1}{2})}{d_2+1}$. These will  be called the ``anti-wedge'' contours.  (See Fig. 1.)    The choice of these contours is determined by the requirement that all moment integrals of the form
     \beq
     \int_{\GH_k}  y^j e^{-V_2(y) +xy }  dy , \quad  \int_{\GC_k }y^k e^{V_2(y) -xy} dy,
     \quad k=1, \dots d_2, \quad j\in \Nbb
     \eeq
    be uniformly convergent in $x\in \Cbb$.
     In the case when the other potential $V_1(x)$ is also a polynomial, of degree $d_1 +1$, the contours $\{\G_k\}_{k=1, \dots d_1}$ in the $x$--plane may be defined similarly.

The ``partition function'' is defined here to be the multiple integral \
\beq
\mathcal Z_N:=\frac 1{N!} \iint_{\varkappa^N} \prod_{j=1}^{N} \d x_j \d y_j \Delta(X) \Delta(Y) \prod_{j=1}^N {\rm e}^{-V_1(x_j) - V_2(y_j) + x_jy_j}
\eeq
where $\Delta(X)$ and $\Delta(Y)$ denote the usual Vandermonde determinants and the factor $\frac 1{N!}$ is chosen for convenience.

Such multiple integral can also be represented as the following determinant \
\beq
\mathcal Z_N = \det[\mu_{ij}]_{0\leq i,j\leq N-1}\ ,\ \ \mu_{ij} := \int_{\varkappa} x^i y^j {\rm e}^{-V_1(x) - V_2(y) + xy}\d x \d y
\eeq
The denomination of ``partition function'' comes from the fact \cite{MS, EynMehta, Harnad} that  when $\varkappa$ coincides with $\mathbb R\times \mathbb R$ then  $\mathcal Z_N$ coincides (up to a normalization for the volume of the unitary group) with the following matrix integral 
\beq
\iint \d M_1 \d M_2 {\rm e}^{-\tr (V_1(M_1) + V_2(M_2) - M_1 M_2)}
\eeq
extended over the space of Hermitean matrices $M_1, M_2$ of size $N\times N$, namely the normalization factor for the measure $\d \mu(M_1,M_2)$ introduced in \ref{101}.

   \begin{figure}
   \begin{center}
\includegraphics[width=.5\textwidth]{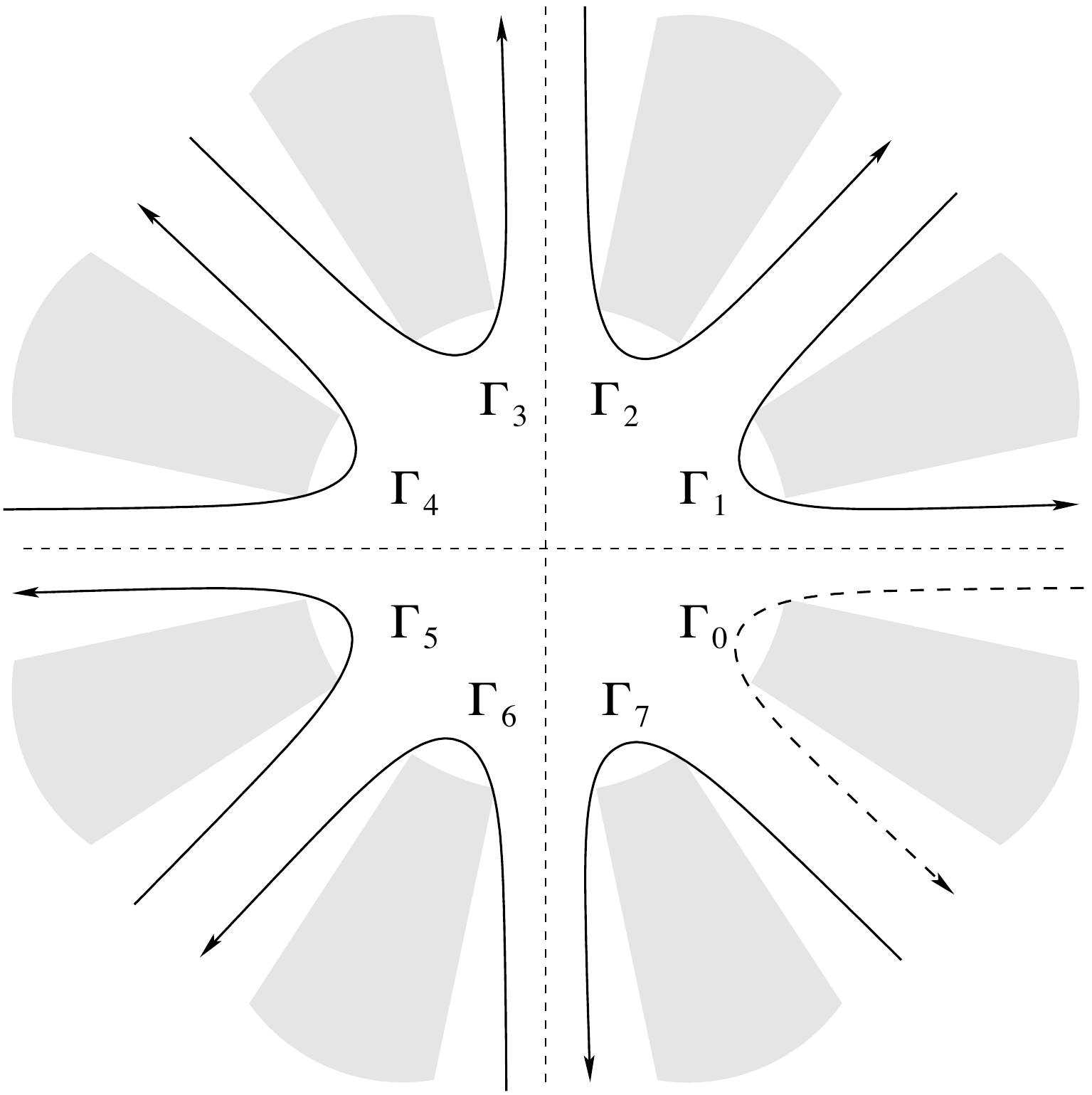}
\caption{ Wedge and anti-wedge contours for $V_2(y)$ of degree $D_2 +1$}
  \end{center}
  \end{figure}

\subsection*{2.1 Riemann--Hilbert characterization for the orthogonal polynomials}
A Riemann--Hilbert characterization of the biorthogonal polynomials is a crucial step towards implementing a steepest--descent analysis. In our context it is also crucial in order to tie the random matrix side to the theory of isomonodromic deformations. 

We first recall the approach given by Kuijlaars and McLaughin (referred to as KM in the rest of the article) in \cite{KMcL}, suitably extended and adapted (in a rather trivial way) to the setting and notation of the present work. 
We quote -paraphrasing and with a minor generalization- their theorem, without proof.
\begin{theorem} [Kuijlaars and McLaughin asymptotic]
\label{KMthm}
The monic bi-orthogonal polynomial $\pi_n(x)$ is the $(1,1)$ entry of the solution $\Gamma(x)$ (if it exists) of the following Riemann-Hilbert problem for $\Gamma(x)$.
\begin{enumerate}
\item The matrix $\Gamma(x)$ is piecewise analytic in $\mathbb C\setminus \bigsqcup \Gamma_j$;
\item the (non-tangential) boundary values of $\Gamma(x)$ satisfy the relations 
\bea
\Gamma(x)_+ = \Gamma(x)_- \left[
\begin{array}{cccc}
 1& w_{j,1}&\dots & w_{j,d_2}\\
  &1&0&0\\
  &&\ddots&\\
  &&&1
\end{array}
\right]\ ,\ \ \ x\in \Gamma_j\\
w_{j,\nu}= w_{j,\nu}(x):= {\rm e}^{-V_1(x)}  \sum_{k=1}^{d_2} \varkappa_{jk}\int_{\GH_k} y^{\nu-1} {\rm e}^{-V_2(y) +xy} \d y 
\eea
\item as $x\to \infty$ we have the following asymptotic expansion 
\beq
\Gamma(x) \sim \left(I_d + \frac {Y_{N,1}}x +\mathcal{O}\left(\frac{1}{x^2}\right)\right)
\begin{pmatrix}
x^{N} & 0 &  0\\
0 & x^{-m_N-1}Id_{r_N}& 0\\
0&0& x^{-m_N}Id_{d_2-r_N} \\
\end{pmatrix}
\label{Gammaexpinf}
\eeq
where we have defined the integers $m_N, r_N$ as follows
    \beq 
    N = m_N d_2 + r_N, \quad m_N, r_N \in \mathbb{N}, \quad 0 \leq r_n \leq d_2-1
    \eeq
\end{enumerate}
\end{theorem}

It follows from \cite{KMcL} that the solution $\Gamma_N(x)$ has the following form 
\bea
&&\Gamma_N(x):= \Gamma(x):= \left[
\begin{array}{cccc}
\pi_{N}(x) & \CC_0(\pi_{N}) & \dots &\CC_{d_2-1} (\pi_{N})\\
p_{N-1}(x) & \CC_0(p_{N-1})&\dots & \CC_{d_2-1}(p_{N-1})\\
\vdots &&&\vdots\\
p_{N-d_2}(x) & \CC_0(p_{N-d_2}) & \dots & \CC_{d_2-1} (p_{N-d_2})
\end{array}
\right]\label{GammaCauchy}\ ,\\
&& \CC_i(f(z)):=\frac{1}{2\pi i}\int\!\!\!\int_\varkappa\frac{f(x)}{x-z}y^i\, e^{-V_1(x)-V_2(y)+xy}dydx
\eea
where the polynomials denoted above by $p_{N-1},\dots,p_{N-d_2}$ are some polynomials of degree not exceeding $N-1$, whose detailed properties are largely irrelevant for our discussion; we refer to \cite{KMcL} for these details.

By a left multiplication of this solution by a suitable constant matrix we can see that the matrix
\beq
\widehat \Gamma_N:= \left[
\begin{array}{cccc}
\pi_{n} & \CC_0(\pi_{n}) & \dots &\CC_{d_2-1} (\pi_{n})\\
\pi_{n-1} & \CC_0(\pi_{n-1})&\dots & \CC_{d_2-1}(\pi_{n-1})\\
\vdots &&&\vdots\\
\pi_{n-d_2} & \CC_0(\pi_{n-d_2}) & \dots & \CC_{d_2-1} (\pi_{n-d_2})
\end{array}
\right]
\eeq
and $\Gamma_N$  are related as 
\beq
 \widehat \Gamma_N(x)=U_N \Gamma_N(x)
 \eeq
where $U_N$ is a constant matrix (depending on $N$ and on the coefficients of the polynomials but not on $x$).
As an immediate consequence, $\widehat \Gamma_N$ solves the same RHP as $\Gamma$ with the exception of the normalization at infinity (\ref{Gammaexpinf}).

The present RHP is not immediately suitable to make the connection to the theory of isomonodromic deformations as described in \cite{JMI, JMII}; we recall that this is the theory that describes the deformations of an ODE in the complex plane which leave  the Stokes' matrices  (i.e. the so--called  {\em extended monodromy data}) invariant.  The solution $\Gamma_N$ (or $\widehat \Gamma_N$) does not solve any ODE as formulated, because the jumps on the contours are non constant. If -however- we can relate $\Gamma_N$ with some other RHP with constant jumps, then its solution can be immediately shown to satisfy a polynomial ODE, which allows us to use the machinery of \cite{JMI,JMII}. This is the purpose of the next section.

\subsection*{2.2 A RHP with constant jumps}

In \cite{Harnad} the biorthogonal polynomials were characterised in terms of an ODE or --which is the same-- of a RHP with constant jumps. In order to connect the two formulations we will use some results and we start by defining some auxiliary quantities:
for $1\leq k \leq d_2$, define the $d_2$ sequences of functions $\{\psi_m^{(k)}(x)\}_{m\in \mathbb{N}}$  as follows:
\beq
 \psi_m^{(k)}(x) := \frac{1}{2\pi i}\int_{\GC_k}ds \int \!\!\!\int_{\varkappa}  dz dw 
\frac{\pi_m(z)e^{-V_1(z)}}{x-z}\frac{V_2'(s)-V_2'(w)}{s-w}  e^{-V_2(w)+V_2(s) +zw - xs }, \quad 1\leq k \leq d_2,
\label{psikmdefAnn}
\eeq
and let
\beq
\psi^{(0)}_m(x):= \pi_m(x)e^{-V_1(x)}.
\eeq

In terms of these define, for $N \ge d_2$, the sequence of $(d_2 + 1) \times (d_2 +1)$
matrix valued functions $\widehat{\PsiN}(x)$ 
\beq
\widehat\PsiN(x):= \left[
\begin{array}{ccc}
\psi_N^{(0)}(x) & \dots& \psi_N^{(d_2)}(x)\\
\vdots &&\vdots\\
\psi_{N-d_2}^{(0)}(x) & \dots& \psi_{N-d_2}^{(d_2)}(x)
\end{array}\right]
\label{hatpsiAnn}
\eeq
The following theorem is easily established using the properties of the bilinear concomitant and it is a very special case of the setting of \cite{MarcoPaths} (Cf. Appendix \ref{B} for a self-contained re-derivation)

\bt[Jump discontinuities in $\widehat{\PsiN}$]
The limits $\widehat{\PsiN}{}_{\pm}$ when approaching the
contours $\G_j$ from the left ($+$) and right($-$) are related by the following
jump discontinuity conditions
\bea
{\widehat \PsiN}{}_{+}(x) &\&= {\widehat \PsiN}{}_{-} (x)\Hb^{(j)}
\label{jumpdiscPsi} \\
\eea
where
\bea
\Hb^{(j)} &\& :=  \Ib  - 2\pi i \eb_0  \kappab^T \cr
\HbT{}^{\!\!(j)}&\&=  (\Hb^{(j)} )^{-1} =  \Ib  + 2\pi i \eb_0  \kappab^T
\label{HHT} \\
 \eb_0&\& := \begin{pmatrix}1 \cr 0 \cr \vdots \cr  0 \end{pmatrix}
\quad \kappab:= \begin{pmatrix} 0 \cr \varkappa_{j 1} \cr \vdots \cr \varkappa_{j d_2}\end{pmatrix}
\eea
\et
{The proof of this theorem is given in Appendix \ref{B}.}
For later convenience we define also 
\beq
\PsiN:= U_N^{-1} \widehat\PsiN
\eeq
The relationship with the matrices $\Gamma_N$,  $\widehat \Gamma_N$ introduced in the previous section  is detailed in the following 
\begin{theorem} [Factorization theorem]
\label{factorization}
\label{thmadditional}
The following identities hold
\beq
{\widehat \PsiN}(x) = \widehat\Gamma_N(x)  V(x) W (x)\ ,\ \ 
\PsiN(x) = \Gamma_N(x) V(x) W(x)  
\eeq
where
\beq
V := \begin{pmatrix} {\rm e}^{-V_1(x)} & 0 \cr
                   0  &  V_0 ,\quad \end{pmatrix}\ ,\qquad  
W(x) :=  \begin{pmatrix} 1 & 0 \cr
                      0 &  W_0(x)\end{pmatrix}                    
\eeq
 and $V_0$, $W_0(x)$ are the $d_2 \times d_2$ matrices with elements

\bea
(V_0)_{jk} &\&= 
\left[
\begin{array}{ccccc}
v_2 &v_3&\dots& & v_{d_2+1}\\
v_3& &&v_{d_2+1}&\\
&&\cdot^{\,\,\ds \cdot^{\,\,\ds \cdot}} &&\\
v_{d_2}&v_{d_2+1} &&&\\
v_{d_2+1} &&&&
\end{array}\right]
= \\
&\&= \begin{matrix} v_{j+k}  & {\rm if} \  j+k \leq d_2 +1\cr 
                                          0 &  {\rm if} \ j+k > d_2 +1, 
                                    \label{V0def}\end{matrix} \\ 
  (W_0(x))_{jk} &\& =\int_{\GC_k} y^{j-1} e^{V_2(y) - xy}  dy, \quad 1\leq j,k \leq d_2   
   \label{W0def}                              
\eea
  \end{theorem}

The proof is a direct verification by multiplication by matrices, noticing that the matrix  $V_0$ is nothing but the matrix representation of $\frac {V_2'(y)-V_2'(s)}{y-s}$ as a quadratic form in the bases $1,y, y^2,\dots, y^{d_2-1}$ and $1,s,s^2, \dots, s^{d_2-1}$ (more details are to be found on appendix \ref{A}, based on \cite{MarcoPaths})
The RHP for $\PsiN$ can be read off from that of $\Gamma_N$ and the fact that the jumps are constants. For convenience we collect the information in the following
\begin{theorem} 
\label{24Ann}
The matrix $\PsiN$ is the unique solution of the following RHP:
\begin{enumerate} 
\item Constant Jumps: \bea
\PsiN{}_{+}(x) &\&= \PsiN{}_{-} (x)\Hb^{(j)} \\
\eea
\item Asymptotic at infinity:
\beq \PsiN(x)\sim\Gamma_N \begin{pmatrix}
x^{N}e^{-V_1(x)} & 0 &  0\\
0 & x^{-m_N-1}Id_{r_N}& 0\\
0&0& x^{-m_N}Id_{d_2-r_N} \\
\end{pmatrix}  \Psi_0(x) 
\eeq
where 
\beq {\Gamma_N}=Id+\frac{{Y}_{N,1}}{x}+...\eeq 
and where $\Psi_0(x):=V(x)W(x)$ will be referred to as  the {\em bare} solution. Its asymptotic at infinity can be computed by steepest descent, but since it is $N$--independent, for the sake of brevity, we do not report on it (details are contained in \cite{BEH2}).
\item $\Psi_N$ has constant jumps 
\item $\Psi_{N}'(x)\Psi_N^{-1}=D_N(x)$ where $D_N(x)$ is a polynomial in $x$
\item $\partial_{u_K}\Psi_{N}(x)\Psi_N^{-1}=U_{K,N}(x)$ is polynomial in $x$.
\item $\partial_{v_J}\Psi_{N}(x)\Psi_N^{-1}=V_{J,N}(x)$ is polynomial in $x$.
\item $\det(\Psi_{N+1}\Psi_N^{-1})=Cste$
\end{enumerate}
\end{theorem}
The points (4,5,6,7) in the above theorem can be found in \cite{BE, BEH2}

In the next section we shall define a proper notion of isomonodromic tau function: it should be pointed out that the definition of \cite{JMI,JMII} cannot be applied as such because --as showed in \cite{BEH2}-- the ODE that the matrix $\PsiN$ (or $\widehat \PsiN$) solves, has a highly degenerate leading coefficient at the singularity at infinity.

In the list, the crucial ingredients are the differential equations (in $x$ or relatively to the parameters $u_K$ and $v_J$). First, the fact that $D_N(x)$ is a polynomial comes from explicit computation (See \cite{BE} for example). The result concerning the determinant of $R_N(x)$ can also be found in \cite{BE} where one has: $\det(\Psi_{N+1}\Psi_N^{-1})=\det(a_N(x))=Cste$.  The properties concerning the differential equations relatively to parameters can be found in \cite{BE} too.
Under all these assumptions, we will show that the proof of Jimbo-Miwa-Ueno can be adapted and that we can define a suitable $\tau$-function in the same way Jimbo-Miwa-Ueno did it.

\section*{3 Definition of the $\tau$-function}
\label{taudef}
In this section, we will place ourselves in a more general context than the one described above; we will show that under few assumptions one can define a good notion of  tau-function. 

 More generally we will denote with $t_a$ the isomonodromic parameters (in our case they are the $u_K$'s and ths $v_J$'s) and a subscript $a$ or $b$ is understood as a derivation relatively to $t_a$ or $t_b$. 
For a function $f$ of the isomonodromic times we will denote by the usual symbol its differential
 \beq
  \d f= \sum_a \partial_{t_a}f \d t_a=\sum_a f_a 
  \d t_a
 \eeq
Our setup falls in the following framework that it is useful to ascertain from the specifics of the case at hands.
Suppose we are given a matrix 
\beq
 \Psi(x) \sim  Y(x)\,  \Xi(x)\ ,
 \ \ Y(x):= \left(\mathbf 1 + \frac{Y_1}x + \frac {Y_2}{x^2}+ \dots \right) x^S
 \eeq 
 where $\Xi(x) = \Xi(x; \mathbf t)$ is some explicit expression (the ``bare'' isomonodromic solution) and $S$ is a matrix independent of the isomonodromic times. This implies that if we define 
 the one--form-valued matrix $\H(x;\mathbf t)$ by 
  \beq
 \H(x; \mathbf t) = \d \Xi(x; \mathbf t)\, \Xi(x;\mathbf t)^{-1}
 \eeq
then $\H(x) = \sum \H_a \d t_a$ (we suppress explicit mention of the $\mathbf t$ dependence henceforth) is some solution of the zero-curvature equations:
\beq
\pa_a \H_b - \pa _b \H_a = [\H_a,\H_b]
\label{barezccAnn}
\eeq
We will {\bf assume} (which is the case in our setting) that all $\H_a$ are {\bf polynomials} in $x$. We will also use that the dressed deformations $\Omega_a$ given by $\Psi_a=\Omega_a \Psi$ are polynomials. Moreover, according to the asymptotic they are given by: 
\beq
\Omega _a = (Y \H_a Y^{-1})_{pol}.
\eeq
In this very general (and generic) setting we can formulate the definition of a ``tau function'' as follows
\begin{definition}
The tau-differential is the one-form defined by 
\beq
\omega:= \sum \omega_a \d t^a:=\sum_a \res{} \tr \left(Y^{-1}Y' \H_a \right) \d t^a
\eeq
\end{definition}
The main point of the matter is that -without any further detail- we can now prove that the tau-differential is in fact closed and hence locally  defines a function.
\bt
\label{closurethm}
The tau-differential is a closed differential and locally defines a $\tau$--function as 
\beq
\d \log \tau = \omega
\eeq
\et
{\bf Proof.} 
We need to prove the closure of the differential.
We first recall the main relations between the bare and dressed deformations
\beq
\pa_a Y = \Omega_a Y - Y \H_a\  \ ; \qquad
Y\H_a Y^{-1} =  \Omega_a - \mathcal R_a\ ;\ \qquad \mathcal R_a:= \pa_a Y Y^{-1}\label{defs}
\eeq
We note that -by construction- $\Omega_a = (Y\H_aY^{-1})_{pol}$ is a polynomial while $\mathcal R_a= \mathcal O(x^{-1})$ {\em irrespectively of the form of $S$}.
We compute the cross derivatives directly 
\bea
\pa_a \omega_b &=&
\restr\bigg(
-Y^{-1} \left(\Omega_a Y - Y \H_a\right) Y^{-1} Y' \H_b + Y^{-1} \left( \Omega_a Y - Y \H_a \right)' \H_b +  Y^{-1} Y' \pa_a \H_b
\bigg) \cr
 &=& \restr \bigg(
\H_a Y^{-1} Y' \H_b + Y^{-1} \Omega_a' Y \H_b  - Y^{-1} Y'  \H_a \H_b - \H_a' \H_b + Y^{-1} Y' \pa_a H_b
\bigg) \cr
&=& \restr \bigg(
Y^{-1} Y' \left([\H_b, \H_a] + \pa_a \H_b\right) + Y^{-1} \Omega_a' Y \H_b - \overbrace{\H_a' \H_b}^{\hbox{polynomial}}\bigg)\cr
&=& \restr \bigg(
Y^{-1} Y' \left([\H_b, \H_a] + \pa_a \H_b\right) - \Omega_a' \mathcal R_b   
\bigg)
\eea
where, in the last step, we have used that $Y  \H_b Y^{-1} = \Omega_b - \mathcal R_b$ and that the contribution coming from $\Omega_b$ vanishes since it is a polynomial.
Rewriting the same with $a\leftrightarrow b$ and subtracting we obtain 
\bea
&\& \pa_a \omega_b - \pa_b \omega_a = \restr \bigg(
2 Y^{-1} Y' [\H_b, \H_a] - \Omega_a' \mathcal R_b + \Omega_b' \mathcal R_a  + Y^{-1}Y' \left(\pa_a \H_b - \pa_b \H _a\right)
\bigg) \cr 
&\&=  \restr \bigg(
 Y^{-1} Y' [\H_b, \H_a] - \Omega_a' \mathcal R_b + \Omega_b' \mathcal R_a  + Y^{-1}Y' \big(\overbrace{\pa_a \H_b - \pa_b \H _a + [\H_b, \H_a]}^{=0\hbox { by the ZCC \ref{barezccAnn}}}\big)
\bigg)\cr
&\& = \restr \bigg(
 Y^{-1} Y' [\H_b, \H_a] - \Omega_a' \mathcal R_b + \Omega_b' \mathcal R_a  \bigg) 
 \label{closure}
\eea
Note that, up to this point, we only used the zero curvature equations for the connection $\nabla = \sum(\pa_a - \H_a )\d t^a$ and the fact that $\H_a$ are polynomials in $x$. 
We thus need to prove that the last quantity in (\ref{closure}) vanishes: this follows from the following computation, which uses once more the fact that $\H_a$ and $\Omega_a$ are all polynomials. Indeed, we have $\restr (\H_a' \H_b) = 0$ and hence (using (\ref{defs}))
\bea
0&\& = \res{}\tr (\H_a' \H_b)=\restr\left( \left(Y \H_a Y^{-1}\right)' Y \H_b Y^{-1}\right) - \restr \left(Y' \H_a \H_b Y^{-1}\right) +  \restr\left(
\H_a Y^{-1} Y' \H_b
\right) \cr
&\& = \restr \left(\left(\Omega_a - \mathcal R_a\right)' \left(\Omega_b - \mathcal R_b\right)\right)  + 
\restr \left(
Y^{-1} Y' \left[\H_b, \H_a\right]\right)  \cr
\&\&=
\restr \bigg( 
\overbrace{\Omega_a'\Omega_b}^{\hbox{poly}}  - \mathcal R_a' \Omega_b - \Omega_a' \mathcal R_b +\overbrace{ \mathcal R_a' \mathcal R_b}^{=\mathcal O(x^{-2})} + Y^{-1}Y' [\H_b,\H_a]
\bigg)
\cr
&\&  = \restr \bigg( 
 - \mathcal R_a' \Omega_b - \Omega_a' \mathcal R_b  + Y^{-1}Y' [\H_b,\H_a]
\bigg)=0
\eea 
Using integration by parts (and cyclicity of the trace) on the first term here above, we obtain precisely the last quantity in (\ref{closure}). The Theorem is proved. {\bf Q.E.D.}

\subsection*{3.1 Application to our problem}
\label{application}
We now apply the general definition above to our setting, with the identifications $\Psi = \Psi_N$, $Y=\Gamma_N$ (as a formal power series at $\infty$) and  $\Xi = \Psi_0$. 
We will write $Y_N$ instead of $\Gamma_N$ in the expressions below to emphasize that we consider its asymptotic expansion at $\infty$
 This reduces the definition of the tau function to the one below
\begin{definition}
\label{tauN}
The $\tau$-function is defined by the following PDE 
\beq  d (\log \tau_N) =\Res_{x \to \infty} \Tr\left(Y_N^{-1}Y_N'  \d(\Psi_0)\Psi_0^{-1}\right) \eeq
where $Y_N$ is the formal  asymptotic expansion of $\Gamma_N$ at infinity
\beq Y_N=\widetilde Y _N\begin{pmatrix} 
x^{N} & 0 &  0\\
0 & x^{-m_N-1}Id_{r_N}& 0\\
0&0& x^{-m_N}Id_{d_2-r_N} \\
\end{pmatrix}\eeq
\end{definition}

\begin{remark}
The matrix $S$ of the previous section  in our case becomes:
\beq S=\begin{pmatrix} 
N & 0 &  0\\
0 & (-m_N-1)\, Id_{r_N}& 0\\
0&0& -m_N\, Id_{d_2-r_N} \\
\end{pmatrix}\eeq
The partial derivatives of $\ln \tau_N$  split into two sets which have different form:
\beq \partial_{u_K} \log \tau_N=-\Res_{x \to \infty} \Tr\left(Y_N^{-1}Y_N' \frac{x^K}{K} \bf{E}_{11} \right)\eeq
\bea 
\partial_{v_J} \log \tau_N&=&\Res_{x \to \infty}\Tr\left(Y_N^{-1}Y_N'  \partial_{v_J}(\Psi_0)\Psi_0^{-1}\right)\eea
where in the last equation the term $\partial_{v_J}(\Psi_0)\Psi_0^{-1}$ has non-zero entries only in the anti-principal minor of size $d_2$.
\end{remark}

One can notice that the situation we are looking at is a generalization of what happen in the one-matrix case. In the 1-matrix model, the matrix $S$ is zero and therefore $Y_N$ are (formal) Laurent series. The matrix $\Psi_0$ matrix is absent in that case since there is only one potential and thus one recovers the usual definition of isomonodromic tau function (see \cite{BEH4}). Note also that in the derivation with respect to $v_J$ we have obtained the second equality using  the block diagonal structure of $\Psi_0$ (first row/column does not play a role). It is remarkable that the two systems are completely decoupled, i.e. that in the first one the matrix $\Psi_0$ (containing all the dependance in $V_2$) disappears and that in the second one the matrix $A_0$ (containing the potential $V_1$) also disappears. 

\subsection*{3.2 Discrete Schlesinger transformation: Tau-function quotient}  
\label{Schlesinger}
In this section we investigate the relationship between the tau-function of Def. \ref{tauN} and the partition function $\mathcal Z_N$ of the matrix model.  

We anticipate that the two object turn out to be the same (up to a nonzero factor that will be explicitly computed, Thm. \ref{main}): the proof relies on two steps, the first of which we prepare in this section. 
These are
\begin{itemize}
\item proving that they satisy the same recurrence relation
\item identifying the initial conditions for the recurrence relation.
\end{itemize}

We start by investigating the relationship between $\tau_N$ and $\tau_{N+1}$; this analysis is essentially identical to the theory developed in \cite{JMII} and used in \cite{BE_mom}, but we report it here for the convenience of the reader.

From the fact that the $\Psi_N$ has constant jumps, we deduce that $\Psi_{N+1}\Psi_N^{-1}$ is an entire function. Moreover asymptotically  it looks like:
\bea \Psi_{N+1}\Psi_N^{-1}&=&\tilde{Y}_{N+1}\begin{pmatrix}
x^{N+1}e^{-V_1(x)} & 0 &  0\\
0 & x^{-m_{N+1}-1}Id_{r_{N+1}}& 0\\
0&0& x^{-m_{N+1}}Id_{d_2-r_{N+1}} \\
\end{pmatrix}  \Psi_0(x) \cr & & \Psi_0(x)^{-1} \begin{pmatrix}
x^{-N}e^{V_1(x)} & 0 &  0\\
0 & x^{m_N+1}Id_{r_N}& 0\\
0&0& x^{m_N}Id_{d_2-r_N} \\
\end{pmatrix}\tilde{Y}_N\eea
\beq \Psi_{N+1}\Psi_N^{-1}=\tilde{Y}_{N+1}\begin{pmatrix}
x & 0 &  0&0\\
0 &Id_{r_N-1} &0&0\\
0& 0&x^{-1}& 0\\
0&0& 0&Id_{d_2-1-r_N} \\
\end{pmatrix}\tilde{Y}_N\eeq
Thus, remembering that $\tilde{Y}_N$ is a series $x^{-1}$, Liouville's theorem states that $\Psi_{N+1}\Psi_N^{-1}$ is a polynomial of degree one, and hence, for some constant matrices $R_N^0, R_N^1$ we must have
\beq
 \Psi_{N+1}\Psi_N^{-1}=R_N(x)=R_N^0+x R_N^1
 \eeq

From the fact that $\det(R_N)$ does not depend on $x$ (last property Thm. \ref{thmadditional}), we know that $R_N^{-1}(x)$ is a  polynomial of degree at most one as well (this is easy if one consider the expression of the inverse of a matrix using the co-matrix).

Comparing the asymptotics of $\Psi_{N+1}$ and $R_N(x) \Psi_N$ term-by-term in the expansion in inverse powers of $x$ and after some elementary algebra one obtains (\cite{JMI} Appendix A):
\beq
 R_N(x)=E_{\alpha_0}x+R_{N,0} \qquad \hbox{and} \qquad R_N^{-1}(x)=E_1 x+R_{N,0}^{-1} 
 \eeq
Here we have introduced the notation $\alpha_0=r_N+1$ which corresponds to the index of the column where the coefficient $x^{-1}$ is to be found in the asymptotic of ${\Psi}_{N+1}{\Psi}_N^{-1}$. This notation is the standard notation used originally by Jimbo-Miwa in a Schlesinger transformation.
The matrix $(R_{N,0})_{\alpha,\beta}$ is given by:
\beq \begin{array}{cccc}
&\beta=\alpha_0&\beta=1 & \beta \neq \alpha_0,1\\
\\
\alpha=\alpha_0 & \frac{-(Y_{N,2})_{\alpha_0,1}+\sum_{\gamma \neq \alpha_0}(Y_{N,1})_{\alpha_0,\gamma}(Y_{N,1})_{\gamma,1}}{(Y_{N,1})_{\alpha_0,1}} & -(Y_{N,1})_{\alpha_0,1}& -(Y_{N,1})_{\alpha_0,\beta} \\
\\
\alpha=1 & \frac{1}{(Y_{N,1})_{\alpha_0,1}} & 0 &0 \\
\\
\alpha \neq \alpha_0, 1 & -\frac{(Y_{N,1})_{\alpha,1}}{(Y_{N,1})_{\alpha_0,1}} & 0 & \delta_{\alpha, \beta} \\
\end{array} \eeq

and $(R_{N,0}^{-1})_{\alpha,\beta}$ is given by:
\beq \begin{array}{cccc}
&\beta=\alpha_0&\beta=1 & \beta \neq \alpha_0,1\\
\\
\alpha=\alpha_0& 0 & (Y_{N,1})_{\alpha_0,1}& 0\\
\\
\alpha=1 & -\frac{1}{(Y_{N,1})_{\alpha_0,1}} & -\frac{-(Y_{N,2})_{\alpha_0,1}}{(Y_{N,1})_{\alpha_0,1}}+(Y_{N,1})_{1,1} & -\frac{(Y_{N,1})_{\alpha_0,\beta}}{(Y_{N,1})_{\alpha_0,1}} \\
\\
\alpha \neq \alpha_0, 1 & 0& (Y_{N,1})_{\alpha,1} & \delta_{\alpha, \beta}\\
\end{array} \eeq

While the formulae above might seem complicated, we will use  the two important observations:
\beq E_{\alpha_0}R_{N,0}^{-1}+R_{N,0}E_1=R_{N,0}^{-1}E_{\alpha_0}+E_1R_{N,0}=0\eeq
\begin{center} $ R_N^{-1}(x)R_N'(x)=R_{N,0}^{-1}E_{\alpha_0}$ does not depend on $x$. \end{center}

The recurrence relation satisfied by the sequence  $\{\tau_N\}$ is derived in the next theorem.
\begin{theorem} 
Up to multiplication by functions that do not depend on the isomonodromic parameters (i.e. independent of the potentials $V_1,V_2$) the following identity holds
\beq  
\frac{\tau_{N+1}}{\tau_N}=(Y_1)_{1,\alpha_0}
\eeq
\end{theorem}
{\bf Proof}
The proof follows \cite{JMII} but we report it here for convenience of the reader.
Consider the following identity
\beq \Psi_{N+1}=Y_{N+1}\Psi_0=R_NY_N\Psi_0\eeq
This implies that 
\beq
 Y_{N+1}=R_NY_N
 \eeq
Taking the derivative with respect to $x$ gives:
\beq Y_{N+1}^{-1}Y_{N+1}'=Y_N^{-1}R_N^{-1}R_N'Y_N+Y_N^{-1}Y_N'\eeq
Therefore we have:
\bea 
\d \log \tau_{N+1}-\d \log \tau_N &=&\Res_{x \to \infty} \Tr( (Y_N^{-1}R_N^{-1}R_N'Y_N+Y_N^{-1}Y_N'-Y_N^{-1}Y_N) \d(\Psi_0 )\Psi_0^{-1}) \cr
&=&\Res_{x \to \infty} \Tr( Y_N^{-1}R_N^{-1}R_N'Y_N\d(\Psi_0 )\Psi_0^{-1}) 
\eea
We now need to ``transfer'' the exterior derivative from $\Psi_0$ to $Y_N$. This can be done using  that $\PsiN =Y_N \Psi_0$, so that
$$\d\PsiN=\d(Y_N) \Psi_0 + Y_N  \d(\Psi_0)$$
Equivalently:
\beq Y_N\d \Psi_0 \Psi_0^{-1}Y_N^{-1}=d(\PsiN)\PsiN^{-1}-dY_NY_N^{-1}\eeq
Inserting these identities  in the tau quotient we obtain the relation
\beq
 d \log \tau_{N+1}-d \log \tau_N =\Res_{x \to \infty} \Tr\left (R_N^{-1}R_N' d(\PsiN)\PsiN^{-1}-R_N^{-1}R_N'dY_NY_N^{-1}\right)\eeq

The first term is residueless at $\infty$ since $\d \PsiN \PsiN^{-1}$ is polynomial in $x$ and $R_N^{-1}R_N'$ does not depend on $x$. Therefore we are left only with:
\beq d \log \tau_{N+1}-d \log \tau_N =-\Res_{x \to \infty} \Tr(R_N^{-1}R_N'dY_NY_N^{-1})\eeq
A direct matrix computation using the explicit form of $R_N$ yields
\beq 
\d \log \tau_{N+1}-\d \log \tau_N =\d \log((Y_{N,1})_{1,\alpha_0})
\eeq
and hence
\beq  
\frac{\tau_{N+1}}{\tau_N}=(Y_1)_{1,\alpha_0}
\eeq
The last equality is to be understood up to a multiplicative constant not depending on the parameters $u_K$ and $v_J$ in $\tau$. {\bf Q.E.D.}\par \vskip 5pt

In order to complete the first step we need to express the entry ${(Y_1)}_{1,\alpha_0}$  in terms of the ratio of two consecutive partition functions. This is accomplished in the following section.

\begin{theorem}
For the matrix $ \Gamma_N $ the asymptotic expansion at infinity (\ref{Gammaexpinf}) is such that 
\beq 
(Y_{N,1})_{1,\alpha_0} = (v_{d_2+1})^ S h_N = (v_{d_2+1})^ S \frac {\mathcal Z_{N+1}}{\mathcal Z_N}
\eeq
where $S$ and $\alpha_0\in \{0,1,\dots, d_2-1\}$ are defined by the following relation
\beq
N = d_2 S + \alpha_0-1 
\eeq
\end{theorem}
{\bf Proof}
In order to compute $(Y_{N,1})_{1,\alpha_0}$ it is sufficient to compute the leading term of the expansion at $\infty$ appearing in the first row of the matrix $\Gamma_N$. Recalling the expression (\ref{GammaCauchy}), we start by the following direct compuation using integration by parts
\bea 
\int\!\!\!\int_\kappa dzdw \,\pi_N(z)z^iw^{k-1}e^{-V_1(z)-V_2(z)+zw} 
\&\& =\int\!\!\!\int_\kappa dzdw \,\pi_N(z)e^{-V_1(z)}w^{k-1}e^{-V_2(w)}\frac{d^i}{dw^i}\left(e^{zw}\right) \cr
&\&=(-1)^i\int\!\!\!\int_\kappa dzdw \,\pi_N(z)e^{-V_1(z)+zw}\frac{d^i}{dw^i}\left(w^{k-1}e^{-V_2(w)}\right)\cr
&\&=\int\!\!\!\int_\kappa dzdw \, \pi_N(z) q_{d_2i+k-1}(w)e^{-V_1(z)-V_2(z)+zw}
\eea
where  $q_{d_2i+k-1}(w)$ is a polynomial of the indicated degree  whose leading coefficient is $v_{d_2+1}^i$. The last RHS is $0$ if $d_2i+k-1<N$ because of orthogonality. If $d_2i+k-1=N$ the integral gives $v_{d_2+1}^i h_N$ by the normality conditions concerning our biorthogonal set. 
This computation allows us to expand the Cauchy transform of $(\Gamma_{N})_{1,\alpha_0}$ near $\infty$ as follows:
\bea \mathcal{C}(p_Nw_{\alpha_0}(x))&=& \frac{1}{2\pi i}\int\!\!\!\int_\kappa dzdw \frac{\pi_N(z)}{z-w} w^{\alpha_0-1}e^{-V_1(z)-V_2(z)+zw} \cr
&=&-\sum_{i=0}^{S-1}\frac{1}{2\pi i}\int\!\!\!\int_\kappa dzdw \pi_N(z)\frac{z^i}{x^{i+1}} w^{\alpha_0-1}e^{-V_1(z)-V_2(z)+zw}\cr
&+& \frac{1}{2\pi i}\frac{1}{x^{S+1}}\int\!\!\!\int_\kappa dzdw \frac{\pi_N(z)}{x-z}z^{S} w^{\alpha_0-1}e^{-V_1(z)-V_2(z)+zw}+ \mathcal O(x^{-S-2})\cr
\eea
By orthogonality the first sum vanishes term-by-term and the leading coefficient of the second term is $v_{d_2+1}^S h_N$. {\bf Q.E.D.}\par\vskip 5pt

Recalling that the $\tau$-function is only defined up to a multiplicative constant not depending on $N$ nor on the coefficients $u_k$and $v_j$, we have
\beq \frac{\tau_{N+1}}{\tau_N}=(v_{d_2+1})^ {S_N} \frac{\mathcal Z_{N+1}}{\mathcal Z_N}\eeq
where $N = d_2 S_N + \alpha_0-1$
Hence for every $n_0$:
\beq \tau_{N}\mathcal Z_{n_0}=\mathcal Z_N \tau_{n_0}(v_{d_2+1})^{\sum_{j=n_0}^{N-1}S_j} \eeq
One would like to take $n_0=0$ because it enables explicit computations. As we will prove now there is a way of extending naturally all the reasoning down to $0$.

%\subsection{Extension down to $N=0$}
The RHP for $\Gamma_N$ (Thm. \ref{KMthm})  is perfectly well--defined for $N=0$ and has solution
\beq 
{\Gamma_0}=\begin{pmatrix}
1& \mathcal C_0(1)&\mathcal C_1(1)& \ldots &\mathcal C_{d_2-1}(1)\\
0&1&0&\ldots&0\\
\vdots&\ddots&\ddots&\ddots&0\\
\vdots&\ddots&\ddots&\ddots&0\\
0&0&\ldots&\ddots&1
\end{pmatrix}\ .
\eeq
Consequently we can take
\beq 
\tau_{N}\mathcal Z_0=(v_{d_2+1})^{\sum_{j=0}^{N-1}S_j} \mathcal Z_N \tau_0 
\eeq
Also note that $\mathcal Z_0\equiv 1$ (by definition).

We can compute $\tau_0$ directly from Def. \ref{tauN} because of the particularly simple and explicit expression of $\mathop{\Psi}_0= \Gamma_0 \Psi_0$.
\beq
\d \ln \tau_0 = \restr\left(Y_0^{-1}Y_0' \d \Psi_0 \Psi^{-1}\right) 
\eeq
We claim that this expression is identically zero (and hence we can define $\tau_0\equiv 1$); indeed,  \beq 
Y_0^{-1}Y_0'=\begin{pmatrix}
0& *&\dots&*\\
0&0& \dots &0\\
\vdots&\vdots& \ddots&0\\
0&0&\dots&0\\
\end{pmatrix}
\eeq 
and 
\beq 
\d \Psi_0(x)\Psi_0^{-1}(x)=\begin{pmatrix}
\star& 0&\dots&0\\
0&\star& \dots &\star\\
\vdots&\vdots& \ddots&\vdots\\
0&\star&\dots&\star\\
\end{pmatrix}
\eeq 
so that the trace of the product is always zero (even before taking the residue). Combining the two results together gives the following theorem:

\begin{theorem} 
\label{main}
The isomodromic $\tau$-function and the partition function are related by:
$$
\forall N \in \mathbb{N}: \mathcal Z_N=(v_{d_2+1})^{\sum_{j=0}^{N-1}S_j} \tau_N \label{PartFuntAndIsoTau}
$$
where we recall that $S_j$ is given by the decomposition of $j+1$ in the Euclidian division by $d_2$: $S_j=E\left[\frac{j+1}{d_2}\right]$. A short computation of the power in $v_{d_2+1}$ gives:
$$
\forall N \in \mathbb{N}: \mathcal Z_N=(v_{d_2+1})^{d_2\frac{\alpha_N(\alpha_N-1)}{2}+\alpha_N(N-\alpha_N d_2)} \tau_N \label{PartFuntAndIsoTau2}$$
where $\alpha_N=E\left[\frac{N}{d_2}\right]$
\end{theorem}

The presence of the power in $v_{d_2+1}$ is due to a bad normalisation of the partition function itself ($\mathcal Z_N$) and can be easily cancelled out by taking $v_{d_2+1}=1$ from the start (it is just a normalization of the weight function). Moreover it is not surprising because in the work of Bergere and Eynard \cite{BgE}, all results concerning the partition function and its derivatives with respect to parameters have special cases for $u_{d_1+1}$ and $v_{d_2+1}$. It also signals the fact that the RHP is badly defined when $v_{d_2+1}=0$ because the contour integrals involved diverge and the whole setup breaks down. Indeed if $v_{d_2+1}=0$ this simply means that $V_{2}$ is a polynomial of lower degree and thus the RHP that we should set up should be of smaller size from the outset.

\section*{Outlook}

In this article, we have restricted ourselves to contours going from infinity to infinity. This allows us to use integration by parts without picking up any boundary term. A natural extension of this work could be to see what happens when contours end in the complex plane, and especially study what happens when the end points moves (models with hard edges). This generalization is important in the computation of the gap probabilities of the Dyson model \cite{TWDyson}, which correspond to a random matrix model with Gaussian potentials but with the integration restricted to intervals of the real axis.

\section*{Acknowledgements}
We would like to thank John Harnad for proposing the problem, Seung Yeop Lee and Alexei Borodin for fruitful discussions. This work was done at the University of Montr\'eal at the departement of mathematics and statistics and the Centre de Recherche Math\'ematique (CRM) and O.M.  would like to thank both for their hospitality.
This work was partly supported
by the Enigma European network MRT-CT-2004-5652,
by the ANR project G\'eom\'etrie et int\'egrabilit\'e en physique 
math\'ematique  ANR-05-BLAN-0029-01,
by the Enrage European network MRTN-CT-2004-005616,
by the European Science Foundation through the Misgam program,
by the French and Japanese governments through PAI Sakurav,
by the Quebec government with the FQRNT.

\section*{Appendix: Factorization of ${\Psi_N}$}
\label{A}
Starting from the definition of  the last $d_2$ columns of $\widehat{\PsiN}$ (\ref{hatpsiAnn})  we observe that
\bea
\psi_m^{(k)}(x) &\& := \frac 1{2i\pi} \int_{\GC_k} \!\!\!{\rm d}s \int\!\!\!\int_{\varkappa} \frac {\pi_m(z)}{x-z} \frac {V_2'(s)- V_2'(w)}{s-w} {\rm e}^{-V_1(z)-V_2(w)+V_2(s) + zw -xs}dwdz \\
&\&= \sum_{p,q}v_{q+p}  \frac 1{2i\pi}  
\int\!\!\!\int_{\varkappa} \frac {\pi_m(z)}{x-z} w^{p-1} {\rm e}^{-V_1(z)-V_2(w)+zw} \int_{\GC_k} {\rm d}s s^{q-1} {\rm e}^{V_2(s) -xs}\cr
&=&\sum_{p,q}(\widehat\Gamma_N)_{m,p}(V_0)_{p,q}(W_0)_{q,k}=(\widehat \Gamma_N \, V_0W_0)_{m,k}
\eea
This proves Thm. \ref{factorization}.

\section*{Appendix: Bilinear concomitant as intersection number}
\label{B}
We recall very briefly the result of \cite{MarcoPaths} stating that
\beq
\frac{V_2'(\pa_x) - V_2'(-\pa_z)}{\pa_x+\pa_z} w(x)f(z)\bigg|_{z=x} =\int_\Gamma \int_{\check \Gamma} \frac {V_2'(\eta)-V_2'(s)}{\eta-s} {\rm e}^{x(\eta-s) -V_2(\eta)+V_2(s)} = 2i\pi \Gamma\# \check \Gamma =\hbox{constant} \ .
\eeq
The last identity is obtained by integration by parts and shows that
the bilinear concomitant is just the intersection number of the
(homology classes) of the contours $\Gamma, \check \Gamma$. 
More precisely we get that:
\bea &&\frac{d}{d x}\int_\Gamma \int_{\check \Gamma}ds d\eta \, \frac {V_2'(\eta)-V_2'(s)}{\eta-s} {\rm e}^{x(\eta-s) -V_2(\eta)+V_2(s)} \cr
&=&\int_\Gamma \int_{\check \Gamma}ds d\eta \, (V_2'(\eta)-V_2'(s)) {\rm e}^{x(\eta-s) -V_2(\eta)+V_2(s)} \cr
&=&\int_\Gamma \int_{\check \Gamma}ds d\eta \, \frac{\partial}{\partial \eta}(-e^{-V_2(\eta)})e^{x\eta}e^{-xs+V_2(s)} -\int_\Gamma \int_{\check \Gamma}d\eta ds \, \frac{\partial}{\partial s}(e^{V_2(s)})e^{-xs}e^{x\eta -V_2(\eta)} \cr
&=&x\int_\Gamma \int_{\check \Gamma}ds d\eta \, e^{x\eta-xs-V_2(\eta)+V_2(s)}-x\int_\Gamma \int_{\check \Gamma}ds d\eta e^{x\eta-xs-V_2(\eta)+V_2(s)}\cr
&=&0
\eea
The matrix expression shows that the pairing is indeed a duality since the determinant is nonzero.
The undressing matrix $\Psi_0$ (that was originally introduced in Thm. \ref{24Ann})  is thus
\beq
\Psi_0 =\left[
\begin{array}{c|c}
1& \\
\hline
& \begin{array}{ccccc}
v_2 &v_3&\dots& & v_{d_2+1}\\
v_3& &&v_{d_2+1}&\\
&&\cdot^{\,\,\ds \cdot^{\,\,\ds \cdot}} &&\\
v_{d_2}&v_{d_2+1} &&&\\
v_{d_2+1} &&&&
\end{array}
\end{array}\right]
\left[ \begin{array}{c|cccc}
1 &&&&\\
\hline
&f_1&f_2 &\dots &f_{d_2}\\
&f_1'&f_2'&\dots &f_{d_2}'\\
 & \vdots &&&\vdots\\
 & f_1^{(d_2-1)}&\dots && f_{d_2}^{(d_2-1)}
\end{array}\right]
\eeq
where the Wronskian subblock in the second term is constructed by
choosing $d_2$ homologically independent contour classes for the
integrations $\check \Gamma$;
\beq
f_k(x):=\int_{\check \Gamma_k} {\rm e}^{-xs+V_2(s)}{\rm d}s\ ,\ \
k=1,\dots,d_2\ .
\eeq
The dressing matrix $\Psi_0$ exhibits a 
  Stokes' phenomenon (of Airy's type) which is the inevitable drawback of removing
the $x$-dependence from the jump matrix.
We can now compute the jumps and see that it does not depend on $x$. For the  $k$-th column we have:
\beq \psi_m^{(k)}(x) := \frac{1}{2\pi i} \int_{\GC_k}ds \int \!\!\!\int_{\varkappa}  dz dw 
\frac{\pi_m(z)e^{-V_1(z)}}{x-z} \frac{V_2'(s)-V_2'(w)}{s-w}  e^{-V_2(w)+V_2(s) +zw - xs }, \quad 1\leq k \leq d_2\eeq
gives:
\bea
\psi_m^{(k)}(x)_+ &=& \psi_m^{(k)}(x)_-+  \psi_m^{(0)}(x)\int\int dsdw 
\frac{V_2'(s)-V_2'(w)}{s-w}  e^{-V_2(w)+V_2(s) +x(w - s) }
\\ 
&=&\psi_m^{(k)}(x)_- + \psi_m^{(0)}(x)\sum_{j=1}^{d_2}\varkappa_{\ell j}( \Gamma^{(y)}_j\# \check \Gamma_k)\ ,
\eea 

%% file: annexef.tex
\selectlanguage{french}
\annexe{Topological expansion of the Bethe ansatz, and non-commutative algebraic geometry} \label{Article[III]}
\selectlanguage{english}

\begin{center}
\vspace{26pt}

\vspace{26pt}

{\sl B.\ Eynard}${}^\dagger$\hspace*{0.05cm}\footnote{ E-mail: bertrand.eynard@cea.fr },
{\sl O.\ Marchal}${}^\dagger\, {}^\ddagger$\hspace*{0.05cm}\footnote{ E-mail: olivier.marchal@cea.fr }\\
\vspace{6pt}
${}^\dagger$ 
Institut de Physique Th\'eorique,\\
CEA, IPhT, F-91191 Gif-sur-Yvette, France,\\
CNRS, URA 2306, F-91191 Gif-sur-Yvette, France.\\
${}^\ddagger$ Centre de recherches math\'ematiques, Universit\'e de Montr\'eal 
C. P. 6128, succ. centre ville, Montr\'eal, Qu\'ebec, Canada H3C 3J7.
\end{center}

\vspace{20pt}
\begin{center}
{\bf Abstract}:

In this article, we define a non-commutative deformation of the "symplectic invariants" (introduced in \cite{OE}) of an algebraic hyperelliptical plane curve. The necessary condition for our definition to make sense is a Bethe ansatz. The commutative limit reduces to  the symplectic invariants, i.e. algebraic geometry, and thus we define non-commutative deformations of some algebraic geometry quantities. 
In particular our non-commutative Bergmann kernel satisfies a Rauch variational formula.
Those non-commutative invariants are inspired from the large N expansion of formal non-hermitian matrix models. Thus they are expected to be related to the enumeration problem of discrete non-orientable surfaces of arbitrary topologies.
%Another application we consider is the geometry of the Gaudin model.

\end{center}

\newpage

%*********************************************************************
%==================== ARTICLE =======================================%******************************************

\section*{1 Introduction}

In \cite{OE}, the notion of symplectic invariants of a spectral curve was introduced. For any given  algebraic plane curve (called spectral curve) of equation:
\beq
0={\cal E}(x,y)= \sum_{i,j} {\cal E}_{i,j}\,\, x^i\, y^j
\eeq
an infinite sequence of numbers
\beq
F^{(g)}({\cal E})
\qquad , \,\, g=0,1,2,\dots,\infty
\eeq
and an infinite sequence of multilinear meromorphic forms $W_n^{(g)}$ (meromorphic on the algebraic Riemann surface of equation ${\cal E}(x,y)=0$) were defined.

Their definition was inspired from hermitian matrix models, i.e. in the case where ${\cal E}={\cal E}_{\rm M.M.}$ is the spectral curve ($y(x)$ is the equilibrium density of eigenvalues) of a formal hermitian matrix integral $Z_{\rm M.M.}=\int dM\, \ee{-N\Tr V(M)}$, the $F^{(g)}$ were such that:
\beq
\ln{Z_{\rm M.M.}} = \sum_{g=0}^\infty N^{2-2g} F^{(g)}({\cal E}_{\rm M.M.})
\eeq
The $F^{(g)}$'s have many remarkable properties (see \cite{OE}), in particular invariance under symplectic deformations of the spectral curve, homogeneity (of degree $2-2g$), holomorphic anomaly equations (modular transformations), stability under singular limits, ...
An important property also, is that the following formal series
\beq
\tau({\cal E})= \ee{\sum_g N^{2-2g} F^{(g)}({\cal E})}
\eeq
is the "formal" $\tau$ function of an integrable hierarchy.

\smallskip
Although those notions were first developed for matrix models, they extend beyond matrix models, and they make sense for spectral curves which are not matrix models spectral curves.
For instance the (non-algebraic) spectral curve ${\cal E}_{\rm WP}(x,y) = (2\pi y)^2 - (\sin{(2\pi \sqrt{x})})^2$ is such that $F^{(g)}({\cal E}_{\rm WP})={\rm Vol}(\overline{\cal M}_{g})$ is the Weyl-Petersson volume of moduli space of Riemann surfaces of genus $g$ (see \cite{EynVolmum, EOVolWP}).
It is conjectured \cite{BKMP} that the $F^{(g)}$'s are deeply related to Gromov-Witten invariants, Hurwitz numbers \cite{BouchardMarino} and topological strings \cite{BKMP}. In particular they are related to the Kodaira-Spencer field theory \cite{DVKS}.

\bigskip

There were many attempts to compute also non-hermitian matrix integrals, and an attempt to extend the method of \cite{OE} was first made in \cite{ChekEynbeta}, and here in this paper we deeply improve the result of \cite{ChekEynbeta}.
The aim of the construction we present here, is to define $F^{(g)}$'s for a "non-commutative spectral curve", i.e. a non commutative polynomial:
\beq
{\cal E}(x,y) = \sum_{i,j} {\cal E}_{i,j}\,\, x^i\, y^j
\virg
[y,x]=\hbar
\eeq
For instance we can view $y$ as $y=\hbar\, {\partial/\partial x}$, and ${\cal E}$ is a differential operator, which encodes a linear differential equation.

In this article we choose ${\cal E}(x,y)$ of degree 2 in the variable $y$, i.e. the case of a second order linear differential equation, i.e. Schroedinger equation, and we leave to a further work the general case.

\bigskip
Here, in this article, we define some $F^{(g)}({\cal E})$, which reduce to those of \cite{OE} in the limit $\hbar\to 0$, and which compute non-hermitian matrix model topological expansions.

For instance consider a formal matrix integral:
\beq
Z=\int_{E_{2\beta,N}} dM \ee{-N\sqrt\beta \Tr V(M)} = \ee{\sum_g N^{2-2g}\, F^{(g)}}
\eeq
where $E_{2\beta,N}$ is one of the Wigner matrix ensembles \cite{Mehta} of rank $N$: $E_{1,N}$ is the set of real symmetric matrices, $E_{2,N}$ is the set of hermitian matrices, and $E_{4,N}$ is the set of self-dual quaternion matrices (see \cite{Mehta} for a review).
We define:
\beq
\hbar={1\over N}\left(\sqrt\beta - {1\over \sqrt\beta}\right)
\eeq
Notice that $\hbar=0$ for hermitian matrices, i.e. the hermitian case is the classical limit $[y,x]=0$.
Notice also that the expected duality $\beta\leftrightarrow 1/\beta$ (cf \cite{mkrtchyan1, Bryc}) corresponds to $\hbar\leftrightarrow -\hbar$, i.e. we expect it to correspond to the duality $x\leftrightarrow y$ (for $\hbar=0$, the $x\leftrightarrow y$ duality was proved in \cite{OE}).

Let us also mention that the topological expansion of non-hermitian matrix integrals is known to be related to the enumeration of unoriented discrete surfaces, and we expect that our $F^{(g)}=\sum_k \hbar^k \,F^{(g,k)}$ can be interpreted as generating functions of such unoriented surfaces.

So, in this article, we provide a method for computing $F^{(g,k)}$ for any $g$ and $k$ (which is more consise than  \cite{ChekEynbeta}).

\subsubsection*{Outline of the article}

\begin{itemize}
\item In section \ref{secdefAnn}, we introduce our recursion kernel $K(x,x')$, and we show that the mere existence of this kernel is equivalent to the Bethe ansatz condition.

\item In section \ref{secdefWngFgAnn}, we define the $W_n^{(g)}$'s and the $F^{(g)}$'s, and we study their main properties, for instance that $W_n^{(g)}$ is symmetric.

\item In section \ref{seclimitclassical}, we study the classical limit $\hbar\to 0$, and we show that we recover the algebro-geometric construction of \cite{OE}.

\item This inspires a notion of non-commutative algebraic geometry in section \ref{secqalgeo}.

\item In section \ref{secMMAnn}, we study the application to the topological expansion of non-hermitian matrix integrals.

\item In section \ref{secGaudin}, we study the application to the Gaudin model.

%\item In section \ref{secnofatgraphs}, we study the application to the enumeration of non-orientable discrete surfaces.

\item Section \ref{secConcl} is the conclusion.

\item All the technical proofs are written in appendices for readability.

\end{itemize}

\section*{2 Definitions, kernel and Bethe ansatz}\label{secdefAnn}

Let $V'(x)$ be a rational function (possibly a polynomial), and we call $V(x)$ the {\bf potential}.
Let $\alpha_i$ be the poles of $V'(x)$ (one of the poles may be at $\infty$). 

\medskip For example, the following potential is called {\bf Gaudin potential} (see section \ref{secGaudin}):
\beq
V_{\rm Gaudin}'(x) = x+ \sum_{i=1}^{\npole} {S_i\over x-\alpha_i}
\eeq
As another example, we will consider formal matrix models in section \ref{secMMAnn}, for which
$V'(x)$ is a polynomial.

However, many other choices can be made.

\subsection*{2.1 The problem}

Our problem is to find $m$ complex numbers $s_1,\dots, s_m$, as well as two functions $G(x_0,x)$ and $K(x_0,x)$ with the following properties:
\begin{enumerate}
\item $G(x_0,x)$ is a rational function of $x$ with poles at $x=s_i$, and a simple pole of residue $+1$ at $x=x_0$, and which behaves as $O(1/x)$ at $x\to\infty$.
\item $G(x_0,x)$ is a rational function of $x_0$ with (possibly multiple) poles at $x_0=s_i$, and a simple pole at $x_0=x$, and $G(x_0,x)$ behaves like $O(1/x_0)$ at $x_0\to\infty$.
\item $B(x_0,x)=-{1\over 2} {\partial\over \partial x}G(x_0,x)$ is symmetric:
$B(x_0,x)=B(x,x_0)$.
\item $K$ and $G$ are related by the following differential equation:
\beq\label{diffeqdefK}
\left(2\hbar \sum_{i=1}^m {1\over x-s_i} - V'(x) - \hbar {\partial\over \partial x}\right)\, K(x_0,x) = G(x_0,x)
\eeq
\item $K(x_0,x)$ is analytical when $x\to s_i$ for all $i=1,\dots,m$.

\end{enumerate}

We shall see below that those 5 conditions determine $K$, $G$, and the $s_i$'s.
In fact condition 5 is the most important one in this list, it amounts to a {\bf no-monodromy condition}, and we shall see below that it implies that the $s_i$'s must obey the {\bf Bethe-ansatz equation}.

\subsection*{2.2 Analytical structure of the kernel $G$}

The 4th and 5th conditions imply that $G(x_0,x)$ has at most simple poles at $x=s_i$. Then condition 3 implies that $G(x_0,x)$ has at most double poles at $x_0=s_i$.

The first 3 conditions imply that there exists a symmetric matrix $A_{i,j}$ such that $G(x_0,x)$ can be written:
\beq
G(x_0,x) = {1\over x-x_0} + 2\sum_{i,j=1}^m {A_{i,j}\over (x-s_i)(x_0-s_j)^2} 
%+  \sum_{i} {c_i\over x_0-s_i}
\eeq
and therefore:
\beq
B(x_0,x) = {1\over 2}\,{1\over (x-x_0)^2} + \sum_{i,j=1}^m {A_{i,j}\over (x-s_i)^2(x_0-s_j)^2} 
\eeq
We will argue in section \ref{secqalgeo}, that $B$ can be viewed as a non=commutative deformation of the algebraic geometry's Bergmann kernel.

\subsection*{2.3 Bethe ansatz and monodromies}\label{sectbetheansatz}

First, we study the conditions under which the differential equation \eq{diffeqdefK} has no monodromies around $s_i$, in other words the condition under which $K(x_0,x)$ is analytical when $x\to s_i$, $\forall i$:
\beq
K(x_0,s_i+\epsilon) = K(x_0,s_i)+\epsilon K'(x_0,s_i)+{\epsilon^2\over 2} K''(x_0,s_i)+{\epsilon^3\over 6} K'''(x_0,s_i)+\dots
\eeq

\medskip
Equating the coefficient of $\epsilon^{-1}$ in \eq{diffeqdefK}, we get:
\beq\label{eqmono1}
\hbar K(x_0,s_i) =  \sum_{j} {A_{i,j}\over (x_0-s_j)^2}
\eeq
equating the coefficient of $\epsilon^{0}$ in \eq{diffeqdefK}, we get:
\beq\label{eqmono2}
\hbar K'(x_0,s_i) = {-1\over x_0-s_i} + V'(s_i) K(x_0,s_i) - 2\hbar\sum_{j\neq i} {K(x_0,s_i)-K(x_0,s_j)\over s_i-s_j} 
%+\sum_j {c_j\over x_0-s_j}
 \eeq
and equating the coefficient of $\epsilon^{1}$ in \eq{diffeqdefK}, we get:
\bea\label{eqmono3}
&& 2\hbar \sum_{j\neq i}{K'(x_0,s_i)\over s_i-s_j} -2\hbar \sum_{j\neq i}{K(x_0,s_i)\over (s_i-s_j)^2} +V''(s_i)K(x_0,s_i)\cr
&=& V'(s_i) K'(x_0,s_i) - {1\over (s_i-x_0)^2} - 2 \sum_{j\neq i} \sum_{k} {A_{j,k}\over (s_i-s_j)^2(x_0-s_k)^2} \cr
\eea
Notice from \eq{eqmono1}, that $K(x_0,s_i)$ has only double poles in $x_0$, with no residue:
\beq
\Res_{x_0\to s_k} K(x_0,s_i)=0
\eeq
Then, taking the residue at $x_0\to s_k$ in \eq{eqmono2}, we see that:
\beq
\hbar \Res_{x_0\to s_k} K'(x_0,s_i) = -\delta_{i,k} 
%+c_k
\eeq
Then, 
%since $\sum_k c_k=1$, we can always choose $k$ such that $c_k\neq \delta_{i,k}$, and 
taking the residue when $x_0\to s_i$ in \eq{eqmono3}, implies that the $s_i$'s are Bethe roots, i.e. they must obey the {\bf Bethe equation}:
\beq\label{Betheeq}
\encadremath{
\forall\, i=1,\dots,m\, , \qquad \quad 2\hbar\,\sum_{j\neq i} {1\over s_i-s_j} = V'(s_i)
}\eeq
Then \eq{eqmono3} becomes:
\beq\label{defA}
{1\over (s_i-x_0)^2}
=  V''(s_i)K(x_0,s_i) + 2\hbar \sum_{j\neq i}{K(x_0,s_i)\over (s_i-s_j)^2}    - 2 \sum_{j\neq i} \sum_{k} {A_{j,k}\over (s_i-s_j)^2(x_0-s_k)^2}
\eeq
i.e. by comparing the coefficient of $1/(x_0-s_k)^2$ on both sides:
\beq\label{Anomono}
\encadremath{
\delta_{i,k} = {1\over \hbar}V''(s_i) A_{i,k} + 2 \sum_{j\neq i} {A_{i,k}-A_{j,k}\over (s_i-s_j)^2}
}\eeq
i.e. $A$ is the inverse of the Hessian matrix $T$:
\beq
A=T^{-1}
\virg
\left\{\begin{array}{l}
T_{i,i} = {1\over \hbar}V''(s_i) + 2 \sum_{j\neq i} {1\over (s_i-s_j)^2} \cr
T_{i,j} = - \, {2\over (s_i-s_j)^2} \cr
\end{array}\right.
\eeq
\beq
T_{i,j} = {1\over \hbar}\,\,{\partial^2\over \partial s_i \partial s_j}\,\,\Big(  \sum_k V(s_k) - \hbar \sum_{k\neq l} \ln{(s_k-s_l)}\Big)
\eeq

Therefore the Bethe ansatz equations \eq{Betheeq} (as well as \eq{Anomono}) are the necessary conditions for $K(x_0,x)$ to be analytical when $x\to s_i$. Those conditions are necessary, but also sufficient conditions, as one can see by solving explicitely the linear ODE for $K$.

\beq\label{solODEK}
K(x_0,x) =  \int^x_{c}\,dx' G(x_0,x')\,\,\ee{{1\over \hbar} (V(x')-V(x))}\,\,\prod_{i} {(x-s_i)^2\over (x'-s_i)^2}
\eeq

\medskip

\br
Notice that $K(x_0,x)$ is not analytical everywhere, it has a logarithmic singularity  at $x=x_0$, and it has essential singularities at the poles of $V'$.
\er

\br
Notice that if one solution of the ODE is analytical near all $s_i$'s, then all solutions have that property.
Indeed, all the solutions differ by a solution of the homogeneous equation, i.e. by:
\beq
\prod_{i} (x-s_i)^2\,\, \ee{-{1\over \hbar} V(x)}
\eeq
which is clearly analytical near the $s_i$'s.

So, for the moment, the requirements 1--5 determine $G(x_0,x)$ uniquely, but $K(x_0,x)$ is not unique.
Let us choose one possible $K(x_0,x)$, and we prove below in theorem \ref{thWngindeptK}, that the objects we are going to define, do not depend on the choice of $K$.

\er

\br
In what follows, it is useful to compute the Taylor expansion of $K$ near a root $s_i$. We write:
\beq
K(x_0,x) = \sum_{k=0}^\infty K_{i,k}(x_0)\,\,(x-s_i)^k
\eeq
The coefficients $K_{i,k}(x_0)$ are themselves rational fractions of $x_0$, and are computed in appendix \ref{appKexpansion}.
\er

\subsection*{2.4 Schroedinger equation}

It is well known  that the Bethe condition can be rewritten as a Schroedinger equation \cite{BBTbook, BabBetGaudin}. We rederive it here for completeness.

Define the wave function:
\beq
\psi(x) = \prod_{i=1}^m (x-s_i) \,\,\, \ee{-{1\over 2\hbar}\, V(x)}
\virg
\om(x) = \hbar \sum_{i=1}^m {1\over x-s_i}
\eeq
\beq
Y(x) = -2 \hbar {\psi'(x)\over \psi(x)} = V'(x)-2\om(x) = V'(x) - 2\hbar \sum_i {1\over x-s_i}
\eeq
then compute:
\bea\label{eqRicattiUY}
U(x) 
&=& Y^2 -2 \hbar Y'(x) = 4\hbar^2 {\psi''(x)\over \psi(x)} \cr
&=& V'(x)^2 - 2\hbar V''(x) + 4(\om(x)^2-V'(x)\om(x)+\hbar \om'(x))
\eea

We have:
\bea
\om(x)^2+\hbar \om'(x) 
&=& \hbar^2 \sum_{i,j} {1\over (x-s_i)(x-s_j)} - \hbar^2 \sum_{i} {1\over (x-s_i)^2}  \cr
&=& \hbar^2 \sum_{i\neq j} {1\over (x-s_i)(x-s_j)}  \cr
\eea
which is a rational fraction with only simple poles at the $s_i$'s.
The residue at $s_i$ is $2\hbar^2 \sum_{j\neq i} {1\over s_i-s_j} = \hbar V'(s_i)$, and thus:
\beq
\om(x)^2+\hbar \om'(x) = \hbar \sum_{i} {V'(s_i)\over (x-s_i)}  
\eeq
which implies:
\beq
\om(x)^2 - V'(x)\om(x) +\hbar \om'(x) = - \hbar \sum_{i} {V'(x)-V'(s_i)\over (x-s_i)}  
\eeq
and thus:
\beq\label{eqUV}
U(x) = V'(x)^2 - 2\hbar V''(x)  - 4 \hbar \sum_{i=1}^m {V'(x)-V'(s_i)\over x-s_i}
\eeq

Therefore $U(x)$ is a rational fraction with poles at the poles of $V'$ (of degree at most those of $V'^2$),  in particular it has no poles at the $s_i$'s.

$U$ is the potential for the Schroedinger equation for $\psi$:
\beq
\encadremath{
4\hbar^2 \psi'' =  U\, \psi
}\eeq

\medskip
As announced in the introduction, this equation can be encoded in a D-module element:
\beq
{\cal E}(x,y) = y^2-{1\over 4}U(x)
\virg
y=\hbar\,{\partial \over \partial x}
\virg [y,x]=\hbar
\eeq
i.e.
\beq
{\cal E}(x,y).\psi = 0
\eeq

Notice that the Schroedinger equation is equivalent to a Ricatti equation for $Y=-2\hbar \psi'/\psi$:
\beq
\encadremath{
Y^2 - 2\hbar Y' = U
}\eeq

\subsection*{2.5 Classical limit}
\label{secclaslim1}

We shall come back in more detail to the classical limit $\hbar\to 0$ in section \ref{seclimitclassical}.
However, let us already make a few comments.

\bigskip
$\bullet$ In the classical limit, the Ricatti equation becomes an algebraic equation (hyperelliptical), which we call the (classical) spectral curve:
\beq
Y_{\rm cl}^2 = U(x) 
\eeq
The function $Y_{\rm cl}(x)=\sqrt{U(x)}$ is therefore a multivalued function of $x$, and it should be seen as a meromorphic function on a branched Riemann surface (branching points are the zeroes of $U(x)$).
We shall see below that in the limit $\hbar\to 0$, the kernel $B(x_0,x)$ tends towards the Bergmann kernel of that Riemann surface.

In other words the classical limit is expressed in terms of {\bf algebraic geometry}.

In fact, in this article we are going to define non-commutative deformations of certain algebraic geometric objects in section \ref{secqalgeo}.

\section*{3 Definition of correlators and free energies}
\label{secdefWngFgAnn}

In this section, we define the quantum deformations of the symplectic invariants introduced in \cite{Eyn1loop, OE}.
The following definitions are inspired from (not hermitian) matrix models. The special case of their application to matrix models will be discussed in section \ref{secMMAnn}.

\subsection*{3.1 Definition of correlators}

\bd\label{defWngAnn}
We define the following functions $W_n^{(g)}(x_1,\dots,x_n)$ (called $n$-point correlation function of "genus"\footnote{here $g$ is any given integer, it has nothing to do with the genus of the spectral curve.} $g$) by the recursion:
\beq
W_1^{(0)}(x) = \om(x) = \hbar \sum_{i=1}^m {1\over x-s_i}
\virg
W_2^{(0)}(x_1,x_2)=B(x_1,x_2)
\eeq
\bea\label{mainrecformulaAnn2}
&& W^{(g)}_{n+1}(x_0,J)  \cr
&=&   \sum_{i=1}^m \Res_{x\to s_i}  K(x_0,x)\, \left( \ovl{W}_{n+2}^{(g-1)}(x,x,J) + \sum_{h=0}^g \sum'_{I\subset J} W_{|I|+1}^{(h)}(x,I) W_{n-|I|+1}^{(g-h)}(x,J/I) \right) \cr
\eea
where $J$ is a collective notation for the variables $J=\{ x_{1},\dots,x_{n} \}$, and where $\sum\sum'$ means that we exclude the terms $(h=0,I=\emptyset)$ and $(h=g,I=J)$, and where:
\beq
\ovl{W}_{n}^{(g)}(x_1,...,x_n) = W_{n}^{(g)}(x_1,...,x_n) - {\delta_{n,2}\delta_{g,0}\over 2}\, {1\over (x_1-x_2)^2}
\eeq

\ed

\vspace{0.5cm}
\br
This is exactly the same recursion as in \cite{OE}, the only difference is that the kernel $K$ is not algebraic, but it is solution of the differential equation \eq{diffeqdefK}.
We shall show in section \ref{seclimitclassical}, that in the limit $\hbar\to 0$, it indeed reduces to the definition of \cite{OE}.
\er

\br
We say that $W_n^{(g)}$ is the correlation function of genus $g$ with $n$ marked points, and sometimes we say that it has characteristics:
\beq
\chi=2-2g-n
\eeq
By analogy with algebraic geometry, we say that $W_n^{(g)}$ is stable if $\chi<0$ and unstable if $\chi\geq 0$.
We see that all the stable $W_n^{(g)}$'s have a common recursive definition  def.\ref{defWngAnn}, whereas the unstable ones appear as exceptions.
\er

\br
In order for the definition to make sense, we must make sure that the behaviour of each term in the vicinity of $x\to s_i$ is indeed locally meromorphic so that we can compute residues, i.e. there must be no log-singularity near $s_i$.
In particular, the requirement of section \ref{sectbetheansatz} for the kernel $K$ is {\bf necessary}.
In other words, a necessary condition for definition eq.\ref{mainrecformulaAnn2} to make sense, is the {\bf Bethe ansatz} !
\er

\subsection*{3.2 Properties of correlators}

The main reason of definition. \ref{defWngAnn}, is because the $W_n^{(g)}$'s have many beautiful properties, which generalize those of \cite{OE}.

We shall prove the following properties:

\bt\label{thpolessiWngAnn}
Each $W_n^{(g)}$ is a rational function of all its arguments. It has poles only at the $s_i$'s (except $W_2^{(0)}$, which also has a pole at $x_1=x_2$).
In particular it has no poles at the $\alpha_i$'s.
Moreover, it vanishes as $O(1/x_i)$ when $x_i\to\infty$.
\et
\proof{ in appendix \ref{approofthpolessiWngAnn}}

\bt\label{thWngPngAnn}
 The $W_n^{(g)}$'s satisfy the loop equation, i.e. Virasoro-like constraints.
This means that the quantity:
\bea\label{loopeqPngAnn2}
 P_{n+1}^{(g)}(x;x_1...,x_n)
 &=&
-Y(x)\overline{W}_{n+1}^{(g)}(x,x_1,...,x_n) + \hbar \partial_{x}{\overline{W}_{n+1}^{(g)}(x,x_1...,x_n)} \cr
&& + \sum_{I\subset J} \ovl{W}_{|I|+1}^{(h)}(x,x_I) \ovl{W}_{n-|I|+1}^{(g-h)}(x,J/I) +
\ovl{W}_{n+2}^{(g-1)}(x,x,J)  \cr
& &+ \sum_{j}
\partial_{x_j} \left( {{\ovl{W}_n^{(g)}(x,J/\{j\})-{\ovl{W}_n^{(g)}(x_j,J/\{j\})}} \over {(x-x_j)}}\right) \cr
\eea
is a rational fraction of $x$ (possibly a polynomial), with no pole at $x=s_i$.
The only possible poles of $P_{n+1}^{(g)}(x;x_1...,x_n)$ are at the poles of $V'(x)$, with degree less than the degree of $V'$.

\et
\proof{ in appendix \ref{approofthWngPngAnn}}

\bt\label{thsymAnn}
 Each $W_n^{(g)}$ is a symmetric function of all its arguments.
\et
\proof{ in appendix \ref{approofthsymAnn}, with the special case of $W_3^{(0)}$ in appendix \ref{approofthW3KrichAnn}.}

\bt\label{thWngindeptK}
The correlation functions $W_n^{(g)}$ are independent of the choice of kernel $K$, provided that $K$ is solution of the equation \eq{diffeqdefK}.
\et
\proof{ in appendix \ref{approofthWngindeptK}}

%\bt\label{thnoressi}
%If $2g-2+n>0$ we have 
%\beq 
%\Res_{x_1\to s_i} W_n^{(g)}(x_1,\dots,x_n) = 0
%\eeq
%\et

\bt\label{thW3KrichAnn}
The 3 point function $W_3^{(0)}$ can also be written:
\beq
W_3^{(0)}(x_1,x_2,x_3) = 4\,\sum_i \Res_{x\to s_i}\,\, {B(x,x_1)B(x,x_2)B(x,x_3)\over Y'(x)}
\eeq
(In section \ref{secqalgeo}, we interpret this equation as a non-commutative version of Rauch variational formula).
\et
\proof{ in appendix \ref{approofthW3KrichAnn}}

\bt\label{thvariationV}
Under an infinitesimal variation of the potential $V\to V+\delta V$, we have:
\beq
\forall n\geq 0, g\geq 0
\,\, , \quad
\delta W_{n}^{(g)}(x_1,\dots,x_n) = - \sum_i \Res_{x\to s_i} W_{n+1}^{(g)}(x,x_1,\dots,x_n)\, \delta V(x)
\eeq
\et
\proof{ in appendix \ref{approofthvariationV}}

This theorem suggest the definition of the "loop operator":
\bd
The loop operator $\delta_x$ computes the variation of $W_n^{(g)}$ under a formal variation $\delta_x V(x')={1\over x-x'}$:
\beq
\delta_{x_{n+1}}\, W_n^{(g)}(x_1,\dots,x_n) = W_{n+1}^{(g)}(x_1,\dots,x_n,x_{n+1})
\eeq
The loop operator is a derivation: $\delta_x (uv) = u\delta_x v+v\delta_x u$, and we have $\delta_{x_1}\delta_{x_2} = \delta_{x_2}\delta_{x_1}$, $\delta_{x_1} \partial_{x_2} =\partial_{x_2}  \delta_{x_1}$.
\ed

\bt\label{thResY} 
For $n\geq 1$, $W_n^{(g)}$ satify the equation:
\beq
 \sum_{i=1}^n {\partial \over \partial x_i}\, \ovl{W}_n^{(g)}(x_1,\dots,x_n)  
= -\sum_i \Res_{x_{n+1}\to s_i}\,\, V'(x_{n+1})\,\,\ovl{W}_{n+1}^{(g)}(x_1,\dots,x_n,x_{n+1})
\eeq
and
\beq
 \sum_{i=1}^n {\partial \over \partial x_i}\, x_i \,\ovl{W}_n^{(g)}(x_1,\dots,x_n)  
= -\sum_i \Res_{x_{n+1}\to s_i}\,\, x_{n+1}\,V'(x_{n+1})\,\,\ovl{W}_{n+1}^{(g)}(x_1,\dots,x_n,x_{n+1})
\eeq
\et
\proof{ in appendix \ref{approofthResY}}

\bt\label{thdilaton} 
For $n\geq 1$, $W_n^{(g)}$ satify the equation:
\beq
(2-2g-n-\hbar {\partial\over \partial \hbar})\, \ovl{W}_{n}^{(g)}(x_1,\dots,x_n)
= - \sum_i \Res_{x_{n+1}\to s_i}\,\, V(x_{n+1})\,\,\ovl{W}_{n+1}^{(g)}(x_1,\dots,x_n,x_{n+1})
\eeq
\et
\proof{ We give a "long" proof in appendix \ref{approofthdilaton}.

There is also a short cut:

If one changes $\hbar\to \l \hbar$, and $V\to \l V$, the $s_i$'s don't change, $B$ and $G$ don't change, and $K$ changes to ${1\over \l}\,K$, thus $W_n^{(g)}$ changes by $\l^{2-2g-n} W_n^{(g)}$.
The theorem is obtained by computing ${\l\partial\over \partial \l} \l^{2g-2+n} W_n^{(g)} = \sum_k {t_k\partial\over \partial t_k} W_n^{(g)} $, and computing the RHS with theorem \ref{thvariationV}, i.e. $\delta V=V$. 

}

\subsection*{3.3 Definition of free energies}

So far, we have defined $W_n^{(g)}$ with $n\geq 1$.
Now, we define $F^{(g)}=W_0^{(g)}$.
\bigskip

Theorem \ref{thvariationV}, and the symmetry theorem \ref{thsymAnn} imply that:
\beq
\delta_{x_1} W_1^{(g)}(x_2) = W_2^{(g)}(x_1,x_2) = W_2^{(g)}(x_2,x_1)
= \delta_{x_2} W_1^{(g)}(x_1) 
\eeq
Thus, the symmetry of $W_2^{(g)}$ implies that there exists a "free energy" $F^{(g)}=W_0^{(g)}$ such that:
\beq
W_{1}^{(g)}(x) = \delta_x F^{(g)} 
\eeq
which is equivalent to saying that for any variation $\delta V$:
\beq
\delta F^{(g)} = - \sum_i \Res_{x\to s_i} W_{1}^{(g)}(x)\, \delta V(x)
\eeq
Therefore, we know that there must exists some $F^{(g)}=W_0^{(g)}$ which satisfy theorem \ref{thvariationV} for $n=0$.

\bigskip

Now, let us give a definition of $F^{(g)}$, inspired from theorem \ref{thdilaton}, and which will be proved to satisfy theorem \ref{thvariationV} for $n=0$.

\bd\label{defallFg}
We define $F^{(g)} \equiv W_0^{(g)}$ by a solution of the differential equation in $\hbar$:
\beq
\forall g\geq 2 \virg
(2-2g-\hbar {\partial\over \partial \hbar})\, F^{(g)} =  - \sum_i \Res_{x\to s_i}\,\, W_{1}^{(g)}(x) \, V(x)
\eeq
more precisely:
\beq\label{defFg}
F^{(g)} =  \hbar^{2-2g}\,\int_0^{\hbar}\, {d\td{\hbar}\over {\td\hbar}^{3-2g}}\,\, \sum_i \Res_{x\to s_i}\,\, V(x)\,\,\, \left. W_{1}^{(g)}(x)\right|_{{\td\hbar}}
\eeq

And the unstable cases $2-2g\geq 0$ are defined by:
\beq
F^{(0)} =   \hbar^2 \sum_{i\neq j} \ln{(s_i-s_j)}  - \hbar \sum_i V(s_i)
\eeq
\beq
F^{(1)} = {1\over 2} \ln{\det A} + \ln{(\Delta(s)^2)} + {F^{(0)}\over \hbar^2} 
\eeq
where $\Delta(s) = \prod_{i>j} (s_i-s_j)$ is the Vandermonde determinant of the $s_i$'s.
\ed
%It is to be noticed that the differential equation in $\hbar$ given in the definition \ref{defFg} does not completely define the $F^{(g)}$'s because they are only defined up to a solution of the homogeneous equation which is $C\,\,\hbar^{2-2g}$. But it is shown in appendix ?? that the constant can be chosen so that $\delta F^{(g)} = \sum_i \Res_{x\to s_i} W_{1}^{(g)}(x)\, \delta V(x)$ making the two definitions compatible.

\bigskip

{\bf Properties of the $F^{(g)}$'s:}

The definition of the $F^{(g)}$'s, is made so that all the theorems for the $W_n^{(g)}$'s, hold for for $n=0$ as well.
Proofs are given in appendices \ref{approofFg}, \ref{approofF0}, \ref{approofF1}.

\bigskip

Explicit computations of the first few $F^{(g)}$'s are given in section \ref{secGaudin} and appendix \ref{appm1}.

\section*{4 Classical limit and WKB expansion}
\label{seclimitclassical}

In the $\hbar\to 0$ limit, all quantities can be expanded formally into powers of $\hbar$:
Write:
\beq
W_n^{(g)}(x_1,\dots,x_n) = \sum_k \hbar^k W_n^{(g,k)}(x_1,\dots,x_n)
\virg
F^{(g)} = \sum_k \hbar^k F^{(g,k)}
\eeq

\subsection*{4.1 Classical limit}

Here we consider the classical limit $\hbar\to 0$.
We noticed in section \ref{secclaslim1}, that in that limit, the Ricatti equation
\beq
Y^2 -2\hbar Y' = U= V'^2 -2\hbar V'' - 4P
\eeq
where $P(x) = \hbar\sum_i {V'(x)-V'(s_i)\over x-s_i}$,
becomes an algebraic hyperelliptical equation:
\beq
\Ycl^2 = U(x) = V'(x)^2 - 4P(x)
\eeq
i.e.
\beq
Y(x) \mathop{\sim}_{\hbar\to 0} \Ycl(x) =  \sqrt{V'(x)^2-4P(x)}
\eeq
$\Ycl(x)$ is a multivalued function of $x$, and it should be seen as a meromorphic function on a 2-sheeted Riemann surface, i.e. there is a Riemann surface $\Sigma$ (of equation $0={\cal E}_{\rm cl}(x,y)=y^2-4U(x)$, such that the solutions of ${\cal E}_{\rm cl}(x,y)=0$ are parametrized by two meromorphic functions on $\Sigma$:
\beq
{\cal E}_{\rm cl}(x,y)=0
\quad\Leftrightarrow\quad
\exists z\in \Sigma \left\{
\begin{array}{l}
x=x(z) \cr
y=y(z)
\end{array}\right.
\eeq

The Riemann surface $\Sigma$ has a certain topology\footnote{This genus $\genus$ has nothing to do with the index $g$ of $F^{(g)}$ or $W_n^{(g)}$. } characterized by its genus $\genus$.
It has a (non-unique) symplectic basis of $2\genus$ non-trivial cycles $\acycle_i\cap \bcycle_j=\delta_{i,j}$.

The meromorphic forms on $\Sigma$ are classified as 1st kind (no pole), 3rd kind (only simple poles), and 2nd kind (multiple poles without residues).

There exists a unique 2nd kind differential $B_{\rm cl}$ on $\Sigma$, called the Bergmann kernel, such that:
$B_{\rm cl}(z_1,z_2)$ has a double pole at $z_1\to z_2$, and no other pole, without residue and normalized (in any local coordinate $z$) as:
\beq
B_{\rm cl}(z_1,z_2) \mathop{\sim}_{z_2\to z_1} {dz_1 dz_2\over (z_1-z_2)^2}+ {\rm reg}
\virg
\forall i=1,\dots,\genus\, , \,\,\, \oint_{\acycle_i} B_{\rm cl} = 0
\eeq
We define a primitive:
\beq
G_{\rm cl}(z_0,z) = -2 \int^z B_{\rm cl}(z_0,z')
\eeq
which is a 3rd kind differential in the variable $z_0$, it is called $dE_{z}(z_0)$ in \cite{OE}.

When $\hbar=0$, the kernel $K(z_0,z)$ satisfies the equation:
\beq
K_{\rm cl}(z_0,z) = -\, {G_{\rm cl}(z_0,z)\over Y_{\rm cl}(z)} = 2\, {\int_c^z B_{\rm cl}(z_0,z')\over Y_{\rm cl}(z)}
\eeq
which coincides with the definition of the recursion kernel in \cite{OE}.

\subsection*{4.2 WKB expansion of the wave function}

When $\hbar$ is small but non-zero, we can WKB expand $\psi(x)$, i.e.:
\beq
\psi(x) \sim \ee{-{1\over 2\hbar} \int^x \Ycl(x')dx'}\,\,{1\over \sqrt{\Ycl(x)}}\,\left(1+\sum_k \hbar^k \psi_k(x)\right)
\eeq
i.e.
\beq
Y \sim \Ycl + \sum_{k=1}^\infty \hbar^k \,Y_k
\eeq
The expansion coefficients $Y_k$ can be easily obtained recursively from the Ricatti equation:
\beq
2\Ycl Y_k  = 2Y_{k-1}' - \sum_{j=1}^{k-1} Y_j\,Y_{k-j}
\eeq
For instance:
\beq
Y_1 = {\Ycl'\over \Ycl}
\virg
Y_2 = {Y'_1\over \Ycl} - {Y_1^2\over 2 \Ycl} = {\Ycl''\over \Ycl^2}-{3\over 2}\,{{\Ycl'}^2\over \Ycl^3}
\virg
\dots\, {\rm etc}
\eeq

\subsection*{4.3 $\hbar$ expansion of correlators and energies}

The kernel $K(x_0,x)$ can also be expanded:
\beq
K(x_0,x) = K_{\rm cl}(x_0,x) + \sum_{k=1}^\infty \hbar^k K_{(k)}(x_0,x)
\eeq
where $K_{(0)}=K_{\rm cl}$ is the kernel of \cite{OE}:
\beq
K_{\rm cl}(x_0,x) = {dE_{x,o}(x_0)\over Y_{\rm cl}(x)}
\eeq

This implies that the correlators $W_n^{(g)}$ can also be expanded:
\beq
 W^{(g)}_{n}(x_1,\dots,x_n)  = \sum_{k=0}^\infty \hbar^k\,\, W^{(g,k)}_{n}(x_1,\dots,x_n) 
\eeq
where the $W^{(g,k)}_{n}$ are obtained by the recursion:
\bea
 W^{(g,k)}_{n+1}(x_0,J) 
&=&   \sum_{l=0}^k \sum_i \Res_{x\to s_i}  K_{(k-l)}(x_0,x)\, \Big[ \overline{W}_{n+2}^{(g-1,l)}(x,x,J)  \cr
&& + \sum_{h=0}^g\sum_{j=0}^{l} \sum_{I\subset J}' W_{|I|+1}^{(h,j)}(x,I) W_{n-|I|+1}^{(g-h,l-j)}(x,J/I) \Big] 
\eea
where $J=\{ x_{1},\dots,x_{n} \}$.

Therefore, we observe that to leading order in $\hbar$, the $\lim_{\hbar\to 0} W^{(g,k)}_{n}=W^{(g,0)}_{n}$ do coincide with the $W^{(g)}_{n}$ computed with only $K_{\rm cl}$, and thus they coincide with the $W^{(g)}_{n}$ of \cite{OE}.

And also, the $\hbar$ expansion must coincide with  the diagrammatic rules of \cite{ChekEynbeta}.

\section*{5 Non-commutative algebraic geometry}
\label{secqalgeo}

We have seen that in the limit $\hbar\to 0$, the correlation functions and the various functions we are considering, are fundamental objects of algebraic geometry.
For instance $B$ is the Bergmann kernel, and $K$ is the recursion kernel of \cite{OE}, which generates the symplectic invariants $F_g$ and the correlators $W_n^{(g)}$ attached to the spectral curve $\Ycl(x)$.

\smallskip

In this paper, when $\hbar\neq 0$, we have defined deformations of those objects, which have almost the same properties as the classical ones, except that they are no longer algebraic functions.

For instance we have:
\begin{itemize}

\item {\bf Spectral curve}

The algebraic equation of the classical spectral curve is replaced by a linear differential equation:
\beq
0={\cal E}(x,y)=\sum_{i,j} {\cal E}_{i,j}\, x^i y^j
\quad \to \quad
0={\cal E}(x,\hbar \partial) \psi=\sum_{i,j} {\cal E}_{i,j}\, x^i (\hbar \partial)^j\,\psi
\eeq
In other words the polynomial ${\cal E}(x,y)$ is replaced by a non-commutative polynomial with $y=\hbar \partial_x$, i.e. $[y,x]=\hbar$.

Here, our non-commutative spectral curve is:
\beq
{\cal E}(x,y) = y^2 - U(x)
\virg y=\hbar \partial_x
\eeq
Notice that it can be factorized as:
\beq
{\cal E}(x,y) = (y-{Y\over 2})\,(y+{Y\over 2})
\eeq
where $Y(x)$ is solution of $Y^2-2\hbar Y'=U$.

\item {\bf Bergmann Kernel} $B(x_1,x_2)$

The non-commutative Bergmann kernel $B(x_1,x_2)$ is closely related to the Inverse of the Hessian $T$, i.e. to $A=T^{-1}$:
\beq
B(x_1,x_2) = {1\over 2(x_1-x_2)^2} + \sum_{i,j} {A_{i,j}\over (x_1-s_i)^2(x_2-s_j)^2}
\eeq
A property of the classical Bergmann kernel $B_{\rm cl}(x_1,x_2)$ is that it computes derivatives, i.e. for any meromorphic function $f(x)$ defined on the spectral curve we have:
\beq
df(x) = -\Res_{x_2\to {\rm poles\, of}\, f}\, B_{\rm cl}(x,x_2)\, f(x_2)
\eeq
Here, this property is replaced by:
for any function $f(x)$ defined on the non-commutative spectral curve (i.e. with poles only at the $s_i$'s), we have:
\beq
f'(x) = -2\sum_i \Res_{x_2\to s_i}\, B(x,x_2)\, f(x_2)\,\, dx_2
\eeq
The factor of $2$, comes from the fact that the interpretation of $x$, and thus of derivatives with respect to $x$, is slightly different.
In the classical case, the differentials are computed in terms of local variables, and $x$ is not a local variable near branch-points. A good local variable near a branchpoint $a$, is $\sqrt{x-a}$.
In the non-commutative case, the role of branchpoints seems to be played by the $s_i$'s, and $x$ is a good local variable near $s_i$.

\item {\bf Rauch variational formula}:
In classical algebraic geometry, on an algebraic curve of equation ${\cal E}(x,y)=\sum_{i,j} {\cal E}_{i,j} x^i y^j=0$, the Bergmann kernel depends only on the location of branchpoints $a_i$. The branchpoints are the points where the tangent is vertical, i.e. $dx(a_i)=0$. Their location is $x_i=x(a_i)$.
The Bergmann kernel is only function of the $x_i$'s, and the classical variational Rauch formula reads:
\beq
{\partial\, B_{\rm cl}(z_1,z_2)\over \partial x_i} = \Res_{z\to a_i}\, {B_{\rm cl}(z,z_1)\,B_{\rm cl}(z,z_2)\over dx(z)}
\eeq
Equivalently, we can parametrize the spectral curve as $x(y)$ instead of $y(x)$, and consider the branchpoints of $y$, i.e. $dy(b_i)=0$, whose location is $y_i=y(b_i)$, and we have:
\beq\label{ClassicalRauch}
{\partial\, B_{\rm cl}(z_1,z_2)\over \partial y_i} = \Res_{z\to b_i}\, {B_{\rm cl}(z,z_1)\,B_{\rm cl}(z,z_2)\over dy(z)}
\eeq

Here, in the non-commutative version, theorem \ref{thW3KrichAnn} and theorem \ref{thvariationV} implies that under a variation of the spectral curve, we have:
\beq
\delta B(x_1,x_2) =-{1\over 2}\,\sum_i \Res_{x\to s_i} {B(x,x_1)B(x,x_2)\over Y'(x)}\,\, \delta Y(x)
\eeq
Consider the branchpoints $b_i$ such that $Y'(b_i)=0$, and define their location as $Y_i=Y(b_i)$, by moving the integration contours we have:
\bea
\delta B(x_1,x_2) 
&=& {1\over 2}\,\sum_i \Res_{x\to b_i} {B(x,x_1)B(x,x_2)\over Y'(x)}\,\, \delta Y(x)\, dx \cr
&=& {1\over 2}\,\sum_i \delta Y_i\, \Res_{x\to b_i} {B(x,x_1)B(x,x_2)\over Y'(x)}\,dx \cr
\eea
i.e.:
\beq
{\partial\, B(x_1,x_2)\over \partial Y_i} = {1\over 2}\,\Res_{x\to b_i}\, {B(x,x_1)\,B(x,x_2)\over Y'(x)}\,\, dx
\eeq
which is thus the quantum version of the Rauch variational formula \eq{ClassicalRauch}.

\end{itemize}

\bigskip
Those properties can be seen as the beginning of a dictionary giving the deformations of classical algebraic geometry into non-commutative algebraic geometry.

\bigskip

{\bf Conjecture about the symplectic invariants}

The $F_g$'s of \cite{OE} are the symplectic invariants of the classical spectral curve,
which means that they are invariant under any cannonical change of the spectral curve which conserves the symplectic form $dx\wedge dy$.
For instance they are invariant under $x\to y,y\to -x$.

Here, we conjecture that we may define some non-commutative $F^{(g)}$'s which are invariant under any cannonical transformation which conserves the commutator $[y,x]=\hbar$.
This duality should also correspond to the expected duality $\beta\to 1/\beta$ in matrix models, cf \cite{mkrtchyan1, Bryc}.

However, to check the validity of this conjecture, one needs to extend our work to differential operators of any order in $y$, and not only order $2$. We plan to do this in a forthcoming work.

\section*{6 Application: non-hermitian Matrix models}
\label{secMMAnn}

The initial motivation for the work of \cite{OE}, as well as this present work, was initially random matrix models. The classical case corresponds to hermitian matrix models, and here, we show that $\hbar\neq 0$ corresponds in some sense to non-hermitian matrix models \cite{BrezNb, Bryc, Dum}.

\smallskip

In this section, we show that non-hermitian matrix models satisfy the loop equation \eq{loopeqappgen} of theorem \ref{thWngPngAnn}.

We define the matrix integral over $E_{m,2\beta}=$set of $m\times m$ matrices of Wigner--type $2\beta$ ($E_{m,1}=$ real symmetric matrices, $E_{m,2}=$ hermitean matrices, $E_{m,4}=$ real quaternion self-dual matrices, see \cite{Mehta}):
\beq
Z = \int_{E_{m,2\beta}} dM\,\, \ee{-{N\sqrt\beta}\, \Tr V(M)}
\eeq
where $N$ is some arbitrary constant, not necessarily related to the matrix size $m$.

It is more convenient to rewrite it in terms of eigenvalues of $M$ (see \cite{Mehta}):
\beq
Z = \int_{{\cal C}^m}\, d\l_1\dots d\l_m\,\, \prod_{i>j} (\l_j-\l_i)^{2\beta}\,\, \prod_i \ee{-{N\sqrt\beta}\, V(\l_i)}
\eeq
This last expression is well defined for any $\beta$, and not only $1/2,1,2$, and for any contour of integration ${\cal C}$ on which the integral is convergent.

We also define the correlators:
\bea
\overline{W}_n(x_1,\dots,x_n) 
&=& <\Tr{1\over x_1-M}\dots \Tr {1\over x_n-M}>_c  \cr
&=& \left(N\sqrt\beta\right)^{-n}\,\,{\partial \over \partial V(x_1)}\dots{\partial \over \partial V(x_n)}\,\ln{Z}
\eea
i.e. in terms of eigenvalues:
\beq
\overline{W}_n(x_1,\dots,x_n) = <\sum_{i_1} {1\over x_1-\l_{i_1}} \dots \sum_{i_n} {1\over x_n-\l_{i_n}}>_c
\eeq
In order to match with the notations of section \ref{secdefWngFgAnn}, we prefer to shift $\overline{W}_2$ by a second order pole, and we define:
\beq
W_n(x_1,\dots,x_n) = \overline{W}_n(x_1,\dots,x_n) + {\delta_{n,2}\over 2(x_1-x_2)^2}
\eeq

\bigskip

We are interested in a case where $Z$ has a large $N$ expansion of the form:
\beq
\ln{Z} \sim \sum_{g=0}^\infty N^{2-2g}\,\, F_g 
\eeq
and for the correlation functions we assume:
\beq\label{Wntopoexp}
W_{n}( x_1,\dots,x_n) = {1\over \beta^{n/2}}\,\,\sum_{g=0}^\infty N^{2-2g-n} W_{n}^{(g)}(x_1,\dots,x_n)
\eeq

\subsection*{6.1 Loop equations}

The loop equations can be obtained by integration by parts, or equivalently, they follow from the invariance of an integral under a change of variable.
By considering the infinitesimal change of variable:
\beq
\l_i \to \l_i + \epsilon{1\over x-\l_i} + O(\epsilon^2)
\eeq
we obtain:
\bea\label{recMMbetaWallkg}
&& {N\sqrt\beta}(V'(x)\, \overline{W}_{n+1}(x, x_1,\dots,x_n) - P_{n+1}(x; x_1,\dots,x_n) ) \cr
&=&  \beta \sum_{J\subset L}  \overline{W}_{1+|J|}(x,J)\,\overline{W}_{1+n-|J|}(x,{L/J})  \cr
&& +  \beta \overline{W}_{n+2}(x,x, x_1,\dots,x_n) \cr
&& - (1-\beta)\,{\partial \over \partial x} \, \overline{W}_{n+1}(x, x_1,\dots,x_n) \cr
&& + \sum_{j=1}^n  {\partial \over \partial x_j} \, {\overline{W}_{n}(x, L/\{x_j\})-\overline{W}_{n}(x_j, L/\{x_j\})\over x-x_j}
\eea
where $P_{n+1}(x; x_1,\dots,x_n) )$ is a polynomial in its first variable $x$, of degree $\delta_{n,1}+\deg V-2$.

If we expand this equation into powers of $N$ using \eq{Wntopoexp}, we have $\forall\, n,g$:
\bea\label{recMMbetaWg}
&& V'(x)\, \overline{W}_{n+1}^{(g)}(x, x_1,\dots,x_n) - P_{n+1}^{(g)}(x; x_1,\dots,x_n) ) \cr
&=&  \sum_{g'=0}^g \sum_{J\subset L}  \overline{W}_{1+|J|}^{(g')}(x,J)\,\overline{W}_{1+n-|J|}^{(g-g')}(x,{L/J})  \cr
&& +  \beta \overline{W}_{n+2}^{(g-1)}(x,x, x_1,\dots,x_n) \cr
&& +\hbar \,{\partial \over \partial x} \, \overline{W}^{(g)}_{n+1}(x, x_1,\dots,x_n) \cr
&& + \sum_{j=1}^n  {\partial \over \partial x_j} \, {\overline{W}^{(g)}_{n}(x, L/\{x_j\})-\overline{W}^{(g)}_{n}(x_j, L/\{x_j\})\over x-x_j}
\eea
where
\beq
\hbar = {\sqrt\beta-{1\over \sqrt\beta}\over N}
\eeq

Those loop equations coincide with 
%\eq{recsurfnoWallk} for the enumeration of unoriented ribbon graphs, and they coincide with 
the loop equations \eq{loopeqPngAnn2} of theorem \ref{thWngPngAnn}.

Moreover we have:
\beq\label{dWngdV}
\overline{W}^{(g)}_n = {\partial \overline{W}_{n-1}^{(g)}\over \partial V}
\eeq
and near $x\to \infty$:
\beq
\sqrt\beta\,\,W_1(x) \sim {m\over x} [N\hbar-\sum_{g=1}^\infty (-1)^g {(2g-2)!\over g! (g-1)!}\,\, (N\hbar)^{1-2g}   ]
\eeq
i.e.
\beq
W^{(0)}_1(x) \sim  {m\hbar \over x} +O(1/x^2)
\virg
W^{(g)}_1(x) \sim - {m\hbar \over x} \,\, \hbar^{-2g}\,\, {(2g-2)!\over g! (g-1)!}+O(1/x^2)
\eeq

One should notice that the loop equations are independent of the contour ${\cal C}$ of integration of eigenvalues.
The contour ${\cal C}$ is in fact encoded in the polynomial $P_{n+1}(x; x_1,\dots,x_n)$.

\subsection*{6.2 Solution of loop equations}

To order $g=0, n=1$ we have:
\beq
 V'(x)\, W_{1}^{(0)}(x) - P_{1}^{(0)}(x)
= W_{1}^{(0)}(x)^2 +\hbar \,{\partial \over \partial x} \, W_{1}^{(0)}(x)
\eeq
which is the same as the Ricatti equation \eq{eqRicattiUY}.

As we said above, the contour ${\cal C}$ is in fact encoded in the polynomial $P^{(0)}_{1}(x) $.
From now on, we choose a contour ${\cal C}$, i.e. a polynomial $P_{1}^{(0)}(x)$ such that the solution of the Ricatti equation is rational:
\beq
W_{1}^{(0)}(x) = \hbar \sum_{i=1}^m {1\over x-s_i}
\eeq
It also has the correct behaviour at $\infty$: $W_{1}^{(0)}(x)\sim {m\hbar\over x}$.
This corresponds to a certain contour ${\cal C}$ which we do not determine here.

Since $W_1^{(0)}(x) =\om(x)$ satisfies the Ricatti equation, i.e. the Bethe ansatz, the kernel $K$ exists, and we can define the functions $K(x_0,x)$, $G(x_0,x)$ and $B(x_0,x)$.

\bigskip

Then, from \eq{dWngdV}, we see that every $\overline{W}_n^{(g)}$ is going to be a rational fraction of $x$, with poles only at the $s_i$'s.
In particular, Cauchy theorem implies:
\beq
\overline{W}_{n+1}^{(g)}(x_0,x_1,\dots,x_n) = \Res_{x\to x_0} G(x_0,x)\,\overline{W}_{n+1}^{(g)}(x,x_1,\dots,x_n)
\eeq
and since both $G(x_0,x)$ and $\overline{W}_{n+1}^{(g)}(x,x_1,\dots,x_n)$ are rational fractions, which vanish sufficientely at $\infty$, we may change the integration contour to the other poles of the integrand, namely:
\bea
&& \overline{W}_{n+1}^{(g)}(x_0,x_1,\dots,x_n) \cr
&=& -\sum_{i} \Res_{x\to s_i} G(x_0,x)\,\overline{W}_{n+1}^{(g)}(x,x_1,\dots,x_n) \cr
&=& -\sum_{i} \Res_{x\to s_i} \overline{W}_{n+1}^{(g)}(x,x_1,\dots,x_n)\,\, (2\om(x)-V'(x)-\hbar\partial_x)K(x_0,x) \cr
&=& -\sum_{i} \Res_{x\to s_i} K(x_0,x)\,\, (2\om(x)-V'(x)+\hbar\partial_x)\overline{W}_{n+1}^{(g)}(x,x_1,\dots,x_n) \cr
\eea

Now, we insert loop equation \eq{recMMbetaWg} in the right hand side, and we notice that the term 
$P_{n+1}^{(g)}$ and ${\partial \over \partial x_j} \, {W^{(g)}_{n}(x_j, L/\{x_j\})\over x-x_j}$ do not have poles at the $s_i$'s, so they don't contribute. We thus get:
\bea
&& \overline{W}_{n+1}^{(g)}(x_0,x_1,\dots,x_n) \cr
&=& \sum_{i} \Res_{x\to s_i}  K(x_0,x)\, \Big(
\overline{W}_{n+2}^{(g-1)}(x,x,x_1,\dots,x_n) \cr
&& + \sum_{g'=0}^g \sum_{J\subset L} W_{1+|J|}^{(g')}(x,J) W_{1+n-|J|}^{(g-g')}(x,L/J) \Big)
\eea
i.e. we find the correlators of def \ref{defWngAnn}.

Special care is needed for $W_2^{(0)}$.
We have:
\bea
&& \overline{W}_{2}^{(0)}(x_0,x_1,\dots,x_n) \cr
&=& -\sum_{i} \Res_{x\to s_i} K(x_0,x)\,\, (2\om(x)-V'(x)+\hbar\partial_x)\overline{W}_{2}^{(0)}(x,x_1) \cr
&=& \sum_{i} \Res_{x\to s_i} K(x_0,x)\,\, {\om(x)\over (x-x_1)^2} \cr
&=& \hbar\,\sum_{i}  {K(x_0,s_i)\over (s_i-x_1)^2} \cr
&=&\sum_{i,j}  {A_{i,j}\over (s_i-x_1)^2(s_j-x_0)^2} \cr
\eea
which also agrees with  def \ref{defWngAnn}.

\section*{7 Application: Gaudin model}
\label{secGaudin}

The Gaudin model's Bethe ansatz is obtained for the potential:
\beq
V_{\rm Gaudin}'(x) = x+ \sum_{i=1}^{\npole} {S_i\over x-\alpha_i}
\eeq
i.e. it corresponds to a Gaussian matrix model with sources:
\beq
Z= \int_{E_{m,2\beta}} dM\,\, \ee{-{N\sqrt\beta \over 2}\Tr M^2}\,\, \prod_i \det(\alpha_i-M)^{-N S_i \sqrt\beta}
\eeq
with $\hbar = {\sqrt\beta-1/\sqrt\beta\over N}$.

$Z$ can also be written in eigenvalues:
\beq
Z = \int d\l_1\dots d\l_m \,\, {\prod_{i=1}^m \ee{-{N\sqrt\beta\over 2}\l_i^2}\over \prod_{i=1}^m\prod_{j=1}^{\ovl{n}} (\alpha_j-\l_i)^{N\sqrt\beta\, S_j}}\,\,\, \prod_{i>j} (\l_i-\l_j)^{2\beta}
\eeq

\subsection*{7.1 Example}

Consider:
\beq
V'(x) = x-{s^2\over x}
\virg
V(x) = {x^2\over 2} - s^2\ln{x}
\eeq
With only 1 root $m=1$, the solution of the Bethe equation $V'(x)=0$ is $x=s$.

Thus we have:
\beq
\om(x) = {\hbar\over x-s}
\eeq
\beq
B(x_1,x_2)={1\over 2(x_1-x_2)^2} + {\hbar\over 2(x_1-s)^2(x_2-s)^2}
\eeq

We find:
\beq
W_3^{(0)}(x_1,x_2,x_3) 
= {\hbar\over 2(x_1-s)^2(x_2-s)^2(x_3-s)^2}\,\left({1\over x_1-s}+{1\over x_2-s}+{1\over x_3-s}+{1\over 2s}\right) 
\eeq
\beq
W_1^{(1)}(x) = {1\over \hbar(x-s)} +{1\over 4s(x-s)^2}+{1\over 2(x-s)^3}
\eeq

For the free energies we have:
\beq
F^{(0)} = {\hbar\,s^2\over 2}\,(\ln{s^2}-1)
\eeq
\beq
F^{(1)} = {1\over 2}\,\ln{({\hbar\over 2})} + {F^{(0)}\over \hbar^2}
\eeq
\beq
F^{(2)} = - {1\over 12 \hbar s^2} -  {F^{(0)}\over \hbar^4}
% -{s^2\over 2\hbar^3}\,(\ln{s^2}-1)
\eeq
\beq
F^{(3)} =  {1\over 12 \hbar^3 s^2} + {2F^{(0)}\over \hbar^6}
%+  {s^2\over \hbar^5}\,(\ln{s^2}-1)
\eeq
and
\beq
Z = \ee{\sum_g N^{2-2g} F^{(g)}} = \ee{-N\sqrt\beta V(s)}\,{1\over \sqrt{2\hbar}}\,\,(1-{1\over 12 s^2 N^2 \hbar^2} + \dots)
\eeq
which is indeed the beginning of the saddle point expansion of:
\beq
Z =  \int dx\,\, \ee{-N\sqrt\beta\,\, V(x)}
\eeq

\section*{8 Conclusion}
\label{secConcl}

In this article, we have defined a special case of non-commutative deformation of the symplectic invariants of \cite{OE}. Many of the fundamental properties of \cite{OE} are conserved or only slightly modified.

The main difference, is that the recursion kernel, instead of beeing an algebraic function, is given by the solution of a differential equation, otherwise the recursion is the same.

\bigskip

The main drawback of our definition, is that it concerns only a very restrictive subset of possible non-commutative spectral curves.
Namely, we considered here only non commutative polynomials ${\cal E}(x,y)=\sum_{i,j} {\cal E}_{i,j}\,\, x^i y^j$ with $y=\hbar \partial_x$, of degree 2 in $y$, and such that the differential equation ${\cal E}(x,\hbar\partial).\psi=0$ has a "polynomial" solution of the form $\psi(x)=\prod_{i=1}^m (x-s_i)\,\, \ee{-V(x)/2\hbar}$.

It should be possible to extend our definitions to other "non-polynomial" solutions $\psi$ (with an infinite number of zeroes $m=\infty$ for instance), and/or to higher degrees in $y$.
In other words, what we have so far, is only a glimpse on more general structure yet to be discovered.

For example, it is not yet clear how our definitions are related to matrix integrals.
We have said that the integration contour for the eigenvalues should be chosen so that the solution of the Schroedinger equation is polynomial of degree $m$, however, it is not known how to find explicitly such integration contours. Conversely, the usual matrix integrals with eigenvalues on the real axis, do probably not correspond to polynomial solutions of the Schroedinger equation.
Similarly, it is not clear what the relationship between our definitions and the number of unoriented ribbon graphs is, for the same reason. The solution of the Schroedinger equation for ribbon graphs, should be chosen such that all the $W_n^{(g,k)}$'s are power series in $t$, and it is not known which integration contour it corresponds to, and which solution of the Schroedinger equation it corresponds to.

\bigskip

Therefore it seems necessary to extend our definitions to arbitrary solutions, i.e. to arbitrary integration contours for the matrix integrals. A possibility could be to obtain non-polynomial solutions as limits of polynomial ones.

\medskip

The extension to higher degree in $y$, can be obtained from multi-matrix integrals, and extension seems rather easy for polynomial solutions again.

\medskip

Finally, like the symplectic invariants of \cite{OE}, we expect those "to be defined" non-commutative symplectic invariants, to play a role in several applications to enumerative geometry, and to topological string theory like in \cite{BKMP}. In other words, we expect our $F^{(g)}$'s to be generating functions for intersection numbers in some non-commutative moduli spaces of unoriented Riemann surfaces, whatever it means...

\section*{Acknowledgments}
We would like to thank O. Babelon, M. Berg\`ere, M. Bertola, L. Chekhov, R. Dijkgraaf, J. Harnad and N. Orantin for useful and fruitful discussions on this subject.
This work is partly supported by the Enigma European network MRT-CT-2004-5652, by the ANR project G\'eom\'etrie et int\'egrabilit\'e en physique math\'ematique ANR-05-BLAN-0029-01, by the Enrage European network MRTN-CT-2004-005616,
by the European Science Foundation through the Misgam program,
by the French and Japaneese governments through PAI Sakurav, by the Quebec government with the FQRNT.

%======================= APPENDICES =================================

\section*{Appendix: Expansion of $K$}\label{appKexpansion}

Since we have to compute residues at the $s_i$'s, we need to compute the Taylor expansion of $K(x_0,x)$ when $x\to s_i$:
\beq\label{TaylexpKikx0}
K(x_0,x) = \sum_{k} (x-s_i)^k\,K_{i,k}(x_0)
\eeq
For instance we find:
\beq
K_{i,0} = {1\over \hbar}\sum_{j}\, {A_{i,j}\over (x_0-s_j)^2}
\eeq
\bea
 \hbar K_{i,1} (x_0)
&=& 
%\sum_j {c_j\over x_0-s_j} 
-{1\over (x_0-s_i)} 
- 2\sum_{a\neq i} \sum_{j} {A_{a,j}\over (s_a-s_i)\,(x_0-s_j)^2}   \cr
\eea
\bea
 \hbar K_{i,3} 
&=& - \hbar\Big(   2\sum_{a\neq i} {1\over (s_a-s_i)^{2}} + {1\over \hbar} V''(s_i) \Big)\, K_{i,1} \cr
&& - \hbar\Big(   2\sum_{a\neq i} {1\over (s_a-s_i)^{3}} + {1\over \hbar}  {V'''(s_i)\over 2}  \Big)\, K_{i,0}  \cr
&& +{1\over (x_0-s_i)^{3}} 
+ 2\sum_{a\neq i} \sum_{j} {A_{a,j}\over (s_a-s_i)^{3}\,(x_0-s_j)^2}   \cr
\eea
Thanks to property \eq{Ki2vanishhyp}, we may assume (but it is not necessary) that:
\beq
K_{i,2}=0
\eeq
Then, we have the recursion for $k\geq 0$:
\bea
&& \hbar\Big( (1-k)K_{i,k+1} - 2\sum_{a\neq i}\sum_{l=0}^{k} {K_{i,k-l}\over (s_a-s_i)^{l+1}} - {1\over \hbar} \sum_{l=0}^k {V^{(l+1)}(s_i)\over l!} K_{i,k-l} \Big) \cr
&=& -{1\over (x_0-s_i)^{k+1}} 
- 2\sum_{a\neq i} \sum_{j} {A_{a,j}\over (s_a-s_i)^{k+1}\,(x_0-s_j)^2}  
%+ \delta_{k,0} \,\sum_j {c_j\over x_0-s_j}  \cr
\eea
This proves that each $K_{i,k}(x_0)$ is a rational fraction of $x_0$, with poles at the $s_j$'s.

\subsection*{Rational fraction of $x_0$}

Thus we write:
\beq\label{TaylexpKikjl}
K_{i,k}(x_0) = \sum_{j,l} {1\over (x_0-s_j)^{k'}}\,\, K_{i,k;j,k'} 
\eeq
For instance we have:
\beq
K_{i,0;j,k'} = {A_{i,j}\over \hbar}\,\delta_{k',2}
\eeq
\beq
 \hbar K_{i,1;j,k'} = 
 - \delta_{k',1}  \delta_{i,j}
%- \delta_{k',1} (c_j  - \delta_{i,j})  
- 2 \delta_{k',2}\, \sum_{a\neq i} {A_{a,j}\over s_a-s_i}   
\eeq
For higher $k$ we have the recursion:
\bea
&& \hbar\Big( (1-k)K_{i,k+1;j,k'} - 2\sum_{a\neq i}\sum_{l=1}^{k} {K_{i,k-l;j,k'}\over (s_a-s_i)^{l+1}} - {1\over \hbar} \sum_{l=1}^k {V^{(l+1)}(s_i)\over l!} K_{i,k-l;j,k'} \Big) \cr
&=& - \delta_{i,j}\, \delta_{k',k+1} 
- 2\delta_{k',2}\,\sum_{a\neq i} {A_{a,j}\over (s_a-s_i)^{k+1}}  
%+ \delta_{k,0}\delta_{k',1}\, c_j  \cr
\eea
In particular, it shows that if $k'>2$, then $K_{i,k;i,k'}$ is proportional to $\delta_{i,j}$.

\subsection*{Generating functions}

We introduce generating functions:
\beq
R_{i;j,k'}(x) = \sum_i K_{i,k;j,k'}\, (x-s_i)^k
\eeq
We have:
\bea
 \hbar\,\Big(  2{\psi'(x)\over \psi(x)} -\partial_x \Big)\, R_{i;j,k'}(x)  
= - \delta_{i,j} (x-s_i)^{k'-1} 
%+ \delta_{k',1} c_j 
+ 2\delta_{k',2}\,\sum_{a} {A_{a,j}\over x-s_a}  
\eea
i.e.
\bea
- \hbar \psi^2(x)\,\partial_x\Big(  { R_{i;j,k'}(x)  \over \psi^2(x)}\Big)
= - \delta_{i,j} (x-s_i)^{k'-1} + \delta_{k',1} c_j
+ 2\delta_{k',2}\,\sum_{a} {A_{a,j}\over x-s_a}  
\eea

In particular with $k'=1$ we find:
\beq\label{Rij1x}
R_{i;j,1}(x) = 
{\delta_{i,j}\over \hbar}\, \psi(x) \phi(x)
%{1\over \hbar}\,(\delta_{i,j}-c_j) \psi(x) \phi(x)
\eeq
where
\beq
\phi(x) = \psi(x)\int^x {dx'\over \psi(x')^2}
\virg
\phi'(x)\psi(x) - \psi'(x)\phi(x) = 1
\eeq

\section*{Appendix: Proof of theorem \ref{thpolessiWngAnn}}
\label{approofthpolessiWngAnn}

{\bf Theorem \ref{thpolessiWngAnn}}
{\it Each $W_n^{(g)}$ is a rational function of all its arguments. If $2g+n-2>0$, it has poles only at the $s_i$'s. In particular it has no poles at the $\alpha_i$'s, and it vanishes as $O(1/x_i)$ when $x_i\to\infty$.
}
\bigskip

{\bf proof:}

It is easy to check that $W_1^{(0)}$, $W_2^{(0)}$ satisfy the theorem. 

We will now make a recursion over $-\chi=2g-2+n$ to prove the result for every $(n,g)$.
We write:
\beq 
W^{(g)}_{n+1}(x_0,x_1,\dots,x_n)
=   \sum_i \Res_{x\to s_i}  K(x_0,x)\,  U^{(g)}_{n+1}(x,x_1,\dots,x_n)
\eeq
where $J=\{ x_1,\dots,x_n\}$, and
\beq\label{defUng1}
U^{(g)}_{n+1}(x,J)
=    \ovl{W}_{n+2}^{(g-1)}(x,x,J) + \sum_{h=0}^g \sum_{I\subset J} W_{|I|+1}^{(h)}(x,I) W_{n-|I|+1}^{(g-h)}(x,J/I)  
\eeq
First, the recursion hypothesis clearly implies that $U_{n+1}^{(g)}(x,x_1,\dots,x_n)$ is a rational fraction in all its variables $x,x_1,...x_n$.

Then we Taylor expand $K(x_0,x)$ as in \eq{TaylexpKikx0} or \eq{TaylexpKikjl}
\bea 
W^{(g)}_{n+1}(x_0,x_1,\dots,x_n)
&=&   \sum_i \Res_{x\to s_i}  K(x_0,x)\,  U^{(g)}_{n+1}(x,x_1,\dots,x_n) \cr
&=&   \sum_i \sum_k K_{i,k}(x_0)\,\, \Res_{x\to s_i}  (x-s_i)^k\,  U^{(g)}_{n+1}(x,x_1,\dots,x_n) \cr
\eea
Since $U_{n+1}^{(g)}(x,x_1,\dots,x_n)$ is a rational fraction of $x$, the sum over $k$ is finite, and
therefore, $W^{(g)}_{n+1}(x_0,x_1,\dots,x_n)$ is a finite sum of rational fractions of $x_0$, with poles at the $s_j$'s, therefore it is a rational fraction of $x_0$ with poles at the $s_j$'s.

It is also clear that $W^{(g)}_{n+1}(x_0,x_1,\dots,x_n)$ is a rational fraction of the other variables $x_1,\dots,x_n$.
 The poles in those variables are necessarily at the $s_j$'s, because as long as the residues can be computed, $W^{(g)}_{n+1}(x_0,x_1,\dots,x_n)$ is finite. The residue cannot be computed everytime an integration contour gets pinched, and since the integration contours are small circles around the $s_i$'s, the only singularities may occur at the $s_i$'s.

\medskip
It remains to prove that each $W_n^{(g)}$ behaves like $O(1/x_i)$ at $\infty$. The proof follows the same line: each $K_{i,k}(x_0)$ behaves like $O(1/x_0)$, and by an easy recursion the result holds for all other variables.
$\square$

\section*{Appendix: Proof of theorem \ref{thWngPngAnn}}
\label{approofthWngPngAnn}

In this subsection we prove theorem \ref{thWngPngAnn}, that all $W_n^{(g)}$'s satisfy
the loop equation.

\medskip
{\bf Theorem \ref{thWngPngAnn}}
{\it 
 The $W_n^{(g)}$'s satisfy the loop equation, i.e. the following quantity $P_{n+1}^{(g)}(x;x_1...,x_n)$
\bea\label{loopeqappgen}
 P_{n+1}^{(g)}(x;x_1...,x_n)
 &=&
-Y(x)\overline{W}_{n+1}^{(g)}(x,x_1,...,x_n) + \hbar \partial_{x}{\overline{W}_{n+1}^{(g)}(x,x_1...,x_n)} \cr
&& + \sum_{I\subset J} \ovl{W}_{|I|+1}^{(h)}(x,x_I) \ovl{W}_{n-|I|+1}^{(g-h)}(x,J/I) +
\ovl{W}_{n+2}^{(g-1)}(x,x,J)  \cr
& &+ \sum_{j}
\partial_{x_j} \left( {{\ovl{W}_n^{(g)}(x,J/\{j\})-{\ovl{W}_n^{(g)}(x_j,J/\{j\})}} \over {(x-x_j)}}\right) \cr
\eea
is a rational fraction of $x$ (possibly a polynomial), with no pole at $x=s_i$.
The only possible poles of $P_{n+1}^{(g)}(x;x_1...,x_n)$ are at the poles of $V'(x)$, and their degree is less than the degree of $V'$.
}
\bigskip

{\bf proof:}

First, from theorem \ref{thpolessiWngAnn}, we easily see that $P_{n+1}^{(g)}(x;x_1...,x_n)$ is indeed a rational function of $x$. 
Moreover it clearly has no pole at coinciding points $x=x_j$.

Then we write Cauchy's theorem for $W_{n+1}^{(g)}$:
\bea
W_{n+1}^{(g)}(x_0,...,x_n)
&=&\Res_{x\to x_0} {1 \over{x-x_0}}\, W_{n+1}^{(g)}(x,x_1,...,x_n)  \cr
&=&\Res_{x\to x_0}  G(x_0,x)\,W_{n+1}^{(g)}(x,x_1,...,x_n) 
\eea
and using again theorem \ref{thpolessiWngAnn}, i.e. that $W_{n+1}^{(g)}$ has poles only at the $s_i$'s, and that both $W_{n+1}^{(g)}$ and $G(x_0,x)$ behave as $O(1/x)$ for large $x$, we may move the integration contours:
\beq
W_{n+1}^{(g)}(x_0,...,x_n)
= -\sum_i \Res_{x\to s_i} G(x_0,x)\,\, W_{n+1}^{(g)}(x,x_1,...,x_n) 
\eeq
Then we use the definition of $K$, and integrate by parts:
 \bea
 W_{n+1}^{(g)}(x_0,...,x_n)   
&=&\sum_i \Res_{x\to s_i} (Y(x)K(x_0,x)+ \hbar K'(x_0,x))W_{n+1}^{(g)}(x,x_1,...,x_n)  \cr
&=& \sum_i \Res_{x\to s_i} K(x_0,x)\,
\Big(Y(x)W_{n+1}^{(g)}(x,x_1,...,x_n)\cr
&-& \hbar \partial_x
W_{n+1}^{(g)}(x,x_1,...,x_n) \Big) \cr
\eea 
From the definition we have also 
\bea 
&& W_{n+1}^{(g)}(x_0,...,x_n) \cr &=& \sum_i
\Res_{x\to s_i} K(x_0,x) \left(\sum_{h=0}^g \sum_{I\subset J}
W_{|I|+1}^{(h)}(x,I) W_{n-|I|+1}^{(g-h)}(x,J/I)
+\ovl{W}_{n+2}^{(g-1)}(x,x,J)\right) \cr
\eea 
then we shift $W_n^{(g)}$ to $\ovl{W}_n^{(g)}$ in the RHS, i.e.: 
\bea &&
W_{n+1}^{(g)}(x_0,...,x_n) \cr &=& \sum_i \Res_{x\to s_i} K(x_0,x)
\Big(\sum_{h=0}^g \sum_{I\subset J} \ovl{W}_{|I|+1}^{(h)}(x,I)
\ovl{W}_{n-|I|+1}^{(g-h)}(x,J/I) + \ovl{W}_{n+2}^{(g-1)}(x,x,J) \cr
&& + \sum_{j=1}^n  {\ovl{W}_n^{(g)}(x,J/\{j\}) \over (x-x_j)^2}
\Big) \cr &=& \sum_i \Res_{x\to s_i} K(x_0,x) \Big(\sum_{h=0}^g
\sum_{I\subset J} \ovl{W}_{|I|+1}^{(h)}(x,I)
\ovl{W}_{n-|I|+1}^{(g-h)}(x,J/I) + \ovl{W}_{n+2}^{(g-1)}(x,x,J) \cr
&& + \sum_{j=1}^n  \partial_{x_j}\,
\left({\ovl{W}_n^{(g)}(x,J/\{j\}) \over x-x_j}\right) \Big) \cr &=&
\sum_i \Res_{x\to s_i} K(x_0,x) \Big(\sum_{h=0}^g \sum_{I\subset J}
\ovl{W}_{|I|+1}^{(h)}(x,I) \ovl{W}_{n-|I|+1}^{(g-h)}(x,J/I) +
\ovl{W}_{n+2}^{(g-1)}(x,x,J) \cr && + \sum_{j=1}^n  \partial_{x_j}\,
\left({\ovl{W}_n^{(g)}(x,J/\{j\})-\ovl{W}_n^{(g)}(x_j,J/\{j\}) \over
x-x_j}\right) \Big) \cr 
\eea 
in the last line we have added for free, the term $\ovl{W}_n^{(g)}(x_j,J/\{j\})$ because it has no pole at $x=s_i$.

Therefore we have:
\bea
%&& c(x_0)  \Res_{x\to \infty} W_{n+1}^{(g)}(x,x_1,...,x_n) \cr
0 &=& \sum_i \Res_{x\to s_i} K(x_0,x)
\Big( -Y(x)W_{n+1}^{(g)}(x,x_1,...,x_n)+ \hbar \partial_x W_{n+1}^{(g)}(x,x_1,...,x_n) \cr
&& + \sum_{h=0}^g \sum_{I\subset J} \ovl{W}_{|I|+1}^{(h)}(x,I)
\ovl{W}_{n-|I|+1}^{(g-h)}(x,J/I) + \ovl{W}_{n+2}^{(g-1)}(x,x,J) \cr
&& + \sum_{j=1}^n  \partial_{x_j}\, \left({\ovl{W}_n^{(g)}(x,J/\{j\})-\ovl{W}_n^{(g)}(x_j,J/\{j\}) \over x-x_j}\right) \Big) \cr
&=& \sum_i \Res_{x\to s_i} K(x_0,x) P_{n+1}^{(g)}(x;x_1,...,x_n) \cr
&=& \sum_i \sum_k K_{i,k}(x_0) \,\, \Res_{x\to s_i} (x-s_i)^k P_{n+1}^{(g)}(x;x_1,...,x_n) \cr
\eea
Notice that this equation holds for any $x_0$.
Since $K_{i,k}(x_0)$ is a rational fraction with a pole of degree $k+1$ in $x_0=s_i$, the $K_{i,k}(x_0)$ are linearly independent functions, and thus we must have:
\beq
\forall k,i \qquad 
0=\Res_{x\to s_i}  (x-s_i)^k\, P_{n+1}^{(g)}(x;x_1,...,x_n) 
\eeq
this means that $P_{n+1}^{(g)}$ has no pole at $x=s_i$.

One easily sees that $P_{n+1}^{(g)}(x;x_1,\dots,x_n)$ is a rational fraction of $x$, and its poles are at most those of $Y(x)$, i.e. at the poles of $V'(x)$.
$\square$

\section*{Appendix: Proof of theorem \ref{thsymAnn}}
\label{approofthsymAnn}

{\bf Theorem \ref{thsymAnn}}
{\it 
Each $W_n^{(g)}$ is a symmetric function of all its arguments.
}
\bigskip

{\bf proof:}

The special case of $W_3^{(0)}$ is proved in appendix \ref{approofthW3KrichAnn} above.
It is obvious from the definition that $W_{n+1}^{(g)}(x_0,x_1,\dots,x_n)$ is
symmetric in $x_1,x_2,\dots,x_n$, and therefore we need to show that (for $n\geq 1$):
\beq
W_{n+1}^{(g)}(x_0,x_1,J)-W_{n+1}^{(g)}(x_1,x_0,J)=0
\eeq
where $J=\{ x_2,\dots,x_{n}\}$.
We prove it by recursion on $-\chi=2g-2+n$. 

Assume that every $W_k^{(h)}$ with $2h+k-2\leq 2g+n$ is symmetric.
We have:
\bea
&& W_{n+1}^{(g)}(x_0,x_1,J) \cr
&=&\sum_i \Res_{x\to s_i} K(x_0,x)\,\, \Big(
W_{n+2}^{(g-1)}(x,x,x_1,J) + 2 \,\,\, B(x,x_1) W_{n}^{(g)}(x,J) \cr
&& + 2 \sum_{h=0}^g\sum'_{I\in J}\,\,\, W_{2+|I|}^{(h)}(x,x_1,I) W_{n-|I|}^{(g-h)}(x,J/I) \Big) \cr
\eea
where $\sum'$ means that we exclude the terms $(I=\emptyset, h=0)$ and $(I=J, h=g)$. Notice also that $\ovl{W}_{n+2}^{(g-1)}=W_{n+2}^{(g-1)}$ because $n\geq 1$.
Then, using the recursion hypothesis, we have:
\bea
&& W_{n+1}^{(g)}(x_0,x_1,J) \cr
&=& 2 \sum_i \Res_{x\to s_i} K(x_0,x)\,\,  B(x,x_1) W_{n}^{(g)}(x,J) \cr
&& + \sum_{i,j} \Res_{x\to s_i} \Res_{x'\to s_j}\,\, K(x_0,x) K(x_1,x')\,\, 
\Big(  W_{n+3}^{(g-2)}(x,x,x',x',J) \cr
&& + 2\sum_h\sum'_{I} W_{2+|I|}^{(h)}(x',x,I) W_{1+n-|I|}^{(g-1-h)}(x',x,J/I) \cr
&& + 2 \sum_h\sum'_{I} W_{3+|I|}^{(h)}(x',x,x,I) W_{n-|I|}^{(g-1-h)}(x',J/I) \cr
&& + 2 \sum_{h}\sum'_{I\in J}\,\,\, W_{n-|I|}^{(g-h)}(x,J/I)
\Big[ W_{3+|I|}^{(h-1)}(x,x',x',I)  \cr
&& + 2 \sum_{h'}\sum'_{I'\subset I} W_{2+|I'|}^{(h')}(x',x,I')   W_{1+|I|-|I'|}^{(h-h')}(x',I/I')  
\Big]\,
\Big) \cr
\eea
Now, if we compute $W_{n+1}^{(g)}(x_1,x_0,J)$, we get the same expression, with the order of integrations exchanged, i.e. we have to integrate $x'$ before integrating $x$.
Notice, by moving the integration contours,  that:
\beq
\Res_{x\to s_i} \Res_{x'\to s_j} - \Res_{x'\to s_j} \Res_{x\to s_i} =
- \delta_{i,j}\Res_{x\to s_i} \Res_{x'\to x} 
\eeq
Moreover, the only terms which have a pole at $x=x'$ are those containing $B(x,x')$.
Therefore:
\bea
&& W_{n+1}^{(g)}(x_0,x_1,J)-W_{n+1}^{(g)}(x_1,x_0,J) \cr
&=& 2 \sum_i \Res_{x\to s_i} \left( K(x_0,x)\,\,  B(x,x_1)  - K(x_1,x)\,\,  B(x,x_0) \right) \, W_{n}^{(g)}(x,J) \cr
&& - 2 \sum_{i} \Res_{x\to s_i} \Res_{x'\to x}\,\, K(x_0,x) K(x_1,x')\,\, B(x,x')\,\,
\Big(     \cr
&& 2W_{1+n}^{(g-1)}(x',x,J)  + 2 \sum_{h}\sum'_{I\in J}\,\, W_{n-|I|}^{(g-h)}(x,J/I)     W_{1+|I|}^{(h)}(x',I)  
\Big) \cr
\eea
The residue $\Res_{x'\to x}$ can be computed:
\bea
&& W_{n+1}^{(g)}(x_0,x_1,J)-W_{n+1}^{(g)}(x_1,x_0,J) \cr
&=& 2 \sum_i \Res_{x\to s_i} \left( K(x_0,x)\,\,  B(x,x_1)  - K(x_1,x)\,\,  B(x,x_0) \right) \, W_{n}^{(g)}(x,J) \cr
&& -  \sum_{i} \Res_{x\to s_i} \,\, K(x_0,x) {\partial \over \partial x'}\Big( K(x_1,x')\,\, 
\Big(     \cr
&& 2W_{1+n}^{(g-1)}(x',x,J)  + 2 \sum_{h}\sum'_{I\in J}\,\, W_{n-|I|}^{(g-h)}(x,J/I)     W_{1+|I|}^{(h)}(x',I)  
\Big)\,\, \Big)_{x'=x} \cr
&=& 2 \sum_i \Res_{x\to s_i} \left( K(x_0,x)\,\,  B(x,x_1)  - K(x_1,x)\,\,  B(x,x_0) \right) \, W_{n}^{(g)}(x,J) \cr
&& -  \sum_{i} \Res_{x\to s_i}\,\, K(x_0,x) K'(x_1,x)\,\, 
\Big(     \cr
&& 2W_{1+n}^{(g-1)}(x,x,J)  + 2 \sum_{h}\sum'_{I\in J}\,\, W_{n-|I|}^{(g-h)}(x,J/I)     W_{1+|I|}^{(h)}(x,I)  
\Big)\,\,  \cr
&& -  \sum_{i} \Res_{x\to s_i} \,\, K(x_0,x) K(x_1,x) {\partial \over \partial x'}\Big(      \cr
&& 2W_{1+n}^{(g-1)}(x',x,J)  + 2 \sum_{h}\sum'_{I\in J}\,\, W_{n-|I|}^{(g-h)}(x,J/I)     W_{1+|I|}^{(h)}(x',I)  
\,\, \Big)_{x'=x} \cr
&=& 2 \sum_i \Res_{x\to s_i} \left( K(x_0,x)\,\,  B(x,x_1)  - K(x_1,x)\,\,  B(x,x_0) \right) \, W_{n}^{(g)}(x,J) \cr
&& -  \sum_{i} \Res_{x\to s_i} \,\, K(x_0,x) K'(x_1,x)\,\, 
\Big(     \cr
&& 2W_{1+n}^{(g-1)}(x,x,J)  + 2 \sum_{h}\sum'_{I\in J}\,\, W_{n-|I|}^{(g-h)}(x,J/I)     W_{1+|I|}^{(h)}(x,I)  
\Big)\,\,  \cr
&& -  {1\over 2} \sum_{i} \Res_{x\to s_i} \,\, K(x_0,x) K(x_1,x) {\partial \over \partial x}\Big(      \cr
&& 2W_{1+n}^{(g-1)}(x,x,J)  + 2 \sum_{h}\sum'_{I\in J}\,\, W_{n-|I|}^{(g-h)}(x,J/I)     W_{1+|I|}^{(h)}(x,I)  
\,\, \Big) \cr
\eea
The last term can be integrated by parts, and we get:
\bea
&& W_{n+1}^{(g)}(x_0,x_1,J)-W_{n+1}^{(g)}(x_1,x_0,J) \cr
&=& 2 \sum_i \Res_{x\to s_i} \left( K(x_0,x)\,\,  B(x,x_1)  - K(x_1,x)\,\,  B(x,x_0) \right) \, W_{n}^{(g)}(x,J) \cr
&& +{1\over 2}  \sum_{i} \Res_{x\to s_i} \,\, \Big( K'(x_0,x) K(x_1,x)-K(x_0,x) K'(x_1,x)\Big)\,\, 
\Big(     \cr
&& 2W_{1+n}^{(g-1)}(x,x,J)  + 2 \sum_{h}\sum'_{I\in J}\,\, W_{n-|I|}^{(g-h)}(x,J/I)     W_{1+|I|}^{(h)}(x,I)  
\Big)\,\,  \cr
\eea
Then we use theorem \ref{thWngPngAnn}:
\bea
&& W_{n+1}^{(g)}(x_0,x_1,J)-W_{n+1}^{(g)}(x_1,x_0,J) \cr
&=& 2 \sum_i \Res_{x\to s_i} \left( K(x_0,x)\,\,  B(x,x_1)  - K(x_1,x)\,\,  B(x,x_0) \right) \, W_{n}^{(g)}(x,J) \cr
&& + \sum_{i} \Res_{x\to s_i} \,\, \Big( K'(x_0,x) K(x_1,x)-K(x_0,x) K'(x_1,x)\Big)\,\, 
\Big(      P_{n}^{(g)}(x,J) \cr
&& + (Y(x) - \hbar \partial_x) W_{n}^{(g)}(x,J)  + \sum_j \partial_{x_j}
\Big( {W_{n-1}^{(g)}(x_j,J/\{x_j\})\over x-x_j} \Big)
\Big)\,\,  \cr
\eea
Since $P_{n}^{(g)}(x,J)$ and $W_{n-1}^{(g)}(x_j,J/\{x_j\})$ have no poles at the $s_i$'s, we have:
\bea
&& W_{n+1}^{(g)}(x_0,x_1,J)-W_{n+1}^{(g)}(x_1,x_0,J) \cr
&=& 2 \sum_i \Res_{x\to s_i} \left( K(x_0,x)\,\,  B(x,x_1)  - K(x_1,x)\,\,  B(x,x_0) \right) \, W_{n}^{(g)}(x,J) \cr
&+& \sum_{i} \Res_{x\to s_i} \,\, \Big( K'(x_0,x) K(x_1,x)-K(x_0,x) K'(x_1,x)\Big)\cr
&&(Y(x) - \hbar \partial_x) W_{n}^{(g)}(x,J)    \cr
\eea
Notice that:
\beq
K'_0 K_1 - K_0 K'_1 = -{1\over \hbar}\,( G_0 K_1 - K_0 G_1)
\eeq
and $B=-{1\over 2}\, G'$, therefore:
\bea
&& W_{n+1}^{(g)}(x_0,x_1,J)-W_{n+1}^{(g)}(x_1,x_0,J) \cr
&=& - \sum_i \Res_{x\to s_i} \left( K_0 G'_1  - K_1 G'_0 \right) \, W_{n}^{(g)}(x,J) \cr
&& -{1\over \hbar}  \sum_{i} \Res_{x\to s_i} \,\, \Big( G_0 K_1- K_0 G_1 \Big)\,\, 
(Y(x) - \hbar \partial_x) W_{n}^{(g)}(x,J)    \cr
\eea
we integrate the first line by parts:
\bea
&& W_{n+1}^{(g)}(x_0,x_1,J)-W_{n+1}^{(g)}(x_1,x_0,J) \cr
&=& \sum_i \Res_{x\to s_i} \left( K'_0 G_1  - K'_1 G_0 \right) \, W_{n}^{(g)}(x,J) \cr
&& + \sum_i \Res_{x\to s_i} \left( K_0 G_1  - K_1 G_0 \right) \, W_{n}^{(g)}(x,J)' \cr
&& -{1\over \hbar}  \sum_{i} \Res_{x\to s_i} \,\, \Big( G_0 K_1- K_0 G_1 \Big)\,\, 
(Y(x) - \hbar \partial_x) W_{n}^{(g)}(x,J)    \cr
\eea
Notice that:
\beq
K'_0 G_1 - G_0 K'_1 = -{Y\over \hbar}\,( K_0 G_1 - G_0 K_1)
\eeq
So we find \beq  W_{n+1}^{(g)}(x_0,x_1,J)-W_{n+1}^{(g)}(x_1,x_0,J)=0\eeq

\section*{Appendix: Proof of theorem \ref{thWngindeptK}}
\label{approofthWngindeptK}

\bt
The correlation functions $W_n^{(g)}$ are independent of the choice of kernel $K$, provided that $K$ is solution of the equation \eq{diffeqdefK}.
\et

\proof{

Any two solutions of  \eq{diffeqdefK}, differ by a homogeneous solution, i.e. by $\psi^2(x)$.
Therefore, what we have to prove is that the following quantity vanishes:
\beq
\sum_i \Res_{x\to s_i} \psi^2(x)\, \Big[ W_{n+2}^{(g-1)}(x,x,J) + \sum_h \sum'_{I\subset J}\, W_{1+|I|}^{(h)}(x,I) W_{1+n-|I|}^{(g-h)}(x,J/I) \Big]
\eeq
Using theorem \ref{thWngPngAnn}, we have:
\bea
&& \Res_{x\to s_i} \psi^2(x)\, \Big[ W_{n+2}^{(g-1)}(x,x,J) + \sum_h \sum'_{I\subset J}\, W_{1+|I|}^{(h)}(x,I) W_{1+n-|I|}^{(g-h)}(x,J/I) \Big] \cr
&=& \Res_{x\to s_i} \psi^2(x) \Big(Y(x)W_n^{(g)}(x,J) - \hbar \partial_x W_n^{(g)}(x,J)+P_n^{(g)}(x;J) \Big) 
\eea
Then we notice that $P_n^{(g)}$ gives no residue, and then we use $Y=-2\hbar \psi'/\psi$, and we integrate by parts:
\bea
&=& -\hbar \Res_{x\to s_i} \psi^2(x) \Big(2\,{\psi'\over \psi}\,W_n^{(g)} + \partial_x W_n^{(g)} \Big) \cr
&=& -\hbar \Res_{x\to s_i} \partial_x\, \Big(\psi^2\,\,  W_n^{(g)} \Big) \cr
&=& 0
\eea
This means that adding to $K(x_0,x)$ a constant times $\psi^2(x)$ doesnot change the $W_n^{(g)}$'s.
In fact we may chose a different constant near each $s_i$, or in other words, we may assume that
\beq\label{Ki2vanishhyp}
K_{i,2}(x_0)=0
\eeq

}

\section*{Appendix: Proof of theorem \ref{thW3KrichAnn}}
\label{approofthW3KrichAnn}

{\bf Theorem \ref{thpolessiWngAnn}}
{\it 
The 3 point function $W_3^{(0)}$ is symmetric and we have:
\beq
W_3^{(0)}(x_1,x_2,x_3) = 4\,\sum_i \Res_{x\to s_i}\,\, {B(x,x_1)B(x,x_2)B(x,x_3)\over Y'(x)}
\eeq
}
\bigskip

{\bf proof:}

The definition of $W_3^{(0)}$ is:
\bea  
&& W_3^{(0)}(x_0,x_1,x_2)\cr 
&=& 2 \sum_i \Res_{x\to s_i} K(x_0,x)B(x,x_1)B(x,x_2) \cr
 &=& {1 \over 2} \sum_i \Res_{x\to s_i}  K_0 \, G_1^{'} \, G'_2 \cr
 &=& {1 \over 2} \sum_i \Res_{x\to s_i} K_0 \left( (\hbar K''_1 +  Y K'_1 +  Y'
 K_1)(\hbar K''_2 + YK'_2 +Y' K_2) \right) \cr
 &=& {1 \over 2} \sum_i \Res_{x\to s_i} K_0 \, (\, \hbar^2 K''_1 K''_2 +  \hbar Y (K'_1
 K''_2+K''_1 K'_2) +  \hbar Y' (K''_1 K_2 +K''_2 K_1) \cr
 && +  Y^2 K'_1 K'_2+  Y Y' (K_1 K'_2+K'_1 K_2)+
{Y'}^2 K_1 K_2 \,) \cr
 \eea
where we have written for short $K_i = K(x_i,x)$, $G_i=G(x_i,x)$, and derivative are w.r.t. $x$.

Since $K(x_i,x)$ has no pole when $x\to s_i$,
the first term vanishes.
Using the Ricatti equation $Y^2 = 2 \hbar Y' + U$ (where $U$ has no pole at $s_i$), we may replace $Y^2$ by $2 \hbar Y'$ and $ Y Y'$ by $\hbar Y''$ without changing the residues, i.e.:
\bea
& & W_3^{(0)}(x_0,x_1,x_2)\cr
 &=& {1 \over 2} \sum_i \Res_{x\to s_i} K_0 \, (  \hbar Y (K'_1
 K''_2+K''_1 K'_2) +  \hbar Y' (K''_1 K_2 +K''_2 K_1) \cr
 && + 2 \hbar Y' K'_1 K'_2+ \hbar Y'' (K_1 K'_2+K'_1 K_2)+
{Y'}^2 K_1 K_2 \,)\ \cr
 &=&  {1\over 2} \sum_i \Res_{x\to s_i} K_0 \, (  \hbar Y (K'_1 K'_2)' +  \hbar Y' (K_1 K_2)'' + \hbar Y'' (K_1 K_2)'+  {Y'}^2 K_1 K_2 \,)\ \cr
 &=&  {1\over 2} \sum_i \Res_{x\to s_i} {Y'}^2 K_0 K_1 K_2 +
 \hbar \big( Y'' K_0 (K_1 K_2)' - (Y K_0)' K'_1 K'_2 - (Y' K_0)' (K_1 K_2)' \big) \cr
 &=&  {1\over 2} \sum_i \Res_{x\to s_i} {Y'}^2 K_0 K_1 K_2 -
 \hbar \big(  (Y K_0)' K'_1 K'_2 + Y' K_0' (K_1 K_2)' \big) \cr
&=&  {1\over 2} \sum_i \Res_{x\to s_i} {Y'}^2 K_0 K_1 K_2 - \hbar Y K'_0 K'_1 K'_2 - \hbar Y' (K_0 K'_1 K'_2+K'_0 K_1 K'_2+K'_0 K'_1 K_2) \cr
\eea
This expression is clearly symmetric in $x_0, x_1, x_2$ as claimed in theorem \ref{thsymAnn}.

Let us give an alternative expression, in the form of the Verlinde or Krichever formula \cite{Krich}:
\beq\label{eq30KricheverAnn}
W_3^{(0)}(x_1,x_2,x_3)=
4 \sum_i \Res_{x\to s_i} \, {B(x,x_1)B(x,x_2)B(x,x_3) \over Y^{'}(x)}
\eeq

{\bf proof:}

In order to prove formula \ref{eq30KricheverAnn}, compute:
\beq
B(x,x_i) = -{1\over 2} G'(x,x_i) = -{1\over 2} G'_i = {1\over 2}(\hbar K''_i +  Y K'_i +  Y' K_i)
\eeq
thus:
\bea
&&  \sum_i \Res_{x\to s_i} \, {B(x,x_1)B(x,x_2)B(x,x_3) \over Y^{'}(x)} \cr
&=&  {1\over 8}\sum_i \Res_{x\to s_i} \, {1\over Y'(x)}\, (\hbar K''_0 +  Y K'_0 +  Y' K_0)(\hbar K''_1 +  Y K'_1 +  Y' K_1)\cr
&& \qquad \qquad (\hbar K''_2 +  Y K'_2 +  Y' K_2)  \cr
&=&  {1\over 8}\sum_i \Res_{x\to s_i} \,
 {\hbar^3\over Y'} K''_0 K''_1 K''_2 + \hbar^2 {Y\over Y'} (K'_0 K''_1 K''_2 + K''_0 K'_1 K''_2 + K''_0 K''_1 K'_2) \cr
&& + \hbar^2  (K_0 K''_1 K''_2 + K''_0 K_1 K''_2 + K''_0 K''_1 K_2) \cr
&& + \hbar {Y^2\over Y'} (K''_0 K'_1 K'_2+K'_0 K''_1 K'_2+K'_0 K'_1 K''_2) \cr
&& + \hbar Y (K_0 K'_1 K''_2 + K_0 K''_1 K'_2 + K'_0 K_1 K''_2 + K'_0 K''_1 K_2 + K''_0 K_1 K'_2 + K''_0 K'_1 K_2) \cr
&& + \hbar Y' (K''_0 K_1 K_2 + K_0 K''_1 K_2 + K_0 K_1 K''_2)
+ {Y^3\over Y'} K'_0 K'_1 K'_2  \cr
&& + Y^2  (K_0 K'_1 K'_2 + K'_0 K_1 K'_2 + K'_0 K'_1 K_2) \cr
&& + Y Y' (K'_0 K_1 K_2 + K_0 K'_1 K_2 + K_0 K_1 K'_2)  + Y'^2 K_0 K_1 K_2  \cr
\eea
Notice that $K_i$ has no pole at the $s_i$'s, and $1/Y'$ has no pole, $Y/Y'$ has no pole, $Y^2/Y'$ has no pole, thus:
\bea
&&  \sum_i \Res_{x\to s_i} \, {B(x,x_1)B(x,x_2)B(x,x_3) \over Y^{'}(x)} \cr
&=&  {1\over 8}\sum_i \Res_{x\to s_i} \,
 \hbar Y (K_0 K'_1 K''_2 + K_0 K''_1 K'_2 + K'_0 K_1 K''_2 + K'_0 K''_1 K_2 + K''_0 K_1 K'_2 \cr
 && + K''_0 K'_1 K_2) 
 + \hbar Y' (K''_0 K_1 K_2 + K_0 K''_1 K_2 + K_0 K_1 K''_2)
+ {Y^3\over Y'} K'_0 K'_1 K'_2  \cr
&& + Y^2  (K_0 K'_1 K'_2 + K'_0 K_1 K'_2 + K'_0 K'_1 K_2) \cr
&& + Y Y' (K'_0 K_1 K_2 + K_0 K'_1 K_2 + K_0 K_1 K'_2)  + Y'^2 K_0 K_1 K_2  \cr
\eea
Notice that $Y^2 = 2\hbar Y' + U$, thus we may replace $Y^3/Y'$ by $2\hbar Y$, and $Y^2$ by $2\hbar Y'$ and $Y Y'$ by $\hbar Y''$, thus:
\bea
&&  \sum_i \Res_{x\to s_i} \, {B(x,x_1)B(x,x_2)B(x,x_3) \over Y^{'}(x)} \cr
&=&  {1\over 8}\sum_i \Res_{x\to s_i} \,
 \hbar Y (K_0 K'_1 K''_2 + K_0 K''_1 K'_2 + K'_0 K_1 K''_2 + K'_0 K''_1 K_2 + K''_0 K_1 K'_2 \cr
 && + K''_0 K'_1 K_2) 
 + \hbar Y' (K''_0 K_1 K_2 + K_0 K''_1 K_2 + K_0 K_1 K''_2)
+ 2\hbar Y K'_0 K'_1 K'_2  \cr
&& + 2 \hbar Y'  (K_0 K'_1 K'_2 + K'_0 K_1 K'_2 + K'_0 K'_1 K_2)
+ \hbar Y'' (K'_0 K_1 K_2 + K_0 K'_1 K_2 + K_0 K_1 K'_2) \cr
&& + Y'^2 K_0 K_1 K_2  \cr
&=&  {1\over 8}\sum_i \Res_{x\to s_i} \,
 \hbar Y (K_0 (K'_1 K'_2)' +  K_1 (K'_0 K'_2)' + K_2 (K'_0 K'_1)') \cr
&& + 2\hbar Y K'_0 K'_1 K'_2  + Y'^2 K_0 K_1 K_2  + \hbar (Y' (K'_0 K_1 K_2 + K_0 K'_1 K_2 + K_0 K_1 K'_2))' \cr
&=&  {1\over 8}\sum_i \Res_{x\to s_i} \,
 \hbar Y (K_0 (K'_1 K'_2)' +  K_1 (K'_0 K'_2)' + K_2 (K'_0 K'_1)') \cr
&& + 2\hbar Y K'_0 K'_1 K'_2  + Y'^2 K_0 K_1 K_2  \cr
&=&  -{1\over 8}\sum_i \Res_{x\to s_i} \, 3 \hbar Y K'_0 K'_1 K'_2 + \hbar Y' (K_0 K'_1 K'_2 + K'_0 K_1 K'_2 + K'_0 K'_1 K_2) \cr
&& - 2\hbar Y K'_0 K'_1 K'_2  - Y'^2 K_0 K_1 K_2  \cr
&=& {1\over 4} W_3^{(0)}(x_0,x_1,x_2)
\eea

\subsection*{Direct computation}

We write 
\bea
&& W_3^{(0)}(z_1,z_2,z_3) \cr
&=& 2 \sum_i \Res_{z\to s_i} K(z_1,z)B(z_2,z)B(z_3,z) \cr
&=& \sum_j\sum_i {A_{i,j}\over (z_2-s_j)^2}\, \Res_{z\to s_i} K(z_1,z){1\over (z-s_i)^2(z_3-z)^2} \,\,+{\rm sym.} \cr
&& + 2 \sum_i\sum_{i'\neq i} \sum_{j,k} {A_{i,j}A_{i',k}\over (z_2-s_j)^2(z_3-s_k)^2}\, \Res_{z\to s_i} K(z_1,z) {1\over (z-s_i)^2(z-s_{i'})^2}  \,\,+{\rm sym.}  \cr
&& + 2 \sum_i \sum_{j,k} {A_{i,j}A_{i,k}\over (z_2-s_j)^2(z_3-s_k)^2}\, \Res_{z\to s_i} K(z_1,z) {1\over (z-s_i)^4}    \cr
&=& \sum_j\sum_i {A_{i,j}\over (z_2-s_j)^2}\, \Big( {K_{i,1}(z_1)\over (z_3-s_i)^2}+{2 K_{i,0}(z_1)\over (z_3-s_i)^3}  \Big)\,\,+{\rm sym.} \cr
&& + 2 \sum_i\sum_{i'\neq i} \sum_{j,k} {A_{i,j}A_{i',k}\over (z_2-s_j)^2(z_3-s_k)^2}\, \Big( {K_{i,1}(z_1)\over (s_{i'}-s_i)^2}+{2 K_{i,0}(z_1)\over (s_{i'}-s_i)^3}  \Big) \,\,+{\rm sym.}  \cr
&& + 2 \sum_i \sum_{j,k} {A_{i,j}A_{i,k}\over (z_2-s_j)^2(z_3-s_k)^2}\, K_{i,3}(z_1)   \cr
&=& \sum_j\sum_i {A_{i,j}\over (z_2-s_j)^2}\, \Big( {K_{i,1}(z_1)\over (z_3-s_i)^2}+{2 K_{i,0}(z_1)\over (z_3-s_i)^3}  \Big)\,\,+{\rm sym.} \cr
&& + 2 \sum_i\sum_{i'\neq i} \sum_{j,k} {A_{i,j}A_{i',k}\over (z_2-s_j)^2(z_3-s_k)^2}\, \Big( {K_{i,1}(z_1)\over (s_{i'}-s_i)^2}+{2 K_{i,0}(z_1)\over (s_{i'}-s_i)^3}  \Big) \,\,+{\rm sym.}  \cr
&& - 2 \sum_i \sum_{j,k} {A_{i,j}A_{i,k}\over (z_2-s_j)^2(z_3-s_k)^2}\, T_{i,i}\, K_{i,1}(z_1)   \cr
&& - 2 \sum_i \sum_{j,k} {A_{i,j}A_{i,k}\over (z_2-s_j)^2(z_3-s_k)^2}\, ({V'''(s_i)\over 2\hbar}+2\sum_{i'\neq i} {1\over (s_{i'}-s_i)^3}) K_{i,0}(z_1)   \cr
&& + {2\over \hbar} \sum_i \sum_{j,k} {A_{i,j}A_{i,k}\over (z_2-s_j)^2(z_3-s_k)^2(z_1-s_i)^3}   \cr
&& + {4\over \hbar} \sum_i\sum_{i'\neq i}\sum_l \sum_{j,k} {A_{i,j}A_{i,k} A_{i',l}\over (z_2-s_j)^2(z_3-s_k)^2(s_{i'}-s_i)^3(z_1-s_l)^2}\,   \cr
&=& {2\over \hbar} \sum_{i,j,k} {A_{i,j}A_{i,k} \over (z_1-s_i)^3(z_2-s_j)^2(z_3-s_k)^2} + {A_{j,i}A_{j,k} \over (z_1-s_i)^2(z_2-s_j)^3(z_3-s_k)^2} \cr
&&+ {A_{k,i}A_{k,j} \over (z_1-s_i)^2(z_2-s_j)^2(z_3-s_k)^3}
 \cr
&& + \sum_{i,j,k} {K_{i,1}(z_1)\over (z_2-s_j)^2(z_3-s_k)^2}\,\Big( A_{j,k}\delta_{i,j}+A_{j,k}\delta_{i,k}  - A_{i,j} \sum_{i'} T_{i,i'} A_{i',k} \cr
&& - A_{i,k} \sum_{i'} T_{i,i'} A_{i',j}  \Big) \cr
&& + 2 \sum_i\sum_{i'\neq i} \sum_{j,k} {A_{i,j}A_{i',k}\over (z_2-s_j)^2(z_3-s_k)^2}\, {2 K_{i,0}(z_1)\over (s_{i'}-s_i)^3}   \,\,+{\rm sym.}  \cr
&& - 2 \sum_i \sum_{j,k} {A_{i,j}A_{i,k}\over (z_2-s_j)^2(z_3-s_k)^2}\, ({V'''(s_i)\over 2\hbar}+2\sum_{i'\neq i} {1\over (s_{i'}-s_i)^3}) K_{i,0}(z_1)   \cr
&& + {4\over \hbar} \sum_i\sum_{i'\neq i}\sum_l \sum_{j,k} {A_{i,j}A_{i,k} A_{i',l}\over (z_2-s_j)^2(z_3-s_k)^2(s_{i'}-s_i)^3(z_1-s_l)^2}\,   \cr
&=& {2\over \hbar} \sum_{l,j,k} {1 \over (z_1-s_l)^2(z_2-s_j)^2(z_3-s_k)^2} \sum_i \Big( {\delta_{i,l} A_{i,j} A_{i,k}\over (z_1-s_i)}+{\delta_{i,j} A_{i,l} A_{i,k}\over (z_2-s_i)}\cr
&& +{\delta_{i,k} A_{i,l} A_{i,j}\over (z_3-s_i)}  \Big) \cr
 && + {4\over \hbar}\sum_{l,j,k}\sum_i \sum_{i'\neq i} {A_{i,j}A_{i,k} A_{i',l} +  A_{i,j}A_{i',k} A_{i,l}  +  A_{i,k}A_{i',j} A_{i,l} -  A_{i,j}A_{i,k}A_{i,l}\over (z_1-s_l)^2(z_2-s_j)^2(z_3-s_k)^2(s_{i'}-s_i)^3}\,\cr
 && - {1\over \hbar^2}\sum_{l,j,k}\sum_i  {A_{i,j}A_{i,k}A_{i,l} V'''(s_i)\over (z_1-s_l)^2(z_2-s_j)^2(z_3-s_k)^2}\,\cr%
\eea

Thus we have:
\bea\label{W30exact}
&& W_3^{(0)}(z_1,z_2,z_3) \cr
&=& {2\over \hbar} \sum_{i,j,k,l} { {\delta_{i,l} A_{i,j} A_{i,k}\over (z_1-s_i)}+{\delta_{i,j} A_{i,l} A_{i,k}\over (z_2-s_i)}+{\delta_{i,k} A_{i,l} A_{i,j}\over (z_3-s_i)}  \over (z_1-s_l)^2(z_2-s_j)^2(z_3-s_k)^2}  \cr
 && + {4\over \hbar}\sum_{l,j,k}\sum_i \sum_{i'\neq i} {A_{i,j}A_{i,k} A_{i',l} +  A_{i,j}A_{i',k} A_{i,l}  +  A_{i,k}A_{i',j} A_{i,l} -  A_{i,j}A_{i,k}A_{i,l}\over (z_1-s_l)^2(z_2-s_j)^2(z_3-s_k)^2(s_{i'}-s_i)^3}\,\cr
 && - {1\over \hbar^2}\sum_{l,j,k}\sum_i  {A_{i,j}A_{i,k}A_{i,l} V'''(s_i)\over (z_1-s_l)^2(z_2-s_j)^2(z_3-s_k)^2}\,\cr%
\eea

\section*{Appendix: Proof of theorem \ref{thvariationV}}
\label{approofthvariationV}

{\bf Theorem \ref{thvariationV}}
{\it 
Under an infinitesimal variation of the potential $V\to V+\delta V$, we have:
\beq
\forall n\geq 0, g\geq 0
\,\, , \quad
\delta W_{n}^{(g)}(x_1,\dots,x_n) = - \sum_i \Res_{x\to s_i} W_{n+1}^{(g)}(x,x_1,\dots,x_n)\, \delta V(x)
\eeq
}
\bigskip

\subsection*{Variation of $\om$}

We have:
\beq
\om(x) = \hbar \sum_i {1\over x-s_i}
\eeq
and
\beq
V'(s_i) = 2\hbar \sum_{j\neq i} {1\over s_i-s_j}
\eeq
Thus taking a variation we have:
\beq
\delta V'(s_i) + \delta s_i V''(s_i)  = -2\hbar \sum_{j\neq i} {\delta s_i-\delta s_j\over (s_i-s_j)^2}
\eeq
i.e.
\beq
\delta V'(s_i) = -\hbar \sum_j T_{i,j} \delta s_j
\eeq
which implies:
\beq
\delta s_i = - {1\over \hbar}\sum_j A_{i,j} \delta V'(s_j)
\eeq
and therefore:
\beq
\delta \om(x) = -\sum_{i,j} {A_{i,j} \delta V'(s_j)\over (x-s_i)^2}
\eeq
which can also be written:
\bea
\delta \om(x) 
&=& - \sum_k \Res_{x'\to s_k} \sum_{i,j} \,\, {A_{i,j} \over (x-s_i)^2\, (x'-s_j)}\,\, \delta V'(x') \cr
&=& - \sum_k \Res_{x'\to s_k} \sum_{i,j} \,\, {A_{i,j} \over (x-s_i)^2\, (x'-s_j)^2}\,\, \delta V(x') \cr
&=& - \sum_k \Res_{x'\to s_k} \, B(x,x')\,\, \delta V(x') 
\eea
and finally we obtain the case $n=1,g=0$ of the theorem:
\beq\label{varomResB}
\encadremath{
\delta \om(x)  = - \sum_k \Res_{x'\to s_k} \, B(x,x')\,\, \delta V(x') 
}\eeq

\subsection*{Variation of $B$}

Consider:
\beq
\ovl{W}_2^{(0)}(x,x') = B(x,x') - {1\over 2}\,{1\over (x-x')^2} =  \sum_{i,j} {A_{i,j}\over (x-s_j)^2(x'-s_i)^2} 
\eeq
Due to \eq{eqmono1} we have:
\bea
\ovl{W}_2^{(0)}(x,x')
&=&  \sum_{i} {\hbar K(x,s_i)\over (x'-s_i)^2} \cr
&=&  \sum_i \Res_{z\to s_i} K(x,z)\,\, {\om(z)\over (z-x')^2} \cr
&=&  {\partial \over \partial x'}\, \sum_i \Res_{z\to s_i} K(x,z)\,\, {\om(z)-\om(x')\over z-x'} \cr
\eea
On the other hand, since $\ovl{W}_2^{(0)}(x,x')$ has poles only at the $s_i$'s we have:
\bea
\ovl{W}_2^{(0)}(x,x')
&=& \Res_{z\to x} G(x,z)\,\, \ovl{W}_2^{(0)}(z,x') \cr
&=& - \sum_i \Res_{z\to s_i} G(x,z)\,\, \ovl{W}_2^{(0)}(z,x') \cr
&=& - \sum_i \Res_{z\to s_i} \left( (2\om(z)-V'(z)+\hbar {\partial_z})K(x,z)\right) \,\, \ovl{W}_2^{(0)}(z,x') \cr
&=& - \sum_i \Res_{z\to s_i} K(x,z)\,\, \left( (2\om(z)-V'(z)-\hbar {\partial_z}) \,\, \ovl{W}_2^{(0)}(z,x') \right)\cr
\eea
This implies that $\forall x$:
\beq
0= - \sum_i \Res_{z\to s_i} K(x,z)\,\, \left( (2\om(z)-V'(z)-\hbar {\partial_z}) \,\, \ovl{W}_2^{(0)}(z,x') + {\partial \over \partial x'} {\om(z)-\om(x')\over z-x'} \right) 
\eeq
and therefore, $\ovl{W}_2^{(0)}(x,x')$ satisfies the loop equation:
\beq\label{apploopeqW20}
(2\om(x)-V'(x)-\hbar {\partial_x}) \,\, \ovl{W}_2^{(0)}(x,x') + {\partial \over \partial x'} {\om(x)-\om(x')\over x-x'} = - P_2^{(0)}(x,x')
\eeq
where $P_2^{(0)}(x,x')$ has no pole at  $x\to s_i$'s.

\medskip
Then we take the variation:
\bea
(2\om(x)-V'(x)-\hbar {\partial_x}) \,\, \delta \ovl{W}_2^{(0)}(x,x') 
&=&
- (2\delta \om(x)-\delta V'(x)) \,\, \ovl{W}_2^{(0)}(x,x')\cr
&& - {\partial \over \partial x'} {\delta \om(x)-\delta \om(x')\over x-x'}  - \delta P_2^{(0)}(x,x') \cr
\eea
$\delta \ovl{W}_2^{(0)}(x,x')$ is a rational fraction of $x$, with poles only at the $s_i$'s, and $\delta P_2^{(0)}(x,x')$ has no pole at  $x\to s_i$'s.
We thus write:
\bea
\delta W_2^{(0)}(x,x')&=&\delta \ovl{W}_2^{(0)}(x,x')\cr
&=& \Res_{z\to x} G(x,z)\,\, \delta \ovl{W}_2^{(0)}(z,x') \cr
&=& - \sum_i \Res_{z\to s_i} G(x,z)\,\, \ovl{W}_2^{(0)}(z,x') \cr
&=& - \sum_i \Res_{z\to s_i} \left( (2\om(z)-V'(z)+\hbar {\partial_z})K(x,z)\right) \,\, \delta \ovl{W}_2^{(0)}(z,x') \cr
&=& - \sum_i \Res_{z\to s_i} K(x,z)\,\, \left( (2\om(z)-V'(z)-\hbar {\partial_z}) \,\, \delta \ovl{W}_2^{(0)}(z,x') \right)\cr
&=&  \sum_i \Res_{z\to s_i} K(x,z)\,\, \Big(  (2\delta \om(z)-\delta V'(z)) \,\, \ovl{W}_2^{(0)}(z,x') \cr
&& + {\partial \over \partial x'} {\delta \om(z)-\delta \om(x')\over z-x'}  + \delta P_2^{(0)}(z,x') \Big)\cr
&=&  \sum_i \Res_{z\to s_i} K(x,z)\,\, \Big(  (2\delta \om(z)-\delta V'(z)) \,\, \ovl{W}_2^{(0)}(z,x') 
 + {\delta \om(z)\over (z-x')^2}   \Big)\cr
&=&  \sum_i \Res_{z\to s_i} K(x,z)\,\,   (2\delta \om(z)-\delta V'(z)) \,\, B(z,x')   \cr
\eea
Then, we use \eq{varomResB}, and we get:
\bea
\delta W_2^{(0)}(x,x')
&=& - 2  \sum_i \Res_{z\to s_i} \sum_k \Res_{x''\to s_k} K(x,z)\,\,  B(z,x'') \delta V(x'')   \,\, B(z,x')   \cr
&& -  \sum_i \Res_{z\to s_i} K(x,z)\,\,   \delta V'(z) \,\, B(z,x')   \cr
&=& -   \sum_i \Res_{z\to s_i} \sum_k \Res_{x''\to s_k} K(x,z)\,\,  G(z,x'') \delta V'(x'')   \,\, B(z,x')   \cr
&& -   \sum_i \Res_{z\to s_i} \Res_{x''\to z} K(x,z)\,\,  G(z,x'') \delta V'(x'')   \,\, B(z,x')   \cr
&=& -   \sum_k \Res_{x''\to s_k} \sum_i \Res_{z\to s_i} K(x,z)\,\,  G(z,x'') \delta V'(x'')   \,\, B(z,x')   \cr
&=& - 2   \sum_k \Res_{x''\to s_k} \sum_i \Res_{z\to s_i} K(x,z)\,\,  B(z,x'') \delta V(x'')   \,\, B(z,x')   \cr
\eea
We thus obtain the case $n=2,g=0$ of the theorem:
\beq\label{varBResW3}
\encadremath{
\delta W_2^{(0)}(x,x') =  - \sum_k \Res_{x''\to s_k} W_3^{(0)}(x,x',x'')\,\, \delta V(x'')    
}\eeq

\subsection*{Variation of other higher correlators}

We prove by recursion on $2g+n$, that:
\beq
\encadremath{
\delta W^{(g)}_{n+1}(x,L) 
= - \sum_k \Res_{x''\to s_k}\,  \delta V(x'')\,\,  W_{n+2}^{(g)}(z,L,x'')  
}\eeq
where $L=\{x_1,\dots,x_n\}$.

\medskip

We write:
\beq\label{defUng2}
U_{n+1}^{(g)}(z,L) = \ovl{W}_{n+2}^{(g-1)}(z,z,L) + \sum_h\sum'_{J\subset L} W^{(h)}_{1+|J|}(z,J)\, W^{(g-h)}_{1+n-|J|}(z,L/J)
\eeq
By definition we have:
\beq
W^{(g)}_{n+1}(x,L) = \sum_i \Res_{z\to s_i} K(x,z)\,\, U^{(g)}_{n+1}(z,L)
\eeq

From the recursion hypothesis, we have:
\bea
\delta U_{n+1}^{(g)}(z,L) 
& = & - \sum_k \Res_{x''\to s_k}\, \delta V(x'')\,\, \Big( W_{n+3}^{(g-1)}(z,z,L,x'') \cr
&& - 2 \sum_h\sum'_{J\subset L}{W}^{(h)}_{2+|J|}(z,J,x'')\, {W}^{(g-h)}_{1+n-|J|}(z,L/J) \Big) \cr
& = & - \sum_k \Res_{x''\to s_k}\, \delta V(x'')\,\, \Big( U_{n+2}^{(g)}(z,L,x'') - 2 B(z,x'') W_{n+1}^{(g)}(z,L)
 \Big) \cr
\eea

Thus:
\bea
&& \delta W^{(g)}_{n+1}(x,L) \cr
&=& \sum_i \Res_{z\to s_i} \delta K(x,z)\,\, U^{(g)}_{n+1}(z,L)  - \sum_i \Res_{z\to s_i} K(x,z)\,\, \sum_k \Res_{x''\to s_k}\, \delta V(x'')\,\, \Big( \cr
&&  U_{n+2}^{(g)}(z,L,x'')  - 2 B(z,x'') W_{n+1}^{(g)}(z,L) \Big) \cr
&=& \sum_i \Res_{z\to s_i} \delta K(x,z)\,\, U^{(g)}_{n+1}(z,L)  - \sum_k \Res_{x''\to s_k}\, \sum_i \Res_{z\to s_i} K(x,z)\,\,  \delta V(x'')\,\, \Big( \cr
&& U_{n+2}^{(g)}(z,L,x'') - 2 B(z,x'') W_{n+1}^{(g)}(z,L) \Big) \cr
&=& \sum_i \Res_{z\to s_i} \delta K(x,z)\,\, U^{(g)}_{n+1}(z,L) \cr
&& +2 \sum_k \Res_{x''\to s_k}\, \sum_i \Res_{z\to s_i} K(x,z)\,\,  \delta V(x'')\,\,  B(z,x'') W_{n+1}^{(g)}(z,L)  \cr
&& - \sum_k \Res_{x''\to s_k}\, \sum_i \Res_{z\to s_i} K(x,z)\,\,  \delta V(x'')\,\,  U_{n+2}^{(g)}(z,L,x'')  \cr
&=& \sum_i \Res_{z\to s_i} \delta K(x,z)\,\, U^{(g)}_{n+1}(z,L) \cr
&& +2 \sum_i \Res_{z\to s_i}\sum_k \Res_{x''\to s_k}\,  K(x,z)\,\,  \delta V(x'')\,\,  B(z,x'') W_{n+1}^{(g)}(z,L)  \cr
&& +2 \, \sum_i \Res_{z\to s_i} \Res_{x''\to z} K(x,z)\,\,  \delta V(x'')\,\,  B(z,x'') W_{n+1}^{(g)}(z,L)  \cr
&& - \sum_k \Res_{x''\to s_k}\,  \delta V(x'')\,\,  W_{n+2}^{(g)}(z,L,x'')  \cr
\eea
We use the loop equation of theorem \ref{thWngPngAnn}, which says that
$U^{(g)}_{n+1}(z,L) + (2\om(z)-V'(z)+\hbar \partial_z)W^{(g)}_{n+1}(z,L) $ has no pole at $z\to s_i$, and thus:
\bea
&& \delta W^{(g)}_{n+1}(x,L) \cr
&=& - \sum_i \Res_{z\to s_i} \delta K(x,z)\,\, (2\om(z)-V'(z)+\hbar \partial_z)W^{(g)}_{n+1}(z,L) \cr
&& +2 \sum_i \Res_{z\to s_i}\sum_k \Res_{x''\to s_k}\,  K(x,z)\,\,  \delta V(x'')\,\,  B(z,x'') W_{n+1}^{(g)}(z,L)  \cr
&& +2 \, \sum_i \Res_{z\to s_i} \Res_{x''\to z} K(x,z)\,\,  \delta V(x'')\,\,  B(z,x'') W_{n+1}^{(g)}(z,L)  \cr
&& - \sum_k \Res_{x''\to s_k}\,  \delta V(x'')\,\,  W_{n+2}^{(g)}(z,L,x'')  \cr
&=& - \sum_i \Res_{z\to s_i} W^{(g)}_{n+1}(z,L) \,\, (2\om(z)-V'(z)-\hbar \partial_z) \delta K(x,z) \cr
&& +2 \sum_i \Res_{z\to s_i}\sum_k \Res_{x''\to s_k}\,  K(x,z)\,\,  \delta V(x'')\,\,  B(z,x'') W_{n+1}^{(g)}(z,L)  \cr
&& +2 \, \sum_i \Res_{z\to s_i} \Res_{x''\to z} K(x,z)\,\,  \delta V(x'')\,\,  B(z,x'') W_{n+1}^{(g)}(z,L)  \cr
&& - \sum_k \Res_{x''\to s_k}\,  \delta V(x'')\,\,  W_{n+2}^{(g)}(z,L,x'')  \cr
\eea
and we have:
\beq
(2\om(z)-V'(z)-\hbar \partial_z) \delta K(x,z)
 = \delta G(x,z) - (2\delta\om(z) - \delta V'(z)) K(x,z)
\eeq

\bea
&& \delta W^{(g)}_{n+1}(x,L) \cr
&=& - \sum_i \Res_{z\to s_i} W^{(g)}_{n+1}(z,L) \,\,  \delta G(x,z) \cr
&& + \sum_i \Res_{z\to s_i} W^{(g)}_{n+1}(z,L) \,\, (2\delta\om(z) - \delta V'(z)) \, K(x,z) \cr
&& +2 \sum_i \Res_{z\to s_i}\sum_k \Res_{x''\to s_k}\,  K(x,z)\,\,  \delta V(x'')\,\,  B(z,x'') W_{n+1}^{(g)}(z,L)  \cr
&& + \, \sum_i \Res_{z\to s_i} K(x,z)\,\,  \delta V'(z)\,\,  W_{n+1}^{(g)}(z,L)  \cr
&& - \sum_k \Res_{x''\to s_k}\,  \delta V(x'')\,\,  W_{n+2}^{(g)}(z,L,x'')  \cr
\eea

We have:
\beq
 \sum_i \Res_{z\to s_i} W^{(g)}_{n+1}(z,L) \,\,  \delta G(x,z) =0
\eeq
because the integrand is a rational fraction, and we have taken the sum of residues at all poles.

Using \eq{varomResB}, we are thus left with:
\beq
\delta W^{(g)}_{n+1}(x,L) 
= - \sum_k \Res_{x''\to s_k}\,  \delta V(x'')\,\,  W_{n+2}^{(g)}(z,L,x'')  
\eeq
which proves the recursion hypothesis for $2g+n+1$.
QED.

\section*{Appendix: Proof of theorem \ref{thResY}}
\label{approofthResY}

{\bf Theorem \ref{thResY}}

For $k=0,1$, 
$W_n^{(g)}$ satify the equation:
\bea
&& \Big(- \sum_{i=1}^n x_i^k{\partial \over \partial x_i}\Big)\, W_n^{(g)}(x_1,\dots,x_n)  \cr
&=& \sum_i \Res_{x_{n+1}\to s_i}\,\, x_{n+1}^k\,V'(x_{n+1})\,\,W_{n+1}^{(g)}(x_1,\dots,x_n,x_{n+1})
\eea

\proof{

Since $W_{n+1}^{(g)}$ has poles only at the $s_i$'s we have (with as usual $J=\{x_1,\dots,x_n\}$):
\bea
&& \sum_i \Res_{x\to s_i}\,\, x^k\,V'(x)\,\,W_{n+1}^{(g)}(J,x)  \cr
&=& \sum_i \Res_{x\to s_i}\,\, x^k\,Y(x)\,\,W_{n+1}^{(g)}(J,x)  \cr
\eea
Then using theorem \ref{thWngPngAnn}, we have:
\bea
&& \sum_i \Res_{x\to s_i}\,\, x^k\,V'(x)\,\,W_{n+1}^{(g)}(J,x)  \cr
&=& \sum_i \Res_{x\to s_i}\,\, x^k\,Y(x)\,\,W_{n+1}^{(g)}(J,x)  \cr
&=& \sum_i \Res_{x\to s_i}\,\, x^k\,
\Big[\hbar \partial_x W_{n+1}^{(g)}(J,x) + U_{n+1}^{(g)}(x,J) - P_{n+1}^{(g)}(x;J) 
-\sum_{j=1}^n \partial_{x_j}\, {W_n^{(g)}(J)\over x-x_j}\Big] \cr
&=& \sum_i \Res_{x\to s_i}\,\, x^k\,
\Big[\hbar \partial_x W_{n+1}^{(g)}(J,x) + U_{n+1}^{(g)}(x,J)  \Big] \cr
\eea
Notice that if $n\geq 1$, $W_{n+1}^{(g)}(J,x)$ behaves like $O(1/x^2)$ at $x\to\infty$, and thus, if $k\leq 1$, $x^k\,\partial_x W_{n+1}^{(g)}(J,x)$ behaves like $O(1/x^2)$. Since we take the residues at all poles, the sum of residues vanish and thus:
\bea
&& \sum_i \Res_{x\to s_i}\,\, x^k\,V'(x)\,\,W_{n+1}^{(g)}(J,x)  \cr
&=& \sum_i \Res_{x\to s_i}\,\, x^k\, U_{n+1}^{(g)}(x,J)  \cr
\eea
Notice that $U_{n+1}^{(g)}(x,J)$ (defined in \eq{defUng2}), behaves at most like $O(1/x^3)$ for large $x$, and thus, if $k\leq 1$, the product $x^k\,U_{n+1}^{(g)}(x,J)$ is a rational fraction, which behaves like $O(1/x^2)$ for large $x$. Its only poles can be at $x=s_i$ or at $x=x_j$. Therefore the sum of residues at $s_i$'s, can be replaced by the sum of residues at $x_j$'s:
\bea
&& \sum_i \Res_{x\to s_i}\,\, x^k\,V'(x)\,\,W_{n+1}^{(g)}(J,x)  \cr
&=& - \sum_{j=1}^n \Res_{x\to x_j}\,\, x^k\, U_{n+1}^{(g)}(x,J)  \cr
\eea
The only terms in $U_{n+1}^{(g)}(x,J)$ which have poles at $x=x_j$, are the terms containing a $B(x,x_j)$, i.e.:
\bea
 \sum_i \Res_{x\to s_i}\,\, x^k\,V'(x)\,\,W_{n+1}^{(g)}(J,x)  
&=& - 2\sum_{j=1}^n \Res_{x\to x_j}\,\, x^k\, B(x,x_j) \, W_n^{(g)}(x,J/\{x_j\})  \cr
&=& - \sum_{j=1}^n \Res_{x\to x_j}\,\, x^k\, {1\over (x-x_j)^2} \, W_n^{(g)}(x,J/\{x_j\})  \cr
&=& - \sum_{j=1}^n {\partial \over \partial x_j}\, \Big( x_j^k\, \, W_n^{(g)}(x_1,\dots,x_n)\, \Big)  \cr
\eea

}

\section*{Appendix: Proof of theorem \ref{thdilaton}}
\label{approofthdilaton}

{\bf Theorem \ref{thdilaton}:}

{\it
For $n\geq 1$, $W_n^{(g)}$ satify the equation:
\beq
(2-2g-n-\hbar {\partial\over \partial \hbar})\, \ovl{W}_{n}^{(g)}(x_1,\dots,x_n)
= - \sum_i \Res_{x_{n+1}\to s_i}\,\, V(x_{n+1})\,\,\ovl{W}_{n+1}^{(g)}(x_1,\dots,x_n,x_{n+1})
\eeq

}

\bigskip

\subsection*{$\hbar$ derivatives for $w(z)$}
We have:
$$V'(s_i)=2\hbar \sum_{\neq i}\frac{1}{s_i-s_j}$$
Taking the derivative with respect to $\hbar$ gives:
$$\hbar  V''(s_i)\partial_\hbar s_i=V'(s_i)-2\hbar^2\sum_{j \neq i} \frac{\partial_\hbar si-\partial_\hbar s_j}{(s_i-s_j)^2}$$
and so
$$V'(s_i)=\hbar\left( V''(s_i)\partial_\hbar s_i+2\hbar\sum_{j \neq i} \frac{\partial_\hbar si-\partial_\hbar s_j}{(s_i-s_j)^2}\right)$$
We recognize the general term of the matrix $T$ and find:
$$V'(s_i)=\hbar^2\sum_{j}T_{i,j}\partial_\hbar s_j$$
Multiplying by the matrix $A$ gives:
\beq \encadremath{\hbar^2 \partial_{\hbar}s_i=\sum_j A_{i,j}V'(s_j)}\eeq
We can use this result to compute:
\bea
\hbar \partial_\hbar \om(x)
&=&\om(x)+\hbar^2\sum_i \frac{\partial_\hbar si}{(x-s_i)^2}\cr
&=& \om(x)+\sum_{i,j}\frac{A_{i,j}V'(s_j)}{(x-s_i)^2} \cr
&=& \om(x)+\sum_k \Res_{x' \to s_k}\sum_{i,j}\frac{A_{i,j}V'(x')}{(x-s_i)^2(x'-s_j)}\cr
&=& \om(x)+\sum_k \Res_{x' \to s_k}\sum_{i,j}\frac{A_{i,j}V(x')}{(x-s_i)^2(x'-s_j)^2}\cr
&=& \om(x)+\sum_k \Res_{x' \to s_k}\ovl{W}_2^{(0)}(x,x')V(x') \cr
&=& \om(x)+\sum_k \Res_{x' \to s_k}W_2^{(0)}(x,x')V(x')\cr
\eea
Thus we have proved the case $n=1, g=0$ of the theorem:
\beq \label{Derivomhbar} \encadremath{\hbar \partial_\hbar \om(x)=\om(x)+\sum_k \Res_{x' \to s_k}W_2^{(0)}(x,x')V(x')}\eeq

\subsection*{$\hbar$ derivatives for $W_2^{(0)}(z)$}

We have seen in appendix \ref{approofthvariationV}, \eq{apploopeqW20}, that $\ovl{W}_2^{(0)}(x,x')$ satisfies the loop equation:
\beq
(2\om(x)-V'(x)+\hbar {\partial_x}) \,\, \ovl{W}_2^{(0)}(x,x') + {\partial \over \partial x'} {\om(x)-\om(x')\over x-x'} = - P_2^{(0)}(x,x')
\eeq
where $P_2^{(0)}(x,x')$ has no pole at  $x\to s_i$'s.

\medskip
Then we take the derivation $\hbar \partial_\hbar$ of this equation:
\bea
(2\om(x)-V'(x)+\hbar \partial_x) \,\, \hbar \partial_\hbar \ovl{W}_2^{(0)}(x,x')+\hbar \partial_x \ovl{W}_2^{(0)}(x,x')+2\hbar \partial_\hbar w(x)\ovl{W}_2^{(0)}(x,x') \cr
=
- {\partial \over \partial x'} {\hbar\partial_\hbar \om(x)-\hbar\partial_\hbar \om(x')\over x-x'}  - \hbar\partial_\hbar P_2^{(0)}(x,x') \cr
\eea
$\hbar \partial_\hbar \ovl{W}_2^{(0)}(x,x')$ is a rational fraction of $x$, with poles only at the $s_i$'s, and $\hbar\partial_\hbar P_2^{(0)}(x,x')$ has no pole at  $x\to s_i$'s.
We thus write:
\bea
&& \hbar \partial_\hbar W_2^{(0)}(x,x')  \cr
&=&\hbar \partial_\hbar \ovl{W}_2^{(0)}(x,x')\cr
&=& \Res_{z\to x} G(x,z)\,\, \hbar \partial_\hbar \ovl{W}_2^{(0)}(z,x') \cr
&=& - \sum_i \Res_{z\to s_i} G(x,z)\,\, \hbar \partial_\hbar \ovl{W}_2^{(0)}(z,x') \cr
&=& - \sum_i \Res_{z\to s_i} \left( (2\om(z)-V'(z)-\hbar {\partial_z})K(x,z)\right) \,\, \hbar \partial_\hbar \ovl{W}_2^{(0)}(z,x') \cr
&=& - \sum_i \Res_{z\to s_i} K(x,z)\,\, \left( (2\om(z)-V'(z)+\hbar {\partial_z}) \,\, \hbar \partial_\hbar \ovl{W}_2^{(0)}(z,x') \right)\cr
&=&  \sum_i \Res_{z\to s_i} K(x,z)\,\, \Big(  (2\hbar \partial_\hbar \om(z)) \,\, \ovl{W}_2^{(0)}(z,x') \cr
&& + {\partial \over \partial x'} {\hbar \partial_\hbar\om(z)+\hbar \partial_\hbar\om(x')\over z-x'}  + \hbar \partial_\hbar P_2^{(0)}(z,x')
+\hbar \partial_z \ovl{W}_2^{(0)}(z,x') \Big)\cr
&=&  \sum_i \Res_{z\to s_i} K(x,z)\,\, \Big(    2\ovl{W}_2^{(0)}(z,x')\,\,\hbar \partial_\hbar \om(z) 
 + {\hbar \partial_\hbar \om(z)\over (z-x')^2}+\hbar \partial_z \ovl{W}_2^{(0)}(z,x')   \Big)\cr
&=&  \sum_i \Res_{z\to s_i} K(x,z)\,\,\Big(2W_2^{(0)}(z,x')\,\hbar \partial_\hbar \om(z)+\hbar \partial_z W_2^{(0)}(z,x')\Big)\cr
\eea
Then, we use \eq{Derivomhbar}, and we get:
\bea
&& \hbar \partial_\hbar W_2^{(0)}(x,x') \cr
&=& \sum_i \Res_{z\to s_i} K(x,z)\,\,\Big(2W_2^{(0)}(z,x')w(z)+\hbar \partial_z W_2^{(0)}(z,x')\Big)\cr
&& + 2\sum_{i,k} \Res_{z\to s_i}\Res_{x'' \to s_k} K(x,z)W_2^{(0)}(z,x')W_2^{(0)}(z,x'')V(x'')\cr
&=& \sum_i \Res_{z\to s_i} W_2^{(0)}(z,x')\,\,\Big(2w(z)-\hbar \partial_z \Big)K(x,z)\cr
&& +\sum_{i,k} \Res_{z\to s_i}\Res_{x'' \to s_k} K(x,z)W_2^{(0)}(z,x')G(z,x'')V'(x'')\cr
&=& \sum_i \Res_{z\to s_i} W_2^{(0)}(z,x')\,\,(G(x,z)+V'(z)K(x,z))\cr
&& +\sum_{i,k} \Res_{z\to s_i}\Res_{x'' \to s_k} K(x,z)W_2^{(0)}(z,x')G(z,x'')V'(x'')\cr
&=& \sum_i \Res_{z\to s_i} W_2^{(0)}(z,x')\,\,G(x,z)\cr
&& +\sum_{i,k} \Res_{z\to s_i}\Res_{x'' \to s_k} K(x,z)W_2^{(0)}(z,x')G(z,x'')V'(x'')\cr
&& +\sum_{i} \Res_{z\to s_i}\Res_{x'' \to z} K(x,z)W_2^{(0)}(z,x')G(z,x'')V'(x'')\cr
&=& \sum_i \Res_{z\to s_i} W_2^{(0)}(z,x')\,\,G(x,z)\cr
&& +\sum_{i,k} \Res_{x'' \to s_k}\,\Res_{z\to s_i} K(x,z)W_2^{(0)}(z,x')G(z,x'')V'(x'')\cr
&=& \sum_i \Res_{z\to s_i} W_2^{(0)}(z,x')\,\,G(x,z)\cr
&& +2 \sum_{i,k} \Res_{x'' \to s_k}\,\Res_{z\to s_i} K(x,z)W_2^{(0)}(z,x')B(z,x'')V(x'')\cr
&=& \sum_i \Res_{z\to s_i} B(z,x')\,\,G(x,z)\cr
&& + \sum_{k} \Res_{x'' \to s_k}\,W_3^{(0)}(x,x',x'')V(x'')\cr
\eea
We now use the fact that $G(x,z)$ and $B(z,x')$ are rational fractions whose only poles are $s_i$'s, as well as $z=x$ and $z=x'$, and we write:
\bea
&& \sum_i \Res_{z\to s_i} B(z,x')\,\,G(x,z)\cr
&=& - \Res_{z\to x} B(z,x')\,\,G(x,z) - \Res_{z\to x'} B(z,x')\,\,G(x,z)\cr
&=& - \Res_{z\to x} B(z,x')\,\,{1\over z-x} - {1\over 2}\,\Res_{z\to x'} {1\over (z-x')^2}\,\,G(x,z)\cr
&=& - \Res_{z\to x} B(z,x')\,\,{1\over z-x} +\,\Res_{z\to x'} {1\over z-x'}\,\,B(x,z)\cr
&=& - B(x,x')+B(x,x')\cr
&=& 0
\eea

So that eventually we have proved the case $n=2,g=0$ of the theorem:
\beq\label{varBResW3hbar}
\encadremath{
\hbar \partial_\hbar W_2^{(0)}(x,x') =  \sum_k \Res_{x''\to s_k} W_3^{(0)}(x,x',x'')\,\, V(x'')    
}\eeq

\subsection*{Recursion for higher correlators}

We proceed by recursion on $2g+n$.

From theorem \ref{thWngPngAnn}, we have that:
\bea
&& (Y(x)-\hbar \partial_x) \hbar\partial_{\hbar} W_{n+1}^{(g)}(x,L) \cr
&=& \hbar\partial_{\hbar} U_{n+1}^{(g)}(x;L) + \hbar \partial_x W_{n+1}^{(g)}(x,L)
- W_{n+1}^{(g)}(x,L) \,  \hbar\partial_{\hbar} Y(x) \cr
&& -  \hbar\partial_{\hbar} \left( P_{n+1}^{(g)}(x;L) +\sum_{x_j\in L} {\partial \over \partial x_j}\, { {\ovl{W}}_{n}^{(g)}(L)\over x-x_j} \right)
\eea
where the term on the last line has no pole at $x=s_i$.
This implies that:
\bea
&& \sum_i \Res_{x\to s_i} K(x_0,x)\,\Big((Y(x)-\hbar \partial_x) \hbar\partial_{\hbar} W_{n+1}^{(g)}(x,L)\Big) \cr
&=& \sum_i \Res_{x\to s_i} K(x_0,x)\,\Big( \hbar\partial_{\hbar} U_{n+1}^{(g)}(x;L) + \hbar \partial_x W_{n+1}^{(g)}(x,L) \cr
&& - W_{n+1}^{(g)}(x,L) \,  \hbar\partial_{\hbar} Y(x)  \Big) 
\eea
We have:
\bea
&& \sum_i \Res_{x\to s_i} K(x_0,x)\,\Big((Y(x)-\hbar \partial_x) \hbar\partial_{\hbar} W_{n+1}^{(g)}(x,L)\Big) \cr
&=& \sum_i \Res_{x\to s_i} \hbar\partial_{\hbar} W_{n+1}^{(g)}(x,L)\,(Y(x)+\hbar \partial_x) K(x_0,x)  \cr
&=& - \sum_i \Res_{x\to s_i} \hbar\partial_{\hbar} W_{n+1}^{(g)}(x,L)\,G(x_0,x)  \cr
&=&  \Res_{x\to x_0} \hbar\partial_{\hbar} W_{n+1}^{(g)}(x,L)\,G(x_0,x)  \cr
&=&   \hbar\partial_{\hbar} W_{n+1}^{(g)}(x_0,L)  
\eea
and therefore:
\bea
&&   \hbar\partial_{\hbar} W_{n+1}^{(g)}(x_0,L)  \cr
&=& \sum_i \Res_{x\to s_i} K(x_0,x)\,\Big( \hbar\partial_{\hbar} U_{n+1}^{(g)}(x;L) + \hbar \partial_x W_{n+1}^{(g)}(x,L)
- W_{n+1}^{(g)}(x,L) \,  \hbar\partial_{\hbar} Y(x)  \Big) \cr
\eea
From the recursion hypothesis we have:
\bea
&& \hbar\partial_{\hbar} U_{n+1}^{(g)}(x;L) \cr
&=& \hbar\partial_{\hbar} W_{n+2}^{(g-1)}(x,x,L)
+ \sum_{k=0}^g\sum'_{J\subset L} W_{1+|J|}^{(k)}(x,J) \hbar\partial_{\hbar} W_{1+n-|J|}^{(g-k)}(x,L/J) \cr
&& + \sum_{k=0}^g\sum'_{J\subset L}  W_{1+n-|J|}^{(g-k)}(x,L/J) \hbar\partial_{\hbar}  W_{1+|J|}^{(k)}(x,J)\cr
&=& (2-2(g-1)-(n+2))  W_{n+2}^{(g-1)}(x,x,L) + \sum_i\Res_{x'\to s_i}  W_{n+3}^{(g-1)}(x,x,L,x')\, V(x') \cr
&& + \sum_{k=0}^g\sum'_{J\subset L} (2-2(g-k)-(1+n-|J|))\,W_{1+|J|}^{(k)}(x,J) \, W_{1+n-|J|}^{(g-k)}(x,L/J) \cr
&& + \sum_{k=0}^g\sum'_{J\subset L}  (2-2k-(1+|J|))\, W_{1+n-|J|}^{(g-k)}(x,L/J) \,  W_{1+|J|}^{(k)}(x,J)\cr
&& + \sum_i \Res_{x'\to s_i} V(x')\sum_{k=0}^g\sum'_{J\subset L} W_{2+|J|}^{(k)}(x,J,x') \, W_{1+n-|J|}^{(g-k)}(x,L/J) \cr
&& + \sum_i \Res_{x'\to s_i} V(x')\sum_{k=0}^g\sum'_{J\subset L} W_{1+|J|}^{(k)}(x,J) \, W_{2+n-|J|}^{(g-k)}(x,L/J,x') \cr
&=& (2-2g-n)\,\,  U_{n+1}^{(g)}(x;L) \cr 
&& + \sum_i \Res_{x'\to s_i} V(x')\,(U_{n+2}^{(g)}(x;x',L) -2B(x,x') W_{n+1}^{(g)}(x,L) )  
\eea

Thus we have:
\bea
&&   \hbar\partial_{\hbar} W_{n+1}^{(g)}(x_0,L)  \cr
&=& (2-2g-n)\sum_i \Res_{x\to s_i} K(x_0,x)\,U_{n+1}^{(g)}(x;L) \cr
&& + \sum_i \Res_{x\to s_i} K(x_0,x)  \sum_j \Res_{x'\to s_j} V(x')\, (U_{n+2}^{(g)}(x;x',L)-2B(x,x') W_{n+1}^{(g)}(x,L) ) \cr
&& + \sum_i \Res_{x\to s_i} K(x_0,x)\,\Big(  \hbar \partial_x W_{n+1}^{(g)}(x,L)
- W_{n+1}^{(g)}(x,L) \,  \hbar\partial_{\hbar} Y(x)  \Big) \cr
&=& (2-2g-n) W_{n+1}^{(g)}(x_0,L) \cr
&& + \sum_j \Res_{x'\to s_j}  \sum_i \Res_{x\to s_i} K(x_0,x)  V(x')\, (U_{n+2}^{(g)}(x;x',L)-2B(x,x') W_{n+1}^{(g)}(x,L) ) \cr
&& + \sum_i \Res_{x\to s_i} K(x_0,x)\,\Big(  \hbar \partial_x W_{n+1}^{(g)}(x,L)
- W_{n+1}^{(g)}(x,L) \,  \hbar\partial_{\hbar} Y(x)  \Big) \cr
&=& (2-2g-n) W_{n+1}^{(g)}(x_0,L)  + \sum_j \Res_{x'\to s_j}   V(x')\, W_{n+2}^{(g)}(x_0,x',L) \cr
&& -2 \sum_j \Res_{x'\to s_j}  \sum_i \Res_{x\to s_i} K(x_0,x)  V(x')\, B(x,x') W_{n+1}^{(g)}(x,L)  \cr
&& + \sum_i \Res_{x\to s_i} K(x_0,x)\,\Big(  \hbar \partial_x W_{n+1}^{(g)}(x,L)
- W_{n+1}^{(g)}(x,L) \,  \hbar\partial_{\hbar} Y(x)  \Big) \cr
&=& (2-2g-n) W_{n+1}^{(g)}(x_0,L)  + \sum_j \Res_{x'\to s_j}   V(x')\, W_{n+2}^{(g)}(x_0,x',L) \cr
&& -2 \sum_i \Res_{x\to s_i} \sum_j \Res_{x'\to s_j}   K(x_0,x)  V(x')\, B(x,x') W_{n+1}^{(g)}(x,L)  \cr
&& -2\sum_i \Res_{x\to s_i} \Res_{x'\to x}   K(x_0,x)  V(x')\, B(x,x') W_{n+1}^{(g)}(x,L)  \cr
&& + \sum_i \Res_{x\to s_i} K(x_0,x)\,\Big(  \hbar \partial_x W_{n+1}^{(g)}(x,L)
- W_{n+1}^{(g)}(x,L) \,  \hbar\partial_{\hbar} Y(x)  \Big) 
\eea

Notice that:
\beq
 \hbar\partial_{\hbar} Y(x)
+ 2\sum_j \Res_{x'\to s_j} B(x,x') V(x')
+ 2\Res_{x'\to x} B(x,x') V(x')
= Y(x)
\eeq
therefore:
\bea
&&   \hbar\partial_{\hbar} W_{n+1}^{(g)}(x_0,L)  \cr
&=& (2-2g-n) W_{n+1}^{(g)}(x_0,L)  + \sum_j \Res_{x'\to s_j}   V(x')\, W_{n+2}^{(g)}(x_0,x',L) \cr
&& + \sum_i \Res_{x\to s_i} K(x_0,x)\,\Big(  \hbar \partial_x W_{n+1}^{(g)}(x,L)
- Y(x) W_{n+1}^{(g)}(x,L)   \Big) \cr
&=& (2-2g-n) W_{n+1}^{(g)}(x_0,L)  + \sum_j \Res_{x'\to s_j}   V(x')\, W_{n+2}^{(g)}(x_0,x',L) \cr
&& - \sum_i \Res_{x\to s_i} W_{n+1}^{(g)}(x,L)\,(  Y(x)+\hbar \partial_x )K(x_0,x)  \cr
&=& (2-2g-n) W_{n+1}^{(g)}(x_0,L)  + \sum_j \Res_{x'\to s_j}   V(x')\, W_{n+2}^{(g)}(x_0,x',L) \cr
&& + \sum_i \Res_{x\to s_i} W_{n+1}^{(g)}(x,L)\,G(x_0,x)  \cr
&=& (2-2g-n) W_{n+1}^{(g)}(x_0,L)  + \sum_j \Res_{x'\to s_j}   V(x')\, W_{n+2}^{(g)}(x_0,x',L) \cr
&& -  \Res_{x\to x_0} W_{n+1}^{(g)}(x,L)\,G(x_0,x)  \cr
&=& (2-2g-n) W_{n+1}^{(g)}(x_0,L)  + \sum_j \Res_{x'\to s_j}   V(x')\, W_{n+2}^{(g)}(x_0,x',L) \cr
&& - W_{n+1}^{(g)}(x_0,L)  \cr
&=& (2-2g-n-1) W_{n+1}^{(g)}(x_0,L)  + \sum_j \Res_{x'\to s_j}   V(x')\, W_{n+2}^{(g)}(x_0,x',L) 
\eea
i.e. we have proved the theorem for $2g+n+1$.

\section*{Appendix: Free Energies}
\label{approofFg}

Here we consider $g\geq 2$.

\medskip

The free energies defined in \eq{defFg}, automatically satisfy theorem \ref{thdilaton}, and thus are homogeneous:
\beq
F^{(g)}(\l V,\l \hbar) = \l^{2-2g}\,\, F^{(g)}( V,\hbar)
\eeq

Here we show that they satisfy theorem \ref{thvariationV}.

\medskip

We start from the definition:
\beq\label{defFg1}
F^{(g)} =  \hbar^{2-2g}\,\int_0^{\hbar}\, {d\td{\hbar}\over {\td\hbar}^{3-2g}}\,\, \sum_i \Res_{x\to s_i}\,\, V(x)\,\,\, \left. W_{1}^{(g)}(x)\right|_{{\td\hbar}}
\eeq
and we compute the loop operator applied to $F^{(g)}$:
\bea
\delta_{x_1}\, F^{(g)} 
&=&  \hbar^{2-2g}\,\int_0^{\hbar}\, {d\td{\hbar}\over {\td\hbar}^{3-2g}}\,\, \sum_i \Res_{x\to s_i}\left(\,\, V(x)\,\,\,  W_{2}^{(g)}(x,x_1)+ \delta_{x_1}\,V(x)\,\,\,  W_{1}^{(g)}(x) \right)_{{\td\hbar}}  \cr
&=&  \hbar^{2-2g}\,\int_0^{\hbar}\, {d\td{\hbar}\over {\td\hbar}^{3-2g}}\,\, \sum_i\Res_{x\to s_i}\,\left( \,  V(x)\,\,\,  W_{2}^{(g)}(x,x_1) + {W_{1}^{(g)}(x)\over x-x_1} \right)_{{\td\hbar}}  \cr
&=&  \hbar^{2-2g}\,\int_0^{\hbar}\, {d\td{\hbar}\over {\td\hbar}^{3-2g}}\,\, \left(   \left(\sum_i\Res_{x\to s_i} V(x)\,\,\,  W_{2}^{(g)}(x,x_1) \right) -  W_{1}^{(g)}(x_1)\right)_{{\td\hbar}}  \cr
&=&  \hbar^{2-2g}\,\int_0^{\hbar}\, {d\td{\hbar}\over {\td\hbar}^{3-2g}}\,\, \left(  \td{h}^{2-2g}\,{d (\td{h}^{2g-1}\,W_{1}^{(g)}(x_1))\over d\td{h}} -W_{1}^{(g)}(x_1) \right)_{{\td\hbar}}  \cr
&=&  \hbar^{2-2g}\,\int_0^{\hbar}\, \left({1\over \td{h}}\,d \left(\td{h}^{2g-1}\,W_{1}^{(g)}(x_1)\right) - {d\td{\hbar}\over {\td\hbar}^{3-2g}}\,\,W_{1}^{(g)}(x_1) \right)_{{\td\hbar}} \cr
\eea
we integrate by parts, and since $2g-2>0$, there is no boundary term coming from the bound at $0$, and thus:
\bea
\delta_{x_1}\, F^{(g)} 
&=& W_1^{(g)}(x_1) + \hbar^{2-2g}\,\int_0^{\hbar}\, \left( \td{h}^{2g-3}\,W_{1}^{(g)}(x_1)- {\td\hbar}^{2g-3}\,\,W_{1}^{(g)}(x_1) \right)_{{\td\hbar}} \, d\td{h}\cr
&=& W_1^{(g)}(x_1)
\eea
Therefore we have proved that the loop operator acting on $F^{(g)}$ is indeed $W_1^{(g)}$, i.e. we have proved theorem \ref{thvariationV}.

\section*{Appendix: $F^{(0)}$}
\label{approofF0}

We have defined $F^{(0)}$ as:
\beq
F^{(0)} =  -\hbar \sum_i  V(s_i) + \hbar^2 \sum_{i\neq j} \ln{(s_i-s_j)}
\eeq

\bigskip
$\bullet$ Proof of theorem \ref{thvariationV} for $F^{(0)}$:

consider a variation $\delta V$, we have: 
\bea 
\delta F^{(0)}
&=& -\hbar \sum_i \delta V(s_i) - \hbar \sum_i V'(s_i) \delta s_i + 2\hbar^2 \sum_{j\neq i} {\delta s_i\over s_i-s_j} \cr
&=& -\hbar \sum_i \delta V(s_i)  \cr
&=& - \sum_i \Res_{x\to s_i} \om(x)\,  \delta V(x)  
\eea

\bigskip
$\bullet$ Proof of theorem \ref{thdilaton} for $F^{(0)}$:

we have:
\bea
\hbar \partial_{\hbar} F^{(0)}
&=& -\hbar \sum_i  V(s_i) + 2\hbar^2 \sum_{i\neq j} \ln{(s_i-s_j)} \cr
&& -\hbar^2 \sum_i {\partial s_i\over \partial \hbar}\,\left(V'(s_i) - 2\hbar\sum_{j\neq i} {1\over s_i-s_j} \right) \cr
&=& -\hbar \sum_i  V(s_i) + 2\hbar^2 \sum_{i\neq j} \ln{(s_i-s_j)} \cr
&=& 2 F^{(0)} + \hbar \sum_i  V(s_i)  \cr
&=& 2 F^{(0)} +  \sum_i \Res_{x\to s_i} \om(x)\, V(x)  
\eea
Therefore:
\beq (2-\hbar\partial_\hbar)F_0=-\sum_i \Res_{x \to s_i}V(x)w(x)\eeq

\section*{Appendix: $F^{(1)}$}
\label{approofF1}

We have defined $F^{(1)}$ as:
\bea
F^{(1)} 
&=& {1\over 2}\,\ln{(\det A)}\, +{F^{(0)}\over \hbar^2} + \ln{(\Delta(s)^2)} \cr
&=& {1\over 2}\,\ln{(\det A)}\, - {1\over \hbar} \sum_i V(s_i) + \sum_{i\neq j} \ln{(s_i-s_j)} + \sum_{i\neq j} \ln{(s_i-s_j)} \cr
&=& {1\over 2}\,\ln{(\det A)}\, - {1\over \hbar} \sum_i V(s_i) + 2\sum_{i\neq j} \ln{(s_i-s_j)}  
\eea

\bigskip
$\bullet$ Proof of theorem \ref{thvariationV} for $F^{(1)}$:

Let us start from $W_1^{(1)}$
\bea
W_1^{(1)}(x) 
&=& \sum_i \Res_{z\to s_i} K(x,z)\,\ovl{W}_2(z,z) \cr
&=& \sum_i \Res_{z\to s_i} K(x,z)\,\Big[{A_{i,i}\over (z-s_i)^4} + 2\sum_{j\neq i} {A_{i,j}\over (z-s_i)^2(z-s_j)^2} \Big] \cr
&=& \sum_i \Res_{z\to s_i} K(x,z)\,{A_{i,i}\over (z-s_i)^4}  \cr
&& +2 \sum_i\sum_{j\neq i}  K'(x,s_i)\, {A_{i,j}\over (s_i-s_j)^2}  \cr
&& -4 \sum_i\sum_{j\neq i}  K(x,s_i)\, {A_{i,j}\over (s_i-s_j)^3}  \cr
\eea

We have:
\bea
&& \sum_i \Res_{z\to s_i} K(x,z)\,{A_{i,i}\over (z-s_i)^4} \cr
&=& {1\over 3} \sum_i \Res_{z\to s_i} K'(x,z)\,{A_{i,i}\over (z-s_i)^3} \cr
&=& {1\over 3} \sum_i \Res_{z\to s_i} ({2\over z-s_i} + 2 \om_i(z) - {1\over \hbar}V'(z))K(x,z) \,{A_{i,i}\over (z-s_i)^3} \cr
&& - {1\over 3\hbar } \sum_i \Res_{z\to s_i} G(x,z) \,{A_{i,i}\over (z-s_i)^3} 
\eea
Therefore:
\bea
&& \sum_i \Res_{z\to s_i} K(x,z)\,{A_{i,i}\over (z-s_i)^4} \cr
&=&  \sum_i \Res_{z\to s_i} ( 2 \om_i(z) - {1\over \hbar}V'(z))K(x,z) \,{A_{i,i}\over (z-s_i)^3} \cr
&& -  {1\over \hbar}\,\sum_i \Res_{z\to s_i} G(x,z) \,{A_{i,i}\over (z-s_i)^3} \cr
&=&  \sum_i \Res_{z\to s_i} \Big[{ 2 \om_i(z) - {1\over \hbar}V'(z)\over z-s_i}\,K(x,z)\Big] \,{A_{i,i}\over (z-s_i)^2} \cr
&& -  {1\over 2\hbar}\,\sum_i \Res_{z\to s_i} G'(x,z) \,{A_{i,i}\over (z-s_i)^2} \cr
&=&  \sum_i A_{i,i}\, \Big[{ 2 \om_i(z) - {1\over \hbar}V'(z)\over z-s_i}\,K(x,z)\Big]'_{z= s_i}  \cr
&& + {1\over \hbar}\,\sum_i \Res_{z\to s_i} B(x,z) \,{A_{i,i}\over (z-s_i)^2} \cr
&=& {1\over 2} \sum_i  ( 2 \om_i''(s_i) - {1\over \hbar}V'''(s_i))K(x,s_i) \,{A_{i,i}} \cr
&&  - \sum_i  K'(x,s_i) \,{A_{i,i} T_{i,i}} \cr
&& +  {1\over \hbar}\,\sum_i \Res_{z\to s_i} B(x,z) \,{A_{i,i}\over (z-s_i)^2} 
\eea

Notice that:
\bea
\Res_{x\to s} K(x,s_i) \delta V(x) 
&=& {1\over \hbar}\, \sum_{j} \Res_{x\to s} {A_{i,j} \delta V(x) \over (x-s_j)^2} \cr
&=& {1\over \hbar}\,\sum_{j}  A_{i,j} \delta V'(s_j) \cr
&=& - \delta s_i 
\eea

\bea
\Res_{x\to s} K'(x,s_i) \delta V(x) 
&=& -{1\over \hbar}\,\sum_j \delta_{i,j}\delta V(s_j) - 2\sum_{j\neq i} {\delta s_j\over s_i-s_j}
\eea

\bea
\Res_{x\to s} \Res_{z\to s_i} {B(x,z)\over (z-s_i)^2}\,\, \delta V(x) 
&=&  \Res_{z\to s_i} \Res_{x\to s}{B(x,z)\over (z-s_i)^2}\,\, \delta V(x)  \cr
&& +  \Res_{z\to s_i} \Res_{x\to z} {B(x,z)\over (z-s_i)^2}\,\, \delta V(x)  \cr
&=&  \Res_{z\to s_i} \Res_{x\to s}{A_{j,l}\over (x-s_l)^2(z-s_j)^2 (z-s_i)^2}\,\, \delta V(x)  \cr
&& +  {1\over 2}\Res_{z\to s_i} \Res_{x\to z} {1\over (x-z)^2(z-s_i)^2}\,\, \delta V(x)  \cr
&=&  {\hbar}\,\Res_{z\to s_i} \Res_{x\to s}{K(x,s_j)\over (z-s_j)^2 (z-s_i)^2}\,\, \delta V(x)  \cr
&& +  {1\over 2}\Res_{z\to s_i}  {1\over (z-s_i)^2}\,\, \delta V'(z)  \cr
&=& - \hbar \, \Res_{z\to s_i} {\delta s_j\over (z-s_j)^2 (z-s_i)^2}  +  {1\over 2}\,\, \delta V''(s_i)  \cr
&=& 2\hbar\,  {\delta s_j\over (s_i-s_j)^3 }  +  {1\over 2}\,\, \delta V''(s_i)  
\eea

That gives:
\bea
&& \Res_x \Res_{z\to s_i} {K(x,z)\, A_{i,i}\over (x-s_i)^4}\, \delta V(x) \cr
&=& - {1\over 2}   ( 2 \om_i''(s_i) - {1\over \hbar}V'''(s_i)) \delta s_i \,{A_{i,i}}  
 + {1\over \hbar}\, \sum_j \delta_{i,j}\delta V(s_j) \,{A_{i,i} T_{i,i}}   \cr
 && + 2\sum_{j\neq i} {\delta s_j\over s_i-s_j}    \,{A_{i,i} T_{i,i}} 
 +  2\hbar\,  {\delta s_j\over (s_i-s_j)^3 }\,A_{i,i}   +  {1\over 2}\,\, \delta V''(s_i) \,A_{i,i} \cr
&=& {1\over 2}  \delta (T_{i,i})  \,{A_{i,i}} 
  + {1\over \hbar}\, \sum_j \delta_{i,j}\delta V(s_j) \,{A_{i,i} T_{i,i}} 
  + 2\sum_{j\neq i} {\delta s_j\over s_i-s_j}    \,{A_{i,i} T_{i,i}} 
\eea

and thus:
\bea
&& \Res_{x\to s} W_1^{(1)}(x) \delta V(x) \cr
&=& \sum_i {1\over 2}  \delta (T_{i,i})  \,{A_{i,i}} 
  + {1\over \hbar}\, \sum_i\sum_j \delta_{i,j}\delta V(s_j) \,{A_{i,i} T_{i,i}} 
  + 2\sum_{j\neq i} {\delta s_j\over s_i-s_j}    \,{A_{i,i} T_{i,i}} \cr
&& -2 \sum_i\sum_{j\neq i}  {{1\over \hbar}\,\sum_l \delta_{i,l}\delta V(s_l)\, \over (s_i-s_j)^2} \,A_{i,j} 
 -4 \sum_i\sum_{j\neq i}\sum_{l\neq i}  {\delta s_l\, \over (s_i-s_l)(s_i-s_j)^2} \,A_{i,j} \cr
&& +4 \sum_i\sum_{j\neq i} { \delta s_i \over (s_i-s_j)^3} \,A_{i,j} \cr
&=& \sum_i {1\over 2}  \delta (T_{i,i})  \,{A_{i,i}} 
  + {1\over \hbar}\, \sum_i\sum_j \sum_l \delta_{i,j}\delta V(s_j) \,{A_{i,l} T_{l,i}} 
  + 2\sum_{j\neq i} {\delta s_j\over s_i-s_j}    \,{A_{i,i} T_{i,i}} \cr
&&  -4 \sum_i\sum_{j\neq i}\sum_{l\neq i}  {\delta s_l\, \over (s_i-s_l)(s_i-s_j)^2} \,A_{i,j} 
 +4 \sum_i\sum_{j\neq i} { \delta s_i \over (s_i-s_j)^3} \,A_{i,j} \cr
&=&  {1\over 2} \Tr A\,\, \delta T 
  + {1\over \hbar}\, \sum_i\sum_j \sum_l \delta_{i,j}\delta V(s_j) \,{A_{i,l} T_{l,i}} 
  + 2\sum_{j\neq i} {\delta s_j\over s_i-s_j}   \cr
&&  +  4\sum_{j\neq i}\sum_{l\neq i} {\delta s_j\over (s_i-s_j)(s_i-s_l)^2}    \,A_{i,l} 
 -4 \sum_{i\neq j\neq l}  {\delta s_l\, \over (s_i-s_l)(s_i-s_j)^2} \,A_{i,j} \cr
&=&  {1\over 2} \Tr A\,\, \delta T 
  + {1\over \hbar}\, \sum_j \delta V(s_j)  
  - \sum_{j\neq i} {\delta s_i-\delta s_j\over s_i-s_j}   \cr
&=&{1\over 2}\delta \ln{\det{T}}+ {1\over \hbar}\, \sum_j\delta(V(s_j)) 
- {1\over \hbar}\, \sum_j V'(s_j)\delta s_j- \sum_{j\neq i} {\delta s_i-\delta s_j\over s_i-s_j}   \cr
&=&{1\over 2}\delta \ln{\det{T}}+ {1\over \hbar}\, \sum_j\delta(V(s_j)) 
- 2\, \sum_j\sum_{i \neq j} {{\delta s_j}\over{s_j-s_i}}- \sum_{j\neq i} {\delta s_i-\delta s_j\over s_i-s_j}   \cr
&=&{1\over 2}\delta \ln{\det{T}}+ {1\over \hbar}\, \sum_j\delta(V(s_j)) 
- \, \sum_j\sum_{i \neq j} {{\delta s_j-\delta s_i}\over{s_j-s_i}}- \sum_{j\neq i} {\delta s_i-\delta s_j\over s_i-s_j}   \cr
&=&{1\over 2}\delta \ln{\det{T}}+ {1\over \hbar}\, \sum_j\delta(V(s_j))-2 \, \sum_{i \neq j} {{\delta s_j-\delta s_i}\over{s_j-s_i}} 
\eea

That implies:
\bea
F_1 
&=& -{1\over 2}\ln{\det{T}}   - {1\over \hbar}\, \sum_j V(s_j) +2 \,\sum_{i \neq j} \ln(s_i-s_j)
\eea

\beq
\encadremath{
F_1  = {1\over 2}\ln{\det{A}}  - {1\over \hbar}\, \sum_j V(s_j) +2 \,\sum_{i \neq j} \ln(s_i-s_j)
}\eeq

\section*{Appendix: Example $m=1$}
\label{appm1}

We choose $s=0$, and $V'(s) = v_2 s + v_3 s^2 + \sum v_{k+1} s^k$.

We have
\beq
\om(x) = {\hbar\over x}
\eeq
\beq
A= {\hbar\over v_2}
\eeq

\beq
K(x_1,x) = \sum_k K_k(x_1)\,\, x^k
\eeq
\beq
K_0 = {1\over v_2 x_1^2}
\virg
K_1=K_2=0
\eeq
\beq
K_3 = {1\over \hbar x_1^3} - {v_3 \over \hbar v_2 x_1^2}
\eeq

\beq
B(x_1,x_2) = {1\over 2(x_1-x_2)^2} + {A\over x_1^2 x_2^2}
\eeq

\bea
W_3^{(0)} 
&=& {2\hbar\over v_2^2\,x_1^2\,x_2^2\,x_3^2}\,({1\over x_1}+{1\over x_2}+{1\over x_3}) - {2\hbar\, v_3\over v_2^3\,x_1^2\,x_2^2\,x_3^2}
\eea

\bea
W_4^{(0)} 
&=& {6\hbar\over v_2^3\,x_1^2\,x_2^2\,x_3^2}\,({1\over x_1^2}+{1\over x_2^2}+{1\over x_3^2}+{1\over x_4^2}) \cr
&& + {8\hbar\over v_2^3\,x_1^2\,x_2^2\,x_3^2}\,({1\over x_1 x_2}+{1\over x_1 x_3}+{1\over x_1 x_4}+{1\over x_2 x_3}+{1\over x_2 x_4}+{1\over x_3 x_4}) \cr
&& -  {12\hbar v_3 \over v_2^4\,x_1^2\,x_2^2\,x_3^2}\,({1\over x_1}+{1\over x_2}+{1\over x_3}+{1\over x_4}) 
+ {12\hbar\, v_3^2\over v_2^5\,x_1^2\,x_2^2\,x_3^2}
- {6\hbar\, v_4\over v_2^4\,x_1^2\,x_2^2\,x_3^2}
\eea

\bea
W_1^{(1)} 
&=& {1\over \hbar x} + {1\over v_2\, x^3} - {v_3\over v_2^2\, x^2}
\eea

\bea
W_2^{(1)} 
&=& {3\over v_2^2\, x_1^2 x_2^2}\,({1\over x_1^2}+{1\over x_2^2}+{2\over 3\, x_1 x_2})
+ {1\over \hbar\,v_2\, x_1^2\,x_2^2} - {4 v_3\over v_2^3\, x_1^2 x_2^2}\,({1\over x_1}+{1\over x_2}) \cr
&& +{4 v_3^2\over v_2^4\, x_1^2 x_2^2}
-{3 v_4\over v_2^3\, x_1^2 x_2^2}
\eea

\bea
W_3^{(1)} 
&=& {12\over v_2^3\, x_1^2 x_2^2 x_3^2}\,({1\over x_1^3}+{1\over x_2^3}+{1\over x_3^3}) \cr
&& + {12\over v_2^3\, x_1^2 x_2^2 x_3^2}\,({1\over x_1^2 x_2}+{1\over x_2^2 x_3}+{1\over x_3^2 x_1}+{1\over x_1 x_2^2}+{1\over x_2 x_3^2}+{1\over x_3 x_1^2})\cr
&& + {8\over v_2^3\, x_1^3 x_2^3 x_3^3}  
+ {2\over \hbar v_2^2  x_1^2 x_2^2 x_3^2}\,({1\over x_1}+{1\over x_2}+{1\over x_3}) \cr
&& - {24 v_3\over v_2^4  x_1^2 x_2^2 x_3^2}\,({1\over x_1^2}+{1\over x_2^2}+{1\over x_3^2}+{1\over x_1 x_2}+{1\over x_2 x_3}+{1\over x_3 x_1})  
- {2 v_3\over \hbar v_2^3  x_1^2 x_2^2 x_3^2} \cr
&& + {32 v_3^2\over v_2^5  x_1^2 x_2^2 x_3^2} \,({1\over x_1}+{1\over x_2}+{1\over x_3})
- {32 v_3^3\over v_2^6  x_1^2 x_2^2 x_3^2}
- {18 v_4\over v_2^4  x_1^2 x_2^2 x_3^2}\,({1\over x_1}+{1\over x_2}+{1\over x_3})\cr
&& + {42 v_3 v_4\over v_2^5  x_1^2 x_2^2 x_3^2}
- {12 v_5\over v_2^4  x_1^2 x_2^2 x_3^2}
\eea

\bea
W_1^{(2)} 
&=& - {1\over \hbar^3 x} 
+ {3\over \hbar\, v_2^2\, x^5} 
- {5 v_3\over \hbar\, v_2^3\, x^4}
+ {5 v_3^2\over \hbar\, v_2^4\, x^3}
- {5 v_3^3\over \hbar\, v_2^5\, x^2}
- {3 v_4\over \hbar\, v_2^3\, x^3} \cr
&& + {8 v_3\, v_4\over \hbar\, v_2^4\, x^2}
- {3 v_5\over \hbar\, v_2^3\, x^2} \cr
\eea

\bea
W_2^{(2)} 
&=& {15\over \hbar\, v_2^3\, x_1^2 x_2^2}\,({1\over x_1^4}+{1\over x_2^4}+{1\over  x_1^2\, x_2^2})
+ {12\over \hbar\, v_2^3\, x_1^2 x_2^2}\,({1\over x_1^3 x_2}+{1\over x_1 x_2^3 })
- {1\over \hbar^3\, v_2\, x_1^2 x_2^2} \cr
&& - {32 v_3\over \hbar\, v_2^4\, x_1^2 x_2^2}\,({1\over x_1^3}+{1\over x_2^3})
 - {30 v_3\over \hbar\, v_2^4\, x_1^2 x_2^2}\,({1\over x_1 x_2^2}+{1\over x_1^2 x_2})
+ {45 v_3^2\over \hbar\, v_2^5\, x_1^2 x_2^2}\,({1\over x_1^2}+{1\over x_2^2})\cr
&& + {40 v_3^2\over \hbar\, v_2^5\, x_1^3 x_2^3} 
 - {50 v_3^3\over \hbar\, v_2^6\, x_1^2 x_2^2}\,({1\over x_1}+{1\over x_2})
+ {50 v_3^4\over \hbar\, v_2^7\, x_1^2 x_2^2}
- {24 v_4\over \hbar\, v_2^4\, x_1^2 x_2^2}\,({1\over x_1^2}+{1\over x_2^2}) \cr
&& - {18 v_4\over \hbar\, v_2^4\, x_1^3 x_2^3}
+ {64 v_3\, v_4\over \hbar\, v_2^5\, x_1^2 x_2^2}\,({1\over x_1}+{1\over x_2})
- {109 v_3^2\, v_4\over \hbar\, v_2^6\, x_1^2 x_2^2}
+ {24 v_4^2\over \hbar\, v_2^5\, x_1^2 x_2^2} \cr
&& - {18 v_5\over \hbar\, v_2^4\, x_1^2 x_2^2}\,({1\over x_1}+{1\over x_2}) 
 + {50 v_3 \, v_5\over \hbar\, v_2^5\, x_1^2 x_2^2}
- {15 v_6\over \hbar\, v_2^4\, x_1^2 x_2^2}
\eea

\bea
W_1^{(3)} 
&=&  {2\over \hbar^5 x} 
+ {15 \over \hbar^2\, v_2^3\, x^7}
- {3 \over \hbar^3\, v_2^2\, x^5} 
- {35 v_3\over \hbar^2\, v_2^4\, x^6}
+ {5 v_3\over \hbar^3\, v_2^3\, x^4}
+ {50 v_3^2\over \hbar^2\, v_2^5\, x^5}
- {5 v_3^2\over \hbar^3\, v_2^4\, x^3} \cr
&& - {60 v_3^3\over \hbar^2\, v_2^6\, x^4}
+ {5 v_3\over \hbar^3\, v_2^5\, x^2} 
+ {60 v_3^4\over \hbar^2\, v_2^7\, x^3}
- {60 v_3^5\over \hbar^2\, v_2^8\, x^2}
- {24 v_4\over \hbar^2\, v_2^4\, x^5}
+ {3 v_4\over \hbar^3\, v_2^3\, x^3} \cr
&& + {75 v_3 v_4\over \hbar^2\, v_2^5\, x^4} 
 - {8 v_3 v_4\over \hbar^3\, v_2^4\, x^2}
- {125 v_3^2 v_4\over \hbar^2\, v_2^6\, x^3}
+ {185 v_3^3 v_4\over \hbar^2\, v_2^7\, x^2}
+ {24 v_4^2\over \hbar^2\, v_2^5\, x^3}
- {99 v_3 v_4^2\over \hbar^2\, v_2^6\, x^2} \cr
&& - {21 v_5\over \hbar^2\, v_2^4\, x^4}
+ {3 v_5\over \hbar^3\, v_2^3\, x^2} 
 + {56 v_3 v_5\over \hbar^2\, v_2^5\, x^3}
- {106 v_3^2 v_5\over \hbar^2\, v_2^6\, x^2}
+ {45 v_4 v_5\over \hbar^2\, v_2^5\, x^2}
- {15 v_6\over \hbar^2\, v_2^4\, x^3} \cr
&& + {50 v_3 v_6\over \hbar^2\, v_2^5\, x^2}
- {15 v_7\over \hbar^2\, v_2^4\, x^2}
\eea

The free energies are:
\beq
F_1 = {1\over 2} \ln{(v_2/\hbar)}
\eeq

\beq
F_2 = -{5 v_3^2\over 6\hbar\, v_2^3} + {3 v_4\over 4\hbar\, v_2^2}
\eeq

\beq
F_3
= {5 v_3^2\over 6\hbar^3\,v_2^3}
- {5 v_3^4\over \hbar^2\,v_2^6}
- {3 v_4\over 4 \hbar^3\,v_2^2}
+ {25 v_3^2 v_4\over 2 \hbar^2\,v_2^5}
- {3 v_4^2\over \hbar^2\,v_2^4}
- {7 v_3 v_5\over \hbar^2\,v_2^4}
+ {5 v_6\over 2 \hbar^2\,v_2^3}
\eeq

%% file: annexeg.tex
\selectlanguage{french}
\annexe{Topological expansion of the Bethe ansatz and quantum algebraic geometry} \label{Article[IV]}
\selectlanguage{english}

\begin{center}
\vspace{26pt}

\vspace{26pt}

{\sl L.\ Chekhov}\hspace*{0.05cm}${}^1$%\footnote{ E-mail: chekhov@mi.ras.ru }
,
{\sl B.\ Eynard}\hspace*{0.05cm}${}^2$%\footnote{ E-mail: bertrand.eynard@cea.fr }
,
{\sl O.\ Marchal}\hspace*{0.05cm}${}^2$%\footnote{ E-mail: olivier.marchal@cea.fr }
\\
\vspace{6pt}
${}^1\,\,$ Steklov Mathematical Institute, ITEP and Laboratoire Poncelet,\\ 
Moscow, Russia\\ 
\vspace{1pt}
${}^2\,\,$ CEA, IPhT, F-91191 Gif-sur-Yvette, France, \\
CNRS, URA 2306, F-91191 Gif-sur-Yvette, France.\\
\end{center}

\vspace{20pt}
\begin{center}
{\bf Abstract}:
In this article, we solve the loop equations of the $\beta$-random matrix model, in a way similar to what was found for the case of hermitian matrices $\beta=1$.
For $\beta=1$, the solution was expressed in terms of algebraic geometry properties of an algebraic spectral curve of equation $y^2=U(x)$.
For arbitrary $\beta$, the spectral curve is no longer algebraic, it is a Schr\"odinger equation $((\hbar\partial)^2-U(x)).\psi(x)=0$ where $\hbar\propto (\sqrt\beta-1/\sqrt\beta)$.
In this article, we find a solution of loop equations, which takes the same form as the topological recursion found for $\beta=1$.
This allows to define natural generalizations of all algebraic geometry properties, like the notions of genus, cycles, forms of 1st, 2nd and 3rd kind, Riemann bilinear identities, and spectral invariants $F_g$, for a quantum spectral curve, i.e. a D-module of the form $y^2-U(x)$, where $[y,x]=\hbar$.
Also, our method allows to enumerate non-oriented discrete surfaces.
\end{center}

%-----------------------------ABSTRACT--------------------------------------
\vspace{26pt}

%*********************************************************************
%==================== ARTICLE =======================================%******************************************

\section*{1 Introduction}

\subsubsection*{Spectral invariants and algebraic geometry}

In \cite{Eyn1loop,OE}, was presented the definition of spectral invariants $F_g$ for any algebraic plane curve, i.e. given by a polynomial equation 
$$
{\cal E}(x,y)=\sum_{i,j} {\cal E}_{i,j}\, x^i y^j=0.
$$ 

Those invariants $F_g({\cal E})$ are defined in terms of algebraic geometry quantities defined on the Riemann surface of equation ${\cal E}(x,y)=0$. Their definition involves residues at branchpoints of some meromorphic forms.
Their definition provides a natural basis of meromorphic forms of 1st, 2nd and 3rd kind, and a natural framework for all algebraic geometry notions.

Moreover, the invariants $F_g$ of \cite{OE} have many nice properties, for instance their deformations under changes of the complex structure of ${\cal E}$ is given by some "special geometry" relations, and provide a natural form-cycle duality.
Also, they are invariants under changes of ${\cal E}$ which conserve the symplectic form $dx\wedge dy$ in $\mathbb C\times \mathbb C$, they have nice modular properties, and finally, they define the tau-function of some dispersionfull integrable system associated to ${\cal E}$.

Also, those invariants $F_g$ have deep relationships with enumerative geometry, for instance they have been related to the Kodaira-Spencer theory \cite{DVKS}, to combinatorics of discrete surfaces (maps), to intersection theory \cite{OE, EynVolmum}, and they are conjectured to be equal to the Gromov-Witten invariants of some toric Calabi-Yau target 3-folds \cite{BKMP}.

\subsubsection*{Algebraic geometry on "quantum" curves}

Here, our goal is to define those notions for a {\bf "quantum curve"}, where ${\cal E}(x,y)$ is a non-commutative polynomial of $x$ and $y$:
\beq\nonumber
{\cal E}(x,y)=\sum_{i,j} {\cal E}_{i,j}\, x^i\, y^j
\qquad , \quad
[y,x]=\hbar.
\eeq
The notion of quantum curve has arised in many ways in the litterature \cite{DDmodules}, and is also called D-modules, i.e. a space of functions quotiented by ${\rm Ker}\, {\cal E}(x,y)$, where $y=\hbar \partial/\partial x$.

In other words, one has to study functions $\psi(x)$ such that:
\beq\nonumber
{\cal E}(x,\hbar \partial_x).\psi(x)=0.
\eeq

In our attempt to define the spectral invariants analogous to those of \cite{OE} for such D-modules, we are naturally led to define all analogous properties of algebraic geometry.
For instance we define the notions of {\bf branch points}, {\bf sheets}, {\bf genus}, {\bf cycles}, {\bf forms}, {\bf Bergman kernel}, and so on...

Because of non-commutativity, some notions like branch-points, cuts and sheets, become "blurred" or "non-localized", i.e. the branchpoint is  no longer a point, but a "region" of the complex plane, and cuts are asymptotic accumulation lines of points. 

But, otherwise, it is surprising to find that almost all relationships of classical algebraic geometry, remained unchanged when $\hbar\neq 0$, for instance the Riemann bilinear identity, the Rauch variational formula, and the topological recursion defining the spectral invariants.

\medskip

Moreover, we shall find, that in order for our quantities to make sense, we must have a "vanishing monodromy" condition, which can be interpreted as a {\bf Bethe ansatz}, and this gives a geometrical interpretation of the Bethe ansatz.

\medskip

Let us also mention that in a previous article \cite{MoiBertrand}, we treated a special case, where the Schr\"odinger potential $U(x)$ was quantized, and we shall see, under the light of this new work, that it was the case of a degenerate quantum surface, with no branchpoints.

\subsubsection*{Hyperelliptical case}

Here, for simplicity, we shall restrict ourselves  to polynomials of degree $2$ in $y$ (called hyperelliptical in algebraic geometry), of the form:
\beq\nonumber
{\cal E}(x,y) = y^2 - U(x)
\virg
[y,x]=\hbar
\eeq
i.e. to the Schr\"odinger equation:
\beq\nonumber
\hbar^2 \psi'' = U \psi.
\eeq
We leave the higher degree case for a further work.

\subsubsection*{Link with $\beta$ matrix models}

The spectral invariants $F_g$ were first introduced for the solution of loop equations arising in the 1-hermitian random matrix model \cite{Eyn1loop, ChekEynFg}. They were later generalized to other hermitian  multi--matrix models \cite{CEO, Chain}.
%Since then they have been used in many different fields like string theory, combinatorics and intersection theory \cite{Krich, BMHurw, Bor, EOVolWP}. 

There exist other matrix models, which are defined with non hermitian matrices. In fact it is well known since Wigner \cite{Mehta} that depending on the symmetry of the problem, it is sometimes interesting to have matrices that are not hermitian. (For example, real-symmetric, unitary, orthogonal or quaternionic, ...). Therefore, it seems reasonable to extend the definition of the spectral invariants for those other models. 
Those other matrix models are often called $\beta$-ensembles, and they are classified by an exponent $\beta$.
The 3 Wigner ensembles (see \cite{Mehta}, and we changed $\beta\to \beta/2$) correspond to $\beta=1$ (hermitian case), $\beta=1/2$ (real symmetric case), $\beta=2$ (real self-dual quaternion case), but it is easy to define a $\beta$ one-matrix model for any other value of $\beta$ (see section \ref{secMM} for more details).

\medskip

In \cite{ChekEynbeta}, a first attempt to generalize the solution of \cite{Eyn1loop} to other matrix models was proposed, but it was not as nice as the topological recursion of \cite{Eyn1loop}.
In \cite{ChekEynbeta}, it was assumed that $\beta=O(1)$ when the size $N$ of the random matrix becomes large, and it was found that all spectral invariants were related to a double series expansion of the form:
\beq\nonumber
\sum_{g,k}\, N^{2-2g-k}\,\,(\sqrt\beta-1/\sqrt\beta)^k F_{g,k}
\eeq
The coefficients $F_{g,k}$ were computed in \cite{ChekEynbeta}.
Here, in this article, we shall work at fixed $\hbar=(\sqrt\beta-1/\sqrt\beta)/N$, instead of fixed $\beta$, i.e. we shall define the resummed $F_g$'s as:
\beq\nonumber
F_g(\hbar) = \sum_k \, \hbar^k\, F_{g,k}.
\eeq
The $F_{g,k}$'s of  \cite{ChekEynbeta} can be recovered by computing the semi-classical small $\hbar$ expansion of $F_g(\hbar)$.
In this article we shall argue that $F_g(\hbar)$ is the natural generalization of the symplectic invariants of \cite{OE} for a "quantum spectral curve" ${\cal E}(x,y)$ with $[y,x]=\hbar$.

\medskip

The tool which we use for studying the $\beta$-matrix model, is the loop equation method. Loop equations are related to the invariance of an integral under change of variable. They can be obtained by integrating by parts. Loop equations for the $\beta$-matrix model have been written many times \cite{Dum, eynbeta}, and here we show how to solve them order by order in $1/N$, at fixed $\hbar$.

\medskip

The $\beta$-matrix model and its loop equations are explained in section \ref{secMM}.

\section*{2 Schr\"odinger equation and Bethe ansatz}\label{secdef}

\subsection*{2.1 Schr\"odinger equation, generalities and notation}
Let:
\beq\label{eqSchroedinger}
\hbar^2 \psi''(x) = U(x)\, \psi(x)
\eeq
be a Schr\"odinger equation with $U(x)$ a polynomial.
Let $U(x)$ be a polynomial of degree $2d$, and define the polynomial "potential" $V(x)$ of degree $d+1$ by its derivative:
\beq\label{defV'Ann}
V'(x) = 2\,(\sqrt{U})_+ = \sum_{k=0}^d t_{k+1}\,x^k
\eeq
where $()_+$ means the polynomial part of the Laurent series at $x\to\infty$.
We also define:
\beq\label{defPAnn}
P(x) = \frac{V'^2(x)}{4} - U(x) -\hbar \frac{V''(x)}{2}
\eeq
so that $P$ is a polynomial of degree $d-1$.

Eventually, we define:
\beq\label{deft0Ann}
t_0 = \mathop{{\rm lim}}_{x\to\infty}\, \frac{xP(x)}{V'(x)}
\eeq

\br
Just in order to give names to those parameters, let us say that
in the language of integrable systems, the coefficients $t_0,t_1,t_2,\dots,t_{d+1}$ are called the ``Casimirs'', and the remaining coefficients of $P$ are the ``conserved charges''. They will play a special role later on in this article.
In matrix model language (see section \ref{secMM}), $t_1,\dots,t_{d+1}$ are called the times associated to the potential $V(x)$, $t_0$ is often called the temperature, and the remaining coefficients of $P$ are called "filling fractions".
In the language of algebraic geometry, parameters $t_k$ with $k\geq 1$ are coupled to 2nd kind meromorphic differential forms, $t_0$ is coupled to 3rd kind, and the remaining coefficients of $P$ are coupled to 1st kind holomorphic differentials, see section \ref{variations} about form-cycle duality.

\er

\subsubsection*{Stokes Sectors}

From the study of the Schr\"odinger equation we know that the function $\psi(x)$ is subject to the Stokes phenomenon, i.e. although $\psi(x)$ is an entire function, its asymptotics look discontinuous near $\infty$. We therefore need to introduce properly the Stokes sectors by defining the following quantities:
Let
\beq\nonumber
\theta_0= {\rm Arg}(t_{d+1})
\eeq
be the argument of the leading coefficient of $V(x)$.

We define the Stokes lines going to $\infty$ as:
\beq
L_k = \left\{x\,\, / \,\, {\rm Arg}(x) =-\frac{\theta_0}{d+1} + \pi \,\frac{k+\frac{1}{2}}{d+1}\, \right\}
\eeq
Those are the lines where $\Re V(x)$ vanishes asymptotically.

We define the sectors:
\beq
S_k =  \left\{{\rm Arg}(x) \in ]-\frac{\theta_0}{d+1}+ \pi \,\frac{k-\frac{1}{2}}{d+1},-\frac{\theta_0}{d+1}+ \pi\,\frac{k+\frac{1}{2}}{d+1}[\, \right\}
\eeq
i.e. $S_k$ is the sector between $L_{k-1}$ and $L_k$.

Notice that in even sectors we have asymptotically $\Re V(x)>0$ and in odd sectors we have $\Re V(x)<0$.

\begin{center} \label{figstokessectors}
	\includegraphics[height=5cm]{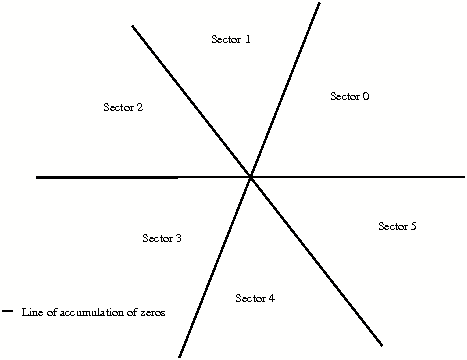}

	Example of sectors for a potential of degree $\deg V=3$, i.e. $d=2$. If $\deg V=d+1$ there are $2d+2$ sectors. 
\end{center}

\subsubsection*{Stokes phenomenon}

Any solution of a linear equation, is analytical where the coefficients of the equation are analytical, and it may possibly have essential singularities where the coefficients are singular. Here, $U(x)$ is an entire function with a singularity (a pole), only at $\infty$, thus, any solution $\psi$ is an entire function with a possible essential singularity at $\infty$. The asymptotics of $\psi$ near $\infty$ are subject to the Stokes phenomenon. This means that, although $\psi$ is analytical in the whole complex plane, its asymptotics at infinity may change from sectors to sectors.

\bigskip

From the study of the Schr\"odinger equation it is known that in each sector $S_k$, $\psi(x)$ has a large $x$ expansion:
\beq\nonumber
\psi(x)  \mathop{{\sim}}_{S_k} \ee{\pm\,{1\over 2\hbar}V(x)}\,x^{C_{k}}\,\, (A_k+\frac{B_k}{x}+\dots)
\eeq
and the sign $\pm$, may jump discontinuously from one sector to another as well as the numbers $A_k, B_k, C_k,\dots$  (and in general, all the coefficients of the series in $\frac{1}{x^j}$ at infinity).

\subsection*{2.3 Decreasing solution}

Let us consider a specific solution $\psi(x)$ of the Schr\"odinger equation which is exponentially decreasing in some even sector at infinity. For writing convenience, \textbf{we will choose $\psi(z)=\psi_0(z)$ a decreasing solution in sector $S_0$. Without further indication, $\psi(z)$ is now understood to be $\psi_0(z)$ in the rest of the article.} Note that this choice is quite arbitrary at the moment, and one should wonder if the quantities we are about to compute depend on this choice, but we are presently not able to answer this question properly, and leave it for further study.

\bigskip

An important and useful result is the Stokes theorem which claims that if the asymptotics of $\psi(x)$ is exponentially small in some sector, then the same asymptotics holds in the two adjacent sectors (and therefore $\psi(x)$ is exponentially large in those two sectors). 

\bigskip

In the general case, (i.e. a generic potential $U(x)$) our solution $\psi(x)$ is decreasing only in sector $0$, and is exponentially large in all other sectors. But if the Schr\"odinger potential $U(x)$ is non-generic (quantized), then there may exist several sectors in which $\psi(x)$ is exponentially small. Due to Stokes theorem, if $\psi$ is exponentially small in some sectors then it must be exponentially large in the adjacent sectors, this implies that there are at most $d+1$ sectors in which $\psi$ is exponentially small.

\medskip
The case studied in \cite{MoiBertrand} was the most degenerate case, such that $\psi$ is exponentially small in $d+1$ sectors.

\subsubsection*{Zeroes of $\psi$}

The main difference with our previous article \cite{MoiBertrand} is that we will not restrict ourselves to the case where $\psi(x)$ is a quasi-polynomial which can only be obtained with very non-generic potential $U(x)$. Here $\psi(x)$ is an entire function with an essential singularity at $\infty$, and with isolated zeroes labelled $s_i$:
\beq
\psi(s_{i})=0
\eeq
\textbf{In particular, the number of zeroes of $\psi$ may be finite or infinite.}

If $\psi(x)$ has an infinite number of zeroes, it is known that the zeroes may only accumulate near $\infty$, and only along the Stokes half-lines $L_j$'s bordering the sectors (see fig.\ref{stokeslines}). In fact, there is an accumulation of zeroes along the half--line $L_j$ if and only if $\psi$ is exponentially large on both sides of the half-line. 

For example in the case where $\psi(x)=\psi_k(x)$ is a solution that exponentially decreases in sector $k$ then it implies that there is no accumulation of zeroes along the half-lines $L_k$ and $L_{k-1}$.

\medskip

If $U(x)$ is generic, then $\psi$ has an infinite number of zeroes, and the zeroes accumulate at $\infty$ along all half-lines $L_j$ with $j\neq 1,2d+1$ (because remember that $\psi$ is implicitely assumed to be $\psi_0$ which decreases in sector $0$), i.e. there are generically $2d$ half-lines of zeroes. The situation is illustrated in fig \ref{stokeslines}.

\begin{center} \label{stokeslines}
	\includegraphics[height=5cm]{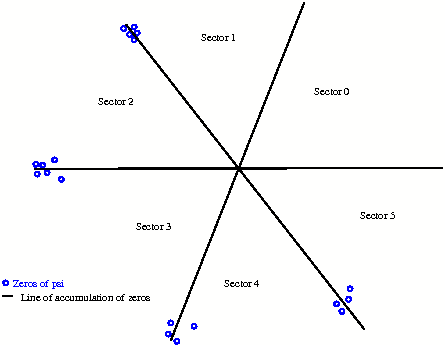}

	The zeroes of $\psi$ accumulate near $\infty$ along the half-lines bordering sectors where $\psi_0$ is exponentially large on both sides. In particular, there is no accumulation of zeroes along $L_0$ and $L_{2d+1}$. 
\end{center}

If $U(x)$ is non-generic (quantized), then there are additional sectors in which $\psi$ is exponentially small, and thus there can be no zeroes accumulating along the two half-lines bordering these sectors. Remember that from Stokes theorem, each time we have a new sector in which $\psi$ is decreasing, we have two half-lines less of zeroes.
Therefore, the number of half-lines of zeroes is always even, and we call it:

\bd The genus $\genus$ of the Schr\"odinger equation is defined by:
\beq
2 {\mathfrak g}+2=\#\,{\hbox{ half-lines\,\, of\,zeroes}}
\eeq
And if $\psi$ has a finite number of zeroes (i.e. there is no half-line of zeroes), we define ${\mathfrak g}=-1$. 
We have
\beq -1\leq \genus\leq d-1\eeq

\ed

\medskip
Note also that the definition of $\genus$ a priori depends on the choice of the solution $\psi=\psi_0$ since two different solutions of the same Schr\"odinger equation may have different numbers of semi-lines of zero accumulation.

An exception, is in the special cases $\genus=-1$ 
%or $\genus=d-1$, 
where it is easy to see that every choice of $\psi=\psi_{2k}$ would give the same value of $\genus$.

Indeed, consider $\genus_{{2k}}$ and $\genus_{{2k'}}$ be the genus defined from the solutions $\psi_{2k}$ exponentially small in sector $S_{2k}$ and $\psi_{2k'}$ exponentially small in sector $S_{2k'}$: 

\smallskip
%$\bullet$ 
if $\genus_{{2k}}=-1$, this means that $\psi_{2k}$ is exponentially small in all even sectors, in particular in sector $S_{2k'}$, and therefore $\psi_{2k}\propto \psi_{2k'}$, and therefore $\genus_{{2k'}}=-1$.

%\smallskip
%$\bullet$ 
%if $\genus_{{2k}}=d-1$, this means that $\psi_{2k}$ is exponentially large in all even sectors except sector $S_{2k}$, in particular $\psi_{2k}$ is exponentially large in  sector $S_{2k'}$.

\subsubsection*{Case $\genus=-1$}

The case $\genus=-1$ was studied in \cite{MoiBertrand}.
This is the case where $\psi$ has only a finite number of zeroes, it is a quasipolynomial:
\beq
\psi(x)\,\ee{{\hbar\over 2}\,V(x)} = {\rm polynomial}.
\eeq

Notice that in order to diminish $\genus$ by $1$, we need to quantize one parameter of $U$, and therefore to reach $\genus$, we need to quantize $d-1-\genus$ parameters.
In particular, to reach $\genus=-1$, we need to quantize $d$ parameters, i.e. $P$ is completely fixed in terms of $V'$, and in particular, $t_0$ is quantized.

In the applications to random matrices, $t_0$ is usually a free parameter (called the temperature) and is never considered quantized, and therefore the case $\genus=-1$ is never obtained in random matrices.

Another way to say that, is that the case $\genus=-1$ has no $\hbar\to 0$ classical limit, and therefore in classical geometry we always have $\genus\geq 0$.

\subsection*{2.4 Resolvent}

The first ingredient of our strategy is to define a resolvent similar to the one in matrix models.

\bd
We define the resolvent for a generic solution $\psi$ by:
\beq
\om(x) = \hbar \frac{\psi'(x)}{\psi(x)} +\frac{V'(x)}{2}
\eeq
\ed

It is clear that this function is analytical except at the zeros of $\psi(x)$ where it has \textbf{simple} poles with residue $\hbar$:
\beq
\om(x) \mathop{{\sim}}_{x\to s_{i}} {\hbar \over x-s_{i}} + {\rm reg}.
\eeq
It also has a possible essential singularity at infinity with the same location of discontinuities as $\psi(x)$. Eventually, note again that the definition of $\om(x)$ depends on the choice of $\psi(x)$.

\subsection*{2.5 Sheets}

In sector $S_k$ we have the asymptotic:
\beq
\psi(x)  \mathop{{\sim}}_{S_k} \ee{{\eta_k\over 2\hbar}V(x)}\,x^{-{\eta_k t_0\over \hbar} - d\,{1+\eta_k\over 2}}\,\, (A_k+\frac{B_k}{x}+\dots)
\eeq
where $\eta_k=\pm 1$. That translates for the resolvent to:
\beq
\om(x) \mathop{{\sim}}_{x\to \infty_k}  {1+\eta_k\over 2}\,(V'(x)-\hbar {d\over x}) - \frac{\eta_k\,t_0}{x}  +O(1/x^2),
\eeq
Therefore it depends if the solution $\psi$ is exponentially big or small in sector $k$ (and of course on the parity of $k$). For a generic $\genus=d-1$ solution which is exponentially big in every sector except $S_0$ (and thus has an alternating sign in the exponential) then $\eta_k= (-1)^k$ (except $\eta_0=-1$). 
%The choice of the $+$ sign in the definition is conventional, it only means in the context of matrix models that the "`physical"' sectors will be the sectors where $\delta_k=0$. 

\bd
We call {\bf "physical sheet"}, the union of sectors where $\eta_k=-1$, in those sectors we have:
\beq
\om(x) \sim {t_0\over x}+O(1/x^2)
\eeq
Notice that the sectors $S_0,S_1$ and $S_{2d+1}$ are always in the physical sheet.

And we call {\bf "second sheet"}, the union of sectors where $\eta_k=+1$, in those sectors we have:
\beq
\om(x) \sim V'(x)+O(1/x)
\eeq

\ed
This definition comes from the analogy with the resolvent in matrix model (see section \ref{secMM} for details).

For a generic potential $U(x)$, all odd sectors are in the physical sheet, and all even sectors except $S_0$ are in the second sheet.

Notice that if $\genus=-1$, there is only the physical sheet, i.e. there is no second sheet.

\subsection*{2.6 The Bethe ansatz}

In the polynomial case studied before \cite{MoiBertrand}, a key ingredient for establishing results was the Bethe ansatz. This ansatz basically deals with the behaviour of $\om(x)$ around zeroes of $\psi$. 
The zeroes of $\psi$ are called "Bethe roots".

The Bethe ansatz can be formulated in many ways. One way to formulate it, is to say that $1/\psi^2$ has no residue at the $s_i$'s:
\beq\label{Betheansatz1}
\Res_{s_i} {1\over \psi^2(x)} = 0
\eeq
 in this way, it will play a key role in defining contour integrals, because all integrals of the type $\int dx/\psi^2(x)$ are insensitive to the exact location integration path with respect to the $s_i$'s, i.e. such integrals will depend only on the homotopy classes of paths.

\smallskip
Equation (\ref{Betheansatz1}) can also be formulated, in a form very similar to the Bethe ansatz in the Gaudin model \cite{Gaudin, BBTbook} as follows:
\bt
The roots $s_i$ of $\psi$ satisfy the Bethe ansatz:
\beq \label{bethe ansatz}
\forall\, i \, , \qquad \quad V'(s_{i}) = 2\, \mathop{{\rm lim}}_{x\to s_i}\,\, \left(\om(x)-{\hbar\over x-s_i}\right).
\eeq

\et

\smallskip

It is a regularized version of the Bethe equation for Gaudin model:
$$
\forall i\, ,\, \qquad V'(s_{i}) \,\, "="\,\, 2\hbar\,\sum_{j\neq i} {1\over s_{i}-s_{j}}
$$
when the number of zeros is infinite and the sum is ill-defined.

\medskip

\proof{This theorem is a classical result and is easy, it just consists in rewriting the Schr\"odinger equation as a Ricatti equation.
We proceed the same way as in \cite{MoiBertrand} and compute:
\bea\label{eqBethe1}
&& V'(x)\om(x) - \om^2(x) - \hbar \om'(x) \cr
&=& V'(x)(\hbar \frac{\psi'(x)}{\psi(x)} +\frac{V'(x)}{2}) - \left({V'(x)^2\over 4}+\hbar V'(x) \frac{\psi'(x)}{\psi(x)}+\hbar^2 \frac{\psi'(x)^2}{\psi^2(x)}\right)\cr
&& -\hbar\left(\hbar{\psi''(x)\over \psi(x)}-\hbar {\psi'^2(x)\over \psi^2(x)} + {V''(x)\over 2}\right) \cr
&=& {V'(x)^2\over 4} - \hbar^2{\psi''(x)\over \psi(x)} -\hbar {V''(x)\over 2} \cr
&=& {V'(x)^2\over 4} - U(x) -\hbar {V''(x)\over 2} \cr
&=& P(x)
\eea
which is a polynomial in $x$, of degree $d-1$.

From its definition, it is clear that $\om^2+\hbar\om'$ has no double pole at the $s_{i}$'s, but it could have simple poles. Consider now a zero $s_{i}$ of $\psi$, and define:
\beq\nonumber
\bar\om_{i}(x) = \om(x) - {\hbar \over x-s_{i}}
\eeq
Then, $\bar\om_{i}(x)$ is regular at $x=s_{i}$, and we may compute $\bar\om_{i}(s_{i})$.
Compute:
\bea
\Res_{x\to s_{i}} \om^2(x)+\hbar \om'(x)
&=& \Res_{x\to s_{i}}\bar\om^2_{i}(x)+2\hbar {\bar\om_{i}(x)\over x-s_{i}} + {\hbar^2\over (x-s_{i})^2}  + \hbar \bar\om'_{i}(x) - {\hbar^2\over (x-s_{i})^2} \cr
&=& \Res_{x\to s_{i}} 2\hbar {\bar\om_{i}(x)\over x-s_{i}} \cr
&=& 2\hbar\, \bar\om_{i}(s_{i})
\eea
On the other hand we have, from \eq{eqBethe1} we have:
\bea
\Res_{x\to s_{i}} \om^2(x)+\hbar \om'(x)
&=& \Res_{x\to s_{i}} V'(x) \om(x) - P(x) \cr
&=& \Res_{x\to s_{i}} V'(x) \om(x)  \cr
&=& \hbar\,V'(s_{i})
\eea

Therefore we find :
\beq\nonumber
\forall\, i \, , \qquad \quad V'(s_{i}) = 2\, \bar\om_{i}(s_{i}).
\eeq
This equation is the Bethe equation for the roots $s_{i}$'s. Note that the potential $V'(x)$ is completely determined by the data of the potential $U(x)$ and does not depend on $\psi$. In particular, in the case where there are only a finite number of $s_{i}$'s we recognize the Bethe equation for Gaudin model \cite{MoiBertrand}:
\beq\nonumber
\forall i\, ,\, \qquad V'(s_{i}) = 2\hbar\,\sum_{j\neq i} {1\over s_{i}-s_{j}}
\eeq
which were completely defining the $s_i$'s.
}

\section*{3 Towards a "Quantum Riemann Surface"}

From the definition of our non-commutative spectral curve (i.e the Schr\"odinger equation), it is tempting to generalize the classical notions kwown in algebraic geometry and Riemann surfaces to our "quantum" case ("quantum" is not to be understood as "quantized" but as "non-commutative" $[y,x]=\hbar$). For a Riemann surface, the central notions are those of cuts, sheets, genus, cycles and meromorphic differentials forms of 1st, 2nd and 3rd kind. In our context, the picture needs a proper adaptation in order to recover the terminology of Riemann surfaces and algebraic geometry. 

In this section we will define the notions of genus, $\acycle$-cycles, $\bcycle$-cycles and the first kind differentials dual to them.
Here, let us assume that $\genus\geq 0$.

\subsection*{3.1 Cuts}

First, we like to think of the 2 sheets, as the sectors which correspond to the 2 possible behaviors of the resolvent at $\infty$: $\om(x) \sim {t_0/x}$ (physical sheet) or $\om(x)\sim V'(x)$ (second sheet).

Then, we consider the cuts as sets of roots $s_i$'s. In some sense, each pair of half lines of accumulation of zeroes can be thought of as a cut.

\bd
We define cuts as pairs of half-lines of zeroes.

There is some arbitrariness in grouping the half-lines of zeroes by pairs.
\ed
There is $\genus+1$ cuts, like in classical algebraic geometry, and notice that the case $\genus=-1$ which has no classical counterpart, has no cuts.

\smallskip

Notice that, contrarily to classical geometry, where the endpoints of the cuts are zeroes of $U(x)$, here the endpoints are somehow blurred, we may move a finite number of $s_i$'s  from one cut to another.

\subsection*{3.2 Cycles}

In standard algebraic geometry, the non-contractible $\acycle$-cycles are often thought of as surrounding cuts in the physical sheet, and their dual $\bcycle$-cycles are going through the cuts, from one sheet to the other, see fig \ref{figclassicalalgebraicgeometry}.

\begin{center} \label{figclassicalalgebraicgeometry}
	\includegraphics[height=5cm]{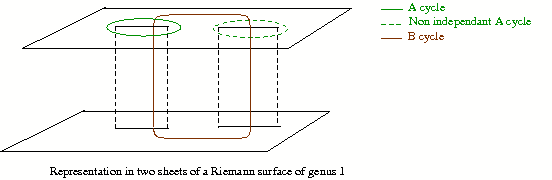}

	\underline{Figure 6}: Representation in two sheets of a Riemann surface of genus $1$.
\end{center}

\subsubsection*{A-Cycles}

Consider the complex plane from which we remove the second sheet (sectors where $\om(x)\sim V'(x)$).
It is clear that it contains $\genus+1$ sectors near $\infty$, and there are $\genus$ homologically linearly independent contours which link them.

\bd
We define $\acycle$-cycles $\acycle_1,\dots,\acycle_{\genus}$ as $\genus$ linearly independent non-contractible contours going from $\infty$ to $\infty$ in the physical sheet.

A choice of $\acycle$-cycles is not unique.
\ed

Remark that this notion really makes sense only for $\genus\geq 1$.

\medskip

Notice that each time $\psi(x)\sim \ee{-V(x)/2\hbar}$ in an even sector, it means it is exponentially small and thus it also behaves like $\ee{-V(x)/2\hbar}$ in the neighboring odd sectors.
That means that we can always choose $\acycle$-cycles going from odd sector to odd sector.

\medskip

Since the first sheet and second sheet are separated by half-lines of accumulations of zeroes, every $\acycle$-cycle surrounds an even number of such half-lines of accumulations of zeroes, i.e. surrounds the cuts in the physical sheet.
Like in standard algebraic geometry, the cuts are identified as pairs of half-lines of zeroes accumulations and the $\cal{A}$ cycles are going enclosing these cuts. 

\subsubsection*{Examples}

In the generic case $\genus=d-1$,  we can define $d$ $\cal{A}$-cycles but only $d-1$ are linearly independent. See picture where $d=7$:
\bigskip

{\mbox{\epsfxsize=14.truecm\epsfbox{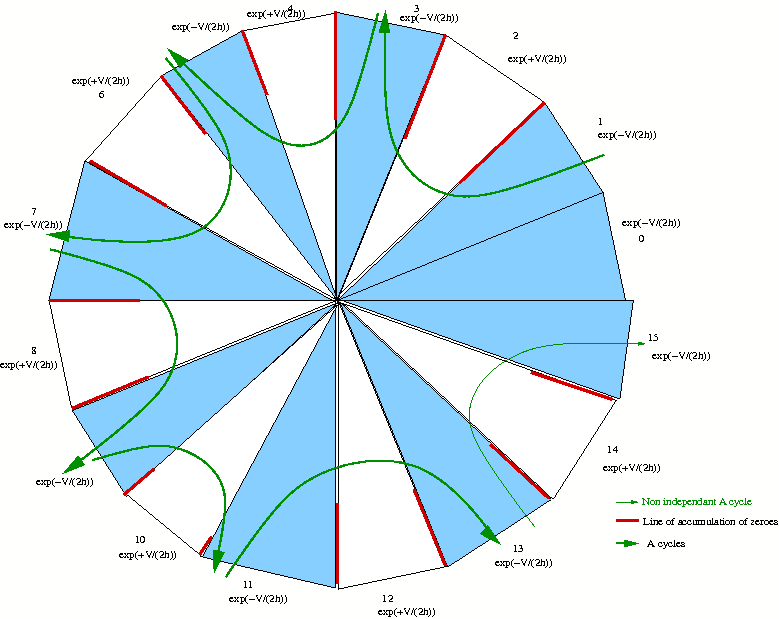}}}

We clearly see that the dashed contour is not linearly independent with the others since the global sum of the contours (dashed included) is contractible in the physical sheet.

For a non-generic case, there are sectors at infinity where $\psi$ is exponentially small. In these cases, the definition of the contours need some adaptations because these sectors correspond to "degenerate" cuts. Here are a few examples of how to deal with these cases. Basically, each time there are two sectors where $\psi$ is small we can replace one of the standard $\cal{A}$ cycle, by a $\hat{ \cal{A}}$ cycle (sometimes called also "degenerate" $\cal{A}$ cycles) that connect them. Here are some examples of the contours in more and more peculiar situations for $d=7$:
\bigskip

{\mbox{\epsfxsize=14.truecm\epsfbox{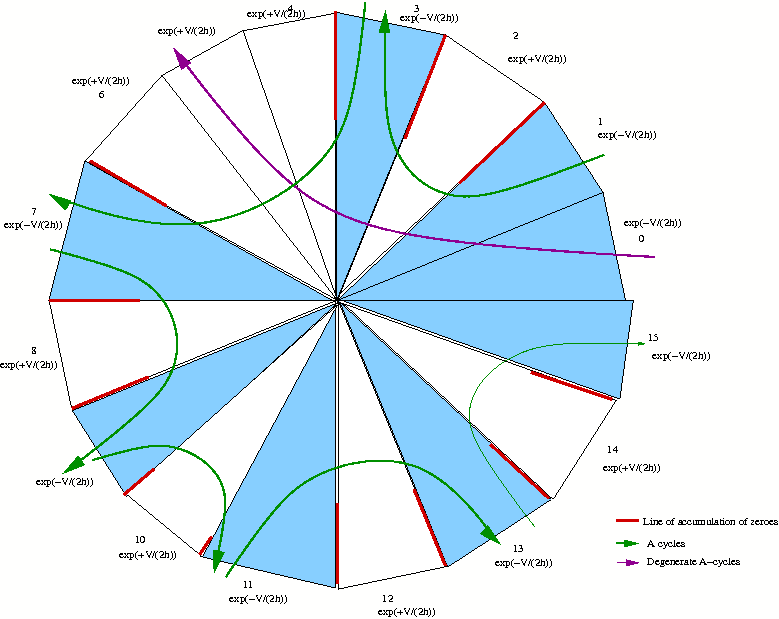}}}

From then it is easy to generalize into more complicated frames:
\bigskip

{\mbox{\epsfxsize=14.truecm\epsfbox{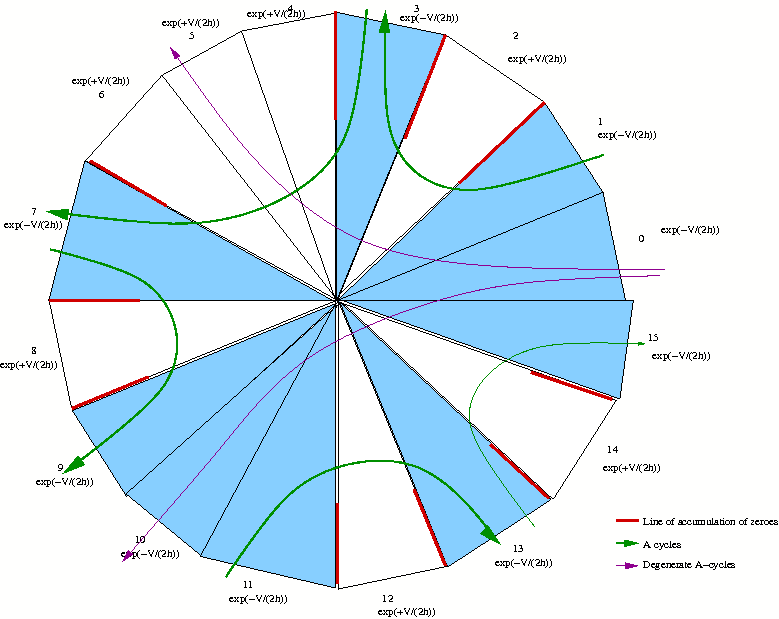}}}

It is then easy to generalize the method in more sophiticated situations. 

In the extreme  case where $\psi$ is exponentially small in all even sectors, there are only $d$ independant  "degenerate" $\hat{ \cal{A}}$ cycles and no $\cal{A}$ cycles, the genus is $\genus=-1$. 
This is the polynomial case studied in \cite{MoiBertrand} where there are no $\cal{A}$ cycles. 

\medskip

From the definitions, it is easy to see that the genus $\genus$ defined above corresponds to the number of independant $\cal{A}$ cycles (we exclude the $\hat{\cal{A}}$ cycles). It is also obvious that the sum of independant $\cal{A}$ and $\hat{ \cal{A}}$ cycles always equals $d-1$.

\subsubsection*{B-Cycles}

As in classical algebraic geometry, it is standard to define the $\cal{B}$ cycles with an origin lying in the non-independant cut. Moreover, although it would be possible to define $\hat{\cal{B}}$ cycles attached to the $\hat{\cal{A}}$ cycles, we prefer limiting ourselves to the definition of $\cal{B}$ cycles attached only to the $\cal{A}$ cycles. Basically, they start from the non-independant cut, goes through their corresponding $\cal{A}$ cycle and end at infinity in the same sector as their corresponding $\cal{A}$ cycle. As there are two sectors in which their corresponding $\cal{A}$ cycle ends, we double them so that one goes into one sector and the other one in the second sector. We also choose the whole so that they intersect only with their corresponding $\cal{A}$-cycles:
\beq
\acycle_\alpha\cap \bcycle_\beta=2 \delta_{\alpha,\beta}
\eeq

This definition is easier understandable with the following pictures: 

Generic case:

{\mbox{\epsfxsize=14.truecm\epsfbox{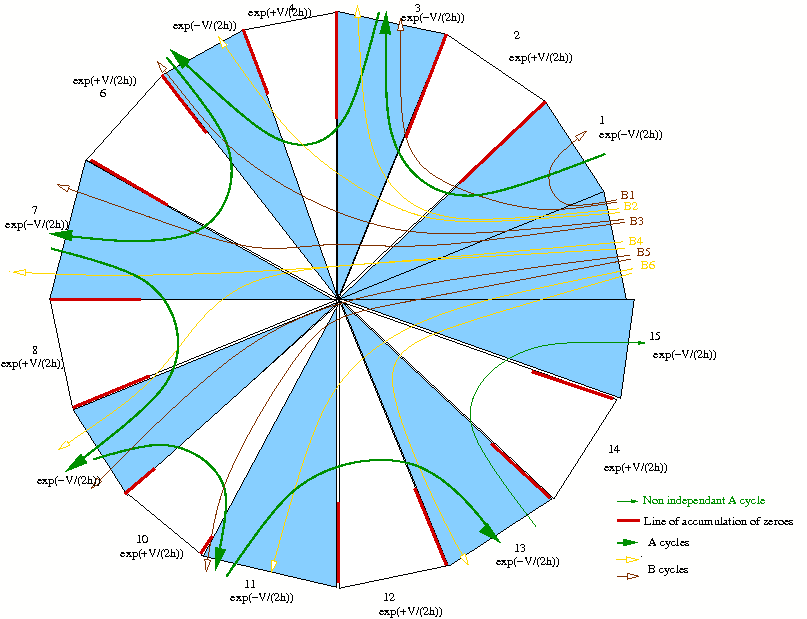}}}

And in a degenerate case:

{\mbox{\epsfxsize=14.truecm\epsfbox{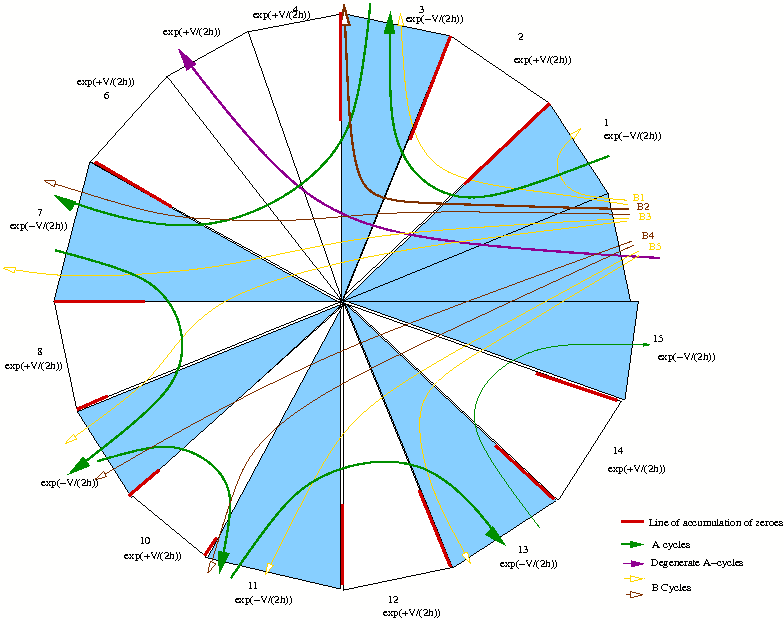}}}

\subsection*{3.3 First kind functions}

After defining the cycles, another important step is to define the equivalent of the first, second and third kind differentials. In this section, we propose a definition of the first kind differentials. 

Let $h_k$, $k=1,\dots,d-1$, be a basis (arbitrary for the moment, but we will choose it orthonormal later on), of the complex vector space of polynomials of degree $\leq d-2$.
To have more convenient notation, we will label the $\hat{\cal{A}}$-cycles as $\cal{A}_\alpha$ , $\genus+1\leq \alpha \leq d-1$ and the standard $\cal{A}$ are labelled $ \cal{A}_\alpha$, $1\leq \alpha \leq \genus$.

Consider the following functions:
\beq
v_k(x) = {1\over \hbar\, \psi^2(x)}\,\int_{\infty_0}^x h_k(x')\,\psi^2(x')\,dx'
\virg
\deg h_k\leq d-2.
\eeq

Notice that, thanks to the Bethe ansatz, $v_{k}(x)$ has double poles with vanishing residues at the $s_{j}$'s (the zeroes of $\psi$), and behaves like $O(1/x^2)$ in sector $S_0$ and in sectors where $\psi$ is exponentially large. (because the polynomial is of degree less than $d-2$).
Therefore, the following integrals are well defined:
\beq
I_{k,\alpha} = \oint_{\acycle_\alpha} v_k(x)\,dx
\virg \alpha=1,\dots,\genus, \, k=1,\dots,d-1.
\eeq

For the degenerate contours $\hat{\cal{A}}_\alpha$, we cannot take the integral since it would not converge. We define instead:
\beq
I_{k,\alpha} = \int_{\hat{\acycle}_\alpha} h_k(x)\, \psi^2(x)\,dx
\virg \alpha=\genus+1,\dots,d-1, \, k=1,\dots,d-1.
\eeq

The matrix $I_{k,\alpha}$ with $k,\alpha=1,\dots,d-1$ is a square matrix, which gives a pairing between the set of paths \{ $\acycle_\alpha,\hat\acycle_\alpha$ \} and the space of polynomials of degree at most $d-2$. Let us choose a basis $h_k$, dual to the $\acycle$-cycles, i.e.:
\beq\label{eqdualvkacycleAnn}
I_{k,\alpha}=\delta_{k,\alpha}.
\eeq

\bigskip

Choosing this set of polynomials gives then the following relations:
\beq \label{orthogonalbasisAnn}
\forall\, i=1,\dots,\genus ,j=1,\dots,d-1\, , \qquad \quad
 \oint_{\acycle_i} v_j(x)\,\, dx = \delta_{i,j}
\eeq
\beq
\forall i=\genus+1,\dots,d-1, \, j=1,\dots,d-1 : \int_{\hat{\acycle}_i} h_j(x)\, \psi^2(x)\,dx =\delta_{j,i}
\eeq

Moreover, from the definitions, we get an asymptotic expression of $v_k(x)$ at infinity:

\bt\label{thvkxlarge}
The functions $v_k(x)$ with $k\leq \genus$ are such that:
\beq
k=1,\dots, \genus, \qquad v_k(x)=O(x^{-2})
\eeq
in all sectors at infinity.

And the functions $v_k(x)$ with $\genus+1 \leq k\leq d-1$ are such that:
\beq
k=\genus+1,\dots, d-1, \qquad v_k(x)=O(x^{-2})
\eeq
in all sectors except in the sector where $\hat\acycle_k$ ends, where we have:
\beq
v_k(x) = {1\over \hbar\psi(x)^2} +O(1/x^2).
\eeq

\et
\proof{
In sector $\infty_0$, we clearly have $v_k(x)\sim O(x^{\deg h_k-d}) = O(x^{-2})$.
%One can multiply in the integral by $V'(x)$ and divide by the same quantity. Integrating by part $V'(x)e^{\frac{\pm V(x)}{2\hbar}}$ gives that in sectors where $\psi$ is exponentially large, we have $v_k(x)\sim O(x^{-2}) + O(1/\psi^2(x)) = O(x^{-2})$. 
And in a sector $S_i$ where $\psi$ is exponentially small we have:
\beq
v_k(x) = {1\over \hbar\psi^2(x)}\,\left[ \int_{\infty_i}^x h_k(x')\,\psi^2(x')\, dx' +  \int_{\infty_0}^{\infty_i} h_k(x')\,\psi^2(x')\, dx'  \right] ,
\eeq
and due to our choice of basis \eq{eqdualvkacycleAnn}, we have
\beq
v_k(x) = {\delta_{i,i_k}\over \hbar\psi(x)^2} +{1\over \hbar\psi^2(x)}\, \int_{\infty_i}^x h_k(x')\,\psi^2(x')\, dx'  = {\delta_{i,i_k}\over \hbar\psi(x)^2} +O(1/x^2),
\eeq
in sector $S_i$.
}

We claim that the function $v_k(x) \, k=1,\dots,\genus$ are the generalization of holomorphic forms (1st kind differentials). 
%Indeed, they have only double poles without residues at the $s_i$'s and behaves as $O(1/{x^2})$ everywhere at infinity, i.e. $v_k(x)dx$ is integrable. Moreover, they are properly normalized on $\cal{A}$ and $\hat{\cal{A}}$ cycles.

\br Classical limit.

The small $\hbar$ BKW expansion $\psi\sim \ee{\pm {1\over 2\hbar}\,\int \sqrt U}$ gives:
\beq
v_k(x) \sim {\pm h_k(x)\over \sqrt{U(x)}} 
\eeq
and $v_k(x)dx$ are indeed the holomorphic forms on the algebraic curve $y^2=U(x)$.

\er

\subsection*{3.4 Riemann matrix of periods}

An interesting quantity in standard algebraic geometry is the Riemann matrix of periods which is the integrations of the holomorphic differentials over $\cal{B}$-cycles. Now that we have defined properly the cycles, we can define a similar ``quantum'' Riemann period matrix $\tau_{i,j}$, $i,j=1,\ldots, \genus$ by:
\beq
\tau_{i,j} \stackrel{{\rm def}}{=} \oint_{\bcycle_i} v_j(x)\,\, dx.
\eeq

Note that this definition makes sense since $v_j(x)$ ($j=1,\dots, \genus$) behaves as $O(1/x^2)$ in the sectors where the $\cal{B}$-cycles go. 
Also, thanks to the Bethe ansatz, $v_j$ has no residue at the roots $s_i$'s, therefore those integrals depend only on the homology class of $\bcycle$-cycles, and not on a representent.

Like for the classical Riemann matrix of periods we have the following property: 

\begin{theorem}
The period matrix $\tau$ is symmetric: $\tau_{i,j}=\tau_{j,i}$. 
\end{theorem}

\proof{
We anticipate on results which shall be proved later, but which don't depend on this theorem.
The proof comes directly from theorem \ref{thBergmanABcycles} below, since:
$$
\oint_{\mathcal{B}_{\beta}} dx \oint_{\mathcal{B}_{\alpha}}B(x,z)dz=2i\pi \oint_{\mathcal{B}_{\beta}} dx v_{\alpha}(x)=2i\pi \tau_{\beta,\alpha}
$$
and from the symmetry theorem \ref{thBergmanSymmetry} for the Bergman kernel $B(x,z)=B(z,x)$:
$$
\oint_{\mathcal{B}_\beta} dx \oint_{\mathcal{B}_\alpha}B(x,z)dz=\oint_{\mathcal{B}_\alpha} dz \oint_{\mathcal{B}_\beta} dx B(x,z)=2i\pi \oint_{\mathcal{B}_\alpha} dz v_{\beta}(z)=2i\pi \tau_{\alpha,\beta}.
$$
}

\subsection*{3.5 Filling fractions}

In random matrices, the notion of filling fractions, is just the $\acycle$-cycle integrals of the resolvent.
%\textcolor{blue}{The notion of filling fractions in random matrices arises from the existence of several cuts in the spectral curve. When several cuts do exist one needs to specify explicitely the repartition of eigenvalues in each cut to completely describe the solution of the problem. This is equivalent to specify the filling fractions that is to say $\acycle$-cycle integrals of the resolvent.}
Here, we easily generalize it by the definition:

\bd
The filling fractions $\epsilon_1,\dots,\epsilon_d$ are defined as follows:
\beq
\alpha=1,\dots,\genus , \qquad
\epsilon_\alpha = {1\over 2i\pi}\,\oint_{\acycle_\alpha} \left(\om(x)-\frac{t_0}{x}\right) +\frac{t_0 n_\alpha}{(d+1)} 
\eeq
where the integer $n_\alpha$ is half the number of Stokes half-lines surrounded by the cycle $\cal{A}_\alpha$. In other words, $\frac{2n_\alpha}{2d+2}$ corresponds to the angular fraction of the complex plane defined by the cycle $\cal{A}_\alpha$.

For $\alpha=\genus+1,\dots,d-1$ we define
\beq
\alpha=\genus+1,\dots,d-1 , \qquad
\epsilon_\alpha =0
\eeq
And for $\alpha=d$, we choose a non-independent $\acycle$-cycle $\acycle_d$, which surrounds all the $s_i$'s which are not surrounded by $\acycle_1,\dots,\acycle_\genus$, and define:
\beq
\epsilon_d = {1\over 2i\pi}\,\oint_{\acycle_d} \left(\om(x)-\frac{t_0}{x}\right) +\frac{t_0 n_d}{(d+1)} 
\eeq

\ed

Note that this definition makes sense because all the cycles $\cal{A_\alpha}$ go from an infinity where $\om(x) -\frac{t_0}{x} \sim O\left(\frac{1}{x^2}\right)$. Note also that this definition depends on the exact locus of the contour $\acycle_\alpha$ and not only on its homotopy class, since $\om(x)$ has simple poles at the $s_i$'s with residue $\hbar$.
If we deform the contour $\acycle_\alpha$, the filling fractions can change by some integer times $\hbar$. 
%In fact the filling fractions can be interpreted as the fraction of $s_i$'s contained in each $\cal{A}$-cycle which defined up to a certain extent the proper end of each cuts (but remember that the choice of the contours $\cal{A}$ is rather arbitrary)

In other words, the filling fractions are  "blurred" when $\hbar\neq 0$, they are defined modulo an integer times $\hbar$. In the classical limit $\hbar\to 0$, they become deterministic.

\medskip
We have:
\bt
\beq
\sum_{\alpha=1}^{d} \epsilon_\alpha = t_0
\eeq
\et

\proof{
When we perform the sum over the contours $\acycle_\alpha$, the contour ${\cal A}_d$ was defined as the "complementary" of the others, i.e. so that the sum is contractible. Since the function $x\to\om(x)- t_0/x$ is integrable at infinity, we find that its global integral is null. With the same argument, it is easy to see that $\sum_{\alpha=1}^{d} n_\alpha=(d+1)$ because we take all Stokes lines once and only once. Therefore we get:
$$
\sum_{\alpha=1}^{d} \epsilon_\alpha=0 + \frac{t_0}{d+1}\sum_{\alpha=1}^{d}n_\alpha=t_0.
$$
Note that it also tells us that only $d-1$ of the epsilon's are independant.
}

\br
In the case $\genus=-1$, the only filling fraction is $\epsilon_d=t_0$, and it is also the sum of residues of $\om$ at the $s_i$'s:
$$
\epsilon_d=t_0=\sum_i \Res_{s_i}\om = \hbar\,\#\{s_i\}
$$
This shows again, that $\genus=-1$ corresponds to a case where $t_0$ is quantized, namely $t_0$ is an integer times $\hbar$:
$$
t_0/\hbar \in \mathbb N.
$$

\er

\section*{4 Kernels}

One of the key geometric objects in \cite{MoiBertrand} and in \cite{OE}, is the "recursion kernel" $K(x,z)$.
It was used in the context of matrix models, to find a solution of loop equations.
Here, it will also allow us to define the 3rd and 2nd kind differentials.

\subsection*{4.1 The recursion kernel $K$}

First we define:
\beq
\hat K(x,z) = {\frac{1}{\hbar}}{1\over \psi^2(x)}\, \int_{\infty_0}^x \psi^2(x')\,{dx'\over x'-z}
\eeq
and for each $\alpha=1,\dots,\genus$, we choose a point $P_\alpha\in \acycle_\alpha$ and  we define:
\beq
\hbar C_\alpha(z) 
=\oint_{\acycle_\alpha}\, {dx''\over \psi^2(x'')}\,\int_{\infty_0}^{P_\alpha} \psi^2(x')\,{dx'\over x'-z}
+\oint_{\acycle_\alpha}\, {dx''\over \psi^2(x'')}\,\int_{P_\alpha}^{x''} \psi^2(x')\,{dx'\over x'-z}
\eeq
where in the last integral, the integration contour between $P_\alpha$ and $x''$, is along $\acycle_\alpha$. This is described in fig.\ref{kernelkdefinition}.

\begin{center} \label{kernelkdefinition}
	\includegraphics[height=5cm]{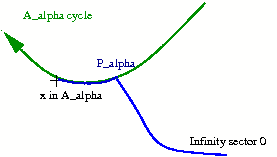}

	Picture of the path of integration used for the definition of the kernel $K(x,z)$.
\end{center}

For each $\alpha=\genus+1,\dots,d-1$, we define:
\beq
C_\alpha(z) 
=\int_{\hat{\cal{A}}_\alpha}\,  \psi^2(x')\,{dx'\over x'-z}.
\eeq

We now need to describe the domain of definition of these functions.

First, one can see that for a fixed $x$, these functions are defined for $z$ outside of some "cuts" (see figure \ref{kernelkdefinition})

$\bullet$ Choose a path between $\infty_0$ and $x$, then $\hat K(x,z)$ is defined for $z$ outside of this path. Across the path $]\infty_0,x]$, $\hat K(x,z)$ has a discontinuity:
\beq
\delta \hat K(x,z) = {2i\pi\over \hbar}\,\,{\psi^2(z)\over \psi^2(x)}
\eeq

$\bullet$ For each $\alpha=1,\dots \genus$, choose a path between $\infty_0$ and $P_\alpha$, then $C_\alpha(z)$ is defined for $z$ outside of this path, and outside $\acycle_\alpha$. 
Across the path $]\infty_0,P_\alpha]$, $C_\alpha(z)$ has a discontinuity:
\beq
\delta C_\alpha(z) = {2i\pi\,\,\psi^2(z)\over \hbar}\,\,\oint_{\acycle_\alpha} {dx''\over \psi^2(x'')}
\eeq
and across the path $\acycle_\alpha$, $C_\alpha(z)$ has a discontinuity:
\beq\label{discKxzAalphaAnn}
\delta C_\alpha(z) = {2i\pi\,\,\psi^2(z)\over \hbar}\,\,\int_{P_\alpha}^z {dx''\over \psi^2(x'')}
\eeq

$\bullet$ For each $\alpha=\genus+1,\dots,d-1$, $C_\alpha(z)$ is defined for $z$ outside of the path $\hat{\acycle}_\alpha$. 
Across the path $\hat{\acycle}_\alpha$, $C_\alpha(z)$ has a discontinuity:
\beq
\delta C_\alpha(z) = {2i\pi\,\,\psi^2(z)}
\eeq

From these remarks, we now define the recursion kernel $K(x,z)$ by:
\bd Definition of the recursion kernel:
\beq
K(x,z) = \hat K(x,z) - \sum_{\alpha=1}^{d-1} v_\alpha(x)\,C_\alpha(z)
\eeq

it is defined for $z$ outside the cuts mentionned above.
\ed

For a fixed $z$, the analytical properties in $x$ of $K(x,z)$ are the same as those of $\hat{K}(x,z)$ since all $v_\alpha(x)$ are analytic. 
For a fixed $z$, the primitive of $\psi^2(x')\,{dx'\over x'-z}$ can be defined locally but not globaly on the complex plane. In fact there is a logarithmic cut to be arbitrarily chosen on $]\infty_0,z]$. Anywhere out of this cut the function $x \to K(x,z)$ is analytic.

\subsubsection*{Properties of kernel $K$}

The definition of the kernel $K(x,z)$ might seem arbitrary at first glance. But in fact, the main reason for the introduction of such kernel is that it has many interesting properties:

It is clear from our definitions that:
\bt\label{thlargeKx}
For a given $z$, the kernel $K$ behaves like:
\beq
K(x,z) \mathop{\sim}\, O(x^{-2})
\eeq
when $x\to \infty$ in all sectors.
\et

\proof{ The result is obvious for sector $S_0$ and for sectors where $\psi$ is exponentially big. When it is not, the fact that we substract $C_\alpha, \alpha=\genus+1,\dots,d-1$ gives the result.}

\bt\label{thKlargez}
We have in all sectors at infinity :
\beq
K(x,z) \mathop{\sim}_{z\to\infty}\, O(z^{-d}).
\eeq
More precisely we have:
\beq
K(x,z) \sim -\,\sum_{k=d-1}^\infty {K_k(x)\over z^{k+1}}
\eeq
with
\beq
\hat K_k(x) = {1\over \hbar \psi^2(x)}\,\int^x_{\infty_0} x'^k\,\psi^2(x')\, dx',
\eeq
and
\beq
K_k(x) = \hat K_k(x) - \sum_{\alpha=1}^{\genus}\, v_\alpha(x)\, \oint_{\acycle_\alpha} \hat K_k(x')\,dx'
- \sum_{\alpha=\genus+1}^{d-1}\, v_\alpha(x)\, \oint_{\acycle_\alpha} \psi^2(x')\,x'^k\,dx'.
\eeq

\et

\proof{
It is clear that
\beq
\hat K(x,z) \sim -\sum_{k=0}^\infty {\hat K_k(x)\over z^{k+1}}
\eeq
where
\beq
\hat K_k(x) = {1\over \hbar \psi^2(x)}\,\int^x_{\infty_0} x'^k\,\psi^2(x')\, dx',
\eeq
and therefore
\beq
K_k(x) = \hat K_k(x) - \sum_{\alpha=1}^{\genus}\, v_\alpha(x)\, \oint_{\acycle_\alpha} \hat K_k(x')\,dx'
- \sum_{\alpha=\genus+1}^{d-1}\, v_\alpha(x)\, \oint_{\acycle_\alpha} \psi^2(x')\,x'^k\,dx'
\eeq
Now, if $k\leq d-2$, notice that $x'^k$ is a polynomial of degree $\leq d-2$, and it is thus a linear combinations of $h_\alpha(x)$'s:
\beq
x'^k = \sum_{\beta=1}^{d-1}\, b_{k,\beta}\, h_\beta(x')
\eeq
This implies:
\beq
\hat K_k(x) = \sum_{\beta=1}^{d-1}\, b_{k,\beta}\, v_\beta(x)
\eeq
Taking now the integral over an $\cal{A}$ cycle and using the normalization choice of $h_k(x)$ gives:
If $\alpha\leq \genus$
\beq
\oint_{\acycle_\alpha} \hat K_k(x')\,dx' = b_{k,\alpha}
\eeq
and if $\alpha> \genus$
\beq
\oint_{\acycle_\alpha} \psi^2(x')\,x'^k\,dx' = b_{k,\alpha}
\eeq
This implies that $K_k(x)=0$ if $k\leq d-2$, and therefore
\beq
K(x,z) = O(z^{-d}).
\eeq

}

\bt\label{thKointAcycles}
Let $\alpha=1,\dots,\genus$, and $z$ on the side of $\acycle_\alpha$ which does not contain $\infty_0$, then:
\beq
\oint_{\acycle_\alpha} K(x,z)\,dx = 0
\eeq
\et

\proof{
 Notice that if $z$ is on that side of $\acycle_\alpha$, we have $C_\alpha(z) = \oint_{\acycle_\alpha} \hat K(x,z)\,dx$, and therefore $\oint_{\acycle_\alpha} K(x,z)\,dx=0$. In fact one can see that the addition of the part with the $C_\alpha(z)$ was just put there to cancel out the $\cal{A}$-cycle integrals.
}

\subsection*{4.2 Third kind differential: kernel $G(x,z)$}

The second important kernel to define is the equivalent of the third kind differential. 
In \cite{MoiBertrand} this kernel was computed from $K$ by derivation, and we use the same definition. 

\bd We define the kernel $G(x,z)$ by:
\beq\label{eqdefGAnn}
G(x,z) =- \hbar\,\psi^2(z)\, \partial_z\, {K(x,z)\over \psi^2(z)}=2\hbar\frac{\psi'(z)}{\psi(z)}K(x,z)-\hbar \partial_z\,K(x,z)
\eeq
\ed

From an easy integration by parts we find:
\bea
G(x,z)
&=& -  {1\over x-z}  + {2\over \psi^2(x)}\,\int^x_{\infty_0} {dx'\over x'-z}\, \psi^2(x') \left( {\psi'(x')\over \psi(x')} - {\psi'(z)\over \psi(z)}  \right) \cr
&& - \hbar \sum_\alpha v_{\alpha}(x)\,  \psi^2(z) \partial_z\, {C_{\alpha}(z)\over \psi^2(z)} \cr
\eea
which shows that near $x=z$ we have $G(x,z)\sim {1\over z-x}$, i.e. there is a simple pole of residue $1$ at $z=x$. Note in particular that ${1 \over x'-z}\left( {\psi'(x')\over \psi(x')} - {\psi'(z)\over \psi(z)}\right)$ has no singularity at $x'=z$ and therefore for a fixed $z$, there is no more any logarithmic cut $]\infty,z]$ as we had for $K(x,z)$.
\smallskip

Note again that a priori, this function of $z$ has the same lines of discontinuity as the kernel $K(x,z)$. 
But notice that the definition of $G$ ensures that all discontinuities of $K$ which are proportional to $\psi^2(z)$ cancel.

\bt\label{thdiscG}

$G(x,z)$ is an analytical function of $x$, with a simple pole at $x=z$ with residue $-1$, and double poles at the $s_{j}$'s (zeros of $\psi(x)$) with vanishing residue, and possibly an essential singularity around $\infty$.

$G(x,z)$ is an analytical function of $z$, with a simple pole at $z=x$ with residue $+1$, simple poles at $z=s_{j}$, and with a discontinuity across $\acycle_\alpha$-cycles with $\alpha=1,\dots,\genus$ (and thus no discontinuity on $\hat{\cal{A}}_\alpha$):
\beq
\delta G(x,z) = -2i\pi \, v_\alpha(x)
\eeq
\et
\proof{
$K(x,z)$ is discontinuous when $z$ crosses either $]\infty_0,x]$, $]\infty_0,P_\alpha]$ or $\acycle_\alpha$. However, the discontinuity of $K(x,z)$ across $]\infty_0,x]$, $]\infty_0,P_\alpha]$, and  $\hat{\cal{A}}_\alpha$ is proportional to $\psi^2(z)$, and this means by derivation that $G(x,z)$ is not discontinuous there.  
Across $\acycle_\alpha$ with $\alpha\leq \genus$, the discontinuity of $K(x,z)$ is given by \eq{discKxzAalphaAnn}, and thus, the discontinuity of $G(x,z)$ is $\delta G(x,z) = -2i\pi \, v_\alpha(x)$.

Since $K(x,z)$ is regular when $z=s_{j}$, then it is clear that $G(x,z)$ has simple poles at $z=s_{j}$, with residue $-2\hbar K(x,s_{j})$.

In the variable $x$, it is clear from the definition and from the Bethe ansatz \ref{bethe ansatz}, that $K(x,z)$ has double poles at $x=s_{j}$ without residue, and this properties follows for $G(x,z)$.

}

\bt\label{thGlargexz}
\beq
G(x,z)=O(1/x^2)
\eeq
when $x\to \infty$ in all sectors. 

And at large $z$ in sector $S_k$:
\beq
\mathop{{\lim}}_{z\to \infty_k}\,\,  G(x,z) = G(x,\infty_k) =  \eta_k\,\,t_{d+1}\,K_{d-1}(x) 
\eeq
where $\eta_k=\pm 1$ is such that $\psi\sim \ee{\eta_k V/2\hbar}$ in sector $S_k$.
\et

\proof{The large $x$ behavior follows from theorem \ref{thlargeKx}. The large $z$ behavior is given by theorem \ref{thKlargez}, i.e. $G(x,z)\sim \eta_k V'(z) K(x,z) \sim \eta_k\, t_{d+1} K_{d-1}(x)$.
The sign depends on the behavior of the solution in this sector. }.

\bt\label{thGointAcycles}
Let $\alpha=1,\dots,\genus$, and $z$ on the side of $\acycle_\alpha$ which does not contain $\infty_0$, then:
\beq
\oint_{\acycle_\alpha} G(x,z)\,dx = 0
\eeq
\et

\proof{
Immediate from theorem \ref{thKointAcycles}}

\subsubsection*{Semi-classical limit}

We claim that this kernel is the quantum version of the third kind differential. Indeed, in classical algebraic geometry a third kind differential is characterized by analyticity except a simple pole with non vanishing residue and a proper normalization on $\acycle$-cycles. Here, apart from the discontinuity along the $\acycle$-cycles which is expected since these contours represent the "quantum cuts", we have analyticity (apart from the $s_i$'s which also define the cuts), a simple pole with residue and a good normalization on $\cal{A}$-cycles. 

\medskip
In the BKW semiclassical expansion we have $\psi \sim \ee{{\pm 1\over \hbar}\int\sqrt{U}}$ and thus
\beq
\hat K(x,z) \sim {2\over x-z}\,{1\over \sqrt{U(x)}}
\eeq
and
\beq
K(x,z) \sim {1\over x-z}\,{1\over 2\sqrt{U(x)}} - \sum_\alpha\, v_\alpha(x)\,C_\alpha(z)
\eeq
and
\beq
G(x,z) \sim 2\sqrt{U(z)} K(x,z) \sim {1\over x-z}\,{\sqrt{U(z)}\over \sqrt{U(x)}} - 2\sum_\alpha\, v_\alpha(x)\,C_\alpha(z)\sqrt{U(z)}
\eeq
The form $G(x,z)dx$ has thus a simple pole at $x=z$, in the physical sheet with residue $+1$ and in the other sheet with residue $-1$, and it is normalized on $\acycle$-cycles $\oint_{\acycle_i} G(x,z)dx=0$.
This is indeed the usual 3rd kind differential in classical algebraic geometry.

\subsection*{4.3 The Bergman kernel $B(x,z)$}

In classical algebraic geometry, the Bergman kernel is the fundamental second kind differential, it is the derivative of the 3rd kind differential, and it is another major tool in classical algebraic geometry. Following the same definition as in \cite{MoiBertrand}, we define:
\beq
B(x,z) = -{1\over 2}\, \partial_z\, G(x,z).
\eeq
The kernel $B$ is going to be called the "quantum" Bergman kernel.

\bt
$B(x,z)$ is an analytical function of $x$, with a double pole at $x=z$ with no residue, and double poles at the $s_{j}$'s with vanishing residues, and possibly an essential singularity around $\infty$.

$B(x,z)$ is an analytical function of $z$, with a double pole at $z=x$ with no residue, and double poles at the $s_{j}$'s with vanishing residues, and possibly an essential singularity around $\infty$.
\textbf{In particular it has no discontinuity along the $\cal{A}$ cycles, it is defined analytically in the whole complex plane except at those double poles.}
\et

\proof{
Those properties follow easily from those of $G(x,z)$ of theorem \ref{thdiscG}. In particular, it is important to notice that the only discontinuity of $G(x,z)$ is along the $\acycle$-cycles, and is independent of $z$, therefore $B(x,z)$ has no discontinuity there.
}

\subsubsection*{Properties of the Bergman kernel}

\bt\label{thBlargexz}
\beq
B(x,z) = O(1/x^2)
\eeq
when $x\to \infty$ in all sectors.

And
\beq
B(x,z) = O(1/z^2)
\eeq
when $z\to \infty$ in all sectors.
\et

\proof{Follows from the large $x$ and $z$ behaviors of $G(x,z)$.}

\bt\label{thloopeqB}
$B$ satisfies the loop equations:
\beq\label{loopeqBxAnn}
(2{\psi'(x)\over \psi(x)}+\partial_x)\,\left(B(x,z)-{1\over 2(x-z)^2}\right) + \partial_z\,{{\psi'(x)\over \psi(x)}-{\psi'(z)\over \psi(z)}\over x-z} = P_2^{(0)}(x,z)
\eeq
where $P_2^{(0)}(x,z)$ is a polynomial in $x$ of degree at most $d-2$.
And
\beq\label{loopeqBzAnn}
(2{\psi'(z)\over \psi(z)}+\partial_z)\,\left(B(x,z)-{1\over 2(x-z)^2}\right) + \partial_x\,{{\psi'(x)\over \psi(x)}-{\psi'(z)\over \psi(z)}\over x-z} =\td{P}_2^{(0)}(z,x)
\eeq
where  $\td{P}_2^{(0)}(z,x)$ is a polynomial in $z$ of degree at most $d-2$.
\et

\proof{This theorem is crucial for all what follows, and its proof is rather non-trivial. Since it is very long and technical, we present the proof in appendix \ref{BergmannLoopEquation}. Those equations are indeed the loop equations for the 2-point function in the $\beta$ matrix model, see section \ref{secMM}.}

\bt\label{thBergmanABcycles}
We have for every $\alpha=1,\dots,\genus$:
\beq
\oint_{\acycle_\alpha} B(x,z)\,dx = 0
\virg
\oint_{\acycle_\alpha} B(x,z)\,dz = 0
\eeq
and
\beq
\oint_{\bcycle_\alpha} B(x,z)\,dz = 2i\pi v_\alpha(x)
\eeq

\et

\proof{
The vanishing of $\acycle$-cycle integrals in the $x$ variable is by construction and can be seen as the consequence of the same result known for $G(x,z)$ on one side of $\cal{A}$ and the fact that $B(x,z)$ has no discontinuity along the $\cal{A}$-cycles. (Therefore, the nullity extend on both sides which no longer need to be treated separately).

For the $z$ variable, notice that if $\acycle_\alpha=]\infty_i,\infty_j[$ goes from $\infty_i$ to $\infty_j$, where both $\infty_i$ and $\infty_j$ are in the physical sheet, we have:
\beq
\oint_{\acycle_\alpha} B(x,z)dz = \int_{\infty_i}^{\infty_j} B(x,z)dz = -{1\over 2}\,(G(x,\infty_j)-G(x,\infty_i))
\eeq
and from theorem \ref{thGlargexz} $G(x,\infty_i) = \eta_i t_{d+1}K_{d-1}(x)$, we get:
\beq
\oint_{\acycle_\alpha} B(x,z)dz = \int_{\infty_i}^{\infty_j} B(x,z)dz = {\eta_i-\eta_j\over 2}\,t_{d+1}\,K_{d-1}(x)
\eeq
and since $\infty_i$ and $\infty_j$ are both in the physical sheet we have $\eta_i=\eta_j=-1$, and therefore
\beq
\oint_{\acycle_\alpha} B(x,z)dz = 0.
\eeq

And similarly, when performing the integral over $\bcycle_\alpha$, the contribution from infinities cancels out since the contour goes in the same sheet. But since $\bcycle_\alpha$ intersects its corresponding $\acycle_\alpha$ (and only this one) where the primitive $-\frac{1}{2}G(x,z)$ is discontinous, the result is the jump of $G(x,z)$ along this $\acycle_\alpha$, that is to say $i\pi v_\alpha(x)$. Eventually, since $\bcycle_\alpha$ and $\acycle_\alpha$ intersect twice, we find \eq{thBergmanABcycles}. 
%\bea
%\oint_{\bcycle_\alpha} B(x,z)\,dz 
%&=& \int_{\infty_0}^{P_{\alpha}-,1}  B(x,z)\,dz + \int^{P_{\alpha}+,1}_{P_{\alpha}+,1}  B(x,z)\,dz + \int^{\infty_{\td\alpha_+,1}}_{P_{\alpha}+,1}  B(x,z)\,dz \cr
%&& +\big(\int_{a}^{P_{\alpha}-,2}  B(x,z)\,dz + \int^{P_{\alpha}+,2}_{P_{\alpha}+,2}  B(x,z)\,dz + \int^{\infty_{\td\alpha_+,2}}_{P_{\alpha}+,2}  B(x,z)\,dz\big) \cr
%&=& -\frac{1}{2} \big(  G(x,\infty_{\td\alpha_+,1})-G(x,\infty_0)+G(x,P_{\alpha}-,1)-G(x,P_{\alpha}+,1)\cr
%&& +G(x,P_{\alpha}-,2)-G(x,P_{\alpha}+,2)+G(x,\infty_{\td\alpha_+,2})-G(x,\infty_0) \big) \cr
%&=& 2i\pi v_\alpha(x) \cr
%\eea
%The key step here is to see that the $2i\pi v_\alpha(x)$ arises because of the discontinuity of the primitive $G(x,z)$ along the $\cal{A}$-cycles. The contributions from infinities cancels out because we are in sectors where the behavior of $G(x,z)$ is always the same. Note that we compensate the factor $\frac{1}{2}$ in the relation between $B$ and $G$, by the fact that the $\cal{A}$ and $\cal{B}$ cycles intersect twice (and so there are $2$ jumps of $i\pi v_\alpha(x)$). This is a little technical difference with the standard algebraic geometry case where the factor $2$ comes directly from the Riemann bilinear identity (which "doubles" all the contours).
}

\medskip

One of our key theorems is:
\bt \label{thBergmanSymmetry}
$B(x,z)$ is symmetric
\beq
B(x,z) = B(z,x)
\eeq

\et

\proof{
The proof relies essentially on the fact that $B(x,z)$ satisfies the loop equation in the two variables. We have:
\bea
&& (2{\psi'(z)\over \psi(z)}+\partial_z)\,(2{\psi'(x)\over \psi(x)}+\partial_x)\, (B(x,z)-{1\over 2(x-z)^2}) \cr
&=& (2{\psi'(z)\over \psi(z)}+\partial_z)\,\Big( P_2^{(0)}(x,z) - \partial_z\,{{\psi'(x)\over \psi(x)}-{\psi'(z)\over \psi(z)}\over x-z} \Big) \cr
&=& (2{\psi'(x)\over \psi(x)}+\partial_x)\,\Big( \td{P}_2^{(0)}(z,x) - \partial_x\,{{\psi'(x)\over \psi(x)}-{\psi'(z)\over \psi(z)}\over x-z} \Big) \cr
\eea
This implies:
\bea
&& (2{\psi'(z)\over \psi(z)}+\partial_z)\, P_2^{(0)}(x,z) -(2{\psi'(x)\over \psi(x)}+\partial_x)\, \td{P}_2^{(0)}(z,x) \cr
&=&  (2{\psi'(z)\over \psi(z)}+\partial_z)\partial_z\,{{\psi'(x)\over \psi(x)}-{\psi'(z)\over \psi(z)}\over x-z}  \cr
&&  - (2{\psi'(x)\over \psi(x)}+\partial_x) \partial_x\,{{\psi'(x)\over \psi(x)}-{\psi'(z)\over \psi(z)}\over x-z}  \cr
&=& 2 {U(x)-U(z)\over (x-z)^2} - {U'(x)+U'(z)\over x-z}
\eea
and therefore:
\bea \label{definitionofR(x,z)}
&& (x-z)^2\,(2{\psi'(z)\over \psi(z)}+\partial_z)\, P_2^{(0)}(x,z) +2U(z)+(x-z)U'(z) \cr
&=& (x-z)^2\,(2{\psi'(x)\over \psi(x)}+\partial_x)\, \td{P}_2^{(0)}(z,x) +2U(x)+(z-x)U'(x) \cr
&\stackrel{{\rm def}}{=}& R(x,z)
\eea
Here, the first line is a polynomial in $x$, whereas the second line is also a polynomial in $z$. Therefore, $R(x,z)$ is a polynomial in both variables, of degree at most $d$ in each variable.
Moreover, we must have:
\beq
R(x,x) = 2U(x)
\eeq
Therefore we must have:
\beq
R(x,z)
= {\frac{1}{\hbar^2}}\left({1\over {2}}\,V'(x)V'(z) -\hbar\,{V'(x) -V'(z)\over x-z} - P(x)-P(z) \right) + (x-z)^2 \td{R}(x,z)
\eeq
where $\td{R}(x,z)$ is a polynomial of both variables of degree at most $d-2$ in each variable.

Putting this back into \ref{definitionofR(x,z)} and using the symmetry $x \leftrightarrow z$ it implies that:
\beq \label{symm} 
 (2{\psi'(z)\over \psi(z)}+\partial_z)\, (P_2^{(0)}(x,z)-\td{P}_2^{(0)}(x,z))
=  \td{R}(x,z)-\td{R}(z,x)
\eeq
Then, we can decompose the r.h.s into the basis $h_\alpha(x)h_\beta(z)$ introduced in \ref{orthogonalbasisAnn}:
\beq \td{R}(x,z)-\td{R}(z,x)=\sum_{\alpha,\beta=1}^{d-1} (\td R_{\alpha,\beta}-\td R_{\beta,\alpha}) h_\alpha(x)h_\beta(z)
\eeq
Integrating the differential equation \eq{symm} then gives:
\beq
P_2^{(0)}(x,z)-\td{P}_2^{(0)}(x,z) = \sum_{\alpha,\beta=1}^{d-1} (\td R_{\alpha,\beta}-\td R_{\beta,\alpha}) h_\alpha(x)\, v_\beta(z) + A_1(x)
\eeq
where $A_1(x)$ is some integration constant.

Then using the loop equations \ref{thloopeqB} we find by substraction that:
\beq \left(2\frac{\psi'(y)}{\psi(y)}+\partial_y\right)\left(B(y,z)-B(z,y)\right)=P_2^{(0)}(y,z)-\td{P}_2^{(0)}(y,z)\eeq
and again, integrating this differential equation we find:
\beq
B(x,z)-B(z,x) = \sum_{\alpha,\beta=1}^{d-1} (\td R_{\alpha,\beta}-\td R_{\beta,\alpha}) v_\alpha(x)\, v_\beta(z) + A(x)+\td A(z)
\eeq
where $(2\psi'/\psi+\partial)A=A_1$, and $\td A(z)$ is some other integration constant.

\smallskip

The large $x$ and large $z$ behavior of $B$ imply that $A(x)=\td A(z)=0$.
We thus get:
\beq \label{difference}
B(x,z)-B(z,x) = \sum_{\alpha,\beta} (\td{R}_{\alpha,\beta}-\td{R}_{\beta,\alpha})\, \td{v}_{\alpha}(x)\td{v}_{\beta}(z)
\eeq

Then, using theorem \ref{thBergmanABcycles}
\beq
\oint_{\acycle_\alpha} B(x,z)dx=0
=\oint_{\acycle_\beta} B(x,z)dz
\eeq
We find:
\beq
\forall \alpha,\beta , \qquad
\td{R}_{\alpha,\beta}=\td{R}_{\beta,\alpha}
\eeq
that is to say by \ref{difference} that the Bergman kernel is symmetric.
}

\bigskip

We claim that all these properties are essential to name this function a "quantum Bergman kernel". Indeed, the symmetry is absolutely necessary and is completely non-trivial. The fact that $B(x,z)$ has no  discontinuity is also essential since in standard algebraic geometry, it is defined everywhere on the Riemann surface. Using all these kernels and their properties, we can then generalize easily the recursion of \cite{Eyn1loop, MoiBertrand} defining the correlation functions.

\subsection*{4.4 Meromorphic forms and properties}

\subsubsection*{Definition of meromorphic forms}

\bd
A meromorphic form ${\cal R}(x)$ is defined as:
\beq
{\cal R}(x) = {1\over \hbar\psi^2(x)}\,\int_{\infty_0}^x\,\, r(x')\,\psi^2(x')\,dx'
\eeq
where $r(x)$ is a rational function of $x$, which behaves at most like $O(x^{d-2})$ at large $x$, and whose poles $r_i$ are such that:
\beq
\Res_{x\to r_i}\, \psi^2(x)\,r(x)=0
\eeq
and for all degenerate $\hat\acycle_\alpha$ cycles
\beq
\int_{\hat\acycle_\alpha} \psi^2(x') r(x')\, dx' = 0.
\eeq

\ed

It is easy to see, that with this definition, the holomorphic forms $v_\alpha(x)$, the kernels $G(x,z)$ and $B(x,z)$ are meromorphic forms of $x$.

\medskip

\subsubsection*{Analiticity properties}

A meromorphic forms ${\cal R}(x)$, has poles at $x=r_i$ the poles of $r(x)$, with degree 1 less than that of $r$, it behaves like $O(x^{-2})$ in all sectors of the physical sheet.
From the Bethe ansatz, it has double poles at the $s_i$'s, with vanishing residues.

In particular, it has an accumulation of poles along the half-lines $L_i$ of accumulations of zeroes of $\psi$.

Also, notice that the following integrals are well defined, and independent of homotopic deformations of $\acycle_\alpha$ (in particular independent of where are the $s_i$'s):
\beq
\oint_{\acycle_\alpha}\, {\cal R}(x)dx.
\eeq

\subsubsection*{\label{integration contour}The integration contours around branch-points}

Let us choose some contour ${\cal C}_i$, such that each ${\cal C}_i$ surrounds (in the trigonometric direction) a half-line $L_i$ of accumulation of zeroes. In other words it surrounds a "branch point".
Let us also assume that $\sum_i {\cal C}_i$ surrounds all roots of $\psi$, i.e. each root of $\psi$ is enclosed in one ${\cal C}_i$.
We also assume that contours ${\cal C}_i$ and $\acycle_\alpha$ do not intersect (they have vanishing intersection numbers):
\beq
\forall i=1,\dots, 2\genus+2,\quad, \forall \alpha=1,\dots,d-1, \qquad \quad
{\cal C}_i\cap \acycle_\alpha = 0
\eeq

\medskip

%Generic case:

{\mbox{\epsfxsize=14.truecm\epsfbox{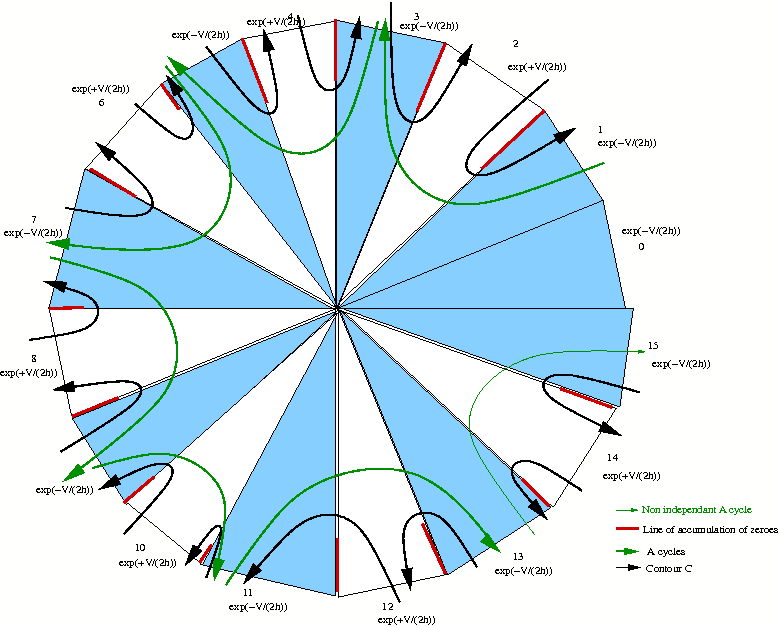}}}

\subsubsection*{Riemann bilinear identity}

For the Riemann bilinear identity, we need the following useful lemma,  which we shall use very often in this article:

\bl \label{NullityOfIntegrals}
For every analytical function $f(x)$ which behaves at infinity at most like $f(x) =O\left(x^{d-2}\right)$ in all directions, and such that it has no singularities inside every contour ${\cal C}_i$ (and thus must be regular at the root $s_{j}$'s) we have, for $x_0$ outside of all $\acycle$-cycles (i.e. on the same side as $\infty_0$) :
$$\forall i\, , \qquad \quad  \,{1\over 2i\pi}\, \oint_{{\cal C}_i}\, dx\,\,  K(x_0,x)\, f(x)=0$$
\el

\proof{
Clearly, the contours ${\cal C}_i$ enclose no singularity of $K(x_0,x)f(x)$ and can be contracted to $0$.
}

\bigskip

Then we can write the bilinear Riemann identity:
\bt {\bf Riemann bilinear identity}

Consider a meromorphic form ${\cal R}(x)$, with poles $r_i$.

Then we have for $x$ outside of all $\acycle$-cycles (i.e. on the same side as $\infty_0$):
\beq
{\cal R}(x) 
=  -\sum_i \Res_{r_i} G(x,z){\cal R}(z)dz +  \sum_{\alpha=1}^{\genus}\, v_\alpha(x)\oint_{\acycle_\alpha} {\cal R}(z)\,dz.
\eeq

\et

\proof{
Since $G(x,z)=1/(z-x)+\dots$, we write Cauchy formula:
\beq
{\cal R}(x) = \Res_{z\to x} G(x,z)\,{\cal R}(z)\,dz
\eeq
and we deform the contour of integration from a small circle around $x$, to contours enclosing all other singularities, i.e. the $r_i$'s and the $s_i$'s.
By doing so, $G(x,z)$ has to cross the $\acycle$-cycles, and picks a discontinuity equal to $2i\pi\,v_\alpha(x)$ i.e. independent of $z$, so the contour integral of the product factorizes for each ${\acycle}_\alpha$.
We thus arrive to:
\bea
{\cal R}(x) 
&=&  -\sum_i \Res_{r_i} G(x,z){\cal R}(z)dz - \sum_i {1\over 2i\pi}\oint_{{\cal C}_i} G(x,z){\cal R}(z)dz \cr
&& +  \sum_{\alpha=1}^{\genus}\, v_\alpha(x)\oint_{\acycle_\alpha} {\cal R}(z)\,dz.
\eea
Then, we need to compute
$$ \oint_{{\cal C}_i} G(x,z){\cal R}(z)dz.$$
Write that $G(x,z)=\psi^2(z)\,\partial_z\, K(x,z)/\psi^2(z)$, and integrate by parts:
$$ \oint_{{\cal C}_i} G(x,z){\cal R}(z)dz = - \oint_{{\cal C}_i} K(x,z)r(z)dz$$
and using lemma \ref{NullityOfIntegrals}, we see that this vanishes.

}

\section*{5 Definition of correlators and free energies}
\label{secdefWngFg}

In this section, we define the quantum deformations of the correlation functions introduced in \cite{Eyn1loop, OE}.
Although the following definitions are inspired from (non hermitian) matrix models (see section \ref{secMM}), they are valid in the present framework of an arbitrary Schr\"odinger equation, not necessarily linked to a matrix model. The special case of their application to matrix models will be discussed in section \ref{secMM}.

\subsection*{5.1 Definition of correlators}

\bd\label{defWng}
We define the following functions $W_n^{(g)}(x_1,\dots,x_n)$ called {\bf $n$-point correlation function of "genus" $g$} by the recursion\footnote{here $g$ is any given integer, it has nothing to do with the genus $\genus$ of the spectral curve.}: 
\beq
W_1^{(0)}(x) = \om(x)
\virg
W_2^{(0)}(x_1,x_2)=B(x_1,x_2)
\eeq
\bea\label{mainrecformulaAnn}
 W^{(g)}_{n+1}(x_0,J)  
&=&   {1\over 2i\pi}\,\sum_{i=1}^{2\genus+2}\, \oint_{{\cal C}_i}\, dx\,\,  K(x_0,x)\, \Big( \ovl{W}_{n+2}^{(g-1)}(x,x,J) \cr
&& + \sum_{h=0}^g\sum'_{I\subset J} {W}_{|I|+1}^{(h)}(x,x_I) {W}_{n-|I|+1}^{(g-h)}(x,J/I) \Big)\cr
%& &+ \sum_{j}
%\partial_{x_j} \left( {{\ovl{W}_n^{(g)}(x,J/\{j\})-{\ovl{W}_n^{(g)}(x_j,J/\{j\})}} \over {(x-x_j)}}\right) \Big) \cr
\eea
where $J$ is a collective notation for the variables $J=\{ x_{1},\dots,x_{n} \}$, and where $\sum\sum'$ means that we exclude the terms $(h=0,I=\emptyset)$ and $(h=g,I=J)$, and where:
\beq
\ovl{W}_{n}^{(g)}(x_1,...,x_n) = W_{n}^{(g)}(x_1,...,x_n) - {\delta_{n,2}\delta_{g,0}\over 2}\, {1\over (x_1-x_2)^2}
\eeq
Here $x_0$ and all the $x_i's$ are outside of the $\acycle$-cycles, i.e. on the same side as $\infty_0$.
The contour ${\cal C}_i$ (defined in section \ref{integration contour}) is a contour which surrounds the branchpoint $L_i$, i.e. a half-line of accumulation of zeroes, and chosen such that every $s_j$ is surrounded by exactly one ${\cal C}_i$, and such that ${\cal C}_i$ doesn't intersect any $\acycle$-cycle.
Very often we shall write
\beq
{\cal C}=\sum_{i=1}^{2\genus+2} {\cal C}_i.
\eeq
\ed

Appart from the precise definition of the kernel $K$, this definition is exactly the same topological recursion as in \cite{OE}, a sum of residues around all branchpoints of the same expression. In other words, the topological recursion is independent of $\hbar$.

\medskip

To shorten equation we will introduce the notation:
\bea
U_n^{(g)}(x,J)&=&\ovl{W}_{n+2}^{(g-1)}(x,x,J)
+ \sum_{I\subset J} \ovl{W}_{|I|+1}^{(h)}(x,x_I) \ovl{W}_{n-|I|+1}^{(g-h)}(x,J/I) \cr
&&+ \sum_{j}
\partial_{x_j} \left( {{\ovl{W}_n^{(g)}(x,J/\{j\})-{\ovl{W}_n^{(g)}(x_j,J/\{j\})}} \over {(x-x_j)}}\right)
\eea

To get:
\bt\label{thWngdefintU}
\beq W^{(g)}_{n+1}(x_0,J) ={1\over 2i\pi}  \oint_{{\cal C}}\, dx\,\,  K(x_0,x)\,U_n^{(g)}(x,J)
\eeq
\et
%This notation is convenient when the precise form of $U(x,J)$ is not required.

\proof{The only difference with the definition, is when we face a term like $B(x,x_j)W_n^{(g)}(x,J/\{j\})$. (note that there are twice this term). It can be split into two terms:
$\ovl{B}(x,x_j)W_n^{(g)}(x,J/\{j\})$ and ${1\over (x-x_j)^2}W_n^{(g)}(x,J/\{j\})$. The second term compensate exactly the $\partial_{x_j}  {{\ovl{W}_n^{(g)}(x,J/\{j\})} \over {(x-x_j)}}$. Thus, the only difference between the two definitions is the term: ${1\over 2i\pi} \oint_{{\cal C}}\, dx\,\,  K(x_0,x)\sum_{j}
\partial_{x_j} {{\ovl{W}_n^{(g)}(x_j,J/\{j\})} \over {(x-x_j)}}$. Therefore the definitions are only the same if these terms are null. This is the case because of Lemma \ref{NullityOfIntegrals}.
}

\subsection*{5.2 Properties of correlators}

The main reason of definition. \ref{defWng}, is because the $W_n^{(g)}$'s have many beautiful properties, which generalize those of \cite{OE}, and in particular they provide a solution of loop equations. We shall prove the following properties:

\bt\label{thpolessiWng}
Each  $W_n^{(g)}(x_1,\dots,x_n)$ with $2-2g-n<0$, is an analytical functions of all its arguments, with poles only when $x_i\to s_{j}$.
Moreover, it vanishes at least as $O\left(1/{x_i^2}\right)$ when $x_i\to\infty$ in all sectors.
It has no discontinuity across $\acycle$-cycles.
\et
\proof{ in appendix \ref{approofthpolessiWng}}

\bt For all $(n,g)\neq (0,0)$ we have
\beq  
\forall \alpha \leq \genus: \label{nullityOfAcycleAnn} \oint_{\cal{A}_\alpha} W_{n+1}^{(g)}(x_0,x_1,...,x_n) dx_1=0
\eeq
\beq  
\forall \alpha \leq \genus: \oint_{\cal{A}_\alpha} W_{n+1}^{(g)}(x_0,x_1,...,x_n) dx_0=0
\eeq
\et

\proof{ We clearly have these properties for $W_2^{(0)}(x_0,x_1)$. By an easy recursion, the first property holds for $x_1,\dots,x_n$. The case of the variable $x_0$ is special and requires explanation. Indeed for fixed values of $x_1,...,x_n$, the dependance in $x_0$ comes from $K(x_0,x)$. The theorem then comes from a permutation of integrals. Indeed, since the contour $\cal{C}$ never crosses any $\cal{A}$-cycles by prescription then we can permute the integrals in $x$ and $x_0$. The nullity of the integral for $K(x_0,x)$ in \ref{thKointAcycles} then gives the result.
}  

\bt\label{thWngPng}
For $2-2g-n<0$, the $W_n^{(g)}$'s satisfy the loop equation, i.e. Virasoro-like constraints.
This means that the quantity:
\bea\label{loopeqPngAnn}
 P_{n+1}^{(g)}(x;x_1...,x_n)
 &=&
2\hbar\frac{\psi'(x)}{\psi(x)}\overline{W}_{n+1}^{(g)}(x,x_1,...,x_n) + \hbar \partial_{x}{\overline{W}_{n+1}^{(g)}(x,x_1...,x_n)} \cr
&& + \sum_{I\subset J} \ovl{W}_{|I|+1}^{(h)}(x,x_I) \ovl{W}_{n-|I|+1}^{(g-h)}(x,J/I) +
\ovl{W}_{n+2}^{(g-1)}(x,x,J)  \cr
& &+ \sum_{j}
\partial_{x_j} \left( {{\ovl{W}_n^{(g)}(x,J/\{j\})-{\ovl{W}_n^{(g)}(x_j,J/\{j\})}} \over {(x-x_j)}}\right) \cr
\eea
is a polynomial in the variable $x$, of degree at most $d-2$.

\et
\proof{ in appendix \ref{approofthWngPng}}

\bt\label{thsym}
 Each $W_n^{(g)}$ is a symmetric function of all its arguments.
\et
\proof{ in appendix \ref{approofthsym}, with the special case of $W_3^{(0)}$ in appendix \ref{approofthW3Krich}.}

\bt\label{thW3Krich}
The 3 point function $W_3^{(0)}$ can also be written:
\beq
W_3^{(0)}(x_1,x_2,x_3) = {4\over 2i\pi}\,\sum_i \oint_{{\cal C}_i}\,\, {B(x,x_1)B(x,x_2)B(x,x_3)\over Y'(x)}\,dx
\eeq
(this can be seen as a quantum version of Rauch variational formula)
\et
\proof{ in appendix \ref{approofthW3Krich}}

\bt\label{thhomogeneity}
For $2-2g-n<0$, $W_n^{(g)}(x_1,\dots,x_n)$ is homogeneous of degree $2-2g-n$:
\beq
\left( \hbar\,{\partial\over \partial \hbar}+ \sum_{j=0}^{d+1} t_j\,{\partial\over \partial t_j} + \sum_{i=1}^\genus \epsilon_i\,{\partial\over \partial \epsilon_i} \right) W_n^{(g)}(x_1,\dots,x_n) = (2-2g-n)\,W_n^{(g)}(x_1,\dots,x_n)
\eeq

\et

\proof{Under a change $t_k\to \l t_k$, $\hbar\to \l \hbar$, $\epsilon_i\to \l\epsilon_i$, the Schr\"odinger equation remains unchanged, and thus $\psi$ is unchanged. The kernel $K$ is changed to $K/\l$ and nothing else is changed. By recursion, $W_n^{(g)}$ is changed by $\l^{2-2g-n}$.}

\section*{6 Deformations \label{variations}}

In this section, we will consider the variations of correlators $W_n^{(g)}$ under infinitesimal variations of the Schr\"odinger potential $U(x)$ or $\hbar$. 
Infinitesimal variations of the resolvent $\om(x)$ can be decomposed on the basis of "meromorphic forms", and forms can be put in duality with cycles. The duality kernel pairing is the Bergman kernel.
We will find in this section, that the classical $\hbar=0$ formulae remain valid for $\hbar\neq 0$, and generalize the corresponding form-cycle duality in special geometry.

\subsection*{6.1 Variation of the resolvent}

Let's consider an infinitesimal polynomial variation:
$$
U\to U+\delta U
\virg
\hbar\to \hbar+\delta \hbar
$$
where $\delta U$ is a polynomial of degree: $\deg \delta U\leq 2d$.
Since we have written $U=V'^2/4 - \hbar V''/2 -P$, we have:
\beq
\delta U = {V'\over 2}\delta V' - {\hbar\over 2}\delta V'' - {\delta \hbar\over 2}\delta V'' - \delta P
\eeq
with
\beq
\delta V'(x) = \sum_{k=1}^{d+1} \delta t_k\,x^{k-1},
\eeq
and $\delta P$ is of degree at most $d-1$:
\beq
\delta P = (t_{d+1} \delta t_0 +t_0 \delta t_{d+1})x^{d-1} + {\rm lower\, degree}.
\eeq

\bigskip

Let us compute $\delta\psi$, or more precisely $f=\delta\ln\psi=\delta\psi/\psi$, let us write it:
\beq
\delta \psi(x)= f(x)\psi(x)
\eeq

The Schr\"odinger equation $\hbar^2\psi''=U\psi$ implies:
\beq\label{deltapsifeq}
\hbar^2 (f\psi)'' - U f \psi = \delta U \psi - \delta\hbar^2\,\psi''
\eeq
i.e.
\beq
\hbar^2(f''\psi+2f'\psi') = (\delta U - 2\,{\delta \hbar\over \hbar}\,U)\,\psi
\eeq
Multiplying by $\psi$ we get:
\beq
\hbar^2(f'\psi^2)' = (\delta U - 2\,{\delta \hbar\over \hbar}\,U)\,\psi^2
\eeq
i.e.:
\beq
\delta (\psi'/\psi) = f'(x) = {1\over \hbar^2\,\psi^2(x)}\,\int_{\infty_0}^x\, \psi^2(x')\, (\delta U(x')- 2\,{\delta \hbar\over \hbar}\,U(x'))\,\, dx'.
\eeq
therefore, since $\om = V'/2+\hbar \psi'/\psi$:
\beq \label{variationw(x)}
\encadremath{
\delta \om(x) = {\delta V'(x)\over 2} + \delta \hbar \, {\psi'(x)\over \psi(x)}  + {1\over \hbar\,\psi^2(x)}\,\int_{\infty_0}^x\, \psi^2(x')\, (\delta U(x')- 2\,{\delta \hbar\over \hbar}\,U(x'))\,\, dx'.
}\eeq

If we write:
\beq
\delta U = {V'\over 2}\delta V'  - {\hbar\over 2}\delta V'' -  {\delta\hbar\over 2} V'' - \delta P
\eeq
where $\delta P$ is of degree at most $d-1$, and $V'/2 = \om-\hbar \psi'/\psi$, we have by integration by parts:
\bea\label{variation2w(x)}
\delta \om(x) 
&=&  {1\over \hbar\,\psi^2(x)}\,\int_{\infty_0}^x\, \psi^2(x')\, \Big(\om(x')\delta V'(x')-\delta P(x') \cr
&& \qquad -\,{\delta \hbar}\,(\om'(x')-{1\over 2}\,V''(x'))\Big)\,\, dx'.
\eea

\subsection*{6.2 Decomposition of variations}

$U(x)$ is a polynomial of degree $2d$, it has $2d+1$ independent coefficients.
If we assume that we have a solution of genus $\genus<d-1$, this means that $U$ is non generic, and satisfies $d-1-\genus$ constraints.
In the space of all possible $U$'s, we shall consider the submanifold corresponding to $U$ of genus $\genus$, which is a submanifold of dimension
\beq
{\rm dim}= d+2+\genus
\eeq
and we shall consider variations of $U$ within that submanifold.
Variations transverse to the genus $\genus$ submanifold, are variations of higher genus and should be computed within a higher genus submanifold.

\medskip

Instead of the $d+2+\genus$ independent coefficients of the polynomial $U$, it is more convenient to choose a system of "flat" coordinates in our genus $\genus$ submanifold, given by:
\beq
t_0,t_1,\dots, t_{d+1},\,\,\epsilon_1,\dots,\epsilon_\genus.
\eeq
We have indeed $d+2+\genus$ coordinates.

\medskip
Let us write the variations as:
\beq
\delta U = \sum_{k=0}^{d+1} U_{t_k}\, \delta t_{k} + \sum_{i=1}^\genus U_{\epsilon_i}\delta \epsilon_i + U_{\hbar} \delta \hbar.
\eeq 

\subsubsection*{Variations relatively to the filling fractions}

For the filling fraction $\delta \epsilon_\alpha$ we have $\delta V'=0$ and thus:
\beq
\delta U(x) = -\delta P(x)
\eeq
where $\deg \delta P\leq d-2$, so we decompose it on the basis of $h_\alpha$'s:
\beq
\delta P(x) = \sum_{\alpha'} c_{\alpha'}\,h_{\alpha'}.
\eeq
and therefore, from \eq{variationw(x)}:
\beq
\delta \om(x) = -\sum_{\alpha'} c_{\alpha'} \,v_{\alpha'}(x).
\eeq
Since $2i\pi \epsilon_{\alpha'} = \oint_{\acycle_{\alpha'}} \om$, we have:
\beq
2i\pi \,\delta_{\alpha,\alpha'} = \oint_{\acycle_{\alpha'}} \delta \om = -\sum_{\alpha''} \oint_{\acycle_{\alpha'}} c_{\alpha''}\,v_{\alpha''}
= -c_{\alpha'}
\eeq
This implies:
\beq
U_{\epsilon_\alpha}(x) = 2i\pi h_\alpha(x)
\eeq
and
\beq \encadremath{
\delta_{\epsilon_i}\om(x) = 2i\pi \, v_\alpha(x) = \oint_{\bcycle_\alpha}\, B(x,z)\,dz.}
\eeq

We shall say that the flat coordinate $\epsilon_\alpha$ is dual to the holomorphic form $v_\alpha$, which is itself dual to the cycle $\bcycle_\alpha$:
\beq
\epsilon_\alpha "=" {1\over 2i\pi}\,\oint_{\acycle_\alpha}\, \om
\qquad , \qquad
\delta_{\epsilon_\alpha} \om = 2i\pi\, v_\alpha = \oint_{\bcycle_\alpha}\, B.
\eeq

\subsubsection*{Variations relatively to $t_0$}

We have:
\beq
\delta U(x) = -\delta P(x) = -t_{d+1}\,x^{d-1} + Q(x)
\eeq
where $\deg Q\leq d-2$.
Using \eq{variationw(x)} we get:
\beq
\delta \om(x) = {1\over \psi^2(x)}\,\int_{\infty_0}^x\, (-t_{d+1}x'^{d-1}+Q(x'))\,\psi^2(x')\,dx'
\eeq
and the polynomial $Q$ is chosen such that $\oint_{\acycle_i} \delta \om=0$ so that when decomposing $Q(x)$ on the basis $v_\alpha(x)$ and performing integrals over $\acycle$-cycles one finds the coefficients of the decomposition as integrals.
Therefore we have:
\bea
\delta \om(x) 
&=& -\,t_{d+1}\,K_{d-1}(x) \cr
&=& -t_{d+1}\,\Big( \hat K_{d-1}(x) - \sum_{\alpha=1}^{\genus}\, v_\alpha(x)\, \oint_{\acycle_\alpha} \hat K_{d-1}(x')\,dx'
- \sum_{\alpha=\genus+1}^{d-1}\, v_\alpha(x)\, \oint_{\acycle_\alpha} \psi^2(x')\,x'^{d-1}\,dx'\Big) \cr
\eea
where
\beq
\hat K_k(x) = {1\over \psi^2(x)}\,\int^x_{\infty_0} x'^k\,\psi^2(x')\, dx',
\eeq
and $K_k(x)$ is the $k^{\rm th}$ term in the large $z$ expansion of $K(x,z)= -\sum_{k=0}^\infty {K_k(x,z)\over z^{k+1}}$ computed in theorem \ref{thKlargez}.
From theorem \ref{thGlargexz} we have
$G(x,\infty_i) = \eta_i t_{d+1}\,K_{d-1}(x) $.
This shows that
\beq
\encadremath{
\delta_{t_0}\om(x)=G(x,\infty_0) = {1\over 2}\,(G(x,\infty_0)-G(x,\infty_-)) = \int_{\infty_0}^{\infty_-} B(x,z)\,dz
}\eeq
where $\infty_0$ is in the physical sheet, and $\infty_-$ is any infinity chosen in the second sheet.

We shall say that the flat coordinate $t_0$ is dual to the 3rd kind meromorphic form $-2G(x,\infty_0)$, which is itself dual to the chain $[\infty_0,\infty_-]$:
\beq \encadremath{
t_0=\Res_{\infty_0} \om
\qquad , \qquad
\delta_{t_{0}} \om = -2G(x,\infty_0) = \int_{\infty_0}^{\infty_-}\, B(x,z)\,dz}
\eeq
where $\Res$ means the coefficient of $1/z$ in the given sector.

\subsubsection*{Variation relatively to $t_k, k=1\dots d$}

For $k=1,\dots, d$ we have:
\beq
U_{t_k}(x) = {V'(x)\over 2}\,x^{k-1} - Q(x)
\virg \deg Q\leq d-2
\eeq
and $Q$ is chosen such that $\oint_{\acycle_i} \delta \om =0$.
Using \eq{variationw(x)} we write:
\beq
\delta \om(x) = \delta \hat\om(x) - \sum_\alpha v_\alpha(x)\,\oint_{\acycle_\alpha} \delta\hat\om(x')\,dx'
\eeq
where
\beq
\delta \hat\om(x)= {1\over \psi^2(x)}\,\int_{\infty_0}^x\, {V'(x')\over 2}\,x'^{k-1}\,\,\psi^2(x')\,dx'
\eeq
Since $V'(x')=\sum_j t_{j+1}\,x'^j$,  we have:
\beq
2 \delta \om(x) =\sum_{j=0}^{d} t_{j+1}\,K_{k+j-1} 
\eeq
Let us compare it with the large $z$ behaviour of $G(x,z)$ in the physical sheet. We have:
\beq
G(x,z) = V'(z)\,K(x,z) +O(z^{-d-1})
\eeq
which means that the large $z$ expansion of $G(x,z)=\sum_k G_k(x)\,z^{-k}$ is given for $k=1,\dots,d$ by:
\beq
G_{k}(x) = -\sum_{j=0}^{d} t_{j+1} K_{k+j-1} 
\eeq
and therefore
\beq
\delta\om(x) = -{1\over 2} G_{k}(x)\,
\eeq
If we write the large $z$ expansion of $B(x,z)$ in the physical sheet, we have
\beq
B(x,z) =\sum_k B_k(x)\,z^{-k-1}\, = -\,{1\over 2}\sum_k k\,G_k(x,z) z^{-k-1}
\eeq
and thus
\beq \encadremath{
\delta_{t_{k}}\om(x) = {1\over k}\,B_{k}(x) = \Res_{\infty_0}\, {z^{k}\over k}\,B(x,z)\,dz}
\eeq

We shall say that the flat coordinate $t_k$ is dual to the 2nd kind meromorphic form ${1\over k}\,B_{k}(x)$, which is itself dual to a residue of $B$.

\subsubsection*{Variations relatively to $t_{d+1}$}

When $k=d+1$, we have a few additional terms of degree $> d-2$:
\beq
U_{t_{d+1}}(x) = {V'(x)\over 2}\,x^{d} -{d\,\hbar\over 2}\,x^{d-1} -t_0 x^{d-1} - Q(x)
\virg \deg Q\leq d-2
\eeq
and $Q$ is chosen such that $\oint_{\acycle_i} \delta \om =0$.
Using \eq{variationw(x)} we write:
\beq
\delta \om(x) = \delta \hat\om(x) - \sum_\alpha v_\alpha(x)\,\oint_{\acycle_\alpha} \delta\hat\om(x')\,dx'
\eeq
where
\beq
\delta \hat\om(x)= {1\over \psi^2(x)}\,\int_{\infty_0}^x\, \left({V'(x')\over 2}\,x'^{d} -{d\,\hbar\over 2}\,x'^{d-1} -t_0 x'^{d-1}\right)\,\,\psi^2(x')\,dx'
\eeq
In other words we have:
\beq
2 \delta \om(x) =\sum_{j=0}^{d} t_{j+1}\,K_{d+j} - d\hbar\, K_{d-1} -2t_0 K_{d-1}
\eeq
Let us compare it with the large $z$ behaviour of $G(x,z)$. We have:
\beq
G(x,z) = (V'(z)-{2t_0\over z})\,K(x,z)-\hbar \partial_z K(x,z) +O(z^{-d-2})
\eeq
which means that the large $z$ expansion of $G(x,z)=\sum_k G_k(x)\,z^{-k}$ is given for $k=d+1$ by:
\beq
G_{d+1}(x) = -\sum_{j=0}^{d} t_{j+1} K_{d+j} + \hbar\,d\,K_{d-1} + 2t_0 K_{d-1} 
\eeq
and therefore
\beq
\delta\om(x) = -{1\over 2} G_{d+1}(x)\,
\eeq
If we write the large $z$ expansion of $B(x,z)$, we have
\beq
B(x,z) =\sum_k B_k(x)\,z^{-k-1}\, = -\,{k\over 2}\sum_k G_k(x,z) z^{-k-1}
\eeq
and thus
\beq
\encadremath{
\delta_{t_{d+1}}\om(x) = {1\over d+1}\,B_{d+1}(x) = \Res_{\infty_0}\, {z^{d}\over d}\,B(x,z)\,dz}
\eeq

We shall say that the flat coordinate $t_{d+1}$ is dual to the 2nd kind meromorphic form ${1\over d+1}\,B_{d+1}(x)$, which is itself dual to a residue of $B$.

\subsection*{6.3 Variation relatively to $\hbar$}

We have:
\beq
\delta_{\hbar} \om(x) 
= -\, {1\over \hbar\,\psi^2(x)}\,\int_{\infty_0}^x\, \psi^2(x')\, \Big( \om'(x')-{1\over 2}\,V''(x')-\delta_\hbar P(x')\Big)\,\, dx'
\eeq
where $\delta_{\hbar} P$ is a polynomial of degree $\leq d-2$ chosen so that $\oint_{\acycle_i} \delta\om=0$.
For the moment, we have not found a good way of writing this expression as an integral with $B$, and we leave that question for a future work.

\subsection*{6.4 Form-cycle duality}

Notice that in all cases, except $\delta_{\hbar}$, there exist a cycle $\delta\om^*$ and a function $\Lambda_{\delta\om}^*$ such that:
\beq
\delta\om(x) = \int_{\delta\om^*}\, B(x,z)\,\Lambda_{\delta\om}^*(z)\,dz.
\eeq
We will use this generic notation later on in order to avoid specifying the $3$ different cases.

Under a suitable reparametrization $z\to z'$ such that $dz'=\Lambda_{\delta\om}^*(z)\,dz$, we say that $\delta\om^*$ in the variable $z'$ is the cycle dual to the "meromorphic form" $\delta\om$.

\subsection*{6.5 Variation of higher correlators}

The following theorem allows to compute the infinitesimal variation of any $W_n^{(g)}$ under a variation of the Schr\"odinger equation.
It tells about the "complex structure deformation" of our quantum Riemann surface.
It can be regarded as special geometry relations.

\bt
Under an infinitesimal deformation $U\to U+\delta U$, 
we have:
\beq
\delta W_n^{(g)}(x_1,\dots,x_n) = \int_{\delta\om^*}\, W_{n+1}^{(g)}(x_1,\dots,x_n,x')\,\Lambda^*(x')\,dx'
\eeq
where $(\delta\om^*,\Lambda_{\delta\om}^*)$ is the dual cycle to the deformation of the resolvent $\om\to \om+\delta \om$.
\et

\proof{
The loop equation  for $W_n^{(g)}(x,J)$ is:
\beq
(2\om(x)-V'(x)+\hbar \partial_x)\,W_n^{(g)}(x,J) + U_n^{(g)}(x,x;J) = P_n^{(g)}(x,J)
\eeq
taking a variation $\delta$ we have:
\bea
&&(2\om(x)-V'(x)+\hbar \partial_x)\,\delta W_n^{(g)}(x,J)
+ (2\delta \om(x)-\delta V'(x)) W_n^{(g)}(x,J)
 + \delta U_n^{(g)}(x,x;J) \cr
 &&=\delta P_n^{(g)}(x,J) 
\eea
notice that $\delta P_n^{(g)}(x,J)$ is a polynomial in $x$, of degree at most $d-2$.

\medskip
On the other hand, consider the loop equation for $W_{n+1}^{(g)}(x,J,x')$ and multiply it by $\Lambda^*(x')$ and integrate $x'$ along $\om^*$, one gets:
\bea
&&(2\om(x)-V'(x)+\hbar \partial_x)\,\int_{\om^*} W_{n+1}^{(g)}(x,J,x')\Lambda^*(x')dx'
 + \int_{\om^*}\delta U_{n+1}^{(g)}(x,x;J,x')\Lambda^*(x')dx' \cr
 &&=  \int_{\om^*}P_{n+1}^{(g)}(x,J,x')\Lambda^*(x')dx'  \cr
\eea

That gives by recursion hypothesis for the computation of $\int_{\om^*}\delta U_{n+1}^{(g)}(x,x;J,x')\Lambda^*(x')dx'$ and using \ref{variationw(x)}:
\bea
&&(2\om(x)-V'(x)+\hbar \partial_x)\left(\int_{\om^*} W_{n+1}^{(g)}(x,J,x')\Lambda^*(x')dx' - \delta W_n^{(g)}(x,J)\right) \cr
 &=& \delta P_n^{(g)}(x,J)-\int_{\om^*}P_{n+1}^{(g)}(x,J,x')\Lambda^*(x')dx' \cr
 &=& \sum_i \alpha_i(J)\, h_i(x)
\eea
where the right hand side is a polynomial of degee at most $d-2$ in $x$, which can be decomposed on the basis $h_i(x)$.

Solving the differential equation gives:
\beq
\int_{\om^*} W_{n+1}^{(g)}(x,J,x')\Lambda^*(x')dx' - \delta W_n^{(g)}(x,J)
= \sum_i \alpha_i(J)\, v_i(x)
\eeq
but since $W_n^{(g)}$ and $W_{n+1}^{(g)}$ are normalized on $\acycle$-cycles, this implies $\alpha_i=0$, i.e.:
\beq
\int_{\om^*} W_{n+1}^{(g)}(x,J,x')\Lambda^*(x')dx' = \delta W_n^{(g)}(x,J)
\eeq

}

\section*{7 Free energies}

We use the variations and theorem \ref{thhomogeneity} to define the $F_g$'s.

Theorem \ref{thhomogeneity} gives:
\beq
(2-2g-n-\hbar\,\partial_{\hbar})\,W_n^{(g)} = \left(t_0\,\partial_{t_0}+\sum_{k=1}^{d+1}\,t_k\,\partial_{t_k}+\sum_{i=1}^\genus \epsilon_i \partial_{\epsilon_i}\right)\,W_n^{(g)}
\eeq
And in the previous section, we have seen how to write the derivatives of $W_n^{(g)}$ as integrals of $W_{n+1}^{(g)}$, that gives:
\beq
(2-2g-n-\hbar\,\partial_{\hbar})\,W_n^{(g)} = \hat H.\,W_{n+1}^{(g)}
\eeq
where $\hat H$ is the linear operator acting as follows:
\beq
\hat H.f(x) = t_0\,\int_{\infty_0}^{\infty_-} f +\sum_{j=1}^{d+1} \Res_{\infty_0}\, {t_{j}\,x^j\over j}\,f + \sum_{i=1}^\genus \epsilon_i\,\oint_{\bcycle_i} f.
\eeq

Those equations allow to define $W_0^{(g)}=F_g$ for $n=0$ and $g\geq 2$ as:

\bd
We define $F_g$ for $g\geq 2$ such that:
\beq
(2-2g-\hbar\,\partial_{\hbar})\,F_g = \hat H.\,W_{1}^{(g)}
\eeq

\ed

\medskip

It would remain to find the correct definitions of $F_0$ (called the prepotential) and $F_1$.
$F_0$ and $F_1$ should be such that under every deformation $\delta=\partial_{t_k},\partial_{t_0}, \partial_{\epsilon_i}$ we should have
\beq
\delta  F_g = H_{\delta}\, W_1^{(g)}.
\eeq
For example $\partial F_0/\partial t_k=\Res x^k \om(x)/k$ i.e. the coefficient of the term $1/x^{k-1}$ in the expansion of $\om(x)$ near $\infty_0$.

\smallskip
We leave the definitions of $F_0$ and $F_1$ for a future work.

\section*{8 Classical and quantum geometry: summary}

Let us summarize the comparison between classical algebraic geometry, and its quantum counterpart introduced here.

\begin{figure}[bth]
\hrule\hbox{\vrule\kern8pt
\vbox{\kern8pt \vbox{
\begin{center}
{\mbox{\epsfxsize=7.truecm\epsfbox{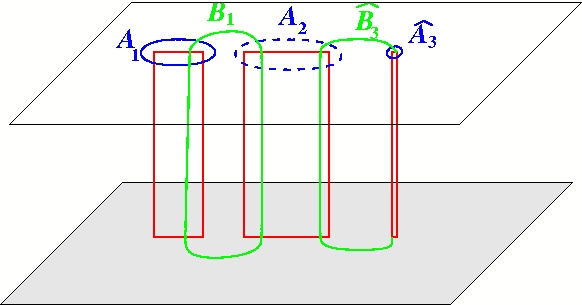}}}
{\mbox{\epsfxsize=7.truecm\epsfbox{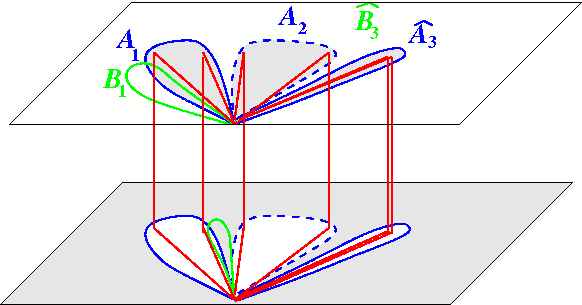}}}
\end{center}
\caption{Classical case $\hbar=0$ of a two sheeted Riemann surface. The branchpoints are paired (in an arbitrary way) to form cuts, and the two sheets are glued along the cuts. Another possibility, is to draw a cut from each branchpoint to $\infty$.
The $\acycle$-cycles surround pairs of branchpoints in the physical sheet.
There are also some degenerate branchpoints, which correspond to cuts of vanishing length.
}
}\kern8pt}
\kern8pt\vrule}\hrule
\end{figure}

\noindent
\begin{tabular}{|l
@{ $\,\, | \,\,$ } l
@{ $\,\, | \,\,$ } 
p{220pt}|}
\hline
\multicolumn{3}{|c|}{
\underline{\bf Summary}} \\
\hline
& {\bf classical $\quad \hbar=0$} & {\bf quantum} \\
\hline
 plane curve:  & 
$E(x,y)=\sum_{i,j} E_{i,j} x^i\,y^j $ &
$E(x,y)=\sum_{i,j} E_{i,j} x^i\,y^j \, , \quad [y,x]=\hbar$ \\
& $E(x,y)=0$ &  $E(x,\hbar \partial_x)\psi=0$  \\
\hline
hyperelliptical  &  $y^2=U(x) $ & $y^2-U(x) \, , \quad [y,x]=\hbar$, 
 \\
plane curve: &  $\deg U=2d$ &
$\hbar^2\psi'' = U\, \psi$  \\
\hline
Potential: &   \multicolumn{2}{c|} { $V'(x)=2(\sqrt{U(x)})_+$ } \\
\hline
2 sheets: & $y\sim_\infty \pm {1\over 2}V'(x) $
& 
$\hbar \psi'/\psi \sim_\infty \pm {1\over 2}V'(x) \sim_{\infty_k}  {\eta_k\over 2}V'(x) $\\
& & choice $\psi=\psi_0\searrow$ in sector $\infty_0$, $\eta_0=-1$\\
\hline
resolvent: & 
$\om(x) = {V'(x)/2} + y $.
& 
$\om(x) = {V'(x)/2} +\hbar {\psi'\over \psi} $.\\
\hline
physical sheet: &  $y\sim_\infty - {1\over 2}V'(x) $, $\om \sim t_0/x$
& 
$\hbar \psi'/\psi \sim_\infty - {1\over 2}V'(x) $, $\om \sim t_0/x$\\
&& sectors where $\psi\sim e^{-{V\over 2\hbar}}$\\
\hline
branchpoints: & simple zeroes of $U(x)$ & half-lines of accumulations \\
& $U(a_i)=0$, $U'(a_i)\neq 0$ & of zeroes of $\psi$ \\
& $i=1,\dots,2\genus+2$ &  $L_i$, $i=1,\dots,2\genus+2$ \\
\hline
genus $\genus$: &  \multicolumn{2}{c|}{$2\genus+2 = \#$ branch points} \\
& \multicolumn{2}{c|}{ $-1\leq \genus\leq d-1$ } \\
\hline
double points: & double zeroes of $U(x)$ & half-lines without accumulations \\
& $U(\hat a_i)=0$, $U'(\hat a_i)= 0$ & of zeroes of $\psi$ \\
\hline
genus $\genus=-1$ & degenerate surface & $\psi\,\ee{V/2\hbar} =$polynomial \\
\hline
$\acycle_\alpha$-cycles  & surround pairs of  & surround pairs of half-lines \\
$\alpha=1,\dots,\genus$  & branchpoints  & of accumulating zeroes \\
\hline
false $\hat\acycle_\alpha$-cycles & surround double  & links 2 sectors \\
$\genus<\alpha<d$ & points  & where $\psi \searrow$ \\
\hline
extra $\acycle_d$-cycle & surrounds last pair of  & surrounds last pair of half-lines  \\
$\alpha=d$ & branchpoints  &  of accumulating zeroes \\
\hline
$\bcycle$-cycles & \multicolumn{2}{c|} { $\acycle_i\cap \bcycle_j=\delta_{i,j}$  } \\
\hline
Holomorphic  & $v_i(x) = {-h_i(x) \over 2\sqrt{U(x)}}\,$ & $v_i(x) = {1\over \hbar \psi^2(x)}\int^x_{\infty_0} \psi^2(x')\,h_i(x')\,dx' $ \\ 
forms, & \multicolumn{2}{c|} {$h_i=$polynomials, $\deg h_i\leq d-2$} \\
1st kind & \multicolumn{2}{c|} {normalized: $\oint_{\acycle_\alpha} v_i(x)\,dx = \delta_{\alpha,i}, \, \alpha=1,\dots,\genus$ }\\
differentials & $h_i(\hat a_\alpha)=-{1\over 2}\delta_{\alpha,i}\,\sqrt{U''(\hat a_\alpha)} $ & $\oint_{\hat\acycle_\alpha} \psi^2 h_i = \delta_{\alpha,i} ,\, \alpha=\genus+1,\dots,d-1$  \\
\hline
Period matrix & \multicolumn{2}{c|} {$\tau_{i,j} = \oint_{\bcycle_j} v_i \quad $, $i,j=1,\dots,\genus$ , $\qquad \tau_{i,j}=\tau_{j,i}$} \\
\hline
Filling fractions & \multicolumn{2}{c|}{ $2i\pi\,\epsilon_\alpha = \oint_{\acycle_\alpha} \om(x)dx\quad $, $\alpha=1,\dots,\genus$, $\qquad \epsilon_{d}=t_0-\sum_{\alpha=1}^\genus \epsilon_\alpha$} \\
\hline
\end{tabular}
\bigskip

\noindent
\begin{tabular}{|l
@{ $\,\, | \,\,$ } l
@{ $\,\, | \,\,$ } 
p{220pt}|}
\hline
\multicolumn{3}{|c|}{
\underline{\bf Summary}} \\
\hline
& classical $\quad \hbar=0$& quantum \\
\hline
3rd kind form &  \multicolumn{2}{c|} {$G(x,z) \sim_{x\to z} 1/(z-x)$} \\
&  \multicolumn{2}{c|} {  $G(x,z) = (2\om(z)-V'(z)-\hbar\partial_z)K(x,z)$  } \\
\hline
Recursion kernel &  \multicolumn{2}{c|} {$K(x,z)$} \\
&  \multicolumn{2}{c|} { $K(x,z) = \hat K(x,z) - \sum_\alpha v_\alpha(x)\,C_\alpha(z)$ } \\
&  \multicolumn{2}{c|} { $C_\alpha(z) = \oint_{\acycle_\alpha}  \hat K(x',z) dx'$ } \\
& $\hat K(x,z) = {1\over z-x}\,{1\over 2\sqrt{U(x)}}$ & $\hat K(x,z) = {1\over \hbar \psi^2(x)}\int^x_{\infty_0} {\psi^2(x')dx'\over x'-z}$ \\
\hline
Bergman kernel & \multicolumn{2}{c|} {$B(x,z) = -{1\over 2}\,\partial_z\, G(x,z)$} \\
2nd kind & \multicolumn{2}{c|} { $B(x,z)\sim 1/2(x-z)^2$}\\
\hline
Symmetry: & \multicolumn{2}{c|} { $B(x,z)=B(z,x)$}\\
\hline  
& \multicolumn{2}{c|} {$\oint_{\acycle_\alpha} B(x,z)dx=0$ } \\
& \multicolumn{2}{c|} {$\oint_{\bcycle_\alpha} B(x,z)dx=2i\pi\, v_\alpha(z)$ } \\
\hline
Meromorphic  & ${\cal R}(x)dx = {r(x)dx\over 2\sqrt{U(x)}}$ & ${\cal R}(x) = {1\over \hbar\psi^2(x)}\,\int_{\infty_0}^x r(x')\,\psi^2(x')\,dx'$ \\
forms & \multicolumn{2}{c|} {$r(x)=$rational with poles $r_i$, $r(x)=O(x^{d-2})$}\\
&  & $\Res_{r_i} r(x')\psi^2(x')=0$\\
\hline
Higher & \multicolumn{2}{c|} {$W_{n+1}^{(g)}(x,J) = \sum_i {1\over 2i\pi}\oint_{{\cal C}_i} K(x,z)\, dz\,\Big( W_{n+2}^{(g-1)}(z,z,J)$ } \\
correlators & \multicolumn{2}{c|} {$\qquad \quad + \sum'_{h+h'=g,\, I\uplus I'=J} W_{1+|I|}^{(h)}(z,I) W_{1+|I'|}^{(h')}(z,I') \Big)$ } \\
& \multicolumn{2}{c|} { where ${\cal C}_i$ surrounds the branchpoint $L_i$} \\
\hline
Symmetry & \multicolumn{2}{c|} {$W_{n}^{(g)}(x_1,x_2,\dots,x_n) = W_{n}^{(g)}(x_{\sigma(1)},x_{\sigma(2)},\dots,x_{\sigma(n)})\, , \qquad\sigma\in S_n$} \\
\hline
Variations and & \multicolumn{2}{c|} {$U(x) \to U(x)+ \delta U(x)$}\\
dual cycle & \multicolumn{2}{c|} {  $\delta U^*$: $\delta \om(x) = \int_{\delta U^*} B(x,x') \,\, \Lambda_{\delta U}(x')\,\,dx'$}\\
\hline
$\delta V'=\sum \delta t_{k}\, x^{k-1}$ & \multicolumn{2}{c|} {$\delta_{t_k} \om(x) = \Res_\infty B(x,x')\,{x'^k\over k}\,dx' $} \\
\hline
variation $\delta t_0$ & \multicolumn{2}{c|} {$\delta_{t_0} \om(x) = \int_{\infty_0}^{\infty_-} B(x,x')\,dx' $} \\
\hline
variation $\delta \epsilon_i$ & \multicolumn{2}{c|} {$\delta_{\epsilon_i} \om(x) = \oint_{\bcycle_i} B(x,x')\,dx' $} \\
\hline
%variation $\delta \hbar$ & \multicolumn{2}{c|} {$\delta_{\hbar} \om(x) = \oint_{?} B(x,x')\,dx' $} \\
%\hline
Variations of & \multicolumn{2}{c|} {$\delta W_{n}^{(g)}(x_1,\dots,x_n) = \int_{\delta U^*} W_{n+1}^{(g)}(x_1,\dots,x_n,x') \,\, \Lambda_{\delta U}(x')\,\, dx'$}\\
higher correlators & \multicolumn{2}{c|} {} \\
\hline
Rauch formula & \multicolumn{2}{c|} { $W_3(x_1,x_2,x_3) = \oint_{{\cal C}} {B(x_1,z)B(x_2,z)B(x_3,z)\over 4 Y'(z)} \,\, dz$ } \\
\hline
\end{tabular}

\section*{9 Application: Matrix models \label{secMM}}

The reason why we introduced those $W_n^{(g)}$'s is because they satisfy the loop equations for $\beta$-random matrix ensembles.

Consider a (possibly formal) matrix integral:
\beq
Z = \int_{E_{N,\beta}}\, dM\,\, \ee{-{N\sqrt\beta\over t_0}\,\tr V(M)}
\eeq
where $V(x)$ is some polynomial, and 
where $E_{N,1}=H_N$ is the set of hermitian matrices of size $N$, $E_{N,1/2}$ is the set of real symmetric matrices of size $N$ and $E_{N,2}$ is the set of quaternion self dual matrices of size $N$ (see \cite{Mehta}).

Alternatively, we can integrate over the angular part and get an integral over eigenvalues only \cite{Mehta}:
\beq
Z= \int d\lambda_1\dots d\lambda_N\,\, \Delta(\lambda)^{2\beta}\,\, \prod_{i=1}^N\, \ee{-{N\sqrt\beta\over t_0}\,V(\lambda_i)}
\eeq
where $\Delta(\lambda)=\prod_{i<j}(\l_j-\l_i)$ is the Vandermonde determinant.

This allows to generalize the matrix model to arbitrary values of $\beta$.
In particular, we shall choose $\beta$ of the form:
\beq
\sqrt \beta = {\hbar N\over 2 t_0}\,\left( 1 \pm \sqrt{1+{4 t_0^2\over \hbar^2\,N^2}}\right)
\eeq
i.e.
\beq
\hbar = {t_0\over N}\,\left(\sqrt\beta-{1\over \sqrt\beta}\right).
\eeq
Notice that $\hbar=0$ correspond to the hermitian case $\beta=1$, and $\hbar\to -\hbar$ corresponds to $\beta\to 1/\beta$.

\subsection*{9.1 Correlators and loop equations}

Then we define the correlators:
\beq
W_k(x_1,\dots,x_k) =  \beta^{k/2}\,\, \left< \sum_{i_1,\dots,i_k} {1\over x_1-\lambda_{i_1}} \dots {1\over x_k-\lambda_{i_k}} \right>_c
\eeq
and
\beq
W_0 = F = \ln{Z}.
\eeq

And we assume (this is automatically true if we are considering formal matrix integrals), that there is a large $N$ expansion of the type (where we assume $\hbar=O(1)$):
\beq\label{WkgdevtopMM}
W_k(x_1,\dots,x_k) = \sum_{g=0}^\infty (N/t_0)^{2-2g-k} W_k^{(g)}(x_1,\dots,x_k)
\eeq
\beq
W_0 = F = \sum_g (N/t_0)^{2-2g} W_0^{(g)} = \sum_g (N/t_0)^{2-2g} F_g.
\eeq
The loop equations are obtained by integration by parts, for example:
\beq
0 = \sum_i \int d\l_1\dots d\l_N {\partial\over \partial \l_i}\left( {1\over x-\l_i}\, \Delta(\l)^{2\beta}\, \prod_j \ee{-{N\sqrt\beta\over t_0}\,V(\l_j)}\right)
\eeq
gives:
\bea
0 &=& \sum_i \left<{1\over (x-\l_i)^2} + 2\beta\sum_{j\neq i}{1\over x-\l_i}{1\over \l_i-\l_j} - {N\sqrt\beta\over t_0}\,{V'(\l_i)\over x-\l_i}\right> \cr
&=& \sum_i \left<{1\over (x-\l_i)^2} + \beta\sum_{j\neq i}{1\over x-\l_i}{1\over x-\l_j} - {N\sqrt\beta\over t_0}\,{V'(\l_i)\over x-\l_i}\right> \cr
&=& \sum_i \left<{1-\beta\over (x-\l_i)^2} + \beta\sum_{j}{1\over x-\l_i}{1\over x-\l_j} - {N\sqrt\beta\over t_0}\,{V'(\l_i)\over x-\l_i}\right> \cr
&=& (\beta-1){1\over \sqrt\beta} W'_1(x) + \beta ({1\over \beta}W_1^2(x) + {1\over \beta}W_2(x,x)) \cr
&& \qquad - {N\sqrt\beta\over t_0}\, \left({1\over\sqrt\beta} V'(x)W_1(x) - \sum_i \left< {V'(x)-V'(\l_i)\over x-\l_i}\right>\right) \cr
\eea
We define the polynomial
\beq
P_1(x) = {\sqrt\beta}\, \sum_i \left< {V'(x)-V'(\l_i)\over x-\l_i}\right> = (V' \, W_1)_+.
\eeq
We thus have:
\beq
W_1^2(x) + \hbar{N\over t_0}\, W_1'(x) + W_2(x,x) = {N\over t_0}\, \left(V'(x)W_1(x)-P_1(x)\right)
\eeq

Using the expansion \eq{WkgdevtopMM}, that gives the Ricatti equation
\beq
{W^{(0)}_1}(x)^2 + \hbar\,{\partial_x} W_1^{(0)}(x)  = V'(x)W^{(0)}_1(x)-P^{(0)}_1(x)
\eeq
which is satisfied by $\om(x)$:
\beq
W_1^{(0)}(x)=\om(x).
\eeq
generalizing to the integration by parts of
\beq
0 = \sum_i \int d\l_1\dots d\l_N {\partial\over \partial \l_i}\Big( {1\over x-\l_i}\sum_{i_1,\dots,i_k}\prod_{j=1}^k {1\over x_j-\l_{i_j}} 
 \quad  \, \Delta(\l)^{2\beta}\, \prod_j \ee{-{N\sqrt\beta\over t_0}\,V(\l_j)}\Big)
\eeq
and using the expansion \eq{WkgdevtopMM} to higher orders in $t_0/N$, one gets the loop equations of theorem \ref{thWngPng},
where
\bea
P_{k+1}(x;x_1,\dots,x_k) 
&=& \sum_g (N/t_0)^{2-2g-k}\, P_{k+1}^{(g)}(x;x_1,\dots,x_k) \cr
&=& \beta^{k/2}\,\left<\sum_i {V'(x)-V'(\l_i)\over x-\l_i}\sum_{i_1,\dots,i_k}\prod_{j=1}^k {1\over x_j-\l_{i_j}} \right>_c.
\eea
In other words, {\bf the correlation functions of $\beta$ matrix models, obey the topological recursion of def. \ref{defWng}}.

\bigskip

{\bf Remark:}

In \cite{MoiBertrand}, a solution of loop equations for the $\beta$-matrix ensemble was proposed, but that solution was such that $U(x)$ was non-generic, corresponding to $\genus=-1$, and that $\psi(x)$ had only a finite number of zeroes. This case implied that $t_0$ was quantized.
Generic matrix models cannot correspond to that situation.

That solution was thus not very useful for actual matrix models.
Here instead, we have the solution for every $U(x)$, i.e. every contour of integration for the $\l_i$'s, and therefore we have the solution of loop equations for the actual matrix model.

\subsection*{9.2 Example: real eigenvalues}

Very often, we are interested in a matrix model with real potential  $V(x)$ of even degree (i.e. $d$ is odd) and such that the eigenvalues are integrated along the real axis.
The resolvent $\om(x)$ is the Stieljes transform of the density of eigenvalues:
\beq
\om(x) = {t_0\sqrt\beta\over N}\, \int_{\mathbb R} {\rho(x')\,dx'\over x-x'}
\virg
\rho(x) = \left< \sum_i \delta(x-\l_i)\right>
\eeq
Let us consider that it is defined by this integral in the upper half-plane for $x\in \mathbb H_+$, and it is extended to the lower half-plane by analytical continuation.

By definition, $\om(x)$ is regular in the upper half-plane, therefore we look for a $\psi(x)$ which has no zero in the upper half-plane, i.e. no zero on the half-lines $L_0, L_1, \dots, L_{d-1}$.
I.e. it has at most $d+1$ half-lines of zeroes , and thus:
$$
\genus\leq {d-1\over 2}.
$$

\section*{10 Non-oriented Ribbon graphs}

\begin{figure}[bth]
\hrule\hbox{\vrule\kern8pt
\vbox{\kern8pt \vbox{
\begin{center}
{\mbox{\epsfxsize=10.truecm\epsfbox{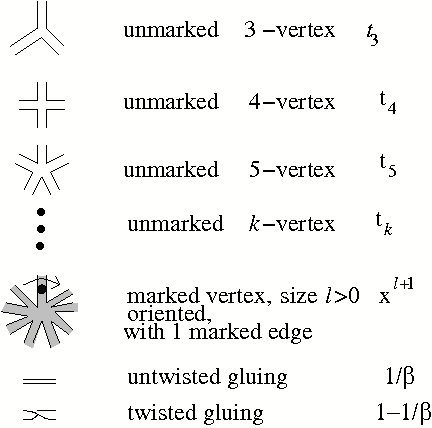}}}
\end{center}
\caption{Consider the set of ribbon graphs obtained by gluing those vertices.
Marked vertices are of degree $l\geq 1$, they are oriented and have one marked half-edge.
Unmarked vertices are unoriented, and are of degree $\geq 3$.
Vertices are glued together by their half-edges, either twisted (with weight $1/\beta$) or untwisted (with weight $1-1/\beta$).\label{figribbonpieces}}
}\kern8pt}
\kern8pt\vrule}\hrule
\end{figure}

\begin{figure}[bth]
\hrule\hbox{\vrule\kern8pt
\vbox{\kern8pt \vbox{
\begin{center}
{\mbox{\epsfxsize=10.truecm\epsfbox{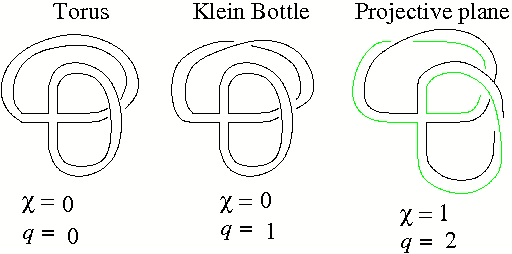}}}
\end{center}
\caption{Examples of ribbon graphs of genus $g=1$.\label{figexribbongraphs}}
}\kern8pt}
\kern8pt\vrule}\hrule
\end{figure}

Consider the set of all closed connected ribbon graphs obtained by gluing the pieces represented in fig. \ref{figribbonpieces}.
Closed means every half-edge is glued to another half-edge.
Connected means every vertex is connected to any other vertex.
See for example fig.~\ref{figexribbongraphs}.

\smallskip

We define the genus of such a ribbon graph ${\cal G}$ as follows.
We replace every twisted edge of ${\cal G}$ by a non-twisted one, we thus obtain another ribbon graph ${\cal G}'$, which is oriented. We define the genus of ${\cal G}$ equal to that of ${\cal G}'$:
$$g({\cal G})=g({\cal G}').$$
The genus of ${\cal G}'$ is computed as usual for oriented ribbon graphs, from the Euler characteristics of ${\cal G}'$:
$$\chi({\cal G}') =2-2g = \#\{{\rm vertices}({\cal G})\}-\#\{{\rm edges}({\cal G})\}+\#\{{\rm single\,lines}({\cal G}')\}$$
where single lines are the lines bordering each side of the ribbon edges. One should follow single lines and see how many connected single lines a graph contains.
Obviously ${\cal G}$ and ${\cal G}'$ have the same number of fat vertices and fat edges (each edge containing two single lines), but they may have different number of single lines.

This defines what we call the genus $g$ of a ribbon graph.

\bigskip

For a given Ribbon graph ${\cal G}$ we call:

$\bullet\,$ $n_i({\cal G})=\#$unmarked vertices of degree $i$, for $3\leq i\leq d+1$,

$\bullet\,$ $l_i({\cal G})=$size of the $i^{\rm th}$ marked vertex, we have $l_i({\cal G})\geq 1$.

$\bullet\,$ $e({\cal G})=\#$edges,

$\bullet\,$ $q({\cal G})=\#$twisted edges,

$\bullet\,$ $v({\cal G})=\#$connected single lines,

$\bullet\,$ $g({\cal G})=$genus,

$\bullet\,$ $\#Aut({\cal G})=$symmetry factor of ${\cal G}$.

\bd
Let $\mathbb M^{(g)}_k(v')$, be the set of ribbon graphs ${\cal G}$ with $k$ marked vertices, $q$ twisted edges, and of genus $g$, and  such that ${\cal G}'$ has $v({\cal G}')=v'$ connected single-lines, .
\ed

\bp
$\mathbb M^{(g)}_k(v')$ is a finite set.
\ep

\proof{
The number of vertices of ${\cal G}'$ is:
$$
\#\{{\rm vertices}\} = k+\sum_{i\geq 3} n_i
$$
The number of edges is twice the number of half-edges, i.e.
$$
2\,\#\{{\rm edges}\} = \sum_{i\geq 3} i\,n_i + \sum_{i=1}^k l_i
$$
That gives:
$$
2-2g = \#\{{\rm vertices}\}-\#\{{\rm edges}\}+v' = k-{1\over 2}\,\sum_{i\geq 3} (i-2) n_i-{1\over 2}\sum_{i=1}^k l_i + v
$$
i.e.
$$
k+v'+2g-2 = {1\over 2}\,\sum_{i\geq 3} (i-2) n_i+{1\over 2}\sum_{i=1}^k l_i 
$$
Since the left hand side is fixed, we see that the number and size of vertices are bounded, so that there is only a finite number of possible oriented ribbon graphs ${\cal G}'$.
Since ${\cal G}'$ has a bounded number of edges, there is only a finite number of possibilities of twisting them, i.e. there are also only a finite number of graphs ${\cal G}$.
}

\subsection*{10.1 Generating functions}

In order to enumerate the sets $\mathbb M^{(g)}_k(v')$, we define the following generating functions:

\bd
We define:
\bea
&& W^{(g)}_k(x_1,\dots,x_k;t_3,\dots,t_{d+1},\beta;t_0) \cr
&=& \beta^{-k/2}\,\sum_{v'\geq 1}\, t_0^{v'}\,\, \sum_{{\cal G}\in \mathbb M^{(g)}_k(v')}
{1\over \#{\rm Aut}({\cal G})}\,\, {t_3^{n_3({\cal G})}\,t_4^{n_4({\cal G})}\,\dots\, t_{d+1}^{n_{d+1}({\cal G})}\over x_1^{l_1({\cal G})}\, x_2^{l_2({\cal G})}\, \dots \, x_k^{l_k({\cal G})}}\,\, \beta^{-e({\cal G})}\,\, (\beta-1)^{q({\cal G})}\,  \cr
&& + {\delta_{k,1}\delta_{g,0}\delta_{q,0}}\,\, {t_0\over x_1}
 + {\delta_{k,2}\delta_{g,0}\delta_{q,0}}\,\, {1\over 2\,(x_1-x_2)^2}. 
\eea
It is a formal series in powers of $t_0$.

Most often, for readability, we shall write only the dependence in the $x_i$'s:
$$ W_k^{(g)}(x_1,\dots,x_k;t_3,\dots,t_{d+1},\beta;t_0) \equiv W_k^{(g)}(x_1,\dots,x_k).$$
Also, for $k=0$ we write
$$
W_0^{(g)}=F_{g}.
$$
\ed

\subsection*{10.2 Tutte's recursive equations}

Tutte's equation is a recursion on the number of edges to construct the ribbon graphs.
It consists in finding a bijection between ribbon graphs of various ensembles, by recursion on the number of edges.
Let
$\mathbb M^{(g)}_{l_1,\dots,l_k}$ be the set of ribbon graphs of genus $g$, and with $k$ marked vertices of size $l_1,\dots,l_k$.

\bigskip

Consider a ribbon graph ${\cal G}\in \mathbb M^{(g)}_{l_0+1,L}$ where $L=\{l_1,\dots,l_k\}$, with marked vertices of degrees $l_0+1,L$.

\begin{figure}[bth]
\hrule\hbox{\vrule\kern8pt
\vbox{\kern8pt \vbox{
\begin{center}
{\mbox{\epsfxsize=7truecm\epsfbox{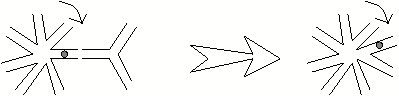}}}$\quad;\quad$
\vspace{2pt}
{\mbox{\epsfxsize=7truecm\epsfbox{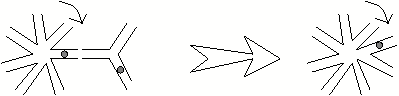}}}
\vspace{2pt}
{\mbox{\epsfxsize=9truecm\epsfbox{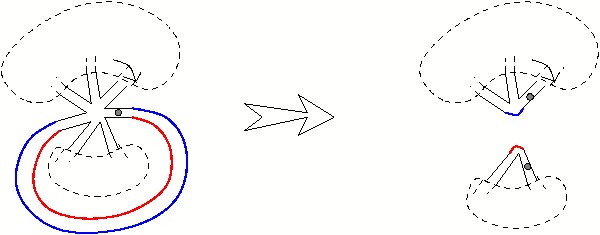}}}
\vspace{2pt}
{\mbox{\epsfxsize=9truecm\epsfbox{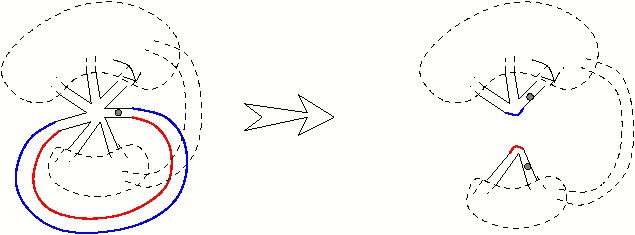}}}
\vspace{2pt}
{\mbox{\epsfxsize=9truecm\epsfbox{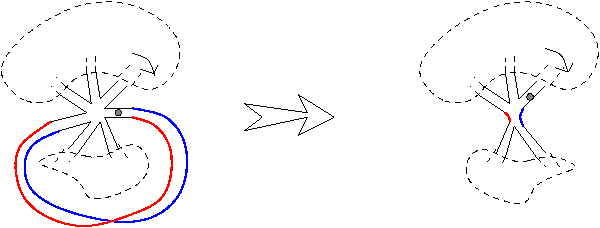}}}
\end{center}
\caption{When we shrink the single lines of a marked edge, several possibilities may occur:
1) the other side is an umarked vertex of order $j+1$, we get a new vertex of order $l_0+j$;
2) the other side is a marked vertex of order $l_i$, we get a new vertex of order $l_0+l_i-1$;
3)4) the other side is the same vertex and the edge is untwisted. Then shrinking the edge splits the vertex into two vertices, this may disconnect the graph or not;
5) the other side is the same vertex and the edge is twisted. Then shrinking the edge doesn't disconnect the vertex.
\label{figtutteseqs}}
}\kern8pt}
\kern8pt\vrule}\hrule
\end{figure}

Consider the marked edge of marked face $0$. It is either twisted or untwisted.
Several mutually exclusive situations may occur (see fig \ref{figtutteseqs}):

$\bullet$ on the other side of the marked edge, there is an unmarked vertex of size $j+1$ with $j\geq 2$.
We then shrink the marked edge to concatenate the two vertices into one marked vertex of degree $l_0+j$. The orientation is inherited from the initial marked vertex, and the marked edge is chosen as the first edge to the left of the shrinked edge. 
It is clear that we don't change the number of single lines in ${\cal G}$ or ${\cal G}'$. We decrease the number of vertices and edges by 1, so we don't change the genus.
We thus get a ribbon graph in $\mathbb \mathbb M^{(g)}_{l_0+j,L}$, and this is weighted with weight $t_{j+1}\,(1/\beta+(1-1/\beta))=t_{j+1}$.

\medskip

$\bullet$ on the other side of the marked edge, there is the marked vertex $i\neq 0$, of size $l_i\geq 1$. We then shrink the marked edge to concatenate the two vertices into one marked vertex of degree $l_0+l_i-1$. The orientation is inherited from the initial marked vertex, and the marked edge is chosen as the first edge to the left of the shrinked edge. It is also clear that we don't change the genus.
Since we forget the marking of the other face, we shall get a symmetry factor $l_i$, corresponding to the $l_i$ places where we glue to the $i^{\rm th}$ marked vertex.
We thus get a ribbon graph in $\mathbb \mathbb M^{(g)}_{l_0+l_i-1,L/\{l_i\}}$, and this is weighted with weight $l_i$.

\medskip

$\bullet$ on the other side of the marked edge, there is the same marked vertex $0$.
Again we shall shrink the marked edge, i.e. shrink the 2 single lines.
Several sub-situations may occur:

\smallskip
$*$ if the edge is untwisted, shrinking the 2 single lines splits the marked vertex of size $l_0+1$ into two vertices of size $l'$ and $l_0-l'-1$. They inherit their orientation and marked edge from the initial marked vertex. We have increased the number of marked vertices by 1.
The two new vertices are either connected together, or not.

$**$ If they are not connected, this means that the number of other marked vertices and  the genus simply add up.
We thus get two ribbon graphs in $\mathbb \mathbb M^{(g')}_{l',L'} \times \mathbb \mathbb M^{(g-g')}_{l_0-l'-1,L/L'}$, and this is weighted with weight $1/\beta$.

$**$ If they are connected, we see that we get a new ribbon graph, with one more vertex, 1 less edge, and we have not changed the connectivity of single lines. The genus has thus decreased by $1$.
We thus get a ribbon graph in $\mathbb \mathbb M^{(g-1)}_{l',l_0-l'-1,L}$, and this is weighted with weight $1/\beta$.

\smallskip
$*$ if the edge is twisted, shrinking the 2 single lines doesn't split the marked vertex.
Instead we get a new vertex of size $l_0-1$.
We assign to it the orientation of the half-vertex situated left of the marked edge, and we mark the edge left of the initial one.
We have decreased $q$ by $1$, and the genus is unchanged.
We thus get a ribbon graph in $\mathbb \mathbb M^{(g)}_{l_0-1,L}$, and this is weighted with weight $(1-1/\beta)\, l_0$ (indeed, there are $l_0$ places where we can glue the marked edge).

%\vfill\eject

\bigskip

For the generating function, those bijections read:
\bea
x\,W_{k+1}^{(g)}(x,X_L) 
&=& \sum_{j=2}^{d}\, t_{j+1}\,x^{j}\,W_{k+1}^{(g)}(x,X_L) \cr
&& + {1\over \sqrt\beta}\,\sum_{i=1}^{k}\, \partial_{x_i}\,\,{W_{k}^{(g)}(x,X_{L/\{x_i\}})-W_{k}^{(g)}(x_i,X_{L/\{x_i\}})\over x-x_i} \cr
&& + {1\over \sqrt\beta}\, \sum_{g',L'\subset L} W_{1+\#L'}^{(g')}(x,X_{L'}) W_{1+k-\#L'}^{(g-g')}(x,X_{L/L'}) \cr
&& + {1\over \sqrt\beta}\,W_{k+2}^{(g-1)}(x,x,X_{L}) \cr
&& + (1-{1\over \beta})\,\partial_x\, W_{k+1}^{(g)}(x,X_L) \cr
&& + {1\over \sqrt\beta}\, P_{k+1}^{(g)}(x,X_L)
\eea
we define
\beq
V'(x) = { \sqrt\beta}\,\left(x-\sum_{j=2}^d t_{j+1} x^j\right)
\eeq
and the last term $P_{k+1}^{(g)}(x;X_L)$ accounts for all the boundary terms, and it is necessarily equal to:
\beq
P_{k+1}^{(g)}(x;X_L) = \left( V'(x)\,\,W_{k+1}^{(g)}(x;X_L)\right)_+.
\eeq
This can be rewritten:
\bea
V'(x)\,W_{k+1}^{(g)}(x,X_L) 
&=& \sum_{i=1}^{k}\, \partial_{x_i}\,\,{W_{k}^{(g)}(x,X_{L/\{x_i\}})-W_{k}^{(g)}(x_i,X_{L/\{x_i\}})\over x-x_i} \cr
&& + \sum_{g',L'} W_{1+\#L'}^{(g')}(x,X_{L'}) W_{1+k-\#L'}^{(g-g')}(x,X_{L/L'}) \cr
&& + W_{k+2}^{(g-1)}(x,x,X_{L}) \cr
&& + \hbar\,\partial_x\, W_{k+1}^{(g)}(x,X_L) \cr
&& + P_{k+1}^{(g)}(x,X_L)
\eea
where
$$
\hbar = {\sqrt\beta-1/\sqrt\beta}.
$$
In other words, the $W_n^{(g)}$'s defined in section \ref{secdefWngFg} provide a solution to Tutte's equations.
They are the generating functions counting our non-oriented ribbon graphs.
One just needs to find the polynomial $P_1^{(0)}(x)$, i.e. $U(x)$, and the choice of $\psi$ which is such that $W_1^{(0)}$ is a formal power series in $t_0$.

\section*{11 Conclusion}
In this article, we have defined some "quantum" versions of quantities known in algebraic geometry and applied them to the resolution of the loop equations in the arbitrary $\beta$-random matrix model case, and in particular the enumeration of some non-orientable ribbon graphs.

Our formalism recovers standard algebraic geometry and the invariants of \cite{MoiBertrand} in the classical limit $\hbar\to 0$.

Instead of an albebraic equation, we have to deal with a differential equation, which we interpreted as a "quantum spectral curve", and we were able to generalize the basic notions arising in classical algebraic geometry, like genus, sheets, branchpoints, meromorphic forms, of 1st kind, 2nd kind, 3rd kind, matrix of periods,...

It is surprising to see that the notion of branchpoints become "blurred", a branchpoint is no longer a point, but an asymptotic accumulation line. Also, there are two sheets, corresponding of the two possible large $x$ asymptotic behaviors of $\psi(x)\sim \exp{\pm V/2\hbar}$, but in fact any solution is a linear combination of these two, so that we could say that we are always in a "linear superposition" of two states like in quantum mechanics.

Another surprising thing, is that, in order for any cohomology theory to make sense, we need the cycle integrals of any forms to depend only on the homology class of the cycles, i.e. we need all forms to have vanishing residues at the $s_i$'s.
This "no-monodromy" condition is equivalent to a Bethe ansatz satisfied by the $s_i$'s, like in the Gaudin model \cite{BabBetGaudin}.
This provides a geometric interpretation of the Bethe ansatz, as the condition for cohomology to make sense.

\medskip

 However, we still lack of a complete understanding of the situation, since most of our results explicitely depend on an initial sector $S_0$ which we choose, whereas in algebraic geometry most of them only depend on the spectral curve and not on its parametrization.
For instance the genus itself depends on a choice of sector.
In some sense, the genus is no longer deterministic. 
 
\medskip

Moreover, we still lack the proper definition of the spectral invariants $F_{g}$, indeed we have defined the $F_g$'s only through solving a differential equation with respect to $\hbar$, which is not as explicit as \cite{OE} or \cite{ChekEynbeta}.
Out of the $F_g$'s, we could expect the possibility to make the link with integrable systems and define a "quantum Tau-function", like in \cite{OE}. 

\medskip
Also, we restricted ourselves to the case of hyperelliptic curves, i.e. second order differential equations, or also a 1-matrix model. 
In a forthcoming paper, we shall generalize all this construction to arbitrary linear differential equations of any order, and generalize to a 2-matrix model. This work is underway, almost finished and the article is being written at this time. As for the hyperelliptical case, the notions of genus, sheets, branchpoints, forms, $W_n^{(g)}$'s ... can be defined. Again there is a Bethe ansatz ensuring a no-monodromy condition so that all cycle integrals depend only on the homology class of cycles. So, there is no qualitative change, the difference is only technical, because the hyperelliptical case has big simplifications due to the involutive symmetry.
The difference between the hyperelliptical case and the general case is comparable to the difference between \cite{Eyn1loop} and \cite{CEO}, i.e. the definition of the kernel $K$ is really more complicated, and there is a rather "big" technical step.

\medskip
Then is would be interesting to see if the $F_g$'s have some sort of symplectic invariance, or more precisely some "canonical invariance", i.e. are unchanged under any change $(x,y)\to (\td x,\td y)$ such that $[\td y,\td x]=[y,x]=\hbar$.

\medskip

Finally, let us mention that we have developped a new geometrical approach to the study of D-modules, and it would be interesting to see how to relate it to more standard approaches, and also to the resurgence theory for studying the Schr\"odinger equation.

\section*{Acknowledgments}
We would like to thank O. Babelon, M. Berg\`ere, G. Borrot, P. Di Francesco, S. Guillermou, V. Pasquier, A. Prats-Ferrer, A. Voros
for useful and fruitful discussions on this subject.
The work of B.E. and O.M. is partly supported by the Enigma European network MRT-CT-2004-5652, ANR project GranMa "Grandes Matrices Al\'eatoires" ANR-08-BLAN-0311-01,  
by the European Science Foundation through the Misgam program,
by the Quebec government with the FQRNT. 
O. Marchal would like to thank the CRM (Centre de recheche math\'ematiques de Montr\'eal, QC, Canada) for its hospitality.

%======================= APPENDICES =================================

\section*{Appendix 1: Proof of the loop equation for $B(x,z)$}
\label{BergmannLoopEquation}

Let's first proove the first loop equation for $B(x,z)$:
Let's define:
\beq
\hat B(x,z) = {1\over 2}\,\partial_z\,(2{\psi'(z)\over \psi(z)}-\partial_z)\, \hat K(x,z)
\eeq
i.e. we have:
\beq
B(x,z) = \hat B(x,z) - \sum_{\alpha=1}^{d-1} v_\alpha(x)\, \oint_{\acycle_\alpha} \hat B(x'',z)dx''
\eeq

Since 
$(2{\psi'(x)\over \psi(x)}+\partial_x)\, v_\alpha(x) = h_\alpha(x)$
 is a polynomial of degree $\leq d-2$, it suffices to prove \eq{loopeqBxAnn} for $\hat B(x,z)$.

Let us compute:
\bea
 (2{\psi'(x)\over \psi(x)}+\partial_x)\,\hat B(x,z)  
&=& {1\over 2}\,\partial_z\,(2{\psi'(z)\over \psi(z)}-\partial_z)\, {1\over x-z}\,  \cr
&=& {1\over 2}\,\partial_z\,(2{\psi'(z)\over \psi(z)(x-z)}-{1\over (x-z)^2})\,  \cr
&=& -{1\over (x-z)^3} + \partial_z\,{\psi'(z)\over \psi(z)(x-z)}   \cr
\eea
and therefore:
\beq
(2{\psi'(x)\over \psi(x)}+\partial_x)\,\left(\hat B(x,z)-{1\over 2(x-z)^2}\right) + \partial_z\,{{\psi'(x)\over \psi(x)}-{\psi'(z)\over \psi(z)}\over x-z} 
= 0
\eeq
This proves \eq{loopeqBxAnn}, with:
\beq
 P_2^{(0)}(x,z) = - \sum_{\alpha=1}^g h_\alpha(x)\, \oint_{\acycle_\alpha} \hat B(x'',z)dx'' -\sum_{\alpha=g+1}^{d-1}\, h_\alpha(x)\, \oint_{\acycle_\alpha} dx''\,\partial_{x''}\,\psi^2(x'') \hat B(x'',z).
\eeq

\bigskip

Let's now proove the second loop equation for $B(x,z)$:
Similarly, let us compute $ (2{\psi'(z)\over \psi(z)}+\partial_z)\,\hat B(x,z)$:
\beq
(2{\psi'(z)\over \psi(z)}+\partial_z)\,\hat B(x,z) = {1\over 2} (2{\psi'(z)\over \psi(z)}+\partial_z)\,\partial_z\,(2{\psi'(z)\over \psi(z)}-\partial_z)\,\hat K(x,z)
\eeq

Notice that the operator $\hat U(z) = {1\over 2} (2{\psi'(z)\over \psi(z)}+\partial_z)\,\partial_z\,(2{\psi'(z)\over \psi(z)}-\partial_z)$, is equal to:
\beq
\hat U(z) = -{1\over 2}\,\partial_z^3 + 2 U(z) \partial_z + U'(z)
\eeq
which is also known in the litterature as the Gelfand-Dikii operator \cite{ZJDFG} (The Gelfand-Dikii differential polynomials $R_k(U)$ are computed recursively by $R_0=1$ and $\partial_z R_{k+1} = \hat U\,.R_k$), which plays a key role in the KdV hierarchy.

However, independently of any relationship with KdV, we get:
\bea
&& (2{\psi'(z)\over \psi(z)}+\partial_z)\,\hat B(x,z) \cr
&=& {1\over \psi^2(x)}\,\int_{\infty_0}^x\, \psi^2(x')\,dx'\,\,\hat U(x'). {1\over x'-z} \cr
&=& {1\over \psi^2(x)}\,\int_{\infty_0}^x\, \psi^2(x')\,dx'\,\,
\Big(-{3\over (x'-z)^4} + {2U(z)\over (x'-z)^2} + {U'(z)\over x'-z} \Big) \cr
\eea
We integrate the first term by parts three times, and we write $Y=\psi'/\psi$ (we have $Y'+Y^2=U$):
\bea
&& (2{\psi'(z)\over \psi(z)}+\partial_z)\,\hat B(x,z) \cr
&=& {1\over (x-z)^3} - {2\over \psi^2(x)}\,\int_{\infty_0}^x\, \psi^2(x')\,dx'\,\,{Y(x')\over (x'-z)^3}  \cr
&& + {1\over \psi^2(x)}\,\int_{\infty_0}^x\, \psi^2(x')\,dx'\,\,
\Big( {2U(z)\over (x'-z)^2} + {U'(z)\over x'-z} \Big) \cr
&=& {1\over (x-z)^3} + {Y(x)\over (x-z)^2} 
 - {1\over \psi^2(x)}\,\int_{\infty_0}^x\, \psi^2(x')\,dx'\,\,{Y'(x')+2Y^2(x')\over (x'-z)^2} \cr
&& + {1\over \psi^2(x)}\,\int_{\infty_0}^x\, \psi^2(x')\,dx'\,\,
\Big( {2U(z)\over (x'-z)^2} + {U'(z)\over x'-z} \Big) \cr
&=& {1\over (x-z)^3} + {Y(x)\over (x-z)^2} 
 + {1\over \psi^2(x)}\,\int_{\infty_0}^x\, \psi^2(x')\,dx'\,\,{Y'(x')\over (x'-z)^2} \cr
&& + {1\over \psi^2(x)}\,\int_{\infty_0}^x\, \psi^2(x')\,dx'\,\,
\Big( {2(U(z)-U(x'))\over (x'-z)^2} + {U'(z)\over x'-z} \Big) \cr
&=& {1\over (x-z)^3} + {Y(x)\over (x-z)^2} 
-{Y'(x)\over x-z} + {1\over \psi^2(x)}\,\int_{\infty_0}^x\, \psi^2(x')\,dx'\,\,{Y''(x')+2Y(x')Y'(x')\over (x'-z)^2} \cr
&& + {1\over \psi^2(x)}\,\int_{\infty_0}^x\, \psi^2(x')\,dx'\,\,
\Big( {2(U(z)-U(x'))\over (x'-z)^2} + {U'(z)\over x'-z} \Big) \cr
&=& {1\over (x-z)^3} - {\partial \over \partial x}\, {Y(x)\over x-z}  
 + {1\over \psi^2(x)}\,\int_{\infty_0}^x\, \psi^2(x')\,dx'\,\,
\Big( {2(U(z)-U(x'))\over (x'-z)^2} + {U'(z)+U(x')\over x'-z} \Big) \cr
\eea
This implies that:
\bea
&& (2{\psi'(z)\over \psi(z)}+\partial_z)\,\Big(\hat B(x,z)-{1\over 2(x-z)^2}\Big)
+ {\partial \over \partial x}\, {Y(x)-Y(z)\over x-z}  \cr
&=& {1\over \psi^2(x)}\,\int_{\infty_0}^x\, \psi^2(x')\,dx'\,\,
\Big( {2(U(z)-U(x'))\over (x'-z)^2} + {U'(z)+U(x')\over x'-z} \Big) \cr
&=& Q(z,x)
\eea
which is clearly a polynomial in $z$.
Taking integrals over $x$ along $\acycle_\alpha$ does not change its structure in $z$, and therefore:
\bea
&& (2{\psi'(z)\over \psi(z)}+\partial_z)\,\Big( B(x,z)-{1\over 2(x-z)^2}\Big)
+ {\partial \over \partial x}\, {Y(x)-Y(z)\over x-z}  \cr
&=& {1\over \psi^2(x)}\,\int_{\infty_0}^x\, \psi^2(x')\,dx'\,\,
\Big( {2(U(z)-U(x'))\over (x'-z)^2} + {U'(z)+U(x')\over x'-z} \Big) \cr
&=& \td{P}_2^{(0)}(z,x)
\eea
is of the required form.

By looking at the behavior of the various terms in the LHS of \eq{loopeqBzAnn} when $z\to\infty$, we find that $\td{P}_2^{(0)}(z,x)$ is a polynomial of degree at most $d-2$ in $z$.

\section*{Appendix 2: Proof of theorem \ref{thpolessiWng}}
\label{approofthpolessiWng}

{\bf Theorem \ref{thpolessiWng}}
{\it Each  $W_n^{(g)}(x_1,\dots,x_n)$ with $2-2g-n<0$, is an analytical functions of all its arguments, with poles only when $x_i\to s_{j}$.
Moreover, it vanishes at least as $O\left(1/{x_i^2}\right)$ when $x_i\to\infty$ in all sectors.
It has no discontinuity across $\acycle$-cycles.}
\bigskip

{\bf proof:}

We proceed by recursion on $2g+n$.
The theorem is true for $W_2^{(0)}$.
Assume it is true up to $2g+n$, we shall prove it for $W_{n+1}^{(g)}(x_0,x_1,\dots,x_n)$.

\medskip

The integrand $U_n^{(g)}$ of theorem \ref{thWngdefintU} is singular only at $x=s_j$'s.
As long as $x_0$ is away from the $s_j$'s, we can continuously deform the $\acycle$-cycles and the contour ${\cal C}$ in order to have $x_0$ outside of the $\acycle$-cycles, and the integral can be evaluated and is analytical in $x_0$.
When $x_0$ approaches $s_i$,  we define $\hat{\cal C}_i$, a contour which surrounds all roots except $s_{i}$, i.e:
\beq
\oint_{{\cal C}} = \oint_{\hat{\cal C}_i} + 2i\pi\,\Res_{s_{i}}
\eeq
The integral over  $\hat{\cal C}_i$ can be evaluated and is convergent, thus it is analytical in $x_0$.

From the recursion hypothesis, all terms in the integrand are meromorphic in the vicinity of $s_{i}$, and thus the residue at $s_{i}$ can be computed by taking a finite Taylor expansion of $K(x_0,x)=\sum_k (x-s_i)^k\,K_{i,k}(x_0)$ in the vicinity of $x\to s_{i}$. The result is a finite sum of terms of the type $K_{i,k}(x_0)$. It is easy to see from the definition of $K$, that each $K_{i,k}(x_0)$ has only poles at $x_0=s_{i}$. Thus we have proved that $W_{n+1}^{(g)}$ has poles at the $s_i$'s in its first variable.

\smallskip

\medskip

In the other variables, the result comes from an obvious recursion.

$\square$

\section*{Appendix 3: Proof of theorem \ref{thWngPng}}
\label{approofthWngPng}

In this subsection we prove theorem \ref{thWngPng}, that all $W_n^{(g)}$'s satisfy
the loop equation.
\bea
 P_{n+1}^{(g)}(x,x_1...,x_n)
 &=&
2\hbar\frac{\psi'(x)}{\psi(x)}\overline{W}_{n+1}^{(g)}(x,x_1...,x_n) + \hbar \partial_{x}{\overline{W}_{n+1}^{(g)}(x,x_1,...,x_n}) \cr
&& + \sum_{I\subset J} \ovl{W}_{|I|+1}^{(h)}(x,x_I) \ovl{W}_{n-|I|+1}^{(g-h)}(x,J/I) +
\ovl{W}_{n+2}^{(g-1)}(x,x,J)  \cr
& &+ \sum_{j}
\partial_{x_j} \left( {{\ovl{W}_n^{(g)}(x,J/\{j\})-{\ovl{W}_n^{(g)}(x_j,J/\{j\})}} \over {(x-x_j)}}\right) \cr
\eea
is a polynomial in $x$ of degree at most $d-2$.

{\bf proof:}

From the definition we have:
\bea
W_{n+1}^{(g)}(x,J)&=&   {1\over 2i\pi} \oint_{{\cal C}}\, dz\,\,  K(x,z)\, U_n^{(g)}(z,J)\cr
&=&{1\over 2i\pi} \oint_{{\cal C}}\, dz\,\,  \hat{K}(x,z)\, U_n^{(g)}(z,J)\cr
&&-\sum_\alpha\, {v_\alpha(x)\over 2i\pi} \oint_{{\cal C}}\, dz\,\,  C_{\alpha}(z)\, U_n^{(g)}(z,J)
\eea
Then, notice that $\hat K(x,z)$ has a logarithmic cut along $]\infty_0,x]$, and the discontinuity across that cut is:
\beq
\delta \hat K(x,z) = {2i\pi\over \hbar}\, {\psi^2(z)\over \psi^2(x)}
\eeq
$U_n^{(g)}$ has no singularity outside of ${\cal C}$, and thus we can deform the contour into a contour enclosing only the logarithmic cut of $\hat K(x,z)$, and therefore:
\beq
{1\over 2i\pi} \oint_{{\cal C}}\, dz\,\,  \hat{K}(x,z)\, U_n^{(g)}(z,J) = -\,{1\over \hbar}\, \int_{\infty_0}^xdz\,\, {\psi^2(z)\over \psi^2(x)}\,U_n^{(g)}(z,J)
\eeq
We then apply the operator: $2\frac{\psi'(x)}{\psi(x)}+\partial_x$, that gives:
\beq
(2\hbar {\psi'(x)\over \psi(x)} + \hbar \partial_x)\,\,{1\over 2i\pi} \oint_{{\cal C}}\, dz\,\,  \hat{K}(x,z)\, U_n^{(g)}(z,J) = - U_n^{(g)}(x,J)
\eeq
and therefore:
\bea
P_{n+1}^{(g)}(x,J)
&=& (2\hbar {\psi'(x)\over \psi(x)} + \hbar \partial_x)\,W_{n+1}^{(g)}(x,J)+U_n^{(g)}(x,J) \cr
&=& -\,(2\hbar {\psi'(x)\over \psi(x)} + \hbar \partial_x)\,\sum_\alpha\, {v_\alpha(x)\over 2i\pi} \oint_{{\cal C}}\, dz\,\,  C_{\alpha}(z)\, U_n^{(g)}(z,J) \cr
&=& -\,\sum_\alpha\, h_\alpha(x) \oint_{{\cal C}}\, dz\,\,  C_{\alpha}(z)\, U_n^{(g)}(z,J)
\eea
which is indeed a polynomial of $x$ of degree at most $d-2$.

$\square$

\section*{Appendix 4: Proof of theorem \ref{thW3Krich}}
\label{approofthW3Krich}

{\bf Theorem \ref{thW3Krich}}
{\it 
The 3 point function $W_3^{(0)}$ is symmetric and we have:
\beq \label{appthW3Krich}
W_3^{(0)}(x_1,x_2,x_3) = {4\over 2i\pi} \oint_{{\cal C}}\, dx\,\, {B(x,x_1)B(x,x_2)B(x,x_3)\over Y'(x)}
\eeq
where $Y(x)=-2\hbar\frac{\psi'(x)}{\psi(x)}$}
\bigskip

%Note that the rhs makes sense because in the direction $\mathcal{C}$, $Y(x)\sim V'(x)$ and $B(x,x_i)=O\left(\frac{1}{x}\right)$

{\bf proof:}

The definition of $W_3^{(0)}$ is:
\bea  
&& W_3^{(0)}(x_0,x_1,x_2)\cr 
&=&  {1\over i\pi}\oint_{{\cal C}}\, dx\,\, K(x_0,x)B(x,x_1)B(x,x_2) \cr
 &=&  {1\over 4i\pi} \oint_{{\cal C}}\, dx\,\,  K_0 \, G_1^{'} \, G'_2 \cr
 &=&  {1\over 4i\pi}\oint_{{\cal C}}\, dx\,\, K_0 \left( (\hbar K''_1 +  Y K'_1 +  Y'
 K_1)(\hbar K''_2 + YK'_2 +Y' K_2) \right) \cr
 &=&  {1\over 4i\pi} \oint_{{\cal C}}\, dx\,\, K_0 \, (\, \hbar^2 K''_1 K''_2 +  \hbar Y (K'_1
 K''_2+K''_1 K'_2) +  \hbar Y' (K''_1 K_2 +K''_2 K_1) \cr
 && +  Y^2 K'_1 K'_2+  Y Y' (K_1 K'_2+K'_1 K_2)+
{Y'}^2 K_1 K_2 \,) \cr
 \eea
where we have written for short $K_p = K(x_p,x)$, $G_p=G(x_p,x)$, and derivative are w.r.t. $x$. Note also that introducing $K_1$ and $K_2$ makes appear some additional and arbitrary logarithmic cuts from $x_1$ to $\infty_{0}$ and from $x_2$ to $\infty_{0}$. But these cuts can be chosen arbitrarily since from the definition of $W_3^{(0)}(x_0,x_1,x_2)$ it should not depend on that. Remember also that to use this definition of $W_3^{(0)}$ we need to assume that $x_1$ and $x_2$ are not circled by the contour $\mathcal{C}$. Therefore we can choose the logarithmic cut of $K_1$ and $K_2$ inside the contour $\mathcal{C}$ like we have done it for $x_0$. We now see that for example $K_0 K_1 K_2$ has no singularity outside $\mathcal{C}$ and thus will not contribute because of theorem \ref{NullityOfIntegrals}. Many other manipulations involving globally defined functions with no singularities outside $\mathcal{C}$ can be done.

For example, using the Ricatti equation $Y_i^2 = 2 \hbar Y_i' + 4U$, we may replace $Y_i^2$ by $2 \hbar Y_i'$ and $ Y_i Y_i'$ by $\hbar Y_i''$.
\bea
& & W_3^{(0)}(x_0,x_1,x_2)\cr
 &=&  {1\over 4i\pi}\oint_{{\cal C}}\, dx\,\, K_0 \, (  \hbar Y (K'_1
 K''_2+K''_1 K'_2) +  \hbar Y' (K''_1 K_2 +K''_2 K_1) \cr
 && + 2 \hbar Y' K'_1 K'_2+ \hbar Y'' (K_1 K'_2+K'_1 K_2)+
{Y'}^2 K_1 K_2 \,)\ \cr
 &=&  {1\over 4i\pi}\oint_{{\cal C}}\, dx\,\, K_0 \, (  \hbar Y (K'_1 K'_2)' +  \hbar Y' (K_1 K_2)'' + \hbar Y'' (K_1 K_2)'+  {Y'}^2 K_1 K_2 \,)\ \cr
 &=&   {1\over 4i\pi}\oint_{{\cal C}}\, dx\,\, {Y'}^2 K_0 K_1 K_2 +
 \hbar \big( Y'' K_0 (K_1 K_2)' - (Y K_0)' K'_1 K'_2 - (Y' K_0)' (K_1 K_2)' \big) \cr
 &=&  {1\over 2} {1\over 4i\pi}\oint_{{\cal C}}\, dx\,\, {Y'}^2 K_0 K_1 K_2 -
 \hbar \big(  (Y K_0)' K'_1 K'_2 + Y' K_0' (K_1 K_2)' \big) \cr
&=&  {1\over 4i\pi}\oint_{{\cal C}}\, dx\,\, {Y'}^2 K_0 K_1 K_2 - \hbar Y K'_0 K'_1 K'_2 - \hbar Y' (K_0 K'_1 K'_2+K'_0 K_1 K'_2+K'_0 K'_1 K_2) \cr
\eea
This expression is clearly symmetric in $x_0, x_1, x_2$ as claimed in theorem \ref{thsym}.

Let us give an alternative expression, in the form of the Verlinde or Krichever formula.
\beq\label{eq30Krichever}
W_3^{(0)}(x_0,x_1,x_2)=
{2\over i\pi}\oint_{{\cal C}}\, dx\,\, {B(x,x_1)B(x,x_2)B(x,x_3) \over Y^{'}(x)}
\eeq

\proof{
In order to prove formula \ref{eq30Krichever}, compute:
\beq
B(x,x_i) = -{1\over 2} G'(x,x_i) = -{1\over 2} G'_i = {1\over 2}(\hbar K''_i +  Y K'_i +  Y' K_i)
\eeq
thus:
\bea
&&  {1\over 2i\pi}\oint_{{\cal C}}\, dx\,\, {B(x,x_1)B(x,x_2)B(x,x_3) \over Y^{'}(x)} \cr
&=&  {1\over 16i\pi}\oint_{{\cal C}}\, dx\,\, {1\over Y'(x)}\, (\hbar K''_0 +  Y K'_0 +  Y' K_0)(\hbar K''_1 +  Y K'_1 +  Y' K_1)(\hbar K''_2 +  Y K'_2 +  Y' K_2)  \cr
&=&  {1\over 16i\pi}\oint_{{\cal C}}\, dx\,\,
 {\hbar^3\over Y'} K''_0 K''_1 K''_2 + \hbar^2 {Y\over Y'} (K'_0 K''_1 K''_2 + K''_0 K'_1 K''_2 + K''_0 K''_1 K'_2) \cr
&& + \hbar^2  (K_0 K''_1 K''_2 + K''_0 K_1 K''_2 + K''_0 K''_1 K_2)
+ \hbar {Y^2\over Y'} (K''_0 K'_1 K'_2+K'_0 K''_1 K'_2+K'_0 K'_1 K''_2) \cr
&& + \hbar Y (K_0 K'_1 K''_2 + K_0 K''_1 K'_2 + K'_0 K_1 K''_2 + K'_0 K''_1 K_2 + K''_0 K_1 K'_2 + K''_0 K'_1 K_2) \cr
&& + \hbar Y' (K''_0 K_1 K_2 + K_0 K''_1 K_2 + K_0 K_1 K''_2)
+ {Y^3\over Y'} K'_0 K'_1 K'_2  \cr
&& + Y^2  (K_0 K'_1 K'_2 + K'_0 K_1 K'_2 + K'_0 K'_1 K_2)
+ Y Y' (K'_0 K_1 K_2 + K_0 K'_1 K_2 + K_0 K_1 K'_2) \cr
&& + Y'^2 K_0 K_1 K_2  \cr
\eea

\bea
&&  {1\over 2i\pi}\oint_{{\cal C}}\, dx\,\, {B(x,x_1)B(x,x_2)B(x,x_3) \over Y^{'}(x)} \cr
&=&  {1\over 16i\pi}\oint_{{\cal C}}\, dx\,\,
 \hbar Y (K_0 K'_1 K''_2 + K_0 K''_1 K'_2 + K'_0 K_1 K''_2 + K'_0 K''_1 K_2 + K''_0 K_1 K'_2 + K''_0 K'_1 K_2) \cr
&& + \hbar Y' (K''_0 K_1 K_2 + K_0 K''_1 K_2 + K_0 K_1 K''_2)
+ {Y^3\over Y'} K'_0 K'_1 K'_2  \cr
&& + Y^2  (K_0 K'_1 K'_2 + K'_0 K_1 K'_2 + K'_0 K'_1 K_2)
+ Y Y' (K'_0 K_1 K_2 + K_0 K'_1 K_2 + K_0 K_1 K'_2) \cr
&& + Y'^2 K_0 K_1 K_2  \cr
\eea
Notice that $Y^2 = 2\hbar Y' + 4U$, thus we may replace $Y^3/Y'$ by $2\hbar Y$, and $Y^2$ by $2\hbar Y'$ and $Y Y'$ by $\hbar Y''$, for the same reasons as before. Thus:
\bea
&&  {1\over 2i\pi}\oint_{{\cal C}}\, dx\,\, {B(x,x_1)B(x,x_2)B(x,x_3) \over Y^{'}(x)} \cr
&=&  {1\over 16i\pi}\oint_{{\cal C}}\, dx\,\,
 \hbar Y (K_0 K'_1 K''_2 + K_0 K''_1 K'_2 + K'_0 K_1 K''_2 + K'_0 K''_1 K_2 + K''_0 K_1 K'_2 + K''_0 K'_1 K_2) \cr
&& + \hbar Y' (K''_0 K_1 K_2 + K_0 K''_1 K_2 + K_0 K_1 K''_2)
+ 2\hbar Y K'_0 K'_1 K'_2  \cr
&& + 2 \hbar Y'  (K_0 K'_1 K'_2 + K'_0 K_1 K'_2 + K'_0 K'_1 K_2)
+ \hbar Y'' (K'_0 K_1 K_2 + K_0 K'_1 K_2 + K_0 K_1 K'_2) \cr
&& + Y'^2 K_0 K_1 K_2  \cr
&=&  {1\over 16i\pi}\oint_{{\cal C}}\, dx\,\,
 \hbar Y (K_0 (K'_1 K'_2)' +  K_1 (K'_0 K'_2)' + K_2 (K'_0 K'_1)') \cr
&& + 2\hbar Y K'_0 K'_1 K'_2  + Y'^2 K_0 K_1 K_2  + \hbar (Y' (K'_0 K_1 K_2 + K_0 K'_1 K_2 + K_0 K_1 K'_2))' \cr
&=&  {1\over 16i\pi}\oint_{{\cal C}}\, dx\,\,
 \hbar Y (K_0 (K'_1 K'_2)' +  K_1 (K'_0 K'_2)' + K_2 (K'_0 K'_1)') \cr
&& + 2\hbar Y K'_0 K'_1 K'_2  + Y'^2 K_0 K_1 K_2  \cr
&=&  -{1\over 16i\pi}\oint_{{\cal C}}\, dx\,\, 3 \hbar Y K'_0 K'_1 K'_2 + \hbar Y' (K_0 K'_1 K'_2 + K'_0 K_1 K'_2 + K'_0 K'_1 K_2) \cr
&& - 2\hbar Y K'_0 K'_1 K'_2  - Y'^2 K_0 K_1 K_2  \cr
&=& {1\over 8i\pi}\oint_{{\cal C}}\, dx\,\, W_3^{(0)}(x_0,x_1,x_2)
\eea
}

\section*{Appendix 5: Proof of theorem \ref{thsym}}
\label{approofthsym}

{\bf Theorem \ref{thsym}}
{\it 
Each $W_n^{(g)}$ is a symmetric function of all its arguments.
}
\bigskip

{\bf proof:}

The special case of $W_3^{(0)}$ is proved in appendix \ref{appthW3Krich} above.

It is obvious from the definition that $W_{n+1}^{(g)}(x_0,x_1,\dots,x_n)$ is
symmetric in $x_1,x_2,\dots,x_n$, and therefore we need to show that (for $n\geq 1$):
\beq
W_{n+1}^{(g)}(x_0,x_1,J)-W_{n+1}^{(g)}(x_1,x_0,J)=0
\eeq
where $J=\{ x_2,\dots,x_n\}$.
We prove it by recursion on $-\chi=2g-2+n$. 

Assume that every $W_k^{(h)}$ with $2h+k-2\leq 2g+n$ is symmetric.
We have:
\bea
&& W_{n+1}^{(g)}(x_0,x_1,J) \cr
&=&\frac{1}{2\pi i} \oint_{\mathcal{C}} \,dx \,\, K(x_0,x)\,\, \Big(
W_{n+2}^{(g-1)}(x,x,x_1,J) + 2 \,\,\, B(x,x_1) W_{n}^{(g)}(x,J) \cr
&& + 2 \sum_{h=0}^g\sum'_{I\in J}\,\,\, W_{2+|I|}^{(h)}(x,x_1,I) W_{n-|I|}^{(g-h)}(x,J/I) \Big) \cr
\eea
where $\sum'$ means that we exclude the terms $(I=\emptyset, h=0)$ and $(I=J, h=g)$. Notice also that $\ovl{W}_{n+2}^{(g-1)}=W_{n+2}^{(g-1)}$ because $n\geq 1$.
Then, using the recursion hypothesis, we have:
\bea
&& W_{n+1}^{(g)}(x_0,x_1,J) \cr
&=& 2 \oint_{\mathcal{C}} \,dx \,\, K(x_0,x)\,\,  B(x,x_1) W_{n}^{(g)}(x,J) \cr
&& + \oint_{\mathcal{C}} \,dx \,\, \oint_{\mathcal{C}} \,dx' \,\, K(x_0,x) K(x_1,x')\,\, 
\Big(  W_{n+3}^{(g-2)}(x,x,x',x',J) \cr
&& + 2\sum_h\sum'_{I} W_{2+|I|}^{(h)}(x',x,I) W_{1+n-|I|}^{(g-1-h)}(x',x,J/I) \cr
&& + 2 \sum_h\sum'_{I} W_{3+|I|}^{(h)}(x',x,x,I) W_{n-|I|}^{(g-1-h)}(x',J/I) \cr
&& + 2 \sum_{h}\sum'_{I\in J}\,\,\, W_{n-|I|}^{(g-h)}(x,J/I)
\Big[ W_{3+|I|}^{(h-1)}(x,x',x',I)  \cr
&& + 2 \sum_{h'}\sum'_{I'\subset I} W_{2+|I'|}^{(h')}(x',x,I')   W_{1+|I|-|I'|}^{(h-h')}(x',I/I')  
\Big]\,
\Big) \cr
\eea
Now, if we compute $W_{n+1}^{(g)}(x_1,x_0,J)$, we get the same expression, with the order of integrations exchanged, i.e. we have to integrate $x'$ before integrating $x$.
Notice, by moving the integration contours,  that:
\beq
\oint_{\mathcal{C}} \,dx \,\, \oint_{\mathcal{C}} \,dx' - \oint_{\mathcal{C}} \,dx' \,\, \oint_{\mathcal{C}} \,dx =
- \oint_{\mathcal{C}} \,dx \frac{1}{2\pi i}\Res_{x'\to x} 
\eeq
Moreover, the only terms which have a pole at $x=x'$ are those containing $B(x,x')$.
Therefore:
\bea
&& W_{n+1}^{(g)}(x_0,x_1,J)-W_{n+1}^{(g)}(x_1,x_0,J) \cr
&=& 2 \oint_{\mathcal{C}} \,dx \,\, \left( K(x_0,x)\,\,  B(x,x_1)  - K(x_1,x)\,\,  B(x,x_0) \right) \, W_{n}^{(g)}(x,J) \cr
&& - 2 \oint_{\mathcal{C}} \,dx \,\, \frac{1}{2i\pi}\Res_{x'\to x}\,\, K(x_0,x) K(x_1,x')\,\, B(x,x')\,\,
\Big(     \cr
&& 2W_{1+n}^{(g-1)}(x',x,J)  + 2 \sum_{h}\sum'_{I\in J}\,\, W_{n-|I|}^{(g-h)}(x,J/I)     W_{1+|I|}^{(h)}(x',I)  
\Big) \cr
\eea
The residue $\Res_{x'\to x}$ can be computed:
\bea
&& W_{n+1}^{(g)}(x_0,x_1,J)-W_{n+1}^{(g)}(x_1,x_0,J) \cr
&=& 2  \oint_{\mathcal{C}} \,dx \,\, \left( K(x_0,x)\,\,  B(x,x_1)  - K(x_1,x)\,\,  B(x,x_0) \right) \, W_{n}^{(g)}(x,J) \cr
&& -  \oint_{\mathcal{C}} \,dx \,\, \,\, K(x_0,x) {\partial \over \partial x'}\Big( K(x_1,x')\,\, 
\Big(     \cr
&& 2W_{1+n}^{(g-1)}(x',x,J)  + 2 \sum_{h}\sum'_{I\in J}\,\, W_{n-|I|}^{(g-h)}(x,J/I)     W_{1+|I|}^{(h)}(x',I)  
\Big)\,\, \Big)_{x'=x} \cr
&=& 2 \oint_{\mathcal{C}} \,dx \,\, \left( K(x_0,x)\,\,  B(x,x_1)  - K(x_1,x)\,\,  B(x,x_0) \right) \, W_{n}^{(g)}(x,J) \cr
&& -   \oint_{\mathcal{C}} \,dx \,\, K(x_0,x) K'(x_1,x)\,\, 
\Big(     \cr
&& 2W_{1+n}^{(g-1)}(x,x,J)  + 2 \sum_{h}\sum'_{I\in J}\,\, W_{n-|I|}^{(g-h)}(x,J/I)     W_{1+|I|}^{(h)}(x,I)  
\Big)\,\,  \cr
&& -  \, \oint_{\mathcal{C}} \,dx \,\, K(x_0,x) K(x_1,x'){\partial \over \partial x'}\Big(      \cr
&& 2W_{1+n}^{(g-1)}(x',x,J)  + 2 \sum_{h}\sum'_{I\in J}\,\, W_{n-|I|}^{(g-h)}(x,J/I)     W_{1+|I|}^{(h)}(x',I)  
\,\, \Big)_{x'=x} \cr
&=& 2  \oint_{\mathcal{C}} \,dx \,\, \left( K(x_0,x)\,\,  B(x,x_1)  - K(x_1,x)\,\,  B(x,x_0) \right) \, W_{n}^{(g)}(x,J) \cr
&& -  \oint_{\mathcal{C}} \,dx \,\, K(x_0,x) K'(x_1,x)\,\, 
\Big(     \cr
&& 2W_{1+n}^{(g-1)}(x,x,J)  + 2 \sum_{h}\sum'_{I\in J}\,\, W_{n-|I|}^{(g-h)}(x,J/I)     W_{1+|I|}^{(h)}(x,I)  
\Big)\,\,  \cr
&& -  {1\over 2}  \oint_{\mathcal{C}} \,dx \,\, K(x_0,x) K(x_1,x) {\partial \over \partial x}\Big(      \cr
&& 2W_{1+n}^{(g-1)}(x,x,J)  + 2 \sum_{h}\sum'_{I\in J}\,\, W_{n-|I|}^{(g-h)}(x,J/I)     W_{1+|I|}^{(h)}(x,I)  
\,\, \Big) \cr
\eea
The last term can be integrated by parts, and we get:
\bea
&& W_{n+1}^{(g)}(x_0,x_1,J)-W_{n+1}^{(g)}(x_1,x_0,J) \cr
&=& 2 \oint_{\mathcal{C}} \,dx \,\, \left( K(x_0,x)\,\,  B(x,x_1)  - K(x_1,x)\,\,  B(x,x_0) \right) \, W_{n}^{(g)}(x,J) \cr
&& +{1\over 2}   \oint_{\mathcal{C}} \,dx \,\, \Big( K'(x_0,x) K(x_1,x)-K(x_0,x) K'(x_1,x)\Big)\,\, 
\Big(     \cr
&& 2W_{1+n}^{(g-1)}(x,x,J)  + 2 \sum_{h}\sum'_{I\in J}\,\, W_{n-|I|}^{(g-h)}(x,J/I)     W_{1+|I|}^{(h)}(x,I)  
\Big)\,\,  \cr
\eea
Then we use theorem \ref{thWngPng}:
\bea
&& W_{n+1}^{(g)}(x_0,x_1,J)-W_{n+1}^{(g)}(x_1,x_0,J) \cr
&=& 2  \oint_{\mathcal{C}} \,dx \,\, \left( K(x_0,x)\,\,  B(x,x_1)  - K(x_1,x)\,\,  B(x,x_0) \right) \, W_{n}^{(g)}(x,J) \cr
&& + \oint_{\mathcal{C}} \,dx \,\, \Big( K'(x_0,x) K(x_1,x)-K(x_0,x) K'(x_1,x)\Big)\,\, 
\Big(      P_{n}^{(g)}(x,J) \cr
&& + (Y(x) - \hbar \partial_x) W_{n}^{(g)}(x,J)  + \sum_j \partial_{x_j}
\Big( {W_{n-1}^{(g)}(x_j,J/\{x_j\})\over x-x_j} \Big)
\Big)\,\,  \cr
\eea
Since $P_{n}^{(g)}(x,J)$ and $W_{n-1}^{(g)}(x_j,J/\{x_j\})$ are entire functions of $x$, we can use the usual thorem \ref{NullityOfIntegrals} to say that they do not contribute. (Note again that we choose the logarithmic cut of $K_1$ inside the contour $\mathcal{C}$, and that we can do that because the contour $\mathcal{C}$ contains $x_1$.)
\bea
&& W_{n+1}^{(g)}(x_0,x_1,J)-W_{n+1}^{(g)}(x_1,x_0,J) \cr
&=& 2 \oint_{\mathcal{C}} \,dx \,\, \left( K(x_0,x)\,\,  B(x,x_1)  - K(x_1,x)\,\,  B(x,x_0) \right) \, W_{n}^{(g)}(x,J) \cr
&+& 2  \oint_{\mathcal{C}} \,dx \,\, \Big( K'(x_0,x) K(x_1,x)-K(x_0,x) K'(x_1,x)\Big)\cr
&&(Y(x) - \hbar \partial_x) W_{n}^{(g)}(x,J)    \cr
\eea
Notice that:
\beq
K'_0 K_1 - K_0 K'_1 = -{1\over \hbar}\,( G_0 K_1 - K_0 G_1)
\eeq
and $B=-{1\over 2}\, G'$, therefore:
\bea
&& W_{n+1}^{(g)}(x_0,x_1,J)-W_{n+1}^{(g)}(x_1,x_0,J) \cr
&=& - 2  \oint_{\mathcal{C}} \,dx \,\, \left( K_0 G'_1  - K_1 G'_0 \right) \, W_{n}^{(g)}(x,J) \cr
&& -{1\over \hbar}  2  \oint_{\mathcal{C}} \,dx \,\, \Big( G_0 K_1- K_0 G_1 \Big)\,\, 
(Y(x) - \hbar \partial_x) W_{n}^{(g)}(x,J)    \cr
\eea
we integrate the first line by parts:
\bea
&& W_{n+1}^{(g)}(x_0,x_1,J)-W_{n+1}^{(g)}(x_1,x_0,J) \cr
&=&  \oint_{\mathcal{C}} \,dx \,\, \left( K'_0 G_1  - K'_1 G_0 \right) \, W_{n}^{(g)}(x,J) \cr
&& +  \oint_{\mathcal{C}} \,dx \,\, \left( K_0 G_1  - K_1 G_0 \right) \, W_{n}^{(g)}(x,J)' \cr
&& -{1\over \hbar}  \oint_{\mathcal{C}} \,dx \,\, \,\, \Big( G_0 K_1- K_0 G_1 \Big)\,\, 
(Y(x) - \hbar \partial_x) W_{n}^{(g)}(x,J)    \cr
\eea
Notice that:
\beq
K'_0 G_1 - G_0 K'_1 = -{Y\over \hbar}\,( K_0 G_1 - G_0 K_1)
\eeq
So we find \beq  W_{n+1}^{(g)}(x_0,x_1,J)-W_{n+1}^{(g)}(x_1,x_0,J)=0\eeq

%% file: annexeh.tex
\selectlanguage{french}
\annexe{A matrix model for the topological string I: Deriving the matrix model} \label{Article[V]}
\selectlanguage{english}

\begin{center}
\vskip 1cm
 {B. Eynard${}^1$, A. Kashani-Poor${}^{2,3}$, O. Marchal${}^{1,4}$}

\vskip.2cm 
{\it ${}^1$ Institut de Physique Th\'eorique,\\
CEA, IPhT, F-91191 Gif-sur-Yvette, France,\\
CNRS, URA 2306, F-91191 Gif-sur-Yvette, France.\\ \vskip0.3cm}

{\it 
$^2$ Institut des Hautes \'Etudes Scientifiques\\
Le Bois-Marie, 35, route de Chartres, 91440 Bures-sur-Yvette, France\\ \vskip0.3cm }

{\it $^3$ Laboratoire de Physique Th\'eorique de l'\'Ecole Normale Sup\'erieure, \\
24 rue Lhomond, 75231 Paris, France \\ \vskip0.3cm}

{\it ${}^4$ Centre de recherches math\'ematiques,
Universit\'e de Montr\'eal \\
C.P. 6128, Succ. centre-ville
Montr\'eal, Qu\'e, H3C 3J7, Canada.\\}

\end{center} 
\vskip 1.5cm

\begin{center}

We construct a matrix model that reproduces the topological string partition function on arbitrary toric Calabi-Yau 3-folds. This demonstrates, in accord with the BKMP ``remodeling the B-model'' conjecture, that Gromov-Witten invariants of any toric Calabi-Yau 3-fold can be computed in terms of the spectral invariants of a spectral curve. Moreover, it proves that the generating function of Gromov-Witten invariants is a tau function for an integrable hierarchy. In a follow-up paper, we will explicitly construct the spectral curve of our matrix model and argue that it equals the mirror curve of the toric Calabi-Yau manifold.
\end{center}

\newpage

\section*{1 Introduction}

In the topological string A-model, the object of study is the moduli space of maps from a Riemann surface $\Sigma_g$ of genus $g$ to a given Calabi-Yau target space ${\mathfrak{X}}$. 
Its partition function is the generating function of Gromov-Witten invariants of ${\mathfrak{X}}$, which roughly speaking count these maps. 

In recent years, deep connections have been unrooted between the topological string on various geometries and random matrix models. A classic result in the field is that intersection numbers, which are related to the Gromov-Witten theory of a point, are computed by the Kontsevich matrix integral \cite{Kontsevich}, see also \cite{Okounkov2}. In the Dijkgraaf-Vafa conjecture \cite{DVmatrix} such a connection is obtained between the topological B-model on certain non-compact Calabi-Yau manifolds and a 1-matrix model. A novel type of matrix model \cite{marino3} inspired by Chern-Simons theory is associated to the topological string in \cite{AKMV_CS}, yielding matrix model descriptions of target spaces obtained from the cotangent space of lens spaces via geometric transition. This work is extended to chains of lens spaces and their duals in \cite{HalmagyiOkuda}.

In the 20 years that have passed since topological string theory was formulated \cite{Witten, String}, various techniques have been developed for computing the corresponding partition function. The topological vertex method \cite{TopologicalVertex} solves this problem completely for toric Calabi-Yau 3-folds at large radius, furnishing the answer as a combinatorial sum over partitions. On geometries with unit first Betti number (the conifold and $\cO(-2) \rightarrow \IC \IP^1 \times \IC$), this formalism yields the partition function as a sum over a single partition with Plancherel measure. In \cite{Partitions2}, such a sum was rewritten as a 1-matrix integral. More complicated examples, such as the topological string on geometries underlying Seiberg-Witten $SU(n)$ theory, can be written as sums over multiple partitions \cite{NekrasovAnn, IK3, MarshakovNekrasov}. 1-matrix integrals that reproduce the corresponding partition functions were formulated in \cite{KlemmSulkowski}. Multi-matrix integrals have arisen in rewriting the framed vertex as a chain of matrices integral \cite{Partitions}. Its Hurwitz-numbers limit (infinite framing of the framed vertex geometry) was shown to be reproduced by a 1-matrix model with an external field in \cite{BorotEynardMulase, MorozovShakirov}.

Here, generalizing the method of \cite{Partitions2}, we are able to formulate a matrix model which reproduces the topological string partition function on a certain fiducial geometry, which we introduce in the next section. Flop transitions and limits in the K\"ahler cone relate the fiducial geometry to an arbitrary toric Calabi-Yau manifold. As we can follow the effect of both of these operations on the topological string partition function, our matrix model provides a description for the topological string on an arbitrary toric Calabi-Yau manifold.

By providing a matrix model realization, we are able to transcribe deep structural insights into matrix models to the topological string setting. E.g., our matrix model involves a chain of matrices, and chain of matrices integrals are always tau functions for an integrable system. Our matrix model realization hence proves integrability of the generating function of Gromov-Witten invariants. Moreover, matrix models satisfy loop equations, which are known to be equivalent to W-algebra constraints. A general formal solution to these equations was found in \cite{Eynard:2005wg}, centered around the introduction of an auxiliary Riemann surface, referred to as the spectral curve of the system. The partition and correlation functions of the matrix model are identified with so-called symplectic invariants of this curve \cite{OE}. The BKMP conjecture \cite{BKMP}, building on work of \cite{marino2Ann}, identifies the spectral invariants of the mirror curve to a toric Calabi-Yau manifold with the topological string partition function with the Calabi-Yau manifold as target space. In a forthcoming publication \cite{topstring2}, we will compute the spectral curve of our matrix model explicitly, thus establishing the validity of this conjecture.

Finally, we would like to emphasize that many different matrix models can yield the same partition function (justifying the choice of indefinite article in the title of this paper). An interesting open problem consists in identifying invariants of such equivalent matrix models. A promising candidate for such an invariant is the symplectic class of the matrix model spectral curve.

The outline of this paper is as follows. In section \ref{the_geometry}, after a very brief review of toric geometry basics, we introduce the fiducial geometry and the notation that we will use in discussing it throughout the paper. We also review the transformation properties of the topological string partition function under flop transitions, which will relate the fiducial to an arbitrary toric geometry, in this section. We recall the topological vertex formalism and its application to geometries on a strip \cite{IqbalKashaniPoor} in section \ref{vertex_calc}. Section \ref{our_matrix_modelAnn} contains our main result: we introduce a chain of matrices matrix model and demonstrate that it reproduces the topological string partition function on the fiducial geometry. By the argument above, we thus obtain a matrix model description for the topological string on an arbitrary toric Calabi-Yau manifold, in the large radius limit. We discuss implications of this result in section \ref{implications}, and point towards avenues for future work in section \ref{conclusions}.

\section*{2 The fiducial geometry and flop transitions} \label{the_geometry}

Toric geometries present a rich class of very computable examples for many questions in algebraic geometry. The topological vertex formalism provides an algorithm for computing the generating function for Gromov-Witten invariants on toric 3 dimensional Calabi-Yau manifolds. These are necessarily non-compact and have rigid complex structure.

The geometry of toric manifolds of complex dimension $d$ can be encoded in terms of a $d$ dimensional fan $\Sigma$, consisting of cones of dimensions 0 to $d$. We denote the set of all $n$ dimensional cones as $\Sigma(n)$. Each such $n$-cone represents the closure of a $(C^*)^{d-n}$ orbit. In particular, 1-cones correspond to hypersurfaces, and for $d=3$, our case of interest, 2-cones correspond to curves.  

The fan for the class of geometries we are interested in is constructed by triangulating a finite connected region of the $\IZ^2$ lattice containing the origin, embedding this lattice in $\IZ^3$ within the $(x,y)$ plane at $z=1$, and defining the cones of the fan via half-lines emanating at the origin and passing through the vertices of this triangulation.\footnote{The canonical class of a toric manifold is given by the sum over all torically invariant divisors. The construction sketched above guarantees that this sum is principal, hence the canonical class trivial: the monomial associated to the 1-cone $(0,0,1)$ generates the class in question. See e.g.  \cite{Fulton}.}

\begin{figure}[h]
 \centering
  \includegraphics[width=5cm]{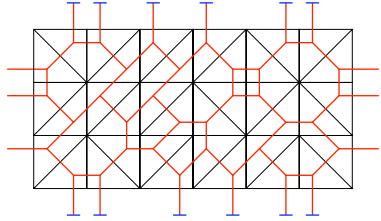}
 \caption{\footnotesize{Example of a box triangulation, corresponding to a 3 dimensional toric fan. The diagram in red is the dual web diagram. Vertices of the triangulations (faces of the web diagram) correspond to 1-cones, edges correspond to 2-cones, and faces (vertices of the dual) correspond to 3-cones.}}
 \label{web}
\end{figure}

We can associate a dual diagram to such toric fans, a so-called web diagram, spanned by lines orthogonal to the projection of 2-cones onto the $\IZ^2$ lattice. In the web diagram, the relation between the dimension of the components of the diagram and the submanifold of the toric geometry they represent coincide: 3-cones (points) correspond to vertices, and 2-cones (curves) to lines, see figure \ref{web}.

\subsection*{2.1 The fiducial geometry}    \label{fiducial}
\begin{figure}[h]
\centering
\includegraphics[width=6cm]{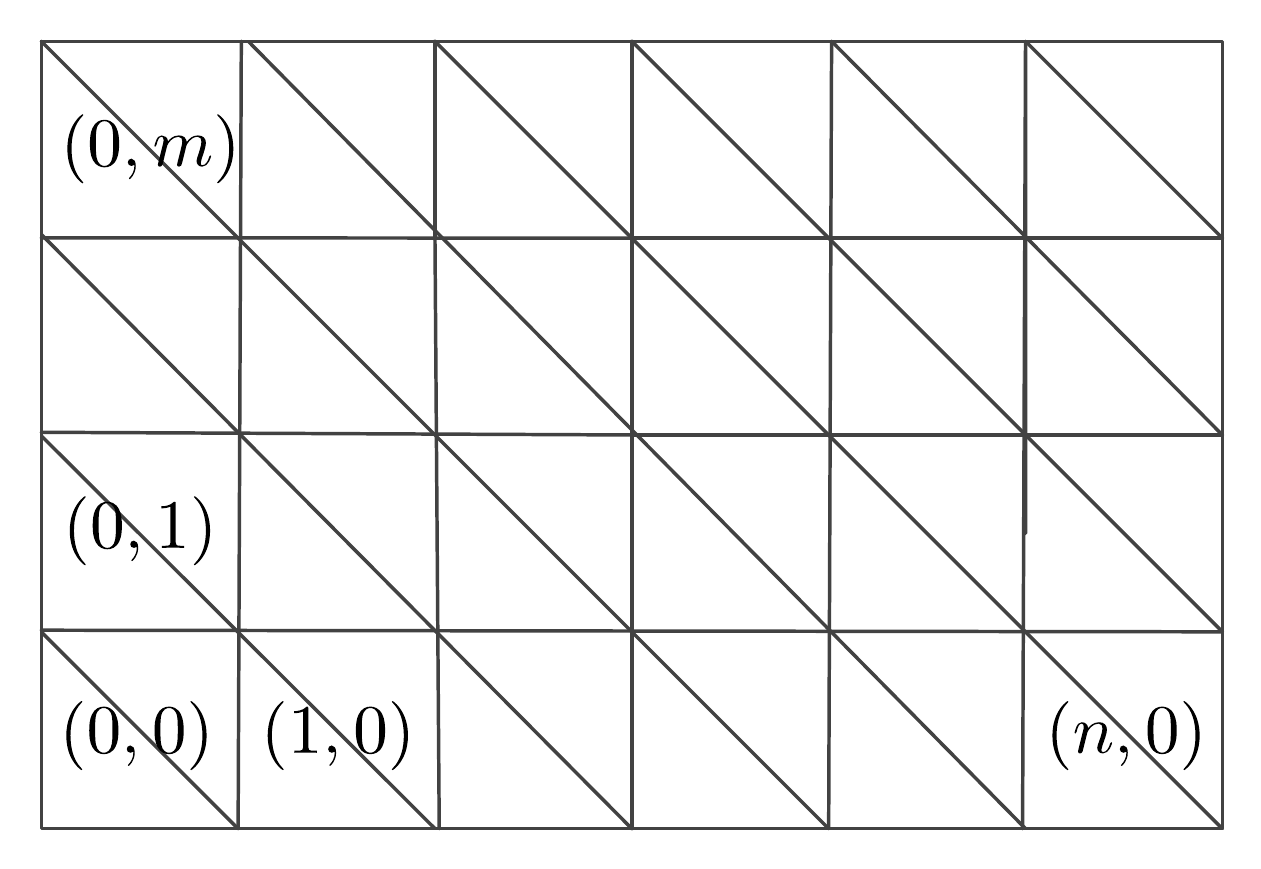}
\caption{\footnotesize{Fiducial geometry $\CYX_0$ with boxes numbered.}}
\label{fiducial_geometry_box}
\end{figure}

The geometry $\CYX_0$ we will take as the starting point of our considerations is depicted in figure \ref{fiducial_geometry_box}. 

\begin{figure}[h]
\centering
\includegraphics[width=3cm]{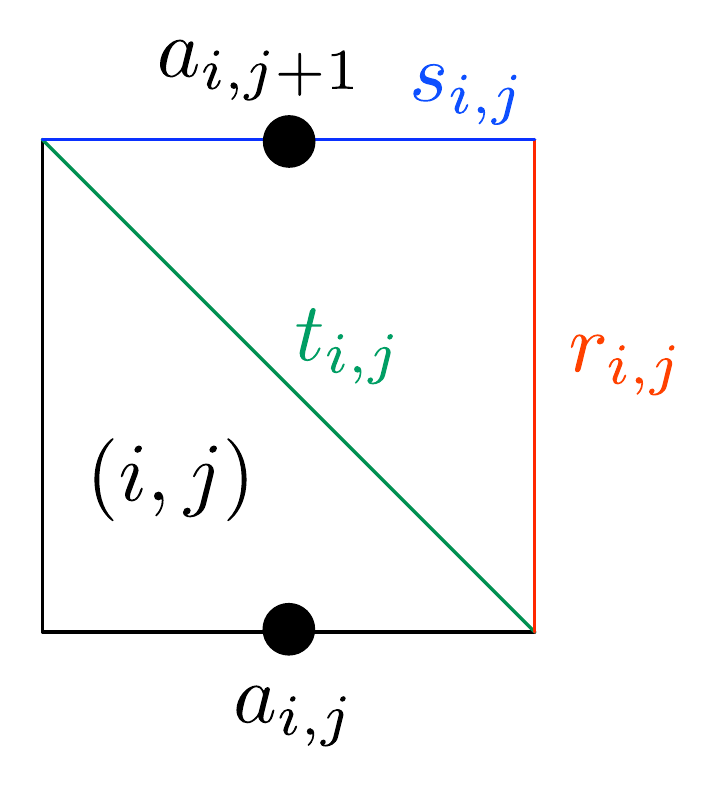}
\caption{\footnotesize{Labeling curve classes, and introducing $a$-parameters.}}
\label{fiducial_labelingAnn}
\end{figure}

Since the torically invariant curves play a central role in our considerations, we introduce a labeling scheme for these in figure \ref{fiducial_labelingAnn}: $(i,j)$ enumerates the boxes as in figure \ref{fiducial_geometry_box}, and we will explain the $a$-parameters further below.

\begin{figure}[h]
 \centering
 \includegraphics[width=8cm]{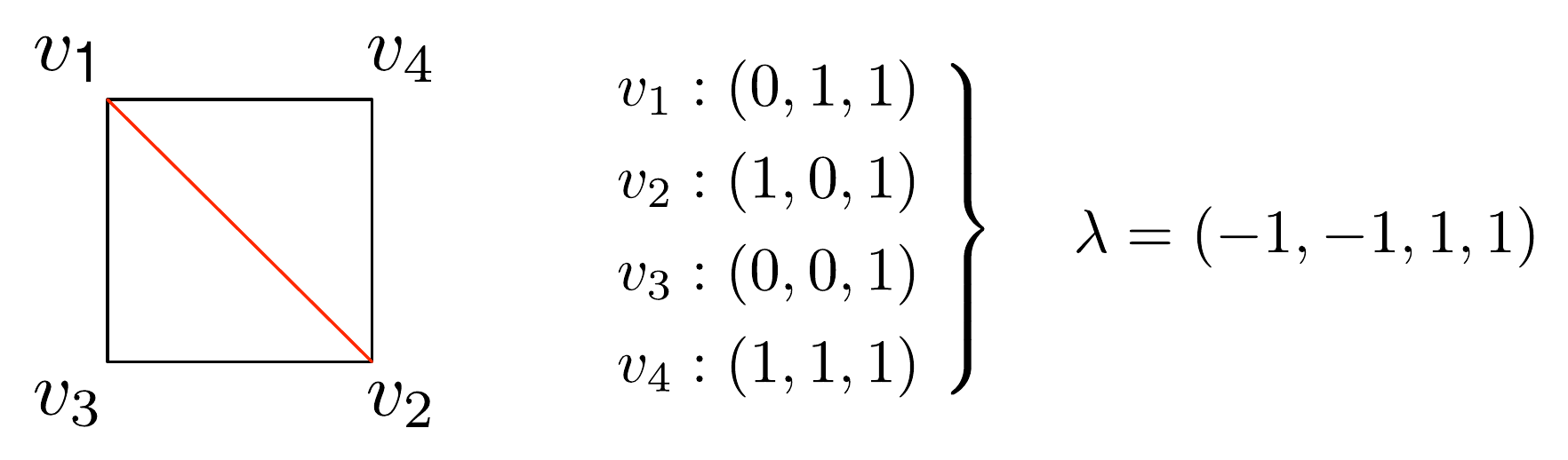}
 \caption{\footnotesize{The 2-cone $\sigma$ corresponds to the relation $\vec{\lambda}$ among 1-cones.}}
 \label{2conerelationcorr}
\end{figure}

In the following, we will, when convenient, use the same notation for a torically invariant curve $\Sigma$, its homology class $[\Sigma] \in H_2(\CYX_0,\IZ)$, and its volume or associated K\"ahler parameter $\int_\Sigma J$, given a K\"ahler form $J$ on $\CYX_0$. The classes of the curves $r_{i,j}, s_{i,j}, t_{i,j}$ introduced in figure \ref{fiducial_labelingAnn} are not independent. To determine the relations among these, we follow \cite[page 39, 40]{CoxKatz}. Consider the integer lattice $\Lambda$ spanned by formal generators $e_\rho$, with $\rho \in \Sigma(1)$ 1-cones of the toric fan,
\beq
\Lambda = \{ \sum_{\rho \in \Sigma(1)} \lambda_\rho e_\rho | \lambda_\rho \in \IZ \} \,.
\eeq
Each torically invariant curve, corresponding to a 2-cone of the fan, maps to a relation between 1-cones, and thus to an element of the lattice $\Lambda$, as follows: a 2-cone $\sigma$ is spanned by two integral generators $v_1$ and $v_2$, and it is contained in precisely two 3-cones, which are each spanned by $v_1, v_2$ and one additional generator $v_3$, $v_4$ respectively. These vectors satisfy the relation $\sum_{i=1}^4 \lambda_{i} v_i =0$, where the $\lambda_{i}$ can be chosen as relatively prime integers, and as $v_3$ and $v_4$ lie on opposite sides of $\sigma$, we can assume that $\lambda_{3}, \lambda_{4} >0$. \cite{CoxKatz} shows that on a smooth variety, the sublattice $\Lambda_h$ generated by the elements $\sum_{i=1}^4 \lambda_{i} e_i$ of $\Lambda$ is isomorphic to $H_2(\CYX_0,\IZ)$. We call this isomorphism $\lambda$,
\beq 
\lambda : H_2(\CYX_0, \IZ) \rightarrow \Lambda_h \,.
\eeq
Figure \ref{2conerelationcorr} exemplifies this map.

\begin{figure}[h]
 \centering
  \includegraphics[width=4cm]{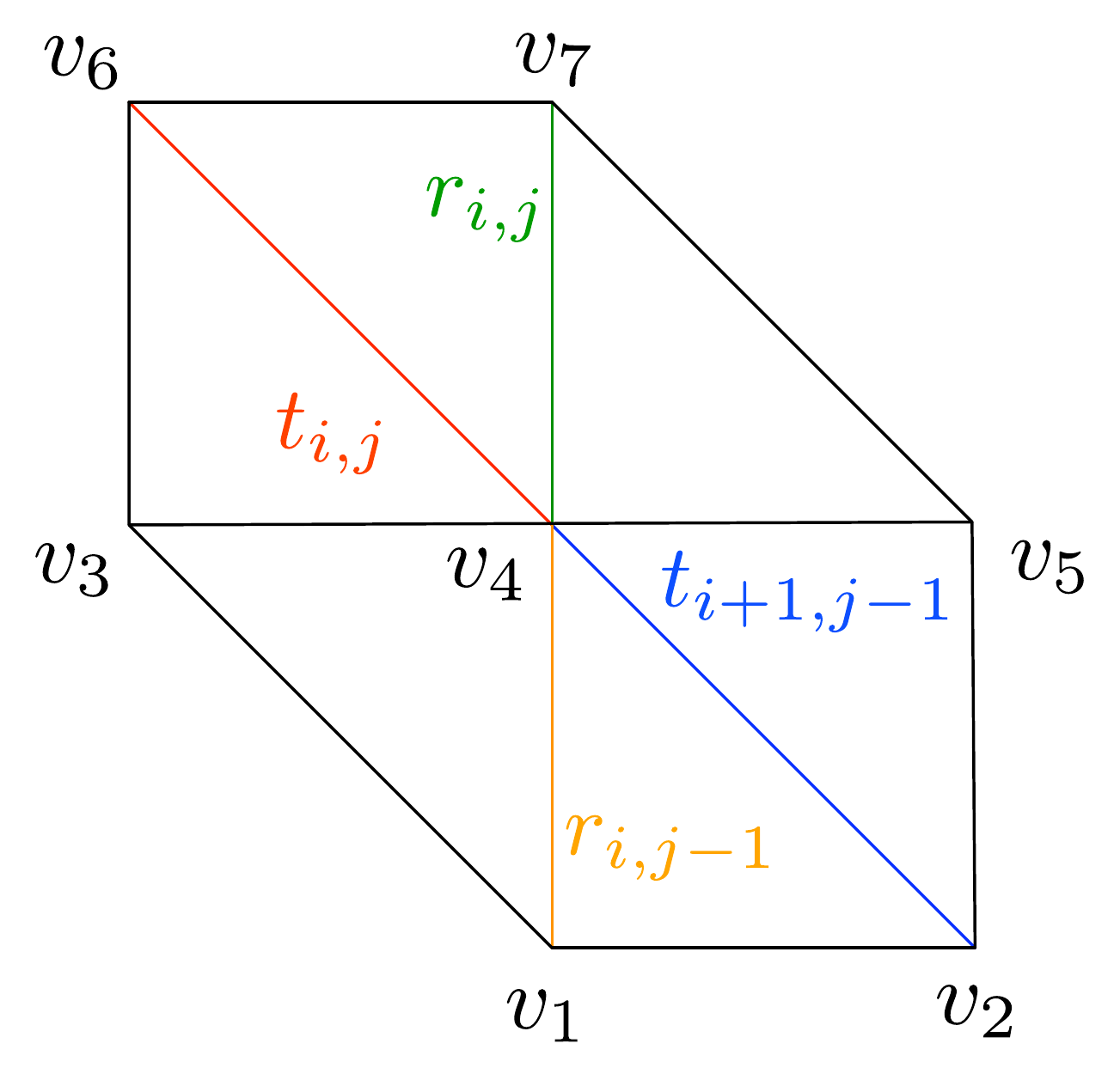}
 \caption{\footnotesize{Determining the relation between curve classes.}}
 \label{rel_bw_curve_classes}
\end{figure}

It allows us to easily work out the relation between the various curve classes. Consider figure \ref{rel_bw_curve_classes}.

\begin{figure}[h]
\centering
\includegraphics[width=12cm]{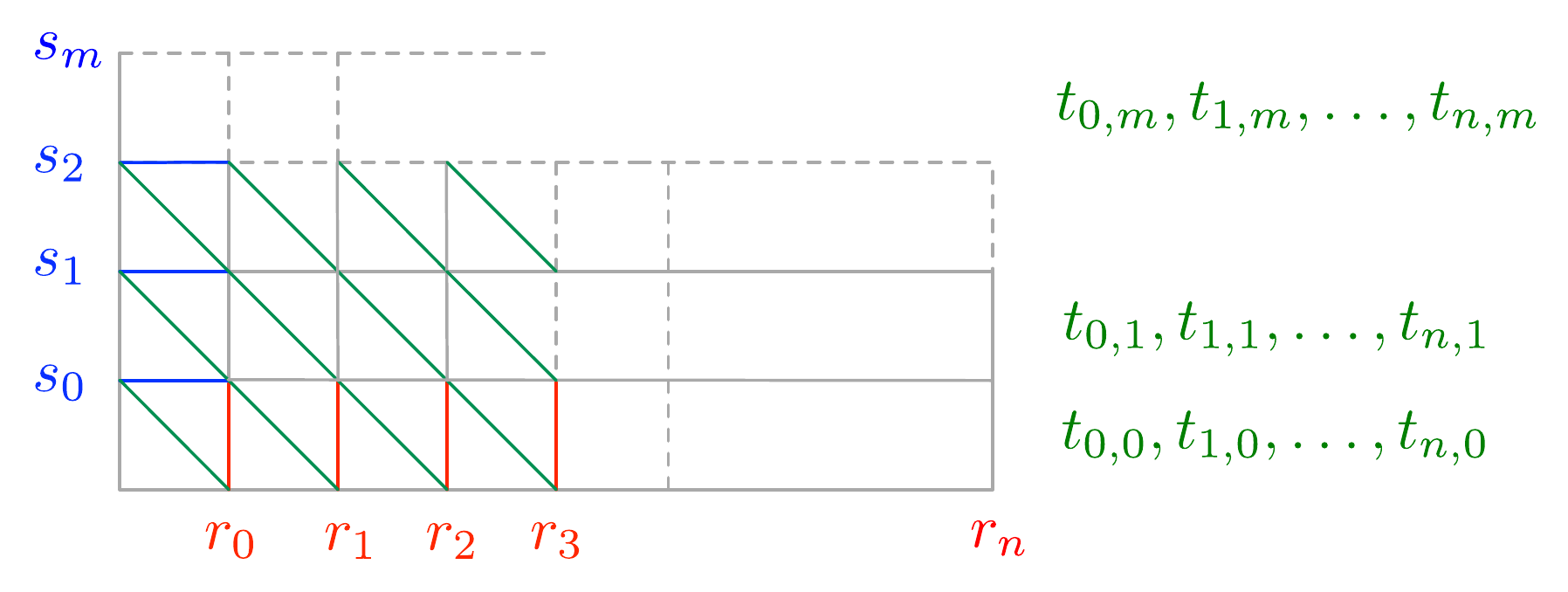}
\caption{\footnotesize{Fiducial geometry with choice of basis of $H_2(\CYX_0,\IZ)$.}}
\label{fiducial_geometryAnn}
\end{figure}

The images of the curve classes depicted there under $\lambda$ are,
\ba
\lambda(r_{i,j}) &=& e_5 + e_6 - e_4 -e_7 \,,\\
\lambda(r_{i,j-1}) &=& e_2 + e_3 -e_1 -e_4\,,\\
\lambda(t_{i,j}) &=& e_3 + e_7 - e_4 - e_6 \,,\\
\lambda(t_{i+1,j-1}) &=& e_1 + e_5 - e_2 - e_4\,.\\
\ea
We read off the relation
\beqn   \label{tr_relation}
t_{i,j}+r_{i,j} = t_{i+1,j-1} + r_{i,j-1} \,.
\eeqn
By symmetry, we also have
\beq
t_{i,j} + s_{i,j-1} = t_{i+1,j-1} + s_{i+1,j-1} \,.
\eeq
A moment's thought convinces us that this constitutes a complete basis for the space of relations. We can solve these in terms of the classes of the curves $r_i, s_i, t_{i,j}$, $i,j = 0,1,\ldots$ depicted in figure \ref{fiducial_geometryAnn}, which hence generate $H_2(\CYX_0,\IZ)$.
The explicit relations are
\ba
r_{i,j} &=& r_i + \sum_{k=1}^j (t_{i+1,k-1} - t_{i,k})\,, \\
s_{i,j} &=& s_j + \sum_{k=1}^i ( t_{k-1,j+1} - t_{k,j})\,.\\
\ea

Our computation for the partition function on $\CYX_0$ will proceed by first considering the horizontal strips in the toric fan describing the geometry, as depicted in figure \ref{fiducial_geometry_box}, individually, and then applying a gluing algorithm to obtain the final result.

For each strip, we find it convenient to write the curve class $w_{IJ} \in H_2(\CYX_0,\IZ)$ of the curve extending between two 3-cones which we label by $I$ and $J$ (recall that 3-cones correspond to vertices in the dual web diagram), with $J$ to the right of $I$, as the difference between two parameters $a_I$ and $a_J$ associated to each 3-cone, 
\beqn   \label{a_parameters}
w_{IJ} = a_I - a_J \,.
\eeqn
We call these parameters, somewhat prosaically, $a$-parameters. It is possible to label the curve classes in this way due to their additivity along a strip. In terms of the notation introduced in figure \ref{fiducial_labelingAnn}, we obtain\beq
t_{i,j}=a_{i,j}-a_{i,j+1} \quad\,, \quad \quad r_{i,j} = a_{i,j+1} - a_{i+1,j} \,.
\eeq

By invoking the relation (\ref{tr_relation}), we easily verify that upon gluing two strips, the curve class of a curve extending between two 3-cones $I$ and $J$ on the lower strip is equal to the class of the curve between the 3-cones $I'$ and $J'$ on the upper strip, where the cones $I$ and $I'$ are glued together, as are the cones $J$ and $J'$,
\beqn
w_{IJ} = w_{I'J'} \,. \label{identify_class}
\eeqn
This allows us to identify the parameters $a_I=a_{I'}$ and $a_J=a_{J'}$ associated to 3-cones glued together across strips.

Note that the basic curve classes $s_{i}$ are not captured by the parameters $a_{i,j}$.

\subsection*{2.2 Flop invariance of toric Gromov-Witten invariants} \label{flops}

\begin{figure}[h]
 \centering
 \includegraphics[width=6cm]{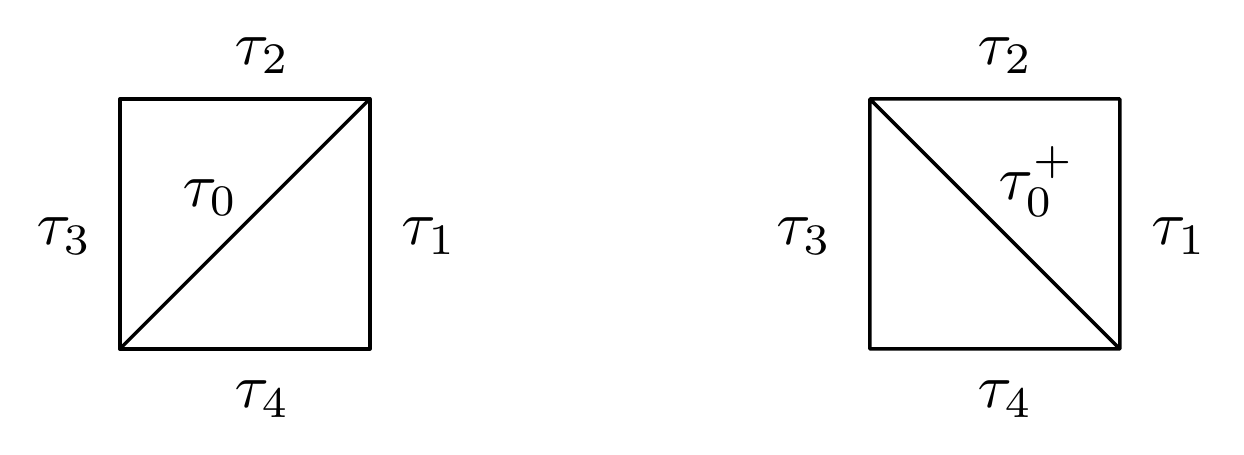}
 \caption{\footnotesize{$\CYX$ and $\CYX^+$ in the vicinity of the (-1,-1) curve.}}
 \label{toric_floppedAnn}
\end{figure}

Under the proper identification of curve classes, Gromov-Witten invariants (at least on toric manifolds) are invariant under flops. Assume $\CYX$ and $\CYX^+$ are related via a flop transition, $\phi: \CYX \rightarrow \CYX^+$. In a neighborhood of the flopped $(-1,-1)$ curve, the respective toric diagrams are depicted in figure \ref{toric_floppedAnn}.

The 1-cones of $\Sigma_\CYX$, corresponding to the toric invariant divisors of $\CYX$, are not affected by the flop, hence can be canonically identified with those of $\CYX^+$. The 2-cones $\tau_i$ in these diagrams correspond to toric invariant 2-cycles $C_i$, $C_i^+$ in the geometry. The curve classes of $\CYX$ push forward to classes in $\CYX^+$ via
\ban\label{floppingAnn}
\phi_*([C_0]) = - [C_0^+] \,, \quad \phi_*([C_i]) =  [C_i^+] + [C_0^+] \,. 
\ean
All other curve classes of $\CYX$ are mapped to their canonical counterparts in $\CYX^+$. Under appropriate analytic continuation and up to a phase factor (hence the $\propto$ in the following formula), the following identity then holds \cite{Witten_Phases,IqbalKashaniPoor, TopologicalVertex2},
\ban
Z_{GW}(\CYX,Q_0,Q_1,\ldots,Q_4,\vec{Q}) \propto Z_{GW}(\CYX^+,1/Q_0, Q_0 Q_1, \ldots, Q_0 Q_4,\vec{Q}) \,,  \label{rel_part_flopped}
\ean
i.e.
\ba
GW_g(\CYX,Q_0,Q_1,\ldots,Q_4,\vec{Q}) = GW_g(\CYX^+,1/Q_0, Q_0 Q_1, \ldots, Q_0 Q_4,\vec{Q}) \,.
\ea

Any toric Calabi-Yau manifold $\CYX$ with K\"ahler moduli $\vec{Q}$ can be obtained from a sufficiently large fiducial geometry $(\CYX_0,\vec{Q}_0)$ upon performing a series of flop transitions and taking unwanted K\"ahler moduli of $\CYX_0$ to $\infty$. Once we obtain a matrix model reproducing the topological string partition function on the fiducial geometry, extending the result to arbitrary toric Calabi-Yau 3-folds will therefore be immediate.

\begin{figure}[h]
 \centering
 \includegraphics[width=10cm]{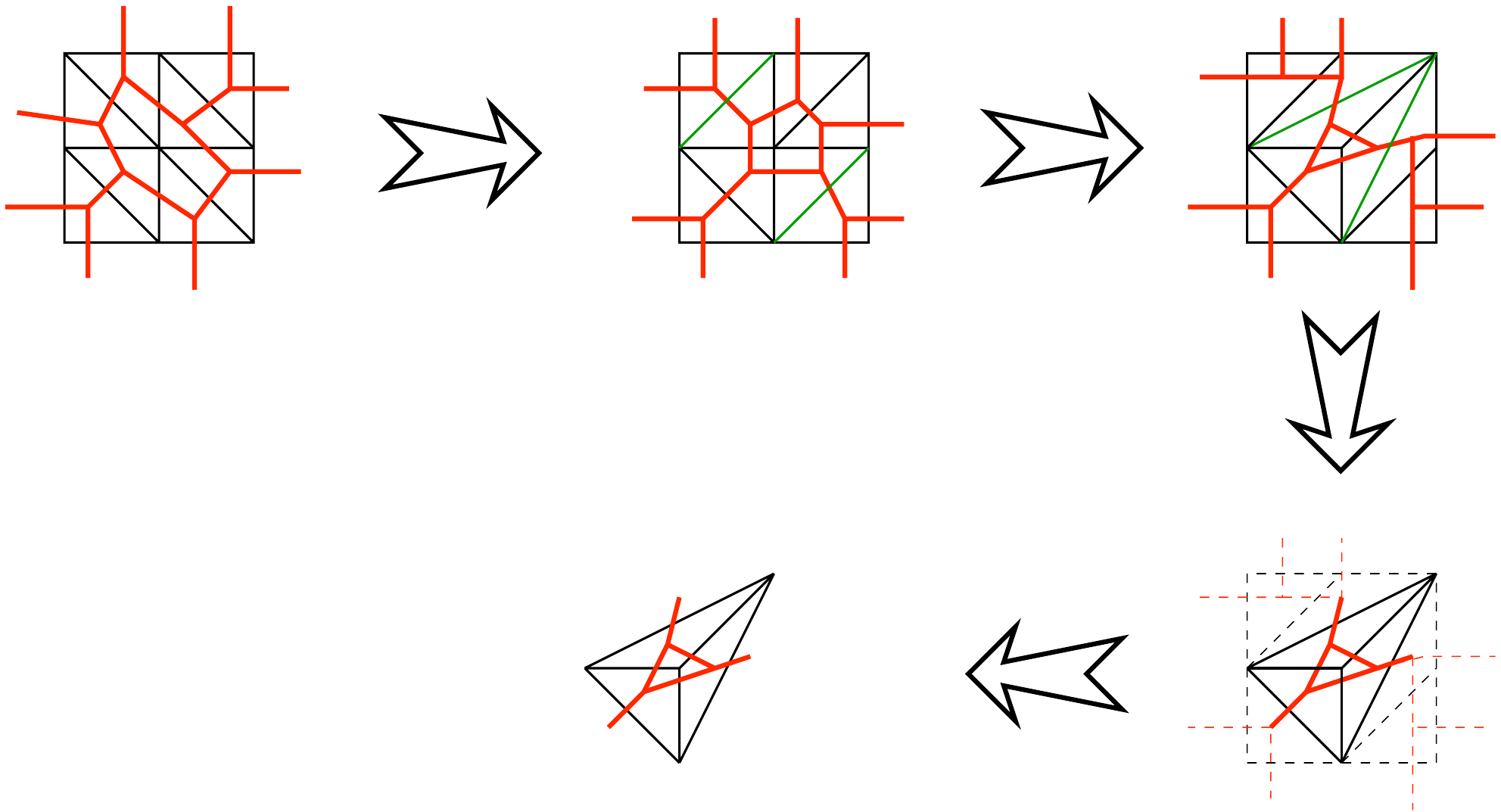}
 \caption{\footnotesize{We obtain local $\mathbb P^2$ from the fiducial geometry with $2\times 2$ boxes by performing five flops and then sending the K\"ahler parameters of the unwanted edges to $\infty$.}}
 \label{flopP2Ann}
\end{figure}

As an example, we show how to obtain the $\mathbb P^2$ geometry from the fiducial geometry with $2\times 2$ boxes in figure \ref{flopP2Ann}.

\section*{3 The partition function via the topological vertex} \label{vertex_calc}
\subsection*{3.1 Gromov-Witten invariants}

Gromov-Witten invariants ${\cal N}_{g,D}(\CYX)$ roughly speaking count the number of maps from a Riemann surface of genus $g$ into the target space $\CYX$, with image in a given homology class $D=(D_1,\dots,D_{k})\in H_2(\CYX,\mathbb Z)$. They can be assembled into a generating series
\beq
GW_g(\CYX,Q) = \sum_{D} {\cal N}_{g,D}(\CYX)\, Q^{D}.
\eeq
Each $GW_g(\CYX,Q)$ is a formal series in powers $Q^D = \prod_i Q_i^{D_i}$ of the parameters $Q=(Q_1,Q_2,\dots,Q_k)$, the exponentials of the K\"ahler parameters. 

\medskip

We can introduce a generating function for Gromov-Witten invariants of all genera by introducing a formal parameter $g_s$ (the string coupling constant) and writing
\beq
GW(\CYX,Q,g_s) = \sum_{g=0}^\infty\,\, g_s^{2g-2}\,\, GW_g(\CYX,Q) \,.
\eeq
%This equality is an equality between formal power series of $Q$, and to each order, the sum over $g$ is finite.

\bigskip

It is in fact more convenient to introduce disconnected Gromov-Witten invariants  ${\cal N}^*_{\chi,D}(\CYX)$, for possibly disconnected surfaces, of total Euler characteristics $\chi$, and to define
\beq
Z_{GW}(\CYX,Q,g_s) = e^{GW(\CYX,Q,g_s)} =  \sum_{D} \, Q^D\,\, \sum_\chi  g_s^{-\chi} \,\, {\cal N}^*_{\chi,D}(\CYX).
\eeq

For toric Calabi-Yau manifolds, an explicit algorithm was presented in \cite{TopologicalVertex} for computing $Z_{GW}$ via the so-called topological vertex formalism, proved in \cite{LLLZ,MOOP}.

\subsection*{3.2 The topological vertex}

In the topological vertex formalism, each vertex of the web diagram contributes a factor $C_q({\alpha, \beta, \gamma})$ to the generating function of GW-invariants, 
where the $\alpha, \beta, \gamma$ are Young tableaux associated to each leg of the vertex, 
and $C_q({\alpha, \beta, \gamma})$  is a formal power series in the variable $q$, where
$$
q=e^{-g_s}.
$$
Topological vertices are glued along edges (with possible framing factors, see \cite{TopologicalVertex}) carrying the same Young tableaux $\alpha$ by performing a sum over $\alpha$, weighted by $Q^{|\alpha|}$, with $Q$ encoding the curve class of this connecting line,
\beq
Z_{\rm vertex}(\CYX,Q,q)=\sum_{{\rm Young\,\, tableaux}\,\alpha_e}\,\, \prod_{{\rm edges}\, e}\, Q_e^{|\alpha_e|}\,\,\, \prod_{{\rm vertices}\, v =(e_1,e_2,e_3)}\,\, C_q(\alpha_{e_1},\alpha_{e_2},\alpha_{e_3}) \,.
\eeq
Note that in practical computations, the sum over representations can ordinarily not be performed analytically. A cutoff on the sum corresponds to a cutoff on the degree of the maps being counted.
 
The equality
\beq
Z_{GW}(\CYX,Q,g_s) = Z_{\rm vertex}(\CYX,Q,q)
\eeq
holds at the level of formal power series in the $Q$'s, referred to as the large radius expansion. It was proved in \cite{MOOP} that the log of the right hand side indeed has a power series expansion in powers of $g_s$.

\subsection*{3.3 Notations for partitions and q-numbers}

Before going further in the description of the topological vertex formula, we pause to fix some notations and introduce special functions that we will need in the following.

\subsubsection*{Representations and partitions} \label{secrepresentations}

Representations of the symmetric group are labelled by Young tableaux, or Ferrer diagrams. For a representation $\gamma$, we introduce the following notation:
\begin{itemize}
\item $\gamma_i$: number of boxes in the $i$-th row of the Young tableau associated to the representation $\gamma$, $\gamma_1\geq \gamma_2\geq \dots \geq \gamma_{d}\geq 0$.

\item The weight $|\gamma|=\sum_i \gamma_i$: the total number of boxes in the corresponding Young tableau.
\item The length $l(\gamma)$: the number of non-vanishing rows in the Young tableau, i.e. $\gamma_i=0$ iff $i>l(\gamma)$.

\item The Casimir $\kappa(\gamma) = \sum_i \gamma_i (\gamma_i - 2i +1 )$.

\item $\gamma^T$ denotes the conjugate representation, which is obtained by exchanging the rows and columns of the associated Young tableau. We have $|\gamma^T|=|\gamma|$, $l(\gamma^T)=\gamma_1$, and $\kappa(\gamma^T)=-\kappa(\gamma)$.

\end{itemize}

\medskip

An integer $d>0$ will denote a cut-off on the length of representations summed over, 
\ba
l(\gamma)\leq d.
\ea
Most expressions we are going to write will in fact be independent of $d$, and we shall argue in \cite{topstring2}, following the same logic as in \cite{Partitions2} based on the arctic circle property \cite{JohanssonAnn}, that our results depend on $d$ only non-perturbatively.

\smallskip
To each representation $\gamma$, we shall associate a parameter $a$ as introduced in (\ref{a_parameters}).

\begin{figure}[h]
 \centering
  \includegraphics[width=14cm]{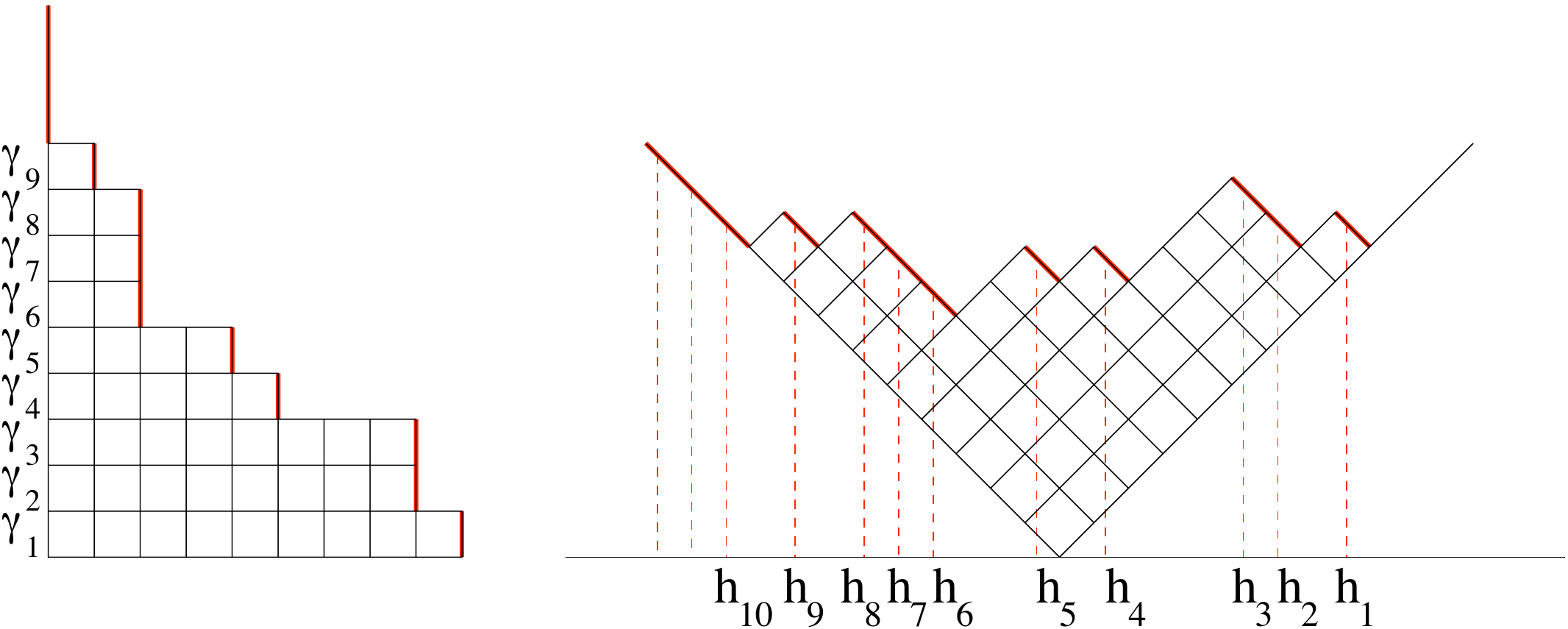}
 \caption{\footnotesize{Relation between a partition $\gamma$ and $h(\gamma)$.}}
 \label{expartition}
\end{figure}

Instead of dealing with a partition $\gamma$, characterized by the condition $\gamma_1 \ge \gamma_2 \ge \ldots \ge \gamma_d \ge 0$, it will prove convenient to define the quantities
\beqn \label{defhAnn}
h_i(\gamma)=\gamma_i-i+d+a \,,
\eeqn
which satisfy instead
\beq
h_1>h_2>h_3>\dots >h_{d}\geq a  \,.
\eeq
The relation between $\gamma$ and $h(\gamma)$, for the off-set $a=0$, is depicted in figure \ref{expartition}.

We finally introduce the functions
\beq 
x_i(\gamma)=q^{h_i(\gamma)}  \,.
\eeq

In terms of the $h_i(\gamma)$, we have
\ba
\kappa(\gamma)=\sum_i h_i^2 - (2d+2a-1)\sum_i h_i + d\,C_{d,a} \,,
\ea
where $C_{d,a}=\frac{1}{3}(d-1)(2d-1)+\,a(a+2d-1)$.

\subsubsection*{q-numbers}

We choose a string coupling constant $g_s$ such that the quantum parameter
$q= e^{-g_s}$ satisfies $|q|<1$. A $q$-number $[x]$ is defined as
\beqn  \label{q_number}
[x] = q^{-\frac{x}{2}} - q^{\frac{x}{2}} = 2 \sinh \frac{x\,g_s}{2} \,.
\eeqn
$q$-numbers are a natural deformation away from the integers; in the limit $q\to 1$, ${1\over g_s}\,[x] \rightarrow  x$.

We also define the $q$-product
\ba
g(x) = \prod_{n=1}^\infty (1-{1\over x}\,q^n) \,.
\ea
The function $g(x)$ is related to the quantum Pochhammer symbol, $g(x)
= [q/x;q]_{\infty}$, and to the $q$-deformed gamma function via $\Gamma_q(x) = (1-q)^{1-x}\,g(1)/g(q^{1-x})$.
$g(x)$ satisfies the functional relation
\ba
g(qx) = (1-{1\over x})\,\, g(x) \,.
\ea
For $\Gamma_q$, this implies $\Gamma_q(x+1) = {1-q^x\over 1-q}\,\, \Gamma_q(x)$, the
quantum deformation of the functional equation $\Gamma(x+1)=x\Gamma(x)$ of the gamma function, which is recovered in the classical limit $q\to 1$.
The central property of $g(x)$ for our purposes is that it vanishes on integer powers of $q$,
\beq
g(q^n)=0 \quad {\rm if}\,\, n\in \mathbb N^*.
\eeq
Moreover, it has the following small $\ln q$ behavior,
\beq
\ln{g(x)} = {1\over \ln q}\,\sum_{n=0}^\infty {(-1)^n\,B_n\over n!}\,(\ln q)^n\,\, \Li_{2-n}(1/x)  \,,
\eeq
where $\Li_n(x) = \sum_{k=1}^\infty {x^k\over k^n}$ is the polylogarithm, and $B_n$ are the Bernouilli numbers
\ba
B_0=1 \,\, , \quad B_1 =-{1\over 2}\,\, , \quad B_2={1\over 6}\,\, , \quad \dots
\ea
$B_{2k+1}=0$ if $k\geq 1$ (see the appendix).

We shall also need the following function $f(x)$,
\ba
{1\over f(x)} 
&=& {g(x) \, g(q/x)\over g(1)^2\,\,\sqrt x}\,\,e^{(\ln{x})^2\over 2\ln q}\,\, e^{-i\pi \ln x\over \ln q}   
%= {e^{-i\pi\ln x\over \ln q}\over \Gamma_q(q/x)\,\,\Gamma_q(x)} 
\cr
&=& {-\ln q\over \theta'({1\over 2} - {i\pi\over \ln q}, -{2i\pi\over \ln q}) }\,\,\, \theta\left({\ln x\over \ln q} +{1\over 2} - {i\pi \over \ln q}, {-2i\pi\over \ln q}\right) ,
\ea
where $\theta$ is the Riemann theta-function for the torus of modulus $-2i\pi/\ln q$.
This relationship is the quantum deformation of the classical gamma function identity $$e^{-i\pi x}/\Gamma(1-x)\Gamma(x) = \sin{(\pi x)}/\pi \,.$$

\subsection*{3.4 The partition function via the vertex\label{secZvertex}}

\begin{figure}[h]
 \centering 
  \includegraphics[width=15cm]{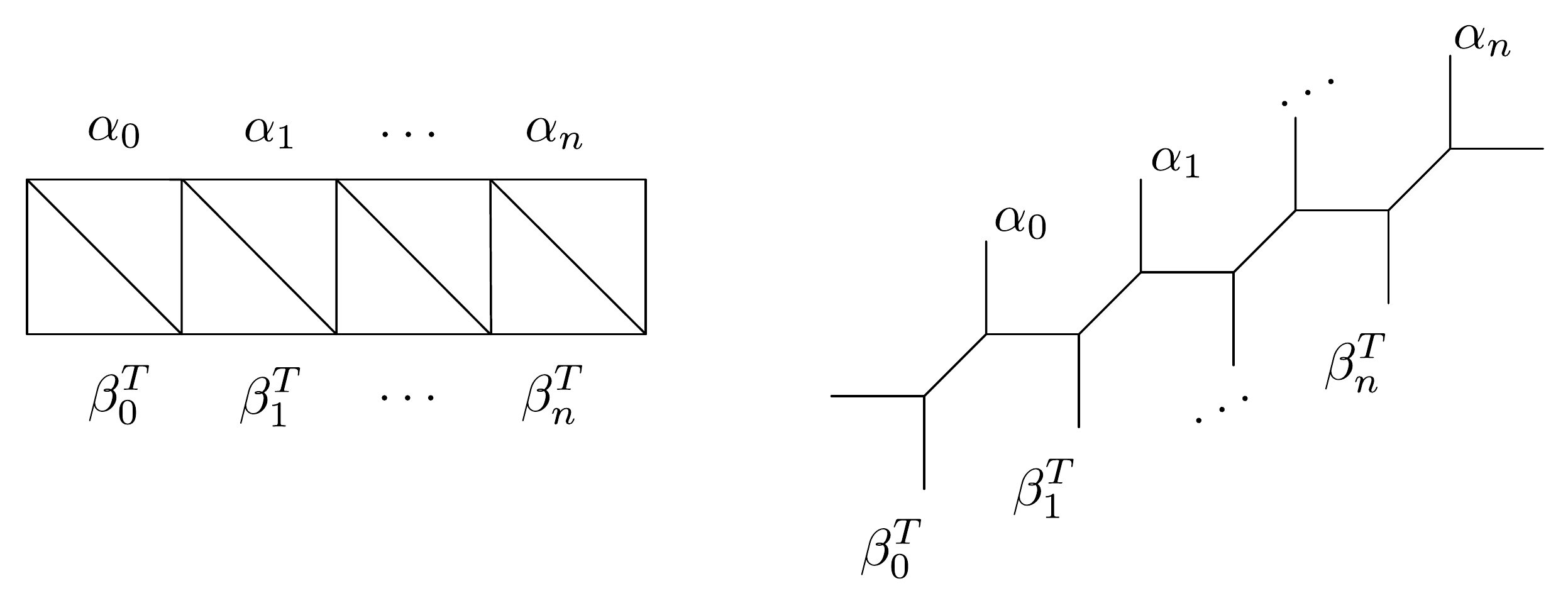}
 \caption{\footnotesize{A horizontal strip of the fiducial geometry and its corresponding web diagram. \label{figstrip}}}
\end{figure}

We begin by considering a single horizontal strip of the fiducial geometry, as depicted in figure \ref{figstrip}.

Of the three legs of the vertex, two point in the direction of the strip and connect the vertex to its neighbors. One leg points out of the strip, either above or below. This leg carries a free representation, $\alpha_i$ or $\beta_i^T$ in the notation of figure \ref{figstrip}. The partition function will hence depend on representations, one per vertex (i.e. face of the triangulation). 

A note on notation: since each 3-cone carries a representation (which up to the final paragraph of this subsection is held fixed) and an a-parameter (see figure \ref{fiducial_labelingAnn}), we will identify the a-parameters by the corresponding representations when convenient.

Using the topological vertex, it was shown in \cite{IqbalKashaniPoor} that the A-model topological string partition function of the strip is given by a product of terms, with the individual factors depending on the external representations and all possible pairings of these. Applied to the fiducial strip, the results there specialize to
\ban
Z_{\rm strip}(\alpha_0;\beta^T)  \label{z_stripAnn}
&=&   \prod_{i=0}^n \frac { [\alpha_i] [\beta^T_i] }{\,\quad \quad [\beta_i,\alpha_i^T]_{Q_{\beta_i,\alpha_i}}} \,\,\,\, {\prod_{i<j} [\alpha_i,\alpha_j^T]_{Q_{\alpha_i,\alpha_j}} \,\,\prod_{i<j} [\beta_i,\beta_j^T]_{Q_{\beta_i,\beta_j}} 
\over \prod_{i<j} [\alpha_i,\beta_j^T]_{Q_{\alpha_i,\beta_j}} [\beta_i,\alpha_j^T]_{Q_{\beta_i,\alpha_j}}
}    
\label{pfcn} \,.
\ean
We explain each factor in turn.

\medskip

$\bullet$ Each vertex $\gamma=\alpha_i$ or $\gamma=\beta_i^T$ contributes a representation dependent factor to the partition function, which we have denoted by $[\gamma]$. It is the $n \rightarrow \infty$ limit of the Schur polynomial evaluated for $x_i = q^{\frac{1}{2}-i}$, $i=1, \ldots, n$, given explicitly by
\ba
[\gamma] 
&=& (-1)^d  q^{\frac{1}{4} \kappa(\gamma)} \prod_{1 \le i < j \le d} \frac{[\gamma_i - \gamma_j + j - i]}{[j-i]} \prod_{i=1}^{d} \prod_{j=1}^{\gamma_i} \frac{1}{[d + j - i]} \\
&=&
\prod_{1\leq i< j\leq d} (q^{h_j}-q^{h_i}) \,\,\prod_{i=1}^d \, \left(\frac{ g(q^{a_\gamma-h_i})}{g(1)}\,\,q^{\frac{1}{2}h_i^2-(a_\gamma+d-1)h_i+\frac{a_\gamma(a_\gamma+d-1)}{2}+\frac{(d-1)(2d-1)}{12}}\right)  \cr
&=& \Delta(X(\gamma))\,\,  e^{-{1\over g_s} \tr U(X(\gamma),a_\gamma)}\,\,e^{-{1\over g_s} \tr U_1(X(\gamma),a_\gamma)}.
\ea
We recall that $h_i(\gamma) = \gamma_i-i+d+a_\gamma$, and we have defined $x_i=q^{h_i}$ and the diagonal matrix $X(\gamma)={\rm diag}(q^{h_1},q^{h_2},\dots,q^{h_d})$. Furthermore, $\Delta(X)$ denotes the Vandermonde determinant of the matrix $X$,
\beq
\Delta(X) = \prod_{1\leq i<j\leq d} (x_j-x_i) \,,
\eeq
and we have written
\beq
U(X,a)  = - g_s \ln{\left( g({q^a\over X})\over g(1)\right)}  \,,
\eeq
\beq
U_1(X,a) =  {(\ln X)^2\over 2} - (a+d-1)\ln{X}\,\ln q  + C(a,d) \,,
\eeq
where $C(a,d) = \frac{a(a+d-1)}{2}+\frac{(d-1)(2d-1)}{12}$.

We have
$$ [\gamma] = q^{\frac{\kappa(\gamma)}{2}} [\gamma^T]\,, \quad \kappa(\gamma^T) = - \kappa(\gamma) \,,$$
and thus
\ba
[\gamma^T] 
&=& \Delta(X(\gamma))\,\,  e^{-{1\over g_s} \tr U(X(\gamma),a_\gamma)}\,\,e^{-{1\over g_s} \tr \td U_1(X(\gamma),a_\gamma)} \,,
\ea
where
\beq
\td U_1(X,a)  = {1\over 2} \ln{X}\,\ln q  + \td C(a,d).
\eeq
$\td C_{a,d}$ is another constant which depends only on $a$ and $d$ and which will play no role for our purposes.

\medskip

$\bullet$ 
In addition, each pair of representations contributes a factor, reflecting the contribution of the curve extended between the respective vertices. In the nomenclature of \cite{IqbalKashaniPoor}, the representations $\alpha_i$ are all of same type, and of opposite type relative to the $\beta_i$. If we take $i<j$, representations of same type (corresponding to (-2,0) curves) contribute a factor of $$[\alpha_i,\alpha_j^T] \quad \mbox{or} \quad [\beta_i^T,\beta_j]\,,$$ whereas representations of different type (corresponding to (-1,-1) curves) contribute a factor of $$\frac{1}{[\alpha_i,\beta_j]}  \quad \mbox{or} \quad  \frac{1}{[\beta_i^T,\alpha_j^T]}\,.$$
The pairing is given by \cite{IK2,IK3,EguchiKanno,IqbalKashaniPoor}
\ban  \label{pairingproduct}
[\gamma, \delta^T]
&=&  Q_{\gamma,\delta}^{-\frac{|\gamma| + |\delta|}{2}} q^{-\frac{\kappa(\gamma) - \kappa(\delta) }{4}} \prod_{i=1}^{d} \prod_{j=1}^{d} \frac{  [ h_i(\gamma)-h_j(\delta)] }{ [ a_\gamma-a_\delta  + j - i] }  \nn\\
&& \times \prod_{i=1}^{d} \prod_{j=1}^{\gamma_i} \frac{1}{[a_\gamma-a_\delta + j - i + d]} \prod_{i=1}^{d} \prod_{j=1}^{\delta_i} \frac{1}{[a_\gamma-a_\delta - j + i - d]} \,\,  \prod_{k=0}^\infty g(Q_{\gamma,\delta}^{-1}q^{-k})\, \nn\\
&=& (-1)^{\frac{d(d-1)}{2}} \prod_{i=1}^d \frac{q^{\frac{1}{2} (h_i(\delta)^2 - h_i(\delta) (2 a_\gamma +2d -1) - a_\delta^2 + 2 a_\gamma a_\delta + (d-2i)a_\gamma + (2i - d - 1) a_\delta )}}{[a_\gamma - a_\delta]^{d}}  \prod_{i=1}^d (-1)^{\delta_i}  \nn \\
&&\prod_{i,j=1}^d (q^{h_j(\delta)}- q^{h_i(\gamma)}) \prod_{i=1}^d \frac{g(q^{a_\gamma-h_i(\delta)})}{g(q^{a_\gamma-a_\delta})} \frac{g(q^{a_\delta-h_i(\gamma)})}{g(q^{a_\delta-a_\gamma})} \nn \\
%&\propto& \Delta(X(\gamma),X(\delta))\,\, 
%\prod_i g(q^{a_\delta-h_i(\gamma)}) \,\, \prod_i g(q^{a_\gamma-h_i(\delta)}) \, q^{{1\over 2}\,h_i(\delta)^2}\,\, q^{-{h_i(\delta)\over 2}(2a_\gamma+2d-1)} \cr
&\propto& \Delta(X(\gamma),X(\delta))\,\, e^{-{1\over g_s} \Tr U(X(\gamma),a_\delta)} e^{-{1\over g_s} \Tr U(X(\delta),a_\gamma)} \,\, e^{-{1\over g_s}\left( \Tr U_2(X(\gamma),a_\delta)+ \Tr \td U_2(X(\delta),a_\gamma)\right)}\, , \nn \\
\ean
where the square brackets on the RHS denote $q$-numbers as defined in (\ref{q_number}), the symbol $\Delta(X(\gamma),X(\delta))$ signifies
\ban \label{doublevan}
\Delta(X(\gamma),X(\delta)) = \prod_{i,j} (X_i(\delta)-X_j(\gamma)) = \prod_{i,j} (q^{h_i(\delta)}-q^{h_j(\gamma)}) \,,
\ean
and
\ba
U_2(X,a)=0 \,,
\ea
\ba
\td U_2(X,a)= {(\ln X)^2\over 2} - (a+d-{1\over 2})\, \ln X\,\ln q + i\pi \ln X.
\ea
The parameter $Q_{\gamma, \delta}$ reflects, given a choice of K\"ahler class $J$ of the metric on $\CYX_0$, the curve class of the curve $\cC$ extended between the vertices labeled by $\gamma$ and $\delta$ via
\beq
w_{\gamma,\delta} = \int_{\cC} J \, \,,\quad \quad Q_{\gamma, \delta} = q^{w_{\gamma,\delta}} \,.
\eeq
By the definition of the a-parameters,
\beq
w_{\gamma,\delta} = a_\gamma-a_\delta.
\eeq

Substituting these expressions into (\ref{z_stripAnn}), we obtain
\ban
\lefteqn{Z_{\rm strip}(\alpha_0,\dots,\alpha_n;\beta_0^T,\dots,\beta_n^T) =}\cr
&& \cr
&=& {\prod_i \Delta(X(\alpha_i))\,\prod_{i<j}\Delta(X(\alpha_i),X(\alpha_j))\,\,\, \prod_i \Delta(X(\beta_i))\,\prod_{i<j}\Delta(X(\beta_i),X(\beta_j))
\over \prod_{i,j} \Delta(X(\alpha_i),X(\beta_j))} \cr
&& \times \prod_i e^{-{1\over g_s}\, \tr (V_{\vec a}(X(\alpha_i))-V_{\vec b}(X(\alpha_i)))}\,\,  \prod_i e^{-{1\over g_s}\, \tr V_i(X(\alpha_i))} \cr
&&\times \prod_i e^{{1\over g_s}\, \tr  (V_{\vec a}(X(\beta_i))-V_{\vec b}(X(\beta_i)))}\,\,  \prod_i e^{-{1\over g_s}\, \tr \td V_i(X(\beta_i))}  \,,  \label{Zstripsum1}
\ean
where we have denoted by $\vec a = (a_0,a_1,\dots,a_n)$ (resp. $\vec b = (b_0,b_1,\dots,b_n)$) the a-parameters of representations on the upper side (resp. lower side) of the strip, and defined
\beqn    \label{partpotAnn}
V_{\vec a}(X) = -g_s\,\sum_{j=0}^n \ln{\left(g(q^{a_j}/X)\right)} \,,
\eeqn
and
\beq
V_i(X) = \ln X\, \ln q\,\, \left({1\over 2} - \sum_{j\leq i} (a_j-b_j)\right)+i\pi \ln X \,,
\eeq
\beq
\td V_i(X) = \ln X\, \ln q\,\, \left({1\over 2} - \sum_{j<i} (b_j-a_j)\right) \,.
\eeq

\subsection*{3.5 Gluing strips}

To obtain the partition function for the full multistrip fiducial geometry $\CYX_0$, we must glue these strips along the curves labelled $s_{i,j}$ in figure \ref{fiducial_labelingAnn}.

Denoting the representations $\alpha_{j,i}$ on line $i$ collectively by
\beq
\vec\alpha_i = (\alpha_{0,i},\alpha_{1,i},\dots,\alpha_{n,i})  \,,
\eeq
this yields
\ban
Z_{\rm vertex}(\CYX_0) = Z_{(n,m)}(\vec\alpha_{m+1},\vec\alpha_0^T) 
=  \sum_{\alpha_{j,i},\, j=0,\dots,n;\, i=1,\dots,m} \quad \,\,
\prod_{i=1}^{m+1} Z_{\rm strip}(\vec\alpha_i,\vec\alpha_{i-1}^T) \,\, \prod_{j=0}^n\prod_{i=1}^m q^{s_{j,i}\,|\alpha_{j,i}|} \,. \nonumber\\  \label{ZvertexprodZstrips}
\ean

Our goal now is to find a matrix integral which evaluates to this sum.

\section*{4 The matrix model}   \label{our_matrix_modelAnn}

\subsection*{4.1 Definition}

Consider the fiducial geometry $\CYX_0$ of size $(n+1)\times (m+1)$, with K\"ahler parameters $t_{i,j}=a_{i,j}-a_{i,j+1}$, $r_{i,j} = a_{i,j+1}-a_{i+1,j}$, and  $s_{i,j}$, as depicted in figures \ref{fiducial_labelingAnn} and \ref{fiducial_geometryAnn}. We write
\beq
\vec a_i  = (a_{0,i},a_{1,i},\dots,a_{n,i}).
\eeq

Assume that the external representations are fixed to $\vec\alpha_{m+1} = (\alpha_{0,m+1},\alpha_{1,m+1},\dots,\alpha_{n,m+1})$ on the upper line, and $\vec\alpha_0 = (\alpha_{0,0},\alpha_{1,0},\dots,\alpha_{n,0})$ on the lower line 
(for most applications, one prefers to choose these to be trivial).

\smallskip

We now define the following matrix integral ${\cal Z}_{\rm MM}$ (${}_{\rm MM}$ for Matrix Model),
\ban
{\cal Z}_{\rm MM}(Q,g_s,\vec\alpha_{m+1},\vec\alpha_0^T)
&=& \Delta(X(\vec \alpha_{m+1}))\,\, \Delta(X(\vec \alpha_0)) \,\, 
\prod_{i=0}^{m+1} \int_{H_N(\Gamma_i)} dM_i \,
 \prod_{i=1}^{m+1}\int_{H_N({\mathbb R}_+)}\,dR_i \nn \\
&& \prod_{i=1}^{m} e^{{-1\over g_s}\,\tr \left[ V_{\vec a_i}(M_i)-V_{\vec a_{i-1}}(M_i) \right]
%+ V_{\vec a_i,\vec a_{i-1}}(X_i) 
} \,\,\,
 \prod_{i=1}^{m} e^{{-1\over g_s}\,\tr \left[V_{\vec a_{i-1}}(M_{i-1})-V_{\vec a_{i}}(M_{i-1}) \right]
%+ \td V_{\vec a_{i-1},\vec a_{i}}(X_{i-1}) 
} \nn \\
&& \prod_{i=1}^{m+1} e^{{1\over g_s} \tr (M_i-M_{i-1})R_i} \,\,\,
 \prod_{i=1}^{m} e^{(S_i+{i\pi\over g_s})\,\tr\, \ln M_i}\,  \nn\\
&& e^{\tr \ln f_{0}(M_0)}\,\,e^{\tr \ln f_{m+1}(M_{m+1})}\,\, \prod_{i=1}^{m} e^{\tr \ln f_{i}(M_i)} \,. \label{m_integralAnn}
\ean
All matrices are taken of size
\beq
N=(n+1)\, d \,,
\eeq
where $d$ is the cut-off discussed in section \ref{secrepresentations}.
We have introduced the notation
\beq
X(\vec \alpha_{m+1})  = {\rm diag} (X(\vec \alpha_{m+1})_i)_{i=1,\dots,N}
\,\, , \qquad
X(\vec \alpha_{m+1})_{j d+k} = q^{h_k(\alpha_{j,m+1})},
\eeq
\beq
X(\vec \alpha_0)  = {\rm diag} (X(\vec \alpha_0)_i)_{i=1,\dots,N}
\,\, , \qquad
X(\vec \alpha_0)_{j d+k} = q^{h_k(\alpha_{j,0})},
\eeq
for $k=1, \ldots, d$, $j=0, \ldots,n$. $\Delta(X)=\prod_{i<j}(X_i-X_j)$ is the Vandermonde determinant. $V_{\vec a_i}(x)$ was introduced in (\ref{partpotAnn}). For $i=1,\dots,m$, we have defined
\beq
f_i(x) =  \prod_{j=0}^n {g(1)^2\,\,e^{({1\over 2}+{i\pi\over \ln q})\, \ln{(x q^{1-a_{j,i}})}}\, \,e^{{(\ln{(x q^{1-a_{j,i}})})^2\over  2 g_s}}\over g(x\,q^{1-a_{j,i}})\, g(q^{a_{j,i}}/x)\,} \,.
\eeq
The denominator of these functions induces simple poles at $x=q^{a_{j,i}+l}$ for $j=0,\dots,n$ and $l\in \mathbb Z$. The numerator is chosen such that they satisfy the relation $f_i(qx)=f_i(x)$. This enforces a simple $l$ dependence of the residues taken at $x=q^{a_{j,i}+l}$, given by a prefactor $q^l$ -- a fact which will be important in the following. These residues are in fact given by
\beqn \label{resfAnn}
 \Res_{q^{a_{j,i}+l}} f_i(x) = q^{a_{j,i}+l}\,\, \hat f_{j,i} =-\, q^{a_{j,i}+l}\,\,   \prod_{k\neq j} {g(1)^2\,\,e^{({1\over 2}+{i\pi\over \ln q})\, (1+a_{j,i}-a_{k,i})\ln q}\, \,e^{{(\ln{(q^{1+a_{j,i}-a_{k,i}})})^2\over  2 g_s}}\over g(q^{a_{j,i}-a_{k,i}})\,(1-q^{a_{k,i}-a_{j,i}}) g(q^{a_{k,i}-a_{j,i}})} \,,
\eeqn
where $\hat f_{j,i}$ is independent of the integer $l$. 

The parameters $S_i$ are defined by 
\beqn \label{siAnn}
S_i =  s_{0,i-1}+t_{0,i-1}= s_{j,i-1} -\sum_{k<j} t_{k,i}+\sum_{k\leq j} t_{k,i-1} \,.
\eeqn
The final equality holds for arbitrary $j$, and can be verified upon invoking (\ref{identify_class}) repeatedly.

For $i=0$ and $i=m+1$, we define
\beq
f_0(x) = {1\over \prod_{j=0}^n \prod_{i=1}^d (x-q^{h_i(\alpha_{j,0})})} \,,
\eeq
\beq
f_{m+1}(x) = {1\over \prod_{j=0}^n \prod_{i=1}^d  (x-q^{h_i(\alpha_{j,m+1})})} \,.
\eeq
Notice that if the representations $\vec\alpha_0$ or $\vec\alpha_{m+1}$ are trivial, i.e. $h_i(\alpha_{j,0})=d-i+a_{j,0}$ or $h_i(\alpha_{j,m+1})=d-i+a_{j,m+1}$, we have
\beq
f_0(x) = \prod_{j=0}^n {g(x\,q^{1-a_{j,0}-d})\over x^d\, g(x\,q^{1-a_{j,0}}) }\,,\hspace{1cm} f_{m+1}(x) = \prod_{j=0}^n {g(x\,q^{1-a_{j,m+1}-d})\over x^d\, g(x\,q^{1-a_{j,m+1}}) }
\eeq
respectively.
The functions $f_0$ and $f_{m+1}$ have simple poles
at $x=q^{h_l(\alpha_{j,0})}$ (resp. $x=q^{h_l(\alpha_{j,m+1})}$) for $l=1,\dots,d$, with residue
\beq
\hat f_{j,0;l} = \Res_{q^{h_l(\alpha_{j,0})}} f_0(x) =   {1\over \prod_{j'\neq j} \prod_{i=1}^d (q^{h_l(\alpha_{j,0})}-q^{h_i(\alpha_{j',0})})} \,{1\over  \prod_{i\neq l} (q^{h_l(\alpha_{j,0})}-q^{h_i(\alpha_{j,0})})} \,,
\eeq
\beq
\hat f_{j,m+1;l} = \Res_{q^{h_l(\alpha_{j,m+1})}} f_{m+1}(x) =   {1\over \prod_{j'\neq j} \prod_{i=1}^d (q^{h_l(\alpha_{j,m+1})}-q^{h_i(\alpha_{j',m+1})})} \,{1\over  \prod_{i\neq l} (q^{h_l(\alpha_{j,m+1})}-q^{h_i(\alpha_{j,m+1})})} \,.
\eeq
The $l$ dependence here is more intricate than above, but this will not play any role since the partitions $\alpha_{j,0}$ and $\alpha_{j,m+1}$ are kept fixed, not summed upon.
  
\bigskip

\begin{figure}[h]
 \centering
 \includegraphics[width=8cm]{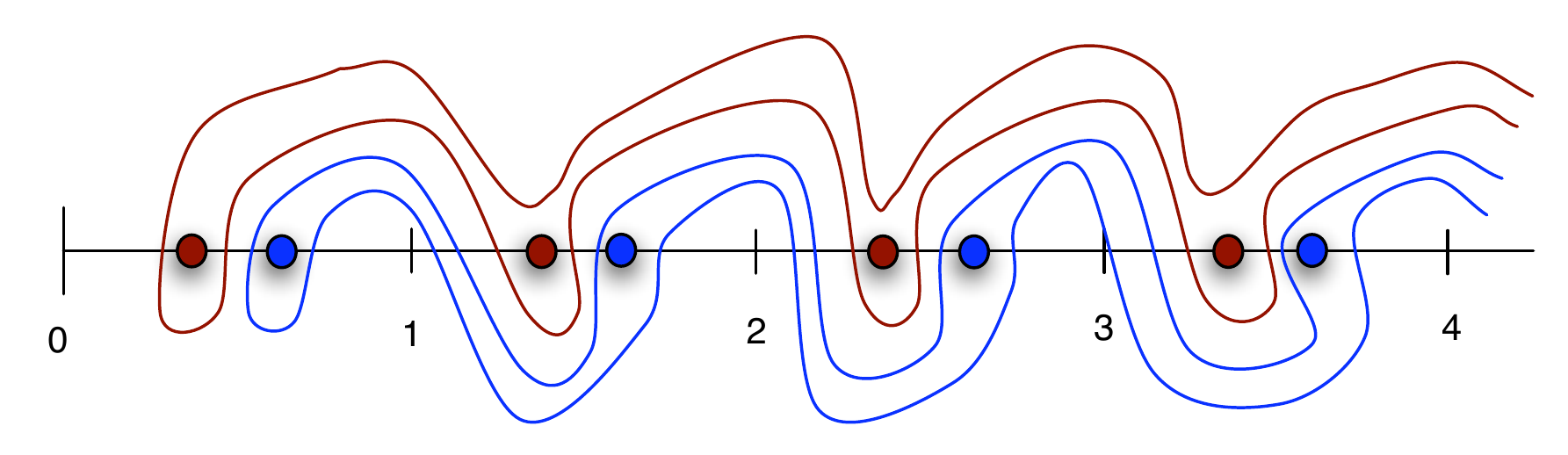}
 \caption{\footnotesize{Two contours surrounding points $a+\mathbb N$ and $b+\mathbb N$, such that  $a-b \notin \mathbb Z$.}}
 \label{contoursAnn}
\end{figure}

The integration domains for the matrices $R_i$ are $H_N(\mathbb R_+^N)$, i.e. the set of hermitian matrices  having only positive eigenvalues. For the matrices $M_i, i=1,\dots,m$, the integration domains are $H_N(\Gamma_i)$, where 
\beq
\Gamma_i = \prod_{j=0}^n\, (\gamma_{j,i})^d \,.
\eeq
$\gamma_{j,i}$ is defined as a contour which encloses all points of the form $q^{a_{j,i}+\mathbb N}$, and does not intersect any contours $\gamma_{k,l}$, $(j,i) \neq (k,l)$. For this to be possible, we must require that the differences $a_{j,i}-a_{j',i'}$ be non-integer. The normalized logarithms of two such contours are depicted in figure \ref{contoursAnn}.

 We have defined
\beq
H_N(\Gamma_i) = \{ M=U\, \Lambda \, U^\dagger \, , \quad U\in U(N)\, , \,\,\, \Lambda={\rm diag}(\lambda_1,\dots,\lambda_N)\, \in \Gamma_i \} \,,
\eeq
i.e. $H_N(\Gamma_i)$ is the set of normal matrices with eigenvalues on $\Gamma_i$. By definition, the measure on $H_N(\Gamma_i)$ is (see \cite{Mehta})
\beqn\label{eqdefdMdUdLAnn}
dM = {1\over N!}\,\, \Delta(\Lambda)^2\,\, dU\,d\Lambda \,,
\eeqn
where $dU$ is the Haar measure on $U(N)$, (normalized not to $1$, but to a value depending only on $N$, such that the Itzykson-Zuber integral evaluates as given in (\ref{eqdefIZ}) with pre\-factor $1$), and $d\Lambda$ is the product of the measures for each eigenvalue along its integration path.

The integration domains for the matrices $M_0$, $M_{m+1}$ are $H_N(\Gamma_0)$, $H_N(\Gamma_{m+1})$ respectively, where 
\beqn  \label{outer_contoursAnn}
\Gamma_0 =  (\sum_{j=0}^n \gamma_{j,0})^N \,,
\qquad \quad
\Gamma_{m+1} =  (\sum_{j=0}^n \gamma_{j,m+1})^N \,.
\eeqn

\bigskip

The goal of the rest of this section is to prove that the matrix integral (\ref{m_integralAnn}) reproduces the topological string partition function for target space the fiducial geometry $\CYX_0$.

 \subsection*{4.2 Diagonalization}

Let us first diagonalize all matrices.
We write
\beq
M_i = U_i \, X_i\, U_i^\dagger \,,
\eeq
\beq
R_i = \td U_i \, Y_i\, \td U_i^\dagger,
\eeq
where $U_i$ and $\td U_i$ are unitary matrices.

By the definition (\ref{eqdefdMdUdLAnn}), the measures $dM_i$ and $dR_i$ are given by
\beq
dM_i = {1\over N!}\, \Delta(X_i)^2\, dU_i\, dX_i
\quad , \quad
dR_i = {1\over N!}\, \Delta(Y_i)^2\, d\td U_i\, dY_i  \,.
\eeq

The matrix integral thus becomes
\ba
{\cal Z}_{\rm MM}(Q,g_s,\vec\alpha_{m+1},\vec\alpha_0^T)
&=& {\Delta(X(\vec \alpha_{m+1}))\,\, \Delta(X(\vec \alpha_0))\over (N!)^{2m+3}} \,\, 
\prod_{i=0}^{m+1} \int_{\Gamma_i} dX_i \,\Delta(X_i)^2\,
 \prod_{i=1}^{m+1}\int_{{\mathbb R}_+^N}\,dY_i\,\Delta(Y_i)^2 \cr
&& \prod_{i=0}^{m+1} dU_i\,\,\prod_{i=1}^{m+1} d \td U_i \cr
&& \prod_{i=1}^{m} e^{{-1\over g_s}\,\tr \left[ V_{\vec a_i}(X_i)-V_{\vec a_{i-1}}(X_i) \right]
%+ V_{\vec a_i,\vec a_{i-1}}(X_i) 
} \,\,\,
 \prod_{i=1}^{m} e^{{-1\over g_s}\,\tr \left[V_{\vec a_{i-1}}(X_{i-1})-V_{\vec a_{i}}(X_{i-1}) \right]
%+ \td V_{\vec a_{i-1},\vec a_{i}}(X_{i-1}) 
} \cr
\cr
&& \prod_{i=1}^{m+1} e^{{1\over g_s} \tr X_i U_i^\dagger \td U_i Y_i \td U_i^\dagger U_i}\,\,e^{{-1\over g_s} \tr X_{i-1} U_{i-1}^\dagger \td U_i Y_i \td U_i^\dagger U_{i-1}} 
\,\, \prod_{i=1}^{m} e^{(S_i+{i\pi\over g_s})\,\tr\, \ln X_i}\, \cr
&& e^{\tr \ln f_{0}(X_0)}\,\,e^{\tr \ln f_{m+1}(X_{m+1})}\,\, \prod_{i=1}^{m} e^{\tr \ln f_{i}(X_i)} \,.
\ea
Next, we introduce the matrices $\hat{U_i}$, $\check{U_i}$, for $i=1, \ldots, m+1$, via
\beq
\hat U_i = U_i^\dagger \td U_i 
\quad , \quad
\check U_i = \td U_i^\dagger  U_{i-1} \,.
\eeq
We can express $U_0, \ldots, U_{m+1}$, and $\tilde{U}_1, \ldots, \tilde{U}_{m+1}$, in terms of these matrices and $U_{m+1}$,
\beq
U_i = U_{m+1}\, \hat U_{m+1}\, \check U_{m+1}\,\hat U_{m}\, \check U_{m} \dots \hat U_{i+1}\, \check U_{i+1} \,,
\eeq
\beq
\tilde U_i = U_{m+1}\, \hat U_{m+1}\, \check U_{m+1}\,\hat U_{m}\, \check U_{m} \dots \hat U_{i+1}\, \check U_{i+1}\, \hat U_i \,.
\eeq

With this change of variables, we arrive at
\ba
{\cal Z}_{\rm MM}(Q,g_s,\vec\alpha_{m+1},\vec\alpha_0^T)
&=& {\Delta(X(\vec \alpha_{m+1}))\,\, \Delta(X(\vec \alpha_0))\over (N!)^{2m+3}} \,\, 
\prod_{i=0}^{m+1} \int_{\Gamma_i} dX_i \,\Delta(X_i)^2\,
 \prod_{i=1}^{m+1}\int_{{\mathbb R}_+^N}\,dY_i\,\Delta(Y_i)^2 \cr
&& \int dU_{m+1}\, \prod_{i=1}^{m+1} d\hat U_i\,\,\prod_{i=1}^{m+1} d \check U_i \cr
&& \prod_{i=1}^{m} e^{{-1\over g_s}\,\tr \left[ V_{\vec a_i}(X_i)-V_{\vec a_{i-1}}(X_i) \right]
} \,\,
 \prod_{i=1}^{m} e^{{-1\over g_s}\,\tr \left[V_{\vec a_{i-1}}(X_{i-1})-V_{\vec a_{i}}(X_{i-1}) \right]
} \cr
&& \prod_{i=1}^{m+1} e^{{1\over g_s} \tr X_i \hat U_i Y_i \hat U_i^\dagger }\,\,e^{{-1\over g_s} \tr X_{i-1} \check U_i^\dagger  Y_i \check U_i } 
\,\, \prod_{i=1}^{m} e^{(S_i+{i\pi\over g_s})\,\tr\, \ln X_i}\, \cr
&& e^{\tr \ln f_{0}(X_0)}\,\,e^{\tr \ln f_{m+1}(X_{m+1})}\,\, \prod_{i=1}^{m} e^{\tr \ln f_{i}(X_i)} \,.
\ea

Notice that the integral over $U_{m+1}$ decouples, and $\int dU_{m+1}={\rm Vol}(U(N))$. 

\subsection*{4.3 Itzykson-Zuber integral and Cauchy determinants}

The $\hat U_i$ and $\check U_i$ appear in the form of Itzykson-Zuber integrals \cite{ItzyksonZuber},
\beqn  \label{eqdefIZ}
I(X,Y) = \int dU\, e^{\tr X U Y U^\dagger} = {\mathop{{\det}}_{p,q} (e^{x_p\,y_q})\over \Delta(X)\,\Delta(Y)} \,,
\eeqn
where $x_p$ and $y_q$ are the eigenvalues of $X$ and $Y$.
We thus have
\ba
{\cal Z}_{\rm MM}(Q,g_s,\vec\alpha_{m+1},\vec\alpha_0^T)
&\propto& {\Delta(X(\vec \alpha_{m+1}))\,\, \Delta(X(\vec \alpha_0))\over (N!)^{2m+3}} \,\, 
\prod_{i=0}^{m+1} \int_{\Gamma_i} dX_i \,\Delta(X_i)^2\,
 \prod_{i=1}^{m+1}\int_{{\mathbb R}_+^N}\,dY_i\,\Delta(Y_i)^2 \cr
&& \prod_{i=1}^{m} e^{{-1\over g_s}\,\tr \left[ V_{\vec a_i}(X_i)-V_{\vec a_{i-1}}(X_i) \right]
} \,\,
 \prod_{i=1}^{m} e^{{-1\over g_s}\,\tr \left[ V_{\vec a_{i-1}}(X_{i-1})-V_{\vec a_{i}}(X_{i-1}) \right]
} \cr
&& \prod_{i=1}^{m+1} I({1\over g_s}X_i,Y_i) \,\, I(-{1\over g_s}X_{i-1},Y_i) 
\,\, \prod_{i=1}^{m} e^{(S_i+{i\pi\over g_s})\,\tr\, \ln X_i}\, \cr
&& e^{\tr \ln f_{0}(X_0)}\,\,e^{\tr \ln f_{m+1}(X_{m+1})}\,\, \prod_{i=1}^{m} e^{\tr \ln f_{i}(X_i)} \cr
&\propto& {\Delta(X(\vec \alpha_{m+1}))\,\, \Delta(X(\vec \alpha_0))\over (N!)^{2m+3}} \,\, 
\prod_{i=0}^{m+1} \int_{\Gamma_i} dX_i \,\,
 \prod_{i=1}^{m+1}\int_{{\mathbb R}_+^N}\,dY_i\, \cr
&& \Delta(X_0)\,\Delta(X_{m+1})\,\,\prod_{i=1}^{m} e^{{-1\over g_s}\,\tr \left[V_{\vec a_i}(X_i)-V_{\vec a_{i-1}}(X_i) \right]
} \cr
&& \prod_{i=1}^{m} e^{{-1\over g_s}\,\tr \left[ V_{\vec a_{i-1}}(X_{i-1})-V_{\vec a_{i}}(X_{i-1}) \right]} \,\,
\,\, \prod_{i=1}^{m} e^{(S_i+{i\pi\over g_s})\,\tr\, \ln X_i}\, 
 \cr
&& \prod_{i=1}^{m+1}  \mathop{{\det}}_{p,q}(e^{{1\over g_s} (X_i)_{p}\, (Y_i)_{q} })
\,\,\, \mathop{{\det}}_{p,q}(e^{{-1\over g_s} (X_{i-1})_{p}\, (Y_i)_{q} })  \cr
&& e^{\tr \ln f_{0}(X_0)}\,\,e^{\tr \ln f_{m+1}(X_{m+1})}\,\, \prod_{i=1}^{m} e^{\tr \ln f_{i}(X_i)} \,,
\ea
where we have dropped an overall sign, powers of $g_s$, and the group volume ${\rm Vol}(U(N))$ which are constant prefactors of no interest to us.

Next, we perform the integrals over $Y_i$ along $\mathbb R_+^N$.
\ba
&& \int_{\mathbb R_+^N} dY \mathop{{\det}}_{p,q}(e^{{1\over g_s} (X_i)_{p}\, (Y)_{q} })
\,\,\, \mathop{{\det}}_{p,q}(e^{{-1\over g_s} (X_{i-1})_{p}\, (Y)_{q} })  \cr
&=& \sum_\sigma \sum_{\td\sigma} (-1)^\sigma (-1)^{\td\sigma}\,\, \prod_{p=1}^N
\int_{0}^\infty dy_p\, e^{{y_p\over g_s}((X_i)_{\sigma(p)}-(X_{i-1})_{\td\sigma(p)})} \cr
&=& \sum_\sigma \sum_{\td\sigma} (-1)^\sigma (-1)^{\td\sigma}\,\, \prod_{p=1}^N {g_s\over (X_{i-1})_{\td\sigma(p)}-(X_i)_{\sigma(p)}} \cr
&=& N!\, g_s^N\,\, \mathop{{\det}}_{p,q}{\left({1\over (X_{i-1})_{p}-(X_i)_{q}}\right)} \,.
\ea
Note that the integral is only convergent for $(X_i)_{\sigma(p)}-(X_{i-1})_{\td\sigma(p)} <0$. For $X_i$ that violate this inequality, we will define the integral via its analytic continuation given in the third line.

An application of the Cauchy determinant formula,
\beq
\det \left( \frac{1}{x_i + y_j} \right)_{1 \le i < j  \le n} = \frac{\prod_{1 \le i < j  \le n} (x_j - x_i) (y_j - y_i)}{\prod_{i,j=1}^n (x_i + y_j)} \,,
\eeq
yields
\beq
\int_{\mathbb R_+^N} dY \mathop{{\det}}_{p,q}(e^{{1\over g_s} (X_i)_{p}\, (Y)_{q} })
\,\,\, \mathop{{\det}}_{p,q}(e^{{-1\over g_s} (X_{i-1})_{p}\, (Y)_{q} })  
= (-1)^{\binom{N}{2}} N!\, g_s^N\,\, {\Delta(X_i)\,\Delta(X_{i-1})\over \Delta(X_{i-1},X_i)} \,,
\eeq
where the notation $\Delta(X_{i-1},X_i)$ was introduced in (\ref{doublevan}). 
Evaluating the $Y_i$ integrals thus, and continuing to drop overall signs and powers of $g_s$, our matrix integral becomes
\ba
{\cal Z}_{\rm MM}(Q,g_s,\vec\alpha_{m+1},\vec\alpha_0^T)
&\propto& {\Delta(X(\vec \alpha_{m+1}))\,\, \Delta(X(\vec \alpha_0))\over (N!)^{m+3}} \,\, 
\prod_{i=0}^{m+1} \int_{\Gamma_i} dX_i \,\,\Delta(X_i)^2 \cr
&& \prod_{i=1}^{m} e^{{-1\over g_s}\,\tr \left[ V_{\vec a_i}(X_i)-V_{\vec a_{i-1}}(X_i) \right]
} \,\,
 \prod_{i=1}^{m} e^{{-1\over g_s}\,\tr \left[V_{\vec a_{i-1}}(X_{i-1})-V_{\vec a_{i}}(X_{i-1}) \right]
} \cr
&& \prod_{i=1}^{m+1}  {1\over \Delta(X_{i-1},X_i)}  \,\, 
\,\, \prod_{i=1}^{m} e^{(S_i+{i\pi\over g_s})\,\tr\, \ln X_i}\, \cr
&& e^{\tr \ln f_{0}(X_0)}\,\,e^{\tr \ln f_{m+1}(X_{m+1})}\,\, \prod_{i=1}^{m} e^{\tr \ln f_{i}(X_i)} \,.
\ea

\subsection*{4.4 Recovering the sum over partitions} \label{recovering_sum}

Following the steps introduced in \cite{KlemmSulkowski} in reverse, we next decompose the diagonal matrix $X_i$ into blocks, $$X_i= {\rm diag}\,(X_{0,i},X_{1,i},\dots,X_{n,i}) \,,$$ where each matrix $X_{j,i}$ is a $d\times d$ diagonal matrix whose eigenvalues are integrated on the contours $\gamma_{j,i}$ surrounding points of the form $q^{a_{j,i}+\mathbb N}$.
We arrive at
\ba
{\cal Z}_{\rm MM}(Q,g_s,\vec\alpha_{m+1},\vec\alpha_0^T)
&\propto& {\Delta(X(\vec \alpha_{m+1}))\,\, \Delta(X(\vec \alpha_0))\over (N!)^{m+3}} \,\, 
\prod_{i=0}^{m+1}\prod_{j=0}^n \int_{(\gamma_{j,i})^d} dX_{j,i} \cr
&& \Delta(X_0)\,\Delta(X_{m+1})\,\, 
\prod_{i=1}^{m+1}  {\Delta(X_{i-1}) \Delta(X_i)\over \Delta(X_{i-1},X_i)}  \cr
&& \prod_{i=1}^{m} e^{{-1\over g_s}\,\tr \left[ V_{\vec a_i}(X_i)-V_{\vec a_{i-1}}(X_i) \right]
} \,\, \prod_{i=1}^{m} e^{{-1\over g_s}\,\tr \left[V_{\vec a_{i-1}}(X_{i-1})-V_{\vec a_{i}}(X_{i-1}) \right]
} \cr
&& e^{\tr \ln f_{0}(X_0)}\,\,e^{\tr \ln f_{m+1}(X_{m+1})}\,\, \prod_{i=1}^{m} e^{\tr \ln f_{i}(X_i)} 
\,\, \prod_{i=1}^{m} e^{(S_i+{i\pi\over g_s})\,\tr\, \ln X_i}\, ,
\ea
with
\beq
 {\Delta(X_{i-1}) \Delta(X_i)\over \Delta(X_{i-1},X_i)}  
= {\prod_{j} \Delta(X_{j,i-1}) \prod_{j} \Delta(X_{j,i}) \, \prod_{j<l} \Delta(X_{j,i-1},X_{l,i-1})\,\,\prod_{j<l} \Delta(X_{j,i},X_{l,i}) \over \prod_{j,l}\Delta(X_{j,i-1},X_{l,i})}  \,.
\eeq

Our next step is to evaluate the $dX_{j,i}$ integrals via Cauchy's residue theorem. The poles of the integrands lie at the poles of $f_i$, and the zeros of $\Delta(X_{i-1}, X_i)$. However, we have been careful to define our contours $\gamma_{j,i}$ in a way that only the poles of $f_i$ contribute. These lie at the points $q^{a_{j,i} + \IN}$. Hence, the integrals evaluate to a sum of residues over the points
\beq
(X_{j,i})_l = q^{a_{j,i}+(h_{j,i})_l} \,,
\eeq
where each $(h_{j,i})_l$ is a positive integer.

Since the integrand contains a Vandermonde of the eigenvalues of $X_{j,i}$, the residues vanish whenever two eigenvalues are at the same pole of $f_i$, i.e. if two $(h_{j,i})_l$ coincide. Moreover, since the integrand is symmetric in the eigenvalues, upon multiplication by $N!$, we can assume that the $(h_{j,i})_l$ are ordered,
\beq
(h_{j,i})_1>(h_{j,i})_2>(h_{j,i})_3> \dots >(h_{j,i})_d \geq 0.
\eeq
The $(h_{j,i})_l$ hence encode a partition $\alpha_{j,i}$ via $(h_{j,i})_l =  (\alpha_{j,i})_l-i+d$, and we have reduced our integrals to a sum over partitions. In terms of the function $h_l(\alpha)$ introduced in (\ref{defhAnn}),
\beq
(X_{j,i})_l = q^{h_l(\alpha_{j,i})}
\,\, , \quad
h_l(\alpha_{j,i}) = (h_{j,i})_l +a_{j,i} \,,
\eeq
\beq
h_1(\alpha_{j,i})>h_2(\alpha_{j,i})>\dots>h_d(\alpha_{j,i})\geq a_{j,i}.
\eeq

\medskip
Notice that unlike $f_i$, $i=1, \ldots, m$, $f_0$ and $f_{m+1}$ only have a finite number of $N=(n+1)d$ poles. Since the $(h_{j,0})_l$, $(h_{j,m+1})_l$ respectively can be chosen pairwise distinct and ordered, $f_0$ and $f_{m+1}$ act as delta functions in the integrals over the $N\times N$ matrices $X_0$ and $X_{m+1}$, and fix these to the prescribed values $X(\vec\alpha_0)$ and $X(\vec\alpha_{m+1})$ respectively.

\medskip

Performing the integrals hence yields
\ba
{\cal Z}_{\rm MM}(Q,g_s,\vec\alpha_{m+1},\vec\alpha_0^T)
&\propto& \Delta(X(\vec \alpha_{m+1}))^2 \Delta(X(\vec \alpha_0))^2  \\ 
&& \hspace{-2cm}\sum_{\{\alpha_{j,i}|j=0,\dots,n;\, i=1,\dots,m+1\}} \,\,\,  \prod_{i=1}^{m+1}  {\Delta(X(\vec\alpha_{i-1})) \Delta(X(\vec\alpha_i))\over \Delta(X(\vec\alpha_{i-1}),X(\vec\alpha_i))}  \cr
&& \prod_{i=1}^{m} e^{{-1\over g_s}\,\tr \left[V_{\vec a_i}(X(\vec\alpha_i))-V_{\vec a_{i-1}}(X(\vec\alpha_i)) \right]
} \,\, \prod_{i=1}^{m} e^{{-1\over g_s}\,\tr \left[V_{\vec a_{i-1}}(X(\vec\alpha_{i-1}))-V_{\vec a_{i}}(X(\vec\alpha_{i-1})) \right]
} \cr
&& \,\, \prod_{i=1}^{m} e^{(S_i+{i\pi\over g_s})\,\tr\, \ln X(\vec\alpha_i)}  \prod_{i=0}^{m+1}\prod_{j=0}^n\,\prod_{l=1}^d  \,\, \left(\Res_{q^{h_l(\alpha_{j,i})}}\, f_i \right)\, .
\ea
Notice that
\beq
\prod_j \prod_l \Res_{q^{h_l(\alpha_{j,0})}}\, f_0   = {1\over  \Delta(X(\vec \alpha_0))^2} \,,
\eeq
\beq
\prod_j \prod_l \Res_{q^{h_l(\alpha_{j,m+1})}}\, f_{m+1}   = {1\over  \Delta(X(\vec \alpha_{m+1}))^2} \,.
\eeq
Furthermore,
\beq
\Res_{q^{h_l(\alpha_{j,i})}}\, f_i   = q^{h_l(\alpha_{j,i})}\,\,\hat f_{j,i} \,,
\eeq
where $\hat f_{j,i}$ computed in  (\ref{resfAnn}) is independent  of $h_l(\alpha_{j,i})$.  
We thus have
\beq
e^{S_i\, \tr \ln X(\vec \alpha_i)} \,\, \prod_{j=0}^n\,\prod_{l=1}^d  \,\, \left(\Res_{q^{h_l(\alpha_{j,i})}}\, f_i \right) = e^{(S_i+1)\tr \ln X(\vec \alpha_i)}\,\,\prod_{j=0}^n\,(\hat f_{j,i})^d\, .
\eeq
Upon substituting the expression (\ref{siAnn}) for $S_i$, we finally arrive at
\ban
&& {\cal Z}_{\rm MM}(Q,g_s,\vec\alpha_{m+1},\vec\alpha_0^T) \cr
&\propto&  \prod_{i=1}^{m}\prod_{j=0}^n\,(\hat f_{j,i})^d\,\,\,
\sum_{\{\alpha_{j,i}|j=0,\dots,n;\, i=1,\dots,m+1\}} \cr  
&& \prod_{i=1}^{m+1} {\prod_{j} \Delta(X(\alpha_{j,i-1})) \prod_{j} \Delta(X(\alpha_{j,i})) \, \prod_{j<l} \Delta(X(\alpha_{j,i-1}),X(\alpha_{l,i-1}))\,\,\prod_{j<l} \Delta(X(\alpha_{j,i}),X(\alpha_{l,i})) \over \prod_{j,l}\Delta(X(\alpha_{j,i-1}),X(\alpha_{l,i}))}   \cr
&& \prod_{i=1}^{m} e^{{-1\over g_s}\,\tr \left[ V_{\vec a_i}(X(\vec\alpha_i))-V_{\vec a_{i-1}}(X(\vec\alpha_i)) \right]
} \,\,  \prod_{i=1}^{m}\prod_{k=0}^n e^{({1\over 2}-\sum_{j\leq k} (a_{j,i}-a_{j,i-1}) -{i\pi\over g_s} )\tr \ln{X(\alpha_{k,i})} } \cr
&& \prod_{i=1}^{m} e^{{-1\over g_s}\,\tr \left[ V_{\vec a_{i-1}}(X(\vec\alpha_{i-1}))-V_{\vec a_{i}}(X(\vec\alpha_{i-1})) \right]} \,\,  \prod_{i=1}^{m} \prod_{k=0}^n e^{({1\over 2}-\sum_{j< k} (a_{j,i}-a_{j,i+1})  )\tr \ln{X(\alpha_{k,i})} } \cr
&& \,\, \prod_{i=1}^{m}\prod_{j=0}^{n} e^{s_{j,i}\tr\ln X(\alpha_{j,i})}\, .\cr
\ean

Comparing to (\ref{Zstripsum1}) and (\ref{ZvertexprodZstrips}), we conclude
\ba
{\cal Z}_{\rm MM}(Q,g_s,\vec\alpha_{m+1},\vec\alpha_0^T) \propto \sum_{\alpha_{j,i}, j=0,\dots,n;\, i=1,\dots,m}\,\,\,
\prod_{i=1}^{m+1} Z_{\rm strip}(\vec\alpha_i,\vec\alpha_{i-1}^T) \,\, \prod_{j=0}^n\prod_{i=1}^m q^{s_{j,i}\,|\alpha_{j,i}|}  \,, 
\ea
i.e.
\beq
{\cal Z}_{\rm MM}(Q,g_s,\vec\alpha_{m+1},\vec\alpha_0^T) \propto Z_{\rm vertex}({\CYX_0}) = e^{\sum_g g_s^{2g-2} \, GW_g(\CYX_0)} \,.
\eeq

Up to a trivial proportionality constant, we have thus succeeded in rewriting the topological string partition function on the fiducial geometry $\CYX_0$ as a {\em chain of matrices} matrix integral. By our reasoning in section \ref{flops}, this result extends immediately to arbitrary toric Calabi-Yau 3-folds as follows. We have argued that any such 3-fold can be obtained from a sufficiently large choice of fiducial geometry via flops and limits. The respective partition functions are related via (\ref{rel_part_flopped}). Upon the appropriate variable identification, we hence arrive at a matrix model representation of the topological string on an arbitrary toric Calabi-Yau 3-fold.

\section*{5 Implications of our result}  \label{implications}

We have rewritten the topological string partition function as a matrix integral. This allows us to bring the rich theory underlying the structure of matrix models to bear on the study of topological string.

\smallskip
The type of matrix integral we have found to underlie the topological string on toric Calabi-Yau 3-folds is a so-called chain of matrices. This class of models has been studied extensively \cite{Mehta1, Mehta}, and many structural results pertaining to it are known.

\subsection*{5.1 Loop equations and Virasoro constraints}

The loop equations of matrix models provide a set of relations among correlation functions. They are Schwinger-Dyson equations; they follow from the invariance of the matrix integral under a change of integration variables, or by an integration by parts argument.

\medskip

Loop equations for a general chain of matrices have been much studied in the literature, in particular in \cite{David1, ZJDFG, Eynchain, Chain}. They can be viewed as W-algebra constraints (a generalization of Virasoro constraints) \cite{MiMo}. Having expressed the topological string partition function as a matrix integral, we can hence conclude that Gromov-Witten invariants satisfy W-algebra constraints.

Moreover, a general formal solution of the loop equations for a chain of matrices matrix model was found in \cite{Chain}, and expressed in terms of so-called symplectic invariants $F_g$ of a spectral curve. The spectral curve for a matrix integral is related to the expectation value of the resolvent of the first matrix in the chain,
\beq
W(x) = \left< \tr {1\over x-M_0}\right>^{(0)} \,.
\eeq
The superscript ${}^{(0)}$ indicates that the expectation value is evaluated to planar order in a Feynman graph expansion. The symplectic invariants $F_g(\curve)$ of an arbitrary spectral curve $\curve$ were defined in \cite{OE}. \cite{Chain} proved that for any chain of matrices integral $Z$, one has
\beq
\ln Z = \sum_g F_g(\curve)
\eeq
with $\curve$ the spectral curve associated to the matrix integral.

\smallskip

Calculating the spectral curve of a chain of matrices matrix model with complicated potentials poses some technical challenges. We will present the spectral curve for our matrix model (\ref{m_integralAnn}) in a forthcoming publication \cite{topstring2}.

\subsection*{5.2 Mirror symmetry and the BKMP conjecture}
The mirror $\hat{{\mathfrak{X}}}$ of a toric Calabi-Yau 3-fold ${\mathfrak{X}}$ is a conic bundle over $\IC^* \times \IC^*$. The fiber is singular over a curve, which we will refer to as the mirror curve ${\cal S}_{\hat {\mathfrak{X}}}$ of $\hat{{\mathfrak{X}}}$. It is a plane curve described by an equation
\beq
{\cal S}_{\hat {\mathfrak{X}}}
 \qquad : \qquad
 H(e^x,e^y)=0 \,,
\eeq
where $H$ is a polynomial whose coefficients follow from the toric data of ${\mathfrak{X}}$ and the K\"ahler parameters of the geometry.

Mirror symmetry is the statement that the topological A-model partition function 
%$$Z_{\mathfrak{X}}(q)=\exp{(\sum_g (\,(\ln{q})^{2g-2} GW_g({\mathfrak{X}}))}$$
with target space ${\mathfrak{X}}$ is equal to the topological B-model partition function with target space $\hat{{\mathfrak{X}}}$.

\smallskip

Extending work of Mari\~no \cite{marino2Ann} proposing a relation between the formalism of \cite{OE} and open and closed topological string amplitudes, Bouchard, Klemm, Mari\~no and Pasquetti (BKMP) conjecture in \cite{BKMP} that
\beq
%\encadremath{
GW_g({\mathfrak{X}}) \stackrel{?}{=} F_g({\cal S}_{\hat{{\mathfrak{X}}}}) \,.
\eeq
Here, the $F_g$'s are the symplectic invariants introduced in \cite{OE}.
The main interest of this conjecture is that it provides a systematic method for computing the topological string partition function, genus by genus, away from the large radius limit, and without having to solve differential equations.

\smallskip

This conjecture was motivated by the fact that symplectic invariants have many intriguing properties reminiscent of the topological string free energies. They are invariant under transformations ${\cal S}\to \td{\cal S}$ which conserve the symplectic form $dx\wedge dy = d\td{x} \wedge d\td{y}$, whence their name \cite{OE}. They satisfy holomorphic anomaly equations \cite{eynhaeq}, they have an integrable structure similar to Givental's formulae \cite{GiventalSemisimple, GiventalHierarchies, AMM1, AMM2, Orantin}, they satisfy some special geometry relations, WDVV relations \cite{CMMV}, and they give the Witten-Kontsevich theory as a special case \cite{OE, EynVolmum}.

\medskip

BKMP succesfully checked their claim for various examples to low genus.

The conjecture was proved for arbitrary genus in \cite{Partitions2} for ${\mathfrak{X}}$ a Hirzebruch rank 2 bundle over $\mathbb P^1$ (this includes the conifold).  Marshakov and Nekrasov \cite{MarshakovNekrasov} proved $F_0=GW_0$ for the family of $SU(n)$ Seiberg-Witten models.  Klemm and Sulkowski \cite{KlemmSulkowski}, generalizing \cite{Partitions2} to Nekrasov's sums over partitions for $SU(n)$ Seiberg-Witten gauge theories, proved the relation for $F_0$, building on work in \cite{NekrasovOkounkov}. In fact, it appears straightforward to extend their computation to arbitrary genus $F_g$. In \cite{Sulkowski}, Sulkowski provided a matrix model realization of $SU(n)$ gauge theory with a massive adjoint hypermultiplet, again using a generalization of \cite{Partitions2} for more general sums over partitions. Bouchard and Mari\~no \cite{countingsurface} noticed that an infinite framing limit of the BKMP conjecture for the framed vertex $\CYX=\mathbb C^3$ implies another conjecture for the computation of Hurwitz numbers, namely that the Hurwitz numbers of genus $g$ are the symplectic invariants of genus $g$ for the Lambert spectral curve $e^x=y\, e^{-y}$. That 
 conjecture was proved recently by another generalization of \cite{Partitions2} using a matrix model for summing over partitions \cite{BorotEynardMulase}, and also by a direct cut and join combinatorial method \cite{EMS}.
The BKMP conjecture was also proved for the framed vertex $\CYX=\mathbb C^3$ in \cite{BKMPcase1, BKMPChen}, using the ELSV formula and a cut and join combinatorial approach.

\medskip

Since we have demonstrated that the topological string partition function is reproduced by a matrix model, we can conclude that the Gromov-Witten invariants coincide with the symplectic invariants 
\beq
\sum_g g_s^{2g-2}\,GW_g = \sum_G F_g(\curve) \,,
\eeq
with $\curve$ the spectral curve of our matrix model. We will compute $\curve$ explicitly in a forthcoming work \cite{topstring2}, and demonstrate that it indeed coincides, up to symplectic transformations, with the mirror curve ${\cal S}_{\hat{{\mathfrak{X}}}}$, thus proving the BKMP conjecture for arbitrary toric Calabi-Yau 3-folds, in the large radius limit.

\subsection*{5.3 Simplifying the matrix model} \label{simplifying_potential}
The matrix models associated to the conifold or to geometries underlying Seiberg-Witten theory have a remarkable property: the spectral curve is the same (perturbatively and up to symplectic transformations) as the one of a simpler matrix model with all $g$-functions replaced by only the leading term in their small $\ln q$ expansion. We will demonstrate in a forthcoming work \cite{topstring2} that this property also holds for our matrix integral (\ref{m_integralAnn}). We can hence simplify the potentials of our matrix model, arriving at
\ba
{\cal Z}_{\rm simp}(Q,g_s,\vec\alpha_{m+1},\vec\alpha_0^T)
&=& \Delta(X(\vec \alpha_{m+1}))\,\, \Delta(X(\vec \alpha_0)) \,\, 
\prod_{i=0}^{m+1} \int_{H_{\bar n_i}(\Gamma_i)} dM_i \,
 \prod_{i=1}^{m+1}\int_{H_{\bar n}({\mathbb R}_+)}\,dR_i \cr
&& \prod_{i=1}^{m} e^{{1\over g_s}\, \tr \sum_{j=0}^n (Li_2(q^{a_{j,i}}/M_i)-Li_2(q^{a_{j,i-1}}/M_{i})) } \cr
&& \,\,\, \prod_{i=0}^{m-1} e^{{1\over g_s}\, \tr \sum_{j=0}^n (Li_2(q^{a_{j,i}}/M_i)-Li_2(q^{a_{j,i+1}}/M_{i})) } \cr
&& \prod_{i=1}^{m+1} e^{{1\over g_s} \tr (M_i-M_{i-1})R_i} \,\,\,
 \prod_{i=1}^{m} e^{(S_i+{i\pi\over g_s})\,\tr\, \ln M_i}\,,
\ea
where the matrix $M_i$ is of size $\bar n_i =\sum_j \bar n_{j,i}$.
\medskip

\subsubsection*{Classical limit}

In the classical limit, the dilogarithm $\Li_2$ becomes the function $x\ln{x}$, and we have
\ba
{\cal Z}_{\rm eff.\, cl}(Q,g_s,\vec\alpha_{m+1},\vec\alpha_0^T)
&=& \Delta(X(\vec \alpha_{m+1}))\,\, \Delta(X(\vec \alpha_0)) \,\, 
\prod_{i=0}^{m+1} \int_{H_{\bar n_i}(\Gamma_i)} dM_i \,
 \prod_{i=1}^{m+1}\int_{H_{\bar n}({\mathbb R}_+)}\,dR_i \cr
&& \prod_{i=1}^{m} e^{{1\over g_s}\, \tr \sum_{j=0}^n (M_i-a_{j,i})\,\ln{(a_{j,i}-M_i)} - (M_i-a_{j,i-1})\,\ln{(a_{j,i-1}-M_i)}
 } \cr
&& \prod_{i=0}^{m-1} e^{{1\over g_s}\, \tr \sum_{j=0}^n (M_i-a_{j,i})\,\ln{(a_{j,i}-M_i)} - (M_i-a_{j,i+1})\,\ln{(a_{j,i+1}-M_i)}
 } \cr
&& \prod_{i=1}^{m+1} e^{{1\over g_s} \tr (M_i-M_{i-1})R_i} \,\,\,
 \prod_{i=1}^{m} e^{(S_i+{i\pi\over g_s})\,\tr\, \ln M_i}\, .
\ea

This model shares features with the Eguchi-Yang matrix model \cite{EguchiYang}, see also \cite{MarshakovNekrasov}.

%, and is often presented as a candidate for an effective matrix model for gauge theories.

\section*{6 Conclusion}  \label{conclusions}

We have rewritten the topological vertex formula for the partition function of the topological A-model as a matrix integral.

Having expressed the topological string in terms of a matrix model, we can bring the immense matrix model toolkit which has been developed since the introduction of random matrices by Wigner in 1951 to bear on questions concerning the topological string and Gromov-Witten invariants. We already started down this path in section \ref{implications} above. Going further, we can apply the method of bi-orthogonal polynomials \cite{Mehta} to our matrix model to unearth the integrable system structure (Miwa-Jimbo \cite{MiwaJimbo1, JMI}) underlying the topological string, at least in the case of toric targets, together with its Lax pair, its Hirota equations (which arise as orthogonality relations), etc. In a related vein, free fermions \cite{Harnad-Orlov2, Kostov2} arise in the theory of matrix models when invoking determinantal formulae to express the matrix model measure \cite{EynMehta}.  It will be very interesting to explore how this is related to the occurrence of free fermions in topological string theory, as studied in \cite{ADKMV, KashaniPoor, DHSV, DDmodules}. More generally, one should study what can be learned about the non-perturbative topological string from its perturbative reformulation as a matrix model, as in the works \cite{MarinoSchiappaWeiss, Marino, EynardMarino,KlemmMarinoRauch}. A recurrent such question, which could be addressed in the matrix model framework (in fact, it was already latently present in the calculations in this work), is that of the quantization of K\"ahler parameters.

On a different note, notice that the matrix model derived in this article, with a potential which is a sum of logs of $q$-deformed $\Gamma$ functions, looks very similar to the matrix model counting plane partitions introduced in \cite{Partitions}. This is a hint that it could be possible to recover the topological vertex formula, corresponding to the topological string with target $\IC^3$ and appropriate boundary conditions, directly from the matrix model approach. Either along these lines or the lines pursued in this paper, it would be interesting to derive a matrix model related to the Nekrasov deformation \cite{NekrasovAnn,IqbalKozcazVafa} of the topological string.

A completely open question is whether the close relation between topological strings and matrix models persists beyond toric target spaces, and more ambitiously yet, whether there exists a general notion of geometry underlying matrix models.

\section*{Acknowledgments}
B.E. and O.M. would like to thank M. Bertola, J. Harnad, V. Bouchard,  M. Mari\~ no, M. Mulase,  H.~ Ooguri, N.~Orantin, B. Safnuk, for useful and fruitful discussions on this subject. A.K. would like to thank V.~Bouchard and I.~Melnikov for helpful conversations. The work of B.E. is partly supported by the Enigma European network MRT-CT-2004-5652, ANR project GranMa "Grandes Matrices Al\'eatoires" ANR-08-BLAN-0311-01,  
by the European Science Foundation through the Misgam program,
by the Quebec government with the FQRNT. 
B.E. would like to thank the AIM, as well as the organizers and all participants to the workshop held at the AIM june 2009.
O.M. would like to thank the CRM (Centre de recheche math\'ematiques de Montr\'eal, QC, Canada) for its hospitality.

\bigskip

%\vfill
%\eject

\section*{Appendix: q-product}
\label{appgq}

The $g$-function, which plays a central role in the definition of our matrix model, is defined as an infinite product,
\beq
g(x) = \prod_{n=1}^\infty (1-{1\over x}\, q^n) \,.
\eeq
It is the quantum Pochhammer symbol $g(x)= [q/x;q]_{\infty}$, and it is related to the $q$-deformed gamma function via $\Gamma_q(x) = (1-q)^{1-x}\,g(1)/g(q^{1-x})$.

The RHS is convergent for $|q|<1$ and arbitrary complex $x \neq 0$. $g(x)$ satisfies the functional equation
\beq
g(qx)=(1-{1\over x})\,g(x) \,.
\eeq
For $n\in \IN$, we have
\beq
g(q^n) = 0
\eeq and
\beq
g'(q^n) = (-1)^{n-1} g(1)\,q^{-{n(n+1)\over 4}}\,\, \prod_{m=1}^{n-1} [m]
 = g(1)\,q^{-{n(n+1)\over 2}}\,\, [n-1]!
= (-1)^{n-1}\,q^{-{n(n+1)\over 2}}\, {g(1)^2\over g(q^{1-n})} \,.
\eeq

Via the triple product representation of the theta function,
\beq
\theta(z;\tau) = \prod_{m=1}^\infty ( 1 - e^{2 \pi i m \tau}) (1 + e^{(2m-1)\pi i \tau + 2 \pi i z})(1 + e^{(2m-1)\pi i \tau - 2 \pi i z})
\,,
\eeq
we obtain the identity
\beq
\theta \left(\frac{1}{2} + \frac{1}{4\pi i} \ln \frac{q}{x^2}\, ;\,\, \frac{\ln q}{2\pi i} \right) = g(x) g(\frac{q}{x}) g(1) \,.
\eeq

We have
\beq
{g(x) \, g(q/x)\over g(1)^2\,\,\sqrt x}\,\,e^{(\ln{x})^2\over 2\ln q}\,\, e^{-i\pi \ln x\over \ln q}   = {-\ln q\over \theta'({1\over 2} - {i\pi\over \ln q}, -{2i\pi\over \ln q}) }\,\,\, \theta\left({\ln x\over \ln q} +{1\over 2} - {i\pi \over \ln q}, {-2i\pi\over \ln q}\right)\,\, 
\eeq
% \beq
% {e^{-i\pi\ln x\over \ln q}\over \Gamma_q(q/x)\,\,\Gamma_q(x)} = {-\ln q\over \theta'({1\over 2} - {i\pi\over \ln q}, -{2i\pi\over \ln q}) }\,\,\, \theta\left({\ln x\over \ln q} +{1\over 2} - {i\pi \over \ln q}, {-2i\pi\over \ln q}\right)\,\, 
% \eeq
where $\theta$ is the Riemann theta-function for the torus of modulus $-2i\pi/\ln q$.
% This relationship is the quantum deformation of $e^{-i\pi x}/\Gamma(1-x)\Gamma(x) = \sin{(\pi x)}/\pi$ for the classical Gamma-function.

At small $\ln q$, the following expansion is valid,
\beq
\ln{g(x)} = {1\over \ln q}\,\sum_{n=0}^\infty {(-1)^n\,B_n\over n!}\,(\ln q)^n\,\, \Li_{2-n}(1/x) \,,
\eeq
where we have used the definition of the Bernoulli numbers $B_n$ as the coefficients in the expansion of $t/(e^t-1)$,
\beq
\frac{t}{e^t -1} = \sum_{n=0}^\infty B_n \frac{t^n}{n!} \,.
\eeq
$\Li_n$ is the polylogarithm function, defined as
\beq
\Li_n(x) = \sum_{k=1}^\infty\, {x^k\over k^n} \,.
\eeq
This is a generalization of the logarithm function, recovered at $n=1$,
\beq
\Li_1(x) = -\ln{(1-x)} \,.
\eeq
It satisfies the functional relation
\beqn
\Li'_n(x) = \frac{1}{x} \Li_{n-1}(x) \,.  \label{derpoly}
\eeqn
Note in particular that this implies that $\Li_n$ is an algebraic function of $x$ for $n \le 0$. E.g.,
\beq
\Li_0(x) = {x\over 1-x} \,.
\eeq
We also define the function
\beq
\psi_q(x) = x\,{g'(x)\over g(x)}  \,.
\eeq
Using the functional equation (\ref{derpoly}) of the polylogarithm, we find its small $\ln(q)$ expansion
\ba
\psi_q(x) &=& -\,{1\over \ln q}\,\sum_{n=0}^\infty {(-1)^n\,B_n\over n!}\,(\ln q)^n\,\, \Li_{1-n}(1/x) \\
&=&  {1\over \ln{q}}\,\Big[
\ln{(1-{1\over x})} -{\ln q\over 2(x-1)}
-\sum_{n=1}^\infty {B_{2n}\over (2n)!}\,\, (\ln q)^{2n}\,\Li_{1-2n}(x) \Big] \,.
\ea
For the second equality, we have used $B_0=1, B_1 = -\frac{1}{2}$, and $B_{2n+1} = 0$ for $n >1$.

\medskip
We have near $x\to\infty$
\beq
\psi_q(x) \sim {q\over 1-q}\,\,{1\over x} + O(x^{-2})
\eeq
and near $x\to 0$:
\beq
\psi(x) \sim {1\over 2}+{i\pi+\ln x\over g_s} +O(x).
\eeq

%% file: annexei.tex
\selectlanguage{french}
\annexe{A matrix model for the topological string II: The spectral curve and mirror geometry} \label{Article[VI]}
\selectlanguage{english}

\begin{center}
\vskip 1cm 
{B. Eynard${}^1$, A. Kashani-Poor${}^{2,3}$, O. Marchal${}^{1,4}$}

\vskip.2cm 
{\it ${}^1$ Institut de Physique Th\'eorique,\\
CEA, IPhT, F-91191 Gif-sur-Yvette, France,\\
CNRS, URA 2306, F-91191 Gif-sur-Yvette, France.\\ \vskip0.3cm}

{\it 
$^2$ Institut des Hautes \'Etudes Scientifiques\\
Le Bois-Marie, 35, route de Chartres, 91440 Bures-sur-Yvette, France\\ \vskip0.3cm }

{\it $^3$ Laboratoire de Physique Th\'eorique de l'\'Ecole Normale Sup\'erieure, \\
24 rue Lhomond, 75231 Paris, France \\ \vskip0.3cm}

{\it ${}^4$ Centre de recherches math\'ematiques,
Universit\'e de Montr\'eal \\
C.P. 6128, Succ. centre-ville
Montr\'eal, Qu\'e, H3C 3J7, Canada.\\}

\end{center} 
\vskip 1.5cm

\begin{center}

In a previous paper, we presented a matrix model reproducing the topological string partition function on an arbitrary given toric Calabi-Yau manifold. Here, we study the spectral curve of our matrix model and thus derive, upon imposing certain minimality assumptions on the spectral curve, the large volume limit of the BKMP ``remodeling the B-model'' conjecture, the claim that Gromov-Witten invariants of any toric Calabi-Yau 3-fold coincide with the spectral invariants of its mirror curve. 
\end{center}

\newpage

\section*{1 Introduction}

In a previous paper \cite{topstring1}, we presented a matrix model that computes the topological string partition function at large radius on an arbitrary toric Calabi-Yau manifold $\CYX$. The goal of this paper is to determine the corresponding spectral curve $\spcurve$. 

That the partition function of a matrix model can be recovered to all genus from its spectral curve was first demonstrated in \cite{Eyn1loop}. \cite{OE} pushed this formalism further, showing that symplectic invariants $F_g(\spcurve)$ can be defined for any analytic affine curve $\spcurve$, with no reference to an underlying matrix model. These invariants coincide with the partition function of a matrix model when $\spcurve$ is chosen as the associated spectral curve. The symplectic invariants $F_g$ satisfy many properties reminiscent of the topological string partition function \cite{eynhaeq, Orantin,CMMV,EynVolmum}, motivating Bouchard, Klemm, Mari\~no, and Pasquetti (BKMP) \cite{BKMP}, building on work of Mari\~no \cite{marino2Ann}, to conjecture that $F_g(\spcurve)$ in fact coincides with the topological string partition function on the toric Calabi-Yau manifold with mirror curve $\spcurve$. BKMP successfully checked their claim for various examples, at least to low genus. The conjecture was subsequently proved in numerous special cases \cite{Partitions2, MarshakovNekrasov, KlemmSulkowski, Sulkowski, BKMPcase1, BKMPChen}.

Bouchard and Mari\~no \cite{BouchardMarino} noticed that an infinite framing limit of the BKMP conjecture for the framed vertex, $\CYX=\mathbb C^3$, implies a conjecture for the computation of Hurwitz numbers, namely that the Hurwitz numbers of genus $g$ are the symplectic invariants of genus $g$ for the Lambert spectral curve $e^x=y\, e^{-y}$. This conjecture was proved recently by a generalization of \cite{Partitions2} using a matrix model for summing over partitions \cite{BorotEynardMulase}, and also by a direct combinatorial method \cite{EMS}. Matrix models and the BKMP conjecture related to toric Calabi-Yau geometries arising from the triangulation of a strip were recently studied in \cite{OSY}.

In this paper, we derive the large radius limit of the BKMP conjecture for {\it arbitrary} toric Calabi-Yau manifolds, but with one caveat: to determine the spectral curve of our matrix model, we must make several minimality assumptions along the way. To elevate our results to a rigorous proof of the BKMP conjecture, one needs to establish a uniqueness result underlying our prescription for finding  the spectral curve to justify these minimal choices. Such a uniqueness result does not exist to date.

Recall that in \cite{topstring1}, we first compute the topological string partition function on a toric Calabi-Yau geometry $\CYX_0$ which we refer to as fiducial. We then present a matrix model which reproduces this partition function. Flops and limits in the K\"ahler cone relate $\CYX_0$ to an arbitrary toric Calabi-Yau 3-fold. As we can follow the action of these operations on the partition function, we thus arrive at a matrix model for the topological string on any toric Calabi-Yau 3-folds. Here, we follow the analogous strategy, by first computing the spectral curve of the matrix model associated to $\CYX_0$, and then studying the action of flops and limits on this curve.

The plan of the paper is as follows. In section \ref{fiducial_section}, we introduce the fiducial geometry $\CYX_0$ and its mirror. The matrix model reproducing the partition function on $\CYX_0$, as derived in \cite{topstring1}, is a chain of matrices matrix model. It is summarized in section \ref{s_our_matrix_model} and appendix \ref{our_matrix_model}. We review general aspects of this class of matrix models and their solutions in section \ref{generalities_on_mm}. In section \ref{the_spectral_curve}, we determine a spectral curve which satisfies all specifications outlined in section \ref{generalities_on_mm}, and demonstrate that it coincides, up to symplectic transformations, with the B-model mirror of the fiducial geometry.  While in our experience with simpler models, the conditions of section \ref{generalities_on_mm} on the spectral curve specify it uniquely, we lack a proof of this uniqueness property. We thus provide additional consistency arguments for our proposal for the spectral curve in section \ref{secproofsp}. Flops and limits in the K\"ahler cone relate the fiducial to an arbitrary toric Calabi-Yau manifold. Following the action of these operations on both sides of the conjecture in section \ref{finishing_proof} completes the argument yielding the BKMP conjecture for arbitrary toric Calabi-Yau manifolds in the large radius limit.  We conclude by discussing possible avenues along this work can be extended.

\section*{2 The fiducial geometry and its mirror} \label{fiducial_section}

\subsection*{2.1 The fiducial geometry}
\begin{figure}[h]
\centering
\includegraphics[width=10cm]{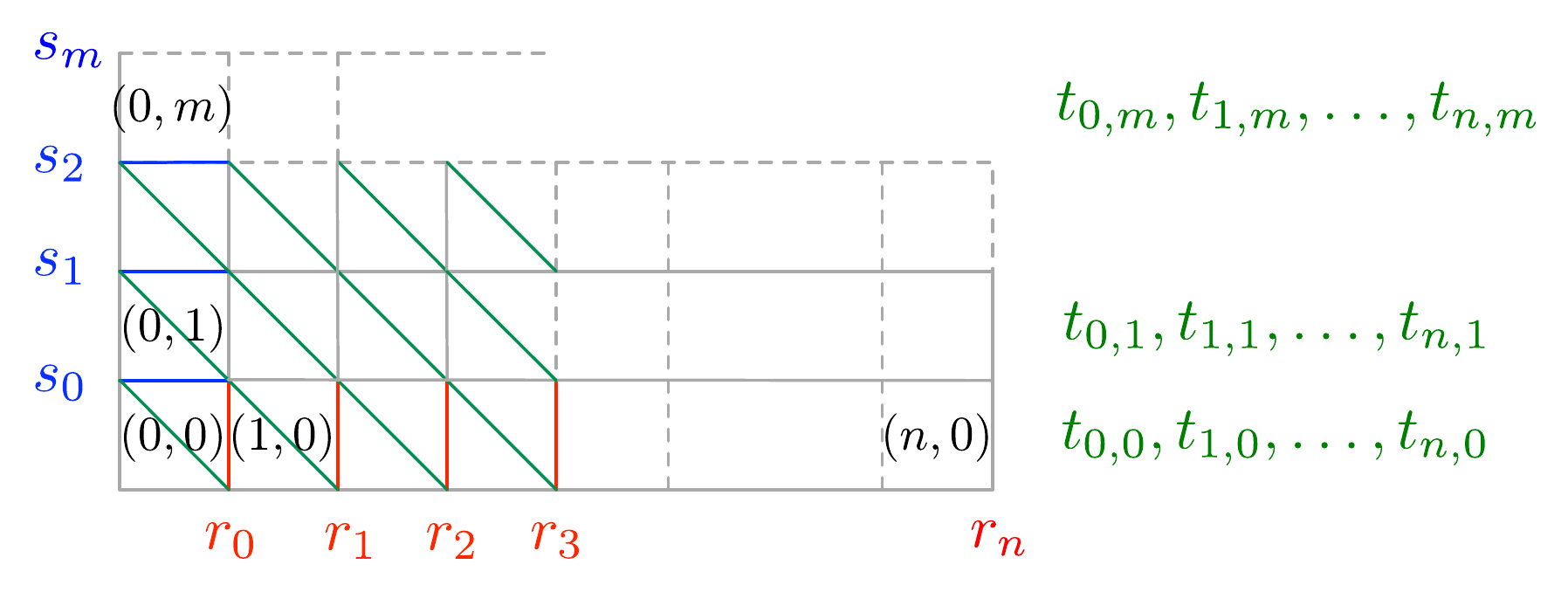}
\caption{\footnotesize{Fiducial geometry $\CYX_0$ with boxes numbered and choice of basis of $H_2(\CYX_0, \IZ)$.}}
\label{fiducial_geometry}
\end{figure}

In \cite{topstring1}, we derived a matrix model reproducing the topological string partition function on the toric Calabi-Yau geometry $\CYX_0$ whose toric fan is depicted in figure \ref{fiducial_geometry}. We refer to $\CYX_0$ as our fiducial geometry; we will obtain the partition function on an arbitrary  toric  Calabi-Yau manifolds by considering flops and limits of $\CYX_0$.

\begin{figure}[h]
\centering
\includegraphics[width=3cm]{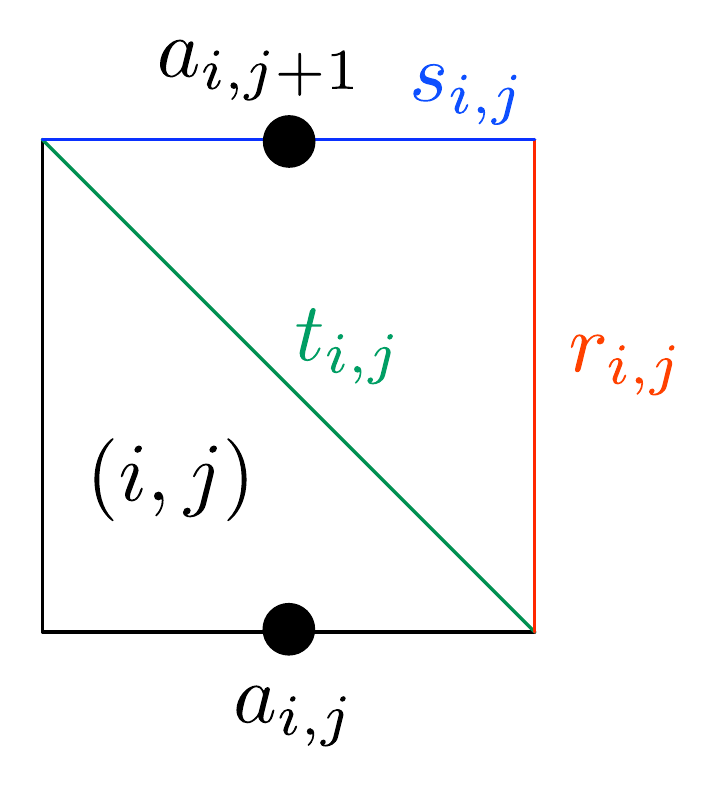}
\caption{\footnotesize{Labeling curve classes, and introducing $a$-parameters.}}
\label{fiducial_labeling}
\end{figure}

We have indicated a basis of $H_2(\CYX_0,\IZ)$ in figure \ref{fiducial_geometry}. Applying the labeling scheme introduced in figure \ref{fiducial_labeling}, the curve classes of our geometry are expressed in this basis as follows,
\ba
r_{i,j} &=& r_i + \sum_{k=1}^j (t_{i+1,k-1} - t_{i,k}) \\
s_{i,j} &=& s_j + \sum_{k=1}^i ( t_{k-1,j+1} - t_{k,j}) \,.
\ea
It proves convenient to express these classes as differences of what we will refer to as $a$-parameters \cite{topstring1}, defined via
\beq
 t_{i,j}=a_{i,j}-a_{i,j+1} \quad\,, \quad \quad r_{i,j} = a_{i,j+1} - a_{i+1,j} \,.
\eeq

\subsection*{2.2 The mirror of the fiducial geometry}
\label{secmirrorcurve}

The Hori-Vafa prescription \cite{HoriVafa} allows us to assign a mirror curve to a toric Calabi-Yau manifold. Each torically invariant divisor, corresponding to a 1-cone $\rho \in \Sigma(1)$, is mapped to a $\IC^*$ variable $e^{-Y_\rho}$. These are constrained by the equation
\beq
\sum_{\rho \in \Sigma(1)} e^{-Y_\rho} = 0 \,.
\eeq
Relations between the 1-cones, as captured by the lattice $\Lambda_h$ introduced in section (2.1) of \cite{topstring1}, map to relations between these variables: for $\sigma \in \Sigma(2)$,
\beqn  \label{y_relations}
\sum_{\rho \in \Sigma(1)} \lambda_\rho(\sigma) Y_\rho = W_\sigma \,.
\eeqn
The $W_\sigma$ are complex structure parameters of the mirror geometry, related to the K\"ahler parameters $w_\sigma= r_{i,j}, s_{i,j}, \ldots$ introduced in the previous subsection via the mirror map, as we will explain in the next subsection.

The Hori-Vafa prescription gives rise to the following mirror curve $\curve_{\CYX_0}$ of our fiducial geometry $\CYX_0$, 
\begin{equation}
\sum_{i=0}^{n+1} \sum_{j=0}^{m+1} x_{i,j}  = 0  \,.  \label{HV}
\end{equation}
We have here labeled the 1-cones by coordinates $(i,j)$, beginning with $(0,0)$ for the cone $(0,0,1)$ in the bottom left corner of box $(0,0)$ as labeled in figure \ref{fiducial_geometry}, and introduced the notation
\beq
x_{i,j} = e^{-Y_{i,j}} \,.
\eeq
Eliminating dependent variables by invoking (\ref{y_relations}) yields an equation of the form
\beqn \label{mirror_curve}
\sum_{i=0}^{n+1} \sum_{j=0}^{m+1} c_{i,j} z_{i,j} = 0 \,.
\eeqn
Here, $$z_{i,j} = x_0^{1-i-j} x_1^i x_2^j \,,$$ where we have defined 
\beq 
x_0=x_{0,0}\,,\,\, x_1=x_{1,0}\,,\,\, x_2=x_{0,1}\,.
\eeq
$(x_0:x_1:x_2)$ define homogeneous coordinates on $\IC \IP^2$. The form of the equation is independent of the choice of triangulation of the toric diagram. What does depend on this choice are the coefficients $c_{i,j}$. It is not hard to write these down for the fiducial geometry $\CYX_0$ with the choice of basis for $H_2(\CYX_0,\IZ)$ indicated in figure \ref{fiducial_geometry}. Explicitly, the relations between the coordinates of the mirror curve (\ref{HV}) are
\begin{equation}
x_{i,0} = \frac{x_{i-1,0} x_{i-1,1}}{x_{i-2,1}} e^{R_{i-2}} \,,\quad x_{0,j}= \frac{x_{0,j-1} x_{1,j-1}}{x_{1,j-2}} e^{S_{j-2}} \,, \quad x_{i,j}=\frac{x_{i-1,j} x_{i,j-1}}{x_{i-1,j-1}} e^{T_{i-1,j-1}} \,.  \label{rel_bw_mirror_variables}
\end{equation}
Solving in terms of $x_0,x_1,x_2$ yields the coefficients $c_{0,0}=c_{0,1}=c_{1,0}=1$, 
\begin{align*}
c_{i,0} &=\exp\left[\sum_{k=1}^{i-1} (i-k) (R_{k-1} + T_{k-1,0})\right] \,, \\
c_{0,j} &=\exp\left[\sum_{l=1}^{j-1} (j-l) (S_{l-1} + T_{0,l-1})\right] \,,\\
\intertext{and for $i,j >0$}
c_{i,j} &= \exp\left[(i+j-1) T_{0,0} + \sum_{k=1}^{i-1} (i-k)(R_{k-1}+T_{k,0}) + \sum_{l=1}^{j-1} (j-l)(S_{l-1}+T_{0,l}) +  \sum_{k=1}^{i-1} \sum_{l=1}^{j-1}  T_{k,l}\right]  \,.
\end{align*}
Note that the number of coefficients $c_{i,j}$, up to an overall rescaling, is equal to the number of independent curve classes $r_i$, $s_j$, $t_{i,j}$.
 
\begin{figure}[h]
 \centering
  \includegraphics[width=6cm]{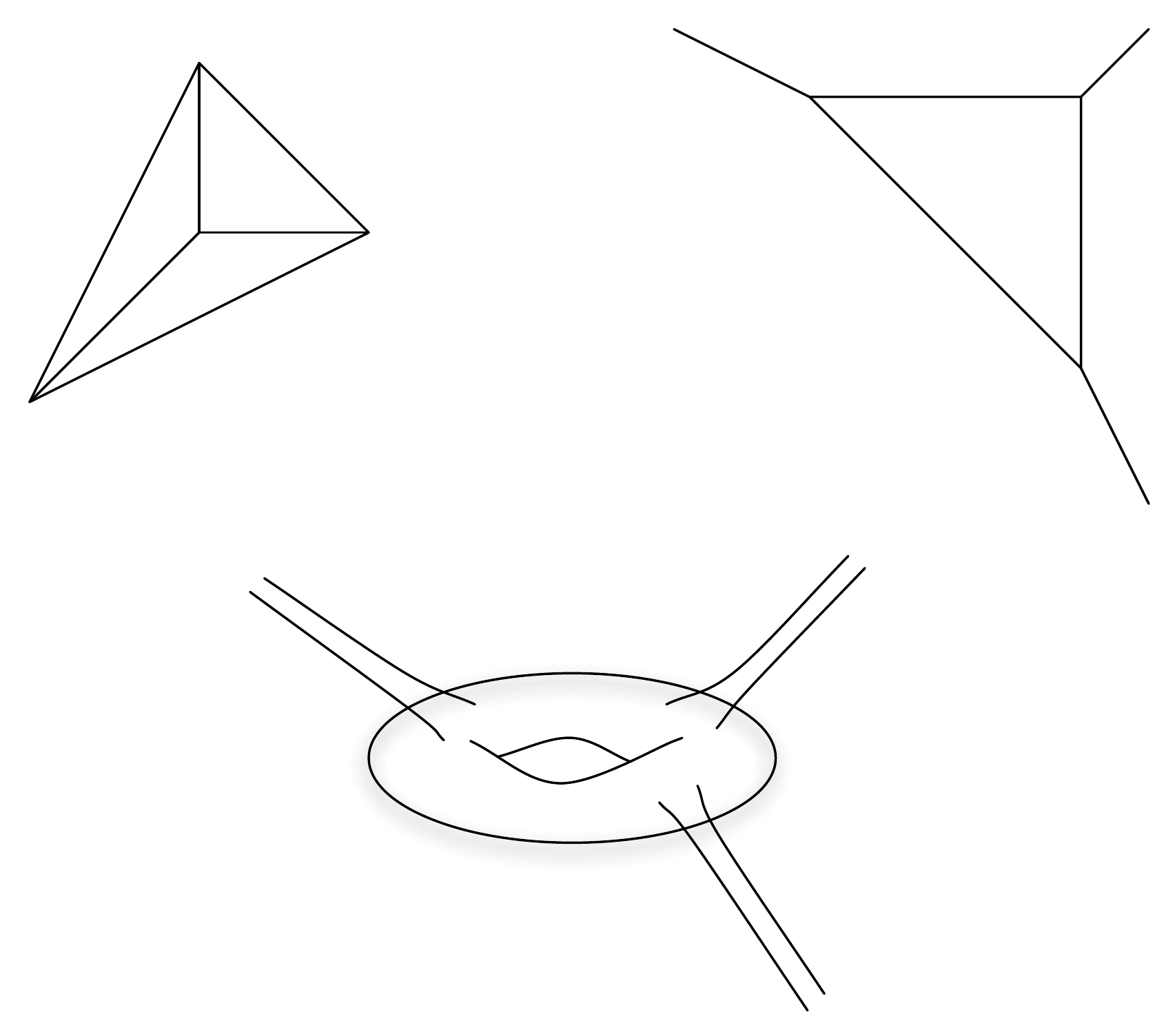}
 \caption{\footnotesize{Example of the thickening prescription: depicted are the fan for $\cO(-3) \rightarrow \IP^2$, the corresponding web diagram, and the mirror curve obtained via the thickening prescription.}}
 \label{thickening}
\end{figure}

In \cite{AKV}, the thickening prescription was put forth for determining the genus and number of punctures of the mirror curve: one is to thicken the web diagram of the original geometry to obtain the Riemann surface of the mirror geometry. The procedure is illustrated in figure \ref{thickening}. We will now verify this procedure by studying the curve (\ref{mirror_curve}) explicitly.

Let's consider the curve (\ref{mirror_curve}) for a single strip (i.e. $n=0$) of length $m+1$,
\begin{multline}
x_0^{m+2} + x_0^{m+1} x_1 + x_0^{m+1} x_2 + c_{1,1} \,x_0^{m} x_1 x_2 + c_{2,0}\, x_0^{m} x_1^2 + c_{2,1} \,x_0^{m-1} x_1^2 x_2 + c_{3,0} \,x_0^{m-1} x_1^3 + \ldots \\+ c_{m+1,0} \,x_0 x_1^{m+1} + c_{m+1,1} \,x_1^{m+1} x_2  = 0 \,. \label{curveonstrip}
\end{multline}
Note that the equation is of degree $m+2$, but the point $(0:0:1)$ is
an $m+1$-tuple point. By choosing the coefficients to be generic, we
can arrange for this singular point to be ordinary. The genus formula
then yields $$ {\frak g}= \frac{(d-1)(d-2)}{2} - \frac{m (m+1)}{2} = 0 \,.$$ In terms of the physical variables $Y_i$, any point on the curve with a vanishing homogeneous coordinate corresponds to a puncture. The punctures on the curve (\ref{curveonstrip}) thus lie at 
\begin{align*}
(0:0:1) \quad &: \quad m+1 \\
(0:1:0) \quad &: \quad 1 \\
(1:x_1^i:0) \quad&: \quad m+1 \\
(1:0:-1) \quad&: \quad 1 \,,
\end{align*}
where $x_1^i$, $i=1, \dots, m+1$, are the solutions of the equation
\ba
1 + x_1 + \sum_{j=1}^{m} d_i x_1^{i+1} &=&0 \,.
\ea
Note that we reproduce the $2m+4$ punctures expected from the thickening prescription of the toric diagram.

For the general case parametrized by $(m,n)$, the degree of the curve is $d=m+n+2$, and we have an ordinary $m+1$-tuple point at $(0:0:1)$ and an ordinary $n+1$-tuple point at $(0:1:0)$. The genus formula now yields $$g=\frac{(m+n)(m+n+1)}{2}-\frac{m(m+1)}{2}-\frac{n(n+1)}{2} = m n\,.$$ The punctures lie at
\begin{align*}
(0:0:1) \quad &: \quad m+1 \\
(0:1:0) \quad &: \quad n+1 \\
(1:x_1^i:0) \quad&: \quad m+1 \\
(1:0:x_2^j) \quad&: \quad n+1  \,,
\end{align*}
with $x_1^i$ the roots of $\sum_{i=0}^{m+1} c_{i,0} x_1^i = 0$ and $x_2^j$ the roots of $\sum_{j=0}^{n+1} c_{0,j} x_2^j = 0$. Again, we see that we reproduce the thickening prescription.

\subsection*{2.3 The mirror map}
Above, we have distinguished between K\"ahler ($A$-model) parameters $w_\sigma$ and complex structure ($B$-model) parameters $W_\sigma$. At large radius/complex structure, these are identified between mirror pairs, but this identification is corrected by the so-called mirror map,\footnote{One could take exception to this nomenclature, arguing that the parameters $W_\sigma$ are the geometric parameters on both sides of the mirror, and refer to the $w_\sigma$ as the instanton or quantum corrected parameters. In such conventions, the curve classes in the various toric diagrams should be labeled by upper case letters.}
\beqn
W_\sigma = w_\sigma + \cO (e^{-w_\sigma}) \,. \label{mmap}
\eeqn
The exponentials of the parameters $W_\sigma$ appear as coefficients in the equation defining the mirror curve. They are global coordinates on the complex structure moduli space of the mirror curve. To compare expressions obtained in the A-model to those obtained in the B-model, all expressions are conventionally expressed in terms of flat coordinates $w_\sigma$. On the A-model side, these coordinates enter (in exponentiated form denoted generically as $Q_{\alpha,\beta}$ below) in the definition of the topological vertex. On the B-model side, they arise as the appropriate periods of a meromorphic one-form $\lambda$, defined in terms of the affine variables $x=\frac{x_1}{x_0}$, $y=\frac{x_2}{x_0}$ in the patch $x_0 \neq 0$ of the curve (\ref{mirror_curve}) as
\beq
\lambda = \log y \frac{dx}{x} \,.
\eeq
By calculating these periods as a function of the coefficients defining the mirror curve, we obtain the mirror map (\ref{mmap}).

The coordinates $w_\sigma$ are not globally defined functions on the complex structure moduli space. In the slightly clearer compact setting, this is due to the fact that the symplectic basis $\{\alpha_A,\beta^A\}$ of $H^3(\CYX,\IZ)$ in which we expand $\Omega$ (the compact analogue of the meromorphic 1-form $\lambda$ introduced above) such that the coefficients of $\alpha_A$ furnish our (local) coordinate system of the complex structure moduli space, undergo monodromy when transported around a singularity in moduli space.\footnote{Note that the symplectic basis makes no reference to complex structure, one might hence be led to believe that a global choice (i.e. one valid for any choice of complex structure) should be possible. This is not so. We consider the family $\pi: \cX \rightarrow \cS$, with $\cS$ the complex structure moduli space. The fiber over each point $w\in \cS$, $\pi^{-1}w = X_w$, is the Calabi-Yau manifold with the respective complex structure. $H^n(X_w,\IC)$ fit together to form a vector bundle $\cF_0$ over $\cS$, with a canonical flat connection, the Gauss-Manin connection. Using this connection, we can parallel transport a symplectic basis of $H^3(X_w,\IC)$ along a curve in $\cS$. As $\cS$ is not generically simply connected (due to the existence of degeneration points of the geometry), this transport may exhibit monodromy. Note that $\Omega$ can be defined as the section of a sheaf in the Hodge filtration of $H^3$ which extends to the singular divisor, hence is single valued. The monodromy in our choice of flat coordinates is therefore entirely due to the choice of symplectic basis.} A good choice of coordinates in the vicinity of a singular divisor $D$ hence involves a choice of basis forms that are invariant under monodromy around that divisor.

\section*{3 Our matrix model} \label{s_our_matrix_model}

We derived a chain of matrices matrix model that reproduces the topological string partition function on $\CYX_0$ in \cite{topstring1}. For $\CYX_0$ of size $(n+1)\times (m+1)$, as depicted in figure \ref{fiducial_geometry}, it is given by
\ban 
{Z}_{\rm MM}(\vec Q,g_s,\vec\alpha_{m+1},\vec\alpha_0^T)
&=& \Delta(X(\vec \alpha_{m+1}))\,\, \Delta(X(\vec \alpha_0)) \,\, 
\prod_{i=0}^{m+1} \int_{H_N(\Gamma_i)} dM_i \,
 \prod_{i=1}^{m+1}\int_{H_N({\mathbb R}_+)}\,dR_i \nn \\
&& \prod_{i=1}^{m} e^{{-1\over g_s}\,\tr \left[ V_{\vec a_i}(M_i)-V_{\vec a_{i-1}}(M_i) \right]
%+ V_{\vec a_i,\vec a_{i-1}}(X_i) 
} \,\,\,
 \prod_{i=1}^{m} e^{{-1\over g_s}\,\tr \left[V_{\vec a_{i-1}}(M_{i-1})-V_{\vec a_{i}}(M_{i-1}) \right]
%+ \td V_{\vec a_{i-1},\vec a_{i}}(X_{i-1}) 
} \nn \\
&& \prod_{i=1}^{m+1} e^{{1\over g_s} \tr (M_i-M_{i-1})R_i} \,\,\,
 \prod_{i=1}^{m} e^{(S_i+{i\pi\over g_s})\,\tr\, \ln M_i}\,  \nn\\
&& e^{\tr \ln f_{0}(M_0)}\,\,e^{\tr \ln f_{m+1}(M_{m+1})}\,\, \prod_{i=1}^{m} e^{\tr \ln f_{i}(M_i)} \,. \label{m_integral}
\ean
We give the explicit expressions for the various functions entering in this definition in appendix \ref{our_matrix_model}. Here, we briefly explain some of its general features. 

The matrix model (\ref{m_integral}) is designed to reproduce the topological string partition function on the toric Calabi-Yau manifold $\CYX_0$ as computed using the topological vertex \cite{TopologicalVertex}. Recall that in this formalism, the dual web diagram to the toric diagram underlying the geometry is decomposed into trivalent vertices. Each such vertex contributes a factor $C(\alpha_i,\alpha_j,\alpha_k$) \cite{TopologicalVertex}, where $\alpha_i$ denote Young tableaux (partitions) of arbitrary size, one associated to each leg of the vertex. Legs of different vertices are glued by matching these Young tableaux and summing over them with appropriate weight.

Aside from the coupling constant $g_s$ and K\"ahler parameters of the geometry, denoted collectively as $\vec Q$, the matrix model (\ref{m_integral}) depends on partitions $\vec\alpha_0$, $\vec\alpha_{m+1}$ associated to the outer legs of the web diagram, which we choose to be trivial in this paper. The two classes of integrals $dR_i$ and $dM_i$ correspond to the two steps in which the topological string partition function on the fiducial geometry $\CYX_0$ can be evaluated: First, the geometry can be decomposed into $m+1$ horizontal strips, with partitions $\alpha_{j,i+1}$ and $\alpha_{j,i}$ associated to the upper and lower outer legs of the associated strip web diagram. $j = 0, \ldots, n$ counts the boxes in figure \ref{fiducial_geometry} in the horizontal direction, $i= 0, \ldots, m+1$ is essentially the strip index. Each such strip has a $dR_i$ integration associated to it. The partition function on such strips was calculated in \cite{IqbalKashaniPoor}. Following \cite{KlemmSulkowski}, we introduce two matrices $M_i$, $M_{i+1}$ per strip. Their eigenvalues encode the partitions $\alpha_{j,i}$ and $\alpha_{j,i+1}$ for all $j$. To work with finite size matrices, we introduce a cut-off $d$ on the number of rows of the Young tableaux we sum over. As we argue in section \ref{arctic}, our matrix model depends on $d$ only non-perturbatively. The strip partition function is essentially given by the Cauchy determinant of the two matrices $M_i$, $M_{i+1}$ \cite{topstring1}, and the $dR_i$ integrals are the associated Laplace transforms. Gluing the strips together involves summing over the partitions $\alpha_{j,i}$. This step is implemented by the $dM_i$ integrations. To obtain a discrete sum over partitions from integration, we introduce functions $f_i(M_i)$ with integrally spaced poles. Integrating $M_i$ along appropriate contours then yields the sum over partitions as a sum over residues, the potentials $V_{\vec{a}_i}$ chosen to provide the proper weight per partition.

\section*{4 Generalities on solving matrix models}  \label{generalities_on_mm}

\subsection*{4.1 Introduction to the topological expansion of chain of matrices}

Chain of matrices matrix models have been extensively studied (see Mehta's book \cite{Mehta} and the review article \cite{ZJDFG}), and the computation of their topological expansion was performed recently in \cite{Eynchain, Chain}.

\smallskip
The solution provided in \cite{Chain} is based on the computation of the spectral curve $\spcurveMM$ of the matrix model. In \cite{Chain, Eynchain}, only the case of potentials whose derivatives are rational functions is considered, and 
similarly to the one matrix model, the planar\footnote{\label{planar}For matrix models with $N$-independent polynomial potentials whose $g_s$ dependence is given by an overall prefactor, the planar limit coincides with the large $N$ limit, but this correspondence can fail if the potential or the integration contours have a non-trivial $N$ or $g_s$ dependence. The planar limit is defined by keeping only planar graphs in the Feynman graph perturbative expansion around an extremum of the potential. However, it is helpful to have in mind the intuitive picture that the planar limit is similar to a large $N$ limit.} expectation value of the resolvent of the first matrix of the chain is shown to satisfy an algebraic equation. The spectral curve is defined to be the solution locus of this equation. 
A general recipe is provided in \cite{Eynchain, Chain} to obtain the spectral curve from algebraic equations and analyticity properties related to rational potentials and integration contours. Here, our potentials contain logs of $g$-functions. As they are not rational, we will have to present a slight extension of the recipe of \cite{Chain} in section \ref{secspcurvegenchain}. This extension from rational potentials to analytical potentials, although not published, is straightforward, and the derivation of these results will appear soon.
In some sense, the derivative of $\ln g(x)$ can be viewed as a rational function with an infinite number of simple poles, i.e. as a limit of a rational function. More precisely, as an expansion in powers of $q$, to each order, it is a rational function. Since the spectral curve can be described by local properties, independent of the number of poles, one can take the limit of the recipe of \cite{Eynchain, Chain}.
This is what we shall do in section \ref{secspcurvegenchain} below.

\smallskip
Having found the spectral curve $\spcurveMM$ of the matrix model, we will compute its symplectic invariants
$$
F_g(\spcurveMM)\, ,\quad g=0,1,2,3,\dots
$$
Symplectic invariants $F_g(\spcurve)$ can be computed for any analytical plane curve $\spcurve$, and thus in particular for $\spcurve=\spcurveMM$. For a general $\spcurve$ they were first introduced in \cite{OE}, as a generalization of the solution of matrix models loop equations of \cite{Eyn1loop}.
Their definition is algebraic and involves computation of residues at branch points of $\spcurve$.
We recall the definition below in section \ref{secdefspinv}.

\subsection*{3.2 Definition of the general chain of matrices}

We consider chain of matrices matrix models of the form
\beqn\label{Zgeneralchain}
Z = \int_{{\cal E}} dM_1\dots dM_L\,\, e^{-{1\over g_s}\Tr \sum_{i=1}^L V_i(M_i)} \,\, e^{{1\over g_s}\Tr \sum_{i=1}^{L-1}  c_i M_{i} M_{i+1}} \,.
\eeqn
Note that aside from the potentials $V_i(M_i)$, the only interactions are between nearest neighbors, whence the name ``chain of matrices.'' Chain of matrices matrix models can be solved when the interaction terms between different matrices are of the form $\Tr M_{i} M_{i+1}$, as is the case here.

${\cal E}$ can be any ensemble of $L$ normal matrices of size $N\times N$, i.e. a submanifold of $\mathbb C^{LN^2}$ of real dimension $LN^2$, such that the integral is convergent. ${\cal E}$ can be many things; for a chain of matrices model, it is characterized by the contours on which eigenvalues of the various normal matrices are integrated (see \cite{MarcoPaths} for the 2-matrix model case). 
For (\ref{Zgeneralchain}) to have a topological expansion, ${\cal E}$ must be a so-called steepest descent ensemble (see \cite{countingsurface}, section 5.5).
For a generic ensemble ${\cal E}$ which would not be steepest descent, $\ln Z$ would be an oscillating function of $1/g_s$, and no small $g_s$ expansion would exist, see \cite{ConvForm1}.

The matrix model introduced in \cite{topstring1} and reproduced in section \ref{s_our_matrix_model} was defined to reproduce the topological string partition function, which is defined as a formal series in $g_s$, and therefore has a topological expansion by construction.

An ensemble ${\cal E}$ is characterized by filling fractions $n_{j,i}$,
\beqn
{\cal E} = \prod_{i=1}^L {\cal E}_i
\quad , \quad
{\cal E}_i  = H_N(\gamma_{1,i}^{n_{1,i}} \times  \gamma_{2,i}^{n_{2,i}} \times \dots \times \gamma_{k_i,i}^{n_{k_i,i}}) \,,   \label{matrix_ensemble}
\eeqn
where $H_N(\gamma_1^{n_1}\times \dots\times \gamma_k^{n_k})$ is the set of normal matrices with $n_1$ eigenvalues on path $\gamma_1$, $n_2$ eigenvalues on path $\gamma_2$, $\dots$, $n_k$ eigenvalues on path $\gamma_k$.

As the filling fractions $n_{j,i}$ must satisfy the relation
\beq
\sum_{j=1}^{k_i} n_{j,i} = N
\eeq
for all $i$, only $\sum_i (k_i-1)$ of them are independent.

We also allow some paths $\gamma_{j,i}$ to have endpoints where $e^{-\Tr \sum_{i=1}^L (V_i(M_i) - M_{i} M_{i+1})}\neq 0$ -- indeed, in our matrix model, the matrices $R_i$ are integrated on $H(\mathbb R_+^N)$.

\subsubsection*{The resolvent}\label{seceigenvalueinterp}
The spectral curve encodes all $W_i(x)$, the planar limits (see footnote \ref{planar}) of the resolvents of the matrices $M_i$,
\beq
W_i(x) = g_s\,\left< \tr {1\over x-M_i}\right>_{planar} \,,
\eeq
see equation (\ref{resolvent}) below. The respective $W_i$ can be expressed as the Stieljes transform
\beq
W_i(x) = \int {\rho_i(x')dx'\over x-x'} 
\eeq
of the planar expectation value of the eigenvalue density $\rho_i(x)$ of the matrix $M_i$,
\beq
\rho_i(x) =g_s\,\left< \tr \delta(x-M_i)\right>_{planar} \,.
\eeq

By general properties of Stieljes transforms, singularities of $W_i(x)$ coincide with the support of the distribution $\rho_i(x)dx$:
\begin{itemize}
\item Simple poles of $W_i(x)$ correspond to delta distributions i.e. isolated eigenvalues.

\item Multiple poles correspond to higher derivatives of delta distributions.

\item Cuts correspond to finite densities, the density being the discontinuity of $W_i(x)$ along the cut,
\beqn  \label{density_as_discontinuity}
\rho_i(x) = {1\over 2i\pi}\,(W_i(x-i0)-W_i(x+i0)) \,.
\eeqn
\end{itemize}

In particular, cuts emerging from algebraic singularities (generically square root singularities) correspond to densities vanishing algebraically (generically as square roots) at the endpoints of the cut. Cuts emerging from logarithmic singularities correspond to constant densities.

\subsubsection*{The spectral curve of the general chain of matrices}
\label{secspcurvegenchain}

When all $V'_i$ are rational, the spectral curve was found in \cite{Eynchain, Chain}, and it is algebraic.
We present here a generalization of this result to more general potentials.

The spectral curve can be obtained by the following procedure:
\medskip

\begin{enumerate}
\item \label{genus_ss} Consider a compact Riemann surface $\curve$ of genus
\beq
{\mathfrak g}= \sum_{i=1}^{L} (k_{i}-1)   \,,
\eeq
where $k_i$ denotes the number of cuts of the $i$-th matrix, as implicitly defined in (\ref{matrix_ensemble}).

\item \label{projections}
Look for $L+2$ functions on $\curve$, 
\beq
x_0(z),\, x_1(z),\, x_2(z),\, \dots , \, x_{L}(z),\, x_{L+1}(z): \curve \rightarrow \IC \IP^1 \,.
\eeq
The $x_i$ are to be holomorphic away from points $z \in \curve$ at which $V'_{i-1}(x_{i-1})$ or $V'_{i+1}(x_{i+1})$ become singular, and satisfy the functional relations
\beqn\label{eqspcurvechainxxV}
c_{i-1}x_{i-1}(z)+c_i x_{i+1}(z) = V'_i(x_i(z)) \,.
\eeqn
Recall that the $c_i$ are the coefficients of the interaction potentials in (\ref{Zgeneralchain}). We have set $c_0=c_{L}=1$.

For each $i=1,\dots,L$, the Riemann surface $\curve$ can be realized as a branched covering of $\mathbb C\IP^1$ by the projection $x_i:\curve\to \mathbb C\IP^1$. A choice of branched covering is not unique: the choice consists in the set of cuts connecting branch points (recall that these are points at which $dx_i(z)=0$). We will determine an appropriate covering below in step \ref{sings}.

\item \label{hard_edges} If some path $\gamma_{j,i}$ has an endpoint $a$ (called ``hard edge'' in the matrix model literature, see \cite{Hardedges}), then choose a pre-image $a_i \in x_i^{-1}(a)$ and require
\beq
dx_i(a_i)=0
\quad {\rm and} \quad
x_{i-1}(z)\,\,{\rm has\, a \,simple\,pole\, at\,} z=a_i.
\eeq
The topological recursion is proved in \cite{Chain} without hard edges, but it is not difficult to see, by mixing the results of \cite{Hardedges}, \cite{ChekhovHardedges} and \cite{Chain}, that the topological recursion continues to hold in the presence of hard edges. The proof will appear in a forthcoming publication. Here, we shall assume that it holds.

\item \label{sings} Choose some contours $\widehat{\cal A}_{j,i}$, $j=1,\dots,k_i$ in $\mathbb C \IP^1$, such that  each $\widehat{\cal A}_{j,i}$ surrounds all points of the contour $\gamma_{j,i}$ (related to the matrix ensemble ${\cal E}_i$ defined in (\ref{matrix_ensemble})) in the clockwise direction and no other contour $\gamma_{j',i}$. 
For $x\in \mathbb C \IP^1$ not enclosed in the contours $\widehat{{\cal A}}_{j,i}$, $j=1,\dots,k_i,$ and given a connected component ${{\cal A}_{j,i}}$ of the pre-image of the contour $\widehat{{\cal A}}_{j,i}$ under $x_i$,
\beq
{\cal A}_{j,i} \subset x_i^{-1}(\widehat{{\cal A}}_{j,i}) \,,
\eeq
 define the function
\beqn  \label{resolvent}
W_i(x) = {c_{i-1}\over 2i\pi}\sum_{j=1}^{k_i}\,\oint_{{\cal A}_{j,i}}\,\, {x_{i-1}(z)\,dx_i(z)\over x-x_i(z)}.
\eeqn

Generalizing \cite{Eynchain} to non-polynomial potentials, we claim that a choice of 
${\cal A}_{j,i}$ exists such that $W_i(x)$ is the planar limit of the resolvent of the matrix $M_i$. In the following, it is this choice that will be referred to as ${\cal A}_{j,i}$.

Notice that not all ${\cal A}_{j,i}$ will be homologically independent on $\curve$. 
We require that we have $\mathfrak g=\sum_{i=1}^L (k_i-1)$ homologically independent ${\cal A}_{j,i}$'s, which coincides with the genus of $\curve$. As a condition on the choice of branched covering, we impose that ${\cal A}_{j,i}$ and $a_i$ lie on the same sheet of $x_i$. This condition, in our experience, uniquely fixes this choice. We will assume that this is the case. We refer to the sheet of $x_i$ containing ${\cal A}_{j,i}$ and $a_i$ as the physical sheet for $x_i$.

\item \label{filling_fraction} In accord with (\ref{density_as_discontinuity}), we consider the discontinuity of $W_i(x)$ along the $j$-th cut. It is given by
\ban \label{disc}
\Disc_j\, W_i(x) &=& \frac{1}{2\pi i} \left( W_i(x_+) - W_i(x_-) \right) \nn \\
&=& \frac{1}{2\pi i} c_{i-1} \,  \Disc_j\, x_{i-1} \,, \label{rho}
\ean
as we explain in figure \ref{discontinuity}.
\begin{figure}[h]
 \centering
  \includegraphics[width=5cm]{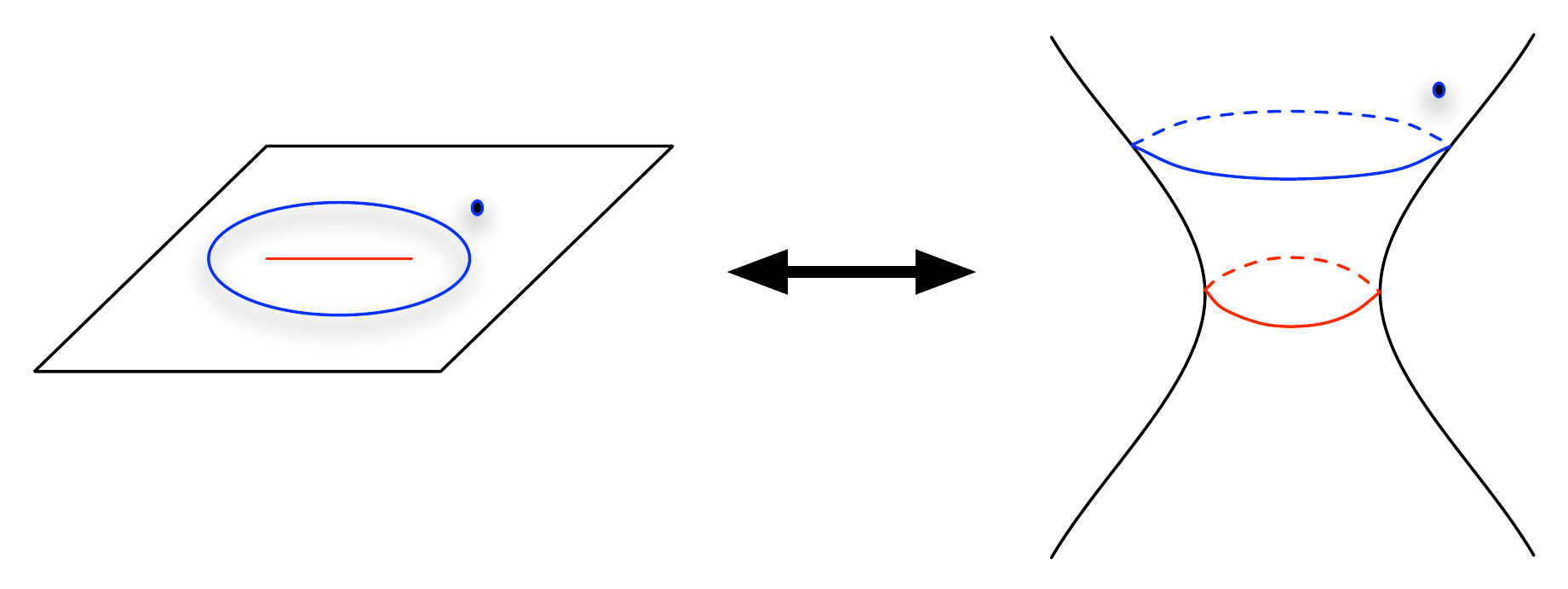}
 \caption{\footnotesize{The integration contour $\widehat{\cal A}_{j,i}$ on the $x$-plane, and its image ${\cal A}_{j,i}$ on $\curve$.}}
 \label{cut}
\end{figure}
\begin{figure}[h]
 \centering
  \includegraphics[width=16cm]{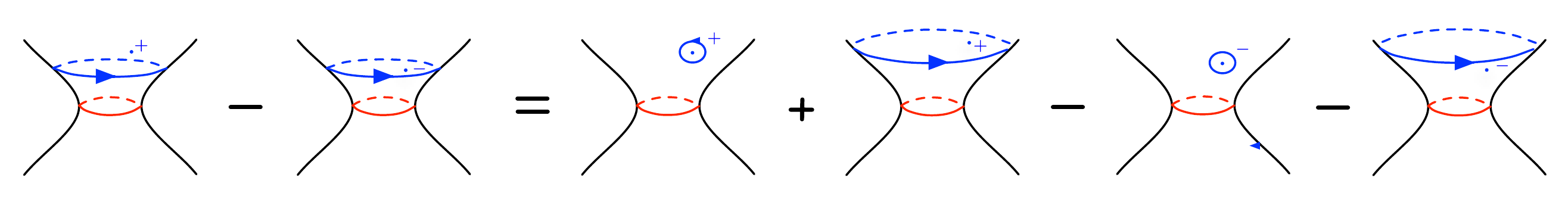}
 \caption{\footnotesize{The preimage of the points $x_+$ and $x_-$ of (\ref{disc}) are depicted as dots in the above diagram, $\widehat{{\cal A}}_{j,i}$ is given by the blue contour, and the preimage of the cut is drawn in red. To take the limit $x_+ \rightarrow x_-$, one must first shift the contours. The second and fourth term on the RHS of the above diagrammatic equation then cancel, yielding the RHS of (\ref{rho}).}}
 \label{discontinuity}
\end{figure}

The definition (\ref{matrix_ensemble}) of the matrix ensemble ${\cal E}_i$ is the condition that there are $n_{j,i}$ eigenvalues of $M_i$ on the contour $\gamma_{j,i}$, hence corresponds to imposing the filling fraction conditions 
\beq
\frac{1}{2\pi i}\oint_{{\cal A}_{j,i}} c_{i-1}x_{i-1} dx_i = g_s\,\, n_{j,i}
\eeq
for $i=1,\dots,L$, $j=1,\dots,k_i$.
\end{enumerate}

\bigskip

In our experience,  the conditions enumerated above have a unique solution and define a unique spectral curve. As emphasized in the introduction, a formal uniqueness proof is however still lacking.

The spectral curve is defined as the data of the Riemann surface $\curve$, and the two functions $x_1(z)$ and $x_2(z)$,
\beqn  \label{spectral_curve}
\encadremath{
\spcurve_{MM} = (\curve,x_1,x_2).
}\eeqn

\subsection*{3.3 Symplectic invariants of a spectral curve}
\label{secdefspinv}

Once we have found the spectral curve $\spcurve_{MM}$ of our matrix model, we can compute the coefficients $F_g$ in the topological expansion of its partition function,
\beq
\ln Z= \sum_{g=0}^\infty g_s^{2g-2} F_g  \,,
\eeq
by computing the symplectic invariants of this curve,
\beq
F_g=F_g(\spcurve_{MM}) \,,
\eeq
 following \cite{Chain}.
\smallskip

Let us recall the definition of these invariants for an arbitrary spectral curve $\spcurve$. 
\bigskip

Let $\spcurve=(\curve,x,y)$ be a spectral curve, comprised of the data of a Riemann surface $\curve$ and two functions $x(z),\, y(z): \curve \rightarrow \IC$, meromorphic on $\curve$ away from a finite set of points (we wish to allow logarithms).\footnote{In fact, the most general setting in which this formalism is valid has not yet been established. We will state it within the generality we need here i.e. we assume that $dx$ is meromorphic forms on $\curve$  (this allows $x$ and $y$ to have logarithms).} We will assume that $dx$ is a meromorphic form on all of $\curve$.

\subsubsection*{Branchpoints}

\begin{figure}[h]
 \centering
  \includegraphics[width=6cm]{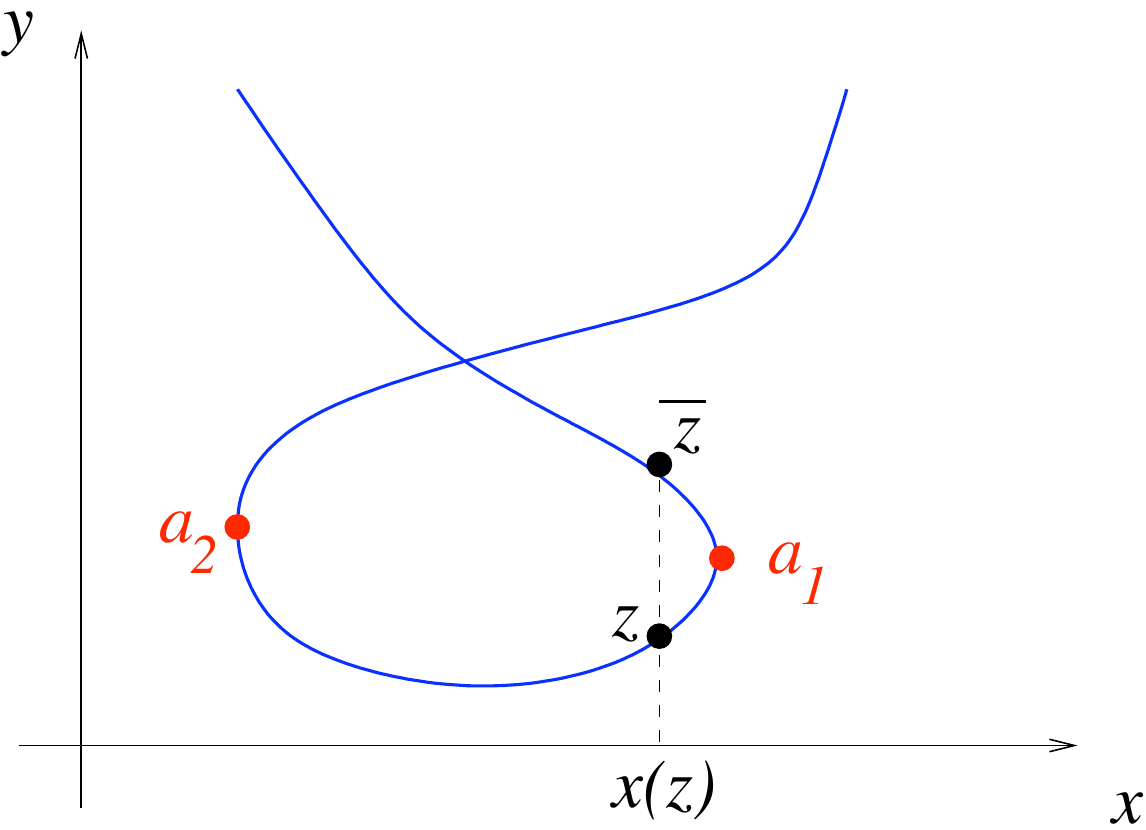}
 \caption{\footnotesize{At a regular branch point $a\in \curve$ of $x$, $y$ as a function of $x$ has a branchcut $y\sim y(a) + C \sqrt{x-x(a)}$. If $z$ is a point on one branch near $a$, we call $\bar z$ the conjugate point on the other branch; it has the same $x$ projection, $x(\bar z)=x(z)$. Notice that $\bar z$ is defined only locally near branch points. If we follow $z$ from $a_1$ to $a_2$, $\bar z$ may have to jump from one branch to another.
 }}
 \label{figspcurvebpzzbar}
\end{figure}

Let $a_i$ be the branch points of the function $x$,
\beq
dx(a_i)=0.
\eeq
We assume that all branch points are simple, i.e. that $dx$ has a simple zero at $a_i$. This implies that in the vicinity of $a_i$, the map $x$ is $2:1$. We introduce the notation $\bar z\neq z$ such that
\beq
x(\bar z)=x(z).
\eeq
$\bar z$ is called the conjugate point to $z$, and it is defined only in the vicinity of branch points, as depicted in figure \ref{figspcurvebpzzbar}.

We also require that the branch points of $x$ and $y$ do not coincide, such that $dy(a_i)\neq 0$ and $y(z)$ therefore has a square-root branchcut as a function of $x$ at $x(a_i)$. If $y$ is finite at $a_i$, its local behavior is hence given by
\beq
y(z) \sim y(a_i)  + C_i\sqrt{x(z)-x(a_i)} \,.
\eeq
If $a_i$ corresponds to a hard edge, we require $y$ to have a pole here. Its local behavior is hence given by 
\beq
y(z) \sim  {C_i\over \sqrt{x(z)-x(a_i)}} \,.
\eeq

\subsubsection*{Bergman kernel}

On a curve $\curve$, there exists a unique symmetric 2-form $B(z_1,z_2)$ with a double pole on the diagonal $z_1=z_2$ and no other poles, with the following normalization on ${\cal A}$-cycles,
\beq
\oint_{z_2\in{\cal A}_{j,i}} B(z_1,z_2)=0.
\eeq
In any local coordinate near $z_1=z_2$, one has 
\beq
B(z_1,z_2) \sim {dz_1\,dz_2\over (z_1-z_2)^2} + {\rm regular}  \,.
\eeq 
$B$ is called the Bergman kernel of $\curve$, or the fundamental 2-form of the second kind \cite{Fay}.

\subsubsection*{Recursion kernel}

We now define the recursion kernel $K$ as
\beq
K(z_0,z) = \frac{\int_{\bar z}^z B(z_0,z')}{2(y(\bar z)-y(z))\,dx(z)} \,.
\eeq
This kernel is a globally defined 1-form in the variable $z_0\in \curve$. In the variable $z$, it is the inverse of a 1-form (that means we have to multiply it with a quadratic differential before computing any integral with it); it is defined only locally near branch points of $x$, such that $K(z_0,\bar z)=K(z_0,z)$. At the branch points, it has simple poles,
\beq
K(z_0,z) \sim -{B(z_0,z)\over 2\,dx(z)\,dy(z)} + {\rm regular}.
\eeq

\subsubsection*{Topological recursion}

Correlation forms $W_n^{(g)}(z_1, \ldots, z_n)$ (not to be confused with the resolvents $W_i(z)$ introduced above) are symmetric $n$-forms defined by
\beq
W_1^{(0)}(z) = -y(z)dx(z) \,,
\eeq
\beq
W_2^{(0)}(z_1,z_2) = B(z_1,z_2) \,,
\eeq
and then by recursion (we write collectively $J=\{z_1,\dots,z_n\}$),
\ba
W_{n+1}^{(g)}(z_0,J) 
&=& \sum_i \Res_{z\to a_i} K(z_0,z)\, \Big[ W_{n+2}^{(g-1)}(z,\bar z,J) \cr
&& + \sum_{h=0}^{g}\sum'_{I\subset J} W_{1+|I|}^{(h)}(z,I)\,W_{1+n-|I|}^{(g-h)}(\bar z,J\setminus I) \Big]
\ea
where $\sum'_I$ is the sum over all subsets of $J$, restricted to $(h,I)\neq (0,\emptyset)$ and $(h,I)\neq (g,J)$.

\medskip

Although it is not obvious from the definition, the forms $W_n^{(g)}$ are symmetric.
For $2-2g-n<0$, they are meromorphic $n$-forms with poles only at branch points. These poles are of degree at most $6g-4+2n$, and have vanishing residues.

For the one matrix model, the $W_n^{(g)}$ coincide with the $n$-point function of the trace of the resolvent at order $g$ in the topological expansion.

\subsubsection*{Symplectic invariants}
\label{secdefspinvs}

Finally, for $g\geq 2$, we define the symplectic invariants $F_g$ (also denoted $W_0(g)$ in \cite{OE}) by
\beq
F_g(\spcurve)  = {1\over 2-2g}\, \sum_i \Res_{z\to a_i} \Phi(z)\,W_1^{(g)}(z) \,,
\eeq
where $\Phi$ is any function defined locally near branch points of $x$ such that $d\Phi=y dx$.

The definitions of $F_1$ and $F_0$ are more involved and we refer the reader to \cite{OE}. $F_0$ is called the prepotential, and $F_1$ is closely related to the determinant of the Laplacian on $\curve$ with metrics $|ydx|^2$, see \cite{KK, EKK}. 

\medskip

The $F_g(\spcurve)$'s depend only on the orbit of $\spcurve$ under the group of transformations generated by
\begin{itemize} 
\item[$\mathfrak{R}:$] $\spcurve\mapsto\td\spcurve=(\curve,x,y+R(x))$ where $R(x)$ is any rational function of $x$,
\item[$\mathfrak{F}:$] $\spcurve\mapsto\td\spcurve=(\curve,f(x),y/f'(x))$ where $f(x)$ is an analytical function of $x$, with $f'$ rational, such that $df=f'dx$ has the same number of zeroes as $dx$,
\item[$\mathfrak{S}:$] $\spcurve\mapsto\td\spcurve=(\curve,y,-x)$.
\end{itemize}
These transformations are symplectic, i.e. they leave $dx\wedge dy$ invariant.

\medskip

The symplectic invariants are homogeneous of degree $2-2g$,
\beqn\label{Fghomogeneous}
F_g(\curve,x,\lambda y) = \lambda^{2-2g}\,F_g(\curve,x,y).
\eeqn
In particular, they are invariant under the parity transformation  $F_g(\curve,x,- y) = F_g(\curve,x,y)$.

\section*{5 The spectral curve for the topological string's matrix model} \label{the_spectral_curve}

Applying the procedure outlined in section \ref{secspcurvegenchain} to our matrix model, we will determine a spectral curve $\spcurve_{MM}(\CYX_0)$ in this section. \cite{Chain} demonstrated that for a chain of matrices, we have
\beq
\ln {Z} = \sum_g g_s^{2g-2} F_g(\spcurveMM) \,,
\eeq
with $F_g$ the symplectic invariants of \cite{OE}. In our case, since we have engineered our matrix model to yield\footnote{As we have here reserved the notation $F_g$ for the symplectic invariants of our matrix model, we refer to the topological string free energies as $GW_g$.} $GW_g(\CYX_0)$ as its partition function, re-computing the partition function via the methods of \cite{Chain} will yield
\beq
GW_g(\CYX_0) = F_g(\spcurveMM)  \,.
\eeq
This relation is already quite interesting, as it allows for explicit computation of the Gromov-Witten invariants. Our goal however will be to go further. We will argue that $\spcurveMM$ is symplectically equivalent to the mirror spectral curve $\spcurve_{\hat \CYX_0}$ of section \ref{secmirrorcurve},
\beq
\spcurveMM \sim \spcurve_{\CYX_0} \,.
\eeq
Since the $F_g$'s are symplectic invariants, this will imply the BKMP conjecture for $\CYX_0$, i.e.
\beq
GW_g(\CYX_0) = F_g(\spcurve_{\CYX_0}).
\eeq

\subsection*{5.1 Applying the chain of matrices rules} \label{applying_the_rules}

We now apply the rules of section \ref{secspcurvegenchain} to the chain of matrices model introduced in section \ref{s_our_matrix_model}.

\medskip
\begin{itemize}
\item Recall that the integration ensembles for the matrices $M_0$ and $M_{m+1}$ are such that for each matrix, all eigenvalues are integrated on the same contour (\ref{outer_contours}). Hence, $k_0=k_{m+1}=1$, and the corresponding filling fractions are equal to $N$.
For $i=1,\dots, m$, the matrix $M_i$ is integrated on $H(\gamma_{0,i}^{d} \times  \gamma_{2,i}^{d} \times \dots \times \gamma_{n,i}^{d})$, where $\gamma_{j,i}$ is a contour which surrounds all points of the form $q^{a_{j,i}+\mathbb N}$. There are thus $k_i=n+1$ filling fractions, each equal to $d$.
The matrices $R_i$ are integrated on $H(\mathbb R_+^N)$. We denote the number of their cuts by $\tilde{k_i}$. Hence, $\td k_i=1$, with the respective filling fraction equal to $N$.

According to condition \ref{genus_ss} of section \ref{secspcurvegenchain}, the genus of the spectral curve $\curve$ is thus given by 
$$
\mathfrak g=\sum_{i=0}^{m+1} (k_i-1)+\sum_{i=1}^{m+1} (\td k_i-1)=nm  \,.
$$

\item Following condition \ref{projections} of section \ref{secspcurvegenchain}, we introduce functions $x_i(z)$, $i=0,\dots,m+1$, associated to the matrices $M_i$, and functions $y_i(z)$, $i=1,\dots,m+1$, associated to the matrices $R_i$, as well as two additional functions $y_0(z)$ and $y_{m+2}(z)$ at the ends of the chain.

They must satisfy the following requirements:
\begin{itemize}

\item Since there is no potential for the matrices $R_i$, equation (\ref{eqspcurvechainxxV}) implies that we have, for $i=1,\dots,m+1$,
\beq
x_{i}(z)-x_{i-1}(z) = 0  \,.
\eeq
We can hence suppress the index $i$ on these functions, $x(z)=x_i(z)$.

\item For $i=1,\dots,m$, equation (\ref{eqspcurvechainxxV}) gives
\beq
y_i(z)-y_{i+1}(z) 
=  2 V'_{\vec a_i}(x(z))-V'_{\vec a_{i+1}}(x(z))-V'_{\vec a_{i-1}}(x(z))   
 - g_s {f'_i(x(z))\over f_i(x(z))} -{g_s \, S_i+i\pi\over x(z)}
\eeq
and
\beq
y_0(z)-y_{1}(z) 
=  V'_{\vec a_0}(x(z))-V'_{\vec a_{1}}(x(z))   - g_s {f'_0(x(z))\over f_0(x(z))} \,,
\eeq
\beq
y_{m+1}(z) -y_{m+2}(z)
=  V'_{\vec a_{m+1}}(x(z))-V'_{\vec a_{m}}(x(z))  - g_s {f'_{m+1}(x(z))\over f_{m+1}(x(z))} \,.
\eeq

More explicitly, in terms of the function $$\psi_q(x)=x g'(x)/g(x)\,,$$ whose small $g_s$ expansion 
\ba
\psi_q(x) &=& -\,{1\over \ln q}\,\sum_{n=0}^\infty {(-1)^n\,B_n\over n!}\,(\ln q)^n\,\, \Li_{1-n}(1/x) \\
&=&  {1\over \ln{q}}\,\Big[
\ln{(1-{1\over x})} -{\ln q\over 2(x-1)}
-\sum_{n=1}^\infty {B_{2n}\over (2n)!}\,\, (\ln q)^{2n}\,\Li_{1-2n}(x) \Big] \,.
\ea
we worked out in appendix A of \cite{topstring1}, we obtain
\ban
\lefteqn{x(z)(y_{i+1}(z)-y_{i}(z) )}  \nn \\
&=& i\pi+ g_s S_i - g_s \sum_j (2\psi_q(q^{a_{j,i}}/x(z)) - \psi_q(q^{a_{j,i+1}}/x(z))-\psi_q(q^{a_{j,i-1}}/x(z))) \nn \\
&& + g_s {x(z)f'_i(x(z))\over f_i(x(z))} \,,    \label{diff_ys}
\ean
as well as
\ba
x(z)(y_1(z)-y_{0}(z) )
&=& -g_s \sum_j \psi_q(q^{a_{j,0}}/x(z))+g_s\sum_j\psi_q(q^{a_{j,1}}/x(z))  \cr
&& -g_s\sum_j\sum_{k=0}^{d-1} \frac{x(z)}{x(z)-q^{a_{j,0}+k}} \,,
\ea
\ba
x(z)(y_{m+2}(z) -y_{m+1}(z))
&=&  -g_s \sum_j \psi_q(q^{a_{j,m+1}}/x(z))+g_s\sum_j\psi_q(q^{a_{j,m}}/x(z)) \cr
&& -g_s\sum_j\sum_{k=0}^{d-1} \frac{x(z)}{x(z)-q^{a_{j,m+1}+k}}
\ea
Note that we have explicitly used the fact that the partitions $\alpha_{j,m+1}$ and $\alpha_{j,0}$ are chosen to be trivial.\\

\item Since the integral over $R_i$ is over $H_N(\mathbb R_+)$, i.e. its eigenvalues are integrated on $\mathbb R_+$, the integration contour has an endpoint (hard edge) at $y_i=0$. Condition \ref{hard_edges} hence requires that at a pre-image $y_i^{-1}(0)$, which we will refer to as $\infty_i$, the following holds
\beq
y_i(\infty_i)=0\, , \quad dy_i(\infty_i)=0\, , \quad x(z){\rm \,\, has\,a \,simple\,pole\,at\,}z=\infty_i \,.
\eeq

Furthermore, introducing a local parameter $z$ in the neighborhood of $\infty_i$, the above translates into 
\beq
y_i(z) \sim z^2 \,, \quad x(z) \sim 1/z  \,.
\eeq
Hence, $\forall \,i=1,\dots,m+1$,
\beq
y_i \sim \cO(1/x^2) \,.
\eeq
\end{itemize}

\item
The relations (\ref{diff_ys}) imply that near $\infty_i$, we have
\beq
x (y_{j+1}-y_j) \mathop{{\sim}}_{z\to\infty_i} i\pi + g_s S_j+ g_s\sum_{l=0}^n (2a_{l,j}-a_{l,j+1}-a_{l,j-1}) + O(1/x) \,.
\eeq
In particular, it follows that $\infty_j\neq \infty_i$. Thus, all points $\{\infty_1,\dots,\infty_{m+1}\}\subset x^{-1}(\infty)$ are distinct , i.e. condition \ref{hard_edges} requires that $x^{-1}(\infty)$ have at least $m+1$ points. 

We will make the minimal assumption that $x^{-1}(\infty)$ has exactly
$m+1$ elements that are simple poles of $x$, and that $x$ has no further singularities, i.e. that $x$ is a meromorphic function of degree  $m+1$ on $\curve$. 

\item By condition \ref{filling_fraction}, since for $i=0,\dots,m+2$ there are $d$ eigenvalues of $M_i$ of the form $q^{a_{j,i}+\mathbb N}$ surrounded by the path $\widehat{\cal A}_{j,i}$ , we have the $(m+2)\times (n+1)$ filling fraction conditions
\beq
{1\over 2i\pi}\oint_{{\cal A}_{j,i}} y_i\,dx = d\, g_s \qquad \mathrm{for}\,\, i=0,\dots,m+1, \,\,\, j=0,\dots,n \,.
\eeq
\end{itemize}

\begin{figure}[!t]
 \centering
  \includegraphics[width=6cm]{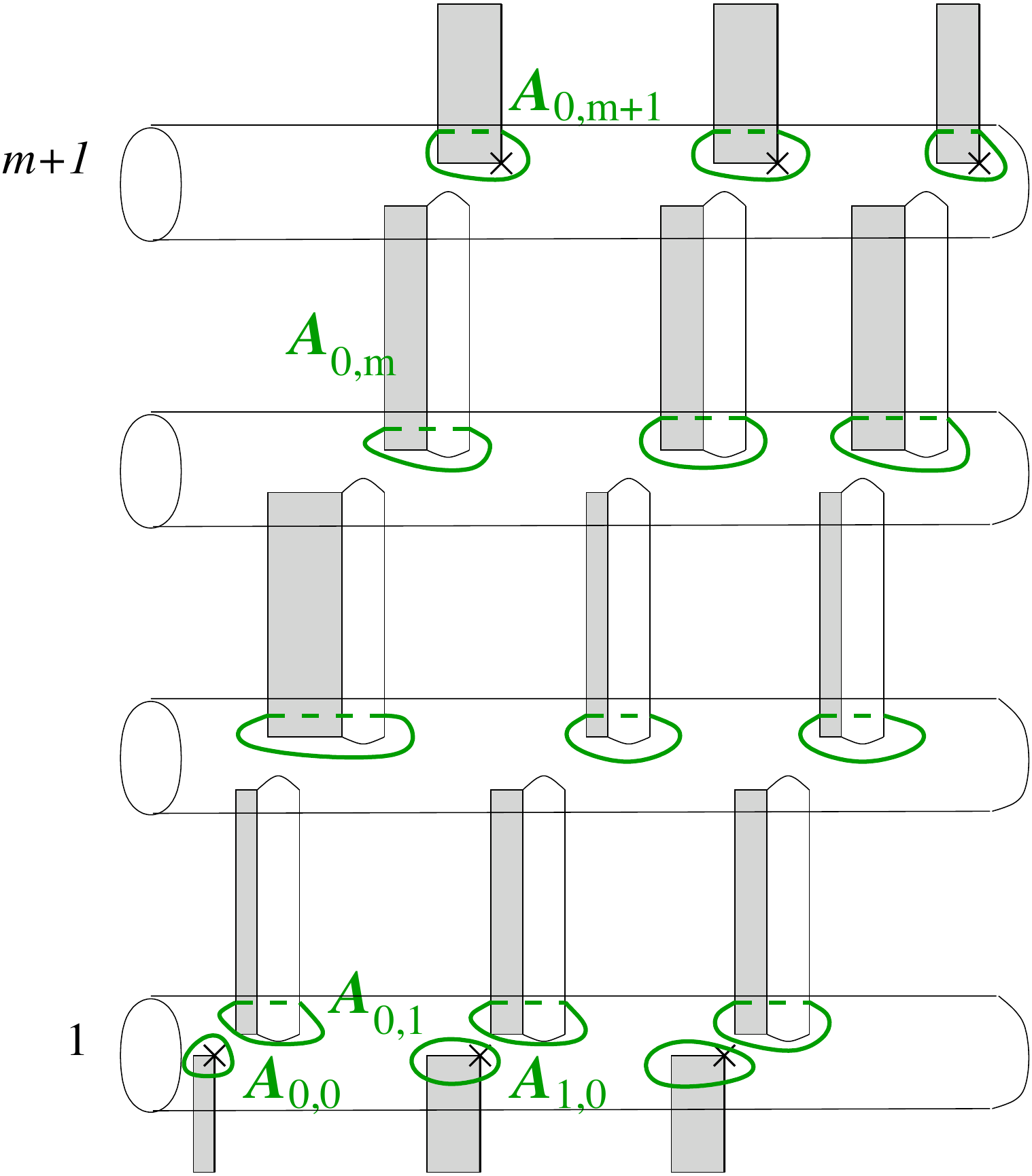}
 \caption{\footnotesize{The spectral curve of our matrix model can be represented as follows. The cover of $\IC \IP^1$ provided by $x$ has $m+1$ sheets. Instead of the projective plane of $x$, we represent the sheets of $\ln x$, which are cylinders.
 Cycles ${\cal A}_{j,i}$ appear in sheets $i-1$ and $i$. They enclose singularities of the resolvent $W_i$. Algebraic cuts are represented as vertical cylinders, and poles and log singularities are represented as grey strips. There is only one cycle ${\cal A}_{j,0}$ (which is in sheet $0$) and one ${\cal A}_{j,m+1}$ (in sheet $m$), and they enclose only poles or log singularities of $y_0$ resp. $y_{m+1}$.
 }}
 \label{figspcurve1}
\end{figure}

$x$ hence defines an $m+1$ sheeted cover of $\IC \IP^1$. Considering the function $\ln x$ instead, with singularities at $x=0$ and $x = \infty$, each sheet of this cover is mapped to a cylinder. We have depicted this covering in figure \ref{figspcurve1}, and indicated the singularities of $y_i$ on each sheet: algebraic cuts are represented by vertical cylinders, and poles and logarithmic cuts by grey strips.

In sheet $i$ we have represented some contours ${\cal A}_{j,i}$ whose image under the projection $x:\curve \to \mathbb C \IP^1$ surrounds all points of type $q^{a_{j,i}+\mathbb N}$. 

For $i=1,\dots,m$, the resolvent $W_i(x)$ of the $i^{\rm th}$ matrix $M_i$ is computed as a contour integral around the sum over $j$ of cycles ${\cal A}_{j,i}$ on sheet $i$,
\beq
W_i(x) = \sum_{j=0}^n\, {1\over 2i\pi}\oint_{{\cal A}_{j,i}}\, {y_i(z')dx(z')\over x-x(z')}  \,.
\eeq

Also, as argued in \cite{topstring1}, the potentials of $M_0$ and $M_{m+1}$ are such that in fact the matrices $M_0$ and $M_{m+1}$ are frozen, and thus their resolvents contain only poles.
In terms of the functions $y_0$ and $y_{m+1}$, we conclude that the singularities of $y_0$ in ${\cal A}_{j,0}$ in sheet $1$ and the singularities of $y_{m+1}$ in ${\cal A}_{j,m+1}$ in sheet $m+1$ can be only poles, not cuts.

Since condition \ref{genus_ss} requires that the genus be $\mathfrak g=nm$, we see that there can be no other cuts than the ones already discussed -- the genus would be higher, otherwise.

\subsection*{5.2 Symplectic change of functions}

The spectral curve of the matrix model is $\spcurve_{MM} = (\curve,x,y_0)$, and our goal is to relate it to the mirror curve described in section \ref{secmirrorcurve}. The mirror curve is described via the algebraic equation (\ref{mirror_curve}) in the two functions $x_1,x_2: \curve_{mirror} \rightarrow \IC \IP^1$  (in the patch $x_0=1$). We wish to obtain a similar algebraic description of $\curve$. Due to log singularities in $y_0$, to be traced to the small $g_s$ behavior of $\psi_q(x)$, an algebraic equation in the variables $(x,y_0)$ cannot exist (recall that $x$ is meromorphic). In this section, we shall, via a series of symplectic transformations on the $y_i$ of the type enumerated in section \ref{secdefspinvs}, arrive at functions $Y_i$ that are meromorphic on $\curve$, and hence each present a viable candidate to pair with $x$ to yield an algebraic equation for $\curve$.

\bigskip

Essentially, we wish to introduce the exponentials of $y_i$. While this will eliminate the log singularities, poles in $y_i$ would be elevated to essential singularities. We hence first turn to the question of eliminating these poles.

\subsubsection*{The arctic circle property} \label{arctic}
On the physical sheet, the interpretation of a pole of $y_i$ is as an eigenvalue of the matrix $M_i$ with delta function support. Such a so-called frozen eigenvalue can arise in the following way:

The sum over all partitions is dominated by partitions close to a typical equilibrium partition, i.e. a saddle point. The typical partition has a certain typical length referred to as its equilibrium length $\bar{n}$. All partitions with a length very different from the equilibrium length contribute only in an exponentially small way (and thus non-perturbatively) to the full partition function. Introducing a cutoff on the length of partitions which is larger than the equilibrium length hence does not change the perturbative part of the partition function. Now recall that when we defined the $h_i(\gamma)$ of a representation $\gamma$ in appendix \ref{our_matrix_model}, we introduced an arbitrary maximal length $d$ such that $l(\gamma)\leq d$ and set
\beq
h_i(\gamma) = a_\gamma + d -i + \gamma_i  \,.
\eeq
Setting $\gamma_i =0$ for $d \ge i > \bar{n}$ yields $h_i$ that do not depend on the integration variables, hence are frozen at fixed values. This behavior is referred to as the arctic circle property \cite{JohanssonAnn}, as all eigenvalues beyond the arctic circle situated at equilibrium length $\bar{n}$ are frozen.

Returning to our matrix model, the eigenvalues of the matrices $M_i$ are given by $q^{(h_{j,i})_l}$. For $d \ge l > n_{j,i}$, they are frozen, and thus contribute poles to $y_i$ by (\ref{resolvent})  (recall that poles of the resolvent correspond to eigenvalues with delta function support) in the physical sheet. We will assume that these are the only poles in the physical sheet and we subtract them to obtain new functions $\td y_i$,
\ba
\td y_0(z) =x(z)y_0(z)- \sum_j \sum_{k=0}^{d-1} {g_sx(z)\over x(z)-q^{a_{j,0}+k}} \,,
\ea
\ba
\td y_{m+2}(z) = x(z)y_{m+2}(z)+ \sum_j \sum_{0}^{d-1} {g_s x(z)\over x(z)-q^{a_{j,m+1}+k}}
\ea
and for $i=1,\dots,m+1$,
\ba
\td y_i(z)=x(z)y_i(z) -\sum_j \sum_{k=0}^{d-n_{j,i}-1} {g_sx(z)\over x(z)-q^{a_{j,i}+k}} + \sum_j \sum_{k=0}^{d-n_{j,i-1}-1} {g_s x(z)\over x(z)-q^{a_{j,i-1}+k}}  \,.
\ea
We have set
\beq
n_{j,0}=0
\,\, , \quad
n_{j,m+1}=0 \,.
\eeq
Notice that at large $x(z)$ in sheet $i$ we have
\beq
\td y_0 \sim  O(1/x(z))
\,\,\, , \qquad
\td y_{m+2} \sim  O(1/x(z))
\eeq
and for $i=1,\dots,m+1$
\beq
\td y_i \sim   g_s \sum_j (n_{j,i}-n_{j,i-1}) + O(1/x(z)).
\eeq

\medskip

As a general property of $\psi_q$, we have for any integer $n_{j,i}\leq d$ 
\beq
\psi_q(q^{a_{j,i}}/x) = \psi_q(q^{a_{j,i}+d-n_{j,i}}/x)+ \sum_{k=0}^{d-n_{j,i}-1} {x\over x-q^{a_{j,i}+k}}   \,.
\eeq
Hence, the loop equations for the new functions $\tilde{y}_i$ read
\ba
\td y_{i+1}(z)-\td y_{i}(z) 
&=& i\pi\, + g_s \, S_i  \cr
&& + g_s \sum_j (2\psi_q(q^{a_{j,i}+d-n_{j,i}}/x(z)) - \psi_q(q^{a_{j,i+1}+d-n_{j,i+1}}/x(z)) \cr
&& -\psi_q(q^{a_{j,i-1}+d-n_{j,i-1}}/x(z)))  + g_s\, {x(z) f_i'(x(z))\over f_i(x(z))}   \,,\cr
\td y_1(z)-\td y_{0}(z) 
&=& g_s \sum_j \psi_q(q^{a_{j,0}+d}/x(z))-g_s\sum_j\psi_q(q^{a_{j,1}+d-n_{j,1}}/x(z)) \,,
\cr
\td y_{m+2}(z) -\td y_{m+1}(z)
&=&  g_s \sum_j \psi_q(q^{a_{j,m+1}+d}/x(z))-g_s\sum_j\psi_q(q^{a_{j,m}+d-n_{j,m}}/x(z)) \,.
\ea

\medskip

\begin{figure}[t]
 \centering
  \includegraphics[width=14cm]{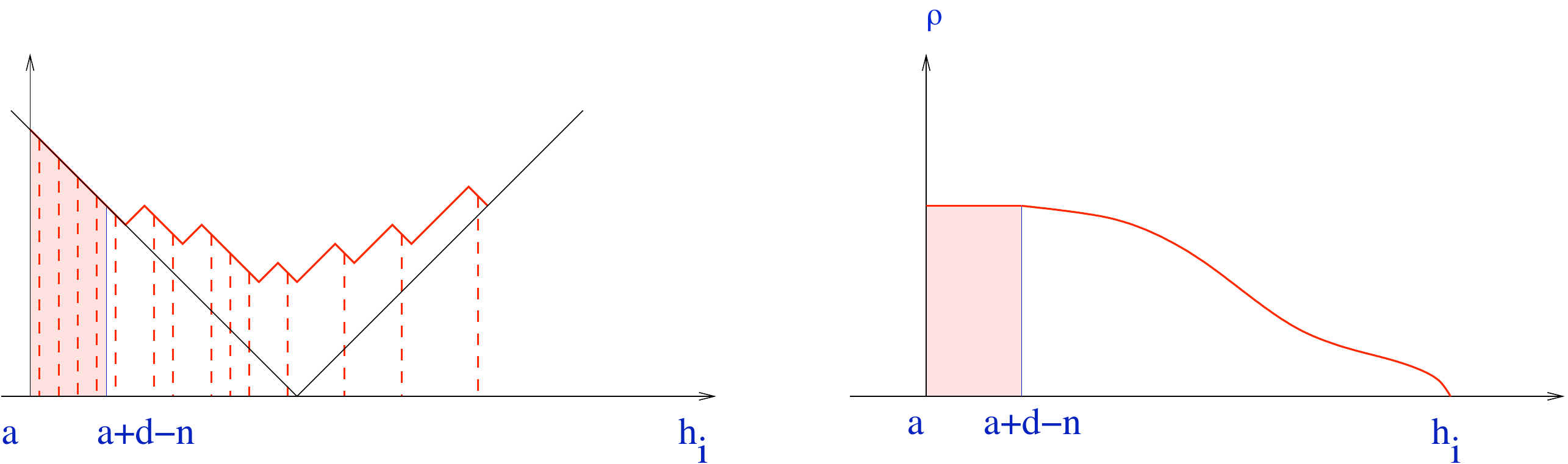}
 \caption{\footnotesize{We shift the cut-off $d$ on the representation lengths, $d\to n_{j,i}$, with $n_{j,i}$ chosen such that frozen eigenvalues in the expected distribution of the $h_i$ are suppressed. In the limit of vanishing spacing ($g_s\to 0$), the equidistant frozen eigenvalues give rise to a constant eigenvalue density region.}}
 \label{figctedensityshiftnij}
\end{figure}

The $n_{j,i}$ in the above definitions are defined as the equilibrium lengths, i.e. by the property that the functions $\tilde{y}_i$ have no poles on their physical sheet. That such a choice of $n_{j,i}$ exists is suggested by the arctic circle property.

Note that the $n_{j,i}$ can also be specified by the fact that $q^{a_{j,i} + d - n_{j,i}}$ be the beginning of the cut encircled by $\gamma_{j,i}$. As we have identified the discontinuities of $y_i$ to lie across branchcuts of $x$, this implies that $x$ has ramification points at the element of $x^{-1}(q^{a_{j,i} + d - n_{j,i}})$ lying on the physical sheets of $y_i$.

Note that the arctic circle property also implies the perturbative independence of our expressions from the arbitrary cut-off $d$. Changing $d$ to $d+d'$ merely introduces $d'$ new frozen eigenvalues $h_i$. This independence from $d$ is important in establishing the equality between the topological string partition function and our matrix integral (\ref{m_integral}), as the topological vertex formulae in fact are formulated in the limit $d\rightarrow \infty$.

\subsubsection*{Obtaining globally meromorphic functions}

We have arrived at functions $\tilde{y_i}$ that have no poles on their physical sheet, and are thus safely exponentiated there. We wish now to use the loop equations to obtain functions which are globally well-behaved. 

To this end, we note that since the Gromov-Witten invariants are defined as a formal power series in $g_s$, we can compute the spectral curve order by order in $g_s$, invoking the following small $\ln q$ expansion \cite{topstring1}:
\ba
\psi_q(q^{a_{j,i}+d-n_{j,i}}/x)
&\sim& -\frac{1}{g_s} \ln(1-\frac{x}{q^{a_{j,i}+d-n_{j,i}}}) +\frac{x}{2(x-q^{a_{j,i}+d-n_{j,i}})} \cr
&& +\frac{1}{g_s}\sum_{n=1}^\infty {B_{2n}\,g_s^{2n}\over (2n)!}\,\Li_{1-2n}(q^{d-n_{j,i}+a_{j,i}}/x)  \,.
\ea
The functions $f'_i/f_i$ are completely non-perturbative; one can
easily check with the above expansion that they can be replaced by $0$ to every order in $g_s$.

Introducing new functions $X(z)$ and $Y_i$ by the formulae
\beq
x(z) = q^d\, X(z) \,,
\eeq
\beq
\td y_0(z) = \ln{Y_0(z)} \,,
\eeq
\beq
\td y_{m+2}(z) = \ln{Y_{m+2}(z)} \,,
\eeq
and for $i=1,\dots,m+1$
\ba
\td y_i(z) 
&=& \ln{Y_i(z)} + \sum_j{X(z) g_s\over 2(X(z)-q^{a_{j,i}-n_{j,i}})} +\frac{1}{g_s} \sum_j \sum_{n=1}^\infty {B_{2n}\, g_s^{2n}\over (2n)!}\, \Li_{1-2n}(q^{a_{j,i}-n_{j,i}}/X(z)) \cr
&& -  \sum_j{X(z) g_s\over 2(X(z)-q^{a_{j,i-1}-n_{j,i-1}})} - \frac{1}{g_s}\sum_j \sum_{n=1}^\infty {B_{2n}\, g_s^{2n}\over (2n)!}\, \Li_{1-2n}(q^{a_{j,i-1}-n_{j,i-1}}/X(z))
\ea

yields loop equations that are algebraic on their right hand side,

\ban \label{ratios} 
\frac{Y_i}{Y_{i+1}}
&=& -\,
e^{-g_s \,S_i}\prod_j \frac {(X-q^{a_{j,i+1}-n_{j,i+1}})(X-q^{a_{j,i-1}-n_{j,i-1}})}{(X-q^{a_{j,i}-n_{j,i}})^2} \cr
&& \qquad \prod_j q^{2(a_{j,i}-n_{j,i})-(a_{j,i+1}-n_{j,i+1})-(a_{j,i-1}-n_{j,i-1})} \,,\cr
\frac{Y_0}{Y_1} &=&\prod_j \frac {(X-q^{a_{j,1}-n_{j,1}})}{(X-q^{a_{j,0}})} \prod_j q^{a_{j,0}-(a_{j,1}-n_{j,1})} \,,\cr
\frac{Y_{m+1}}{Y_{m+2}} &=&\prod_j \frac {(X-q^{a_{j,m}-n_{j,m}})}{(X-q^{a_{j,m+1}})} \prod_j q^{a_{j,m+1}-(a_{j,m}-n_{j,m})} \,,
\ean
i.e.
\ba
{Y_i\over Y_0} 
= e^{g_s(S_1+\dots+S_{i-1})}\,\prod_j q^{(a_{j,i}-n_{j,i})-(a_{j,i-1}-n_{j,i-1})}\, \prod_j {X-q^{a_{j,i-1}-n_{j,i-1}}\over X-q^{a_{j,i}-n_{j,i}}}  \,.
\ea
Since we have argued that the $Y_i$ are holomorphic on their physical sheet, and the ratio $Y_i/Y_{i+1}$ is purely algebraic, we conclude that the $Y_i$ are meromorphic functions on all of $\curve$. This was the goal we had set out to achieve.

Note that the above changes of variables have modified the asymptotics at infinity and the integrals over the $\cal{A}$-cycles. More precisely, we have
\beqn \label{asymptotics} 
\forall i \in [1,m+1]: \, \ln Y_i \mathop{\sim}_{\infty_i}  \td y_i \mathop{\sim}_{\infty_i} g_s\sum_j (n_{j,i}-n_{j,i-1}) +O(\frac{1}{x})\,,
\eeqn
\beq \ln Y_0 = \td y_0 \mathop{\sim}_{\infty} O(\frac{1}{x})\,,
\eeq
\beq \ln Y_{m+2} = \td y_{m+2} \mathop{\sim}_{\infty} O(\frac{1}{x}) \,.
\eeq
The filling fraction equation reads
\beqn \label{integrals} \frac{1}{2i\pi} \int_{{\cal{A}}_{j,i}} \frac{\td y_i(z)}{x(z)}dx(z)=g_s(d-(d-n_{j,i}))= g_s\,n_{j,i} \,.
\eeqn
In terms of $Y_0$, these conditions can be rewritten as
\beq
\ln Y_0 \mathop{\sim}_{\infty_i}   -g_s(S_1+\dots+S_{i-1}) + g_s\sum_{j=0}^n (a_{j,i}-a_{j,i-1}) +O(\frac{1}{x})
\eeq
and
\ba
\frac{1}{2i\pi} \int_{{\cal{A}}_{j,i}} \ln {Y_0}\, \frac {dX}{X}
&=& \frac{1}{2i\pi} \int_{{\cal{A}}_{j,i}} \ln {Y_i}\, \frac {dX}{X} + \frac{1}{2i\pi} \int_{{\cal{A}}_{j,i}} \ln {X}\,\, d \ln \left(\frac{Y_i}{Y_0} \right) \cr
&=& \frac{1}{2i\pi} \int_{{\cal{A}}_{j,i}} \ln {Y_i}\, \frac {dX}{X} + \frac{1}{2i\pi} \int_{{\cal{A}}_{j,i}} \ln {X}\,\, d \left( \sum_{k=1}^i \ln \frac{Y_k}{Y_{k-1}} \right) \cr
&=& \frac{1}{2i\pi} \int_{{\cal{A}}_{j,i}} \ln {Y_i}\, \frac {dX}{X} - \frac{1}{2i\pi} \int_{{\cal{A}}_{j,i}} \ln {X}\,\,\left( \sum_l {dX\over X-q^{a_{l,i}-n_{l,i}}} - {dX\over X-q^{a_{l,i-1}-n_{l,i-1}}}\right) \cr
&=& \frac{1}{2i\pi} \int_{{\cal{A}}_{j,i}} \ln {Y_i}\, \frac {dX}{X} + g_s(a_{j,i}-n_{j,i}) \cr
&=& g_s n_{j,i}+ g_s\,(a_{j,i}-n_{j,i}) \cr
&=& g_s a_{j,i}  \,.
\ea

\subsection*{5.3 Recovering the mirror curve}

We have argued above that $X$ and $Y_i$, and hence in particular
$Y_0$, are meromorphic functions on $\curve$. There must hence exist a
polynomial $H(X,Y)$ such that (see e.g. Theorem 5.8.1 in \cite{Jost})
\beq
H(X,Y_0)=0  \,.
\eeq
The facts that $X$ provides an $m+1$ sheeted cover of $\IC \IP^1$ and that $Y_0$ may have $n+1$ poles in its physical sheet imply that the polynomial $H$ has degrees at least $(n+1,m+1)$. As above, we shall choose the minimal hypothesis that it has exactly these degrees. Thus,
\beqn  \label{mirror_curve_H}
H(X,Y) = \sum_{i=0}^{m+1}\sum_{j=0}^{n+1} H_{i,j} X^j \, Y^i.
\eeqn
As we saw in section \ref{secmirrorcurve}, projectivizing a generic polynomial of these degrees (yielding a homogeneous polynomial of degree $m+n+2$) indeed gives rise to a curve of genus ${\mathfrak g}=nm$.

We now need to determine the $(n+2)(m+2)-1$ unknown coefficients of $H$ ($H$ is defined up to a global multiplicative constant).

\bigskip

The cycle integrals 
\beq
\oint_{{\cal A}_{j,i}} \ln{Y_0}\, \, {dX\over X} = 2i\pi g_s\,\, a_{j,i}
\eeq
provide $(n+1)m$ constraints on the coefficients of $H$.
We also have $m+1$ constraints for the behavior at $\infty_i$, $i=1, \ldots, m+1$,
\beq
\Res_{\infty_i} \ln{Y_0}\, \, {dX\over X} = g_s(S_1+\dots+S_{i-1}) - g_s\sum_{j=0}^n (a_{j,i}-a_{j,i-1})  \,.
\eeq
Finally, requiring that $Y_0$ has poles at $q^{a_{j,0}}$ and $Y_{m+2}$ has zeroes at $q^{a_{j,m+1}}$ gives another $2(n+1)$ constraints, which we may write as
\beq
\Res_{q^{a_{j,0}}} \ln{X}\, \, {dY_0\over Y_0} = g_s a_{j,0} \,,
\eeq
\beq
\Res_{q^{a_{j,m+1}}} \ln{X}\, \, {dY_{m+2}\over Y_{m+2}} = g_s a_{j,m+1}  \,.
\eeq
This gives enough equations to completely determine $H$.
Knowing $H$, we know the location of branch points as functions of $a_{j,i}$'s and $S_i$'s, and can hence determine the $n_{j,i}$ by requiring that $q^{a_{j,i}-n_{j,i}}$ be a branch point.

\medskip
Notice that we can choose to express the period integrals in any linear combination of ${\cal A}$-cycles. In particular,
\beq
\oint_{{\cal A}_{j,i+1}-{\cal A}_{j,i}} \ln{Y_0}\, \, {dX\over X} = 2i\pi g_s\,\, (a_{j,i+1}-a_{j,i}) = 2i\pi\,\, g_s\, t_{j,i}  \,,
\eeq
\beq
\oint_{{\cal A}_{j,i+1}-{\cal A}_{j+1,i}} \ln{Y_0}\, \, {dX\over X} = 2i\pi g_s\,\, (a_{j,i+1}-a_{j+1,i}) = 2i\pi\,\, g_s\, r_{j,i}  \,.
\eeq
Similarly, we may also take linear combinations of ${\cal A}$-cycles together with circles surrounding the poles or zeroes of $x$ in order to get the $s_{j,i}$ classes. We hence conclude that the periods of the curve $H(X,Y_0)=0$ yield the quantum corrected K\"ahler parameters of the fiducial toric geometry $\CYX_0$, allowing us to identify it with the corresponding mirror curve.

\begin{figure}[h]
 \centering
  \includegraphics[width=5cm]{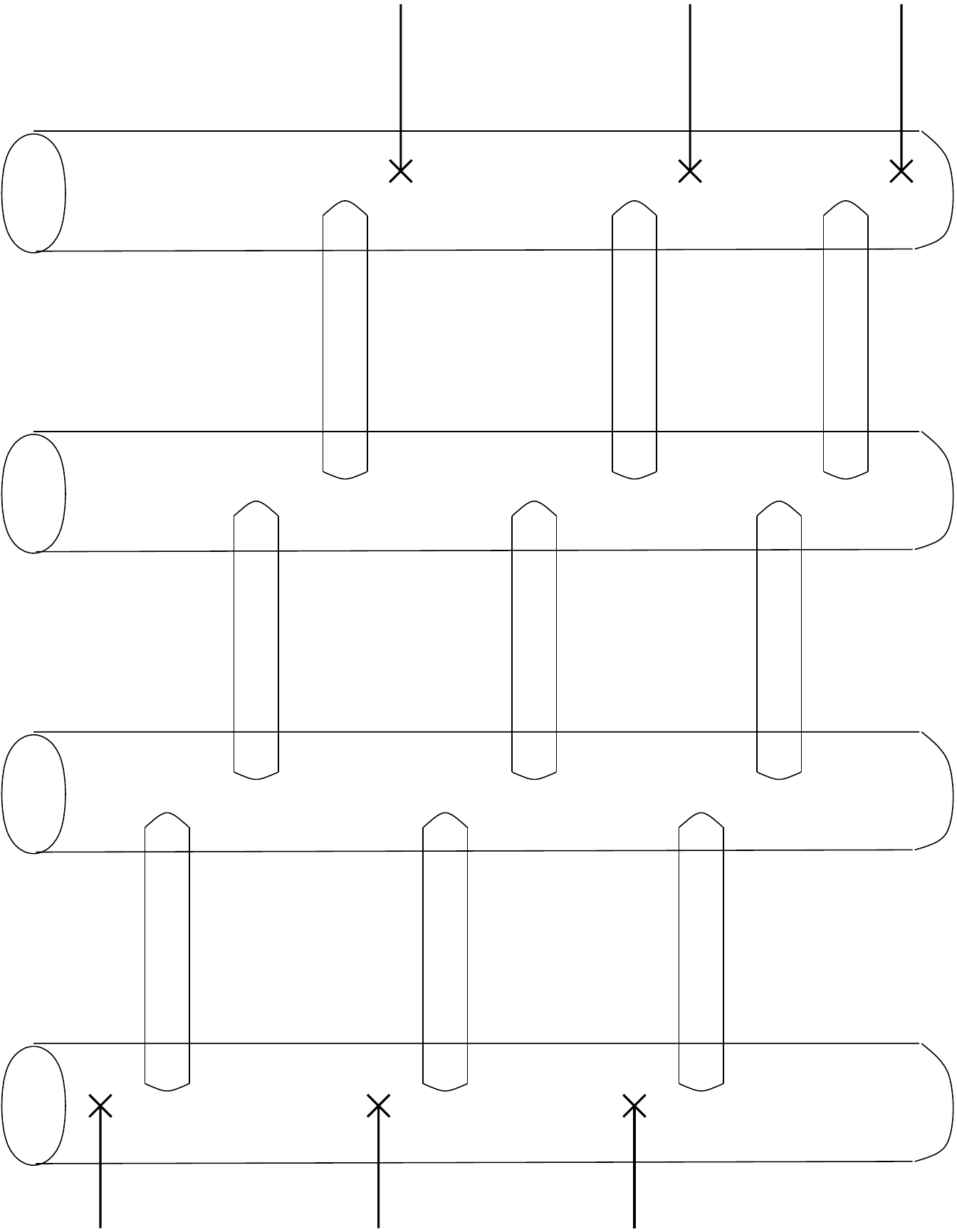}
 \caption{\footnotesize{The spectral curve $(X,{1\over X}\ln{(Y_0)})$ has the following structure:
 $X(z)$ is a meromorphic function of degree $m+1$ on a curve of genus ${\frak{g}}=nm$. Therefore it has $m+1$ poles and $m+1$ zeroes. It provides a branched covering of $\IC \IP^1$. We prefer to represent $\ln X$ instead of $X$, and thus we have $m+1$ copies of the $\ln X$-cylinder. In each sheet there is one zero and one pole of $X$. $Y_0$ is a meromorphic function of degree $n+1$, so that it has $n+1$ zeroes in sheet $0$, and $n+1$ poles in sheet $m+1$.
We recognize the mirror curve ${\cal S}_{\hat\CYX_0}$,  which is a thickening of the toric web diagram.
 }}
 \label{figspcurveMM1}
\end{figure}

\subsection*{5.4 Topological expansion and symplectic invariants}

Following \cite{Chain}, we obtained
\beq
\spcurve_{MM}=(\curve,x,y_0) 
\eeq
as the spectral curve of our matrix model at the end of section \ref{applying_the_rules}.

As reviewed in section \ref{secdefspinv}, we can compute the corresponding symplectic invariants $F_g(\check\spcurve_{MM})$, which assemble to yield the matrix model partition function \cite{Chain},
\beq
\ln Z = \sum_g g_s^{2g-2}\, F_g(\spcurve_{MM}) \,.
\eeq
The symplectic transformation $\mathfrak{R}$ of section \ref{secdefspinvs} maps $(\curve, x, y_0)$ to $(\curve,x,\frac{1}{x}\ln Y_0)$ order by order in $g_s$. $\mathfrak{F}$ maps this to $(\curve,X,{1\over X}\ln Y_0)$, and a second application of $\mathfrak{F}$ yields
\beq
\hat\spcurve_{MM}=(\curve,\ln{X},\ln Y_0) \,.
\eeq
By the symplectic invariance of the $F_g$, we therefore have, order by order in powers of $g_s$,
\beq
F_g(\spcurve_{MM}) = F_g(\hat\spcurve_{MM}) \,.
\eeq
Since our matrix model was engineered to reproduce the Gromov-Witten invariants of $\CYX_0$, we have arrived at
\beq
GW_g(\CYX_0) = F_g(\curve,\ln{X},\ln Y_0)  \,,
\eeq
with $X$ and $Y_0$ obeying the algebraic equation
\beq
H_0(X,Y_0)=0
\eeq
which coincides with the equation (\ref{mirror_curve}) describing the mirror curve of $\CYX_0$.

Given our minimality assumptions on the spectral curve, we have thus derived the BKMP conjecture for the fiducial geometry $\CYX_0$.

\subsection*{5.5 The small $q$ limit and the thickening prescription}
\label{secproofsp}

The above derivation of the spectral curve for the matrix model is not fully rigorous, as we have relied on making minimal assumptions along the way. Although the spectral curve we have found here satisfies all the constraints of section \ref{secspcurvegenchain}, to prove that it is the spectral curve of our matrix model requires a uniqueness result which we currently do not have.

In this section, we provide a heuristic argument that the qualitative behavior of the spectral curve and the mirror curve coincide at small $q$.

\begin{figure}[h]
 \centering
  \includegraphics[width=7cm]{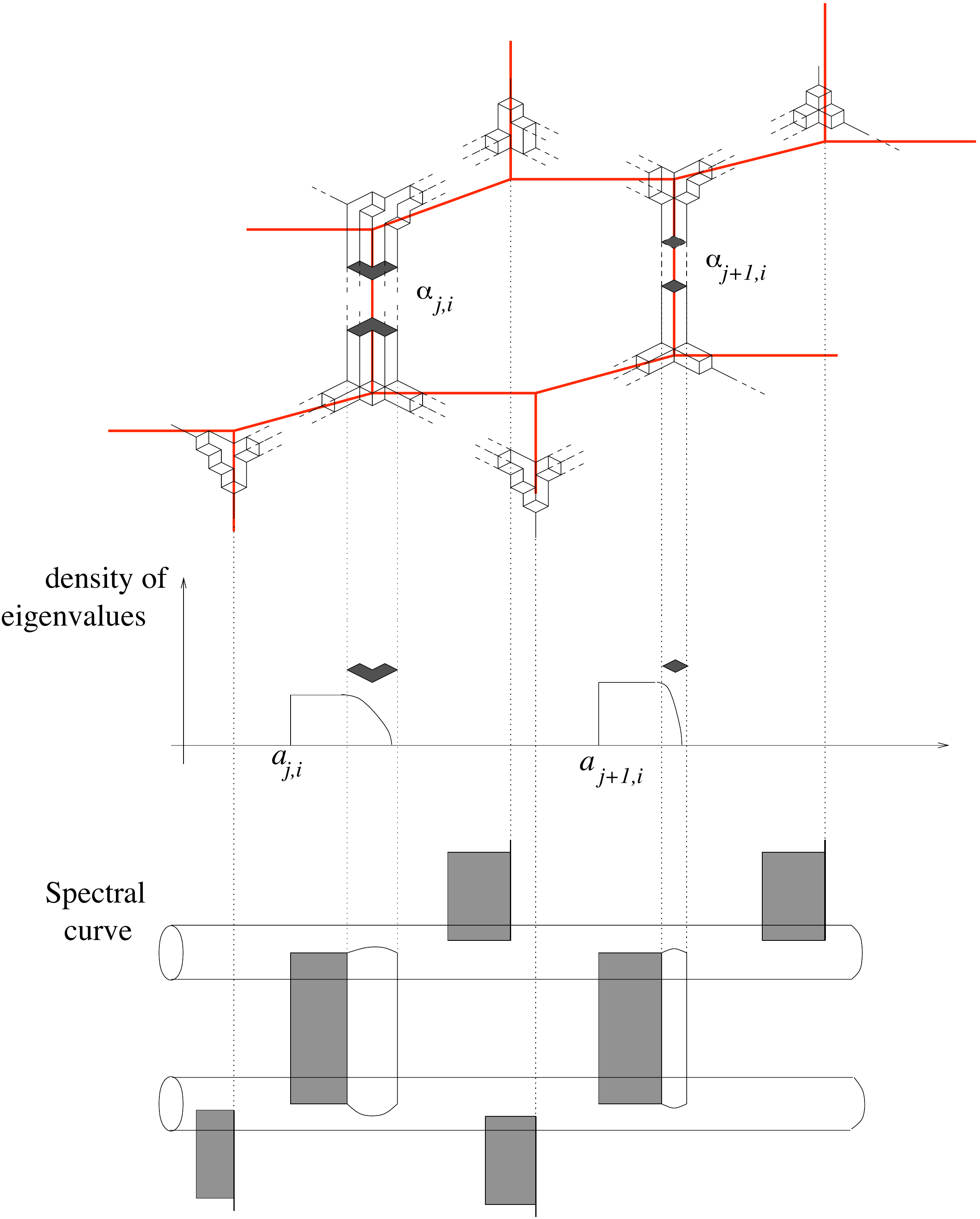}
 \caption{\footnotesize{ In the small $q$ limit, only very small partitions contribute to the matrix integral, therefore the density of eigenvalues of $M_i$ tends to the flat density (a Dirac comb of equidistant delta functions), the non-flat part, which reflects the cuts of the spectral cut, shrinks to zero.}}
 \label{figsmallqpplane}
\end{figure}

At small $q$, only very small partitions contribute to the matrix integral. Almost all eigenvalues of $M_i$ are frozen to the values $q^{a_{j,i}+d-l}$. By the arguments in section \ref{arctic}, the resolvent $W_i(x)$ hence behaves at small $q$ as
\beq
W_i(x) \sim \sum_{j=0}^n \sum_{l=1}^d\,\, {1\over x-q^{a_{j,i}-l+d}} + {\rm small\, cut\, near\,} q^{a_{j,i}+d}  \,.
\eeq
Pictorially, the size of the cuts is shrinking in this limit, replacing the spectral curve by its skeleton, see figure \ref{figsmallqpplane}.

On the other hand, the mirror curve is a priori a tree level quantity, hence does not depend on $q=e^{-g_s}$. However, recall that we have defined the K\"ahler parameters $Q$ associated to a curve $\curve$ as 
\beq
Q= q^{\int_{\curve} J} \,.
\eeq
The large $q$ limit hence corresponds to the large curve class limit, i.e. the distance between the vertices of the pairs of pants out of which the mirror curve is constructed is taken to infinity. Just as the spectral curve, the mirror curve thus collapses to its skeleton in the $q \rightarrow 0$ limit.
\section*{6 The general BKMP conjecture}  \label{finishing_proof}

So far, we have obtained the BKMP conjecture only for the fiducial geometry $\CYX_0$. Studying the behavior of the partition function under flop transitions will allow us to extend our argument to arbitrary toric geometries.

\subsection*{6.1 Flop invariance of toric Gromov-Witten invariants}

\begin{figure}[h]
 \centering
 \includegraphics[width=6cm]{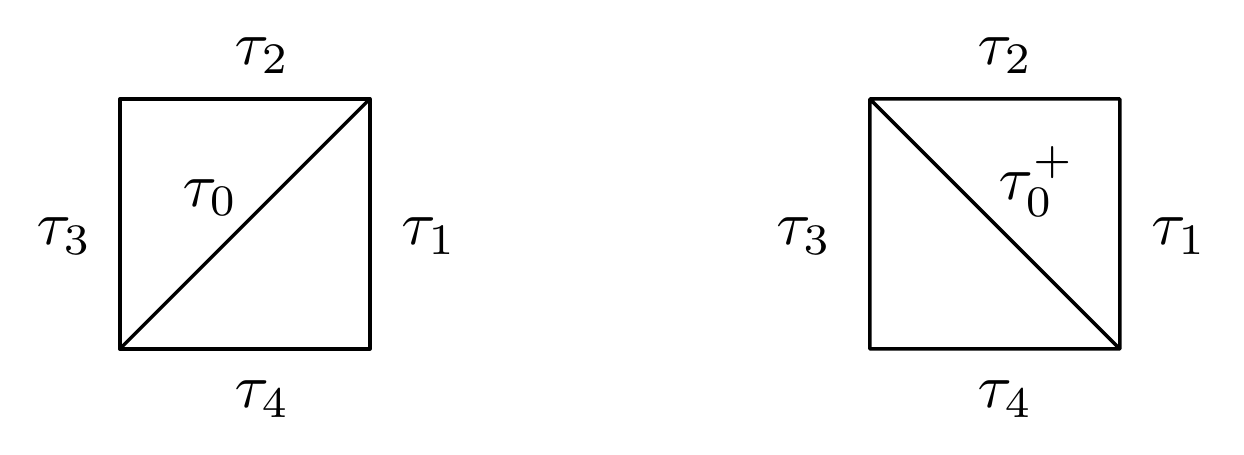}
 \caption{\footnotesize{$\CYX$ and $\CYX^+$ in the vicinity of the (-1,-1) curve.}}
 \label{toric_flopped}
\end{figure}

Under the proper identification of curve classes, Gromov-Witten invariants (at least on toric manifolds) are invariant under flops. Assume the toric Calabi-Yau manifolds $\CYX$ and $\CYX^+$ are related via a flop transition, $\phi: \CYX \rightarrow \CYX^+$. In a neighborhood of the flopped $(-1,-1)$ curve, the respective toric diagrams are depicted in figure \ref{toric_flopped}.

The 1-cones of $\Sigma_\CYX$, corresponding to the toric invariant divisors of $\CYX$, are not affected by the flop, hence can be canonically identified with those of $\CYX^+$. The 2-cones $\tau_i$ in these diagrams correspond to toric invariant 2-cycles $C_i$, $C_i^+$ in the geometry. The curve classes of $\CYX$ push forward to classes in $\CYX^+$ via
\ban\label{flopping}
\phi_*([C_0]) = - [C_0^+] \,, \quad \phi_*([C_i]) =  [C_i^+] + [C_0^+] \,. 
\ean
All other curve classes $\vec{C}$ of $\CYX$ are mapped to their canonical counterparts in $\CYX^+$. Under appropriate analytic continuation and up to a phase factor (hence the $\propto$ in the following formula), the following identity then holds \cite{Witten_Phases,IqbalKashaniPoor, TopologicalVertex2},
\ba
Z_{GW}(\CYX,Q_0,Q_1,\ldots,Q_4,\vec{Q}) \propto Z_{GW}(\CYX^+,1/Q_0, Q_0 Q_1, \ldots, Q_0 Q_4,\vec{Q}) \,,
\ea
i.e.
\ba
GW_g(\CYX,Q_0,Q_1,\ldots,Q_4,\vec{Q}) = GW_g(\CYX^+,1/Q_0, Q_0 Q_1, \ldots, Q_0 Q_4,\vec{Q}) \,.
\ea

\subsection*{6.2 Proof of flop invariance via mirror symmetry}
Flop invariance of Gromov-Witten invariants upon the identification (\ref{flopping}) is immediate upon invoking mirror symmetry, as (\ref{flopping}) maps the mirror curve of $\CYX$ to that of $\CYX^+$. The proof is a simple computation.

\begin{figure}[h]
 \centering
 \includegraphics[width=6cm]{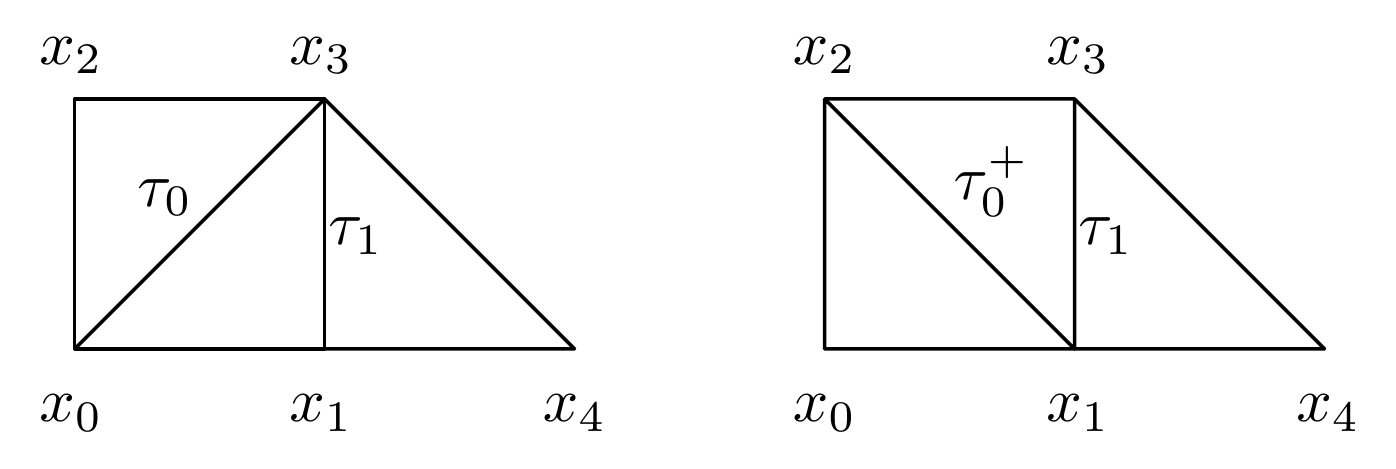}
 \caption{\footnotesize{$\CYX$ and $\CYX^+$ in the vicinity of the (-1,-1) curve.}}
 \label{mirror_flopped}
\end{figure}
Let us introduce the notation $t_0, t_1, t_0^+, t_1^+$ for the K\"ahler volume of the curve classes $C_i,C_i^+$ corresponding to the respective 2-cones.
In terms of these, we obtain for the mirror curve of $\CYX$
\beq
x_0 + x_1 + x_2 + \frac{x_1 x_2}{x_0} e^{T_0} + \frac{x_1^2}{x_0} e^{-T_1} = 0 \,,
\eeq
while the mirror curve of $\CYX^+$ is given by
\beq
x_0 + x_1 + x_2 + \frac{x_1 x_2}{x_0} e^{-T_0^+} + \frac{x_1 x_3}{x_2} e^{-T_1^+} = 0\,.
\eeq
Upon invoking $x_3 = \frac{x_1 x_2}{x_0}e^{-T_0^+}$, we easily verify that the identification (\ref{flopping}) maps these curves and their associated meromorphic 1-forms $\lambda$ into each other. 

\subsection*{6.3 The BKMP conjecture}
\begin{figure}[h]
 \centering
 \includegraphics[width=10cm]{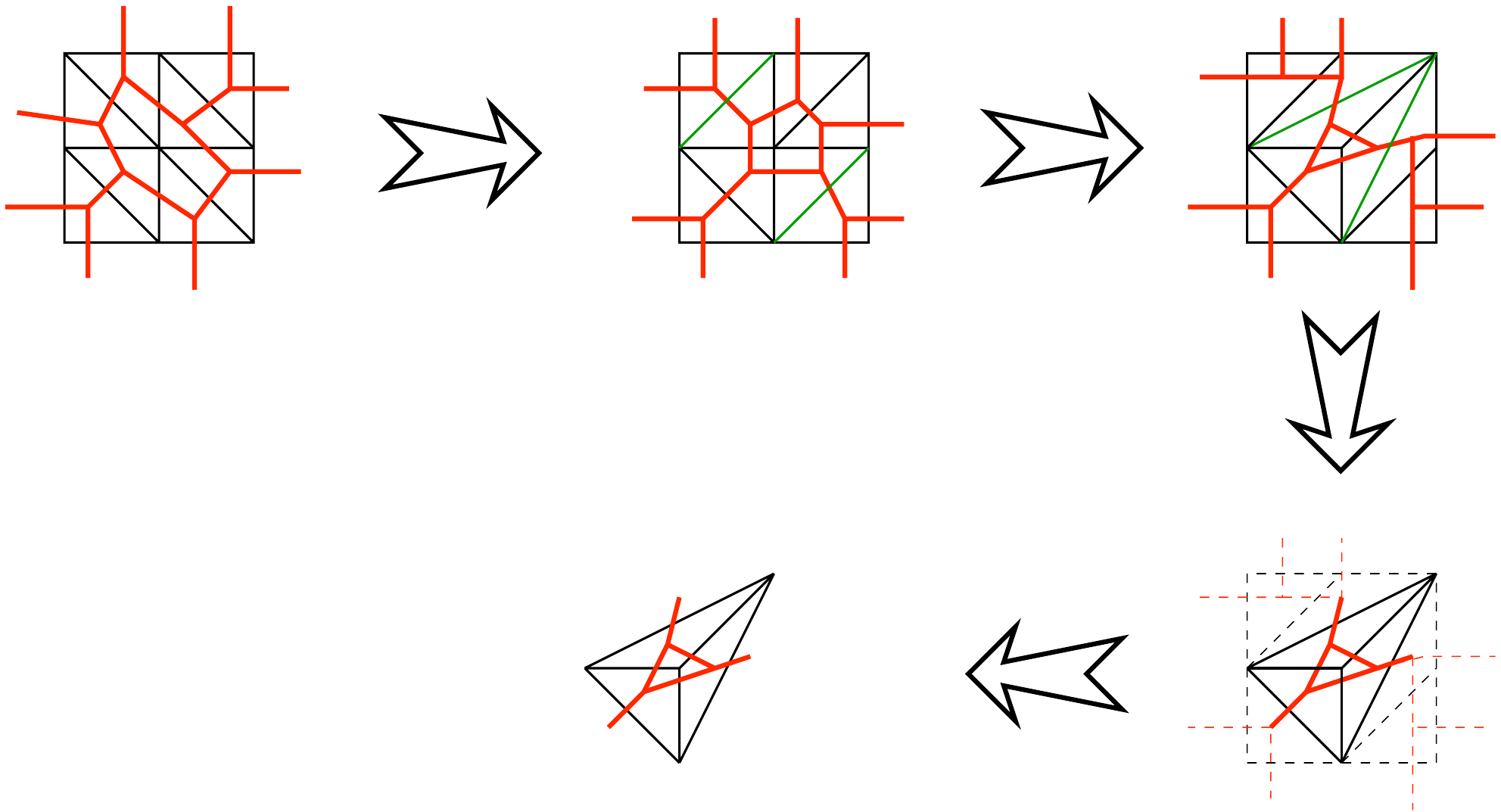}
 \caption{\footnotesize{Example: We obtain local $\mathbb P^2$ from the fiducial geometry with $2\times 2$ boxes by performing five flops and then sending the K\"ahler parameters of the unwanted edges to $\infty$.}}
 \label{flopP2}
\end{figure}

Any toric Calabi-Yau manifold $\CYX$ with K\"ahler moduli $\vec{Q}$ can be obtained from a sufficiently large fiducial geometry $(\CYX_0,\vec{Q}_0)$ upon performing a series of flop transitions and taking unwanted K\"ahler moduli of $\CYX_0$ to $\infty$, see figure \ref{flopP2} for an example.

The K\"ahler moduli of $\CYX$ are related to those of $\CYX_0$ by some relation $\vec{Q}=f(\vec{Q}_0)$. We have just argued that the mirror curves of $\CYX_0$ and $\CYX$ are equal upon this identification,
\beq
\spcurve_{\CYX,\vec{Q}} = \spcurve_{\CYX_0,\vec{Q_0}} \,,
\eeq
as are the respective Gromov-Witten invariants,
\beq
GW_g(\CYX,\vec{Q}) = GW_g(\CYX_0,\vec{Q}_0) \,.
\eeq
Given the BKMP conjecture for the fiducial geometry,
\beq
GW_g(\CYX_0,\vec{Q}_0)
=F_g(\spcurve_{\CYX_0,\vec{Q}_0}) \,,
\eeq
its validity thus follows for any toric Calabi-Yau manifold:
\beq
\encadremath{
GW_g(\CYX,\vec{Q})
=F_g(\spcurve_{\CYX,\vec{Q}})
}\eeq

\section*{7 Conclusion}
Taking our matrix model from \cite{topstring1} as a starting point and imposing certain minimality conditions on the spectral curve, we have thus derived the BKMP conjecture, for closed topological strings, for all toric Calabi-Yau manifolds in the large radius limit. As we have emphasized throughout, elevating our procedure to a formal proof of the conjecture requires a more rigorous derivation of the spectral curve of our matrix model.

It should also be possible to extend our argument to
open Gromov-Witten invariants by invoking loop operators, which relate
closed to open invariants. In \cite{OE}, such an operator was defined in the matrix model context. An analogous operator should also exist in the theory of Gromov-Witten invariants. Establishing the equivalence of these two loop operators would allow us to conclude that the $W_n^{(g)}$'s of the spectral curve $\spcurve_{\CYX}$ are the open Gromov-Witten invariants of $\CYX$.

Finally, our treatment of the BKMP conjecture took place at large radius. One should study the behavior of the matrix model as one moves away from large radius e.g. to orbifold points, and see whether the phase transitions of the topological string are captured accurately by the matrix model. Of course, the main tool on the topological string side employed in this work, the topological vertex, is no longer applicable in these regions of moduli space.

\section*{Acknowledgments}
B.E. and O.M. would like to thank M. Bertola, J. Harnad, V. Bouchard,  M. Mari\~ no, M. Mulase,  H.~ Ooguri, N.~Orantin, B. Safnuk, for useful and fruitful discussions on this subject. A.K. would like to thank Vincent Bouchard and Ilarion Melnikov for helpful conversations. The work of B.E. is partly supported by the Enigma European network MRT-CT-2004-5652, ANR project GranMa "Grandes Matrices Al\'eatoires" ANR-08-BLAN-0311-01,  
by the European Science Foundation through the Misgam program,
by the Quebec government with the FQRNT. 
B.E. would like to thank the AIM, as well as the organizers and all participants to the workshop held at the AIM june 2009.
O.M. would like to thank the CRM (Centre de recheche math\'ematiques de Montr\'eal, QC, Canada) for its hospitality.

\section*{Appendix: The matrix model}   \label{our_matrix_model} 
In this appendix, which is mainly a reprint of section 4 of \cite{topstring1}, we present the matrix model which reproduces the topological string partition function on the fiducial geometry $\CYX_0$, and whose spectral curve we derive in the text.

SConsider the fiducial geometry $\CYX_0$ of size $(n+1)\times (m+1)$, with K\"ahler parameters $t_{i,j}=a_{i,j}-a_{i,j+1}$, $r_{i,j} = a_{i,j+1}-a_{i+1,j}$, and  $s_{i,j}$, as depicted in figure \ref{fiducial_geometry}. We write
\beq
\vec a_i  = (a_{0,i},a_{1,i},\dots,a_{n,i}).
\eeq

Assume that the external representations are fixed to $\vec\alpha_{m+1} = (\alpha_{0,m+1},\alpha_{1,m+1},\dots,\alpha_{n,m+1})$ on the upper line, and $\vec\alpha_0 = (\alpha_{0,0},\alpha_{1,0},\dots,\alpha_{n,0})$ on the lower line. For the most part, we will choose these to be trivial.

\bigskip

We now define the following matrix integral ${\cal Z}_{\rm MM}$ (${}_{\rm MM}$ for Matrix Model),
\ban
{\cal Z}_{\rm MM}(Q,g_s,\vec\alpha_{m+1},\vec\alpha_0^T)
&=& \Delta(X(\vec \alpha_{m+1}))\,\, \Delta(X(\vec \alpha_0)) \,\, 
\prod_{i=0}^{m+1} \int_{H_N(\Gamma_i)} dM_i \,
 \prod_{i=1}^{m+1}\int_{H_N({\mathbb R}_+)}\,dR_i \nn \\
&& \prod_{i=1}^{m} e^{{-1\over g_s}\,\tr \left[ V_{\vec a_i}(M_i)-V_{\vec a_{i-1}}(M_i) \right]
%+ V_{\vec a_i,\vec a_{i-1}}(X_i) 
} \,\,\,
 \prod_{i=1}^{m} e^{{-1\over g_s}\,\tr \left[V_{\vec a_{i-1}}(M_{i-1})-V_{\vec a_{i}}(M_{i-1}) \right]
%+ \td V_{\vec a_{i-1},\vec a_{i}}(X_{i-1}) 
} \nn \\
&& \prod_{i=1}^{m+1} e^{{1\over g_s} \tr (M_i-M_{i-1})R_i} \,\,\,
 \prod_{i=1}^{m} e^{(S_i+{i\pi\over g_s})\,\tr\, \ln M_i}\,  \nn\\
&& e^{\tr \ln f_{0}(M_0)}\,\,e^{\tr \ln f_{m+1}(M_{m+1})}\,\, \prod_{i=1}^{m} e^{\tr \ln f_{i}(M_i)} \,. 
\ean
All matrices are taken of size
\beq
N=(n+1)\, d \,.
\eeq
$d$ denotes a cut-off on the size of the matrices, on which, as discussed in section \ref{arctic}, the partition function depends only non-perturbatively.
We have introduced the notation
\beq
X(\vec \alpha_{m+1})  = {\rm diag} (X(\vec \alpha_{m+1})_i)_{i=1,\dots,N}
\,\, , \qquad
X(\vec \alpha_{m+1})_{j d+k} = q^{h_k(\alpha_{j,m+1})},
\eeq
\beq
X(\vec \alpha_0)  = {\rm diag} (X(\vec \alpha_0)_i)_{i=1,\dots,N}
\,\, , \qquad
X(\vec \alpha_0)_{j d+k} = q^{h_k(\alpha_{j,0})},
\eeq
for $k=1, \ldots, d$, $j=0, \ldots,n$, where
\beqn \label{defh}
h_i(\gamma)=\gamma_i-i+d+a \,.
\eeqn
$\Delta(X)=\prod_{i<j}(X_i-X_j)$ is the Vandermonde determinant. The potentials $V_{\vec a_i}(x)$ are given by
\beqn    \label{partpot}
V_{\vec a}(X) = -g_s\,\sum_{j=0}^n \ln{\left(g(q^{a_j}/X)\right)} 
\eeqn
in terms of the $q$-product
\ba
g(x) = \prod_{n=1}^\infty (1-{1\over x}\,q^n) \,.
\ea
For $i=1,\dots,m$, we have defined
\beq
f_i(x) =  \prod_{j=0}^n {g(1)^2\,\,e^{({1\over 2}+{i\pi\over \ln q})\, \ln{(x q^{1-a_{j,i}})}}\, \,e^{{(\ln{(x q^{1-a_{j,i}})})^2\over  2 g_s}}\over g(x\,q^{1-a_{j,i}})\, g(q^{a_{j,i}}/x)\,} \,.
\eeq
The denominator of these functions induces simple poles at $x=q^{a_{j,i}+l}$ for $j=0,\dots,n$ and $l\in \mathbb Z$. The numerator is chosen such that they satisfy the relation $f_i(qx)=f_i(x)$. This enforces a simple $l$-dependence of the residues taken at $x=q^{a_{j,i}+l}$, given by a prefactor $q^l$ -- a fact which will be important in the following. These residues are in fact given by
\beqn \label{resf}
 \Res_{q^{a_{j,i}+l}} f_i(x) = q^{a_{j,i}+l}\,\, \hat f_{j,i} =-\, q^{a_{j,i}+l}\,\,   \prod_{k\neq j} {g(1)^2\,\,e^{({1\over 2}+{i\pi\over \ln q})\, (1+a_{j,i}-a_{k,i})\ln q}\, \,e^{{(\ln{(q^{1+a_{j,i}-a_{k,i}})})^2\over  2 g_s}}\over g(q^{a_{j,i}-a_{k,i}})\,(1-q^{a_{k,i}-a_{j,i}}) g(q^{a_{k,i}-a_{j,i}})} \,,
\eeqn
where $\hat f_{j,i}$ is independent of the integer $l$. 

The parameters $S_i$ are defined by 
\beqn \label{si}
S_i =  s_{0,i-1}+t_{0,i-1}= s_{j,i-1} -\sum_{k<j} t_{k,i}+\sum_{k\leq j} t_{k,i-1} \,.
\eeqn
The final equality holds for arbitrary $j$ \cite{topstring1}.

For $i=0$ and $i=m+1$, we define
\beq
f_0(x) = {1\over \prod_{j=0}^n \prod_{i=1}^d (x-q^{h_i(\alpha_{j,0})})} \,,
\eeq
\beq
f_{m+1}(x) = {1\over \prod_{j=0}^n \prod_{i=1}^d  (x-q^{h_i(\alpha_{j,m+1})})} \,.
\eeq
Notice that if the representations $\vec\alpha_0$ or $\vec\alpha_{m+1}$ are trivial, i.e. $h_i(\alpha_{j,0})=d-i+a_{j,0}$ or $h_i(\alpha_{j,m+1})=d-i+a_{j,m+1}$, we have
\beq
f_0(x) = \prod_{j=0}^n {g(x\,q^{1-a_{j,0}-d})\over x^d\, g(x\,q^{1-a_{j,0}}) }\,,\hspace{1cm} f_{m+1}(x) = \prod_{j=0}^n {g(x\,q^{1-a_{j,m+1}-d})\over x^d\, g(x\,q^{1-a_{j,m+1}}) }
\eeq
respectively.
The functions $f_0$ and $f_{m+1}$ have simple poles
at $x=q^{h_l(\alpha_{j,0})}$ (resp. $x=q^{h_l(\alpha_{j,m+1})}$) for $l=1,\dots,d$, with residue
\beq
\hat f_{j,0;l} = \Res_{q^{h_l(\alpha_{j,0})}} f_0(x) =   {1\over \prod_{j'\neq j} \prod_{i=1}^d (q^{h_l(\alpha_{j,0})}-q^{h_i(\alpha_{j',0})})} \,{1\over  \prod_{i\neq l} (q^{h_l(\alpha_{j,0})}-q^{h_i(\alpha_{j,0})})} \,,
\eeq
\beq
\hat f_{j,m+1;l} = \Res_{q^{h_l(\alpha_{j,m+1})}} f_{m+1}(x) =   {1\over \prod_{j'\neq j} \prod_{i=1}^d (q^{h_l(\alpha_{j,m+1})}-q^{h_i(\alpha_{j',m+1})})} \,{1\over  \prod_{i\neq l} (q^{h_l(\alpha_{j,m+1})}-q^{h_i(\alpha_{j,m+1})})} \,.
\eeq
The $l$ dependence here is more intricate than above, but this will not play any role since the partitions $\alpha_{j,0}$ and $\alpha_{j,m+1}$ are kept fixed, and not summed upon.
  
\bigskip

\begin{figure}[h]
 \centering
 \includegraphics[width=8cm]{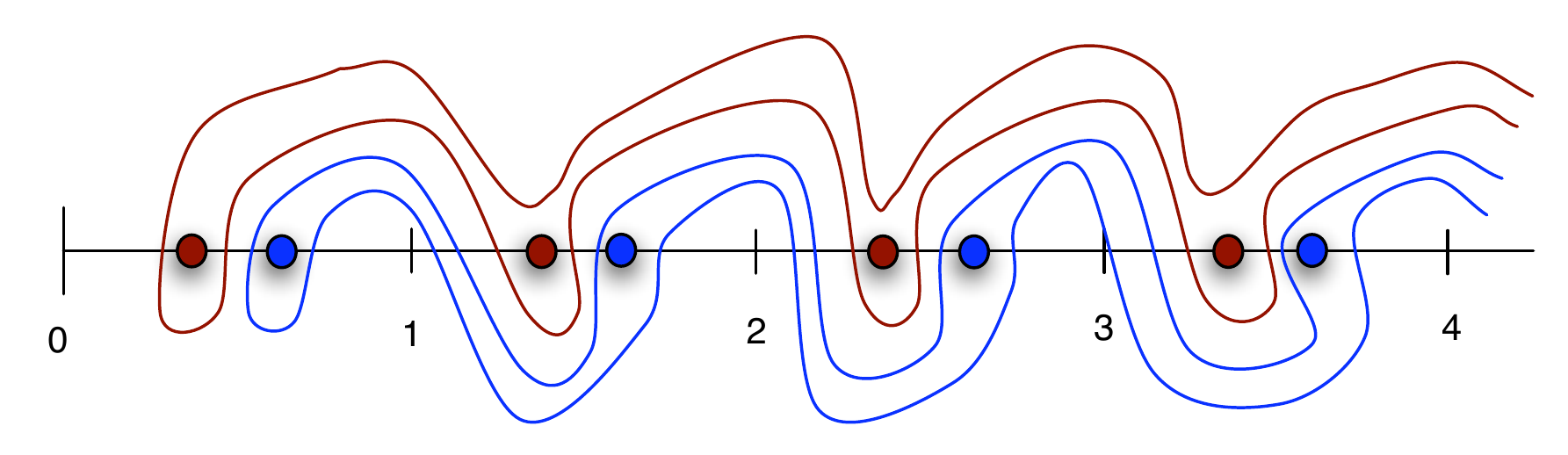}
 \caption{\footnotesize{Two contours surrounding points $a+\mathbb N$ and $b+\mathbb N$, such that  $a-b \notin \mathbb Z$.}}
 \label{contours}
\end{figure}

The integration domains for the matrices $R_i$ are $H_N(\mathbb R_+^N)$, i.e. the set of hermitian matrices  having only positive eigenvalues. For the matrices $M_i, i=1,\dots,m$, the integration domains are $H_N(\Gamma_i)$, where 
\beq
\Gamma_i = \prod_{j=0}^n\, (\gamma_{j,i})^d \,.
\eeq
$\gamma_{j,i}$ is defined as a contour which encloses all points of the form $q^{a_{j,i}+\mathbb N}$, and does not intersect any contours $\gamma_{k,l}$, $(j,i) \neq (k,l)$. For this to be possible, we must require that the differences $a_{j,i}-a_{j',i'}$ be non-integer. The normalized logarithms of two such contours are depicted in figure \ref{contours}.

 We have defined
\beq
H_N(\Gamma_i) = \{ M=U\, \Lambda \, U^\dagger \, , \quad U\in U(N)\, , \,\,\, \Lambda={\rm diag}(\lambda_1,\dots,\lambda_N)\, \in \Gamma_i \} \,,
\eeq
i.e. $H_N(\Gamma_i)$ is the set of normal matrices with eigenvalues on $\Gamma_i$. By definition, the measure on $H_N(\Gamma_i)$ is (see \cite{Mehta})
\beqn\label{eqdefdMdUdL}
dM = {1\over N!}\,\, \Delta(\Lambda)^2\,\, dU\,d\Lambda \,,
\eeqn
where $dU$ is the Haar measure on $U(N)$, and $d\Lambda$ is the product of the measures for each eigenvalue along its integration path.

The integration domains for the matrices $M_0$, $M_{m+1}$ are $H_N(\Gamma_0)$, $H_N(\Gamma_{m+1})$ respectively, where 
\beqn  \label{outer_contours}
\Gamma_0 =  (\sum_{j=0}^n \gamma_{j,0})^N \,,
\qquad \quad
\Gamma_{m+1} =  (\sum_{j=0}^n \gamma_{j,m+1})^N \,.
\eeqn

%% file: These.bbl
\begin{thebibliography}{0}
\expandafter\ifx\csname natexlab\endcsname\relax\def\natexlab#1{#1}\fi
\expandafter\ifx\csname url\endcsname\relax
  \def\url#1{{\tt #1}}\fi

\end{thebibliography}


\begin{thebibliography}{99}

\bibitem{Hist} H. Hotelling, ``Analysis of a complex of statistical variables into principal components'', \textit{Jour. Educ. Psych.}, Vol \textbf{24},  417-441, 1933.

\bibitem{Hist2} A. James, ``Distributions of matrix variates and latent roots derived from normal samples'', \textit{Annals of Mathematical Statistics}, Vol \textbf{35}, 475-501, 1964.

\bibitem{VieuxWigner} E. Wigner, A.M. Weinberg, ``Physical Theory of Neutron Chain Reactors'', \textit{University of Chicago Press.}, ISBN 0-226-88517-8, 1958.

\bibitem{VieuxDyson} F.J. Dyson, ``Correlations between the eigenvalues of a random matrix'', \textit{Commun. Math. Phys.}, Vol \textbf{19}, 1970.

\bibitem{FirkMiller} F.W. Firk, S.J. Miller, ``Nuclei, Primes and the Random Matrix Connection''. \textit{Symmetry}, Vol \textbf{1}, 64-105, 2009.

\bibitem{DerrienLeal} H. Derrien, L. Leal, N. Larson, ``Status of new evaluation of the neutron resonance parameters of 238U at ORNL'', \textit{PHYSOR}, Amer. Nucl. Soc. LaGrange Park IL, 2004.

\bibitem{Od1} A. Odlyzko, ``On the distribution of spacings between the zeros the zeta function'', \textit{Math. Comp.}, Vol \textbf{48}, 273-308, 1987.

\bibitem{Od2} A. Odlyzko, M. van Frankenhuysen, M. L. Lapidus, ``The 1022-nd zero of the Riemann zeta function, Proc. Conference on Dynamical, Spectral and Arithmetic Zeta-Functions'', Amer. Math. Soc., \textit{Contemporary Math. series}, 2001.

\bibitem{RNA} G. Vernizzi, H. Orland, A. Zee,``Enumeration of RNA structures by Matrix Models'', \textit{Phys. Rev. Lett.}, Vol \textbf{94}, 168103, 2005.

\bibitem{RNA2} I. Garg, N. Deo, ``RNA matrix models with external interactions and their asymptotic behavior'', \textit{Phys. Rev. E}, Vol \textbf{79}, 061903, 2009.

\bibitem{RNA3} M. Bon, G. Vernizzi, H Orland, A Zee, ``Topological classification of RNA structures'', \textit{Journal of Molecular Biology}, Vol. \textbf{379}, Issue 4, 900-911, 2008.

\bibitem{signal} R. Muller, H. Hofstetter, ``A Random Matrix Model of Communication Via Antenna Arrays'', \textit{IEEE Transactions on information theory}, Vol. \textbf{48}, No. 9, 2002.

\bibitem{protein} WJ. Wilbur, ``On the PAM matrix model of protein evolution'', \textit{Mol. Biol. Evol.}, Vol \textbf{5}, 434-47, 1985.

\bibitem{Erdos} L. Erdos, ``Universality of Wigner Random Matrices: a Survey of Recent Results'', \textit{arXiv}:1004.0861v1 [math-ph].

\bibitem{Erdos2} L. Erdos, H. Yau, J. Yin, ``Universality for generalized Wigner matrices with Bernoulli distribution'', \textit{arXiv}:1003.3813v3 [math-ph].

\bibitem{Guillonet} A. Guionnet, ``Large deviation upper bounds and central limit theorems for band matrices'', \textit{Ann. Inst. H. Poincare Probab. Statist.}, Vol \textbf{38}, 341-384, 2002.

\bibitem{Johansson} K. Johansson, ``Universality of the local spacing distribution in certain ensembles of Hermitian Wigner matrices'', \textit{Comm. Math. Phys.}, Vol \textbf{215}, No 3, 683-705, 2001.

\bibitem{Bessis} D. Bessis, ``Orthogonal polynomials, Padé approximation and Julia sets'', \textit{OrthogonalPolynomials},
\textit{NATO Adv. Sci. Inst. Ser. C Math. Phys. Sci.}, Vol \textbf{294}, 55-97, Kluwer Acad. Publ., Dordrecht, 1990.

\bibitem{Mehta} M. L. Mehta, ``Random matrices (3e édition)'', \textit{Pure and Applied Mathematics Series}, Vol \textbf{142}, Elsevier, ISBN 0120884097, 2004.

\bibitem{BPS} A. Boutet de Monvel, L. Pastur, M. Shcherbina, ``On the statistical mechanics approach in the random matrix theory: Integrated density of states'', \textit{J. Statist. Phys.}, Vol \textbf{79}, 585-611, 1995.

\bibitem{JOH} K. Johansson, ``On fluctuations of eigenvalues of random matrices'', \textit{Duke Math. Jour.}, Vol \textbf{91}, 151-204, 1988.

\bibitem{BleherEynard} P. Bleher, B. Eynard, ``Double scaling limits of random matrices models and a nonlinear hierarchy of differential equations'', \textit{J. Phys. A: Math. Gen.}, Vol \textbf{36}, 3085, 2003.

\bibitem{OE}  N.Orantin, B. Eynard  ``Invariants of algebraic curves and topological expansion'', \textit{Comm. in Number Theory and Physics}, Vol \textbf{1}, No 2, 2007.

\bibitem{Kazakov} E. Brézin, V.A. Kazakov, ``Exactly solvable field theories of closed strings'' \textit{Phys. Lett. B}, Vol \textbf{236}, 144-150, 1990. 

\bibitem{determinantalformulae} M. Berg\`{e}re, B. Eynard, ``Determinantal formul\ae{} and loop equations'', \textit{arXiv}:0901.3273v2 [math-ph], 2009.

\bibitem{Nonperturbative} B. Eynard, ``Large N expansion of convergent matrix integrals, holomorphic anomalies, and background independence'', \textit{JHEP}, 0903:003, 2009.

\bibitem{Staudacher} M. Staudacher, ``The Yang-Lee edge singularity on a dynamical planar random surface'', \textit{Nuclear Physics B}, Vol \textbf{336}, 349-362, 1990.

\bibitem{Staudacher2} W. Krautha, M. Staudacher, ``Eigenvalue distributions in Yang-Mills integrals'', \textit{Physics Letters B}, Vol \textbf{453}, 253-257, 1999.

\bibitem{BergereEynard} M. Berg\`{e}re, B. Eynard, ``Universal scaling limits of matrix models, and $(p,q)$ Liouville gravity'', \textit{arXiv}:math-ph/0909.085v1

\bibitem{MoiMattia} O. Marchal, M. Cafasso, ``Double scaling limits of random matrices and minimal $(2m,1)$ models: the merging of two cuts in a degenerate case'', \textit{arXiv}:1002.3347v2 [math-ph]

\bibitem{AlvarezAlonsoMedina} G. Alvarez, LM Alonso, E. Medina, ``Phase transitions in multi-cut matrix models and matched solutions of Whitham hierarchies'', \textit{J. Stat. Mech.}, \textbf{P03023}, 2010.

\bibitem{BE} M. Bertola, B. Eynard, ``The PDEs of biorthogonal polynomials arising in the two-matrix model'', \textit{Mathematical Physics, Analysis and Geometry}, Vol \textbf{9}, No 1, 23-52, 2006.

\bibitem{BEH} M. Bertola, B. Eynard,  J. Harnad, ``Duality of spectral curves arising in two-matrix models'' 
{\it  Theor. Math. Phys.}, Vol {\bf 134}, 27-38  2003. 

\bibitem{BEH1} M. Bertola, B. Eynard, J. Harnad, ``Duality, Biorthogonal Polynomials and Multi-Matrix Models'',
\textit{Commun. Math. Phys.}, Vol {\bf 229}, 73-120, 2002.

\bibitem{BEH2} M. Bertola, B. Eynard, J. Harnad, ``Differential systems for biorthogonal polynomials appearing in
2-matrix models and the associated Riemann-Hilbert problem'', \textit{Commun. Math. Phys.}, Vol {\bf 243}, 193-240, 2003.

\bibitem{KMcL}  A. B. J. Kuijlaars, K. T-R McLaughlin ``A Riemann-Hilbert problem for biorthogonal polynomials'',  \textit{Journ. computational and applied mathematics}, Vol \textbf{178}, no 1-2, 313-320, 2005.

\bibitem{Harnad}  M. Bertola, B. Eynard, J. Harnad ``Partition functions for matrix models and isomonodromic tau functions'' \textit{Journal of Physics A}, Vol \textbf{36}, 3067-3083, 2003. 

\bibitem{JMI} M. Jimbo, T. Miwa, K. Ueno ``Monodromy preserving deformation of linear ordinary differential equations with rational coefficients'', \textit{ Physica D}, Vol {\bf 2}, 306-352, 1981.

\bibitem{JMII} M. Jimbo, T. Miwa, K. Ueno ``Monodromy preserving deformation of linear ordinary differential equations with rational coefficients II'', \textit{Physica D}, Vol {\bf 2}, 407-448, 1981.

\bibitem{JMIII} M. Jimbo, T. Miwa, K. Ueno ``Monodromy preserving deformation of linear ordinary differential equations with rational coefficients III'', \textit{Physica D}, Vol {\bf 4}, Issue 1, 26-46, 1981.

\bibitem{MoiMarco} M. Bertola, O. Marchal ``The partition function of the two-matrix model as an isomonodromic tau-function'', \textit{J. Math. Phys.} \textbf{50}, 013529, 2009. 

\bibitem{NotesBleher} P. Bleher ``Lectures on random matrix models. The Riemann-Hilbert approach'', \textit{arXiv}:0801.1858v2 [math-ph], 2008.

\bibitem{Angular} M. Berg\`{e}re, B. Eynard,  ``Some properties of angular integrals'', \textit{J. Phys. A: Math. Theor.}, Vol \textbf{42}, 2009.

\bibitem{MoiBertrand} B. Eynard, O. Marchal, ``Topological expansion of the Bethe ansatz, and non-commutative algebraic geometry'', \textit{JHEP} 0903:094, 2009.

\bibitem{MoiLeonidBertrand} L. Chekov, B. Eynard, O. Marchal, ``Topological expansion of the Bethe ansatz, and quantum algebraic geometry'' arXiv:0911.1664v2 [math-ph], 2009.

\bibitem{Vershik} A.M. Vershik, ``Limit distribution of the energy of a quantum ideal gas from the viewpoint of the theory of partitions of natural numbers'', \textit{Uspekhi Mat. Nauk}, Vol \textbf{52} 139-146, 1997. 

\bibitem{David} F. David, ``Randomly triangulated surfaces in $2$ dimensions'', \textit{Phys. Lett. B}, Vol \textbf{159}, 303-306, 1985.

\bibitem{David2} F. David. ``Random matrices and two-dimensional gravity'', Ed. Elsevier, Eighth International Summer School on Fundamental Problems in Statistical Mechanics, Altenberg, Germany, 1994.

\bibitem{Okunkov} A. M. Vershik, A. Okunkov,  ``A new approach to representation theory of symmetric groups II'', \textit{J. Math. Sci.}, Vol \textbf{2}, 5471-5494, 2005.

\bibitem{Douglas} M.R. Douglas, V.A. Kazakov, ``Large $N$ Phase Transition in Continuum QCD$_2$'', \textit{Phys. Lett. B}, Vol \textbf{319}, 1993.

\bibitem{BiPZ} E. Brézin, C. Itzykson, G. Parisi, J.B. Zuber, ``Planar diagrams'', \textit{Comm. Math. Phys.}, Vol \textbf{59}, 35-51, 1978.

\bibitem{Nekrasov} Piotr Sulkowski ``Matrix models for $\beta$-ensembles from Nekrasov partition functions'', \textit{JHEP}, Vol \textbf{2010}, No 4, 1-29, 2010.

\bibitem{Nekrasov2} A. Mironov, A. Morozov, Sh. Shakirov, ``Matrix Model Conjecture for Exact BS Periods and Nekrasov Functions'', \textit{JHEP}, Vol \textbf{2010}, No 2, 1-26, 2010.

\bibitem{Nekrasov3} Hiroshi Itoyama, Takeshi Oota, ``Method of Generating q-Expansion Coefficients for Conformal Block and N=2 Nekrasov Function by beta-Deformed Matrix Model'', \textit{Nucl. Phys. B}, Vol \textbf{838}, 298-330, 2010. 

\bibitem{AGT} L. Alday, D. Gaiotto, Y. Tachikawa, ``Liouville Correlation Functions from Four-dimensional Gauge Theories'', \textit{Letters in math. physics}, Vol \textbf{91}, No 2, 167-197, 2010.

\bibitem{toprecint} B. Eynard, N. Orantin, ``Geometrical interpretation of the topological recursion, and integrable string theories'', \textit{arXiv}:0911.5096v1 [math-ph], 2009.

\bibitem{BorotEynardMulase} G. Borot, B. Eynard, M. Mulase, B. Safnuk, ``A matrix model for simple Hurwitz numbers, and topological recursion'', \textit{arXiv}:0906.1206v1 [math-ph], 2009.

\bibitem{countingsurface} B. Eynard, N. Orantin,``Algebraic methods in random matrices and enumerative geometry'', \textit{arXiv}:0811.3531v1 [math-ph], 2008.

\bibitem{countingsurface2} B. Eynard, ``Formal matrix integrals and combinatorics of maps'', \textit{CRM Series in Mathematical Physics}, arXiv:math-ph/0611087v3, 2007.

\bibitem{IZ} C. Itzykson, J.-B. Zuber, ``Combinatorics of the Modular Group II: the Kontsevich integrals'', \textit{Int. J. Mod. Phys. A}, Vol \textbf{7}, 5661-5705, 1992.

\bibitem{IZ2} C. Itzykson, J.-B. Zuber, ``Addendum to the paper Combinatorics of the modular group II: the Kontsevich integrals'', \textit{arXiv}:hep-th/9201055v1, 1992. 

\bibitem{doublelimiteechelle} M.R. Douglas, S.H. Shenker, ``Strings in less than one dimension'', \textit{Nucl. Phys. B}, Vol \textbf{335}, 635-654, 1990. 

\bibitem{doublelimiteechelle2} D.J. Gross, A.A. Migdal, ``Nonperturbative two-dimensional quantum gravity'', \textit{Phys. Rev. Lett.}, Vol \textbf{64}, 127-130, 1990.

\bibitem{TracyWidom} C. Tracy, H. Widom, ``Universality of the Distribution Functions of Random Matrix Theory'', \textit{CRM Series in Mathematical Physics}, Vol \textbf{26}, 1999.

\bibitem{TW} C. Tracy, H. Widom, ``Level spacing distributions and the Airy kernel'', \textit{Commun. Math. Phys.}, Vol \textbf{159} ,151-174, 1994.

\bibitem{TW2}  C. Tracy, H. Widom, ``On orthogonal and symplectic ensembles'', \textit{Commun. Math. Phys.}, Vol \textbf{177}, 727-754, 1996.

\bibitem{JMMS} M. Jimbo, T. Miwa, Y. Môri, M. Sato, ``Density matrix of an impenetrable Bose gas and the fifth Painlevé transcendent'',\textit{ Physica 1D}, 80-158, 1980.

\bibitem{topstring1} B. Eynard, A. Kashani-Poor, O. Marchal, ``A matrix model for the topological string I: Deriving the matrix model'', \textit{arXiv}:1003.1737v2 [hep-th].

\bibitem{topstring2} B. Eynard, A. Kashani-Poor, O. Marchal, ``A matrix model for the topological string II: The spectral curve and mirror geometry'', \textit{arXiv}:1007.2194v1 [hep-th].

\bibitem{GromovWitten} McDuff, Dusa, Salamon, Dietmar ``J-Holomorphic Curves and Symplectic Topology'', \textit{American Mathematical Society colloquium publications}, ISBN 0-8218-3485-1, 2004.

\bibitem{Marino} M. Mari\~{n}o, ``Les Houches lectures on matrix models and topological strings'', \textit{arXiv}:hep-th/0410165v3, 2004

\bibitem{corde1} M. Mari\~{n}o ``Chern-Simons theory and topological strings'', \textit{Rev. Mod. Phys.}, Vol \textbf{77}, 675-720, 2005.

\bibitem{Marino2} M. Mari\~{n}o, ``Chern-Simons theory, the 1/N expansion, and string theory'', \textit{arXiv}:1001.2542v3 [hep-th], 2010.

\bibitem{Vonk} M. Vonk, ``A mini-course on topological strings'', \textit{arXiv}:hep-th/0504147v1, 2005.

\bibitem{Klemm} A. Klemm, ``Introduction to Topological String Theory on Calabi-Yau manifolds'', disponible à l'adresse http://www.math.ist.utl.pt/~strings/AGTS/topstrings.pdf.

\bibitem{TopologicalVertex} M. Aganagic, A. Klemm, M. Marino, C. Vafa, ``The topological vertex'', \textit{Commun. Math. Phys.}, Vol \textbf{254}, 425-478, 2005.

\bibitem{TopologicalVertex2} Y. Konishi, S. Minabe ``Flop invariance of the topological vertex'', \textit{Int. Journ. Math.}, Vol \textbf{19}, 27-45, 2008.

\bibitem{Witten} E. Witten, ``Mirror manifolds and topological field theory'', \textit{arXiv}:hep-th/9112056, 1991.

\bibitem{String} M. Bershadsky, S. Cecotti, H. Ooguri, C Vafa, ``Kodaira-Spencer theory of gravity and exact results for quantum string amplitudes'', \textit{Commun. Math. Phys.}, Vol \textbf{165}, 311-428, 1994.

\bibitem{BKMPcase1} Jian Zhou, ``Local Mirror Symmetry for One-Legged Topological Vertex'' \textit{arXiv}:0910.4320v2 [math.AG], 2009.

\bibitem{BKMP} V. Bouchard, A. Klemm, M. Mari\~{n}o, S. Pasquetti. ``Remodeling the B-model'', \textit{Commun. Math. Phys.}, Vol \textbf{287}, 117-178, 2009.

\bibitem{BKMPChen} L. Chen, ``Bouchard-Klemm-Marino-Pasquetti conjecture for C3'', \textit{arXiv}:0910.3739v1 [math.AG], 2009.

\bibitem{Chain} B. Eynard, A. Prats Ferrer, ``Topological expansion of the chain of matrices'', \textit{JHEP}, Vol \textbf{7}, 2009.

\bibitem{Partitions} B. Eynard, ``A Matrix model for plane partitions'', \textit{Journ. Stat. Mech.}, P10011, 2009.

\bibitem{Partitions2} B. Eynard, ``All orders asymptotic expansion of large partitions'', \textit{Journ. Stat. Mech.}, P07023, 2008.

\bibitem{Hardedges} B. Eynard, ``Loop equations for the semiclassical 2-matrix model with hard edges'',  \textit{J. Stat. Mech.}, P10006, 2005.

\bibitem{Hardedges2} L. Chekhov, ``Solving matrix models in the 1/N-expansion'', \textit{Russ. Math. Surv.}, Vol \textbf{61}, No 3, 2006.

\bibitem{AAM} G. Alvarez, l. Martinez-Alonso, E. Medina, ``Phase transitions in multi-cut matrix models and matched solutions of Whitham hierarchies'', 	\textit{Journ. Stat. Mech.}, P03023, 2010.

\bibitem{BBTbook} O. Babelon, D. Bernard, M. Talon, ``Introduction to Classical Integrable Systems'', Cambridge University Press, 1999.

\bibitem{RHP} P. Bleher, A. Its, ``Double scaling limit in the matrix model: the Riemann-Hilbert approach'', \textit{Comm. Pure Appl. Math.}, Vol \textbf{56}, no. 4, 433-516, 2003.

\bibitem{CIK} T. Claeys, A. Its, I. Krasovsky, ``Higher order analogues of the Tracy-Widom distribution and the Painleve II hierarchy'', \textit{Commun. pure and applied math.},  Vol \textbf{63}, No 3, 362-412, 2010.

\bibitem{mKDV}  P. Clarkson, N. Joshi, M. Mazzocco ``The Lax pair for the mKdV hierarchy'', \textit{S\'{e}minaires et Congr\`{e}s, Soci\'{e}t\'{e} Math\'{e}matique de France}, Vol \textbf{14}, 53-64. ISBN 978-2-85629-229-7, 2006.

\bibitem{MM} M. Mazzocco, M.Y. Mo ``The Hamiltonian structure of the second Painlev\'e hierarchy'', \textit{Nonlinearity}, Vol \textbf{20}, No 12, 2845-2882, 2007. 

\bibitem{FN} H. Flashka, A.C. Newell ``Monodromy and spectrum-preserving deformations I'', \textit{Comm. Math. Phys}, Vol \textbf{76}  No. 1, 65-116, 1980.

\bibitem{ConvForm1} G. Bonnet, F. David, B. Eynard, ``Breakdown of universality in multi-cut matrix models'', \textit{J. Phys. A: Math. Gen.}, Vol \textbf{33}, 6739, 2000.

\bibitem{BabBetGaudin} O. Babelon, D. Talalaev, ``On the Bethe Ansatz for the Jaynes-Cummings-Gaudin model'', \textit{J. Stat. Mech.}, P06013, 2007.

\bibitem{Bryc} W. Bryc, V. Pierce, ``Duality of real and quaternionic random matrices'', \textit{Electronic Journal of Probability}, Vol \textbf{14}, 2009.

\bibitem{ChekEynbeta} L. Chekhov, B. Eynard, ``Matrix eigenvalue model: Feynman graph technique for all genera'', \textit{JHEP}, Vol \textbf{26}, 2006.

\bibitem{ChekEynFg} L. Chekhov, B. Eynard, ``Hermitian matrix model free energy: Feynman graph technique for all genera'', 
\textit{JHEP}, Vol \textbf{03}, 2006.

\bibitem{CEO} L. Chekhov, B. E., N. Orantin, ``Free energy topological expansion for the 2-matrix model'', \textit{JHEP}, 0612, 2006.

\bibitem{ZJDFG} P. Di Francesco, P. Ginsparg, J. Zinn-Justin, ``2D Gravity and Random Matrices'', \textit{ Phys. Rep.} {\bf 254}, 1 (1995).

\bibitem{DDmodules} R. Dijkgraaf, L. Hollands, P. Sulkowski, ``Quantum Curves and D-Modules'', \textit{JHEP}, Vol \textbf{11}, 2009. 

\bibitem{DVKS} R. Dijkgraaf,  C. Vafa, ``Two Dimensional Kodaira-Spencer Theory and Three Dimensional Chern-Simons Gravity'', \textit{arXiv}:0711.1932, 2007.

\bibitem{Dum} I. Dumitriu, A. Edelman, ``Matrix models for beta ensembles'', \textit{J. Math. Phys.}, Vol \textbf{43}, 2002.

\bibitem{Eyn1loop} B. Eynard, ``Topological expansion for the 1-hermitian matrix model correlation functions'', \textit{JHEP}, Vol \textbf{2004}, 2004.

\bibitem{eynbeta} B. Eynard, ``Asymptotics of skew orthogonal polynomials'', \textit{J. Phys A.}, Vol \textbf{34}, 2001.

\bibitem{EynVolmum} B. Eynard, ``Recursion between Mumford volumes of moduli spaces'',  \textit{arXiv}:0706.4403 [math-ph], 2007.

\bibitem{Farkas} H.M. Farkas, I. Kra, ``Riemann surfaces'' 2nd edition, Springer Verlag, 1992.

\bibitem{Fay} J.D. Fay, ``Theta functions on Riemann surfaces'', Springer Verlag, 1973.

\bibitem{Gaudin} M. Gaudin, ``La fonction d'onde de Bethe'', ed. Masson, ISBN:2225796076, 1997.

\bibitem{Mulase} M. Mulase, A.Waldron, ``Duality of Orthogonal and Symplectic Matrix Integrals and Quaternionic Feynman Graphs'',
\textit{Comm. Math. Phys.}, Vol \textbf{240}, 553-586, 2003. 

\bibitem{hezari1} H. Hezari, ``Complex zeros of eigenfunctions of 1D Sch\"odinger operators'', \textit{International mathematics research notices}, ISSN  1687-0247, 2008 .

\bibitem{EGS} A. Eremenko, A. Gabrielov, B. Shapiro, ``Zeros of eigenfunctions of some anharmonic oscillators'', \textit{Annales de l'Institut Fourier}, Vol \textbf{58}, No 2, 603-624, 2008.

\bibitem{EGS2}  A. Eremenko, A. Gabrielov, B. Shapiro, ``High energy eigenfunctions of one-dimensional Scr\"odinger operators with polynomial potentials'', \textit{Computational Methods and Function Theory}, Vol \textbf{8}, No 2, 513-529, 2008.

\bibitem{Kontsevich} M. Kontsevich, ``Intersection theory on the moduli space of curves and the matrix Airy function'',
\textit{ Commun. Math. Phys.}, {\bf 147}, 1-23, 1992.

\bibitem{Okounkov2} A.Okounkov, R. Pandharipande, ``Gromov-Witten theory, Hurwitz numbers, and matrix models I'', \textit{Proc. Symposia Pure Math.}, Vol \textbf{80}, 325-414, 2009.

\bibitem{DVmatrix} R. Dijkgraaf, C. Vafa, ``Matrix models, topological strings, and supersymmetric gauge theories'', \textit{Nucl. Phys. B}, Vol {\bf 644}, 3-20, 2002.

\bibitem{marino3} M. Mari\~{n}o, ``Chern-Simons theory, matrix integrals, and perturbative three-manifold invariants'', \textit{ Commun. Math. Phys.}, Vol {\bf 253}, 25-49, 2004.

\bibitem{AKMV_CS} M. Aganagic, A. Klemm, M. Mari\~{n}o, C. Vafa, ``Matrix model as a mirror of Chern-Simons theory'', \textit{JHEP}, Vol {\bf 02},  010, 2004.

\bibitem{HalmagyiOkuda} N. Halmagyi, T. Okuda, ``Bubbling Calabi-Yau geometry from matrix models'', \textit{JHEP}, Vol {\bf 03}, 028, 2008.

\bibitem{NekrasovAnn} N.A. Nekrasov, ``Seiberg-Witten Prepotential From Instanton Counting'', \textit{Adv. Theor. Math. Phys.}, Vol {\bf 7}, 831-864, 2004.

\bibitem{IK3} A.Iqbal, A.K. Kashani-Poor, ``SU(N) geometries and topological string amplitudes'', \textit{Adv. Theor. Math. Phys.}, Vol {\bf 10}, 1-32, 2006.

\bibitem{MarshakovNekrasov} A. Marshakov, N. Nekrasov, ``Extended Seiberg-Witten theory and integrable hierarchy'', \textit{JHEP}, Vol {\bf 01}, 2007.

\bibitem{KlemmSulkowski} A. Klemm, P. Sulkowski, ``Seiberg-Witten theory and matrix models'', \textit{Nucl. Phys. B}, Vol {\bf 819},  400-430, 2009.

\bibitem{MorozovShakirov} S. Shakirov, A. Morozov, ``On equivalence of two Hurwitz matrix models'', \textit{Mod. Phys. Lett. A}, Vol {\bf 24}, 2659-2666, 2009.

\bibitem{Eynard:2005wg} B. Eynard, A.P. Ferrer, ``2-matrix versus complex matrix model, integrals over the unitary group as triangular integrals'', \textit{Commun. Math. Phys.}, Vol {\bf 264},  115-144, 2006.

\bibitem{marino2Ann} M.Marino, ``Open string amplitudes and large order behavior in topological string theory'', \textit{JHEP}, Vol {\bf 03}, 060, 2008.

\bibitem{IqbalKashaniPoor} A. Iqbal, A.K. Kashani-Poor, ``The vertex on a strip'', \textit{Adv. Theor. Math. Phys.}, Vol {\bf 10},  317--343, 2006. 

\bibitem{Fulton} W.Fulton, ``Introduction to toric varieties'', \textit{Annals of Mathematics Studies}, Vol \textbf{131}, Princeton University Press, Princeton, NJ, 1993.

\bibitem{CoxKatz} D.A. Cox,S.~Katz, ``Mirror symmetry and algebraic geometry'', \textit{Mathematical Surveys and Monographs}, Vol \textbf{68}, American Mathematical Society, Providence, RI, 1999.

\bibitem{Witten_Phases} E. Witten, ``Phases of N = 2 theories in two dimensions'' \textit{Nucl. Phys. B}, Vol {\bf 403}, 159-222, 1993.

\bibitem{LLLZ} J. Li, C.M. Liu, K. Liu, J. Zhou, ``A mathematical theory of the topological vertex'', \textit{Geom. Topol.}, Vol {\bf 13}, No 1, 527-621, 2009.

\bibitem{MOOP} D. Maulik, A. Oblomkov, A. Okounkov, R.Pandharipande, ``Gromov-Witten/Donaldson-Thomas correspondence for toric 3-folds'', \textit{arXiv}:0809.3976 [math.AG], 2008.

\bibitem{JohanssonAnn} K. Johansson, ``The arctic circle boundary and the Airy process'', \textit{Ann. Probab.}, Vol {\bf 33}, No. 1, 1-30, 2005.

\bibitem{IK2} A. Iqbal, A.K. Kashani-Poor, ``Instanton counting and Chern-Simons theory'', \textit{Adv. Theor. Math. Phys.}, Vol {\bf 7}, 457-497, 2004.

\bibitem{EguchiKanno} T. Eguchi, H. Kanno, ``Topological strings and Nekrasov's formulas'', \textit{JHEP}, Vol {\bf 12}, 2003.

\bibitem{Hooft} G. 't Hooft, ``A two-dimensional model for mesons'',  \textit{Nucl. Phys. B}, Vol {\bf 75}, 461-853, 1974.

\bibitem{Hooft2} G. 't Hooft, ``A planar diagram theory for strong interactions'', \textit{Nucl. Phys. B}, Vol {\bf 72}, 461-470, 1974.

\bibitem{HarishChandra} M. Harish-Chandra ``Differential operators on a semi-simple Lie algebra'' \textit{Am. J. Math.},  79-85, 1957.

\bibitem{ItzyksonZuber} C. Itzykson, J.B. Zuber, ``The Planar Approximation. 2'', \textit{J. Math. Phys.}, Vol {\bf 21}, 1980.

\bibitem{Mehta1} M.L. Mehta, ``A method of integration over matrix variables'', \textit{Comm. Math. Phys.}, Vol {\bf 79}, No 2, 327-340, 1981. 

\bibitem{David1} F. David, ``Loop equations and nonperturbative effects in two-dimensional quantum gravity'',
\textit{Mod. Phys. Lett. A}, Vol {\bf 5}, 1019-1030, 1990.

\bibitem{Eynchain} B.Eynard, ``Master loop equations, free energy and correlations for the chain of matrices'', \textit{JHEP}, Vol {\bf 11}, 2003.

\bibitem{MiMo} A. Mironov, A. Morozov, ``On the origin of Virasoro constraints in matrix models: Lagrangian approach'',
\textit{Phys. Lett. B}, Vol {\bf 252}, 47-52, 1990.

\bibitem{eynhaeq} B. Eynard, M. Marino, N. Orantin, ``Holomorphic anomaly and matrix models'', \textit{JHEP}, Vol {\bf 6}, 2007.  

\bibitem{GiventalSemisimple} A.B. Givental, ``Semisimple Frobenius structures at higher genus'', \textit{Inter. Math. Res. Notices}, No 23, 1265-1286, 2001.

\bibitem{GiventalHierarchies} A. Givental, ``{$A\sb {n-1}$} singularities and {$n$}{K}d{V} hierarchies,''
 \textit{Mosc. Math. J.}, Vol {\bf 3}, No. 2, 475-505, 2003.

\bibitem{AMM1} A.S. Alexandrov, A. Mironov, A. Morozov, ``Solving Virasoro constraints in matrix models'', \textit{Fortsch. Phys.}, Vol {\bf 53}, 512-521, 2005.  

\bibitem{AMM2} A.S. Alexandrov, A. Mironov, A. Morozov, ``M-theory of matrix models'', \textit{Teor. Mat. Fiz.}, Vol {\bf 150}, 179--192, 2007.

\bibitem{Orantin} N. Orantin, ``Symplectic invariants, Virasoro constraints and Givental decomposition'', \textit{arXiv}:0808.0635 [math-ph], 2008.

\bibitem{CMMV} L. Chekhov, A.Marshakov, A. Mironov, D. Vasiliev, ``DV and WDVV'', \textit{Phys. Lett. B}, Vol {\bf 562}, 323-338, 2003. 

\bibitem{NekrasovOkounkov} N. Nekrasov, A. Okounkov, ``Seiberg-Witten theory and random partitions'',
\textit{arXiv}:hep-th/0306238, 2003.

\bibitem{Sulkowski} P. Sulkowski, ``Matrix models for 2* theories'', \textit{arXiv}:0904.3064 [hep-th], 2009.

\bibitem{BouchardMarino} V. Bouchard,  M.Marino, ``Hurwitz numbers, matrix models and enumerative geometry'',
\textit{arXiv}:0709.1458 [math.AG], 2007.

\bibitem{EMS} B. Eynard, M. Mulase, B. Safnuk, ``The Laplace transform of the cut-and-join equation and the Bouchard-Marino conjecture on Hurwitz numbers'', \textit{arXiv}:0907.5224 [math.AG], 2009.

\bibitem{EguchiYang} T. Eguchi, S.K. Yang, ``The Topological CP1 model and the large N matrix integral'',  \textit{Mod. Phys. Lett.}, Vol {\bf A9}, 2893-2902, 1994.

\bibitem{MiwaJimbo1} M. Jimbo, T. Miwa, ``Solitons and infinite dimensional lie algebras'', \textit{Publ. Res. Inst. Math. Sci. Kyoto}, Vol {\bf 19}, 943,  1983.

\bibitem{Harnad-Orlov2} J. Harnad , A.Y. Orlov, ``Fermionic construction of tau functions and random processes'',
\textit{arXiv}:0704.1157 [math-ph], 2007.

\bibitem{Kostov2} I.K. Kostov, ``Conformal field theory techniques in random matrix models'', \textit{arXiv}:hep-th/9907060, 1999.

\bibitem{EynMehta} B. Eynard, M.L. Mehta, ``Matrices coupled in a chain: I. Eigenvalue correlations'', \textit{Journal of Physics A: Mathematical and General}, Vol {\bf 31}, No 19, 4449-4456, 1998.

\bibitem{ADKMV} M. Aganagic, R. Dijkgraaf, A. Klemm, M. Marino, C. Vafa, ``Topological strings and integrable hierarchies'', \textit{Commun. Math. Phys.}, Vol {\bf 261}, 451-516, 2006. 

\bibitem{KashaniPoor} A.K. Kashani-Poor, ``The Wave Function Behavior of the Open Topological String Partition Function on the Conifold'', \textit{JHEP}, Vol {\bf 04}, 2007.

\bibitem{DHSV} R. Dijkgraaf, L. Hollands, P. Sulkowski, C. Vafa, ``Supersymmetric Gauge Theories, Intersecting Branes and Free Fermions'', \textit{JHEP}, Vol {\bf 02}, 2008.

\bibitem{MarinoSchiappaWeiss} M. Marino, R. Schiappa, M. Weiss, ``Nonperturbative Effects and the Large-Order Behavior of Matrix Models and Topological Strings'', \textit{arXiv}:0711.1954 [hep-th], 2007.

\bibitem{MarinoAnn} M. Marino, ``Nonperturbative effects and nonperturbative definitions in matrix models and topological strings'', \textit{JHEP}, Vol {\bf 12}, 2008.

\bibitem{EynardMarino} B. Eynard, M. Marino, ``A holomorphic and background independent partition function for matrix models and topological strings'', \textit{arXiv}:0810.4273 [hep-th],2008.

\bibitem{KlemmMarinoRauch} A. Klemm, M. Marino, M. Rauch, ``Direct Integration and Non-Perturbative Effects in Matrix Models'', \textit{arXiv}:1002.3846 [hep-th], 2010.
 
\bibitem{IqbalKozcazVafa} A. Iqbal, C. Kozcaz, C. Vafa, ``The refined topological vertex'', \textit{arXiv}:hep-th/0701156, 2007.

\bibitem{OSY} H. Ooguri, P. Sulkowski,M. Yamazaki, ``Wall Crossing As Seen By Matrix Models'', \textit{arXiv}:1005.1293 [hep-th], 2010.

\bibitem{HoriVafa} K. Hori, C. Vafa, ``Mirror symmetry'', \textit{arXiv}:hep-th/0002222, 2000.

\bibitem{AKV} M. Aganagic, A. Klemm, C.Vafa, ``Disk instantons, mirror symmetry and the duality web'', \textit{Z. Naturforsch. A}, Vol {\bf 57}, 1-28, 2002.

\bibitem{MarcoPaths} M. Bertola, ``Biorthogonal polynomials for 2-matrix models with semiclassical potentials'', \textit{J. Approx. Theory}, Vol {\bf 2}, 162, 2007.

\bibitem{ChekhovHardedges} L. Chekhov, ``Matrix models with hard walls: Geometry and solutions'', \textit{J. Phys. A}, Vol {\bf 39}, 8857-8894, 2006.  

\bibitem{KK} A. Kokotov, D. Korotkin, ``Tau-function on Hurwitz spaces'', \textit{Math. Phys., Analysis and Geometry}, Vol {\bf 7}, 47-96, 2004.

\bibitem{EKK} B. Eynard, A. Kokotov, D. Korotkin, ``Genus one contribution to free energy in hermitian two- matrix model'', \textit{Nucl. Phys. B}, Vol \textbf{694}, 443-472, 2004.

\bibitem{Jost} J. Jost, ``Compact Riemann surfaces'', \textit{Springer-Verlag}, Berlin, second ed., 2002.

\bibitem{BrezNb} E. Brezin, H. Neuberger, ``Large-N scaling limits of symmetric-matrix models as systems of fluctuating unoriented surfaces'', \textit{Phys. Rev. Lett.}, Vol \textbf{65}, 2098-2101, 1990.

\bibitem{EOVolWP} B. Eynard, N. Orantin, ``Weil-Petersson volume of moduli spaces, Mirzakhani's recursion and matrix models'',  math-ph: arXiv:0705.3600v1.

\bibitem{Krich} I. Krichever, ``The tau-function of the universal Witham hierarchy, matrix models and topological field theories'', \textit{Commun. Pure Appl. Math.}, Vol {\bf 47} 437, 1992.

\bibitem{mkrtchyan1} RL. Mkrtchyan, ``The equivalence of $Sp(2N)$ and $SO(2N)$ gauge theories'', \textit{Physics Letters B}, 1981.

\bibitem{AvM} M. Adler, P. Van Moerbeke, ``The Spectrum of Coupled Random Matrices'', {\it Ann. Math.}, Vol {\bf 149}, 921-976, 1999.

\bibitem{BgE} M. Berg\`{e}re, B. Eynard ``Mixed correlation function and spectral curve for the 2-matrix model'', \textit{J. Phys. A}, Vol {\bf 39}, 15091-15134, 2006.

\bibitem{BE_mom} M. Bertola, ``Moment determinants as isomonodromic tau functions'', \textit{Non-linearity}, Vol  \textbf{22}, No 1, 29-50, 2009.

\bibitem{BertoGekhtman} M. Bertola, M. Gekhtman, ``Biorthogonal Laurent polynomials, T\"oplitz determinants, minimal Toda orbits and isomonodromic tau functions'', \textit{Constr. Approx.}, Vol {\bf 26}, No 3, 383-430, 2007.

\bibitem{BM} M. Bertola, M.Y. Mo, ``Isomodromic deformation of resonant rational connections'', \textit{IMRP}, Vol {\bf 11}, 565, 2005.

\bibitem{BEH4} M. Bertola, B. Eynard, J. Harnad, ``Semiclassical orthogonal polynomials, matrix models and isomonodromic tau functions'', {\it Commun. Math. Phys.}, Vol {\bf 263}, 401-437, 2006.

\bibitem{DKK} J.M. Daul, V. Kazakov, I.K. Kostov, ``Rational Theories of 2D Gravity from the Two-Matrix Model'', {\it Nucl. Phys. B} Vol {\bf 409}, 311-338, 1993.

\bibitem{ErM} N. M. Ercolani, K. T.-R. McLaughlin ``Asymptotics and integrable structures for biorthogonal polynomials associated to a random two-matrix model'', {\it Physica D},  Vol {\bf 152-153}, 232-268, 2001.

\bibitem{FIK} A. Fokas, A. Its, A. Kitaev, ``The isomonodromy approach to matrix models in 2D quantum gravity'', {\it Commun. Math. Phys.}, Vol {\bf 147}, 395-430, 1992.

\bibitem {ITW} A. R. Its, A. Tracy, H. Widom, ``Random words, Toeplitz determinants and integrable systems II'', \textit{Advances in nonlinear mathematics and science, Phys. D}, Vol {\bf 152-153}, 199-224, 2001.

\bibitem{Kap} A. A. Kapaev, ``The Riemann--Hilbert problem for the bi-orthogonal polynomials'', \textit{J. Phys. A}, Vol {\bf 36}, 4629-4640, 2003.

\bibitem{Kaz} V.A. Kazakov, ``Ising model on a dynamical planar random lattice: exact solution'', {\it Phys Lett. A}, Vol {\bf 119}, 140-144, 1986.

\bibitem{MS} M. L. Mehta, P. Shukla, ``Two coupled matrices: eigenvalue correlations and spacing functions'', {\it J. Phys. A}, Vol {\bf 27}, 7793-7803, 1994.

\bibitem{TWDyson} C. Tracy, H. Widom, ``Differential equations for Dyson processes'', \textit{Comm. Math. Phys.}, Vol {\bf 252},  No 1-3, 7-41, 2004.

\bibitem{Conformal} P. Di Francesco, P. Mathieu, D. Sénéchal, ``Conformal Field Theory'', \textit{Springer-Verlag}, ISBN 0-387-94785-X, 1997.

\bibitem{blouh} A. Its, R. Novokshenov, V. Yu, ``The isomonodromic deformation method in the theory of Painlevé equations, Lecture Notes in Mathematics'', \textit{Springer-Verlag}, MR851569, ISBN 978-3-540-16483-8, 1986. 

\bibitem{conffieldtheory} A. Belavin, A. Polyakov, A. Zamolodchikov, ``Infinite conformal symmetry in two-dimensional quantum field theory'', \textit{Nucl. Phys. B}, Vol \textbf{241}, 333-380, 1984.

\end{thebibliography}
